%% file: main.tex
\documentclass[12pt,oneside]{thesis} 
\newcommand{\tablefont}{\fontsize{10}{12}\selectfont}
\usepackage[
    letterpaper,
    left=1in,
    right=1in,
    top=1in,
    bottom=1in
]{geometry}
\usepackage{supertabular}
\usepackage{amsmath}
\usepackage{amssymb}
\usepackage{tabls}
\usepackage{titlesec}
    \titleformat{\chapter}
        {\normalfont\large}{Chapter \thechapter:}{1em}{}
\usepackage{times}
\usepackage[USenglish]{babel}
\usepackage{subcaption}
\usepackage{graphicx}
\usepackage{cite}
\usepackage{lscape}
\usepackage{indentfirst}
\usepackage{latexsym}
\usepackage{multirow}
\usepackage{epstopdf}
\usepackage{wrapfig}
\usepackage{slashbox}
\usepackage{float}
\usepackage{caption}
\usepackage[onehalfspacing]{setspace} 
\usepackage{bibentry} 
\nobibliography*
\usepackage{array}
\usepackage{booktabs}
\usepackage{needspace}
\usepackage{tabularx}
\usepackage{adjustbox}
\usepackage{makecell}

\usepackage[colorlinks=true,urlcolor=black,linkcolor=blue,citecolor=blue]{hyperref}

\usepackage{tocloft,calc}
\renewcommand{\cftlottitlefont}{\hspace*{\fill}\large}
\renewcommand{\cftafterlottitle}{\hspace*{\fill}}
\renewcommand{\cftloftitlefont}{\hspace*{\fill}\large}
\renewcommand{\cftafterloftitle}{\hspace*{\fill}}
\renewcommand{\cftchapaftersnum}{:\ }
\renewcommand{\cftchappresnum}{\chaptername\space}
\makeatletter
\g@addto@macro\appendix{%
  \addtocontents{toc}{%
    \protect%
  }%
}
\renewcommand{\baselinestretch}{2}
\newcommand{\ket}[1]{\left|#1\right\rangle}

\usepackage{subfiles}

\begin{document}
\pagestyle{empty}
\include{Abstract} 
\include{Titlepage} 
\include{Copyright} 

\pagestyle{plain}
\pagenumbering{roman}
\setcounter{page}{2}

\phantomsection 
\addcontentsline{toc}{chapter}{Acknowledgements}
\include{Acknowledgements} 

\phantomsection 
\vspace*{-2.5cm}
\addcontentsline{toc}{chapter}{Table of Contents}
    \renewcommand{\contentsname}{Table of Contents}
\renewcommand{\baselinestretch}{1}
\small\normalsize
\tableofcontents 
\newpage

\phantomsection 
\addcontentsline{toc}{chapter}{List of Tables}
    \renewcommand{\contentsname}{List of Tables}
\listoftables 

\clearpage
\phantomsection 
\addcontentsline{toc}{chapter}{List of Figures}
    \renewcommand{\contentsname}{List of Figures}
\listoffigures 

\clearpage
\phantomsection 
\addcontentsline{toc}{chapter}{List of Abbreviations}
\input{Abbreviations-supertabular}

\clearpage
\setlength{\parskip}{0em}
\renewcommand{\baselinestretch}{2}
\small\normalsize

\setcounter{page}{1}
\pagenumbering{arabic}
\clearpage
\subfile{Chapter1}
\clearpage
\subfile{Chapter2}
\clearpage
\subfile{Chapter3}
\clearpage
\subfile{Chapter4}
\clearpage
\subfile{Chapter5}
\clearpage
\subfile{Chapter6}
\clearpage
\subfile{Chapter7}
\clearpage
\subfile{Chapter8}
\clearpage
\subfile{Chapter9}
\appendix
\clearpage
\subfile{AppendixA}
\clearpage
\subfile{AppendixB}
\clearpage
\subfile{AppendixC}
\clearpage
\subfile{AppendixD}
\clearpage

\renewcommand{\baselinestretch}{1}
\small\normalsize
\vspace*{-1in}
\bibliographystyle{unsrt}
\bibliography{Dissertation}

\end{document}

%% file: Abstract.tex

\hbox{\ } \vspace{.7in}
\renewcommand{\baselinestretch}{1}
\small \normalsize

\begin{center}
\large{{ABSTRACT}}

\vspace{3em}

\end{center}
\hspace{-.15in}
\begin{tabular}{ll}
Title of Dissertation:    & {\large  LOW VIBRATION CLOSED-CYCLE FLOW CRYOSTAT}\\
&                     {\large  FOR ION-CAVITY EXPERIMENTS} \\
\ \\
&                          {\large  Andrew Laugharn} \\
&                           {\large Doctor of Philosophy, 2026} \\
\ \\
Dissertation Directed by: & {\large Dr. Joseph W. Britton} \\
&               {\large Department of Physics } \\
&               {\large CCDC Army Research Laboratory} \\
&               {\large Quantum Technology Center} \\
\end{tabular}

\vspace{3em}

\renewcommand{\baselinestretch}{2}
\large \normalsize

Cryogenic operation below 10 K suppresses trap-electrode and dielectric heating of trapped ions while cryopumping residual gases to ultra-high vacuum (UHV) without requiring a bakeout. These advantages are particularly important for cavity-integrated ion traps, where nearby dielectric surfaces can increase motional heating and rapid design-build-test iterations benefit from eliminating bakeout cycles. We have developed a custom 4 K cryostat continuously refrigerated by a ColdEdge Technologies (CET) Stinger closed-cycle $^{4}\text{He}$ flow cryogenic system. Two such cryostats are now in operation at the University of Maryland to study various ion trap optical cavity structures designed to serve as a photonic interconnect between remote, compute-capable trapped $^{171}\text{Yb}^{+}$ ion quantum processors, enabling long baseline entanglement distribution $(>10\text{ km})$. In one of these systems, we integrate an Alpine Quantum Technologies (AQT) Paul trap and Fabry-Pérot optical cavity at the 4 K stage. The cooling capacities of the cryostat 40 K and 4 K stages were measured to be 5 W at 45 K and 0.5 W at 6.1 K, respectively. Vibrations of the 4 K stage relative to the optical table are less than $12\text{ nm}_{\text{RMS}}$ in three dimensions, as measured using a Michelson interferometer. An ultra-high vacuum (UHV) is inferred by studying the rate of hydrogen-$\mathrm{Yb}^{+}$ ion collisions, modeled using classical Langevin theory. Ion storage times exceeding 5 days without laser cooling have been reported and the position of the trapped ions are reproducible within $\sim10$ \textmu{}m following repeated thermal cycles to room temperature. The uptime of this system is 72 days and offers convenient access to the 4 K stage for in-vacuum maintenance and system modifications. Using this platform, we demonstrate spatial overlap and interaction between a trapped $^{174}\text{Yb}^{+}$ ion and the optical cavity mode on the 1345 nm transition, establishing preliminary experimental validation of the integrated cryogenic ion-cavity structure.

%% file: Titlepage.tex

\thispagestyle{empty} \hbox{\ }
\vspace{.5in}

\renewcommand{\baselinestretch}{1}
\small\normalsize
\begin{center}

\large{{LOW VIBRATION CLOSED-CYCLE FLOW CRYOSTAT \\
FOR ION-CAVITY EXPERIMENTS}}\\
\ \\
\ \\
\large{by} \\
\ \\
\large{Andrew Laugharn}
\ \\
\ \\
\ \\
\ \\
\normalsize
Dissertation submitted to the Faculty of the Graduate School of the \\
University of Maryland, College Park in partial fulfillment \\
of the requirements for the degree of \\
Doctor of Philosophy \\
2026
\end{center}

\vspace{7.5em}

\noindent Advisory Committee: \\
\hbox{\ }\hspace{.5in}Professor Norbert Linke, Chair \\
\hbox{\ }\hspace{.5in}Dr. Joseph W. Britton, Co-Chair/Advisor \\
\hbox{\ }\hspace{.5in}Professor Steven Rolston\ \\
\hbox{\ }\hspace{.5in}Assistant Professor Timothy Koeth \\
\hbox{\ }\hspace{.5in}Professor Bao Yang

%% file: Copyright.tex

\thispagestyle{empty}
\hbox{\ }

\vfill
\renewcommand{\baselinestretch}{1}
\small\normalsize

\begin{center}
\large{\copyright \hbox{ }Copyright by\\
Andrew Laugharn  
\\
2026}
\end{center}

\vfill

\newpage

%% file: Acknowledgements.tex

\renewcommand{\baselinestretch}{2}

\small\normalsize
\hbox{\ }

\vspace*{-.85in}

\begin{center}
\large{Acknowledgements}
\end{center}

\vspace{1ex}

First and foremost, I would like to thank my advisor, Dr. Joseph W. Britton, for his guidance and for giving me the opportunity to build these cryostats from the ground up. Throughout my PhD, I was trusted with designing, building, and commissioning a wide range of systems and was given the freedom to make mistakes and learn from them. This process was exciting, rewarding, and at times frustrating, but it ultimately shaped me into the experimentalist and problem solver I am today.

I would also like to thank my labmates, Connor Goham, Wance Wang, and Roame Hildebrand, with whom I had the privilege of building an atomic physics laboratory from the ground up. Connor was always willing to answer my questions about ARTIQ and trapped-ion physics. It was a pleasure working alongside Wance, especially during the integration of the optical cavity into the cryostat. We made a great team and solved countless problems together. I also thank Roame for his help designing and building the water leak detection interlocks for the helium compressors.

I am grateful to the University of Maryland ion trapping community, including Dr. Norbert Linke, Dr. Qudsia Quraishi, Dr. Alaina Green, Dr. Yu Liu, and Dr. Chris Monroe, for their support and for fostering the collaborative environment that made my time at UMD so rewarding.

I would like to thank Alessandro Restelli for allowing me to use the 3D printers to fabricate many parts for the cryostats and for the many enjoyable discussions we had about electronics.

Finally, I would like to thank the IREAP machine shop staff, Thomas Weimar and Nolan Ballew, for training me in the machine shop and for their willingness to help fabricate countless parts for these cryostats. They taught me how to machine components independently, and I especially appreciate their patience as I inevitably broke more than a few end mills while learning.

%% file: Abbreviations-supertabular.tex

\renewcommand{\baselinestretch}{1}
\small\normalsize
\hbox{\ }

\vspace{-.6in}

\begin{center}
\large{List of Abbreviations}
\end{center}

\vspace{1em}

\begin{supertabular}{ll}
QCCD & Quantum Charge-Coupled Device \\
DI-QKD & Device-Independent Quantum Key Distribution \\
HOM & Hong-Ou-Mandel \\
CG & Clebsch–Gordan \\
FP & Fabry-P\'erot \\
MRI & Magnetic Resonance Imaging \\
LNG & Liquefied Natural Gas \\
DCT & Deep Cryogenic Treatment \\
CERN & Conseil Europ\'een pour la Recherche Nucl\'eaire \\
UHV & Ultra-High Vacuum \\
ITER & International Thermonuclear Experimental Reactor \\
CTE & Coefficient of Thermal Expansion \\
BEC & Boise-Einstein Condensate \\
TES & Transition-Edge Sensors \\
ADR & Adiabatic Demagnetization Refrigerators \\
SNSPD & Superconducting Nanowire Singe Photon Detector \\
GM & Gifford-McMahon \\
CET & ColdEdge Technologies \\
JT & Joule-Thomson \\
AQT & Alpine Quantum Technologies \\
PZT & Piezoelectric Transducer \\
ARTIQ & Advanced Real-Time Infrastructure for Quantum Physics \\
HX & Heat Exchanger \\
RGC & Recirculating Gas Cryocooler \\
STP & Standard Temperature and Pressure \\
SLPM & Standard Liter Per Minute \\
SCCM & Standard Cubic Centimeters Per Minute \\
NPT & National Pipe Thread \\
SOP & Standard Operating Procedure \\
CF & Conflat Flange \\
KF & Klein Flansche \\
SDS & Safety Data Sheet \\
REDOX & Reduction-Oxidation \\ 
HV & High Voltage \\
FPGA & Field-Programmable Gate Array \\
DAC & Digital-to-Analog Converter \\
ADC & Analog-to-Digital Converter \\
DDS & Direct Digital Synthesis \\
CW & Continuous-Wave \\
AOM & Acousto-Optic Modulator \\
EOM & Electro-Optic Modulator \\
VHSIC & Very High-Speed Integrated Circuit \\
VHDL & VHSIC Hardware Description Language \\
RPC & Remote Procedure Call \\
LAN & Local Area Network \\
GUI & Graphical User Interface \\
PMT & Photomultiplier Tube \\
NDSP & Network Device Support Package \\
I/O & Input/Output \\
DRTIO & Distributed Real-Time I/O \\
ECDL & External-Cavity Diode Laser \\
DC & Direct Current \\
RF & Radio Frequency \\
EMI & Electromagnetic Interference \\
PCB & Printed Circuit Board \\
ENIG & Electroless Nickel Immersion Gold \\
AWG & American Wire Gauge \\
MOT & Magneto-Optical Trap \\
FEL & Free-Electron Laser \\
RKF & Runge–Kutta–Fehlberg \\
EMCCD & Electron-Multiplying Charge-Coupled Device \\
NA & Numerical Aperture \\
WD & Working Distance \\ 
PSF & Point Spread Function \\
DOF & Depth of Focus \\
SM & Single-Mode \\
COM & Center of Mass \\
SNR & Signal-to-Noise \\
RGA & Residual Gas Analyzer \\
CAD & Computer-Aided Design \\
PD & Photodiode \\
PID & Proportional-Integral-Derivative \\
ENBW & Equivalent Noise Bandwidth \\
TIG & Tungsten Inert Gas \\

\end{supertabular}

%% file: Chapter1.tex
 
\chapter{Introduction}
\label{chap:Introduction}

\section{Quantum Computing}
\label{sec:Quantum-Computing}

Quantum computers operate on fundamentally different principles than classical computers by harnessing the laws of quantum mechanics to process information. As a result, they are expected to solve certain classes of problems that are intractable for modern classical computers. One of the earliest envisioned applications of quantum computers was the simulation of physical systems governed by quantum mechanics, such as molecular structure and chemical reaction dynamics \cite{feynmanQuantumMechanicalComputers,reiherElucidatingReactionMechanisms2017,aspuru-guzikSimulatedQuantumComputation2005,abramsSimulationManyBodyFermi1997}. Because nature itself is governed by quantum mechanics, quantum computers can, in principle, simulate these systems more efficiently than classical computers. Another landmark development came in the 1990s when Peter Shor introduced a quantum algorithm for factoring large integers that provides an exponential speedup over the best known classical algorithms \cite{ShorFactoring}. Since the security of many widely used public-key cryptographic protocols relies on the computational difficulty of integer factorization, fault-tolerant quantum computers would have profound implications for modern cryptography.

Shor's algorithm also demonstrated that quantum computers may provide advantages for a broader class of computational problems. Although large-scale fault-tolerant quantum computers remain a long-term goal, near-term quantum devices may still provide advantages for selected applications, including optimization and quantum simulation \cite{preskillQuantumComputingNISQ2018}. The computational power of quantum computers arises primarily from two uniquely quantum phenomena: superposition and entanglement. 

Classical computers encode information in bits that take values of either 0 or 1. Quantum computers instead use quantum bits, or qubits, which may occupy the basis states $\ket{0}$ and $\ket{1}$ as well as arbitrary superpositions of the two. A general single-qubit state is

\begin{equation}
\ket{\Psi}=\alpha\ket{0}+\beta\ket{1}
\label{eq:General-quantum-state}
\end{equation}
where $\alpha$  and $\beta$  are complex amplitudes satisfying $|\alpha|^{2}+|\beta|^{2}=1$. For a system of $N$ qubits, the quantum state is described by $2^{N}$ complex amplitudes, whereas only $N$ numbers are required to describe a classical $N$-bit string. This exponential scaling of complexity in quantum systems is quite distinct from classical systems. For example, a full specification of a 100-qubit quantum state could require on the order of $2^{100}\sim10^{30}$ complex numbers.

However, quantum superpositions are inherently fragile and collapse upon measurement, yielding $\ket{0}$ with probability $|\alpha|^{2}$ or $\ket{1}$ with probability $|\beta|^{2}$. Although, a quantum computer evolves a superposition over exponentially many states, $2^{N}$, measurement returns only an $N$-bit string sampled from the total quantum state. This apparent limitation is overcome through quantum interference. By carefully designing quantum operations, amplitudes corresponding to undesired outcomes can interfere destructively, while those corresponding to desired solutions interfere constructively. The goal of a quantum algorithm is therefore to engineer interference patterns that result in output states that encode information of interest.

Quantum entanglement is a uniquely quantum correlation with no classical analogue. Entanglement led Albert Einstein to question the completeness of quantum mechanics \cite{einsteinCanQuantumMechanicalDescription1935}, though subsequent experiments testing John Bell's inequality have supported the predictions of quantum theory \cite{bellEinsteinPodolskyRosen1964}. One example of an entangled state is

\begin{equation}
\ket{\Psi}=\frac{1}{\sqrt{2}}\left(\ket{00}+\ket{11}\right)
\label{eq:prototypical-entangled-state}
\end{equation}
In this state, measuring one qubit immediately determines the state of the other, regardless of spatial separation. Entanglement is an essential resource for many quantum algorithms and quantum communication protocols.

Quantum computers therefore do not simply provide a uniform speedup over classical computation. Rather, they enable fundamentally different computational strategies that exploit the principles of quantum mechanics. While several compelling applications have already emerged, it is likely that some of the most impactful uses of quantum computing remain to be discovered.

\subsection{Trapped Ion Quantum Computing}
\label{subsec:Trapped-Ion-Quantum-Computing}

There are several competing qubit technologies being developed to create large-scale quantum computers. These technologies include trapped ions \cite{debnathDemonstrationSmallProgrammable2016a}, neutral atoms \cite{bluvsteinFaulttolerantNeutralatomArchitecture2025}, superconducting circuits \cite{devoretSuperconductingCircuitsQuantum2013}, spins in semiconductor quantum dots \cite{awschalomQuantumSpintronicsEngineering2013}, color centers in solids \cite{bradleyTenQubitSolidStateSpin2019,borettiAdvancementsChallengesColor2025}, and all-photonic systems \cite{kokLinearOpticalQuantum2007}. At this point, various domestic companies are pursuing quantum computing using these approaches: 1.) Trapped Ions: IonQ, Quantinuum, 2.) Neutral Atoms: QuEra, Infleqtion, 3.) Superconducting Circuits: Google, IBM, Rigetti, 4.) Quantum Dots: Intel, 5.) Color Centers: Quantum Brilliance, 6.) Photonic: PsiQuantum, Xanadu. Despite some early successes, all these platforms face significant challenges on the path to scalability.

Superconducting circuits, quantum dots, and color centers rely on fabricated structures, and therefore suffer from device-to-device variability arising from imperfections in fabrication. While photonic qubits are based on identical particles of light, scalable implementations depend on integrated optical components - such as waveguides and interferometers - whose performance is likewise limited by fabrication tolerances and optical loss. Consequently, for these “hardware-defined” platforms, scaling is closely tied to advances in high-precision manufacturing and materials control. In contrast, atomic qubits - such as trapped ions and neutral atoms - are defined by fundamental constants of nature and are therefore intrinsically identical. As a result, the primary challenges shift away from qubit reproducibility and toward the classical control infrastructure required to manipulate them. This includes precise trapping potentials, laser-based control, and the ability to position, transport, and interact individual atoms with high fidelity.

Trapped ion systems, in particular, have demonstrated all of the criteria outlined by David Divicenzo for the physical implementation of a quantum computer \cite{divincenzoPhysicalImplementationQuantum2000}. The central remaining challenge is to scale these systems to sizes relevant for practical applications. A leading architecture for scaling trapped ion quantum computers is the Quantum Charge-Coupled Device (QCCD)\cite{kielpinskiArchitectureLargescaleIontrap2002}. In this approach, ions are confined in segmented traps that support multiple chains. Entangling operations are performed within a chain via shared motional modes, after which ions can be separated, transported through the trap, and recombined with other chains as required by the quantum algorithm. In practice, such operations introduce motional excitation, necessitating the use of sympathetic cooling ions to remove excess energy and maintain high-fidelity gate operations \cite{barrettSympatheticCooling92003}. While the QCCD architecture provides a clear path toward systems with tens to potentially hundreds of qubits per chip, further scaling will require interconnecting multiple modules. This motivates a modular architecture in which separate ion trap systems are linked via photonic interconnects \cite{brownCodesigningScalableQuantum2016,monroeLargescaleModularQuantumcomputer2014}. Such quantum networking approaches establish entanglement between distant modules and provide a pathway toward large-scale distributed quantum computers. 

\section{Quantum Networking}
\label{sec:Quantum-Networking}

A quantum network connects spatially separated quantum systems, or quantum nodes, enabling the transmission of quantum information between them. The most powerful applications of such networks rely on the ability to generate and distribute entanglement across distant nodes. This distributed entanglement underpins Device-Independent Quantum Key Distribution (DI-QKD) \cite{vaziraniFullyDeviceIndependentQuantum2014}, often regarded as the ultimate form of cryptographic security, and forms the foundation of a future quantum internet \cite{kimbleQuantumInternet2008}. It also enables globally synchronized atomic clocks \cite{jozsaQuantumClockSynchronization2000}, sensing beyond the standard quantum limit \cite{zwierzGeneralOptimalityHeisenberg2010,qianHeisenbergscalingMeasurementProtocol2019}, and distributed quantum computing, where $M$ quantum processors each containing $N$ qubits are entangled to access a joint Hilbert space of dimension $2^{NM}$ \cite{ciracDistributedQuantumComputation1999}. 

\subsection{Long-Baseline Quantum Networking}
\label{subsec:Long-Baseline-Quantum-Networking}

The central objective of our research is long baseline entanglement distribution $(>10\text{ km})$ between remote, compute-capable quantum nodes. While such capabilities support applications including networked atomic clocks and sensing beyond the standard quantum limit, I'm personally most excited by their potential for distributed quantum computing. In this scheme each node has $^{171}\text{Yb}^{+}$ ions with a long-lived hyperfine ground state qubit and laser-mediated gates. To enable entanglement between nodes, we are developing optical cavities that function as ion-photon interfaces that are compatible with commercial optical fiber networks. 

Contemporary approaches to long-baseline ion-based quantum networking involve co-trapping two ion species. For example, systems based on $^{43}\text{Ca}^{+}$ and $^{88}\text{Sr}^{+}$ ions \cite{mainDistributedQuantumComputing2025} or $^{171}\text{Yb}^{+}$ and $^{138}\text{Ba}^{+}$ ions \cite{carterIonTrapInvacuum2024} use one species as a long-lived qubit and the other as a photon emitter. While this strategy has demonstrated success, it introduces additional complexity, including the need for quantum frequency conversion to telecom wavelengths and interspecies swap gates to transfer quantum information between the emitter and qubit. In contrast, our approach investigates the use of a single ion species, $^{171}\text{Yb}^{+}$, to serve both roles. This leverages a distinctive feature of several high-mass ions, which support optical, metastable, and ground-state qubits within the same species - an architecture colloquially known as OMG \cite{allcockOmgBlueprintTrapped2021}. 

\begin{figure}[!!h]
\centering
\includegraphics[width=\textwidth]{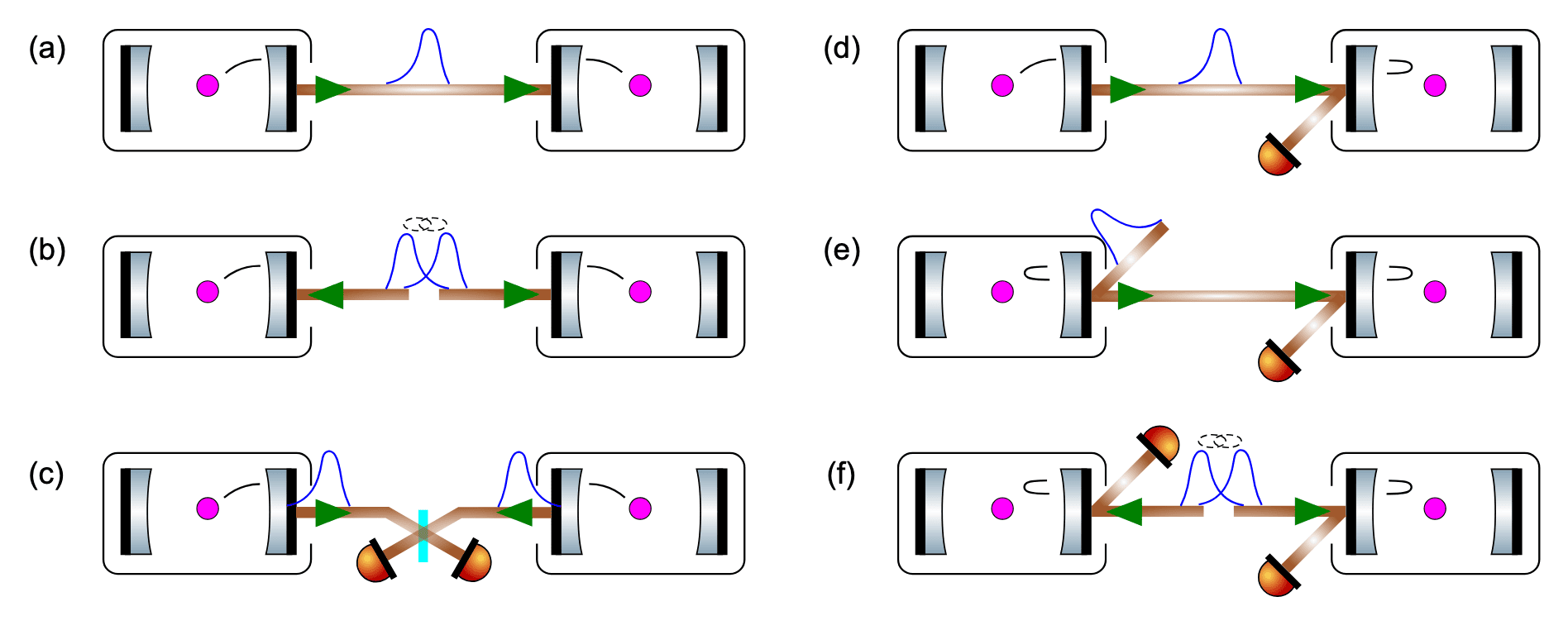}
\caption[Entanglement Distribution Schemes]{Schematic illustration of several entanglement distribution schemes. (a-f) Representative schemes for distributing entanglement between two spatially separated quantum nodes. Pink circles represent matter qubits, while the two-mirror optical cavity denotes a matter-photon interface. Each quantum node is enclosed by a black rectangular outline. Blue curves represent single-photon wave packets, with black curves indicating photon emission, absorption, or reflection from the cavity. Brown lines denote optical fibers serving as quantum channels, and green arrows indicate the direction of photon propagation. Overlapping dashed black circles represent entanglement between photonic qubits. Cyan rectangles denote non-polarizing beam splitters, and orange semicircles represent single-photon detectors.}
\label{fig:entanglement-distribution-schemes}
\end{figure}

Regardless of the choice of qubit architecture, long-distance entanglement ultimately requires the efficient conversion of stationary qubits into flying photonic qubits. This is accomplished through an ion-photon interface, in which the internal quantum state of a trapped ion becomes entangled with a single emitted photon. Photons generated at two remote nodes are transmitted through optical fibers to a central station, where they interfere on a beam splitter. When the photons are indistinguishable in all degrees of freedom except the encoded quantum information, Hong-Ou-Mandel (HOM) interference erases the which-path information, allowing a Bell-state measurement on the photons to herald the creation of entanglement between the distant ions. Figure \ref{fig:entanglement-distribution-schemes} illustrates the general classes of entanglement distribution protocols. Our work falls within the category of cavity-mediated ion-photon interfaces, where a trapped ion is strongly coupled to a high-finesse optical cavity that directs emitted photons into a single spatial mode for efficient coupling into optical fibers.

A straightforward approach to collecting photons is to image the ion fluorescence using a high numerical aperture objective. Because spontaneous emission is nearly isotropic, however, even large numerical aperture optics intercept only a modest fraction of the emitted photons. Furthermore, the collected light occupies many spatial modes, limiting the achievable coupling efficiency into single-mode optical fibers. Optical cavities overcome these limitations by modifying the electromagnetic mode density surrounding the ion through the Purcell effect. Rather than emitting into free space, the ion preferentially emits into the cavity mode, producing photons in a well-defined Gaussian spatial mode that can be efficiently coupled into optical fiber. Although cavity integration introduces significant experimental complexity, the resulting increase in photon collection efficiency makes optical cavities a promising approach for scalable trapped-ion quantum networking.

Integrating optical cavities with trapped ions is particularly challenging for ultraviolet transitions. Mirror coatings exhibit higher optical losses and are more susceptible to laser-induced degradation at ultraviolet wavelengths than in the visible or infrared. In addition, the small cavity mode volumes required for strong ion-cavity coupling necessitate positioning dielectric mirror surfaces within a few hundred micrometers of the ion, increasing sensitivity to stray electric fields, charging, and contamination. These challenges have motivated the development of networking architectures that leverage longer-wavelength optical transitions, where optical components are considerably more mature. Our work follows this approach by exploiting the multiple qubit encodings available within a single $^{171}\text{Yb}^{+}$ ion through the OMG architecture. Local quantum computation is performed using the ground-state hyperfine qubit, while remote entanglement is generated using a metastable optical qubit that emits single photons at the telecommunications wavelength of 1650 nm. This architecture enables direct transmission through low-loss commercial optical fiber without quantum frequency conversion and eliminates the need to co-trap multiple ion species and implement interspecies swap gates, substantially simplifying the experimental architecture while preserving a compute-capable trapped-ion node.

\begin{figure}[!!h]
\centering
\includegraphics[width=\textwidth]{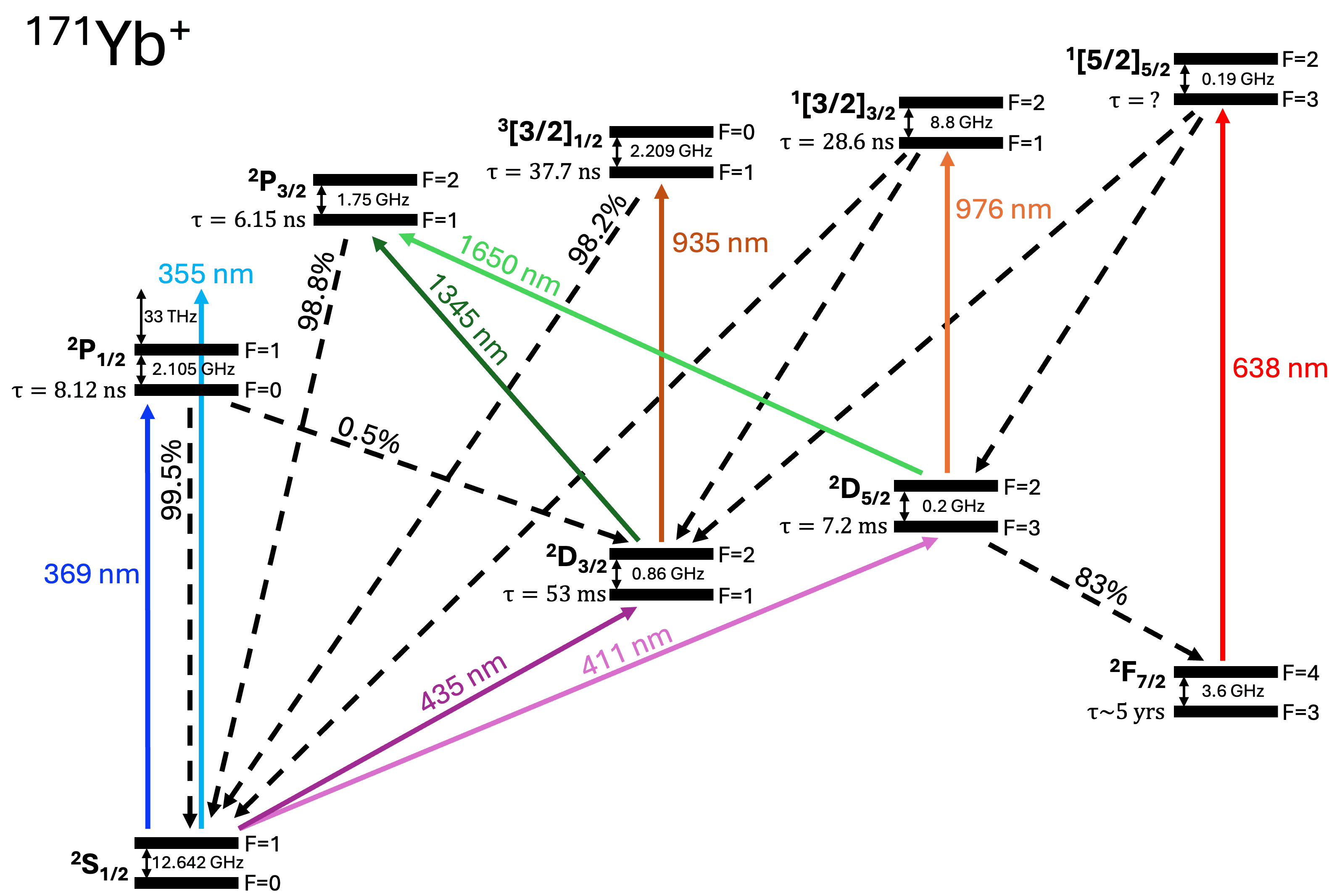}
\caption[$^{171}\text{Yb}^{+}$ level diagram 1]{$^{171}\text{Yb}^{+}$ level diagram. The nuclear spin of $^{171}\text{Yb}^{+}$ is $\text{I}=\frac{1}{2}$. The diagram is shown in the absence of an external magnetic field so the $m_{\text{F}}$ states are degenerate. The branching ratios from the $^{1}[5/2]_{5/2}$ and $^{1}[3/2]_{3/2}$ states are unknown.}
\label{fig:171Yb-level-diagram-Intro}
\end{figure}

The $^{171}\text{Yb}^{+}$ ion provides a mature experimental platform that supports all of the fundamental operations required for quantum information processing, see level diagram in Figure \ref{fig:171Yb-level-diagram-Intro}. Laser cooling is performed on the $^{2}\text{S}_{1/2}-^{2}\text{P}_{1/2}$ transition after which optical pumping prepares a well-defined hyperfine qubit state. State detection is performed using state-dependent fluorescence on the same cycling transition. Coherent microwave pulses implement single-qubit rotations, while laser-mediated stimulated Raman (355 nm) interactions couple the ions' internal states to their shared motional modes to realize high-fidelity entangling gates. Together, these capabilities enable universal quantum computation within each trapped-ion node.

\begin{figure}[!!h]
\centering
\begin{subfigure}{0.7\textwidth}
    \includegraphics[width=\textwidth]{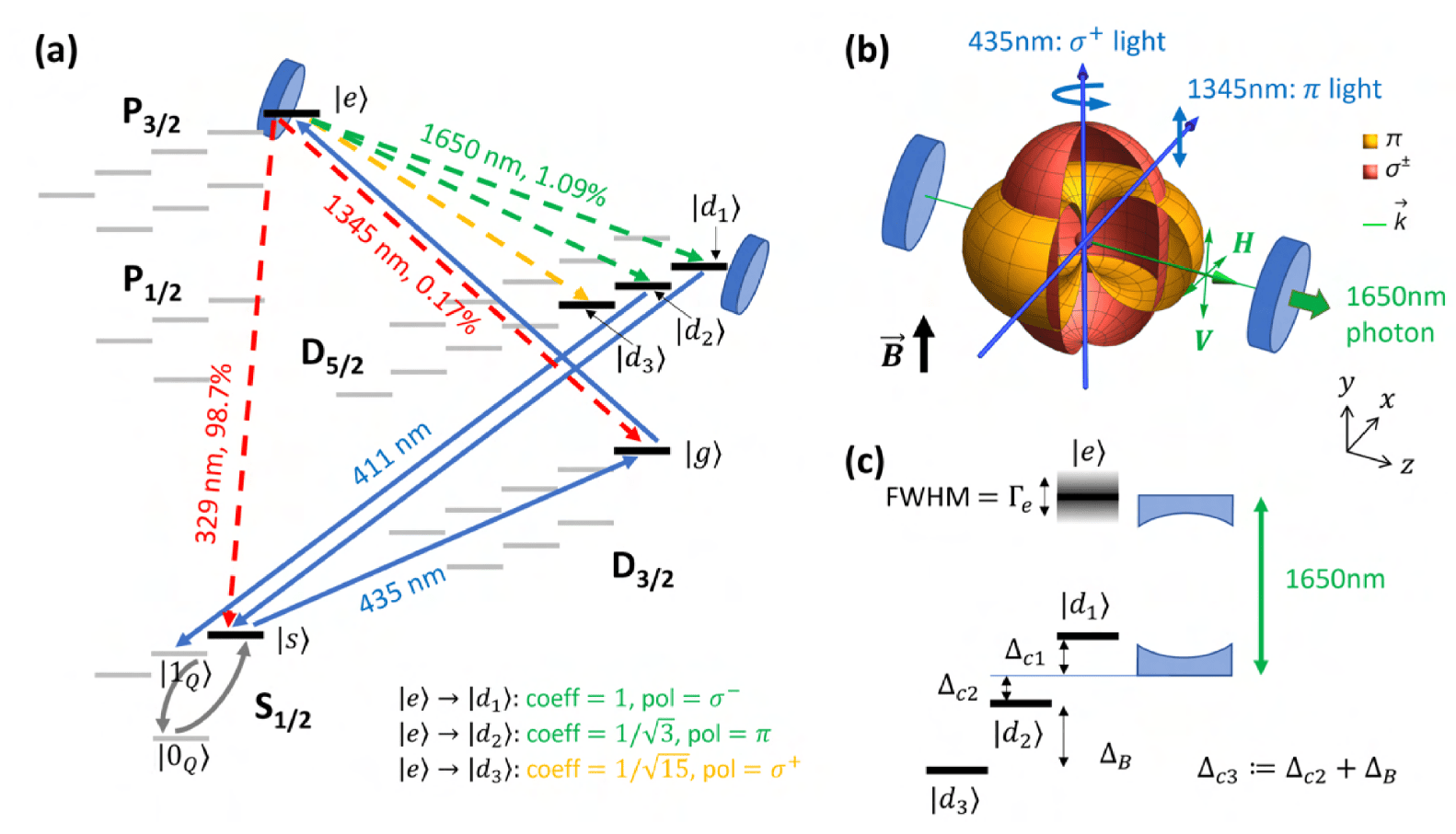}
\end{subfigure}
\begin{subfigure}{0.28\textwidth}
    \includegraphics[width=\textwidth]{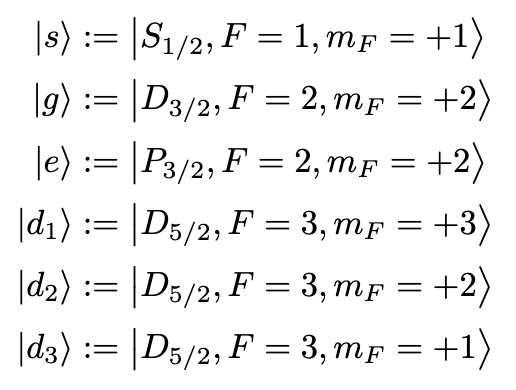}
\end{subfigure}
\caption[Quantum Networking Scheme]{$^{171}\mathrm{Yb}^{+}$ photon entanglement scheme at 1650 nm (adapted from Wance Wang's thesis). (a) Laser-driven transitions and decay pathways used to generate ion-photon entanglement. The six relevant hyperfine states are shown in bold and labeled using ket notation. Gray curved arrows denote microwave transitions, blue arrows denote laser excitation, and dashed arrows represent spontaneous and cavity-enhanced emission. Red, green, and yellow dashed arrows indicate decay within the nearly closed optical cycle, the desired single-photon generation pathway, and undesired decay channels, respectively. A Fabry-Pérot (FP) cavity is resonant with the 1650 nm transition. Clebsch–Gordan coefficients and emitted-photon polarizations are labeled. (b) Experimental geometry showing the laser beam orientations, cavity axis, and photon polarizations. The dipole radiation patterns for the $\sigma^{\pm}$ and $\pi$ transitions are shown, with the cavity supporting the horizontal ($H$) and vertical ($V$) polarization modes corresponding to the $\sigma^{\pm}$ and $\pi$ components, respectively. (c) Cavity detunings, $\Delta_{c,i}$, relative to the three effective two-level systems ${|e\rangle,|d_i\rangle}$ ($i=1,2,3$). The Zeeman splitting of the $^{2}\mathrm{D}_{5/2}$, $F=3$ manifold for a magnetic field of $\vec{B}=5~\mathrm{G}$ is denoted by $\Delta_B$. The natural linewidth (FWHM) of the excited state $\ket{e}$ is also indicated.} 
\label{fig:Quantum-networking-scheme}
\end{figure}

The quantum networking protocol developed in our laboratory by Wance Wang extends the capabilities of the $^{171}\mathrm{Yb}^{+}$ platform by exploiting the multiple qubit encodings available within the OMG architecture. As illustrated in Figure \ref{fig:Quantum-networking-scheme}, local quantum computation is performed using the long-lived ground-state hyperfine qubit, while remote entanglement is generated through an optical qubit encoded in the metastable $^{2}\mathrm{D}_{5/2}$ manifold. Following local gate operations, the hyperfine qubit is coherently mapped to selected $^{2}\mathrm{D}_{5/2}$ states. A two-photon excitation, consisting of simultaneous 435 nm and 1345 nm lasers, drives the ion to the $^{2}\text{P}_{3/2}$ manifold, from which it spontaneously emits a photon on the $^{2}\text{P}_{3/2}-{}^{2}\mathrm{D}_{5/2}$ transition at 1650 nm.

An optical cavity resonant with the 1650 nm transition enhances emission into a single spatial cavity mode, increasing the probability that the emitted photon is coupled into an optical fiber. The cavity is oriented such that the $\sigma^{\pm}$ and $\pi$  decay channels couple to orthogonal cavity polarizations, generating an entangled state between the polarization of the emitted photon and the ion's $^{2}\mathrm{D}_{5/2}$ Zeeman state. Because the photon is generated directly at a telecommunications wavelength, it can be efficiently coupled into standard single-mode optical fiber for low-loss transmission without requiring quantum frequency conversion.

Photons generated by two remote nodes are subsequently interfered on a beam splitter. When the photons are indistinguishable, HOM interference erases the which-path information, allowing a Bell-state measurement to herald the successful creation of entanglement between the distant ions. Finally, the quantum information stored in the metastable $^{2}\mathrm{D}_{5/2}$ manifold is coherently mapped back to the ground-state hyperfine qubit using the 411 nm quadrupole transition, where its long coherence time enables subsequent local quantum gate operations and distributed quantum computation.

\section{Going Cryogenic}
\label{sec:Going-Cryogenic}

In this section, I'll provide a brief survey of the field of cryogenics followed by the design goals and summary of the cryogenic ion trap apparatus built in this thesis.

\subsection{Cryogenics}
\label{subsec:Cryogenics}

The field of cryogenics plays a vital role in the world economy spanning the medical, energy, industrial, aerospace, and quantum computing sectors as well as basic science. In the medical sector, biological samples can be preserved using liquid nitrogen (77 K) or dry ice (195 K) and bulk storage tanks of liquid oxygen (90 K) are housed at hospitals enabling the provision of oxygen to patients when required. Storing oxygen as a liquid state rather than a gas reduces its volume by a factor of $\sim900$ making it more efficient to store and transport. Additionally, the superconducting magnets required to make Magnetic Resonance Imaging (MRI) work would not function without cooling to 4-80 K. In the energy sector, the trade of natural gas (methane, $\text{CH}_{4}$) between countries with no existing pipeline is enabled by liquefaction plants. Liquefying methane (112 K), colloquially known as LNG, reduces its volume by a factor of $\sim600$ enabling more efficient bulk shipment of methane between countries via a tanker. In the industrial sector, Deep Cryogenic Treatment (DCT) is cold thermal process used to increase the yield strength and corrosion resistance of stainless steels, cast iron, aluminum, and copper by submerging the metals in liquid nitrogen (77 K) for days to change their crystal structure followed by a slow warming back to room temperature. Additionally, cryogenic shrink fitting between mechanical parts (e.g., a shaft and housing) is routinely performed by first machining the shaft and housing to have an interference fit. Utilizing the thermal contraction at cryogenic temperatures, the shaft is submerged in liquid nitrogen (77 K) and assembled with the room temperature housing so that as the shaft equilibrates to room temperature, it expands and creates a tight, high-strength interference fit with the housing \cite{CryogenicsBradshaw}. In the aerospace industry in particular for the SpaceX Raptor engines for Starship, methane is used as the fuel while oxygen is used at the oxidizer enabling combustion. Storing them as liquids (Methalox: liquid oxygen (90 K) and liquid methane (112 K)) - which requires liquefaction plants - is more efficient as the volume is reduced enabling higher density storage in a rocket and also serves to cool the engine utilizing their heat of vaporization before entering the combustion chamber as a gas, known as regenerative cooling.\footnote{SpaceX: spacex.com/.} 

In basic science, the field of superconductors in general would not be able to be explored without cryogenics. In particular for the Large Hadron Collider at CERN, immense cryogenic systems are required to cool the thousands of superconducting magnets stretching around the 25 km accelerator ring to $\sim2\text{ K}$. This cooling also creates an Ultra-High Vacuum (UHV) environment for the beam line to make the particle accelerator function.\footnote{CERN: home.cern/.} Another large scale ongoing project known as the International Thermonuclear Experimental Reactor (ITER) will also require vast $\sim4\text{ K}$ cryogenic systems to cool its superconducting magnets and to create UHV environments in order to make its tokamak fusion device work.\footnote{ITER: iter.org/.} Yet another large scale project known as the James Webb Space Telescope requires continuous cryogenic cooling of its optics and detectors to $\sim4\text{ K}$ to reduce infrared background noise and Johnson noise in order to detect faint infrared signals from distant space objects.\footnote{James Webb Telescope: science.nasa.gov/mission/webb/cryocooler/.} Cryogenic cooling has also played a vital role in the development of ultrastable lasers used in precision measurement science, e.g., where lasers are locked to a optical cavity cooled to $\sim4\text{ K}$ which reduces thermal noise and relaxes temperature stability requirements as the slope of the coefficient of thermal expansion (CTE) for its materials drastically decreases below $\sim80\text{ K}$ \cite{zhangUltrastableSiliconCavity2017a}. In addition, the lowest reported temperature of a bulk solid-state object was achieved with precooling to $\sim10\text{ mK}$ using a $^{3}\text{He}/^{4}\text{He}$ dilution refrigerator followed by adiabatic demagnetization of nuclear spins in a 6 milligram rhodium sample ($\sim10^{19}$ atoms), yielding a temperature of 100 pK \cite{knuuttilaPolarizedNucleiNormal2001}.\footnote{The lowest temperature reported for a Boise-Einstein Condensate (BEC) is 38 pK. This did not require any cryogenic cooling, but only consisted of $\sim10^{5}$ rubidium atoms \cite{deppnerCollectiveModeEnhancedMatterWave2021}.}

In the quantum computing sector, cryogenic cooling is essential to several approaches. For solid-state based quantum computing technologies such as superconducting circuits \cite{devoretSuperconductingCircuitsQuantum2013} or charges and spins in semiconductor quantum dots \cite{awschalomQuantumSpintronicsEngineering2013}, $\sim10\text{ mK}$ temperatures are required, provided by $^{3}\text{He}/^{4}\text{He}$ dilution refrigerators, to initialize the systems in their ground state and to avoid errors due to thermal excitation during operation. The key players in industry using superconducting circuit technology are Google, IBM, and Rigetti while Intel is using semiconductor quantum dots. For photonic based quantum computing technologies \cite{kokLinearOpticalQuantum2007}, cryogenic cooling is exclusively used to maintain the superconductivity required for its single photon detectors to function in the temperature range of 0.05-2 K (depending on the detector and wavelength), rather than to cool the photonic qubits themselves. Transition-Edge Sensors (TES) require base temperatures in the range of 50-100 mK which can be achieved using $^{3}\text{He}/^{4}\text{He}$ dilution refrigerators or Adiabatic Demagnetization Refrigerators (ADR). Another common detector is the Superconducting Nanowire Singe Photon Detector (SNSPD) which requires base temperatures in the range of 0.8-2 K depending on the wavelength. Temperatures of $\sim2\text{ K}$ can be achieved using closed-cycle 2 K cryocoolers \footnote{Quantum Opus SNSPD: quantumopus.com/web/.} while temperatures $<2\text{ K}$ require closed-cycle 2-4 K cryocoolers together with evaporative cooling driven by cryosorption pumping of $^{4}\text{He}$.\footnote{Photon Spot SNSPD: photonspot.com/.} The key players in industry at this point using the photonic qubit technology are Xanadu and PsiQuantum. While the trapped ion technology has shown to be viable at room temperature \cite{debnathDemonstrationSmallProgrammable2016a}, cryogenic cooling of ion trap systems to $<10\text{ K}$ - most commonly achieved using closed-cycle 4 K cryocoolers - has shown to reduce a range of perturbations to ion motion as will be discussed in more detail in the following section. The key players in industry at this point building cryogenic trapped ion systems are IonQ and Quantinuum. 

\subsection{Design Goals and Cryogenic Ion Trap}
\label{subsec:Design-Goals-and-Cryogenic-Ion-Trap}

Quantum coherent control of trapped atomic ions \cite{winelandNobelLectureSuperposition2013} has applications including scalable quantum information processing \cite{debnathDemonstrationSmallProgrammable2016a,monzRealizationScalableShor2016}, quantum simulation \cite{monroeProgrammableQuantumSimulations2021}, frequency metrology \cite{ludlowOpticalAtomicClocks2015,brewer27QuantumLogicClock2019a} and entanglement distribution \cite{stephensonHighRateHighFidelityEntanglement2020b,krutyanskiyLightmatterEntanglement502019}. Laser-mediated ground state cooling of ions’ motional states is key to many of these applications and is routinely accomplished in room temperature apparatuses. Nevertheless, cryogenic cooling of an ion trap apparatus to $<10\text{ K}$ reduces a range of perturbations to ion motion including reduced anomalous heating \cite{turchetteHeatingTrappedIons2000a,deslauriersScalingSuppressionAnomalous2006,brownnuttIontrapMeasurementsElectricfield2015a} and fewer, lower energy residual hydrogen gas collisions \cite{paganoCryogenicTrappedionSystem2018a,wangProbingExtremehighVacuum2025,spiveyHighStabilityCryogenicSystem2022,dubielzigUltralowvibrationClosedcycleCryogenic2021}. The residual gases (nominally $\text{H}_{2}\text{O}$, $\text{CO}_{2}$, $\text{N}_{2}$, $\text{O}_{2}$, $\text{H}_{2}$) in a well prepared vacuum chamber will all be cryocondensed $<20\text{ K}$, with the exception of hydrogen. While hydrogen is only cryocondensed $<3\text{ K}$, cooling to $<10\text{ K}$ increases the efficiency of cryosorption pumping, reducing the residual hydrogen pressure \cite{baglinCryopumpingVacuumSystems2020,benvenutiINFLUENCETHERMALRADIATION}. These benefits enable long-lived hundred-ion ion chains \cite{paganoCryogenicTrappedionSystem2018a}, higher fidelity phonon-mediated two-qubit gates \cite{ballanceHighFidelityQuantumLogic2016a}, and robust operation of shallow-potential-well surface-electrode ion traps \cite{chiaveriniSurfaceElectrodeArchitectureIonTrap2005,seidelinMicrofabricatedSurfaceElectrodeIon2006,brownCoupledQuantizedMechanical2011}. In optical atomic clock applications, a cryogenic enclosure reduces uncertainty due to unknown Doppler shifts and blackbody radiation shifts \cite{poitzschCryogenicLinearIon1996a,ludlowOpticalAtomicClocks2015,itanoShift121982}. Lastly, this cryogenic cooling in general relaxes the typical in-vacuum material constraints of ion trap systems as cryopumping creates UHV conditions without performing a bakeout.

In this work, our research goal is to explore various $^{171}\text{Yb}^{+}$ ion trap optical cavity structures designed to ultimately serve as a photonic interconnect between multiple $^{171}\text{Yb}^{+}$ ion traps. Substantial work in coupling ions to optical cavities has been done but has proven to be difficult \cite{takahashiStrongCouplingSingle2020,brandstatterIntegratedFibermirrorIon2013,northupIoncavityInterfaceQuantum2015a}. With an aim towards building small mode volume optical cavities for ion-cavity QED to reach the strong coupling regime, the ion cavity distance needs to be small. This is challenging as the dielectric surfaces that comprise an optical cavity are known to produce electric field noise that disturbs the motion of ions, increasing heating rates and compromising the fidelity of gate operations for quantum computing \cite{tellerHeatingTrappedIon2021}. This dielectric-induced heating is predicted to scale as $\frac{T}{d^{3}}$ \cite{tellerHeatingTrappedIon2021} where $d$ and $T$ are the ion-cavity distance and temperature respectively, giving a strong motivation for building a cryogenic ion-cavity structure. 

Building optical cavities around ion traps will require many design iterations as follows: design, build, experiment, update design, and repeat. This creates a practical challenge for room temperature ion trap systems as they require a bakeout that can take weeks to recreate UHV conditions after every design iteration - i.e., after breaking vacuum - causing a long turnaround time. Work has been done to skip the bakeout in ion trap systems where the chamber was backfilled with nitrogen to reduce the accumulation of adsorbed air. Successful ion trapping was then reported after breaking vacuum this way without performing a bakeout at the cost of having ion lifetimes on the order of $\sim10\text{ s}$. An alternate technique for mitigating the long turnaround time of the bakeout was applied to work in neutral atom-cavity experiments known as the cavity load lock apparatus \cite{yinCavityLoadLock2023}. This apparatus consists of a science chamber and load lock chamber which are vacuum isolated by a gate valve while the cavity structure can be moved from the science chamber to the load lock chamber via a translator. In short, design iterations are performed by moving the cavity to the load lock chamber, closing the gate valve, and breaking vacuum in the load lock chamber. Once the new cavity design is installed, a low temperature $410\text{ K}$ bakeout is performed in the load lock chamber pumping the residual gas with a turbo pump and ion pump. Once the pressure in the load lock chamber is on the level of $10^{-10}\text{ Torr}$, the temperature is ramped down, the gate valve is opened, and the cavity structure is repositioned in the science chamber. This prevents the science chamber from being exposed to air, allows the ambient optical setups to remain in place during the entire process, and has a one week turnaround time. This is a clever technique, but is still a complex system that is bulky, requires two large ion pumps, requires an initial bakeout of the science chamber, and only achieves a vacuum pressure on the $10^{-10}\text{ Torr}$ level. In addition, it's not clear how many times the load lock chamber can be effectively evacuated as operating an ion pump with only a $410\text{ K}$ bakeout temperature might prematurely saturate the ion pump, reducing its pumping speed over time. Alternatively, cryogenically cooling an ion trap to $<10\text{ K}$ creates UHV conditions without a bakeout as the residual gases, with the exception of hydrogen, will be completely condensed at this temperature enabling rapid ion trap and cavity design iteration motivating yet another reason to go cryogenic. 

The advantages of cryogenic cooling come with added technical complexity as well as expensive liquid cryogens or cryocoolers. Thus, the first design decision is to determine what type of cooling technology to use. Historically, researchers thermally coupled cryostats to dewars filled with liquid helium to cool the trap to $4\text{ K}$. These setups exhibited low vibration and excellent temperature stability \cite{brownCoupledQuantizedMechanical2011,brandlCryogenicSetupTrapped2016a,diederichObservingSingleHydrogenlike1998}, but due to a world-wide scarcity of helium this approach is no longer practical in an academic lab setting. Contemporary approaches include semi-closed-cycle cryocoolers where cryogenic gases are purified and re-liquefied \cite{wangCryogenRecyclerPulse2015,ballSiteresolvedImagingBeryllium2019} and fully closed-cycle cryocoolers that utilize closed-loop gas circulation that can achieve cooling to $4\text{ K}$ without using liquid cryogens \cite{paganoCryogenicTrappedionSystem2018a,niedermayrCryogenicSurfaceIon2014b,wangProbingExtremehighVacuum2025,qinQinEvaluationSuppressionVibrational2025,mickeClosedcycleLowvibration42019,vittoriniModularCryostatIon2013,dubielzigUltralowvibrationClosedcycleCryogenic2021,spiveyHighStabilityCryogenicSystem2022}. In general, cooling to $3\text{ K}$ is desirable as this ensures complete cryocondensation of residual hydrogen in the chamber. However, this becomes impractical for RF Paul traps given the limited cooling power available at $3\text{ K}$ for these commercial close-cycle cryocoolers. Additionally, the commercially available $^{3}\text{He}/^{4}\text{He}$ dilution refrigerators are not designed to have large cooling power at $3\text{ K}$. In principle, more cooling power could be created at 3 K but will require an expensive custom cryogenic cooling system - e.g., the ITER 4 K stage used for cooling its superconducting magnets has a cooling power of $\sim75\text{ kW}$.\footnote{ITER: iter.org/.} These closed-cycle cryocoolers (typically Gifford-McMahon (GM), Pulse Tube, or Stirling) are mechanical refrigeration systems that can achieve cooling to 4 K but produce significant vibrations on the order of $10\text{ \ensuremath{\mu}m}$ \cite{tomaruVibrationAnalysisCryocoolers2004}. Such vibrations can interfere with the tightly focused laser beams used to cool and manipulate ions, substantially increasing the technical complexity of obtaining the benefits of cryocooling referenced earlier in this section. 

Extensive work has gone into building low vibration 4 K cryogenic ion trap systems as well as cavities for optical atomic clocks using closed-cycle cryocoolers. Many custom cryogenic ion trap systems have been built utilizing passive vibration isolation techniques where the thermal link between the ion trap and the cryocooler is a helium gas exchange region \cite{paganoCryogenicTrappedionSystem2018a,dubielzigUltralowvibrationClosedcycleCryogenic2021,moritzCryogenicPaulTrap2025,hartsellDesignCharacterizationCryogenic2025}, a flexible copper braid \cite{wangProbingExtremehighVacuum2025,niedermayrCryogenicSurfaceIon2014b}, or a long copper inertial pendulum \cite{qinQinEvaluationSuppressionVibrational2025,mickeClosedcycleLowvibration42019}. These systems were particularly advantageous as they exhibited a cooling power of $\sim1\text{ W}$ at 4 K, but did not consistently report low vibration levels. In contrast, ion traps \cite{spiveyHighStabilityCryogenicSystem2022} and optical cavities for atomic clocks \cite{zhangUltrastableSiliconCavity2017a} have been installed in plug-and-play type cryostats completely engineered by Montana Instruments Corporation - the thermal link and vibration isolation mechanism between the cryocooler and cryostat 4 K stage are not publicly known. These systems have the benefit of outsourcing the complex cryogenic engineering and vibration isolation, and are nicely designed to enable high optical and maintenance access to the 4 K stage while the cryocooler sits besides the optics table. However, these systems exhibited lower cooling powers in the range of 0-250 mW at 4 K. 

Alternatively, another cryogenic cooling system built by ColdEdge Technologies (CET), known as the Stinger, which integrates a Gifford-McMahon (GM) cryocooler with a closed-cycle helium recirculation loop, where precooled helium undergoes Joule-Thomson (JT) expansion to cool remote cryostat stages. In this case, the thermal link between the 4 K stage of the cryostat and the cryocooler is the recirculating helium via a long flexible transfer line which provides passive vibration isolation. This system is advantageous as it is specced to have a cooling power of $>200$ mW at 4 K and a vibration level of $<10\text{ nm}_{\text{RMS}}$. In addition, the cryostat is not tightly constrained to the GM cryocooler and the transfer line can be easily disconnected from the cryostat. This contrasts with the custom cryogenic systems as well as the plug-and-play type Montana Instruments systems reference earlier in this section. This functionality provides more room on the optics table, high optical access to the 4 K stage, and enables doing in-vacuum maintenance on the cryostat without adjusting the position of the cryocooler. In addition, the cryostat can be removed from the optics table by simply disconnecting the transfer line to perform more complicated in-vacuum maintenance and perform a bakeout if necessary. In light of this, we decided to build a custom cryostat that interfaces with the CET Stinger system giving us more control of the cryostat design. It is also worth noting that during the course of my PhD, successful custom cryogenic neutral atom systems \cite{zhangHighOpticalAccess2025,jinExtendedRydbergLifetimes2026} have been built that are cooled using the CET Stinger system. 

I'll now present the design goals for building a 4 K cryostat suitable for our ion-cavity research that interfaces with the CET Stinger system. This boils down to the following constraints: 1.) high optical access to the 4 K stage, 2.) low vibration of the 4 K stage relative to the optics table, 3.) bakeable to 500 K , 4.) built with non-magnetic materials, and 5.) fast design iteration turnaround time. I'll summarize these design goals one by one in the following paragraphs. 

\noindent\textbf{High Optical Access}

High optical access to the ion trap and cavity at the 4 K stage of this cryostat is required for ion trapping, implementing the ion-cavity protocol, and applying entangling gate operations on the ions. This boils down to optical access for the photoionization, Doppler cooling, repumper, quadrupole, cavity, and Raman beams as well as optical access to collect photons from the ions for imaging and detection. This motivates designing the cryostat that enables many beam orientations to the ion trap with appropriately oriented windows, a high numerical aperture (NA) for imaging and detection, and quantization axis (external magnetic field) tunability for driving quadrupole transitions. 

\noindent\textbf{Low Vibration}

Movement of the 4 K stage relative to the optics table affects laser-ion interactions by creating relative motion between the ions and interrogating lasers producing intensity fluctuations and a Doppler shift of the laser frequency in the rest frame of the ion. These intensity fluctuations and Doppler shifts can manifest as Rabi frequency errors and detuning errors for Raman beams creating gate errors for quantum computation \cite{ballanceHighFidelityQuantumLogic2017a}, see Subsection \ref{subsec:vibration-impact-on-ion-laser-interactions-and-cavity-length-fluctuations}. In particular for ion-cavity interactions, these vibrations can perturb the cavity length which degrades the ion-cavity cooperativity, see Subsection \ref{subsec:vibration-impact-on-ion-laser-interactions-and-cavity-length-fluctuations}. This motivates careful design of the cryostat cold stages to isolate the vibrations from the GM cryocooler.

\noindent\textbf{Bakeable to 500 K}

Although a bakeout is not required to achieve UHV, the cold surfaces of the cryostat become getters for the residual gases (nominally $\text{H}_{2}\text{O}$, $\text{CO}_{2}$, $\text{N}_{2}$, $\text{O}_{2}$, $\text{H}_{2}$). In particular, cryodeposit on the optical cavity mirrors and trap electrodes can affect ion heating rates. Cryodeposit on the cavity mirrors or trap electrodes can trap stray charges, forming patch potentials that create static electric fields that shift the ion's position \cite{turchetteHeatingTrappedIons2000a}. In particular for polar $\text{H}_{2}\text{O}$ and the most polarizable molecule $\text{CO}_{2}$ of the nominal residual gases, the electric fields from the trap electrodes can rearrange these molecules which can further alter the effective electric field at the ions. While anomalous ion heating has been shown to decrease at cryogenic temperatures \cite{turchetteHeatingTrappedIons2000a,deslauriersScalingSuppressionAnomalous2006,brownnuttIontrapMeasurementsElectricfield2015a}, the efficacy of baking a cryostat to further reduce anomalous ion heating has not yet been reported. With the long term goal of building small mode volume optical cavities for ion-cavity QED, this motivates designing a cryostat with low outgassing materials that is bakeable to $500\text{ K}$ to ensure clean cryogenic trap electrode and optical cavity mirror surfaces to mitigate ion heating. 

\noindent\textbf{Non-Magnetic Materials}

Having magnetic materials in the cryostat can create local magnetic field gradients across the ion trap resulting in a position-dependent frequency shift. As a corollary to that, vibrations of the 4 K stage (i.e., the ions) relative to an interrogating laser can create a time varying magnetic field at the ion if there is a field gradient causing decoherence of qubit operations. In addition, a local magnetic material can become magnetized by an external magnetic field (e.g., the quantization axis), and could change over time due to hysteresis (and temperature) of the material making it necessary to periodically measure and compensate for field drift using an ion as the field probe. For our quantum networking protocol using $^{171}\text{Yb}^{+}$, the hyperfine ground state qubit $S_{1/2}$ is defined by states with $F=0,m_{\text{F}}=0$ and $F=1,m_{\text{F}}=0$ and is thus first-order insensitive to magnetic field fluctuations. In contrast, the protocol involves driving narrow quadrupole transitions and generating 1650 nm photons entangled with atomic hyperfine states which are all highly sensitive to magnetic field fluctuations as covered in Subsection \ref{subsec:Long-Baseline-Quantum-Networking}. This motivates designing the cryostat with non-magnetic materials.

\noindent\textbf{Fast Turnaround Time}

To enable fast design iteration turnaround time of ion-cavity structures, the 4 K stage must be easily accessible and the cryostat must be thermally cycled quickly. Extremely low background pressures on the level of $10^{-16}\text{ Torr}$ have been reported in 4 K cryogenic ion trap systems that have a vacuum sealed 4 K enclosure \cite{gabrielseThousandfoldImprovementMeasured1990,diederichObservingSingleHydrogenlike1998}. This can be beneficial as the in-vacuum regions can be divided into a UHV clean and bakeable 4 K enclosure surrounded by a high-vacuum 40 K and 300 K vacuum space relaxing material constraints \cite{dubielzigUltralowvibrationClosedcycleCryogenic2021,vittoriniModularCryostatIon2013}. However, accessing the 4 K stage in these systems creates a long turnaround time as the 4 K enclosure needs to be resealed, pumped, and rebaked \cite{dubielzigUltralowvibrationClosedcycleCryogenic2021}. This motivates building a cryostat with a 4 K stage that is vacuum continuous with the rest of the chamber. This places harder material constraints on the cryostat as all in-vacuum materials must have low outgassing rates and be bakeable to 500 K as discussed earlier, but enables easier access to the 4 K stage for ion-cavity design iterations. 

\begin{figure}[!!h]
\centering
\includegraphics[width=\textwidth]{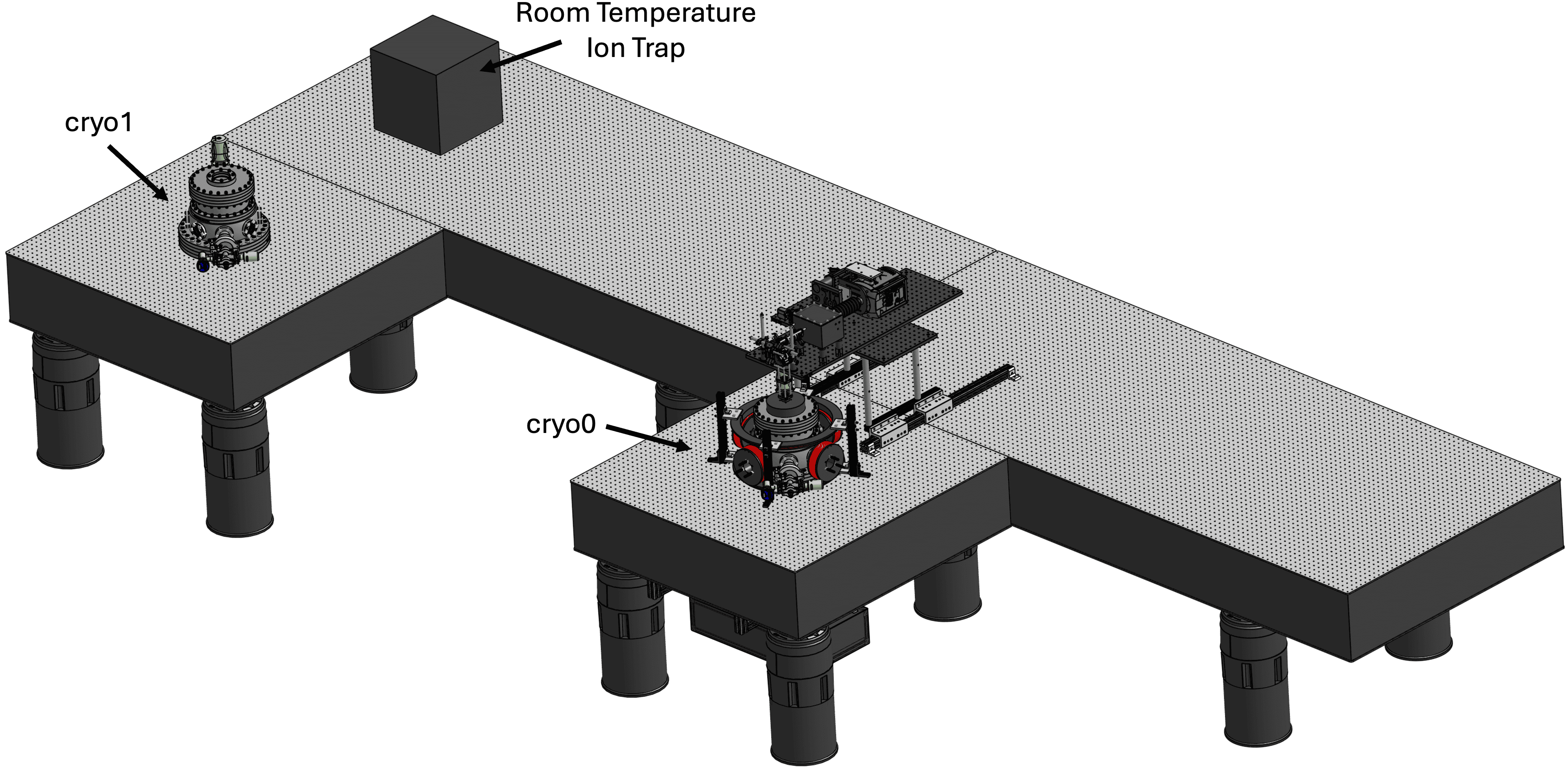}
\caption[Britton Lab Experiment Optical Table]{CAD drawing of the Britton Lab experiment optical table consisting of two cryostats (cryo0 and cryo1) and a room temperature ion trap. I designed and built the two cryostats while the room temperature ion trap experiments were conducted by Connor Goham which are covered in his thesis. }
\label{fig:Britton-Lab-Ion-Traps}
\end{figure}

In this thesis, I designed, built, and commissioned two cryostats (cryo0 and cryo1) that interface with the CET Stinger system to enable the ion-cavity research for our lab. Both systems are composed of three identical vacuum continuous regions: a 300 K stage, a 40 K stage, and a 4 K stage. Each cryostat is designed to mate with and protrude through the bore of a 12 inch optics table with an ion trap situated at the 4 K stage as shown in Figure \ref{fig:Britton-Lab-Ion-Traps}. This design allows UHV vacuum components, electrical feedthroughs, and the CET Stinger cryogen feedthroughs to interface the cryostat at the 300 K stage from beneath the table, maximizing the available optical access on the top surface. This configuration enables laser access both from beneath the table and at a height of 4 inches above the table surface, while allowing for photon collection from above. The cold stages were designed to be mechanically stiff to isolate vibrations communicated to the cryostat from the GM cryocooler. This design also enables easy access to the 4 K stage by simply removing the top conflat flange and radiation shield lids. In addition, the cryostat can be cooled down to its base temperature in $\sim17\text{ h}$ and can be warmed back up to room temperature in $\sim5\text{ h}$ using heaters mounted at the cold stages, enabling a fast design iteration turnaround time. The cryostats were built exclusively with non magnetic materials when possible due to commercially available parts. The cryostats were not baked in this work due to time constraints. Thus, the cryostats were built with mainly low outgassing materials that are bakeable to 500 K, but some non-bakeable parts were installed for convenience. 

The primary focus of this thesis involved the development of cryo0 (formerly Frosty), which houses an Alpine Quantum Technologies (AQT) RF Paul trap and Fabry-Pérot (FP) optical cavity at the 4 K stage. The optical cavity was designed and built by Wance Wang and is covered in his thesis. For this cryostat, I designed and built an in-vacuum objective providing a 0.5 NA for imaging and detection, an oven serving as the atomic source of $\text{Yb}$, and an in-vacuum resonator to create the radially confining potential for the ion trap all installed at the 4 K stage. I also designed and built all of the in-vacuum wiring including an DC trap electrode filter board installed at the 4 K stage. I built magnetic field coils to generate an arbitrary quantization axis in 3D relaxing the optical access constraints for quadrupole transition beams. I also built the laser cooling system and optomechanical systems used to deliver the beams to the ions and collect photons for imaging and detection. I set up the experimental control using ARTIQ to enable running ion trapping experiments and characterized the performance of the trap including secular frequency measurements, micromotion compensation, photon collection efficiency, and ion lifetime. I also characterized the cryostat performance by measuring the cooling capacity, vibration level, and vacuum level and built a helium purification device that increased the uptime of the cryostat from 30 days to 72 days. Lastly, Wance Wang and I installed an FP optical cavity (cavity designed by Wance) around the AQT Pine trap and demonstrated spatial overlap and interaction between a trapped $^{174}\mathrm{Yb}^{+}$ ion and the cavity mode on the 1345 nm transition.

A second system, cryo1 (formerly Crystal), was built and commissioned to verify the reproducibility of the design, but it was not used for the ion trapping experiments conducted in this work. In addition, I built a bakeout station designed to perform a bakeout of the cryostat and other UHV chambers.

\section{Thesis Outline}
\label{sec:Thesis-Outline}

I will begin this thesis discussing the CET Stinger system and the helium purification circuit I built in Chapter \ref{chap:Cryocooler} followed by the construction of the cryostat I built in Chapter \ref{chap:Cryostat}. I'll then the present the various subsystems built that enable cryogenic ion trapping experiments in Chapter \ref{chap:cryogenic-ion-trap-system} followed by the ion trap characterization including the secular frequency measurement, micromotion compensation, construction of the resonator, filtering circuits, magnetic field coils, and ion-cavity mode overlap in Chapter \ref{chap:ion-trap}. The construction and characterization of the atomic source and imaging and detection system are then presented in Chapters \ref{chap:atomic-source} and \ref{chap:imaging-and-detection}. Lastly, the characterization of the cryostat including the cooldown and warmup procedures, cooling capacity measurements, vibration level measurements, and vacuum level measurements are presented in Chapter \ref{chap:cryostat-characterization}. 

Supplemental material is covered in the Appendices. Information pertaining to vacuum physics and the bakeout station are covered in Appendix \ref{chap:vacuum-processing}. Information about materials, connectors, and machine shops used in this work are covered in Appendix \ref{chap:manufacturing}. Details about the in-vacuum oven construction are covered in Appendix \ref{chap:pulse-arc-and-resistance-welder}. Lastly, information about estimating power spectral densities is covered in Appendix \ref{chap:estimating-vibration-amplitude-spectra}.

\ifSubfilesClassLoaded{%
  \bibliographystyle{unsrt}
  \bibliography{Dissertation}
}{}

\end{document}

%% file: Chapter2.tex
\chapter{Cryocooler}
\label{chap:Cryocooler}

In this chapter, I will present the ColdEdge Technologies (CET) Stinger system used to cool the cryostats cryo0 and cryo1. Two identical Stinger systems were used to cool each cryostat, but majority of the work in this thesis was done on the system cooling cryo0. A high-level explanation of the Stinger as well as its various subsystems are covered in Section \ref{sec:ColdEdge-Technologies-Stinger}. I'll discuss the thermodynamics of the cooling system and present a simple thermal model of the cryostat cold stage heat exchangers in Section \ref{sec:Thermal-Model}. Next, I present the helium purification device I built to increase the uptime of the system in Section \ref{sec:helium-purification}. Lastly, I'll present the standard operating procedure I used to ensure optimal performance of the system in Section \ref{sec:standard-operating-procedure}.

\section{ColdEdge Technologies Stinger}
\label{sec:ColdEdge-Technologies-Stinger}

\begin{figure}[!!h]
\centering
\includegraphics[width=\textwidth]{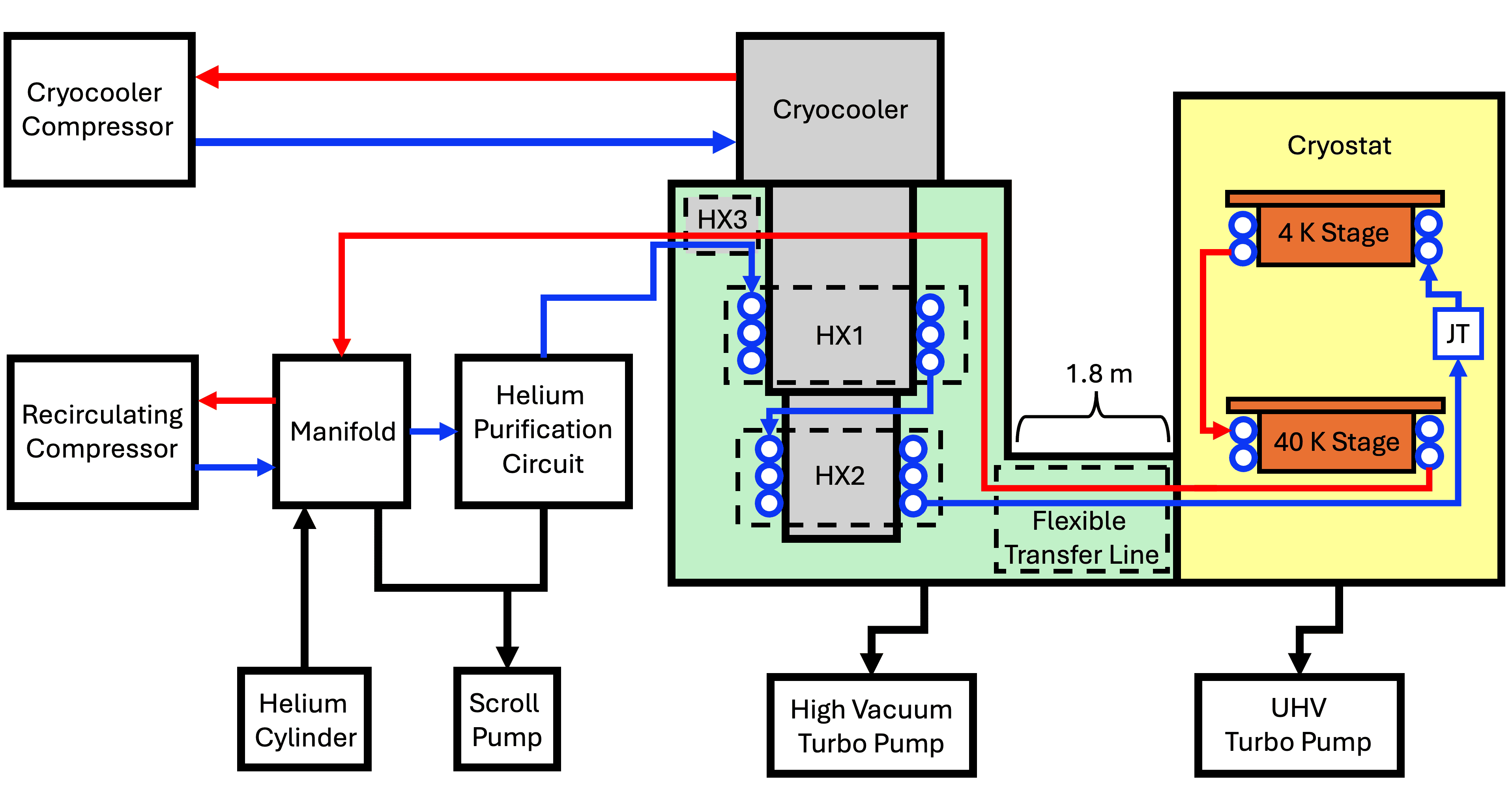}
\caption[ColdEdge Technologies Stinger Cryogenic System]{Schematic of the CET Stinger closed-cycle flow cryocooler and how it interfaces with the cryostat (not to scale). The red and blue arrows depict the supply and return sections of each loop, respectively. The GM cryocooler cold stages are depicted in the shaded gray regions with heat exchangers HX1 and HX2. A third heat exchanger (HX3) between the supply and return of the recirculating loop is depicted in a separate shaded gray region. The cryocooler vacuum space, depicted by the shaded green region, is evacuated using a turbo pump. The cryostat vacuum space, depicted by the shaded yellow region, is isolated from the cryocooler vacuum space and has a dedicated turbo pump. A scroll pump is used to evacuate the helium in the precooling stages and transfer line to the JT capillary and to regenerate the getters in the helium purification circuit.}
\label{fig:Schematic-of-CET-Stinger-Cyrogenic-system}
\end{figure}

The system consists of two independent closed-cycle $^{4}\text{He}$ loops: a Gifford-McMahon (GM) cryocooler loop and a separate recirculation loop shown in Figure \ref{fig:Schematic-of-CET-Stinger-Cyrogenic-system}. In the recirculation loop, high-pressure helium at roughly 110 psig is first precooled via heat exchangers (HX1 and HX2) at the GM cryocooler cold stages. The precooled helium is then transported through a 1.8 meter vacuum insulated flexible transfer line where it undergoes Joule-Thomson (JT) expansion through a capillary restriction and cools the cryostat cold stages. The expanded helium returns through the transfer line at roughly 0 psig and cools the incoming high pressure helium at another heat exchanger (HX3) before returning to the recirculating compressor, completing the closed loop. The helium purification circuit in series with the recirculating loop consists of three getters used to mitigate contamination by removing trace impurities that would otherwise condense or solidify within the JT capillary and is discussed in more detail in Section \ref{sec:helium-purification}. The gas manifold is equipped with valves to isolate the getters, precooling stages, and transfer line to the JT capillary from the recirculating compressor. In addition, it includes a pressure regulator for flow control, a vacuum port for evacuating the loop, and a charging port for helium recharge.

For this CET Stinger system, a GM cryocooler (Sumitomo, PN: RDE-418D4) was used together with a compressor (Sumitomo, PN: F-70L) which is specced with cooling capacities of 2 W at 4.2 K and 50 W at 50 K while a second compressor (Sumitomo, PN: HC-4E1) was used for the recirculating helium loop. The main challenge for maintaining long uptimes using this CET Stinger closed-cycle flow system is due to the JT expansion technology. Impurities in the recirculating helium loop can precipitate in solid form and accumulate in the small capillary creating a clog which decreases the flow of helium and thus the cooling power at the 4 K stage of the cryostat. These clogs can become so bad causing the 4 K stage temperature to rise at which point the system needs to be completely warmed back up to room temperature so that the precipitants can be evacuated from the recirculating helium line. This downtime is simply annoying for running any atomic physics experiments. In this work, I spent considerable time battling this issue and engineered a helium purification device that improved the uptime of the cryostat from 30 days to 72 days which is covered in Section \ref{sec:helium-purification}.

There are similar commercially available recirculating gas cryogenic systems built by Lake Shore Cryotronics (Infinite Helium system and the Recirculating Gas Cryocooler (RGC) system) and Cryo Industries of America (Universal Cryocooler) where a recirculating helium gas loop is precooled by a GM cryocooler and then cools a sample region in a separate vacuum space. The Lake Shore Infinite Helium and RGC systems do not incorporate JT expansion and are specced to have low vibration at the sample region of $<10\text{ nm}_{\text{RMS}}$ and cooling powers of 450 mW and 250 mW at 5 K, respectively. The Cryo Industries of America Universal Cryocooler system also does not incorporate JT expansion at the sample region, but its vibration level and cooling capacity are not specced. To my knowledge, these Lake Shore systems were not yet commercially available and I didn't know about the company Cryo Industries of America at the beginning of my PhD so none of these systems were considered. 

\subsection{Gas Manifold}
\label{subsec:Gas-Manifold}

\begin{figure}[!!h]
\centering
\includegraphics[width=\textwidth]{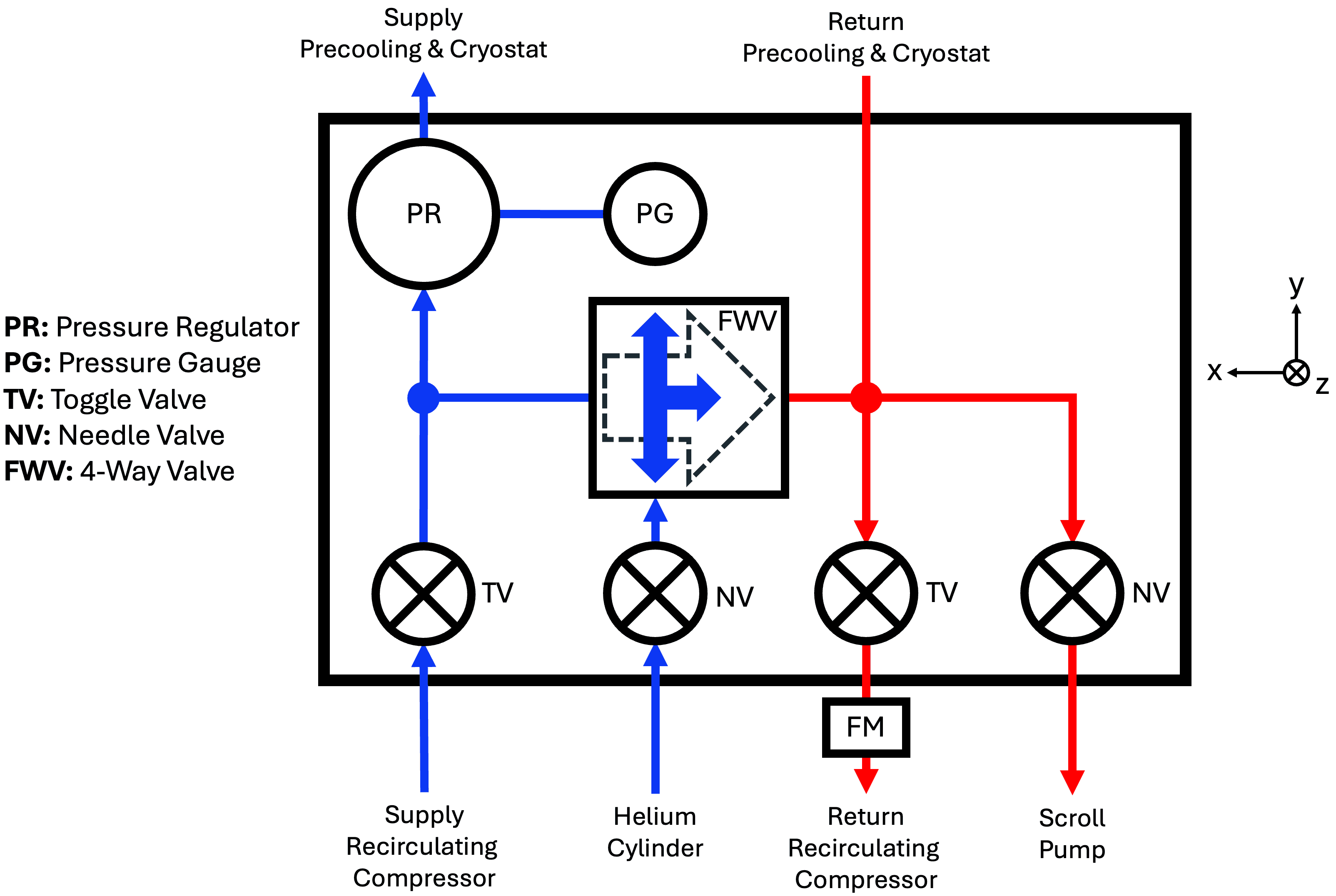}
\caption[Gas Manifold]{Schematic of gas manifold in the CET Stinger system while it is cooling the cryostat. The 4-way valve pointing in the $-\hat{x}$ direction (depicted by the dashed black arrow), the toggle valves for the recirculating helium are open, and the needle valves at the charge and vacuum ports are closed. The blue and red arrows depict the supply and return of the recirculating helium.}
\label{fig:Gas-Manifold}
\end{figure}

The gas manifold is equipped with valves to isolate the getters, precooling stages, and transfer line to the JT capillary from the recirculating compressor as shown in Figure \ref{fig:Gas-Manifold}. In addition, it includes a pressure regulator for flow control, a flow meter, a vacuum port for evacuating the loop, and a charging port for helium recharge. The internal plumbing of the gas manifold consists of welded stainless steel tubes and Swagelok connectors as well as a multiple pressure relief valves that are not shown for simplicity. The schematic is shown in the configuration while the Stinger system is cooling the cryostat with the 4-way valve pointing in the $-\hat{x}$ direction (depicted by the dashed black arrow), the toggle valves for the recirculating helium are open, and the needle valves at the charge and vacuum ports are closed. More details about various operational configurations of the gas manifold are covered in Section \ref{sec:standard-operating-procedure}. 

We use a mass flow meter (Omega, PN: FMA1824A) to monitor the flow of the helium at the return side of the gas manifold when the helium is around 0-1 psig and 300 K as shown in Figure \ref{fig:Gas-Manifold}. This is an invaluable tool as it is the only device in the system that can directly detect a clog at the JT capillary. The device works by applying a known amount of power, $\dot{Q}$, to the gas and measures the temperature gradient, $\Delta T$, at two points in the meter. The mass flow is then calculated using the specific heat, $C$, of the gas. 

\begin{equation}
    \dot{m}=\frac{\dot{Q}}{C\Delta T}
    \label{eq:mass-flow-rate-flow-meter}
\end{equation}
However, the volume flow rate, $\dot{V}=\frac{\dot{m}}{\rho}$, is displayed on the meter in units of standard liter per minute (SLPM), by convention, and is calibrated with nitrogen at standard temperature and pressure (STP), 1 atm and 273 K.\footnote{Another conventional unit commonly used is standard cubic centimeters per minute (SCCM). 1\text{ SLPM=1000\text{ SCCM}}.} This can be converted to the volume flow rate of helium in SLPM using the following equation

\begin{equation}
    \dot{V}_{\text{He}}=\dot{V}_{\text{N}_{2}}\frac{\rho_{\text{N}_{2}}C_{\text{N}_{2}}}{\rho_{\text{He}}C_{\text{He}}}=\dot{V}_{\text{N}_{2}}\times1.45
    \label{eq:volume-flow-rate-calibration-for-He-SLPM}
\end{equation}
where $\rho$  and $C$ are the densities and specific heats of nitrogen and helium at STP. However, the mass flow rate in $\frac{\text{kg}}{\text{s}}$ or $\frac{\text{mol}}{\text{s}}$ is a more convenient unit used in calculations and will be used solely through this work. For example using mass conservation, the mass flow measured at the meter at 300 K equals the mass flow through the JT capillary at 6 K, $\dot{m}_{\text{He,FM,300\,K}}=\dot{m}_{\text{He,JT,6\,K}}$ where the mass flow is given as $\dot{m}=\rho Av$. As the temperature changes, the density and velocity change accordingly to keep the mass flow constant for a fixed cross-sectional area $A$. 

\subsection{Water Cooling}
\label{subsec:Water-cooling}

\begin{figure}[!!h]
\centering
\includegraphics[width=\textwidth]{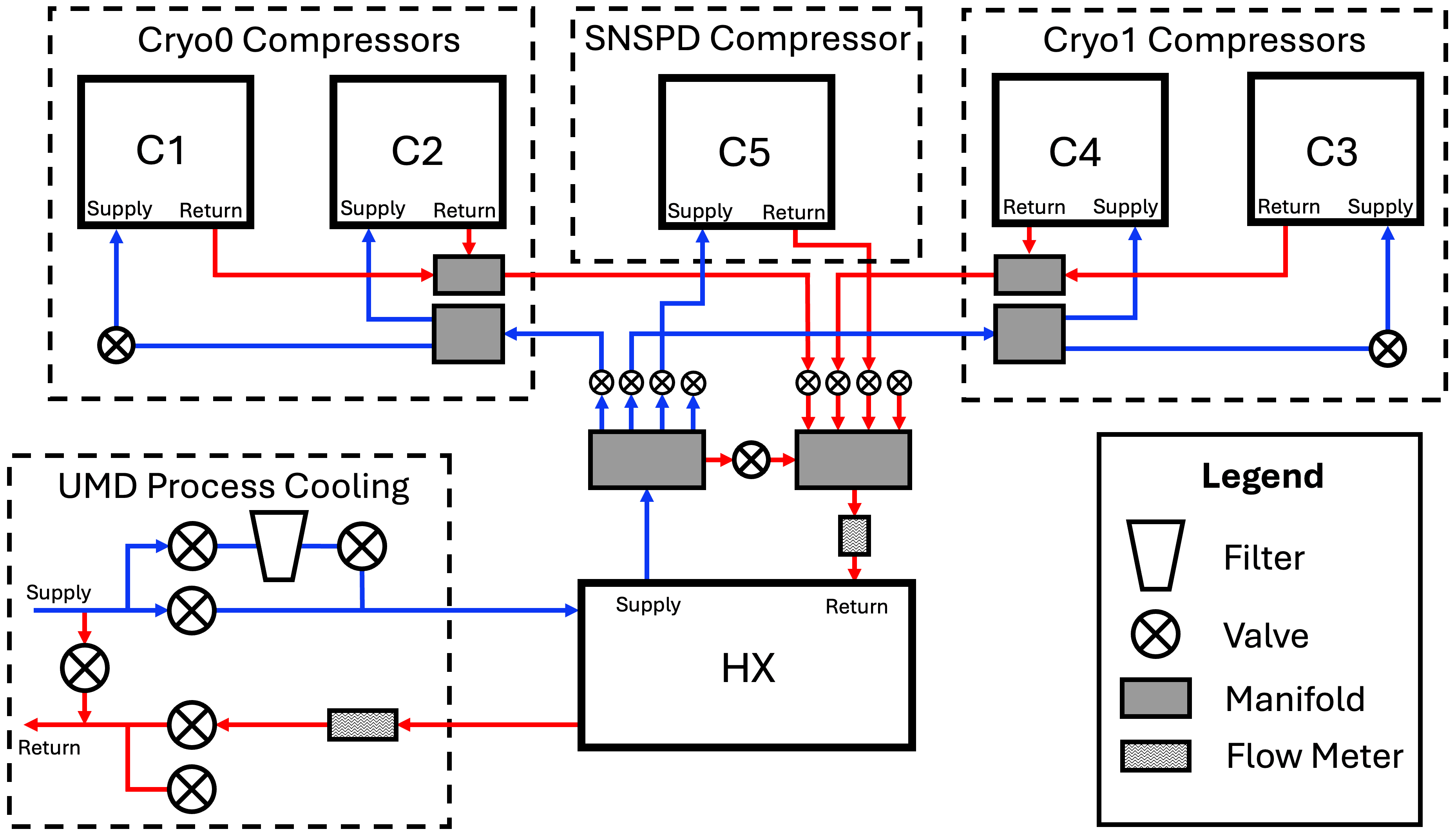}
\caption[Compressor Closet Water Cooling Circuit]{Schematic of the compressor closet water cooling circuit. The parts in the UMD process cooling loop are connected with NPT-3/4 fittings. The parts in the local cooling loop are connected with NPT-1, NPT-3/4, and NPT-1/2 fittings. The materials used for all fittings in each loop are copper, brass, bronze, or stainless steel 300 series to mitigate galvanic corrosion. A part list for this schematic is given in Table \ref{tab:part-list-for-compressor-water-cooling-circuit}.}
\label{fig:Schematic-of-compressor-water-cooling-circuit}
\end{figure}

In order for the GM cryocooler (in general any cryocooler) to work as well as to recirculate the helium in the CET Stinger system requires compressors. Compressing helium generates intense heat requiring additional cooling loops and in this work we used water cooled helium compressors. As the purity of the UMD process cooling water is not compatible with the helium compressor purity requirements, we opted to water cool the compressors driven by a local pump that is cooled by the UMD process cooling water via a heat exchanger as shown in Figure \ref{fig:Schematic-of-compressor-water-cooling-circuit}.

\begin{table}[!!h]
\centering
\tablefont
\caption[Parts List for Compressor Closet Water Cooling Circuit]{Parts list for compressor closet water cooling circuit in Figure \ref{fig:Schematic-of-compressor-water-cooling-circuit}}
\label{tab:part-list-for-compressor-water-cooling-circuit}
\begin{tabularx}{\textwidth}{>{\centering\arraybackslash}p{0.49\textwidth}>{\centering\arraybackslash}p{0.44\textwidth}}
\toprule
\textbf{Schematic Label} & \textbf{Part} \\
\midrule
C2, C4 & \makecell{GM Cryocooler Compressor\\Sumitomo, PN: F-70L} \\
\midrule
C1, C3 & \makecell{Recirculating Compressor\\Sumitomo, PN: HC-4E1} \\
\midrule
C5 & \makecell{Photon Spot SNSPD Cryocooler Compressor\\Sumitomo, PN: HC-4E2} \\
\midrule
HX & \makecell{Water Cooling Heat Exchanger\\Thermo Fisher, PN: SYS4 D3 CP100 230/60/3 AR} \\
\bottomrule
\end{tabularx}
\end{table}

\section{Thermal Model}
\label{sec:Thermal-Model}

\begin{figure}[!!h]
\centering
\includegraphics[width=\textwidth]{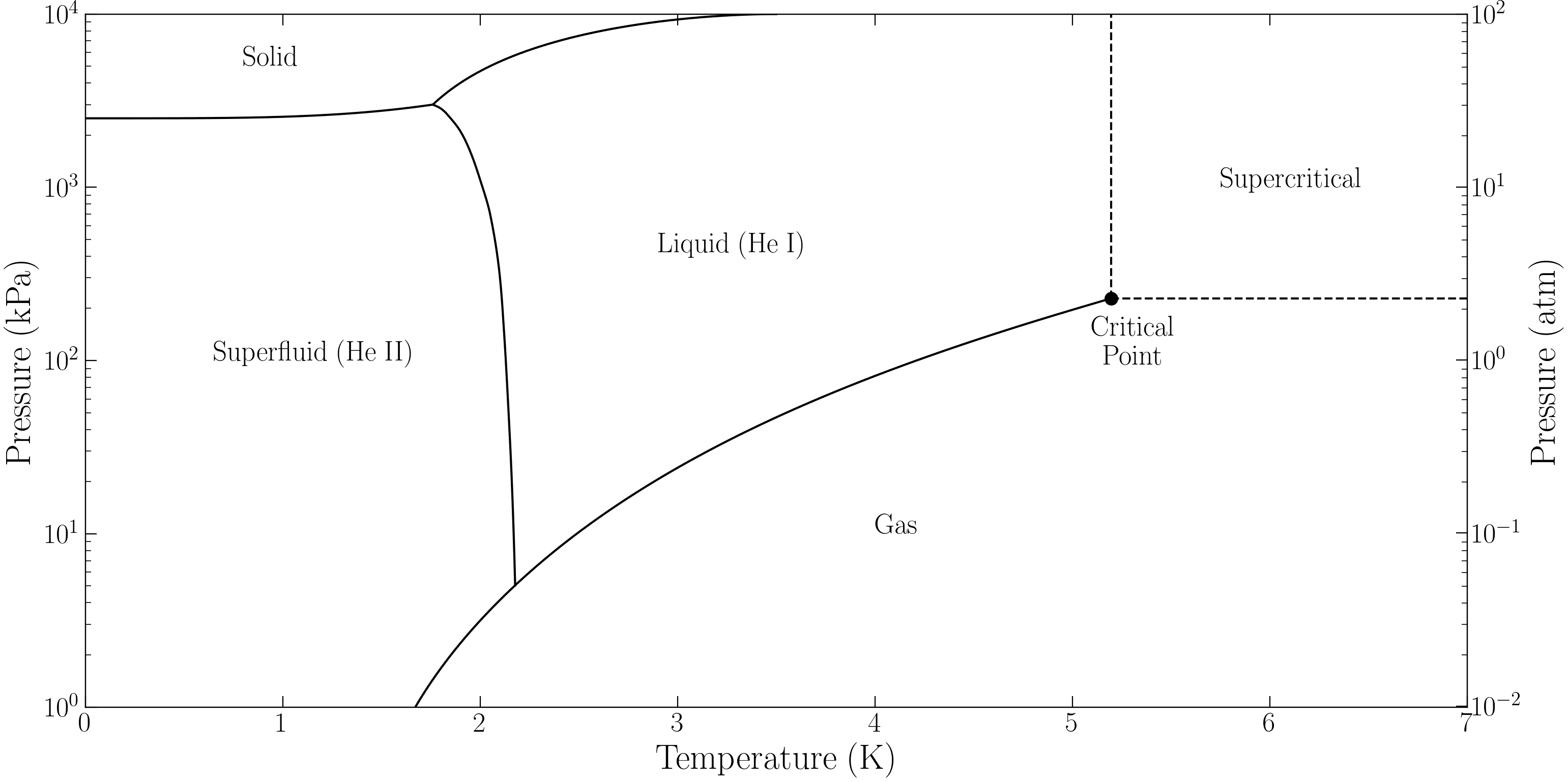}
\caption[$^{4}\text{He}$ Phase Diagram in $PT\text{-space}$]{Phase diagram of $^{4}\text{He}$ in $PT$-space.}
\label{fig:Helium4-phase-diagram}
\end{figure}

In this section I'll present the thermodynamics of the JT expansion for $^{4}\text{He}$ as well as a simulation of the cooling capacity of the 4 K and 40 K heat exchangers. The phase diagram for $^{4}\text{He}$ is given for reference in Figure \ref{fig:Helium4-phase-diagram}.

The JT expansion of a gas can be modeled as a constant enthalpy process where the enthalpy state function is defined as follows,

\begin{equation}
    H=U+PV=\text{const.}
    \label{eq:enthalpy-state-function}
\end{equation}
where $U$ is the internal energy, $P$ is the pressure, and $V$ is the volume. As we are interested in how the gas changes temperature when we reduce the pressure at constant enthalpy, we define the JT coefficient $\mu_{\text{JT}}=\left(\frac{\partial T}{\partial P}\right)_{H}$ \cite{sciverHeliumCryogenics2012}. Using thermodynamic relations, we can write the JT coefficient as follows,

\begin{equation}
    \mu_{\text{JT}}=\frac{1}{C_{\text{P}}}\left[T\left(\frac{\partial V}{\partial T}\right)_{P}-V\right]
    \label{eq:JT-coefficient}
\end{equation}
The change in temperature from pressure $P_{1}$ to $P_{2}$ following a JT expansion is,

\begin{equation}
    \Delta T=\int_{P_{1}}^{P_{2}}\mu_{\text{JT}}dP=\int_{P_{1}}^{P_{2}}\frac{1}{C_{\text{P}}}\left[T\left(\frac{\partial V}{\partial T}\right)_{P}-V\right]dP
    \label{eq:JT-temperature-change}
\end{equation}
Since $\text{d}H=T\text{d}S+V\text{d}P=0$, the entropy change is $\Delta S=-\int_{P_{1}}^{P_{2}}\frac{V}{T}\text{d}P$. Thus, cooling of the gas at constant enthalpy when the pressure is reduced is only obtained if $\mu_{\text{JT}}>0$ \cite{sciverHeliumCryogenics2012}. Also note that $\mu_{\text{JT}}=0$ for an ideal gas. Thus, a more accurate equation of state must be used to model a JT expansion. In this analysis, I will use the empirical equation of state for $^{4}\text{He}$ from CoolProp/REFPROP (NIST) \cite{bellPurePseudopureFluid2014CoolProp,lemmon2010nistREFPROP}. 

\begin{figure}[!!h]
\centering
\includegraphics[width=\textwidth]{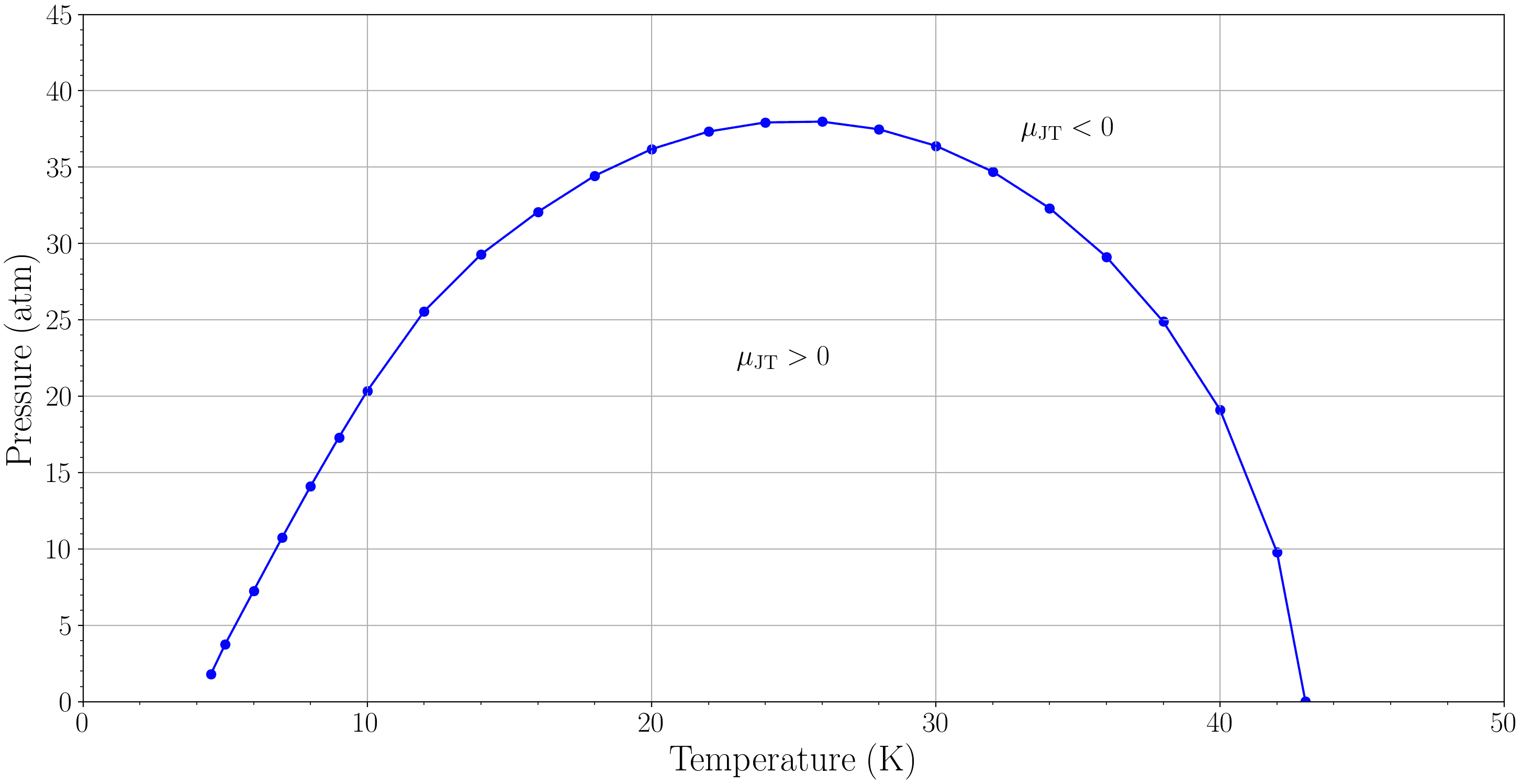}
\caption[$^{4}\text{He}$ JT Inversion Curve in $PT\text{-space}$]{JT inversion curve for $^{4}\text{He}$ in $PT$-space. The curve indicates the temperatures and pressures where $\mu_\text{JT}=0$. The equation of state is taken from Coolprop/REFPROP (NIST) \cite{bellPurePseudopureFluid2014CoolProp,lemmon2010nistREFPROP}. }
\label{fig:Helium4-JT-inversion-curve}
\end{figure}

The inversion curve in $PT$ space showing for what temperatures and pressures $^{4}\text{He}$ can cool via JT expansion shown in Figure \ref{fig:Helium4-JT-inversion-curve}. The maximum inversion temperature for 110 psig helium is $\sim42\text{ K}$ \cite{bellPurePseudopureFluid2014CoolProp,lemmon2010nistREFPROP}. Thus helium must be precooled to at least this temperature which explains the need for the GM cryocooler. 

\begin{figure}[!!h]
\centering
\includegraphics[width=\textwidth]{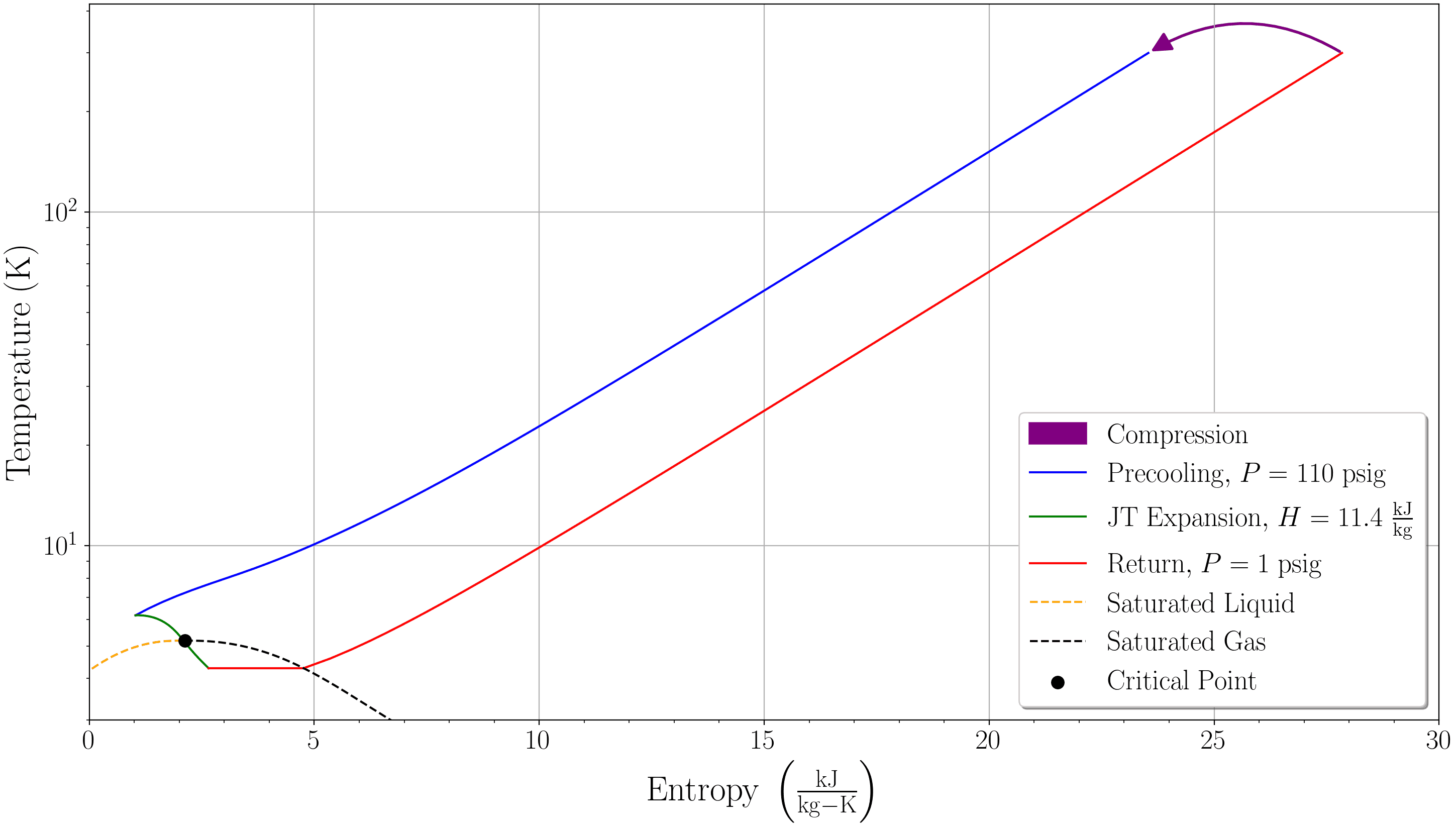}
\caption[ColdEdge Stinger JT Cooling Cycle for $^{4}\text{He}$ in $TS\text{-space}$]{Stinger JT cooling cycle for $^{4}\text{He}$ in $TS$-space, using the CoolProp/REFPROP equation of state \cite{bellPurePseudopureFluid2014CoolProp,lemmon2010nistREFPROP}. Cooling cycle: Compression to 110 psig, 110 psig isobaric cooling, JT expansion at constant enthalpy, 1 psig isobaric return. Not, the entropy does not actually zero, this is just due to the defined entropy reference point in CoolProp/REFPROP.}
\label{fig:Helium4-JT-cooling-cycle-in-TS-space}
\end{figure}

I will now summarize the thermodynamic cycle of the Stinger recirculating helium loop as shown in Figure \ref{fig:Helium4-JT-cooling-cycle-in-TS-space}. First, helium is pressurized to 110 psig indicated by the purple arrow. Next, helium is precooled at a constant pressure of 110 psig via the GM cryocooler and counterflow heat exchanger with the return helium to 6.185 K, indicated by the blue isobar. Next, the helium undergoes a JT expansion at a constant enthalpy of $11.4\,\frac{\text{kJ}}{\text{kg}}$ to 1 psig reaching a temperature of 4.295 K where a fraction of the gas is liquefied indicated by the green arrow. Finally, the helium is returned at a constant pressure of 1 psig to the compressor at which point it is recompressed. 

The JT expansion at constant enthalpy can be written as a weighted sum of the saturated liquid enthalpy, $H_{\text{l}}(T_{2},P_{2})$, and the saturated gas enthalpy, $H_{\text{g}}(T_{2},P_{2})$,

\begin{equation}
    H(T_{1},P_{1})=H(T_{2},P_{2})=yH_{\text{l}}(T_{2},P_{2})+(1-y)H_{\text{g}}(T_{2},P_{2})
    \label{eq:JT-expansion-constant-enthalpy-liquefaction}
\end{equation}
where the liquefaction yield is,
\begin{equation}
    y=\frac{H_{\text{g}}(T_{2},P_{2})-H(T_{1},P_{1})}{H_{\text{g}}(T_{2},P_{2})-H_{\text{l}}(T_{2},P_{2})}
    \label{eq:JT-expansion-liquefaction-yield}
\end{equation}
and the denominator is the latent heat of vaporization $\Delta H_{\text{vap}}=H_{\text{g}}(T_{2},P_{2})-H_{\text{l}}(T_{2},P_{2})$ \cite{sciverHeliumCryogenics2012}. As the input temperature and pressure to the JT capillary is known, we can calculate the enthalpy for the JT expansion, $H(6.185\text{\,K},110\text{\,psig})=11.4\,\frac{\text{kJ}}{\text{kg}}$. Since we know the output pressure of the JT capillary, we can solve for the temperature after the capillary, $T_{2}=4.295\,\text{K}$, assuming constant enthalpy  $H(4.295\text{\,K},1\text{\,psig})=11.4\,\frac{\text{kJ}}{\text{kg}}$. We can now use the saturated liquid enthalpy, $H_{\text{l}}(6.185\text{\,K},1\text{\,psig})=0.382\,\frac{\text{kJ}}{\text{kg}}$, and saturated gas enthalpy, $H_{\text{g}}(6.185\text{\,K},1\text{\,psig})=20.5\,\frac{\text{kJ}}{\text{kg}}$, to get a liquefaction yield of $y=0.451$ using Equation \eqref{eq:JT-expansion-liquefaction-yield}. 

I will now estimate the cooling capacity of the cryostat cold stage heat exchangers. I'll assume that all the pressure is dropped during the JT expansion such that the helium is at a constant pressure of $1\text{ psig}$ through each heat exchanger (this is a reasonable assumption). For the 4 K stage heat exchanger, I'll take the input temperature of the helium to be $T_{\text{He}}=4.295\text{ K}$ and the temperature of the 4 K stage to be $4.936\text{ K}$ (base temperature under these conditions). I'll then compute the heat transferred to the helium along heat exchanger length to determine the temperature of the helium just as it exits the heat exchanger. Dividing the heat exchanger into 1000 segments, I compute the heat transferred to the helium at each segment and add it to the enthalpy, recompute the gas quality and corresponding new temperature. This encapsulates the two-phase mixture and the vaporization of the liquid helium.

The differential equation to solve is given as follows,
\begin{equation}
    \frac{\text{d}H}{\text{d}z}=\frac{G^{\prime}(z)}{\dot{m}}(T_{\text{stage}}-T_{\text{He}}(z))
    \label{eq:HX-enthalpy-gain-method}
\end{equation}
where $G'(z)$ is the conductance per unit length that encapsulates the forced-convection heat transfer from the helium to the helium tube sidewall $dG_{\text{conv}}=h_{n}\pi D_{\text{i}}dz$ where $h_{n}$ is the forced convection heat transfer coefficient computed at each segment $n$, the conduction heat transfer through the tube to the heat exchanger $dG_{\text{tube}}=\frac{2\pi k_{\text{Cu}}dz}{\ln(D_{\text{o}}/D_{\text{i}})}$, and the conduction from the heat exchanger to the annular plate $G_{\text{annulus}}=\frac{2\pi k_{\text{Cu}}t}{\ln(r_{\text{o}}/r_{\text{i}})}$. At each segment, the new enthalpy and gas quality $x=1-y$ is computed using Equation \eqref{eq:JT-expansion-liquefaction-yield} to determine whether or not the helium is a two-phase mixture or just a gas. This is important as it changes the forced-convection heat transfer coefficient which is computed as follows.

\begin{figure}[!!h]
\centering
\begin{subfigure}{0.49\textwidth}
    \includegraphics[width=\textwidth]{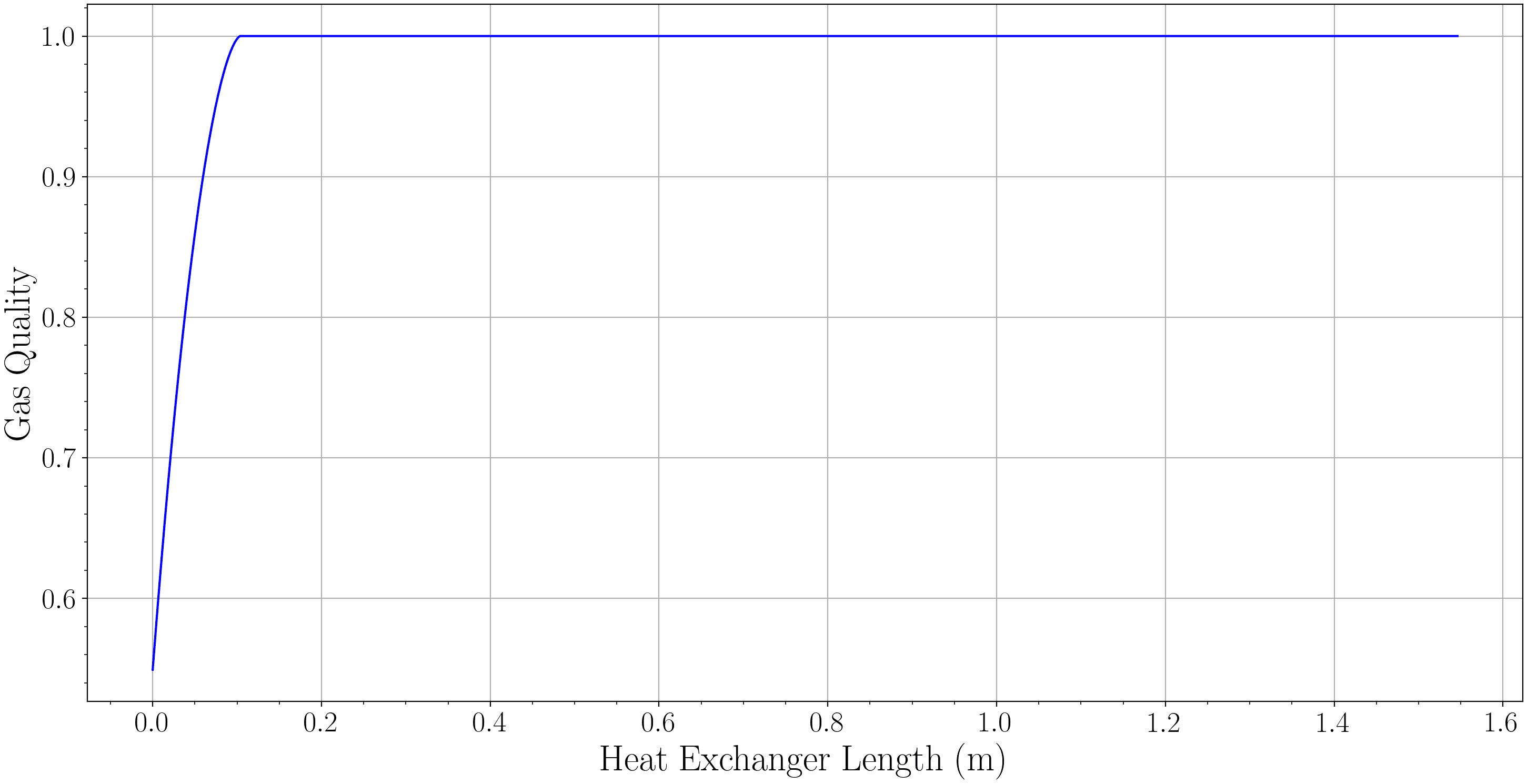}
\end{subfigure}
\begin{subfigure}{0.49\textwidth}
    \includegraphics[width=\textwidth]{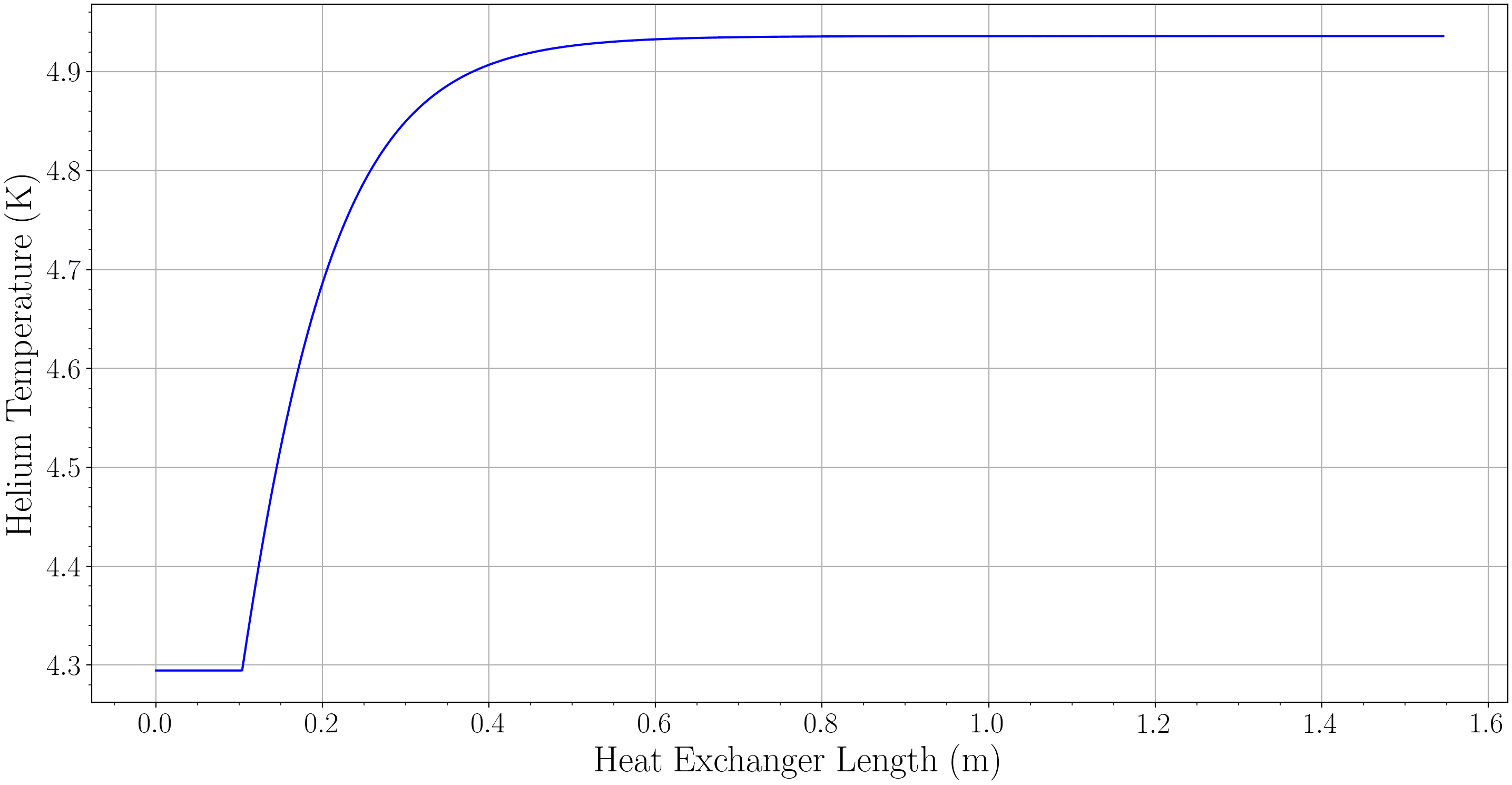}
\end{subfigure}
\caption[4 K Stage HX Model]{4 K stage heat exchanger model.}
\label{fig:4K-HX-analysis}
\end{figure}

\begin{figure}[!!h]
\centering
\begin{subfigure}{0.49\textwidth}
    \includegraphics[width=\textwidth]{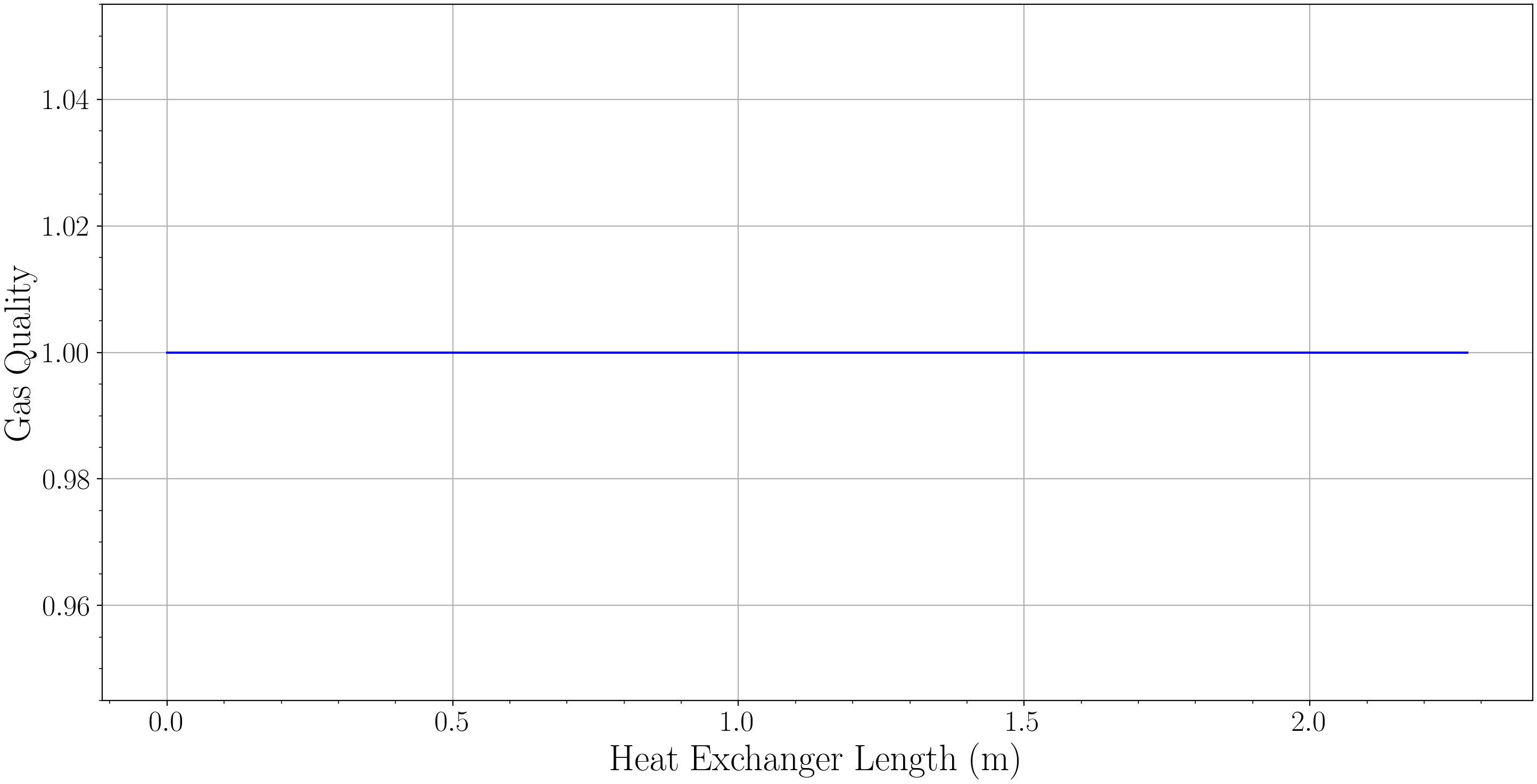}
\end{subfigure}
\begin{subfigure}{0.49\textwidth}
    \includegraphics[width=\textwidth]{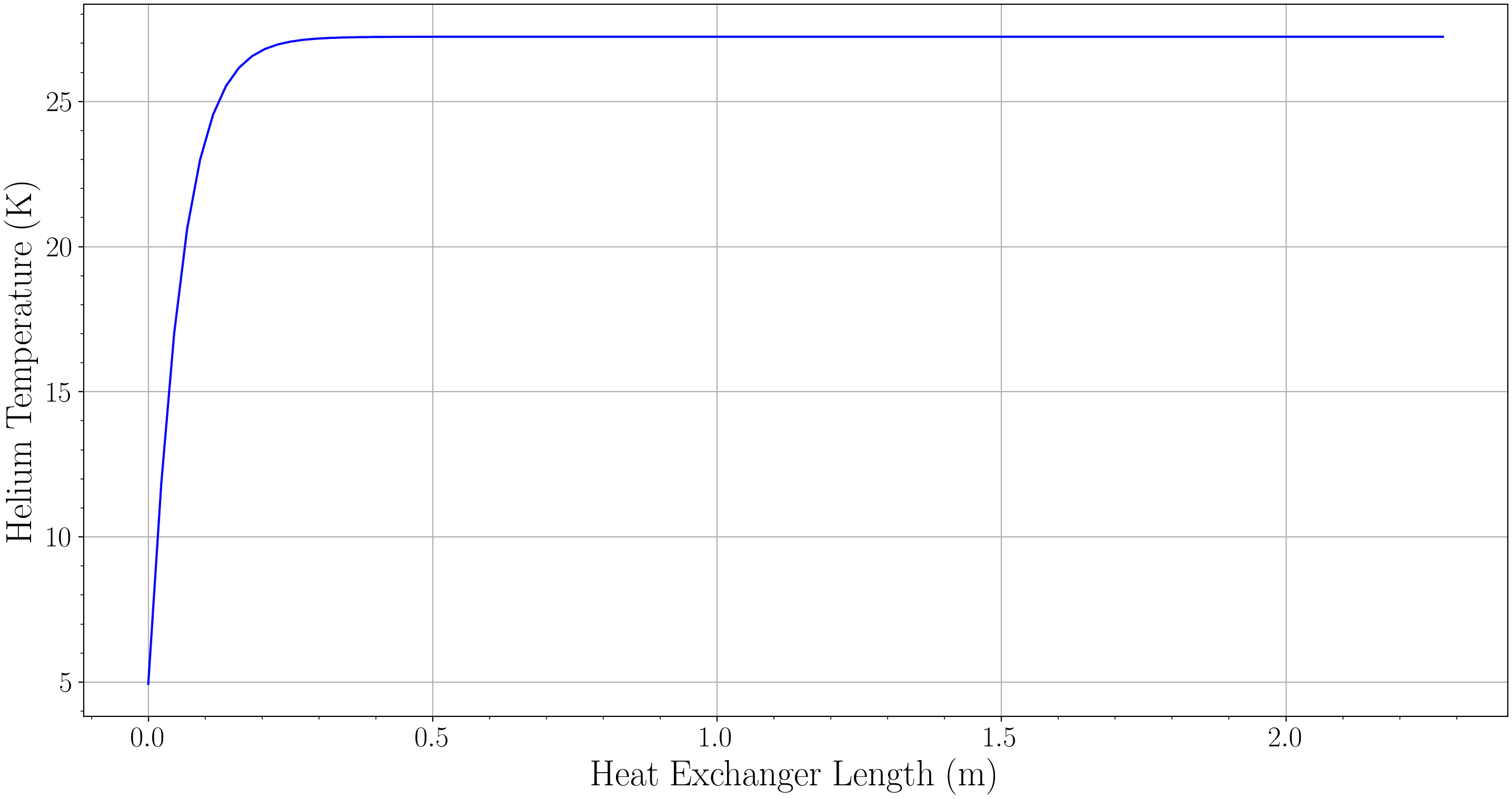}
\end{subfigure}
\caption[40 K Stage HX Model]{40 K stage heat exchanger model.}
\label{fig:40K-HX-analysis}
\end{figure}

First, the Reynolds number is computed for a fluid or gas using,
\begin{equation}
    \text{Re}=\frac{\rho_{\text{f}}v_{\text{f}}D_{\text{i}}}{\mu_{\text{f}}}=\frac{4\dot{m}_{\text{f}}}{\pi D_{\text{i}}\mu_{\text{f}}}
    \label{eq:Reynolds-number}
\end{equation}
Then the Prandtl number is computed for a liquid or gas using,
\begin{equation}
    \text{Pr}=\frac{\mu_{\text{f}}C_{\text{P,f}}}{k_{\text{f}}}
    \label{eq:Prandtl-number}
\end{equation}
The Nusselt number for gaseous helium is
\begin{equation}
    \text{Nu}=0.023\text{Re}^{4/5}\text{Pr}^{2/5}
    \label{eq:Nusselt-number-gaseous-helium}
\end{equation}
and Nusselt number for two-phase helium can be computed using
\begin{equation}
    \text{Nu}=0.023\text{Re}_{\text{l}}^{0.8}\text{Pr}_{\text{l}}^{0.4}(1-x)^{0.8}5.4\left[\left(\frac{\rho_{\text{g}}}{\rho_{\text{l}}}\right)^{0.6}\left(\frac{\mu_{\text{l}}}{\mu_{\text{g}}}\right)^{0.1}\left(\frac{1-x}{x}\right)\right]^{-0.385}
    \label{eq:Nusselt-number-two-phase-helium}
\end{equation}
Finally, the forced convection heat transfer coefficient can be computed at each segment to be,
\begin{equation}
    h_{\text{conv}}=\frac{\text{Nu}k_{\text{f}}}{D_{\text{i}}}
    \label{eq:forced-convection-heat-transfer-coefficient}
\end{equation}
where if the helium is a two-phase mixture $k_{\text{f}}=xk_{\text{g}}+yk_{\text{l}}$ and if it's a gas, then $k_{\text{f}}=k_{\text{g}}$ \cite{sciverHeliumCryogenics2012}. Once, the the final temperature of the helium is determined after each stage, the cooling power can be computed as follows,
\begin{equation}
    \dot{Q}_{\text{cool}}=\dot{m}\left(H_{\text{f}}(T_{\text{f}},1\,\text{psig})-H_{\text{i}}(T_{\text{i}},1\,\text{psig})\right)
    \label{eq:HX-cooling-power}
\end{equation}
which is the enthalpy gained by the helium via the heat exchanger times the mass flow rate, $\dot{m}=51\times10^{-6}\,\frac{\text{kg}}{\text{s}}$ \cite{sciverHeliumCryogenics2012}. The simulated cooling power of the 4 K stage compared with the measured cooling power is summarize in Table \ref{tab:4K-HX-cooling-power} and a plot showing the temperature and gas quality of helium through the heat exchanger for the 4 K stage temperature of 4.936 K is shown in Figure \ref{fig:4K-HX-analysis}. Similarly, the simulated cooling power of the 40 K stage compared with the measured cooling power is summarize in Table \ref{tab:40K-HX-cooling-power} and a plot showing the temperature and gas quality of helium through the heat exchanger for the 40 K stage temperature of 27.229 K is shown in Figure \ref{fig:40K-HX-analysis}. 

\begin{table}[!!h]
\centering
\tablefont
\caption[Simulated Cooling power of 4 K stage HX]{I measured a cooling power of 0.5 W at 6.1 K for the 4 K stage. Taking the 4 K stage temperature to be 6.1 K, I simulate a 0.4 W cooling power relative to the base temperature, 4.9 K.}
\label{tab:4K-HX-cooling-power}
\begin{tabularx}{\textwidth}{>{\centering\arraybackslash}p{0.49\textwidth}>{\centering\arraybackslash}p{0.44\textwidth}}
\toprule
\textbf{Temperature (K)} & \textbf{Cooling Power (W)} \\
\midrule
4.9 & 0.7 \\
\midrule
6.1 & 1.1 \\
\bottomrule
\end{tabularx}
\end{table}

\begin{table}[!!h]
\centering
\tablefont
\caption[Simulated Cooling power of 40 K stage HX]{I measured a cooling power of 5 W at 45 K for the 40 K stage. Taking the 40 K stage temperature to be 45 K, I simulate a 4.7 W cooling power relative to the base temperature, 27.2 K.}
\label{tab:40K-HX-cooling-power}
\begin{tabularx}{\textwidth}{>{\centering\arraybackslash}p{0.49\textwidth}>{\centering\arraybackslash}p{0.44\textwidth}}
\toprule
\textbf{Temperature (K)} & \textbf{Cooling Power (W)} \\
\midrule
27.2 & 6.1 \\
\midrule
45 & 10.8 \\
\bottomrule
\end{tabularx}
\end{table}

\section{Helium Purification}
\label{sec:helium-purification}

The Stinger is a closed-cycle flow cryostat that relies on JT cooling of helium via a small capillary. Impurities in the helium precipitate in solid form and accumulate in the small capillary, causing a clog and force system warmup. This motivates an engineering solution to purify helium and lab best practices to prevent contamination.

The main impurities in helium from an AirGas cylinder are $\text{N}_{2}$, $\text{O}_{2}$, $\text{H}_{2}\text{O}$, $\text{CO}_{2}$, $\text{CH}_4$ and $\text{CO}$. Although it's not specced, trace amounts of $\text{H}_{2}$ have also been measured in helium gas delivery systems \cite{gabalHydrogenFreeLiquidHeliumRecovery2016}. In addition, hydrogen contamination can also be introduced from oil degradation in the helium compressors and from permeation out of the stainless steel helium lines. Hydrogen is particularly challenging to filter as its cryocondensation temperature is 3 K, thus its adsorption potentials are very weak at temperatures above 33 K - hydrogen's critical temperature. Consequently, the best way to filter hydrogen above 33 K is via catalytic conversion where hydrogen is oxidized creating water \cite{gabalHydrogenFreeLiquidHeliumRecovery2016}. Molecular sieves or liquid nitrogen traps can then be placed after this hydrogen filter to easily filter out water. 

I have determined empirically that hydrogen is probably contaminating the recirculating line. After multiple weeks of steady uptime, I measured a clog in the system via a decrease in mass flow. I then heated the system up to 20 K - hydrogen's boiling point - and was able to relieve the clog. 

\begin{figure}[!!h]
\centering
\includegraphics[width=\textwidth]{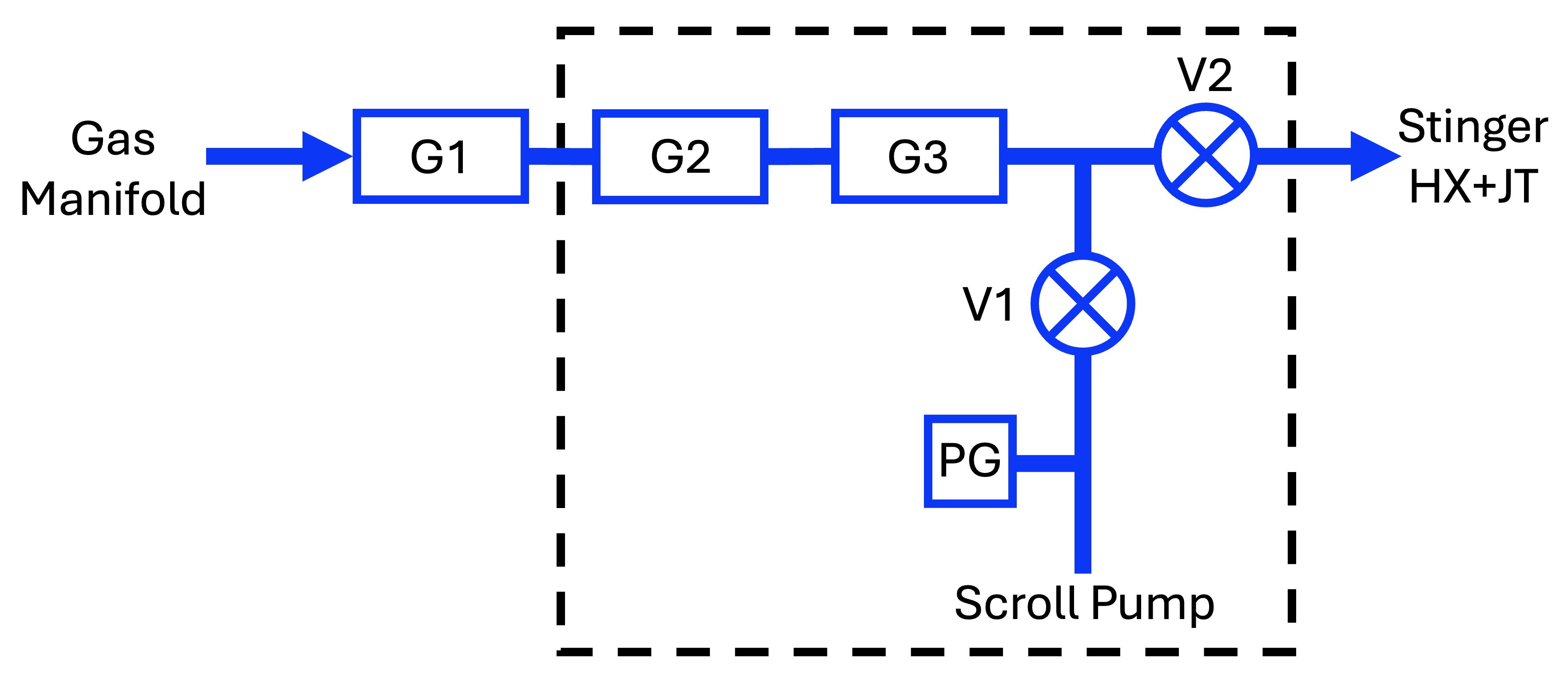}
\caption[Helium Purification Circuit]{Schematic of helium purification circuit installed in series with the recirculating loop between the gas manifold and Stinger heat exchangers (HX) and JT capillary. The dashed box indicates the part of the circuit that I built. Each part was connected with Swagelok-1/2 and Aeroquip-8 fittings all made of 316 stainless steel. The parts list is given in Table \ref{tab:part-list-for-helium-purification-circuit}.}
\label{fig:helium-purification-circuit}
\end{figure}

After many discussions with engineers at CET and Sumitomo, I built a helium purification circuit that consists of a charcoal getter heat sunk at 77 K using liquid nitrogen in series with a molecular sieve and helium purifier as shown in Figure \ref{fig:helium-purification-circuit}. The charcoal getter and helium purifier was provided by CET and the molecular sieve was provided by Sumitomo. I built the device using swagelok and aeroquip connectors and is placed in series with the recirculating helium loop between the gas manifold and Stinger precooling heat exchangers. It is rigidly mounted to an 8020 structure with permanent heater tape installed allowing in situ regeneration of the filters via a 100 C bakeout while evacuating with a scroll pump. Implementing this circuit increased the system uptime from 30 days to 72 days. 

The charcoal getter is heat sunk at 77 K - provided by a dewar filled with liquid nitrogen - and is efficient at cryocondensing water and carbon dioxide and cryosorption pumping of methane. However, cryosorption pumping of carbon monoxide, nitrogen, and oxygen are less efficient. The molecular sieve is designed to be efficient at sorption pumping of water at 300 K - as told by Sumitomo. The helium purifier is specced to efficiently filter carbon dioxide, carbon monoxide, methane, water, and oxygen. Although, the helium purifier is not specced to filter hydrogen, the Safety Data Sheet (SDS) lists 10-20\% copper (II) oxide, 1-5\% tricobalt tetraoxide, and 1-5\% nickel oxide which can all form REDOX reactions with hydrogen as shown in Equation \eqref{eq:REDOX-reactions-with-hydrogen} - hydrogen is oxidized and forms water. I can't be sure without gas chromatography, but this might be the reason for the increased system uptime.

\begin{equation}
\begin{aligned}
    \text{H}_{2}+\text{CuO} & \rightarrow \text{Cu}+\text{H}_{2}\text{O} \\
    \text{H}_{2}+\text{NiO} & \rightarrow \text{Ni}+\text{H}_{2}\text{O} \\
    \text{H}_{2}+\text{Co}_{3}\text{O}_{4} & \rightarrow 3\text{CoO}+\text{H}_{2}\text{O}
\end{aligned}
\label{eq:REDOX-reactions-with-hydrogen}
\end{equation}

\begin{table}[!!h]
\centering
\tablefont
\caption[Parts List for Helium Purification Circuit]{Parts list for helium purification circuit in Figure \ref{fig:helium-purification-circuit}.}
\label{tab:part-list-for-helium-purification-circuit}
\begin{tabularx}{\textwidth}{>{\centering\arraybackslash}p{0.49\textwidth}>{\centering\arraybackslash}p{0.44\textwidth}}
\toprule
\textbf{Schematic Label} & \textbf{Part} \\
\midrule
G1 & \makecell{Getter \\Liquid Nitrogen Trap \\(manufactured by CET)} \\
\midrule
G2 & \makecell{Getter \\Molecular Sieve Filter \\Sumitomo, PN: SET\_00088} \\
\midrule
G3 & \makecell{Getter \\Helium Purifier \\Supelco, PN: 27601-U} \\
\midrule
V1 & \makecell{Needle Valve \\Hy-Lok, PN: NV4-H-8T-S316} \\
\midrule
V2 & \makecell{Toggle Valve \\Hy-Lok, PN: TG3-H-8T-S316} \\
\midrule
PG & \makecell{Pressure Gauge \\InstruTech, PN: CVM211GBL} \\
\bottomrule
\end{tabularx}
\end{table}

\section{Standard Operating Procedure}
\label{sec:standard-operating-procedure}
In this section, I'll present the standard operating procedure (SOP) I used to ensure optimal performance of the Stinger system. 

\subsection{Evacuate Recirculating Loop}
\label{subsec:evacuate-recirculating-loop-SOP}

\noindent\textbf{At the gas manifold:} \ref{fig:Gas-Manifold}
\begin{enumerate}
    \item Turn on scroll pump.
    \item Turn 4-way valve to point in the $-\hat{y}$ direction.
    \item Close recirculating compressor supply and return toggle valves.
    \item Open scroll pump needle valve.
\end{enumerate}
\noindent\textbf{At the helium purification circuit:} \ref{fig:helium-purification-circuit}
\begin{enumerate}
    \item Open valves 1 and 2.
    \item Turn on heaters at getters 1, 2, and 3. Bakeout temperature is $100\text{ C}$. Monitor pressure of recirculating loop using gauge.
\end{enumerate}

\subsection{Cooldown}
\label{subsec:cooldown-SOP}
Start these steps after evacuating the recirculating helium loop in \ref{subsec:evacuate-recirculating-loop-SOP} for $\sim 12$ hours. Cryostat vacuum space should also be evacuated to $<1\times10^{-4}$ Torr. Turn on overpressure vacuum space scroll pump (Figure \ref{fig:cryostat-overpressure-vacuum-space}) only after cryostat has been evacuated.

\noindent\textbf{At the helium purification circuit:} \ref{fig:helium-purification-circuit}
\begin{enumerate}
    \item Close valve 1. Keep valve 2 open.
    \item Turn off heaters at getters 1, 2, and 3.
    \item Disconnect getter 1 from scroll pump and submerge in LN2 dewar.
\end{enumerate}

\noindent\textbf{At the compressor closet:} \ref{fig:Schematic-of-compressor-water-cooling-circuit}
\begin{enumerate}
    \item Turn on HX.
    \item Turn on recirculating compressor (either C1 or C3).
\end{enumerate}

\noindent\textbf{At the gas manifold:}\ref{fig:Gas-Manifold}
\begin{enumerate}
    \item Turn 4-way valve to point in the $\hat{x}$ direction. Keep scroll pump needle valve open.
    \item Purge helium cylinder line at $\sim50$ psig.
    \item Reconnect helium cylinder line to manifold and increase pressure to $100-110$ psig.
    \item Open helium cylinder needle valve and monitor the pressure increase using gauge.
    \item You should hear the scroll pump make a more throaty noise. This indicates the JT capillary is not clogged. If you don't hear this, evacuate helium and return to  procedure \ref{subsec:evacuate-recirculating-loop-SOP}. 
    \item Close helium cylinder and scroll pump needle valves. Turn off scroll pump and close valve at helium cylinder.
    \item Turn 4-way valve to point in the $-\hat{x}$ direction.
    \item Disconnect recirculating compressor flexible helium line at the supply and connect it to getter 1 submerged in LN2. Connect the other end of getter 1 to the supply.
    \item Open recirculating compressor return toggle valve and wait $\sim30$ s.
    \item Open recirculating compressor supply toggle valve. 
    \item Mass flow meter should display $0.5-1$ SLPM (nitrogen calibration).
    \item Turn on helium interface heater (Figure \ref{fig:helium-interface-heater}).
\end{enumerate}

\noindent\textbf{At the compressor closet:} \ref{fig:Schematic-of-compressor-water-cooling-circuit}
\begin{enumerate}
    \item Turn on GM cryocooler compressor (either C2 or C4).
\end{enumerate}

Flow should increase throughout the cooldown as in Figure \ref{fig:cryostat-helium-mass-flow-vs-time-during-cooldown}. If flow increases faster, the Stinger bayonet is not seated correctly causing the helium to bypass the JT capillary. If flow plateaus at a lower value or starts decreasing, the JT capillary is clogged.  After a successful cooldown or before the LN2 in the dewar boils off, proceed as follows:

\noindent\textbf{At the gas manifold:}\ref{fig:Gas-Manifold}
\begin{enumerate}
    \item Close recirculating compressor supply toggle valve. 
    \item Disconnect getter 1 from supply. Reconnect recirculating compressor flexible helium line to supply.
    \item Open recirculating compressor supply toggle valve.
\end{enumerate}

\subsection{Warmup}
\label{subsec:warmup-SOP}

\noindent\textbf{At the gas manifold:}\ref{fig:Gas-Manifold}
\begin{enumerate}
    \item Turn on scroll pump.
    \item Close recirculating compressor supply toggle valve. 
    \item Turn 4-way valve to point in the $-\hat{y}$ direction. Monitor helium pressure decrease to $\sim1$ psig using gauge.
    \item Close recirculating compressor return toggle valve.
    \item Incrementally open scroll pump needle valve to evacuate remaining helium and impurities. You should hear the scroll pump make a more throaty noise. Continue incrementally opening the scroll pump needle valve to prevent damaging the scroll pump.
    \item Turn off helium interface heater.
\end{enumerate}

\noindent\textbf{At the compressor closet:} \ref{fig:Schematic-of-compressor-water-cooling-circuit}
\begin{enumerate}
    \item Turn off GM cryocooler compressor (either C2 or C4).
    \item Turn off recirculating compressor (either C1 or C3).
\end{enumerate}

It is recommended to the turn the turbo pump on. When the cold stages reach $\sim120$ K, open the cryostat vacuum valve to pump on the cryostat vacuum space. See Section \ref{subsec:warmup} for actively heating the cryostat cold stages during warmup. Once the cryostat cold stages have successfully warmed up to $300$ K, turn off overpressure vacuum space scroll pump.

\subsection{Disconnect Stinger Bayonet}
\label{subsec:disconnect-stinger-bayonet}
The recirculating line is evacuated at this point following procedure \ref{subsec:evacuate-recirculating-loop-SOP}.

\noindent\textbf{At the gas manifold:}\ref{fig:Gas-Manifold}
\begin{enumerate}
    \item Turn 4-way valve to point in the $\hat{x}$ direction.
\end{enumerate}

\noindent\textbf{At the helium purification circuit:} \ref{fig:helium-purification-circuit}
\begin{enumerate}
    \item Close valves 1 and 2.
\end{enumerate}

The Stinger bayonet can now be removed. This procedure prevents venting the getters in Figure \ref{fig:helium-purification-circuit} to atmosphere.

\subsection{Connect Stinger Bayonet}
\label{subsec:connect-stinger-bayonet}
Ensure the procedure in \ref{subsec:disconnect-stinger-bayonet} has been followed. Then connect the Stinger bayonet.

\noindent\textbf{At the gas manifold:}\ref{fig:Gas-Manifold}
\begin{enumerate}
    \item Turn on scroll pump. Open scroll pump needle valve.
    \item Turn 4-way valve to point in the $-\hat{y}$ direction.
\end{enumerate}

\noindent\textbf{At the helium purification circuit:} \ref{fig:helium-purification-circuit}
\begin{enumerate}
    \item Open valves 1 and 2.
\end{enumerate}

\ifSubfilesClassLoaded{%
  \bibliographystyle{unsrt}
  \bibliography{Dissertation}
}{}

\end{document}

%% file: Chapter3.tex
\chapter{Cryostat}
\label{chap:Cryostat}

\begin{figure}[!!h]
\centering
\includegraphics[width=\textwidth]{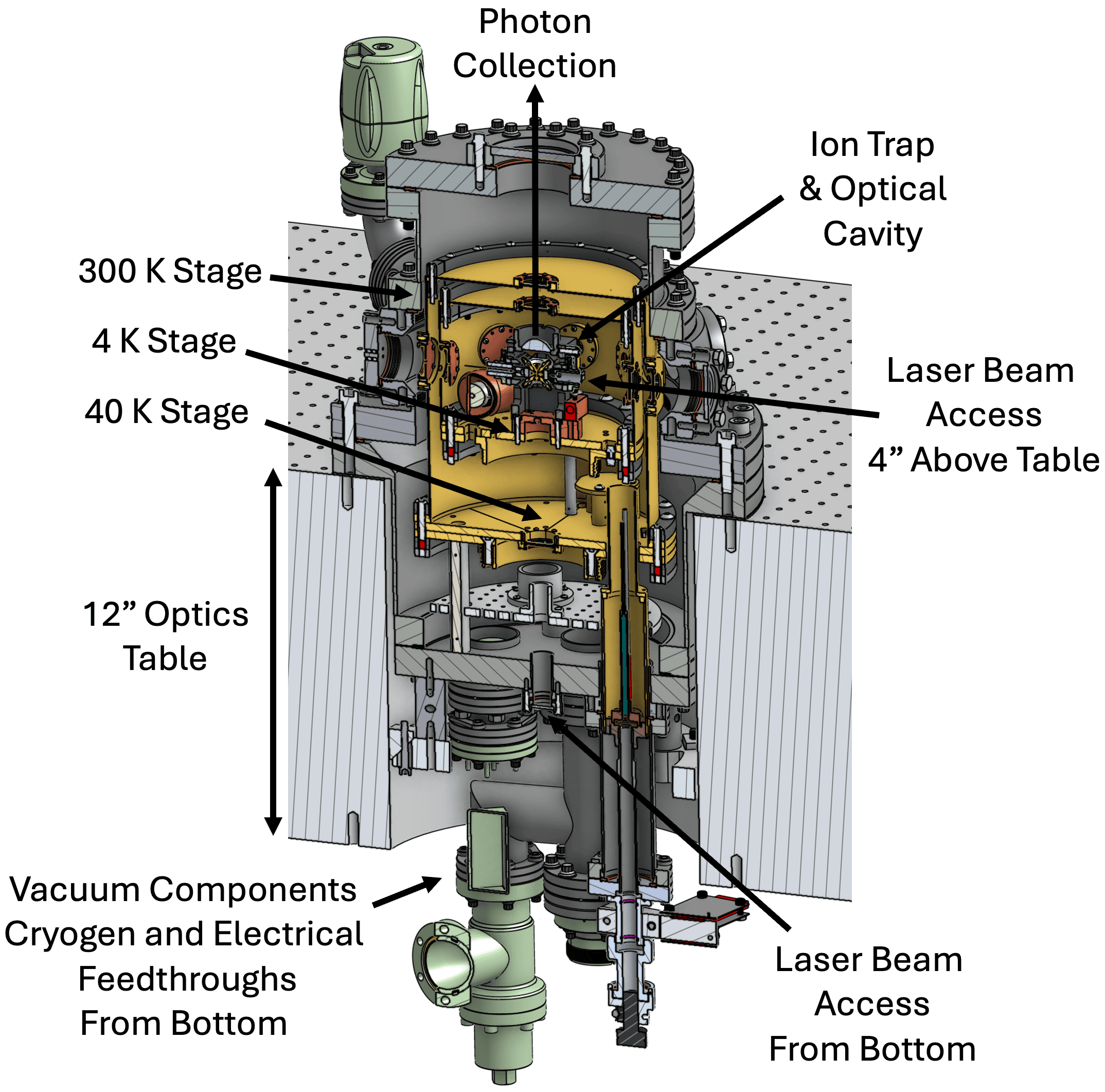}
\caption[Cryostat in Optics Table]{CAD drawing of cryostat, cryo0, in optics table.}
\label{fig:Cryostat-in-optics-table}
\end{figure}

This chapter presents the design and construction of two cryostats developed during this work. Both systems are composed of three identical vacuum continuous regions: a 300 K stage, a 40 K stage, and a 4 K stage. Each cryostat is designed to mate with and protrude through the bore of a 12 inch optics table with an ion trap situated at the 4 K stage. This design allows UHV vacuum components, electrical feedthroughs, and cryogen feedthroughs to interface the cryostat at the 300 K stage from beneath the table, maximizing the available optical access on the top surface. This configuration enables laser access both from beneath the table and at a height of 4 inches above the table surface, while allowing for photon collection from above. The primary focus of this thesis involved the development of cryo0 (formerly Frosty), which houses an ion trap and optical cavity at the 4 K stage as shown in Figure \ref{fig:Cryostat-in-optics-table}. A second system, cryo1 (formerly Crystal), was built and commissioned to verify the reproducibility of the design, but it was not used for the ion trapping experiments conducted in this work. Detailed discussions regarding the mechanical and thermal design, components, and wiring of the stages are provided in Sections \ref{sec:300K-stage}, \ref{sec:4K-and-40K-stages}, and \ref{sec:heat-load-budget}. Information about materials, connectors, and machine shops used in this work are covered in Appendix \ref{chap:manufacturing}.

 \section{300 K Stage}
 \label{sec:300K-stage}

 The purpose of the 300 K stage is to provide a vacuum enclosure for the cryostat and to provide a central interface for vacuum components, electrical feedthroughs, and cryogen feedthroughs. The sub-regions of the 300 K stage are presented in this section. 

 \subsection{300 K CF-10 Flange}
 \label{subsec:300K-F-10-Flange}

\begin{figure}[!!h]
\centering
\includegraphics[width=\textwidth,height=0.45\textheight,keepaspectratio]{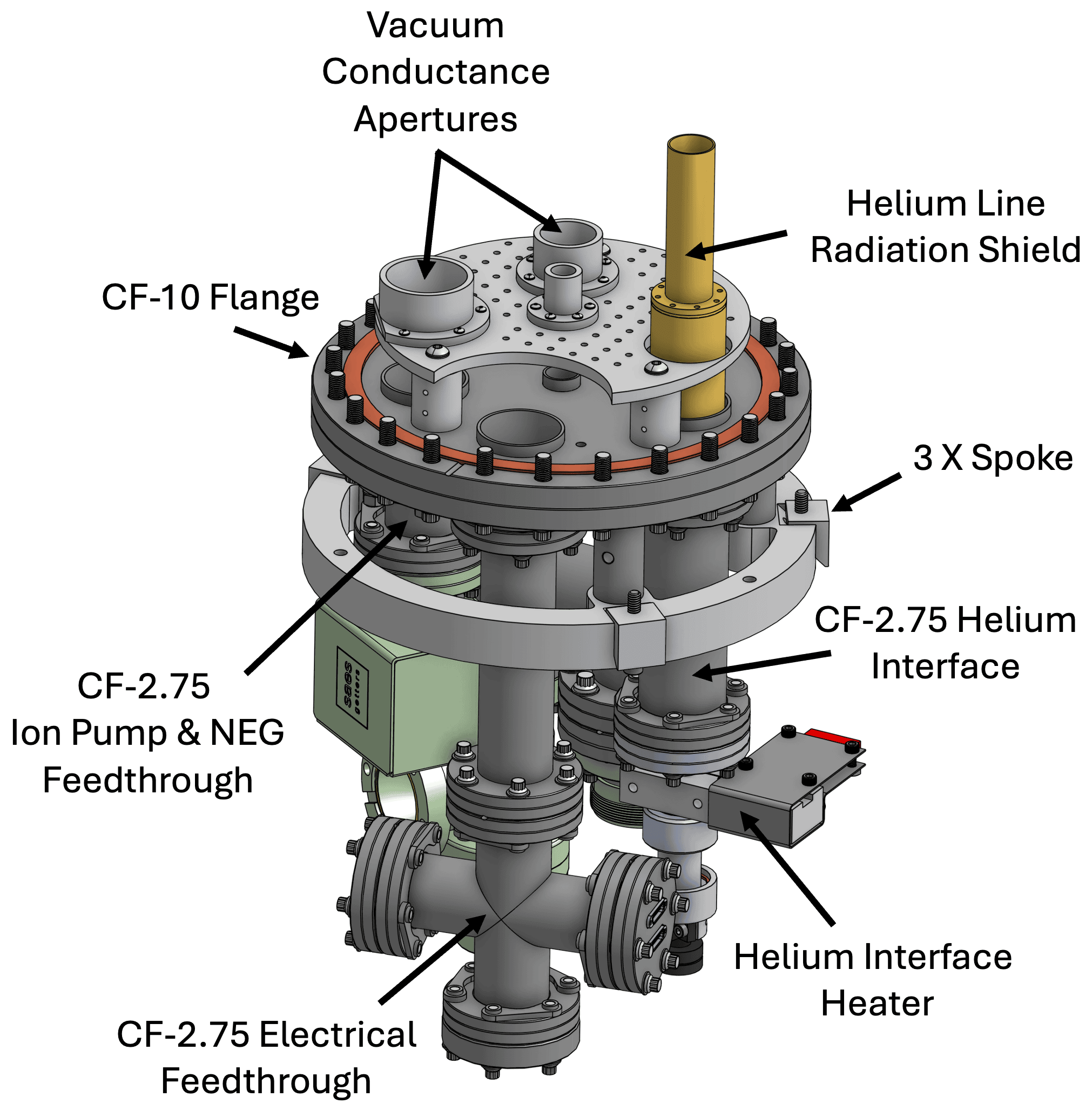}
\newline
\includegraphics[width=\textwidth,height=0.45\textheight,keepaspectratio]{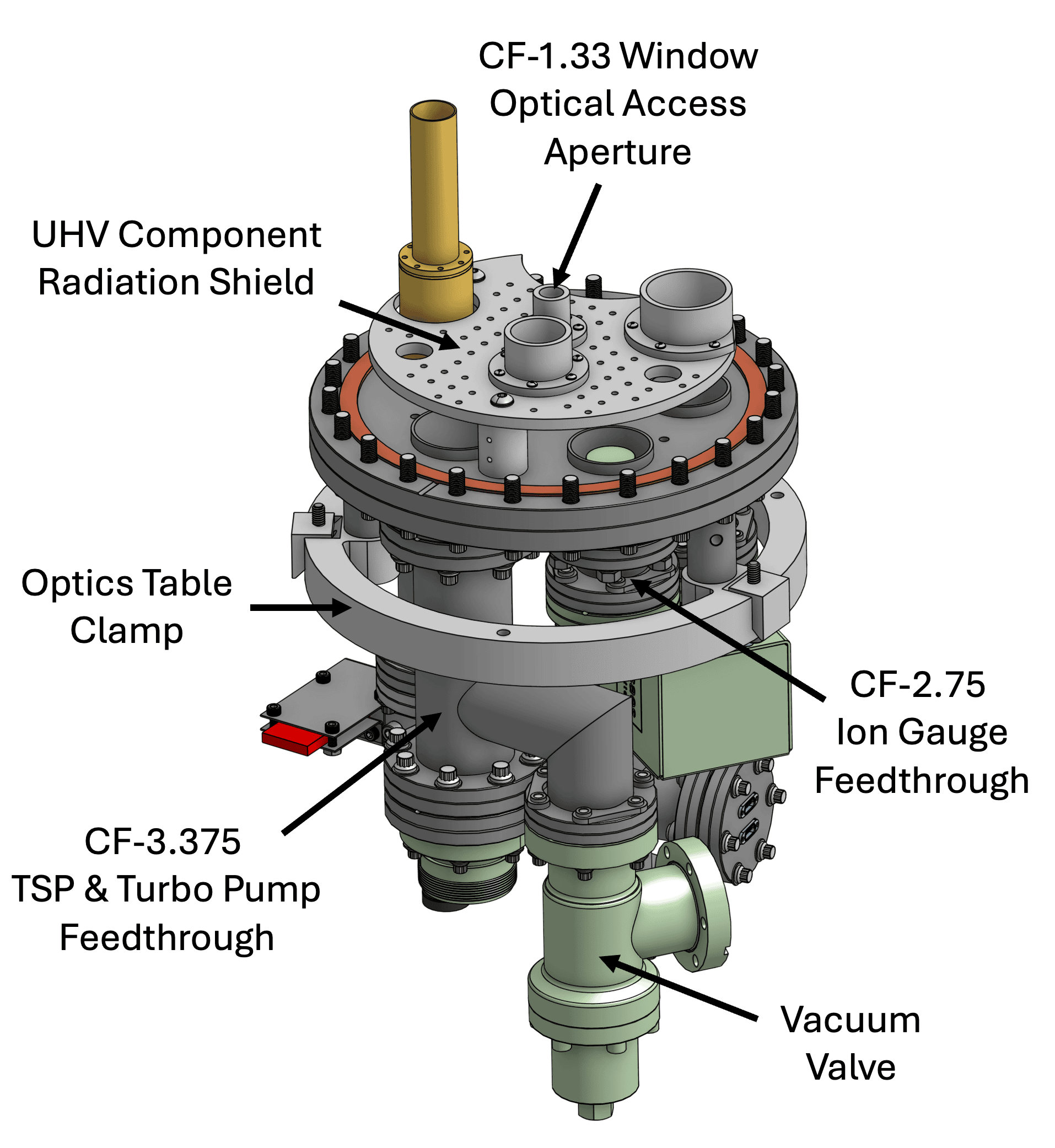}
\caption[300 K CF-10 Flange Assembly]{CAD drawings of 300 K CF-10 flange. A parts list in summarized in Table \ref{tab:part-list-for-300K-CF10-flange-assembly}.}
\label{fig:300K-CF10-flange-assembly}
\end{figure}

To maximize optical access on the top of the optics table, the UHV vacuum components, electrical feedthroughs, and helium recirculating loop interface enter the cryostat at a single CF-10 flange as shown in Figure \ref{fig:300K-CF10-flange-assembly}. It is desirable to use non-magnetic materials such as titanium or SS316 for all the 300 K vacuum components to minimize perturbations to ion qubits. However, it becomes impractical to manufacture as titanium and SS316 are typically non-standard material choices from vacuum component vendors, increasing the pice and lead time. As a compromise, all of the flanges, nipples, and fasteners at the 300 K CF-10 flange are made of SS304 as they are farthest away from the ion trap $(B\propto r^{-3})$ and it is cheaper to manufacture. The 300 K CF-10 flange was manufactured by machining apertures into a standard CF-10 flange and then UHV welding standard half nipples at the vacuum side of the flange. The recirculating helium loop interfaces with the cryostat at the CF-2.75 Helium Interface where the helium undergoes a Joule-Thomson (JT) expansion through a capillary restriction and subsequently cools the cryostat cold stages, see Section \ref{sec:4K-and-40K-stages}. A radiation shield encloses the in-vacuum recirculating helium line. The turbo pump interfaces with the cryostat at the vacuum valve and the conductance path from there is provided by the CF-3.375 TSP \& Turbo Pump Feedthrough. The larger CF-3.375 nipple enclosing the TSP is utilized to maximize the surface area for the titanium plating, increasing the effective TSP pumping speed. An ion pump \& NEG assembly and an ion gauge interface with the CF-10 flange at separate feedthroughs. All the DC and RF signals enter the cryostat at the CF-2.75 Electrical Feedthrough. An Al6061T6 radiation shield is mounted above the CF-10 flange to prevent direct-line-of sight radiation from the hot ion gauge. Aperture tubes are mounted on the shield to provide vacuum conductance and optical access unobstructed from the in-vacuum DC and RF wires. Additional 1/4-20 taps are machined in the radiation shield to mount other parts in the future. The cryostat is centered on the bore of the optics table by a clamp connected at the ambient side of the CF-10 flange. The clamps consist of three spokes installed in a camming mechanism that press against the sidewalls of the optics table bore and thus centers the cryostat, locks its position, and increases its stiffness. 

\begin{table}[!!h]
\centering
\tablefont
\caption[Parts List for 300 K CF-10 Flange Assembly]{Parts list for UHV vacuum components, electrical feedthroughs, and cryogenic penetrations at the 300 K CF10 flange in Figure \ref{fig:300K-CF10-flange-assembly}.}
\label{tab:part-list-for-300K-CF10-flange-assembly}
\begin{tabularx}{\textwidth}{>{\centering\arraybackslash}p{0.42\textwidth}c}
\toprule
\textbf{Schematic Label} & \textbf{Part} \\
\midrule
CF-2.75 Helium Interface & \makecell{Recirculating Helium Loop Interface (CET) \\SS304 Nipple (CET) \\Helium Interface Heater (See Subsection \ref{subsec:helium-interface-heater}) }\\
\midrule
CF-3.375 TSP \& Turbo Pump Feedthrough & \makecell{TSP (Gamma Vacuum, PN: 360819) \\Vacuum Valve (VAT, PN: 54132-GE02) \\Custom SS304 Tee (Kurt J. Lesker)} \\
\midrule
CF-2.75 Ion Pump \& NEG Feedthrough & \makecell{Ion pump \& NEG (SAES, PN: NEXTorr Z 100)\\Custom SS304 Nipple (Kurt J. Lesker)} \\
\midrule
CF-2.75 Ion Gauge Feedthrough & \makecell{Ion gauge (SRS, PN: NR-F-UHV) \\Custom SS304 Nipple (Kurt J. Lesker) }\\
\midrule
CF-2.75 Electrical Feedthrough & \makecell{DC $2\times25$ Pin Micro-D (MPF, PN: A8978-4-W) \\DC $2\times15$ pin Micro-D (MPF, PN: A8978-4-W) \\RF $4\times$  SMA (MPF, PN: A12377-4-CF) \\Standard SS304 Nipple and Cross (Kurt J. Lesker)} \\
\midrule
UHV Component Radiation Shield & \makecell{Machined Al6061T6 (Zero Hour Parts) \\Electropolished (CET)} \\
\midrule
CF-10 Flange & \makecell{Custom SS304 Flange and Half Nipples (Kurt J. Lesker)} \\
\midrule
CF-1.33 Window & Custom SS304 (MPF) \\
\bottomrule
\end{tabularx}
\end{table}

\subsection{Helium Interface Heater}
\label{subsec:helium-interface-heater}

\begin{figure}[!!h]
\centering
\includegraphics[width=0.75\textwidth]{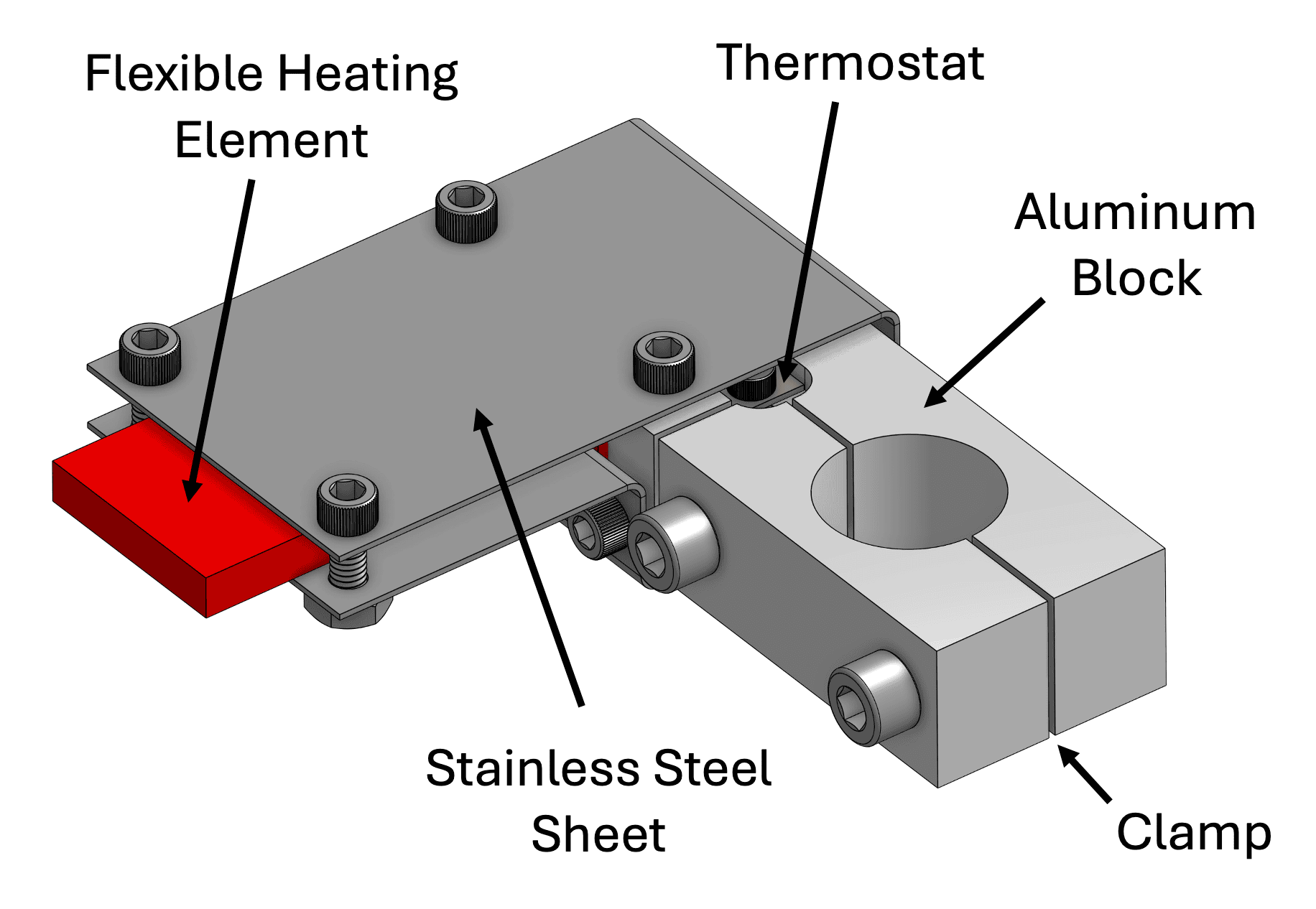}
\caption[Helium Interface Heater Assembly]{Helium interface heater CAD drawing. The thermostat is bolted directly to the aluminum block. The flexible heating element is held in place on the aluminum block using stainless steel sheet. The clamp to the CF-2.75 helium interface consists of the aluminum block that has 1/4-20 taps and a separate piece of aluminum has 1/4 inch through holes. Pressure to the clamp is supplied by tightening the 1/4-20 bolts.}
\label{fig:helium-interface-heater}
\end{figure}

\begin{figure}[!!h]
\centering
\includegraphics[width=0.75\textwidth]{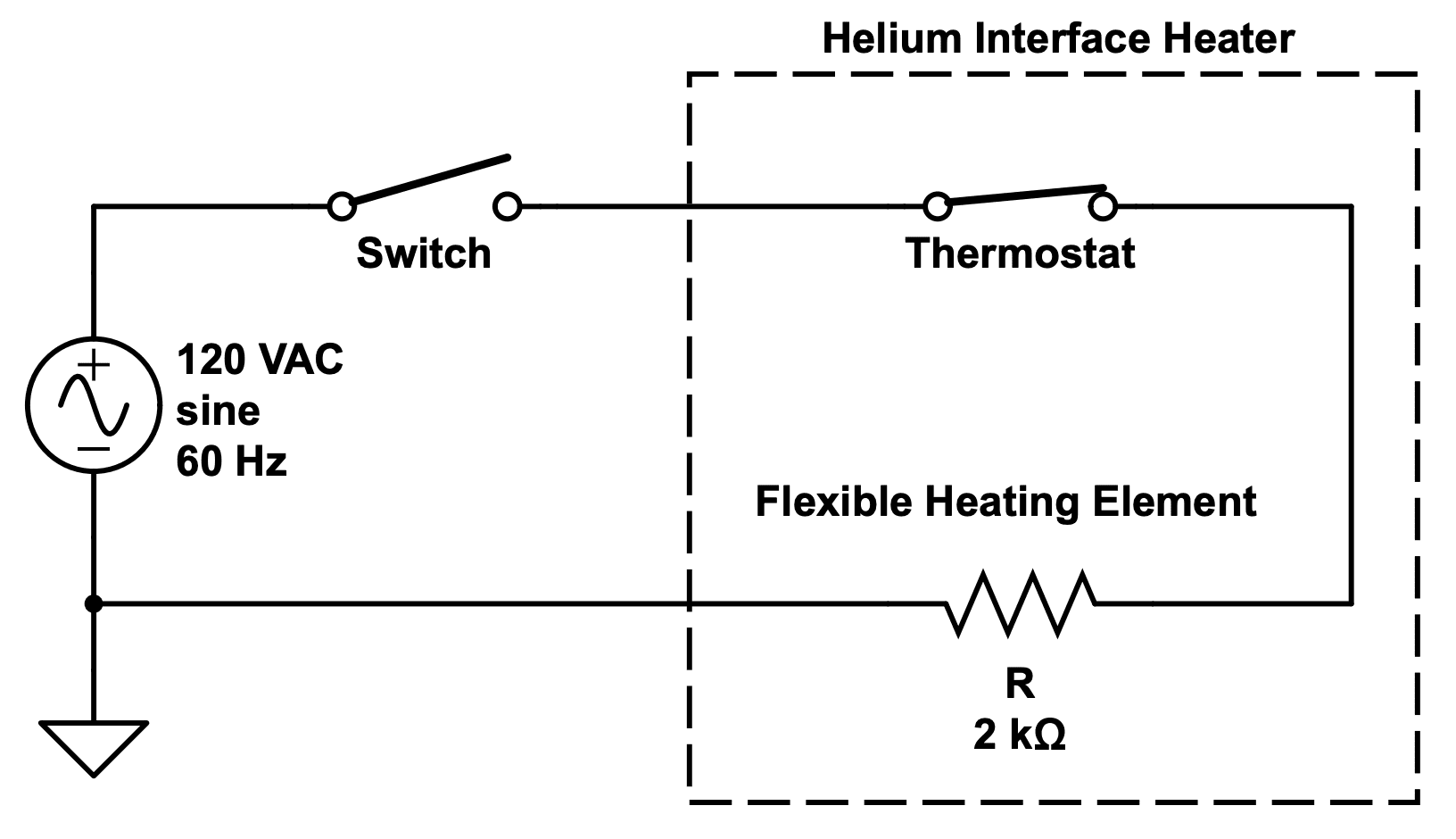}
\caption[Helium Interface Heater Circuit]{Helium interface heater circuit. Line voltage is supplied to the circuit using an external switch. The thermostat is a normally closed switch controlling current to the resistive heating element, provided the temperature of the thermostat is $<$ 323 K. Once the temperature reaches 323 K, the thermostat switches open.}
\label{fig:helium-interface-heater-circuit}
\end{figure}

I built a heater that attaches to the CF-2.75 helium interface at the cryostat to prevent condensation as shown in Figure \ref{fig:helium-interface-heater}. The heater consists of an aluminum block that is servoed to 323 K via a thermostat (Sensata, PN:67L050P) that controls current to a 6 W flexible resistive heating element (Watlow, PN:010050C20007M) as shown in Figure \ref{fig:helium-interface-heater-circuit}. The flexible heating element is held in place on the aluminum block using 30 mil thick stainless steel sheet and the aluminum block is clamped to the CF-2.75 helium interface to transfer the heat via conduction. The parts were designed and machined by myself at UMD. 

\subsection{300 K Chamber}
\label{subsec:300K-chamber}

\begin{figure}[!!h]
\centering
\includegraphics[width=\textwidth]{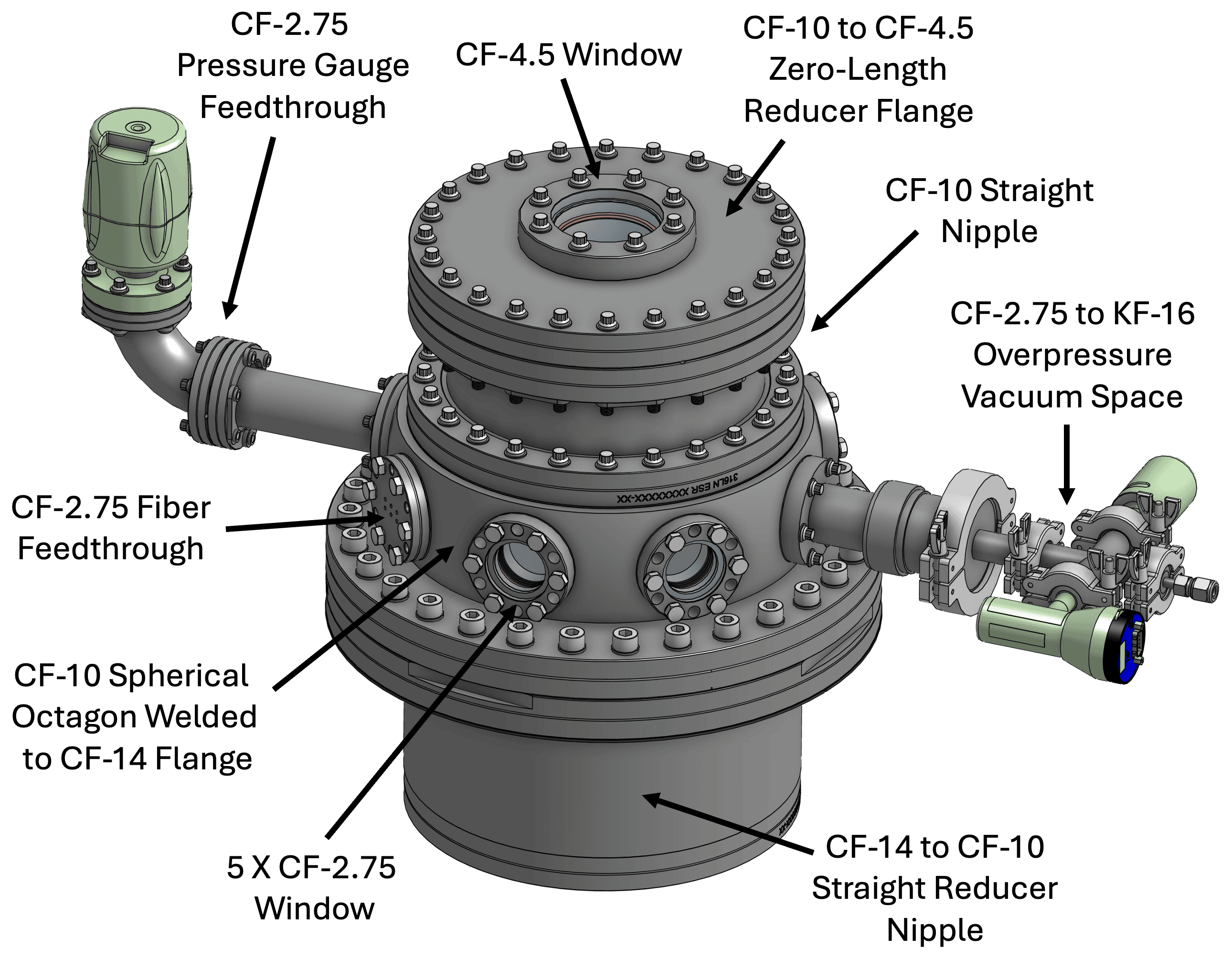}
\caption[300 K Chamber Assembly]{CAD drawing of 300 K chamber. A parts list is summarized in Table \ref{tab:part-list-for-300K-chamber-assembly}.}
\label{fig:300K-chamber-assembly}
\end{figure}

The 300 K chamber is the section of the 300 K stage that encloses the cold stages and is bolted to the 300 K CF-10 flange. The main parts of this section consist of a CF-14 to CF-10 straight reducer nipple, a CF-10 spherical octagon welded to a CF-14 flange, a CF-10 straight nipple, and a CF-10 to CF-4.5 zero-length reducer flange as shown in Figure \ref{fig:300K-chamber-assembly}. The CF-14 to CF-10 straight reducer nipple snares the top of the optics table on top of nylon spacers to provide electrical isolation. These spacers also reduce the friction between the flange and optics table making it easy to rotate the cryostat while its fully seated. Access to the 4 K stage for in-vacuum maintenance is achieved by removing either the CF-10 to CF-4.5 zero-length reducer flange or the CF-10 straight nipple. Additionally, the cryostat can be lifted out of the optics table using an adapter that interfaces with the CF-10 straight nipple. All of the flanges, nipples, and fasteners at this section are made out of SS316 as it directly encloses the ion trap $(B\propto r^{-3})$. The fasteners are all silver plated to prevent galling of the threads after repeated assembly/disassembly of the system. There is a CF-4.5 window attached at the top of the chamber and five CF-2.75 windows with double density through holes attached on the sides of the CF-10 spherical octagon. The details of the window coatings are discussed in Subsection \ref{subsec:ion-trap-optical-access}. The optical fiber for the cavity penetrates the cryostat at the CF-2.75 Fiber Feedthrough and is also machined with double density through holes. All the taps for each flange in the CF-10 spherical octagon are double density enabling mounting structures directly to the chamber with threaded rods. A wide-range-gauge is mounted onto the chamber with CF-2.75 components and an overpressure vacuum space is integrated with the chamber and is discussed in Section \ref{subsec:cryostat-overpressure-vacuum-space}. 

\begin{table}[!!h]
\centering
\tablefont
\caption[Parts List for 300 K Chamber Assembly]{Parts list for 300 K chamber in Figure \ref{fig:300K-chamber-assembly}.}
\label{tab:part-list-for-300K-chamber-assembly}
\begin{tabularx}{\textwidth}{>{\centering\arraybackslash}p{0.38\textwidth}>{\centering\arraybackslash}p{0.55\textwidth}}
\toprule
\textbf{Schematic Label} & \textbf{Part} \\
\midrule
CF-14 to CF-10 Straight Reducer Nipple & Custom SS316 (Kurt J. Lesker) \\
\midrule
\makecell{CF-10 Spherical Octagon\\ Welded to CF-14 Flange} & \makecell{Custom SS316 (Kimball Physics) \\Double Density Taps on Octagon} \\
\midrule
CF-10 Straight Nipple & Custom SS316 (Kurt J. Lesker) \\
\midrule
CF-10 Zero-Length Reducer Flange & Custom SS316 (Kurt J. Lesker) \\
\midrule
CF-4.5 Window & Custom SS316 (MPF) \\
\midrule
$5\times$ CF-2.75 Window & \makecell{Custom SS316 (MPF) \\Double Density Through Holes} \\
\midrule
CF-2.75 Pressure Gauge Feedthrough & \makecell{Wide-Range-Gauge (Edwards Vacuum, PN: WRG-S-DN40CF) \\Custom SS316 Nipple and Right Angle (Kurt J. Lesker)} \\
\midrule
\makecell{CF-2.75 to KF-16\\ Overpressure Vacuum Space} & See Subsection \ref{subsec:cryostat-overpressure-vacuum-space}\\
\midrule
Fasteners & Custom Silver-Plated SS316 (UC Components) \\
\bottomrule
\end{tabularx}
\end{table}

\subsection{Cryostat Overpressure Vacuum Space}
\label{subsec:cryostat-overpressure-vacuum-space}

An engineering solution is presented here that protects the cryostat from overpressure - a positive pressure relative to atmosphere. This is an issue for this cryostat as it contains seven windows that are not designed for large positive pressures ($>15\text{ psi}$)\footnote{Confirmed by MPF Products.}, leading to a hazardous explosion of glass in the lab. There are four ways that overpressure can occur in this system, all occurring when the system is cryogenic: 1.) there is a leak from the in-vacuum helium recirculating loop into the chamber, 2.) there is a leak from atmosphere into the chamber, 3.) the non-evaporable getter (NEG) is accidentally activated releasing previously adsorbed gases, and 4.) too much gas is intentionally leaked into the cryostat during warmup to create a thermal short between the cold stages and 300 K stage (see Subsection \ref{subsec:warmup}).

In case 1.), we can estimate the the worst case positive pressure assuming all the helium in the recirculating loop leaks into the cryostat. Taking the pressure of the helium in the loop to be $P_{\text{L}}=125\text{ psi}$\footnote{This is absolute pressure. i.e., $110\text{ psig}+15\text{ psi}=125\text{ psi}$.}, the volume of the loop to be $V_{\text{L}}=6.4\text{ L}$, the pressure of the cryostat when cryogenic to be $P_{\text{cryostat}}\approx0\text{ psi}$, and the volume of the cryostat to be $V_{\text{cryostat}}=18\text{ L}$, the pressure in the cryostat after a leak can be computed using the following equation,

\begin{equation}
    P_{\text{cryostat}}=\frac{P_{\text{L}}V_{\text{L}}}{V_{\text{L}}+V_{\text{cryostat}}}
    \label{eq:cryostat-pressure-due-to-JT-leak}
\end{equation}
We get that the final helium pressure in the cryostat is $P_{\text{cryostat}}=33\text{ psi}$ which is a 18 psi positive pressure relative to atmosphere. This will cause overpressure as the 4 K stage is not cold enough to condense helium as shown in Table \ref{tab:vapor-pressures-for-various-gas-molecules}. For case 2.), air will leak into the cryostat and condense at the cold stages which will eventually lead to the stages warming up due to the increase in emissivity of ice \cite{pepperAbsorptionInfraredRadiation}, but will not immediately cause overpressure. Similarly for case 3.), gases released from the NEG will condense, but will not immediately cause overpressure. However, if the cryostat warms up there is now a new gas load that was not initially evacuated with the turbo pump before cooldown and could create a hazardous positive pressure. Lastly for case 4.), overpressure simply occurs if too much gas is intentionally leaked into the cryostat. 

\begin{figure}[!!h]
\centering
\includegraphics[width=0.75\textwidth]{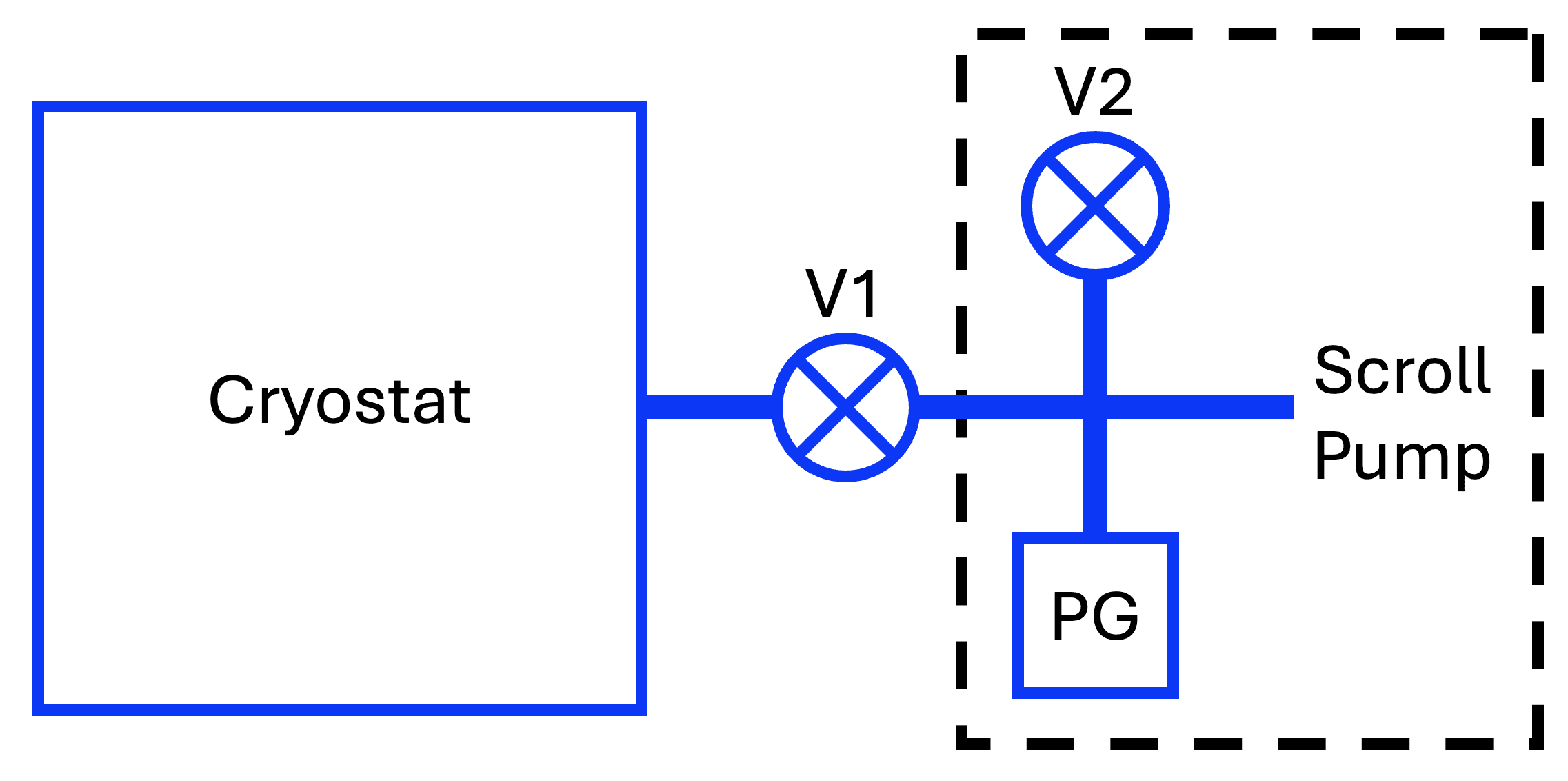}
\caption[Cryostat Overpressure Vacuum Space Assembly]{Schematic of the cryostat overpressure vacuum space depicted inside the dashed box. All parts to the right of V1 are connected with KF-16 and Swagelok-1/4 components. V1 is connected to the cryostat with a CF-2.75 flange. A parts list for this schematic is summarized in Table \ref{tab:part-list-for-cryostat-overpressure-vacuum-space-assembly}.}
\label{fig:cryostat-overpressure-vacuum-space}
\end{figure}

We built an overpressure vacuum space that is designed to engage when there is an overpressure event as shown in Figure \ref{fig:cryostat-overpressure-vacuum-space}. It consists of a burst disc (V1) that will rupture when it senses a pressure difference of 5 psi. To the right of the burst disc (V1) is a rough vacuum space with a relief valve (V2) and pressure gauge (PG) evacuated with a scroll pump to a pressure of $10^{-1}\text{ Torr}$ ($2\times10^{-3}\text{ psi}$). So if there is overpressure, the burst disc will rupture when the cryostat vacuum space is $260\text{ Torr}$ (5 psi) and prevent the windows from experiencing a positive pressure relative to atmosphere. The gas load is then pumped away with the scroll pump. There will be substantial back-streaming from the scroll pump depending on the temperatures of the cold stages, but this is a lower level priority problem than exploding windows in the lab. In the case that an overpressure scenario occurs during a prolonged power outage, the scroll pump will eventually turn off after the UPS battery dies and a relief valve (V2), with a cracking pressure of 0.5 psi, will vent the gas load to atmosphere. Lastly, the standard operating procedure before a cooldown is to ensure that the cryostat vacuum space is evacuated before the overpressure vacuum space to prevent the burst disc from rupturing. 

\begin{table}[!!h]
\centering
\tablefont
\caption[Parts List for Cryostat Overpressure Vacuum Space Assembly]{Parts list for cryostat overpressure vacuum space schematic in Figure \ref{fig:cryostat-overpressure-vacuum-space}.}
\label{tab:part-list-for-cryostat-overpressure-vacuum-space-assembly}
\begin{tabularx}{\textwidth}{>{\centering\arraybackslash}p{0.4\textwidth}c}
\toprule
\textbf{Schematic Label} & \textbf{Part} \\
\midrule
V1 & \makecell{Burst Disc with Gas Recovery Port (5 psi differential) \\ANCORP, PN: BD-LP-CF275XQF50-GRP} \\
\midrule
V2 & \makecell{Relief Valve (0.5 psi differential) \\Accu-Glass Products, PN: 113155 }\\
\midrule
PG & \makecell{Pressure Gauge (Rough Vacuum) \\InstruTech, PN: CVM211GBL} \\
\bottomrule
\end{tabularx}
\end{table}

\section{4 K and 40 K Stages}
\label{sec:4K-and-40K-stages}

\begin{figure}[!!h]
\centering
\includegraphics[width=\textwidth]{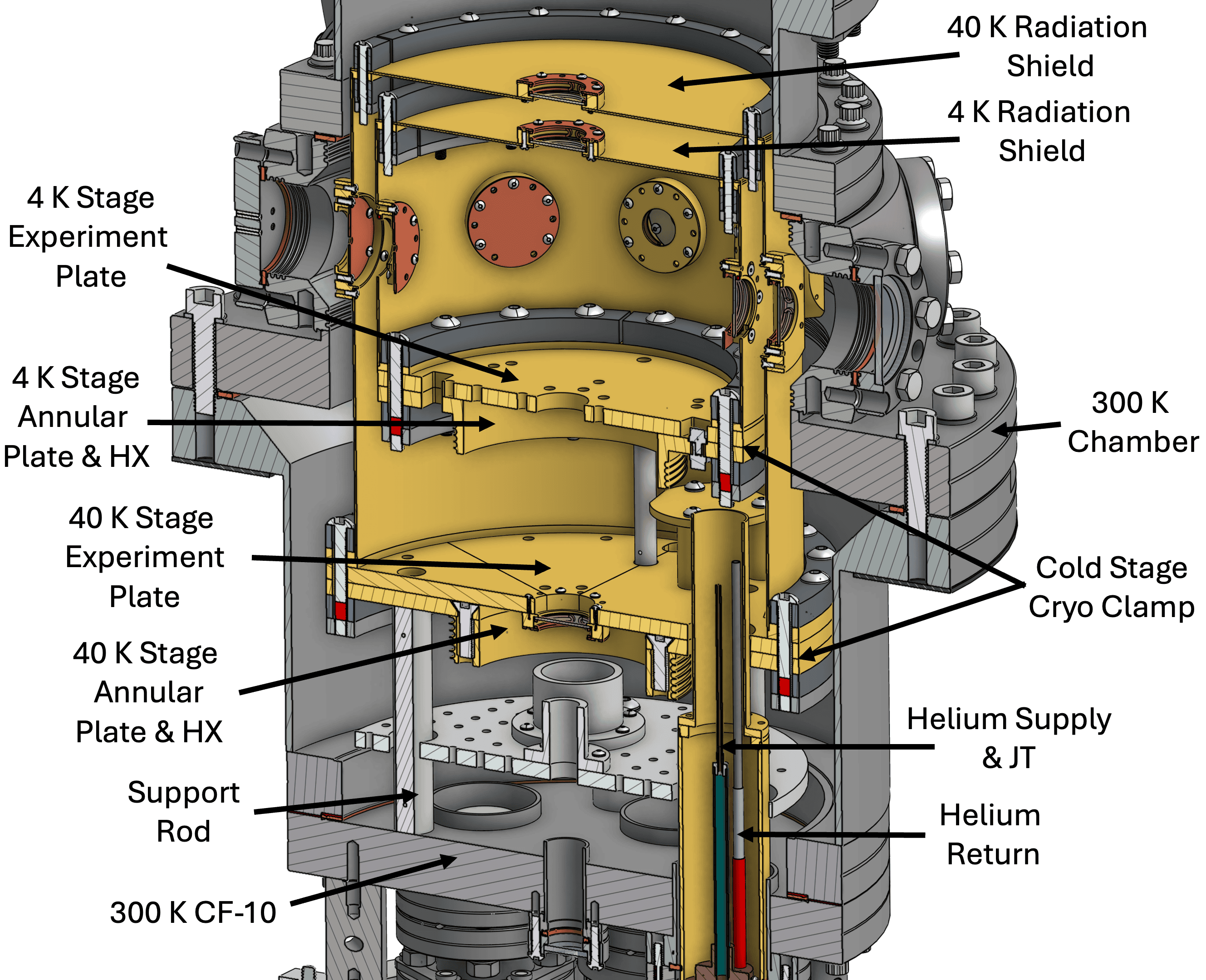}
\caption[Cryostat Cold Stages Assembly]{CAD drawing of cryostat cold stages.}
\label{fig:cryostat-cold-stages-assembly}
\end{figure}

The cryostat 4 K and 40 K cold stages are thermally isolated from each other and the 300 K stage via PEEKGF30 support rods as shown in Figure \ref{fig:cryostat-cold-stages-assembly}. As there is limited cooling power available at the 4 K stage due to the thermodynamics of helium at 4 K (see Section \ref{sec:Thermal-Model}), a 40 K stage is required that has more cooling power and consumes the bulk of the heat load from the 300 K stage. The 40 K stage encloses the 4 K stage, where the ion trap is positioned, and each stage has a radiation shield. 

After the helium undergoes a JT expansion at the CF-2.75 Helium Interface, it is first routed to the 4 K stage and then to the 40 K stage. After that, it is routed back through the CF-2.75 Helium Interface exiting the cryostat. The heat exchanger (HX) between the in-vacuum helium lines and the cold stages consists of a helical corrugated surface to which the helium lines are brazed. This heat exchanger is interfaced with an annular plate that serves as the cold bath for each stage. The helium lines between each stage and the CF-2.75 Helium Interface are connected using VCR-1/4 fittings enabling simple assembly/disassembly. The electrical feedthroughs, helium line feedthroughs, and vacuum conductance apertures are machined into the annular plates. Separate experiment plates are clamped to the annular plates and are machined to accommodate the feedthroughs and apertures in the annular plates. The details of the clamp are discussed in Subsection \ref{subsec:the-cryo-clamp}. This enables cheaper design iteration of the experiment plates without changing the annular plates and heat exchangers. Lastly, all in-vacuum bolts in the cryostat are made of SS316 or silver-plated SS316 and the bolts that are tapped into C10100, Al6061T6, or PEEKGF30 have TiG5 helicoils (McMaster Carr or MSC Industrial Supply). This prevents galling of the threads after repeated assembly/disassembly of the system. 

\subsection{4 K and 40 K Stage Plates}
\label{subsec:4K-and-40K-stage-plates}

\begin{figure}[!!h]
\centering
\includegraphics[width=0.49\textwidth]{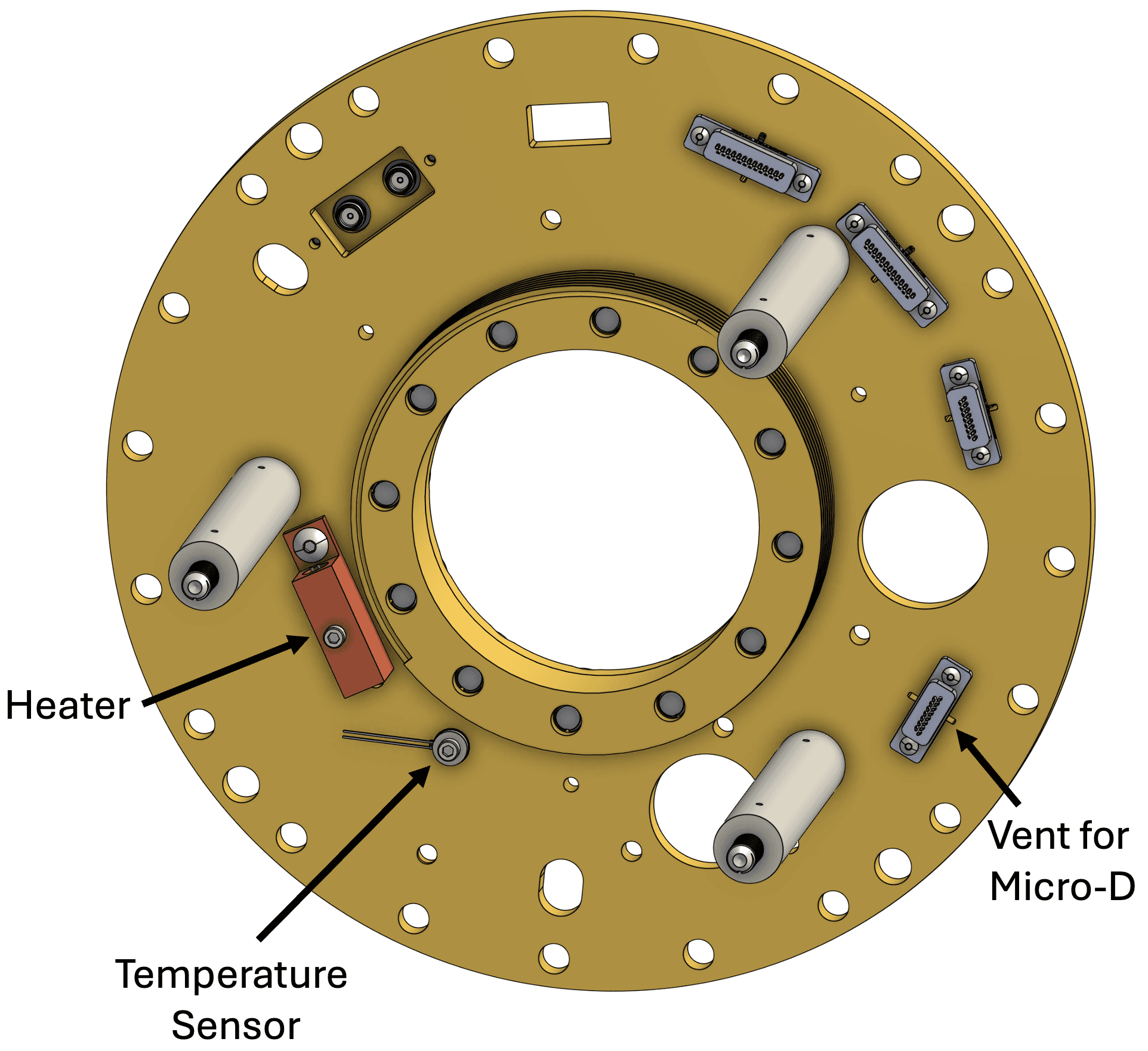}
\includegraphics[width=0.49\textwidth]{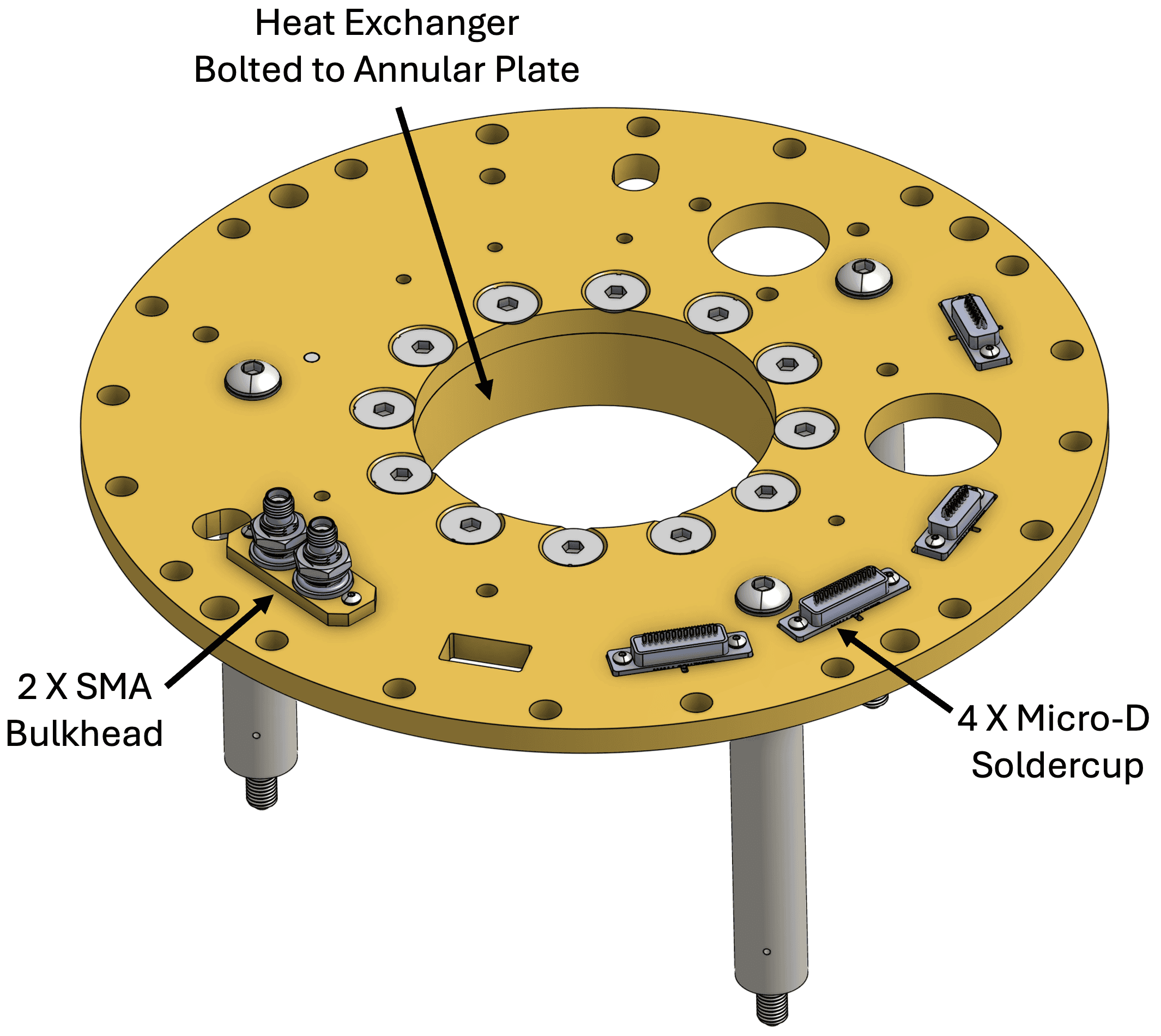}
\newline
\includegraphics[width=0.49\textwidth]{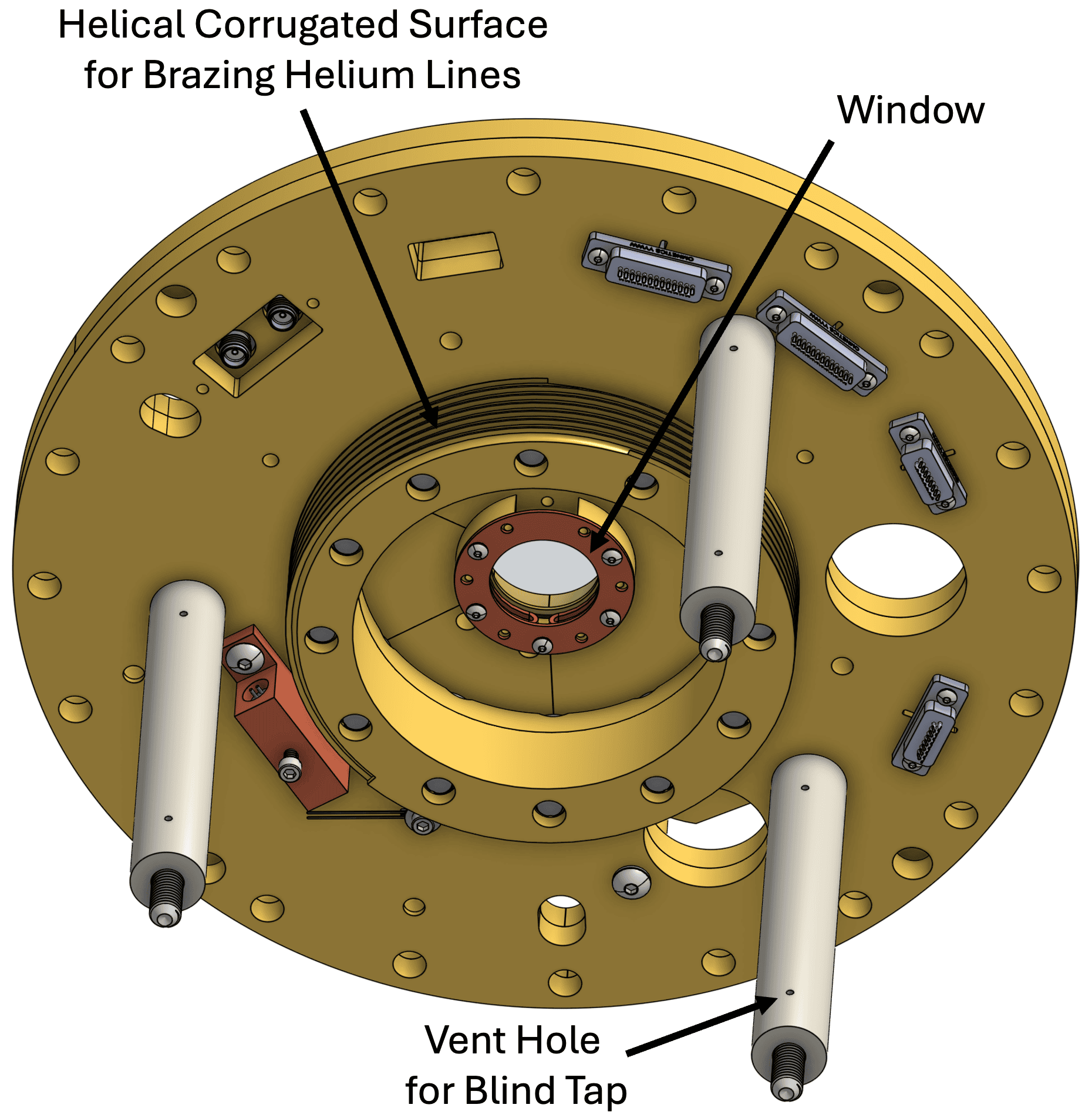}
\includegraphics[width=0.49\textwidth]{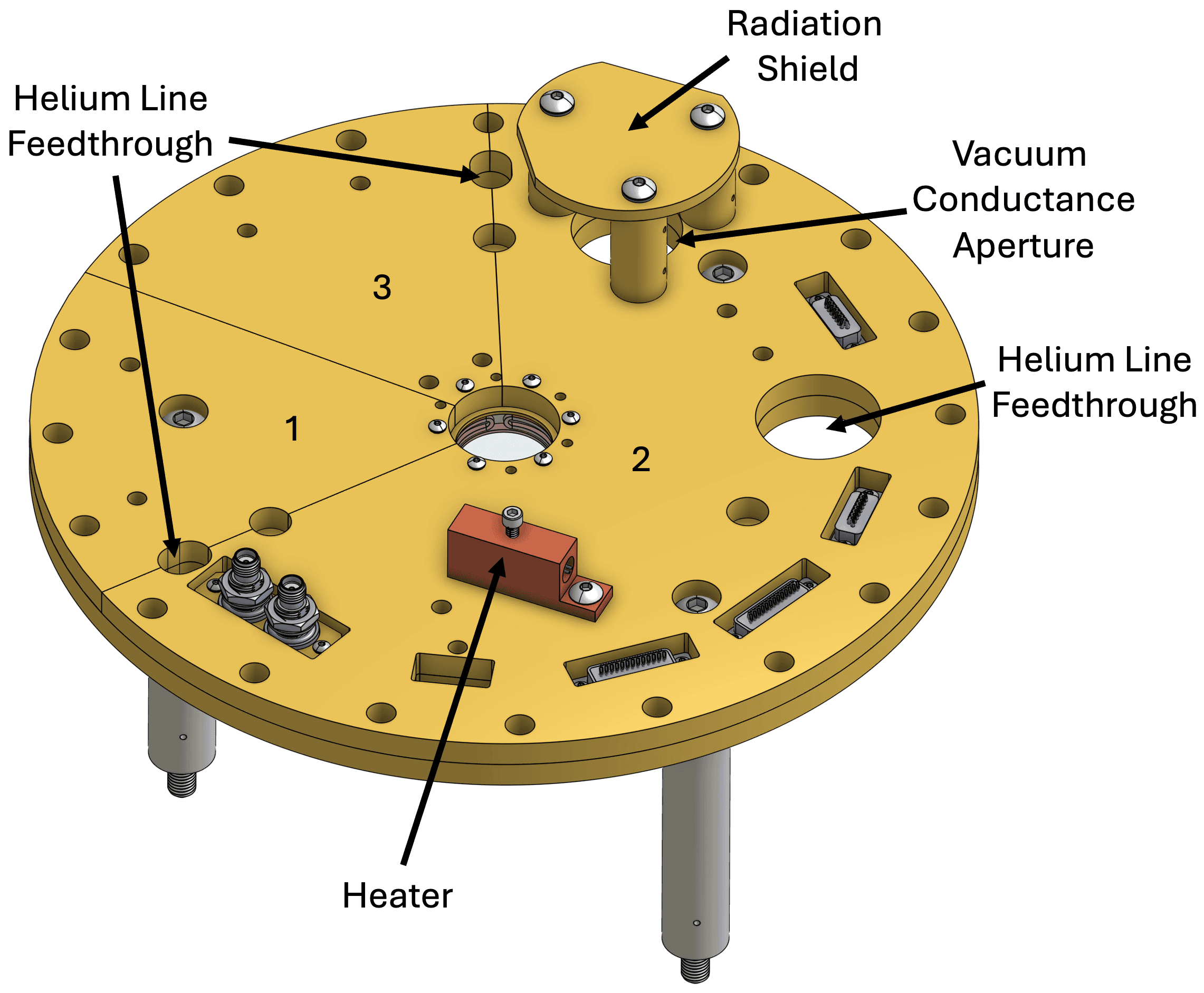}
\caption[Cryostat 40 K Stage Plates Assembly]{CAD drawing of 40 K stage plates. (Top) Top and bottom views of 40 K annular plate. (Bottom) Top and bottom views of 40 K stage annular and experiment plates.}
\label{fig:40K-stage-annulus-plates-CAD}
\end{figure}

The 40 K stage consists of an annular plate and 3 experiment plates as shown in Figure \ref{fig:40K-stage-annulus-plates-CAD}. The experiments plates enable alternate designs for a cryogenic ion trap with an oven and resonator located at the 40 K stage. The heat exchanger needed to be bolted to the annular plate to prevent warping of the large annular plate during brazing - confirmed by CET. A heater and temperature sensor are mounted on the bottom of the annular plate and another heater is mounted on the top of experiment plate 2. Four Micro-D soldercups are directly mounted to the annular plate with milled slots for venting gas. The Micro-Ds are heat sunk to the annular plate to provide a thermal sink for the wires. Optical access is provided through a window attached to the experiment plates to prevent direct line-of-sight radiation from the 300 K to 4 K stage, see Table \ref{tab:heat-load-budget}. A vacuum conductance aperture is machined into both plates giving a vacuum conductance path between the 40 K stage and 300 K stage. A radiation shield is mounted above these apertures to prevent direct line-of-sight radiation from the 300 K to 4 K stage, see Table \ref{tab:heat-load-budget}.  Lastly, additional apertures are machined into both plates for the helium feedthroughs.

\begin{figure}[!!h]
\centering
\includegraphics[width=0.49\textwidth]{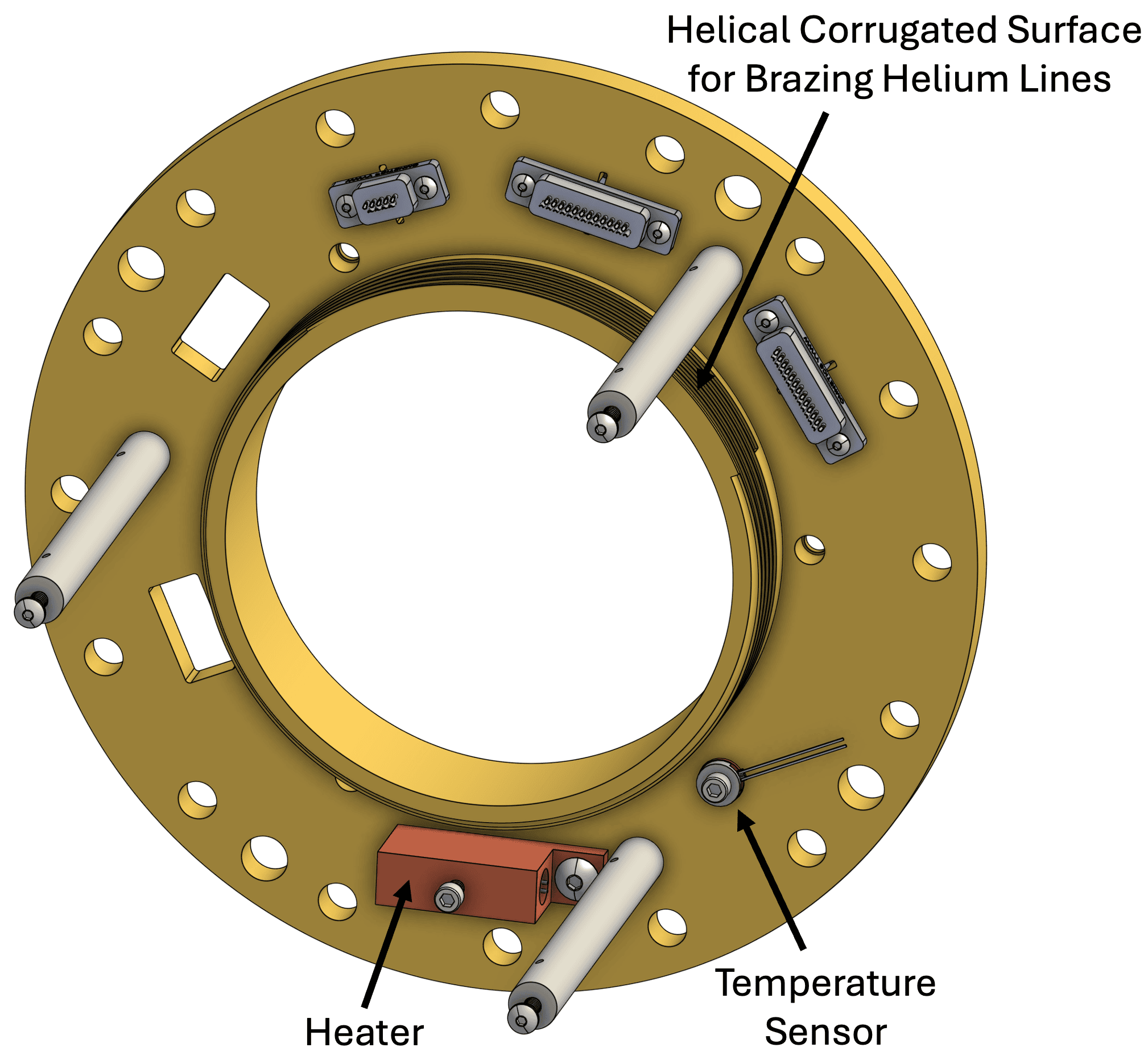}
\includegraphics[width=0.49\textwidth]{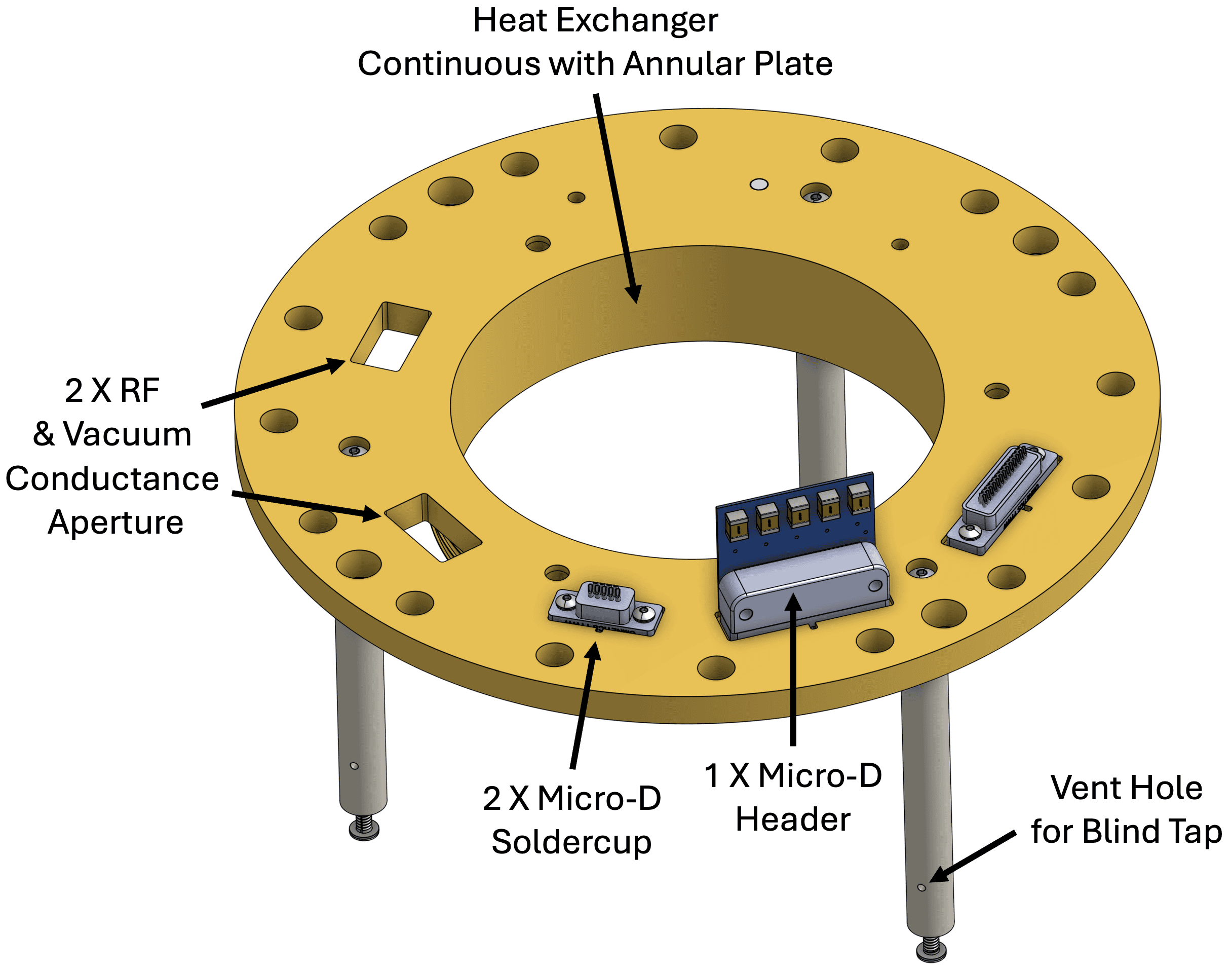}
\newline
\includegraphics[width=0.49\textwidth]{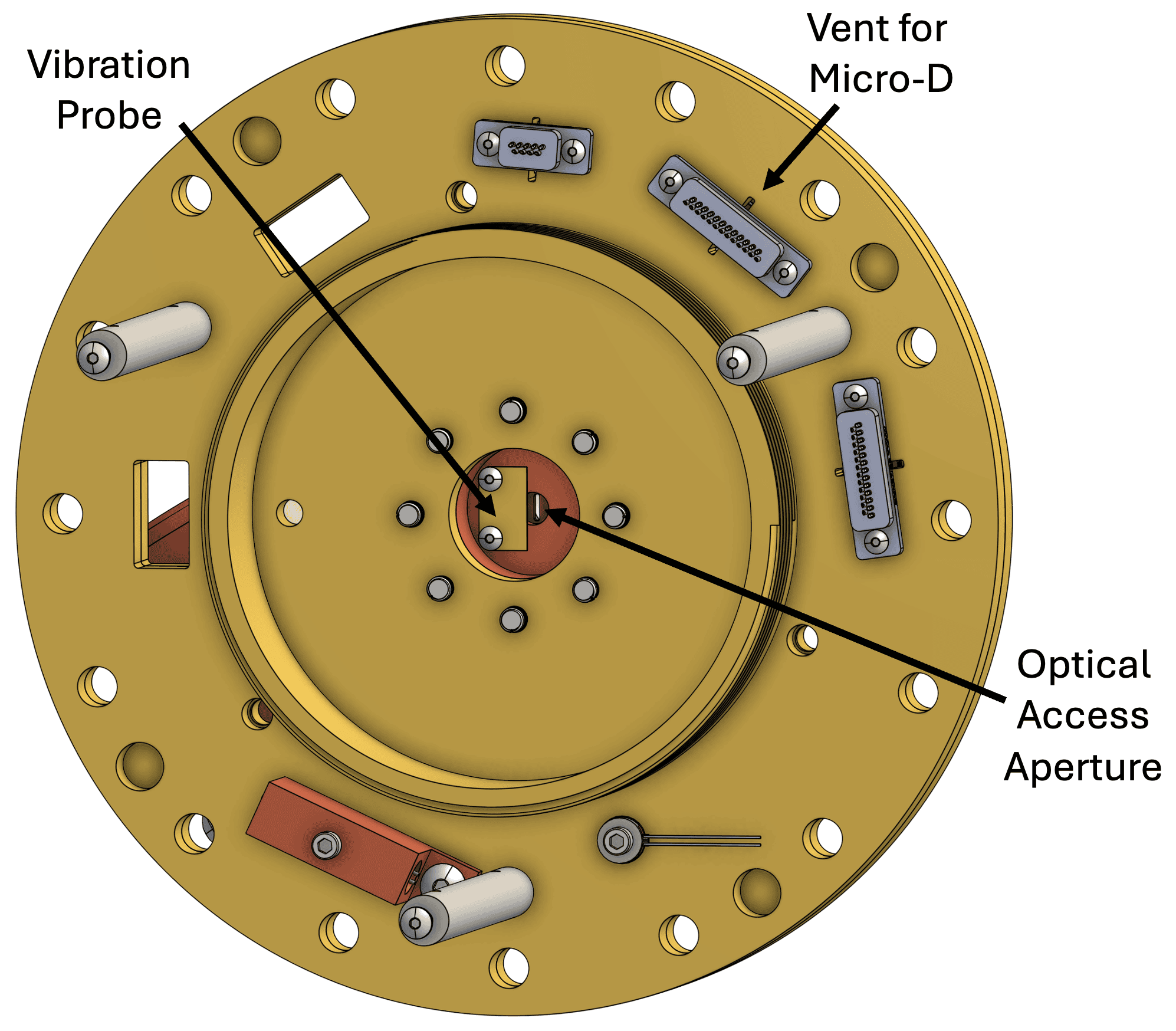}
\includegraphics[width=0.49\textwidth]{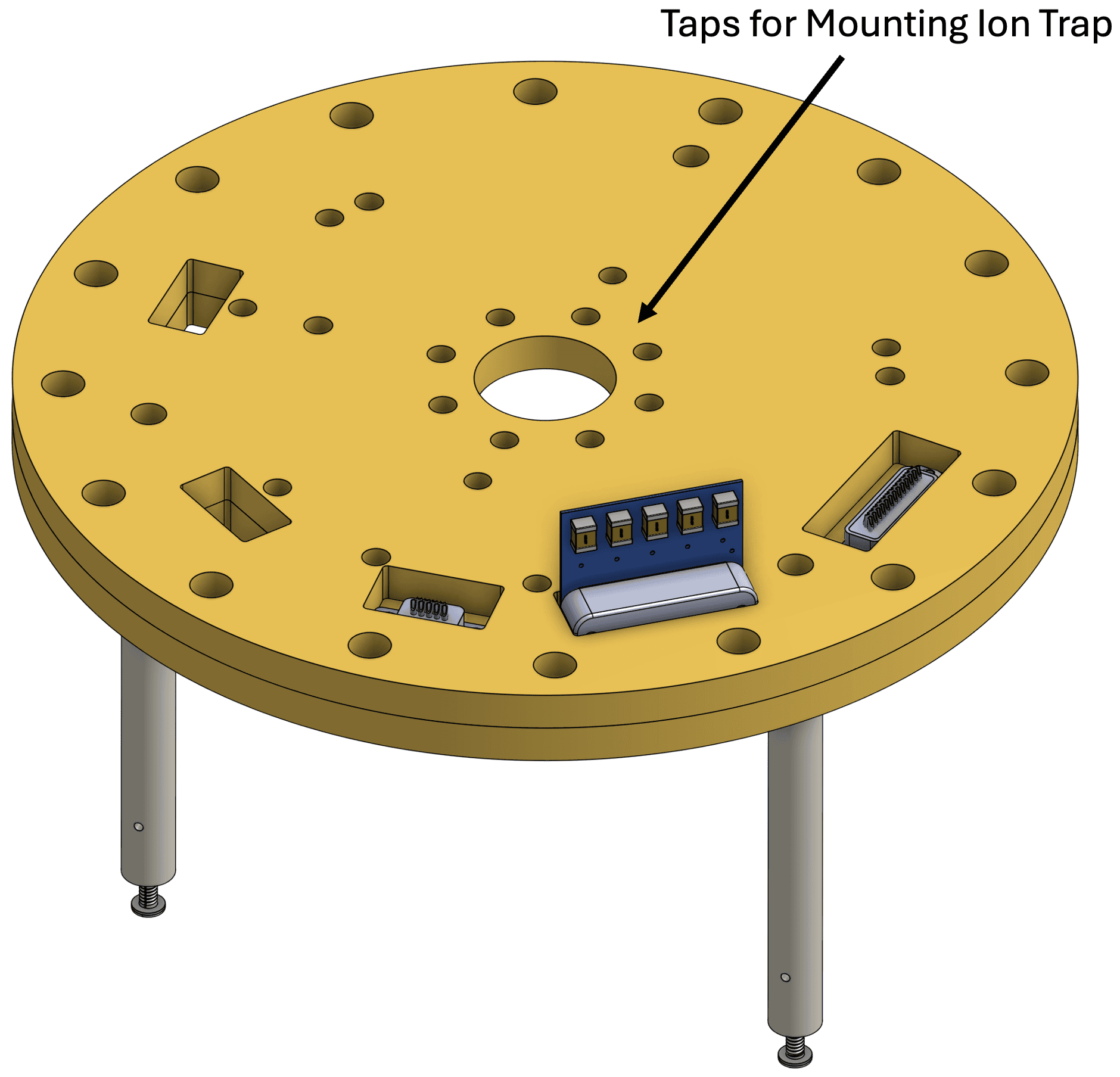}
\caption[Cryostat 4 K Stage Plates Assembly]{CAD drawing of 4 K stage plates. (Top) Top and bottom views of 4 K annular plate. (Bottom) Top and bottom views of 4 K stage annular and experiment plates.}
\label{fig:4K-stage-annulus-plates-CAD}
\end{figure}

The 4 K stage consists of an annular plate and 1 experiment plate, where the ion trap is mounted, as shown in Figure \ref{fig:4K-stage-annulus-plates-CAD} The heat exchanger was able to be continuous with the annular plate at this stage as this smaller geometry was easier to braze - confirmed by CET. A heater and temperature sensor are mounted on the bottom of the annular plate and another heater and temperature sensor are mounted at the Trap Riser and Trap Mount respectively. Two Micro-D soldercups and one Micro-D header are directly mounted to the annular plate with milled slots for venting gas. Optical access is provided through a small aperture milled into the Trap Riser. Two slots are milled into the annular plate to provide vacuum conductance between the 4 K stage and 300 K stage and to provide a feedthrough for RF signals. The ion trap, helical resonator, optical cavity, and oven are all mounted at the experiment plate and is discussed in Chapter \ref{chap:ion-trap}. 

\subsection{Radiation Shields}
\label{subsec:radiation-shields}

\begin{figure}[!!h]
\centering
\includegraphics[width=0.8\textwidth]{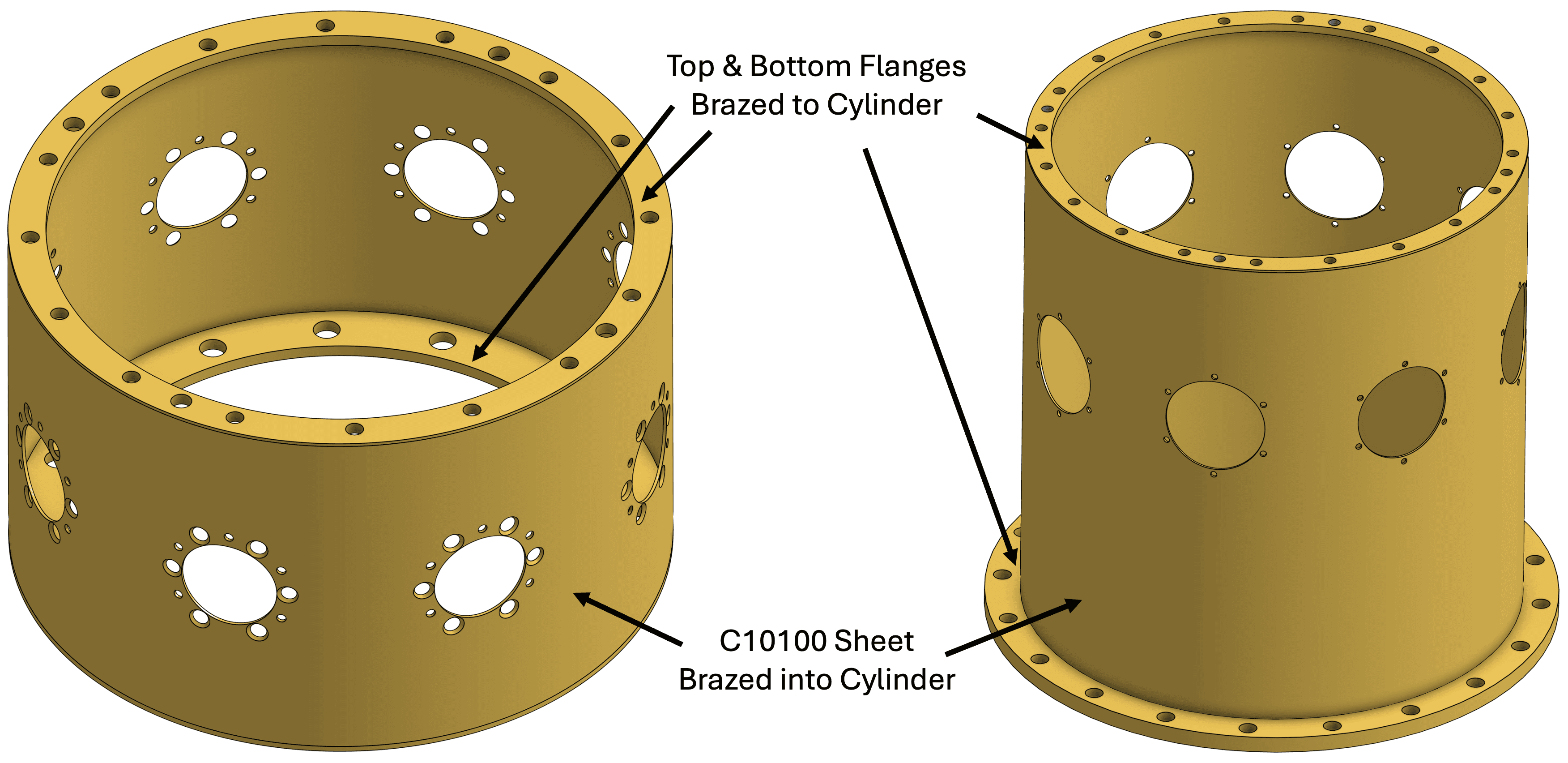}
\newline
\includegraphics[width=0.8\textwidth]{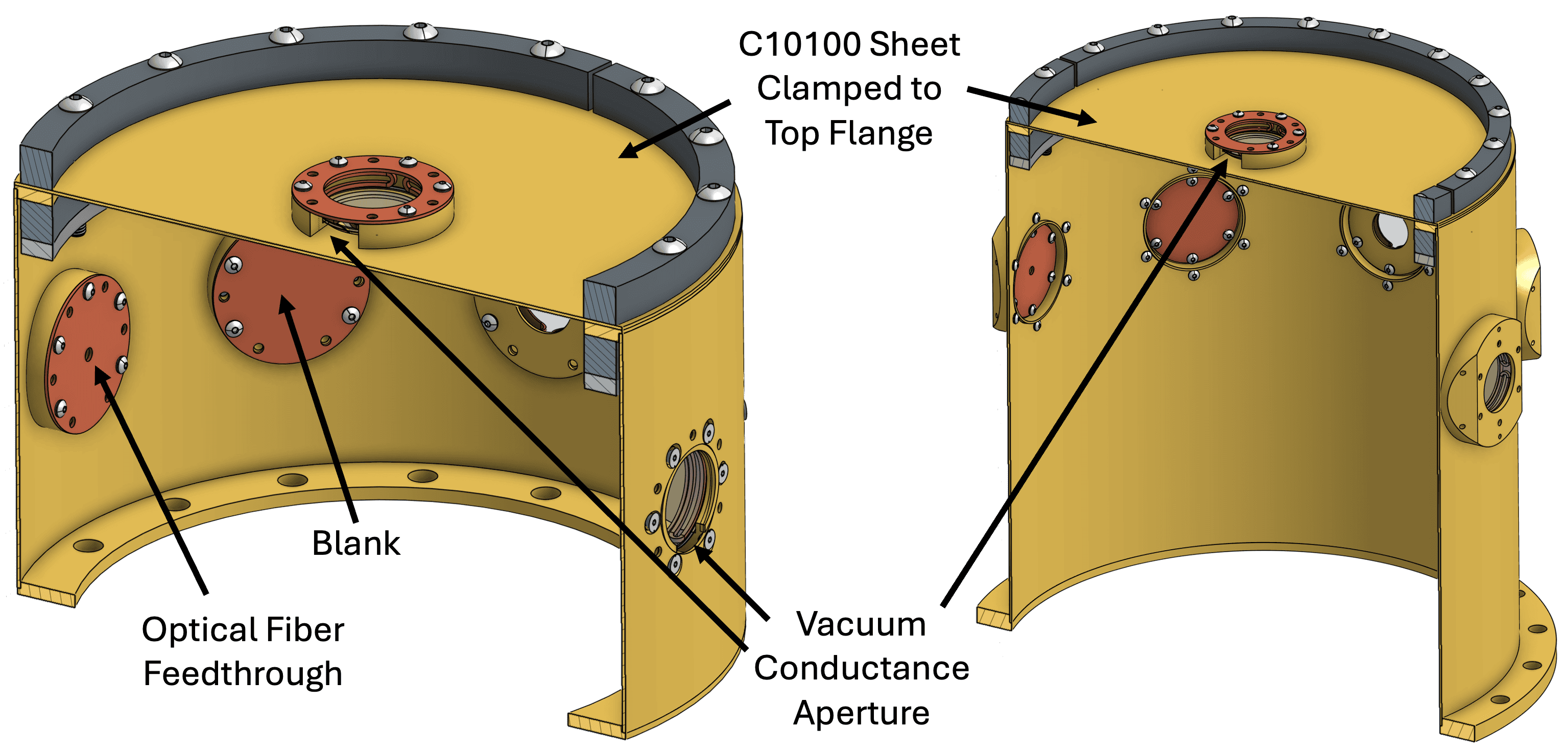}
\caption[Cryostat Cold Stage Radiation Shields Assembly]{CAD drawings of cold stage radiation shields.}
\label{fig:radiation-shields-assmebly}
\end{figure}

The cold stage radiation shields were manufactured by brazing 0.032 in (0.8 mm) thick C10100 sheet into a cylinder, then brazing the cylinder to a top and bottom flange as shown in Figure \ref{fig:radiation-shields-assmebly}. Each shield is clamped to the cold stage plates with the bottom flange and a lid is clamped to the top flange of each shield. Window holders mate with the apertures machined in the shield side walls and lids, discussed in Subsection \ref{subsec:window-holders}. The window holders mounted at the lids and 4 K sidewalls have a vacuum conductance aperture. 

\subsection{Window Holders}
\label{subsec:window-holders}

\begin{figure}[!!h]
\centering
\includegraphics[width=0.5\textwidth]{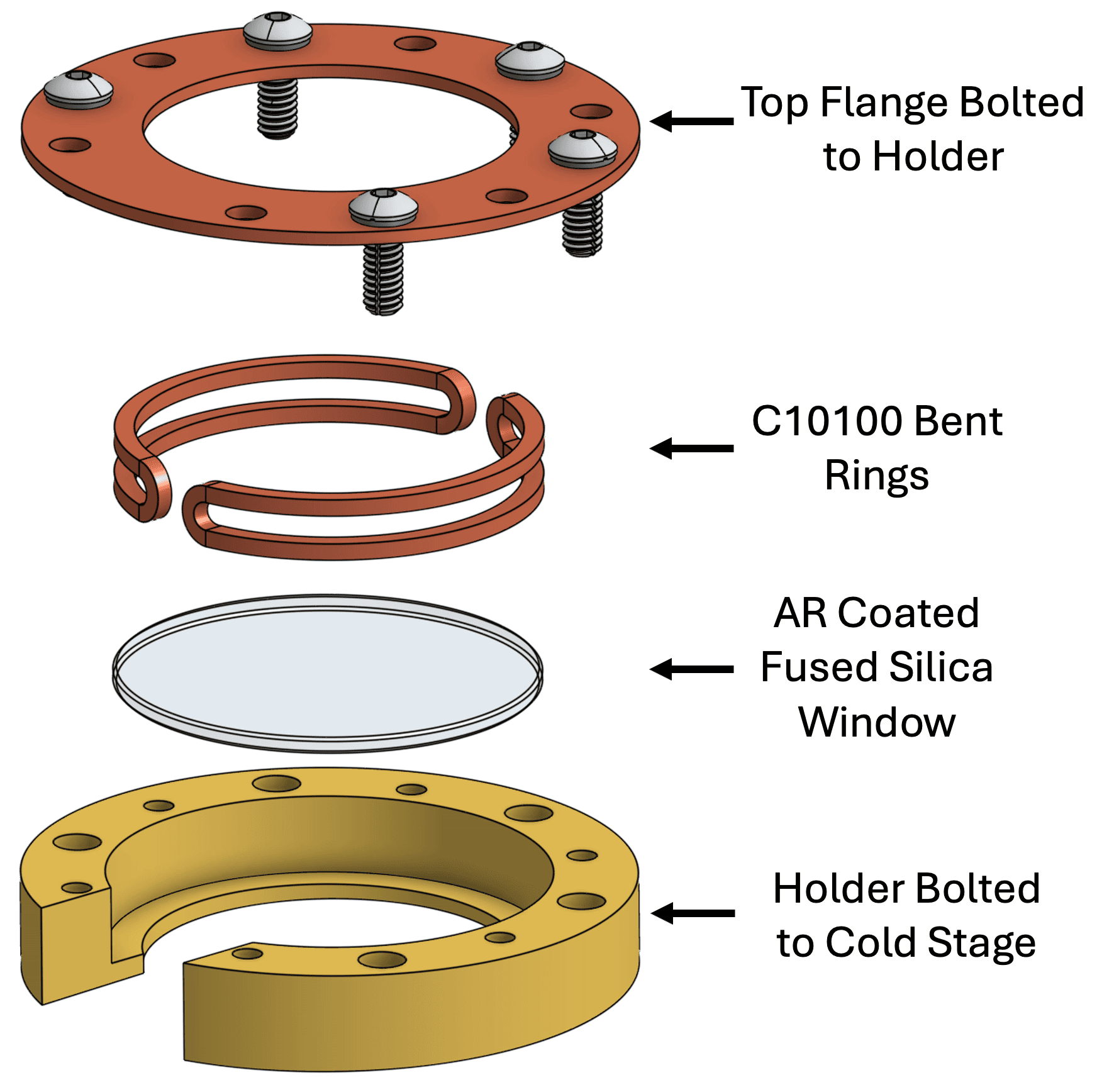}
\caption[Cryostat Cold Stage Window Holders Assembly]{Exploded CAD drawing of the window holder design.}
\label{fig:window-holders}
\end{figure}

The window holders were designed to accommodate the differential thermal contraction of C10100 and fused silica. This was done by machining C10100 rings and bending them to create springs. The 1 mm thick windows and springs are then clamped between two pieces of C10100 - the top flange and holder - so that the net strain due to differential thermal contraction will be in the spring and not the fused silica as shown in Figure \ref{fig:window-holders}. The exact geometry of the holder is slightly different between the 4 K and 40 K stages, but the design principle is the same. Additionally, the size of the aperture of the top flange and optical coating of the fused silica varies depending on the optical access required in that direction and is discussed more in Subsection \ref{subsec:ion-trap-optical-access}. 

\subsection{The Cryo Clamp}
\label{subsec:the-cryo-clamp}

\begin{figure}[!!h]
\centering
\includegraphics[width=0.75\textwidth]{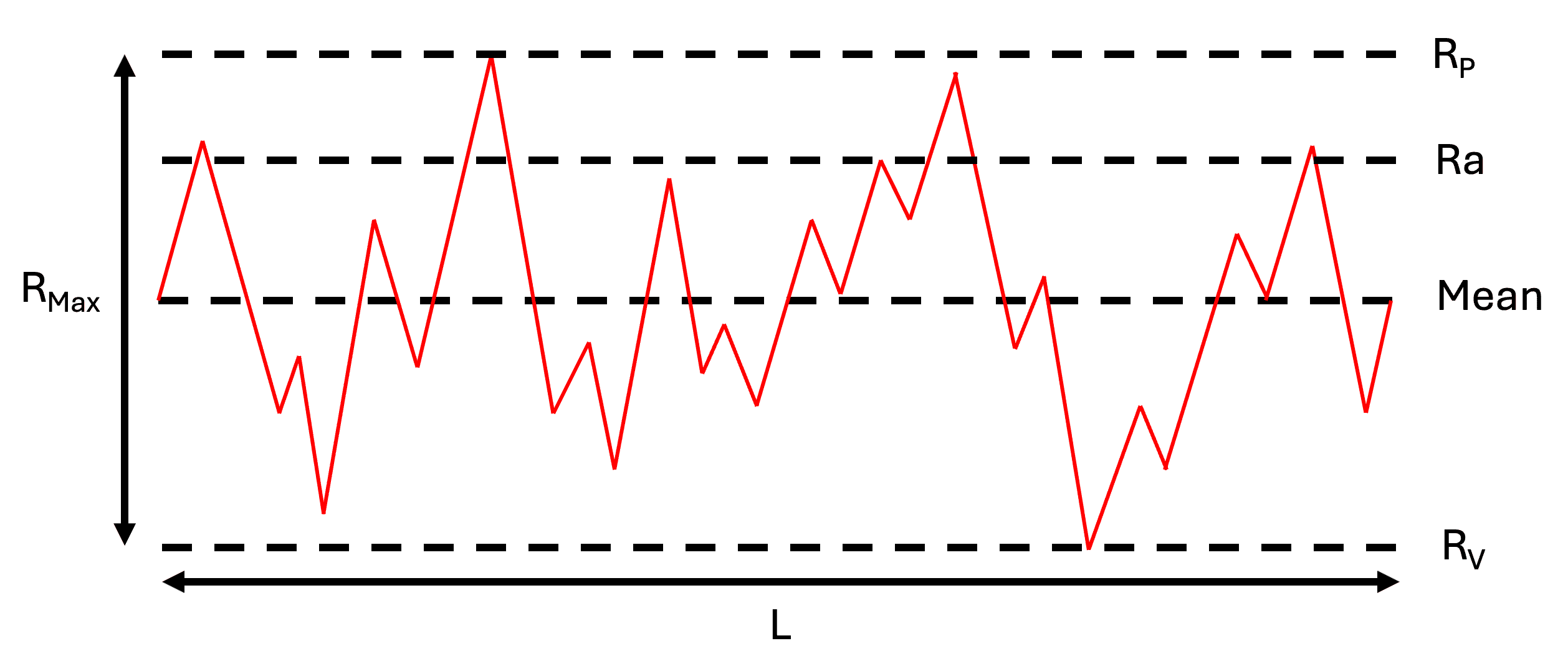}
\caption[Schematic of Average Surface Roughness]{Schematic of a material surface of length $L$ showing how the average roughness, Ra, is defined in Equation \eqref{eq:average-roughness-Ra}. $R_{P}$ and $R_{V}$ indicate the locations of the largest asperity (Peak) and the deepest interstice (Valley) relative to the mean with a maximum roughness, $R_{Max}$, and average roughness, Ra.}
\label{fig:surface-roughness-schematic}
\end{figure}

There is a resistance to heat flow at the interface of two materials due to the surface roughness of the materials. A very common surface roughness specification, which was used to machine the parts in this cryostat, is known as the average roughness, Ra, which is the average between peaks and valleys on a surface as shown in Figure \ref{fig:surface-roughness-schematic}. The formula for average roughness is as follows, 

\begin{equation}
    \text{Ra}=\frac{1}{L}\int_{0}^{L}|y(x)|dx
    \label{eq:average-roughness-Ra}
\end{equation}
where $|y(x)|$ is the vertical deviation from the mean line at a distance $x$ along the surface. The default, as-machined, average roughness for most machine shops is $3.2\text{ \ensuremath{\mu}m}$ ($128\text{ \ensuremath{\mu}in}$), while lower finishes will need to be specified in a mechanical drawing.\footnote{Imperial unit is in micro inches. 1\text{ \ensuremath{\mu}m}=40\text{ \ensuremath{\mu}in}.} 

Consequently, the real contact area between the materials is only due to the asperities at the surface and can be as low as 1\% of the nominal surface area of the materials. Thus, the heat transfer at this interface is via conduction through the asperities and radiation between the remaining gaps at the surface. This decreases the thermal contact conductance and creates a temperature drop at the interface of the two materials. This thermal contact conductance can be increased by three mechanisms: 1.) applying a force between the materials, 2.) reducing the surface roughness of the materials, or 3. ) placing an interstitial material in between the two main contact materials. However, at cryogenic temperatures the thermal contact conductance for a fixed load decreases because the yield strength of metals slightly increases (40\% for copper \cite{simonPropertiesCopperCopper1992}) making the asperities even harder to deform, further motivating the importance of low surface roughness. Additionally, for temperatures below 20 K, Kapitza resistance or thermal boundary resistance becomes significant as as it scales as $T^{-3}$. This phenomenon arises from Acoustic Mismatch Theory (AMT) where heat is primarily transported by phonons (quantized lattice vibrations) behaving as acoustic waves. As the phonon wavelength increases with decreasing temperature, the interface acts as a barrier to energy transmission, resulting in a temperature drop at the interface \cite{pollackKapitzaResistance1969}. 

The thermal contact conductance across metal-metal interfaces has been measured to increase as the surface roughness decreases and vary linearly with applied contact force in the temperature range of 4-20 K \cite{dillonThermalInterfaceMaterial2017,bermanThermalContactLow1958a,dhuleyPressedCopperGoldplated2019a} and at 300 K \cite{tariqExperimentalInvestigationThermal2016}. In particular, the thermal contact conductance across C10100 copper-copper interfaces was compared with different interstitial materials (bare copper, indium foil, Apiezon-N grease, and gold) at 4 K given a fixed load and surface roughness \cite{dillonThermalInterfaceMaterial2017}. Indium and gold are used because they are soft metals relative to copper and fill the interstices at the surface of the copper with an applied contact force that increases the thermal contact conductance.\footnote{Yield Strength of Indium, Gold, and Copper respectively: 4.5 MPa, 100 MPa, and 300 MPa \cite{simonPropertiesCopperCopper1992,ekinExperimentalTechniquesLowtemperature2006a}.} Apiezon-N grease can also be used because it fills the interstices completely without an applied contact force and exhibits the same effect. However, indium or Apiezon-N grease are not suitable for this cryostat as it is designed to be bakeable to 500 K and contain UHV compatible materials. The thermal contact conductance of a gold-plated C10100 copper-copper interface with a surface roughness of $1.6\text{ \ensuremath{\mu}m}$ was measured at 4 K to be $2155\text{ }\frac{\text{W}}{\text{m}^{2}\text{-K}}$ while the bare C10100 copper-copper interface was measured to be $129\text{ }\frac{\text{W}}{\text{m}^{2}\text{-K}}$ given a 458 N contact force and \cite{dillonThermalInterfaceMaterial2017}. By contrast, the thermal contact conductance of a bare copper-copper interface with a surface roughness of $1.6\text{ \ensuremath{\mu}m}$ was measured with the same force at 300 K to be $40\text{ }\frac{\text{kW}}{\text{m}^{2}\text{-K}}$, a factor of 300 greater, showing the strong temperature effect \cite{tariqExperimentalInvestigationThermal2016}. Recognizing that the thermal contact conductance scales linearly with applied force we can write an empirical expression to calculate the thermal contact conductance for an arbitrary force at 4 K, 
\begin{equation}
    h_{\text{c}}(F)=h_{\text{c,458\,N}}\frac{F}{458\,\text{N}}
    \label{eq:thermal-contact-conductance-vs-applied-force}
\end{equation}
where $F$ is the contact force in N and $h_{\text{c,458\,N}}$ is the thermal contact conductance in $\frac{\text{W}}{\text{m}^{2}\text{-K}}$ measured with a 458 N load \cite{dillonThermalInterfaceMaterial2017}. 

In our cryostat, the circulating helium cools the 4 K annular plate and serves as a cold bath for the 4 K experiment plate that houses our ion trap. Therefore, we need to generate a high thermal contact conductance between the 4 K annular-experiment plate joint to enable optimal heat extraction. Recognizing that the thermal contact conductance between a gold-plated C10100 copper-copper interface is a factor of 17 higher than that of bare copper-copper interface at 4 K \cite{dillonThermalInterfaceMaterial2017}, we chose to gold plate the annular and experiment plates. As the thermal contact conductance is sensitive to surface degradation, the gold plating has an added benefit of preventing copper-oxide layers forming that will degrade the thermal contact conductance over time after repeated exposure to air. 

We will now present a simple thermal model to determine the required force at the 4 K annular-experiment plate joint to enable optimal heat extraction. We model the radial heat flow through the experiment plate and then across the 4 K annular-experiment plate joint as follows,
\begin{equation}
    \dot{Q}=\left(\frac{G_{\text{EP}}G_{\text{c}}}{G_{\text{EP}}+G_{\text{c}}}\right)\Delta T
    \label{eq:4K-plate-conductance-with-contact}
\end{equation}
$G_{\text{EP}}=\frac{2\pi t_{\text{Cu}}k_{\text{Cu}}}{\ln(\frac{r_{\text{o}}}{r_{\text{i}}})}$ is the radial conductance of the 4 K experiment plate where $r_{\text{o}}=8.4\times10^{-2}\text{ m}$ and $r_{\text{i}}=1.1\times10^{-2}\text{ m}$ are the outer and inner radii of the plate respectively, $t_{\text{Cu}}=6\text{ mm}$ is the thickness of the plate, and $k_{\text{Cu}}=320\text{ }\frac{\text{W}}{\text{m-K}}$ is the thermal conductivity of C10100 at 4 K. $G_{\text{c}}=h_{\text{c}}A_{\text{c}}$ is the conductance across the 4 K annular-experiment plate joint parametrized by the thermal contact conductance $h_{\text{c}}$ in $\frac{\text{W}}{\text{m}^{2}\text{-K}}$ and the contact surface area of the joint $A_{\text{c}}$ in $\text{m}^{2}$. The term in parentheses represents the total thermal conductance of the system $G_{\text{tot}}$ in $\frac{\text{W}}{\text{K}}$. Thus, the design goal is to determine the force across the joint to make $G_{\text{c}}\gg G_{\text{EP}}$ such that $G_{\text{tot}}\approx G_{\text{EP}}$ - i.e., the radial heat flow through the 4 K experiment plate is not limited by the contact conductance. Given this geometry, $G_{\text{EP}}=6\text{ \ensuremath{\frac{\text{W}}{\text{K}}}}$ meaning that $G_{\text{c}}\approx1000\text{ \ensuremath{\frac{\text{W}}{\text{K}}}}$ will suffice.

\begin{figure}[!!h]
\centering
\includegraphics[width=0.75\textwidth]{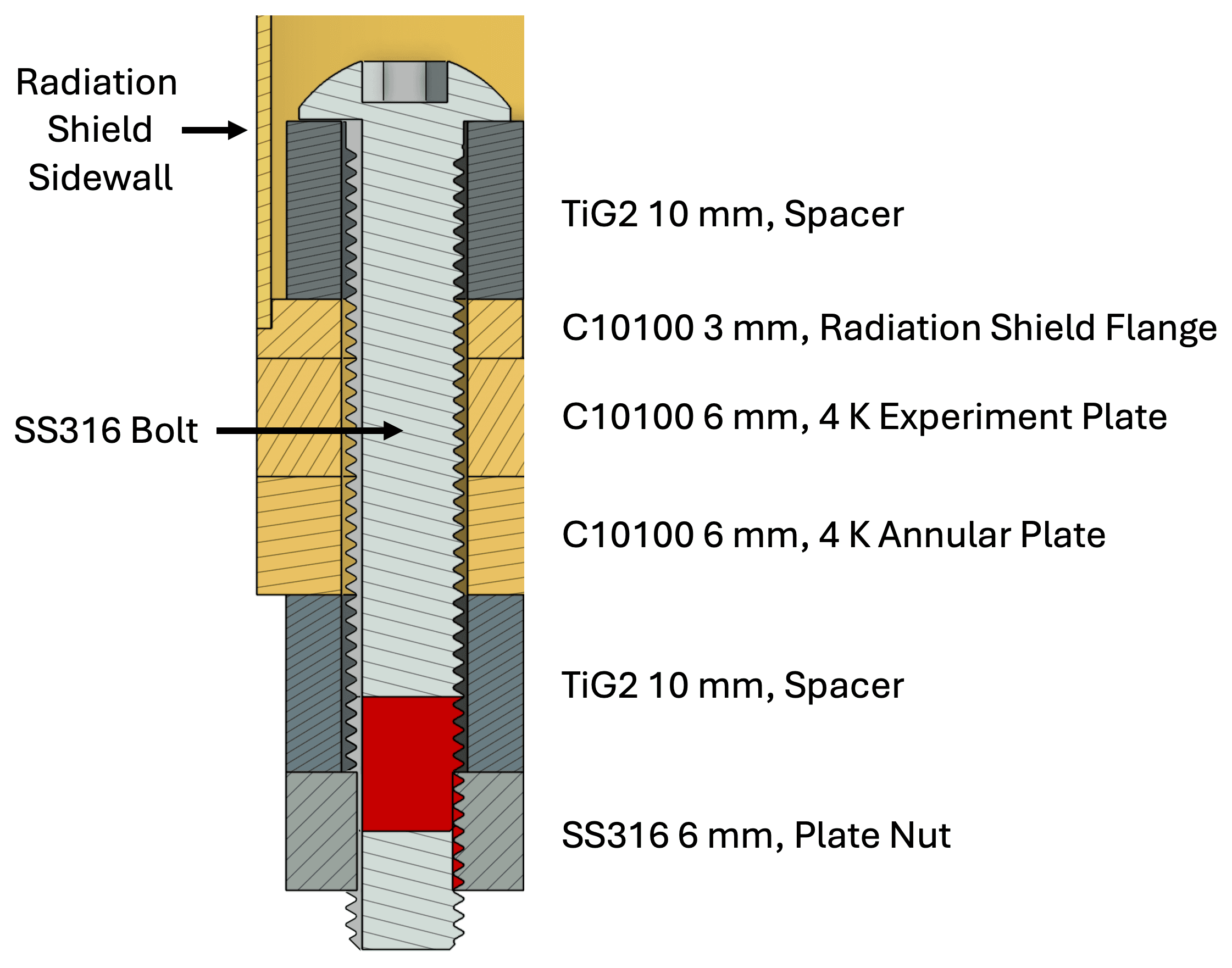}
\caption[Cryo Clamp Assembly]{CAD drawing of the cryo clamp. }
\label{fig:cryo-clamp}
\end{figure}

We will now discuss the cryo clamp design built in this cryostat that reliably generates sufficient force to create a conductance, $G_{\text{c}}\approx1000\text{ \ensuremath{\frac{\text{W}}{\text{K}}}}$. Generating a force can be accomplished by clamping the two copper plates together with stainless steel bolts, but the differential thermal contraction between stainless steel and copper from 300 K to 4 K will cause the bolts to loosen. Thus, one needs to apply additional prestress to the stainless steel bolts at room temperature to ensure that the joint remains tight at 4 K. Lock washers can also be implemented to help mitigate this effect. However, repeatedly applying high prestress to bolts can easily lead to stripping threads and galling especially if the thread is in a softer metal such as copper or aluminum, motivating the use of helicoils and silver plating the bolts for lubrication. Another common choice of bolt material is brass because it has a favorable differential contraction from 300 K to 4 K relative to copper such it will self clamp at 4 K, but it is not UHV compatible. We remedy this issue by engineering a self-clamping mechanism (the cryo clamp) that utilizes the differential thermal contraction of UHV compatible stainless steel, titanium, and copper from 300 K to 4 K and eliminates the need for helicoils.\footnote{Thermal contractions of TiG2, SS316, C10100, and Brass (35\% Zn) from 300 K to 4 K, $\frac{\Delta L}{L}$ (\%): 0.151, 0.297, 0.324, 0.384 \cite{ekinExperimentalTechniquesLowtemperature2006a}.} The the part stack of the cryo clamp consists of the C10100 4 K annular plate, 4 K experiment plate, and 4 K radiation shield flange sandwiched between two TiG2 spacers by 16 equally spaced 1/4-20$\times$ 0.875 silver-plated SS316 bolts as shown in Figure \ref{fig:cryo-clamp}. The bolts snare the top titanium spacer and are tapped into a SS316 plate nut held beneath the bottom titanium spacer. Corresponding through holes are machined in the C10100 and TiG2 parts for the SS316 bolts.

\begin{figure}[!!h]
\centering
\includegraphics[width=0.75\textwidth]{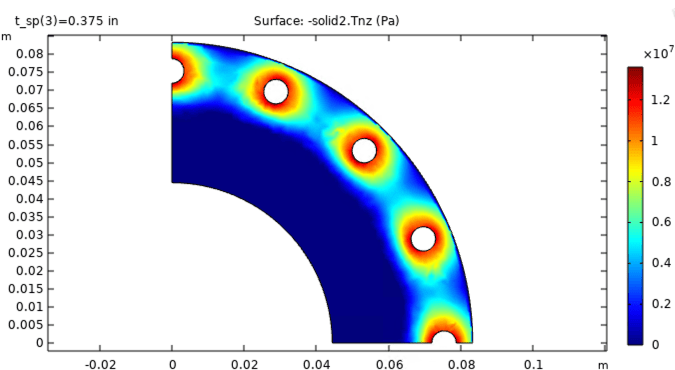}
\caption[Cryo Clamp Contact Force Simulation]{COMSOL simulation of the contact pressure distribution of the 4 K annular-experiment plate joint driven by the differential thermal contraction of C10100, TiG2, and SS316 from 300 K to 4 K from 300 K to 4 K. The real contact area of the joint is localized around the bolts to be $A_{\text{c}}=5.3\times10^{-3}\text{ m}^{2}$. Integrating over this area yields a total force of $4\times10\text{ kN}=40\text{ kN}$ that produces a thermal contact conductance to of $h_{\text{c}}(40\,\text{kN})=188\text{ \ensuremath{\frac{\text{kW}}{\text{m}^{2}\text{-K}}}}$ using Equation \eqref{eq:thermal-contact-conductance-vs-applied-force}, that ensures $G_{\text{c}}=h_{\text{c}}A_{\text{c}}=1000\text{ \ensuremath{\frac{\text{W}}{\text{K}}}}$.}
\label{fig:contact-force-simulation}
\end{figure}

Using COMSOL Finite-Element-Analysis (FEM) software, we simulated the pressure distribution at the 4 K annular-experiment plate joint driven by the differential thermal contraction of C10100, TiG2, and SS316 from 300 K to 4 K and determined an optimal TiG2 spacer thickness of 10 mm as shown in Figure \ref{fig:contact-force-simulation}. To improve computational efficiency and reduce simulation runtime, the COMSOL model took advantage of the system's rotational symmetry. By modeling only a quadrant of the plate and applying symmetry boundary conditions, the mesh density was optimized without sacrificing the accuracy of the full-scale solution. We see that the real contact area between the 4 K annular plate and experiment plate is $5.3\times10^{-3}\text{ m}^{2}$ - a factor of 2.6 lower than the nominal surface area of the joint and is localized around the bolt circle. The contact force at the joint is calculated by integrating over this area yielding a force of $4\times10\text{ kN}=40\text{ kN}$. Taking the real surface area to be the contact surface area $A_{\text{c}}=5.3\times10^{-3}\text{ m}^{2}$ and the thermal contact conductance to be $h_{\text{c}}(40\,\text{kN})=188\text{ \ensuremath{\frac{\text{kW}}{\text{m}^{2}\text{-K}}}}$ using Equation \eqref{eq:thermal-contact-conductance-vs-applied-force} with $h_{\text{c,458\,N}}=2155\text{ \ensuremath{\frac{\text{W}}{\text{m}^{2}\text{-K}}}}$ for gold-plated C10100, we get that $G_{\text{c}}=1000\text{ \ensuremath{\frac{\text{W}}{\text{K}}}}$. If instead we used bare C10100, where $h_{\text{c}}(40\,\text{kN})=11\text{ \ensuremath{\frac{\text{kW}}{\text{m}^{2}\text{-K}}}}$ using Equation \eqref{eq:thermal-contact-conductance-vs-applied-force} with $h_{\text{c,458\,N}}=129\text{ \ensuremath{\frac{\text{W}}{\text{m}^{2}\text{-K}}}}$, we get that $G_{\text{c}}=60\text{ \ensuremath{\frac{\text{W}}{\text{K}}}}$, which only limits the radial heat flow through the experiment plate by 10\%, but is at risk of becoming worse overtime to due oxide layer surface degradation.

The optimal TiG2 spacer thickness was simulated without any prestress so in principle the bolts can be hand tightened. In practice, I use lock washers with these bolts and tighten with a wrench until the lock washer is seated on the TiG2 spacer. The recommended tensile force per SS316 1/4-20 bolt is $\approx4.4\text{ kN}$, so the 16 bolts provide enough density such that each bolt is loaded to $\approx\frac{40\text{ kN}}{16}\approx2.5\text{ kN}$. The same clamp design is implemented on the 40 K annular-experiment plate joint as well as both 4 K and 40 K radiation shield lids to ensure sufficient contact conductance. The relative thickness of the TiG2 spacers is not different between the 4 K and 40 K stages as the thermal expansion coefficients for the materials do not change significantly in this temperature range.\footnote{Thermal contractions of TiG2, SS316, C10100, and Brass (35\% Zn) from 300 K to 40 K, $\frac{\Delta L}{L}$ (\%): 0.150, 0.296, 0.322, 0.380 \cite{ekinExperimentalTechniquesLowtemperature2006a}.} 

\subsection{Wiring}
\label{subsec:wiring}

As the wires in the cryostat are heat sunk at each stage, there is a design compromise between having wires with a low thermal conductivity and a low resistivity. The best alternative to C10100 is phosphor bronze (CuSnP alloy) as its thermal conductivity is roughly two orders of magnitude lower and its resistivity is roughly an order of magnitude higher. Manganin (CuMnNi alloy) is the next best alternative with a thermal conductivity similar to phosphor bronze but its resistivity is roughly two orders of magnitude larger than C10100. However, phosphor bronze and manganin are not suitable as the cryostat is designed to have low outgassing and non-magnetic materials. SS316 is alternative wire choice material with a thermal conductivity that is lower than C10100, phosphor bronze, and manganin, but has the highest resistivity out of all these materials. I engineered SS316 wire assemblies, but they ended up being impractical due to the difficulty in soldering/welding the SS316 wires to the copper Micro-D feedthrough soldercups. To circumvent this issue, C10100 wire assemblies were built at the cost of having extremely long wire lengths (1-3 m) between each stage of the cryostat to reduce thermal loading. 

The original design for the cryostat was to have 50 DC signals and 4 RF signals cables routed to the 4 K stage for operating an ion trap and optical cavity. The DC signals are used for the trap electrodes and cavity PZTs while the RF signals are used for the trap RF confinement and $^{171}\text{Yb}^{+}$ qubit control. An additional 22 DC signals are routed to the cold stages for the oven, heaters, and temperature sensors. The temperature sensors are an imperative diagnostic tool for operating the cryostat. The heaters together with the temperature sensors enable servoing the cryostat cold stage temperatures. Additionally, the heaters are used to quickly warm up the JT capillary to 20 K to release condensed hydrogen, which unclogs the capillary, and also to warm up the cryostat to 300 K in 5 hours as discussed in Subsection \ref{subsec:warmup}. The signals are divided into 4 categories: 1.) High Current Signals (S\_HC), 2.) Temperature Signals (S\_T), 3.) High Voltage Signals (S\_HV), and 4.) RF Signals (S\_RF). All the DC signals are routed in the cryostat with 24-34 AWG silver-plated C10100, Kapton coated wires (Accu-Glass Products) and are heat sunk at each cold stage using Micro-D feedthroughs. In particular, filters for the DC trap electrode signals are installed on a PCB connected to a Micro-D header. The wires for the DC signals are soldered or welded at the Micro-D feedthrough soldercups. The RF signals are routed in the cryostat with coaxial cable (Elspec) and are either heat sunk to the cold stages via SMA bulkhead or directly routed to the 4 K stage. The electrical components and wire routing for this original cryostat design are summarized in Tables \ref{tab:part-list-for-electrical-components-in-cryostat} and \ref{tab:wire-routing-original-design}. 

In this work, I installed the Alpine Quantum Technologies (AQT) Pine Trap at the 4 K experiment plate. This involved routing 5 DC signals and 1 RF signal for the trap electrodes and 6 DC signals for the cavity PZTs. A filter board for the DC electrodes was installed on a Micro-D header and is discussed in Section \ref{sec:trap-electrode-filtering-circuits}. An additional 22 DC signals were routed to the cold stages for the oven, heaters, and temperature sensors. This is cryo0 and the electrical components and wire routing are summarized in Tables \ref{tab:part-list-for-electrical-components-in-cryostat} and \ref{tab:wire-routing-cryo0}. For the second cryostat, 22 DC signals were routed to the cold stages for heaters and temperature sensors. This is cryo1 and the electrical components and wire routing are summarized in Tables \ref{tab:part-list-for-electrical-components-in-cryostat} and \ref{tab:wire-routing-cryo1}.

\begin{table}[!!h]
\centering
\tablefont
\caption[Parts List of Electrical Components in Cryostat]{Parts list of the electrical components in the cryostat.}
\label{tab:part-list-for-electrical-components-in-cryostat}
\begin{tabularx}{\textwidth}{>{\centering\arraybackslash}p{0.38\textwidth}c}
\toprule
\textbf{Schematic Label} & \textbf{Part} \\
\midrule
Temperature Sensor & Diode Sensor (Lake Shore Cryotronics, PN: DT-670B-SD) \\
\midrule
Micro-D Soldercup Feedthrough & \makecell{25 Pin M/F (Omnetics, PN: MDLS-025-P00-SS-RH) \\15 Pin M/F (Omnetics, PN: MDLS-015-P00-SS-RH) \\9 Pin M/F (Omnetics, PN: MMDS-009-N00-SS-HT)} \\
\midrule
Micro-D Header Feedthrough & 25 Pin M (Milnec PN:MDP4-N25X3-H) \\
\midrule
SMA Bulkhead Feedthrough & SMA (Pasternack, PN: PE9184NH) \\
\midrule
DC Wires & \makecell{24 AWG-34 AWG (Accu-Glass Products) \\Ag-plated C10100, Kapton Coated} \\
\midrule
Solder and Flux & \makecell{Solder (Accu-Glass Products, PN: 110796) \\Flux (Accu-Glass Products, PN: 110797)} \\
\midrule
RF Wires & Coaxial Cable (Elspec, PN: SF047 CRYO) \\
\bottomrule
\end{tabularx}
\end{table}

\begin{table}[!!h]
\centering
\tablefont
\caption[Wire Routing for Original Cryostat Design]{Wire routing for original design of cryostat. Blank cells in feedthrough columns indicate that the signal was not connected to that feedthrough. }
\label{tab:wire-routing-original-design}
\begin{tabularx}{\textwidth}{>{\centering\arraybackslash}p{0.15\textwidth}cccc}
\toprule
\textbf{Signals} & \textbf{300 K Feedthrough} & \textbf{40 K Feedthrough} & \textbf{4 K Feedthrough} & \textbf{Destination/Purpose}\\
\midrule
S\_HV1\_\{1-25\} & 300K\_FT\_HV1 & 40K\_FT\_HV1 & 4K\_FT\_HV1 & \makecell{4 K Filter Board for\\DC Trap Electrodes}\\
\midrule
S\_HV2\_\{1-25\} & 300K\_FT\_HV2 & 40K\_FT\_HV2 & 4K\_FT\_HV2 & \makecell{4 K Filter Board for\\DC Trap Electrodes\\\& 4 K Cavity PZTs}\\
\midrule
S\_HC\_\{1,2\} & 300K\_FT\_HC & - & - & \makecell{Heater on 40 K \\Annular Plate }\\
\midrule
S\_HC\_\{3,4\} & 300K\_FT\_HC & 40K\_FT\_HC & - & \makecell{Heater on 40 K \\Experiment Plate} \\
\midrule
S\_HC\_\{5,6\} & 300K\_FT\_HC & 40K\_FT\_HC & - & \makecell{Heater on 4 K \\Annular Plate} \\
\midrule
S\_HC\_\{7,8\} & 300K\_FT\_HC & 40K\_FT\_HC & 4K\_FT\_HCT & \makecell{Heater on 4 K \\Experiment Plate} \\
\midrule
S\_HC\_\{9,10\} & 300K\_FT\_HC & 40K\_FT\_HC & - & Oven at 40 K Stage \\
\midrule
S\_HC\_\{11,12\} & - & - & - & Extra Signals \\
\midrule
S\_T\_\{1-4\} & 300K\_FT\_T & - & - & \makecell{Temperature Sensor on \\40 K Annular Plate} \\
\midrule
S\_T\_\{5-8\} & 300K\_FT\_T & 40K\_FT\_T & - & \makecell{Temperature Sensor on \\4 K Annular Plate} \\
\midrule
S\_T\_\{9-12\} & 300K\_FT\_T & 40K\_FT\_T & 4K\_FT\_HCT & \makecell{Temperature Sensor \\on Ion Trap Mount} \\
\midrule
S\_RF\_\{1,2\} & 300K\_FT\_RF & 40K\_FT\_RF & 4K\_FT\_RF & \makecell{4 K Resonator \\and Pickoff} \\
\midrule
S\_RF\_\{3,4\} & 300K\_FT\_RF & 40K\_FT\_RF & 4K\_FT\_RF & \makecell{4 K Antenna for \\ $^{171}\text{Yb}^{+}$ Qubit Control} \\
\bottomrule
\end{tabularx}
\end{table}

\begin{table}[!!h]
\centering
\tablefont
\caption[Wire Routing for Cryo0]{Wire routing for cryo0 in this work. Blank cells in feedthrough columns indicate that the signal was not connected to that feedthrough.}
\label{tab:wire-routing-cryo0}
\begin{tabularx}{\textwidth}{>{\centering\arraybackslash}p{0.15\textwidth}cccc}
\toprule
\textbf{Signals} & \textbf{300 K Feedthrough} & \textbf{40 K Feedthrough} & \textbf{4 K Feedthrough} & \textbf{Destination/Purpose}\\
\midrule
S\_HV1\_\{1-5\} & 300K\_FT\_HV1 & 40K\_FT\_HV1 & 4K\_FT\_HV1 & \makecell{4 K Filter Board for\\DC Trap Electrodes}\\
\midrule
S\_HV2\_\{1-6\} & 300K\_FT\_HV2 & 40K\_FT\_HV2 & 4K\_FT\_HV2 & 4 K Cavity PZTs\\
\midrule
S\_HC\_\{1,2\} & 300K\_FT\_HC & - & - & \makecell{Heater on 40 K \\Annular Plate} \\
\midrule
S\_HC\_\{3,4\} & 300K\_FT\_HC & 40K\_FT\_HC & - & \makecell{Heater on 40 K \\Experiment Plate} \\
\midrule
S\_HC\_\{5,6\} & 300K\_FT\_HC & 40K\_FT\_HC & - & \makecell{Heater on 4 K \\Annular Plate}\\
\midrule
S\_HC\_\{7,8\} & 300K\_FT\_HC & 40K\_FT\_HC & 4K\_FT\_HCT & \makecell{Heater on 4 K \\Trap Riser} \\
\midrule
S\_HC\_\{9,10\} & 300K\_FT\_HC & 40K\_FT\_HC & - & Oven at 4 K Stage \\
\midrule
S\_T\_\{1-4\} & 300K\_FT\_T & - & - & \makecell{Temperature Sensor on \\40 K Annular Plate} \\
\midrule
S\_T\_\{5-8\} & 300K\_FT\_T & 40K\_FT\_T & - & \makecell{Temperature Sensor on \\4 K Annular Plate} \\
\midrule
S\_T\_\{9-12\} & 300K\_FT\_T & 40K\_FT\_T & 4K\_FT\_HCT & \makecell{Temperature Sensor \\on Ion Trap Mount} \\
\midrule
S\_RF\_\{1\} & 300K\_FT\_RF & - & - & 4 K Resonator \\
\bottomrule
\end{tabularx}
\end{table}

\begin{table}[!!h]
\centering
\tablefont
\caption[Wire Routing for Cryo1]{Wire routing for cryo1 in this work. Blank cells in feedthrough columns indicate that the signal was not connected to that feedthrough. }
\label{tab:wire-routing-cryo1}
\begin{tabularx}{\textwidth}{>{\centering\arraybackslash}p{0.15\textwidth}cccc}
\toprule
\textbf{Signals} & \textbf{300 K Feedthrough} & \textbf{40 K Feedthrough} & \textbf{4 K Feedthrough} & \textbf{Destination/Purpose}\\
\midrule
S\_HC\_\{1,2\} & 300K\_FT\_HC & - & - & \makecell{Heater on 40 K \\Annular Plate} \\
\midrule
S\_HC\_\{3,4\} & 300K\_FT\_HC & 40K\_FT\_HC & - & \makecell{Heater on 40 K \\Experiment Plate} \\
\midrule
S\_HC\_\{5,6\} & 300K\_FT\_HC & 40K\_FT\_HC & - & \makecell{Heater on 4 K \\Annular Plate} \\
\midrule
S\_HC\_\{7,8\} & 300K\_FT\_HC & 40K\_FT\_HC & 4K\_FT\_HCT & \makecell{Heater on 4 K \\Experiment Plate} \\
\midrule
S\_T\_\{1-4\} & 300K\_FT\_T & - & - & \makecell{Temperature Sensor on \\40 K Annular Plate} \\
\midrule
S\_T\_\{5-8\} & 300K\_FT\_T & 40K\_FT\_T & - & \makecell{Temperature Sensor on \\4 K Annular Plate} \\
\midrule
S\_T\_\{9-12\} & 300K\_FT\_T & 40K\_FT\_T & 4K\_FT\_HCT & \makecell{Temperature Sensor on\\4 K Experiment Plate} \\
\bottomrule
\end{tabularx}
\end{table}

\section{Heat Load Budget}
\label{sec:heat-load-budget}

A thermal analysis is presented in this section to calculate the heat load on the cryostat cold stages. Heat is exchanged between the 300 K stage and the cold stages via conduction and radiation
\begin{equation}
    \dot{Q}=\dot{Q}_{\text{cond}}+\dot{Q}_{\text{rad}}
    \label{eq:heat-transfer-cond-conv-heat-budget}
\end{equation}
and we assume that the heat transferred via conduction and convection through the vacuum is negligible and is not included in the model. The heat transferred via conduction through a solid material is given as 
\begin{equation}
   \dot{Q}_{\text{cond}}=\bar{k}\frac{A}{L}\Delta T
    \label{eq:Fouriers-law-solids-heat-budget}
\end{equation}
where $\bar{k}$ is the mean thermal conductivity of the material in $\frac{\text{W}}{\text{m-K}}$, $A$ is the cross-sectional area of the material in $\text{m}^{2}$, $L$ is the length of the material in $\text{m}$, and $\Delta T$ is the temperature gradient through the material in $\text{K}$. The mean thermal conductivity is used here because the thermal conductivity, $k(T)$, is temperature dependent and is given as $\bar{k}=\frac{1}{\Delta T}\int_{T_{1}}^{T_{2}}k(T)dT$ where $\Delta T=T_{2}-T_{1}$. The heat load on the cold stages via conduction is due to the support rods and cryostat wiring and is summarized in Table \ref{tab:heat-load-budget}. We assume the support rod or wire with cross sectional area $A$ and length $L$ is heat sunk to each stage, then calculate the mean thermal conductivity of the material \cite{ruleLowtemperatureThermalConductivity1989,simonPropertiesCopperCopper1992} over that temperature gradient to compute the heat load using Equation \eqref{eq:Fouriers-law-solids-heat-budget}.\footnote{RF coaxial cable is Elspec, PN: SF047 CRYO. Thermal conductance is stated in datasheet.}

The relevant theory for radiative heat transfer is presented here to calculate the thermal load on the cold stages. When an object at a given temperature receives electromagnetic radiation, it is either absorbed, reflected, or transmitted. In general, it is a combination of all three processes and by conservation of energy we can write,
\begin{equation}
    \alpha+\beta+\tau=1
    \label{eq:absorb-reflect-transmit-energy-conservation}
\end{equation}
where $\alpha $, $\beta$ , and $\tau$  is the absorptivity, reflectivity, and transmissivity respectively. If $\tau=0$, all the energy is either absorbed or reflected and the object is called opaque. If $\tau=0$ and $\beta=0$, then we get that $\alpha=1$ and the object absorbs all the energy and is called a blackbody. The radiative power emitted from an object is given as,
\begin{equation}
    \dot{Q}_{\text{rad}}=\epsilon A\sigma T^{4}
    \label{eq:power-emitted-from-blackbody}
\end{equation}
where $\sigma=5.67\times10^{-8}\frac{\text{W}}{\text{m}^{2}\text{-K}^{4}}$ is the Stefan-Bolztmann constant, $A$ is the surface area of the object in $\text{m}^{2}$, $T$ is the temperature of the surface in $\text{K}$, and $\epsilon$  is the emissivity of the object. The emissivity of an object quantifies how well it emits thermal radiation and is expressed as a fraction of emitted power of the blackbody ($\epsilon=1$). In addition, the peak wavelength of radiation emitted from a blackbody is given by the Wien's displacement law as follows, 
\begin{equation}
    \lambda_{\text{peak}}=\frac{2.89\times10^{-3}\text{\,m-K}}{T}
    \label{eq:Wien-displacement-law}
\end{equation}
where $T$ is the temperature of the blackbody. The radiative heat transfer from object 1 to object 2 is given as
\begin{equation}
    \dot{Q}_{\text{rad,1-2}}=\frac{A_{1}\sigma(T_{1}^{4}-T_{2}^{4})}{\frac{1-\varepsilon_{1}}{\varepsilon_{1}}+\frac{1}{F_{12}}+\frac{1-\varepsilon_{2}}{\varepsilon_{2}}\frac{A_{1}}{A_{2}}}
    \label{eq:radiative-heat-transfer-heat-budget}
\end{equation}
where $T_{1}$ and $T_{2}$ are the temperatures of objects 1 and 2 in $\text{K}$, $A_{1}$ and $A_{2}$ are the surface areas of objects 1 and 2 in $\text{m}^{2}$, $\epsilon_{1}$ and $\epsilon_{2}$ are the emissivities of objects 1 and 2, and $F_{12}$ is the fraction of radiative heat due to object 1 that is communicated to object 2, denoted the view factor which is dependent on solid angle. View factors for various geometries can be derived or found in heat transfer textbooks and must satisfy $A_{1}F_{12}=A_{2}F_{21}$ by conservation of energy and $F_{12},F_{21}\in[0,1]$ \cite{brewsterThermalRadiativeTransfer1992}. 

By examining Equation \eqref{eq:radiative-heat-transfer-heat-budget}, the radiative heat load on the cold stages can be decreased by engineering radiation shields with low emissive surfaces. The emissivity depends on the material of the surface, the peak wavelength of radiation presented to the surface, and the surface roughness. I will first present how the emissivity for the metals relevant for this cryostat can be calculated - copper, gold, stainless steel, and aluminum. The Drude model describes the response of a metal's free-electron gas due to incidence electromagnetic radiation. The radiation relevant for heat transfer in this cryostat are the peak wavelengths emitted from 300 K and 40 K surfaces which are $10\text{ \ensuremath{\mu}m}$ and $72\text{ \ensuremath{\mu}m}$ respectively using Equation \eqref{eq:Wien-displacement-law}. At these wavelengths, the frequency is much lower than the plasma frequency of these metals which results in a very small skin depth. Consequently, these metals are opaque to this radiation, $\tau=0$, and so $\alpha+\beta=1$ using Equation \eqref{eq:absorb-reflect-transmit-energy-conservation}. If the participating surfaces in the radiative heat exchange are isothermal over the entire surface we can assume that the emissivity is equal to the absorptivity, $\epsilon=\alpha$ , which is known as Kirchhoff's Law \cite{brewsterThermalRadiativeTransfer1992}. Thus, the emissivity in this case can be determined by the reflectivity, $\epsilon=1-\beta$ . The reflectivity can be calculated using the normal incidence Fresnel equation together with the complex refractive index derived from the Drude model,
\begin{equation}
    \beta(\lambda)=\frac{(n-1)^{2}+k^{2}}{(n+1)^{2}+k_{}^{2}}
    \label{eq:reflectivity-Fresnel-Drude-normal-indcidence}
\end{equation}
where $n$ and $k$ are the optical constants of the material and are the real and imaginary parts of the complex refractive index, $\tilde{n}(\lambda)=n(\lambda)+ik(\lambda)$. The empirical optical constants for most materials are well documented \cite{palikHandbookOpticalConstants1985,brewsterThermalRadiativeTransfer1992}. For the wavelengths $10\text{ \ensuremath{\mu}m}$ and $72\text{ \ensuremath{\mu}m}$ (frequencies much lower than the plasma frequency), the emissivity can be conveniently written as follows, 
\begin{equation}
    \epsilon(\lambda)\approx0.365\sqrt{\frac{\rho}{\lambda}}
    \label{eq:Hagen-Ruben}
\end{equation}
where $\rho$  is the resistivity of the material in $\Omega\text{-m}$ and $\lambda$  is the radiation wavelength and is called the the Hagen-Rubens equation \cite{brewsterThermalRadiativeTransfer1992,ekinExperimentalTechniquesLowtemperature2006a}. Thus, it is advantageous to use metals with low resistivities for radiation shields. As a corollary to that, the emissivity of metals will decrease with temperature as the resistivity also decreases with temperature. 

We will now present how the surface roughness of the radiation shield influences its effective reflectivity, and thus the emissivity since $\epsilon=1-\beta$. A surface is considered electromagnetically smooth if the phase difference between light rays reflecting from the top of a peak and the bottom of a valley on the surface is small, maintaining constructive interference and enabling specular reflection. We can write this phase difference as $\Delta\phi=\frac{2\pi}{\lambda}2\text{Ra}\cos\theta$  where $\lambda$  is the wavelength of the radiation, and $2\text{Ra}\cos\theta$  is the average path difference where Ra is the average roughness defined in Equation \eqref{eq:average-roughness-Ra} and $\theta$  is the angle of incidence. Imposing that this phase difference be $\Delta\phi<\frac{\pi}{2}$, we get that
\begin{equation}
    \text{Ra}\ll\frac{\lambda}{8\cos\theta}
    \label{eq:Rayleigh-roughness-criterion}
\end{equation}
which is known as the Rayleigh roughness criterion \cite{1987ah...book.....B}. If this inequality is not satisfied, three effects occur that reduce the effective reflectivity of the surface: 1.) the waves can destructively interfere, 2.) the waves will reflect in many different directions, known as diffuse reflection, and 3.) the waves can get trapped in an interstice bouncing multiple times where energy is absorbed on each bounce. The peak wavelengths, respectively, emitted from 300 K and 40 K surfaces are $10\text{ \ensuremath{\mu}m}$ and $72\text{ \ensuremath{\mu}m}$ using Equation \eqref{eq:Wien-displacement-law} which means that the 40 K and 4 K surfaces should have a surface roughness of less than $1.25\text{ \ensuremath{\mu}m}$ and $9\text{ \ensuremath{\mu}m}$ using Equation \eqref{eq:Rayleigh-roughness-criterion}. Standard machine shop tools can reliably achieve surface roughnesses down to $0.8\text{ \ensuremath{\mu}m}$. To reach $<0.8\text{ \ensuremath{\mu}m}$ requires secondary processes like grinding, buffing, or electropolishing. It is important to note that the apparent reflectivity of a surface observed by eye is not necessarily a good proxy to determine if it will be reflective at $10\text{ \ensuremath{\mu}m}$ and $72\text{ \ensuremath{\mu}m}$. Taking the peak wavelength of light emitted from room lights to be $\approx600\text{ nm}$, the surface will be electromagnetically smooth if the roughness is less than 75 nm using Equation \eqref{eq:Rayleigh-roughness-criterion}. Thus, it might not be discernible by eye if a surface will be reflective at $10\text{ \ensuremath{\mu}m}$ and $72\text{ \ensuremath{\mu}m}$.

We will now discuss the material design choice for the radiation shields in this cryostat and also present the estimated radiative heat load on the cold stages. While the ideal emissivity for metals can be calculated using Equation \eqref{eq:Hagen-Ruben}, in reality it also depends on the surface finish so the emissivities used to the calculate the heat load on the cold stages are empirical and are summarized in Table \ref{tab:emissivities-of-various-materials}. These emissivities are reported at 300 K as there is more data for these materials at this temperature with a specified surface finish. A good material choice for the cold stages should not only have a low emissivity, but also needs to be resistant to surface degradation. In particular, polished copper can have an emissivity of 0.02, but when an oxide layer is formed after prolonged exposure to air its emissivity can increase to 0.6 \cite{ekinExperimentalTechniquesLowtemperature2006a}. Polished gold also has an emissivity of 0.02, but does not oxidize \cite{ekinExperimentalTechniquesLowtemperature2006a}. To accommodate the design goal of frequent vacuum cycling, the copper radiation shields were gold-plated to prevent oxidization and maintain low surface emissivity. The thickness of the plating is $\approx1\text{ \ensuremath{\mu}m}$, confirmed by CET, which is 300 times the skin depth of gold for $10\text{ \ensuremath{\mu}m}$ and $72\text{ \ensuremath{\mu}m}$ thermal radiation onto the 40 K and 4 K cold stages respectively. However, surface degradation can still occur on these gold plated surfaces due to fingerprints/contamination, excess scratching, or gross accumulation of adsorbed residual gases. This motivates standard operating procedures of UHV cleanliness and care while handling the radiation shields as well as properly evacuating the cryostat prior to cooldown, see Section \ref{sec:cooldown-and-warmup}.

The vacuum chamber is made of stainless steel which is a standard material choice for manufacturing and has a typical emissivity 0.2, but can be reduced to 0.1 by electropolishing \cite{ekinExperimentalTechniquesLowtemperature2006a,brewsterThermalRadiativeTransfer1992}. The Kimball Physics spherical octagons are delivered electropolished, but we chose not to electropolish the remaining stainless steel parts. First, the factor of two decrease in emissivity only decreases the radiative heat load on the 40 K stage by 8\% as it is more sensitive to the emissivity of the 40 K stage surface. Second, if we eventually air bake the stainless steel parts to $\approx700\text{ K}$ to intentionally create an oxide layer to reduce hydrogen diffusion, the emissivity will be degraded anyway \cite{odakaDependenceOutgassingRate1996a}. The emissivity for polished oxidized stainless steel is not well documented, but if we assume it increases to 0.6, this will only increase the radiative heat load on the 40 K stage by 8\%. The upshot is that electropolishing the stainless steel versus using bare stainless steel will not make a significant difference to the heat load at the 40 K stage. The radiative heat load on the gold plated 40 K stage due to the stainless steel 300 K stage is 1.6 W. The heat load on the gold plated 4 K stage due to the gold plated 40 K stage is only 0.2 mW and is summarized in Table \ref{tab:heat-load-budget}.

The aluminum 300 K radiation shield built for shielding the 40 K stage from the hot ion gauge was electropolished creating an emissivity of 0.03 \cite{brewsterThermalRadiativeTransfer1992,ekinExperimentalTechniquesLowtemperature2006a}. Taking the normal operating temperature of the ion gauge to be 500 K, the radiative heat load on the 40 K stage is 600 mW and the equilibration temperature of the aluminum shield is 430 K. Without the shield, the heat load on the 40 K stage would be 1.3 W. The load could be further decreased by placing a stack of aluminum shields in between the ion gauge and 40 K stage where the load is reduced by a factor of $\approx N+1$ where $N$ is the number of shields, but was not done in this work. 

The emissivity of the fused silica windows was estimated by calculating the reflectivity, transmissivity, and absorptivity of $10\text{ \ensuremath{\mu}m}$ and $72\text{ \ensuremath{\mu}m}$ radiation using documented optical constants. We can calculate the transmissivity through a material with thickness $d$ using,
\begin{equation}
    \tau=\frac{I}{I_{0}}=(1-\beta)^{2}e^{\mu_{\text{a}}d}
    \label{eq:transmissivity-Beers-Law}
\end{equation}
where $\mu_{\text{a}}=\frac{4\pi k}{\lambda}$ is the absorption coefficient, $\beta$  is the reflectivity calculated using Equation \eqref{eq:reflectivity-Fresnel-Drude-normal-indcidence}, and $I_{0}$ and $I$ is the incident intensity and final intensity respectively.\footnote{The skin depth can also be calculated from the optical constant $k$, $\delta=\frac{\lambda}{2\pi k}$. However, this is not the pertinent parameter in this case.} Additionally, we can calculate the absorptivity as whatever light did not survive the trip through the glass,
\begin{equation}
    \alpha=(1-\beta)\left(1-e^{\mu_{\text{a}}d}\right)
    \label{Absorptivity}
\end{equation}
The optical constants at $10\text{ \ensuremath{\mu}m}$ and $72\text{ \ensuremath{\mu}m}$ are not specced for our fused silica windows (Corning 7980). However, the optical constants at these wavelengths have been empirically determined for samples of silicon dioxide ($\text{SiO}_{2}$) glass at 300 K, which we will use for this analysis \cite{brewsterThermalRadiativeTransfer1992,palikHandbookOpticalConstants1985,kitamura2007optical}. At $10\text{ \ensuremath{\mu}m}$ and $72\text{ \ensuremath{\mu}m}$, we find that $\beta=0.1$, $\tau\approx0$, and $\alpha=0.9$ meaning that fused silica is opaque at these wavelengths and we will take the emissivity to be $\epsilon=0.9$. The infrared absorption of fused silica (Corning 7980) is fundamentally governed by the vibrational modes of the $\text{SiO}_{2}$ lattice, making it consistent across other substrates. At $10\text{ \ensuremath{\mu}m}$, the material is opaque due to the resonance with the stretching of the silicon-oxygen bond. As the wavelength increases to $72\text{ \ensuremath{\mu}m}$, the mechanism shifts from internal bond stretching to acoustic modes of the $\text{SiO}_{2}$ lattice \cite{kitamura2007optical}. A common impurity in $\text{SiO}_{2}$ glass manufacturing is hydroxyl (OH) ions which create an absorption band around $2.7\text{ \ensuremath{\mu}m}$, as seen in the Corning 7980 datasheet, but does not effect the absorption defined by the silicon-oxygen lattice itself \cite{brewsterThermalRadiativeTransfer1992,kitamura2007optical}. Consequently, fused silica remains an absorber for both 300 K ($10\text{ \ensuremath{\mu}m}$) and 40 K ($72\text{ \ensuremath{\mu}m}$) thermal radiation. However, for 4 K ($724\text{ \ensuremath{\mu}m}$) thermal radiation we find that $\beta=0.1$, $\tau=0.7$, and $\alpha=0.1$ \cite{palikHandbookOpticalConstants1985}. Thus, fused silica is transparent to thermal radiation of cold stages $\leq$  4 K ($\geq724\text{ \ensuremath{\mu}m}$) as the frequency is too low to excite these molecular and lattice-scale resonances. This is not a concern for this cryostat, but may be for sub 4 K cryostats depending on the application. The radiative heat loads on the 40 K and 4 K stage due to the windows are 600 mW and 0.2 mW respectively and are documented in Table \ref{tab:heat-load-budget}.

In summary, the 40 K stage is required as it consumes the bulk of the heat load (3.2 W) from the 300 K stage, effectively shielding the 4 K stage with only 21 mW of power consumed. Gold is a good plating metal as it does not oxidize, preserving the low emissivity after many cycles of breaking vacuum. The windows contribute the most radiative heat load per unit surface area as the emissivity is $\epsilon=0.9$, so the window apertures should be small. Lastly, direct line-of-sight radiation from the 300 K stage to the 4 K stage should be avoided due to the strong ($T_{1}^{4}-T_{2}^{4}$) dependence in Equation \eqref{eq:radiative-heat-transfer-heat-budget}. The only direct line-of-sight radiation from the 300 K stage to the 4 K stage in this cryostat is through the 3 mm diameter aperture fiber feedthroughs in each radiation shield, but it is still $\approx2\text{ mW}$. This could be improved by machining the apertures off center relative to one another up to the minimum bend radius of the fiber optic cable, but was not implemented in this work. Lastly, it is clear that the 4 K radiation shield is not needed from a heat transfer perspective, however it provides greater cold surface area for cryosorption pumping of hydrogen as discussed in Subsection \ref{subsec:pressure-estimate-using-cryopuming-speed-and-cryosorption-capacity}.

\begin{table}[!!h]
\centering
\tablefont
\caption[Emissivities of Various Materials]{Emissivities, measured at 300 K, of relevant materials used to calculate the radiative heat transfer in the cryostat. The emissivities for the metals are from \cite{ekinExperimentalTechniquesLowtemperature2006a,brewsterThermalRadiativeTransfer1992} while the emissivity for fused silica is calculated using tabulated optical constants at $10\text{ \ensuremath{\mu}m}$ and $72\text{ \ensuremath{\mu}m}$ for silicon dioxide ($\text{SiO}_{2}$) glass at 300 K \cite{brewsterThermalRadiativeTransfer1992,palikHandbookOpticalConstants1985,kitamura2007optical}.}
\label{tab:emissivities-of-various-materials}
\begin{tabularx}{\textwidth}{>{\centering\arraybackslash}p{0.3\textwidth}>{\centering\arraybackslash}p{0.3\textwidth}c}
\toprule
\textbf{Material} & \textbf{Emissivity} & \textbf{Comments} \\
\midrule
Stainless Steel & 0.2 (0.6) & Typical (Oxidized) \\
\midrule
Copper & 0.02 (0.6) & Polished (Oxidized) \\
\midrule
Gold & 0.02 & Polished \\
\midrule
Aluminum & 0.03 (0.6) & Electropolished (Oxidized) \\
\midrule
Fused Silica & 0.0 & Silicon Dioxide ($\text{SiO}_{2}$) glass \\
\bottomrule
\end{tabularx}
\end{table}

\begin{table}[!!h]
\centering
\tablefont
\caption[Heat Load Budget]{Heat load budget of the cryostat cold stages in this work calculated using Equations \eqref{eq:Fouriers-law-solids-heat-budget} and \eqref{eq:radiative-heat-transfer-heat-budget}.}
\label{tab:heat-load-budget}
\begin{tabularx}{\textwidth}{>{\centering\arraybackslash}p{0.3\textwidth}>{\centering\arraybackslash}p{0.3\textwidth}c}
\toprule
\textbf{Heat Source} & \textbf{Load on 40 K Stage (mW)} & \textbf{Load on 4 K Stage (mW)} \\
\midrule
Support Rods & 300 & 5 \\
\midrule
DC Wiring & 70 & 11 \\
\midrule
RF Wiring & - & 3 \\
\midrule
Radiation (Shields) & 1600 & 0.2 \\
\midrule
Radiation (Windows) & 600 & 0.2 \\
\midrule
Radiation (Ion Gauge) & 600 & - \\
\midrule
\makecell{Radiation\\(direct line-of-sight from 300 K)} & - & 2 \\
\midrule
Total & 3170 & 21.4 \\
\bottomrule
\end{tabularx}
\end{table}

\ifSubfilesClassLoaded{%
  \bibliographystyle{unsrt}
  \bibliography{Dissertation}
}{}
\end{document}

%% file: Chapter4.tex
\chapter{Cryogenic Ion Trap System}
\label{chap:cryogenic-ion-trap-system}

There are multiple subsystems that comprise this cryogenic ion trap system. There is the cryostat which provides the cryogenic temperatures and vacuum environment and is discussed in detail in Chapter \ref{chap:Cryostat}. There is also the ion trap, atomic source, and imaging and detection system which provide the electrodynamic confinement of $\text{Yb}^{+}$ ions as well as ion position monitoring and state detection which are covered in Chapters \ref{chap:ion-trap}, \ref{chap:atomic-source}, and \ref{chap:imaging-and-detection}. In this chapter, I will present the additional subsystems that enable conducting cryogenic trapped ion-cavity experiments: 1.) the experimental control system covered in Section \ref{sec:experimental-control-system}, the laser systems covered in Section \ref{sec:laser-systems}, the optomechanical systems covered in Section \ref{sec:optomechanical-systems}, and the in-vacuum maintenance system covered in Section \ref{sec:in-vacuum-maintenance-system}. 

\section{Experimental Control System}
\label{sec:experimental-control-system}

There are many hardware components in a trapped-ion experiment that must be coordinated with precise timing for reliable operation. At long timescales, system parameters such as ion lifetime, laser power, and voltage calibrations must be monitored and updated. At intermediate timescales, continuous-wave (CW) laser beams must be switched to perform Doppler cooling, state preparation, and fluorescence detection. At the shortest timescales, RF frequencies and phases must be rapidly controlled to execute quantum operations.

The experiment is controlled by a field-programmable gate array (FPGA) mounted on a custom breakout board that provides digital input/output lines, analog-to-digital converters (ADCs), digital-to-analog converters (DACs), and direct digital synthesizer (DDS) channels. The DDS channels generate programmable single-frequency RF signals used to drive acousto-optic modulators (AOMs) and perform experiments such as Rabi oscillations, Ramsey spectroscopy, and frequency scans.

The FPGA is programmed using the ARTIQ control framework, which compiles high-level Python code running on a host computer into deterministic FPGA timing sequences. This approach combines the nanosecond timing resolution of FPGA hardware with the accessibility of Python, eliminating the need for low-level hardware description languages such as Verilog or VHDL. Although FPGAs are well suited for deterministic timing, they are less efficient at computationally intensive tasks such as floating-point arithmetic. ARTIQ addresses this limitation through remote procedure calls (RPCs), allowing the FPGA to temporarily delegate complex calculations to the host computer before resuming execution.

ARTIQ also provides a distributed hardware architecture in which device drivers may run on multiple computers connected through a local area network (LAN). Rather than routing all experimental hardware to a single control computer, individual subsystems communicate through a client-server architecture, improving scalability and simplifying laboratory integration. In addition, ARTIQ includes graphical user interfaces, experiment scheduling, and centralized databases for storing experimental data and calibration parameters, enabling automated experiment execution while maintaining awareness of the current state of the system.

In this work, I setup an ARTIQ crate that provides experimental control to cryo0. The breakout board provides 20 digital output lines, 4 digital input lines, 8 DAC channels, 8 ADC channels, and control over 8 DDS chips. This crate provides enough experimental control for laser cooling, micromotion compensation, secular frequency measurements, and ion position measurements for $\text{Yb}^{+}$ ions: 1.) running the 369 nm and 935 nm laser systems, 2.) counting photons from the PMT, 3.) applying DC voltages to trap electrodes, and 4.) applying RF voltages to trap electrodes. For running more complicated experiments that require control of the 638 nm, 976 nm, 411 nm, 435 nm, 1345 nm, and 1650 nm laser systems an additional satellite ARTIQ crate was built by Connor Goham. Instantiating the cryo0 ARTIQ crate into DRTIO mode enables control of this satellite ARTIQ crate - i.e., control of the 638 nm, 976 nm, 411 nm, 435 nm, 1345 nm, and 1650 nm laser systems. This is a compromise as Connor also requires control over these laser systems to conduct experiments on our lab's room temperature ion trap. So instead of building duplicate optical setups, we can simply share these laser systems at the cost of not being able to run experiments simultaneously. 

The main ARTIQ code base for ion-control in our lab was designed and written by Connor and is covered in his thesis. The variables for the devices (ARTIQ crate devices and NDSP devices) in the code were given aliases so that I can simply update the cryo0 device database file to assign the local cryo0 devices to the correct aliases. This allows the code for all experiments common to the room temperature and cryo0 ion trap systems to remain the same: laser cooling, detection, micromotion compensation, secular frequency probing, Rabi flopping, Ramsey sequences, etc. The experiment to measure the position of the ion-cavity mode was written locally for the cryo0 crate. 

\section{Laser Systems}
\label{sec:laser-systems}

\begin{figure}[!!h]
\centering
\includegraphics[width=\textwidth]{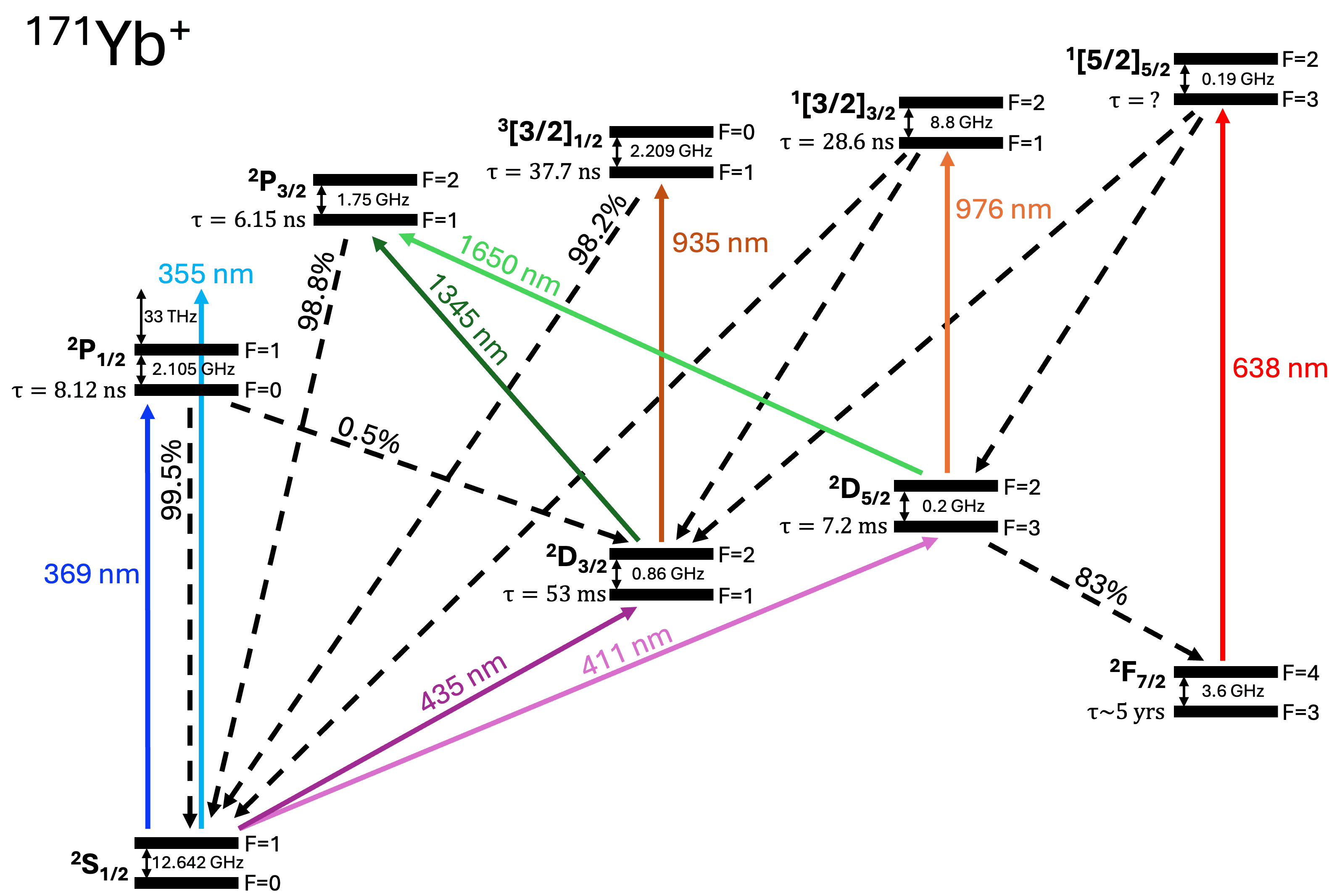}
\caption[$^{171}\text{Yb}^{+}$ level diagram 2]{$^{171}\text{Yb}^{+}$ level diagram. The nuclear spin of $^{171}\text{Yb}^{+}$ is $\text{I}=\frac{1}{2}$. The diagram is shown in the absence of an external magnetic field so the $m_{\text{F}}$ states are degenerate. The branching ratios from the $^{1}[5/2]_{5/2}$ and $^{1}[3/2]_{3/2}$ states are unknown.}
\label{fig:171Yb-level-diagram-Cryogenic-ion-trap-system}
\end{figure}

In this section, I'll discuss the relevant laser systems I built for $\text{Yb}^{+}$ trapped ion control. For the experiments conducted in this work, I trapped solely $^{174}\text{Yb}^+$ ions, but the laser systems were designed to accommodate trapping of $^{171}\text{Yb}^+$ ions for our quantum networking scheme.  The level diagram for $^{171}\text{Yb}^+$ is shown in Figure \ref{fig:171Yb-level-diagram-Cryogenic-ion-trap-system}. Laser light at 369 nm and 935 nm are required for laser cooling. Cooling is performed on the $^{2}\text{S}_{1/2}-^{2}\text{P}_{1/2}$ transition at 369 nm, but the optically excited $^{171}\text{Yb}^{+*}$ ions in the $^{2}\text{P}_{1/2}$ state leak into the $^{2}\text{D}_{3/2}$ state and are repumped back to $^{2}\text{S}_{1/2}$ via the $^{2}\text{D}_{3/2}-^{3}[3/2]_{1/2}$ transition at 935 nm. Additionally, state preparation and imaging and state detection is performed on the $^{2}\text{S}_{1/2}-^{2}\text{P}_{1/2}$ transition at 369 nm. Optically excited $^{171}\text{Yb}^{+*}$ ions in the $^{2}\text{P}_{1/2}$ and $^{2}\text{D}_{3/2}$ states can undergo hydrogen collision-induced shelving to the $^{2}\text{F}_{7/2}$ state or simply decay to the the $^{2}\text{F}_{7/2}$ state from the $^{2}\text{D}_{5/2}$. These ions can be repumped back to the $^{2}\text{S}_{1/2}$ state via the $^{2}\text{F}_{3/2}-^{1}[5/2]_{5/2}$ transition at 638 nm together with the $^{2}\text{D}_{5/2}-^{1}[3/2]_{3/2}$ transition at 976 nm and the $^{2}\text{D}_{3/2}-^{3}[3/2]_{1/2}$ transition at 935 nm. The quantum networking protocol to generate 1650 nm photons with polarization entangled with the atomic states in the $\text{F}=3$ hyperfine manifold of the $^{2}\text{D}_{5/2}$ level involves the $^{2}\text{S}_{1/2}-^{2}\text{D}_{3/2}$ transition at 435 nm, the $^{2}\text{D}_{3/2}-^{2}\text{P}_{3/2}$ transition at 1345 nm, and the $^{2}\text{D}_{5/2}-^{2}\text{P}_{3/2}$ transition at 1650 nm. The atomic component of the entangled ion-photon state is mapped back to the ground state via the $^{2}\text{S}_{1/2}-^{2}\text{D}_{5/2}$ transition at 411 nm. Lastly, entangling gate operations can be performed using the beat note of two Raman beams with center frequency at 355 nm. 

\subsection{Cooling, Detection, State Preparation, and Photoionization Lasers}
\label{subsec:Cooling-Detection-State-Preparation-and-photoionization-Lasers}

\begin{figure}[!!h]
\centering
\includegraphics[width=\textwidth]{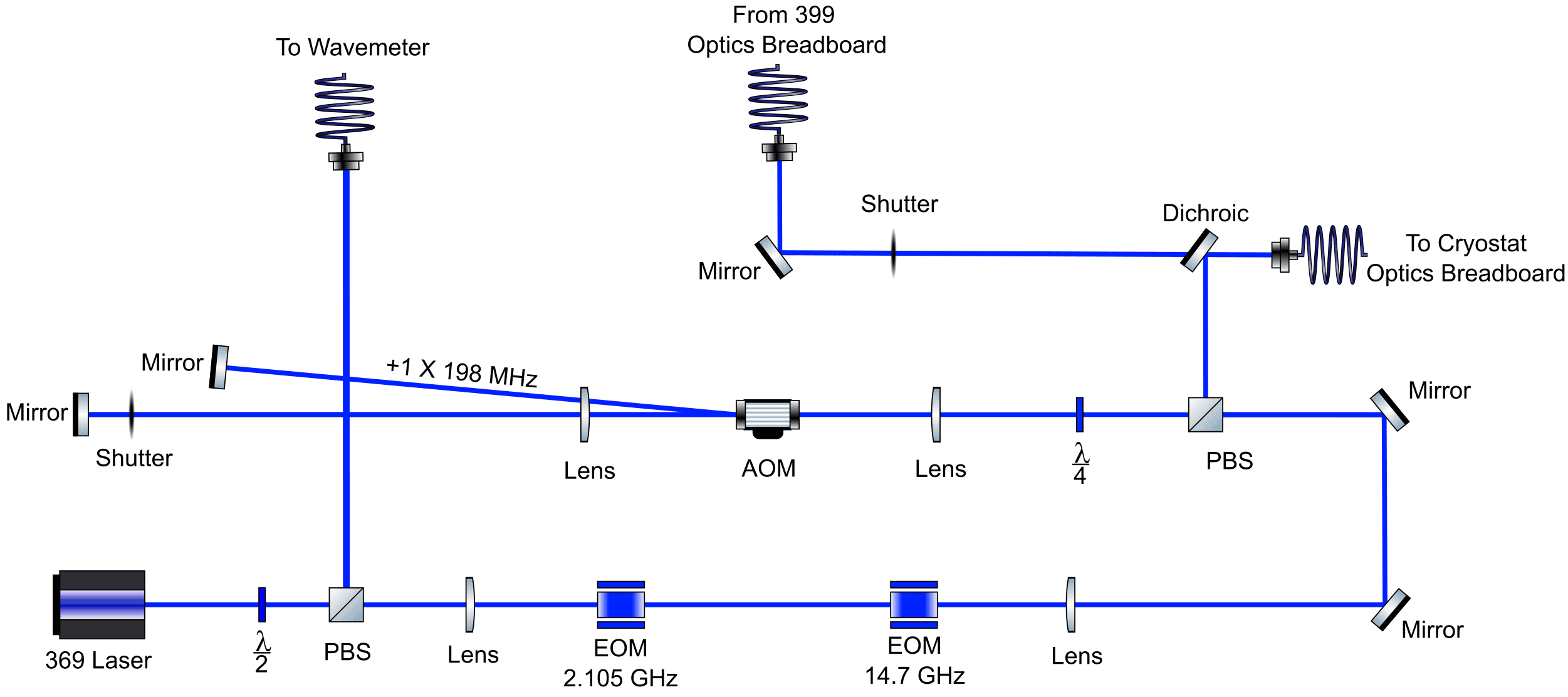}
\caption[Schematic of 369+399 Optics Breadboard.]{Schematic of 369+399 optics breadboard.}
\label{fig:369-399-optics-breadboard}
\end{figure}

There is a dedicated 369 nm laser and optical setup for cryo0 and is shown in Figure \ref{fig:369-399-optics-breadboard}, while optical power for all other lasers is shared between the room temperature trap and cryo0 experiments. The frequency of this 369 nm ECDL is servoed by using a wavemeter to measure the laser's output frequency, calculating the difference from a target frequency, and sending a feedback voltage to a PZT that alters the diffraction grating angle. Resonant free-space EOMs at 2.105 GHz and 14.7 GHz enable complete depopulation of the $^{171}\text{Yb}^{+}$ hyperfine states for efficient laser cooling on the $^{2}\text{S}_{1/2}(\text{F}=1)-^{2}\text{P}_{1/2}(\text{F}=0)$ and state preparation to the $^{2}\text{S}_{1/2}(\text{F}=0)$ state. The 369 nm laser frequency is servoed $\sim379\text{ MHz}$ red of the $^{2}\text{S}_{1/2}(\text{F}=1)-^{2}\text{P}_{1/2}(\text{F}=0)$ transition such that the first order diffracted double pass AOM frequency shift is detuned $-\Gamma/2$ for optimal laser cooling. The double pass frequency detuning of the AOM $2\times(+1\times198\text{ MHz})$ is also a integer multiple of the RF trap drive frequency $\sim22\text{ MHz}$ to ensure cooling of micromotion sidebands which is covered in Subsection \ref{subsec:micromotion-compensation}. A more detailed discussed of Doppler cooling is covered in Subsection \ref{subsec:doppler-cooling}. We call the first order diffracted beam the resonant beam and the zeroth order beam the far-detuned beam. The far-detuned beam $\sim60\text{ \ensuremath{\mu}W}$ is used for cooling very hot atoms as the cooling force is most efficient when the laser detuning matches the Doppler-broadened velocity distribution of the ions. As the ions slow down, they move closer to the atomic resonance where cooling is most efficient with the resonant beam $\sim5\text{ \ensuremath{\mu}W}$. The far detuned beam can be toggled off with an optical shutter to increase signal to noise for imaging and detection. For the two-stage photoionization of neutral $\text{Yb}$ using 399 nm and 369 nm light covered in Section \ref{sec:photoionization-of-Yb}, both the resonant and far-detuned beams are toggled on and overlapped into the same optical fiber with the 399 nm beam. This light is routed to the cryostat optics breadboard as discussed in Subsection \ref{subsec:cryostat-optics-breadboard}. The 399 nm light comes from a separate board where optical power is shared between the cryo0 and room temperature experiments. The 399 nm beam can be toggled off with an optical shutter after photoionization and loading into the trap to increase signal to noise for imaging and detection.

\subsection{Repumpers}
\label{subsec:repumpers}

\begin{figure}[!!h]
\centering
\includegraphics[width=\textwidth]{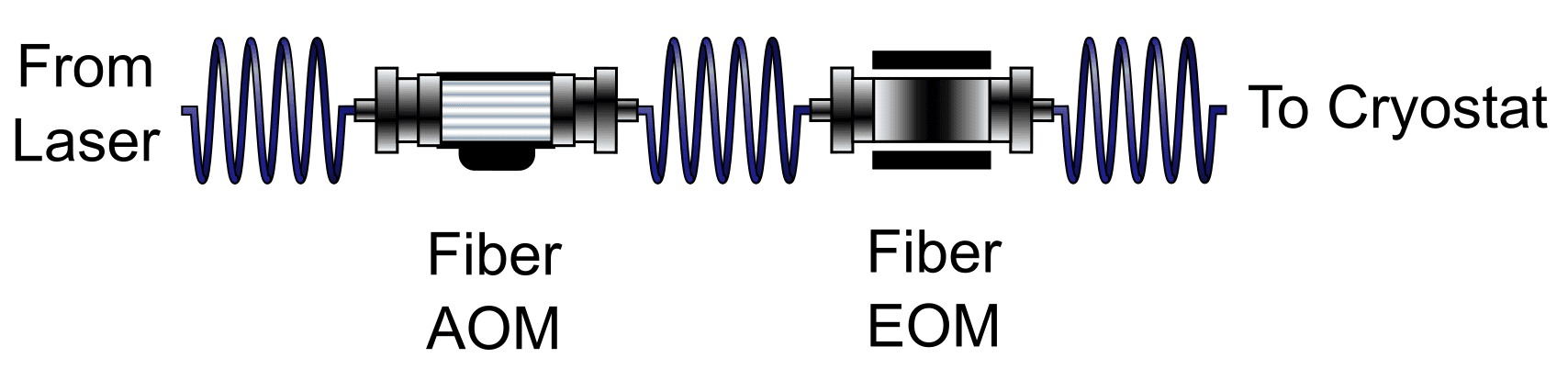}
\caption[Schematic of General Modulation Breadboard for Repumpers.]{General modulation breadboard design used for repumpers 638 nm, 976 nm, 1345 nm, and 1650 nm. }
\label{fig:general-modulation-breadboard}
\end{figure}

\begin{figure}[!!h]
\centering
\includegraphics[width=\textwidth]{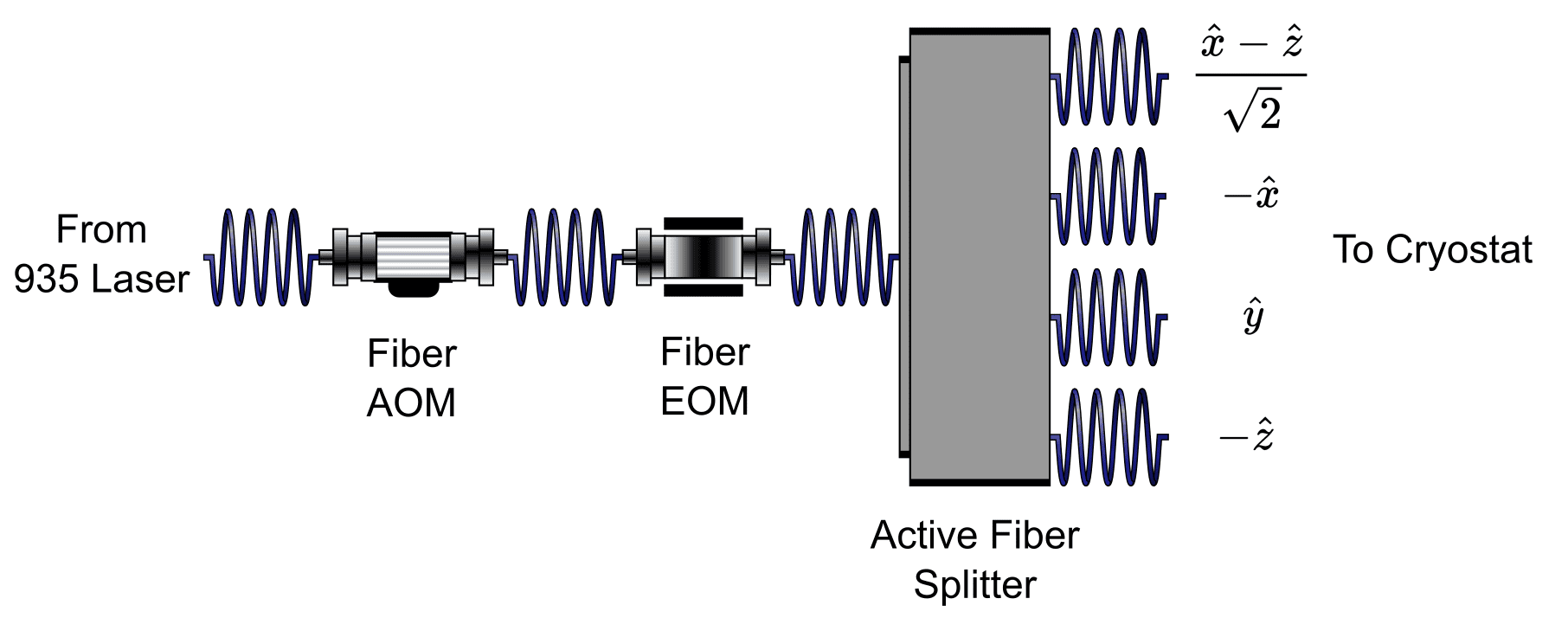}
\caption[Schematic of Modulation Breadboard for the 935 nm Repumper.]{Modulation breadboard design used for the 935 nm repumper. The design is identical to the general modulation breadboard in Figure \ref{fig:general-modulation-breadboard} but with an additional active fiber splitter enabling 935 nm beam access to the ions along the cooling direction $\frac{\hat{x}-\hat{z}}{\sqrt{2}}$ as well as $\hat{x},\hat{y},\hat{z}$.}
\label{fig:935-modulation-breadboard}
\end{figure}

In general, the main repumpers for $^{171}\text{Yb}^{+}$ ions are the 935 nm, 638 nm, 976 nm, 1345 nm, and 1650 nm lasers. To enable toggling these beams on or off as well as depopulating all hyperfine states, an AOM and EOM are installed in the beam paths. For the 638 nm, 976 nm, 1345 nm, and 1650 nm lasers appropriate fiber AOMs and EOMs are used which is shown in Figure \ref{fig:general-modulation-breadboard}. For the 935 nm beam, an additional active fiber splitter is used enabling 935 nm beam access to the ions along the cooling direction $\frac{\hat{x}-\hat{z}}{\sqrt{2}}$ as well and the $\hat{x},\hat{y},\hat{z}$ directions which is shown in Figure \ref{fig:935-modulation-breadboard}. This design enables probing micromotion in 3D as discussed in Subsection \ref{subsec:micromotion-compensation} and easy alignment of optomechanics in 3D by monitoring ion fluorescence. i.e., after loading an ion using the 935 nm beam along cooling direction $\frac{\hat{x}-\hat{z}}{\sqrt{2}}$, the optomechanical system alignment along $\hat{x},\hat{y},\hat{z}$ can be optimized by toggling the active fiber splitter and maximizing ion fluorescence. 

\subsection{Quadrupole and Raman Lasers}
\label{subsec:quadrupole-and-raman-lasers}

All the optical setups and lasers locks for the quadrupole (411 nm, 435 nm) and Raman (355 nm) lasers were setup by Connor Goham and are documented in his thesis.

\subsection{Cavity Lasers}
\label{subsec:cavity-lasers}

 All the optical setups and laser locks for the cavity lasers (1345 nm, 1650 nm) were setup by Wance Wang and are documented in his thesis.

\section{Optomechanical Systems}
\label{sec:optomechanical-systems}

 In this section, I will present the optical access to the ion trap at the 4 K stage of the cryostat and the optomechanical systems used to the align the laser beams to the ions, vibration probes, and ablation targets as well as the optomechanical system used to collect photons from the ions for imaging and detection.

\subsection{Ion Trap Optical Access}
\label{subsec:ion-trap-optical-access}

\begin{figure}[!!h]
\centering
\includegraphics[width=0.49\textwidth]{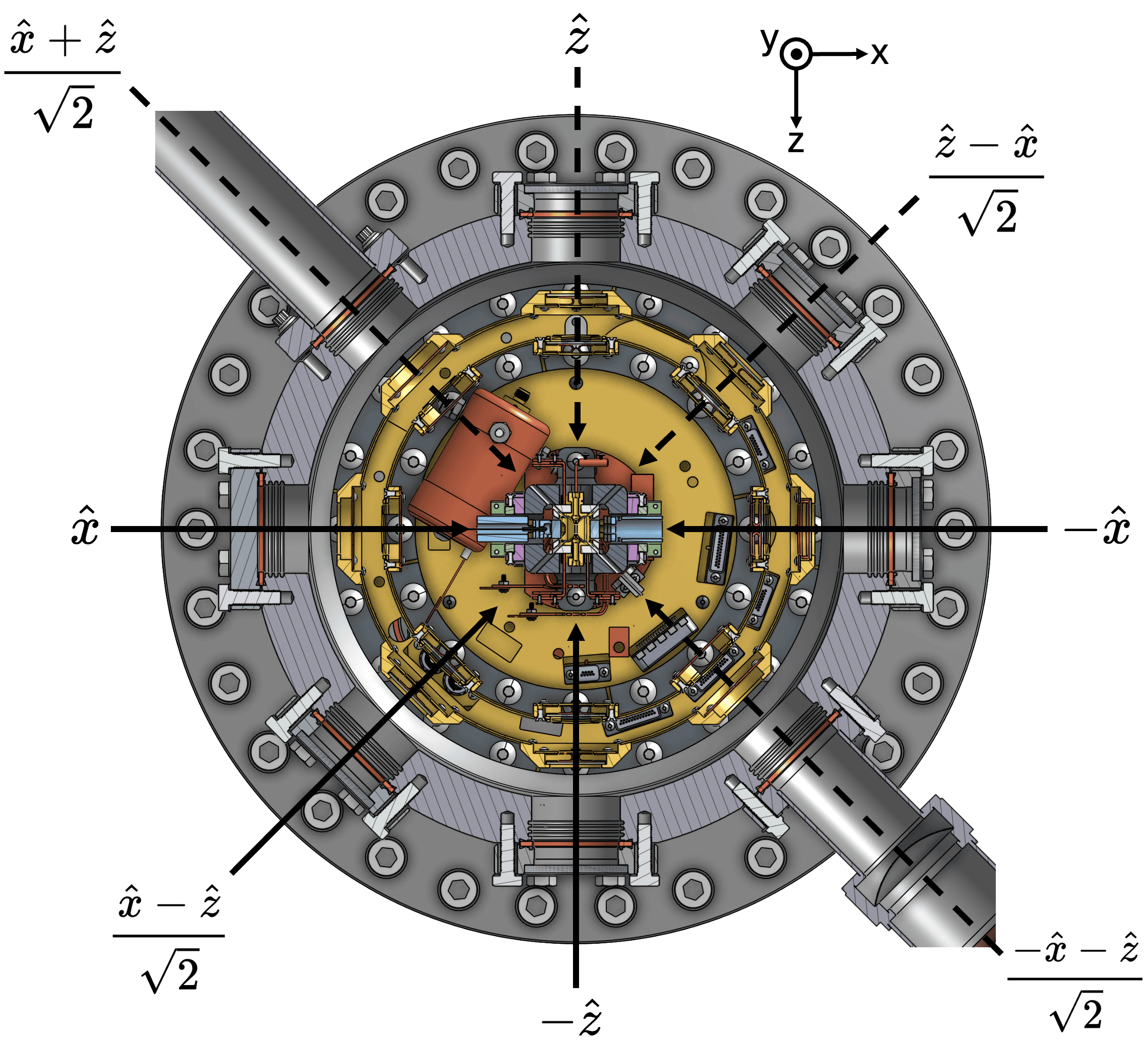}
\includegraphics[width=0.49\textwidth]{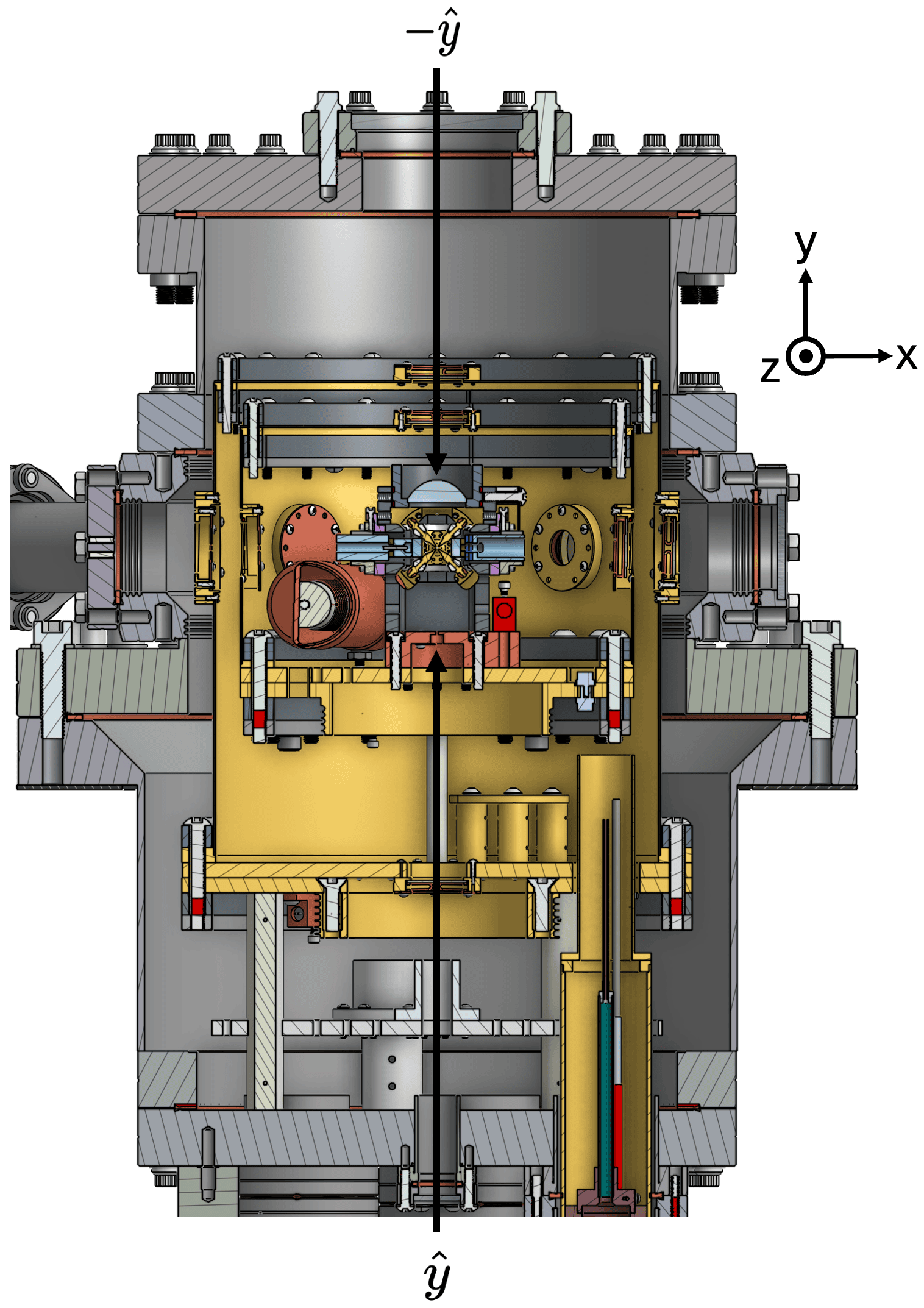}
\caption[Ion Trap Optical Access]{CAD drawings cryo0 showing optical access to the ion trap at the 4 K stage. Optical access that was implemented in this work are depicted with solid black arrows. Optical access that could be used but was not implemented in this work are depicted with dashed black arrows. Detailed descriptions of each beam direction are summarized in Table \ref{tab:ion-trap-optical-access-summary}.}
\label{fig:ion-trap-optical-access}
\end{figure}

The optical access to the ion trap in the $xz$-plane and $xy$-plane is summarized in Figure \ref{fig:ion-trap-optical-access} and Table \ref{tab:ion-trap-optical-access-summary}.

\begin{table}[!!h]
\centering
\tablefont
\caption[Ion Trap Optical Access Summary]{Summary of optical access to the ion trap including the windows used along each beam direction.}
\label{tab:ion-trap-optical-access-summary}
\begin{tabularx}{\textwidth}{>{\centering\arraybackslash}p{0.07\textwidth}cccc}
\toprule
\textbf{Direction} & \makecell{\textbf{Optical Access}\\ \textbf{in This Work}} & \makecell{\textbf{Optical Access}\\\textbf{in Future Work}} & \makecell{\textbf{Numerical}\\\textbf{Aperture to Ions}} & \textbf{Windows} \\
\midrule
$\frac{\hat{x}-\hat{z}}{\sqrt{2}}$ & \makecell{369 nm, 399 nm\\935 nm, 638 nm\\411 nm, 435 nm} & 355 nm & 0.05 & \makecell{\textbf{300 K:} MPF, PN: A22316-1\\(AR coating: 335 nm - 435 nm)\\\textbf{40K:} Thorlabs, PN: WG41010-A\\\textbf{4K:} Thorlabs, PN: WG41010-A}\\
\midrule
$\frac{\hat{z}-\hat{x}}{\sqrt{2}}$ & - & 355 nm & 0.05 & Same as $\frac{\hat{x}-\hat{z}}{\sqrt{2}}$ direction \\
\midrule
$\frac{-\hat{x}-\hat{z}}{\sqrt{2}}$ & - & \makecell{355 nm, 411 nm\\435 nm} & 0.05 & \makecell{CF-2.75 Tee Needed with window\\Oven Installed Below Trap\\\textbf{300K:} MPF, PN: A22317-1\\(AR coating: 411 nm, 435 nm,\\1345 nm, 1650 nm)\\ \textbf{40K:} Thorlabs, PN: WG41010-A\\ \textbf{4K:} Thorlabs, PN: WG41010-A} \\
\midrule
$\frac{\hat{x}+\hat{z}}{\sqrt{2}}$ & - & \makecell{355 nm, 411 nm\\435 nm} & 0.05 & Same as $\frac{-\hat{x}-\hat{z}}{\sqrt{2}}$ direction \\
\midrule
$-\hat{z}$ & \makecell{935 nm, 976 nm\\1650 nm, 1345 nm} & - & 0.06 & \makecell{\textbf{300 K:} MPF, PN: A22317-1\\(AR coating: 411 nm, 435 nm,\\1345 nm, 1650 nm)\\\textbf{40K:} Thorlabs, PN: WG41010-A\\\textbf{4K:} Thorlabs, PN: WG41010-A}\\
\midrule
$\hat{z}$ & - & 411 nm, 435 nm & 0.06 & Same as $-\hat{z}$ direction \\
\midrule
$\hat{x}$ & 1345 nm, 1650 nm & - & \makecell{Governed by Cavity\\(Wance Wang's Thesis)} & \makecell{\textbf{300 K:} MPF, PN: A22315-1\\(AR coating: 1345 nm, 1650 nm)\\\textbf{40K:} Thorlabs, PN: WG41010-C\\\textbf{4K:} Thorlabs, PN: WG41010-C} \\
\midrule
$-\hat{x}$ & 1650 nm & - & \makecell{Governed by Cavity\\(Wance Wang's Thesis)} & \makecell{Fiber Feedthrough\\(Wance Wang's Thesis)} \\ 
\midrule
$\hat{y}$ & 650 nm, 935 nm & 532 nm, 1345 nm & 0.01 & \makecell{\textbf{300 K:} MPF, PN: A22318-1\\(AR coating: 532 nm)\\\textbf{40 K:} Thorlabs, PN: WG41010-A} \\
\midrule
$-\hat{y}$ & 369 nm & - & \makecell{0.49, in-vacuum\\objective\\(See Subsection \ref{subsec:in-vacuum-aspheric-lens-optomechanics})} & \makecell{\textbf{300 K:} MPF, PN: A22319-1\\(AR coating: 350 - 435 nm,\\1345 nm, 1650 nm)\\\textbf{40K:} Thorlabs, PN: WG41010-A\\\textbf{4K:} Thorlabs, PN: WG41010-A} \\
\bottomrule
\end{tabularx}
\end{table}

\subsection{Cryostat Optics Breadboard}
\label{subsec:cryostat-optics-breadboard}

\begin{figure}[!!h]
\centering
\includegraphics[width=\textwidth]{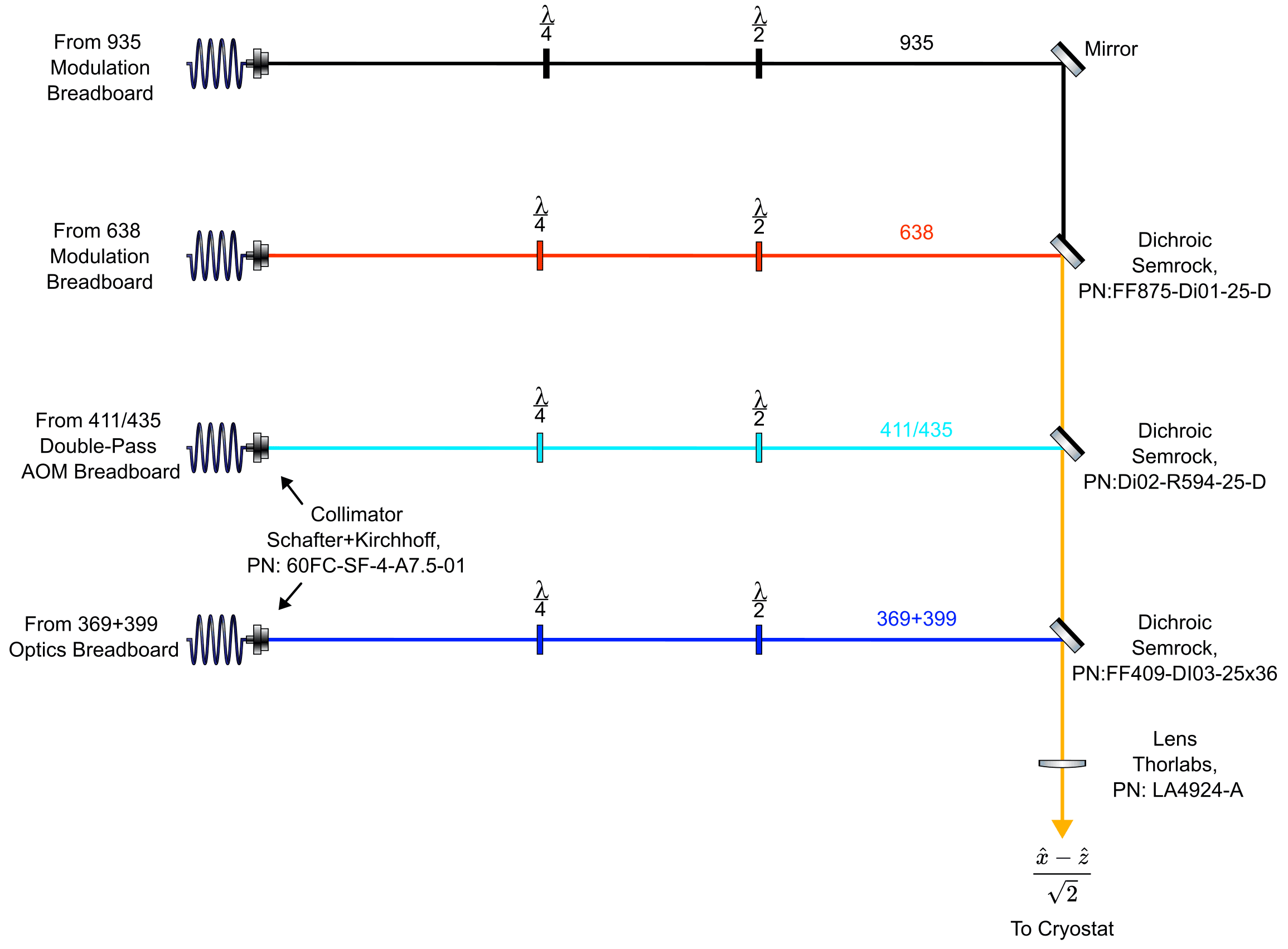}
\caption[Cryostat Optics Breadboard]{Schematic of cryostat optics breadboard.}
\label{fig:cryostat-optics-breadboard}
\end{figure}

The purpose of the cryostat optics breadboard, see Figure \ref{fig:cryostat-optics-breadboard}, is to align the 399 nm and 369 nm photoionization beams, the 369 nm Doppler cooling beam, the 935 nm and 638 nm repumper beams, and the 411 nm and 435 nm quadrupole beams along the $\frac{\hat{x}-\hat{z}}{\sqrt{2}}$ direction to the ions. The 355 nm Raman beam could also be applied in this breadboard in between the 369+399 nm dichroic and the focusing lens in Figure \ref{fig:cryostat-optics-breadboard} using an appropriate fiber, collimator, wave plates, and dichroic (Semrock, PN:DI02-R355-25.4-D) but was not done in this work. Sending the beams along the $\frac{\hat{x}-\hat{z}}{\sqrt{2}}$ direction generates a component of the light along each principal axis of the trap $\left(\frac{\hat{x}+\hat{y}}{\sqrt{2}}, \frac{\hat{x}-\hat{y}}{\sqrt{2}}, \hat{z}\right)$ enabling cooling of all motional modes. This ensures efficient Doppler cooling using the 369 nm beam and ground state cooling using either the 411 nm, 435 nm, or 355 nm beams. The final focusing lens focal length was chosen to to be 175 mm which creates a $25\text{ \ensuremath{\mu}m}$ beam waist for the 369 nm cooling beam at the ions. A similar beam waist at the ions will be created for the 399 nm, 355 nm, 411 nm, and 435 nm beams as they are close in wavelength while the 935 nm and 638 nm beam waists vary more but are not as important because they are repumpers.

\subsection{Repumper Optomechanics}
\label{subsec:repumper-optomechanics}

\begin{figure}[!!h]
\centering
\includegraphics[width=\textwidth]{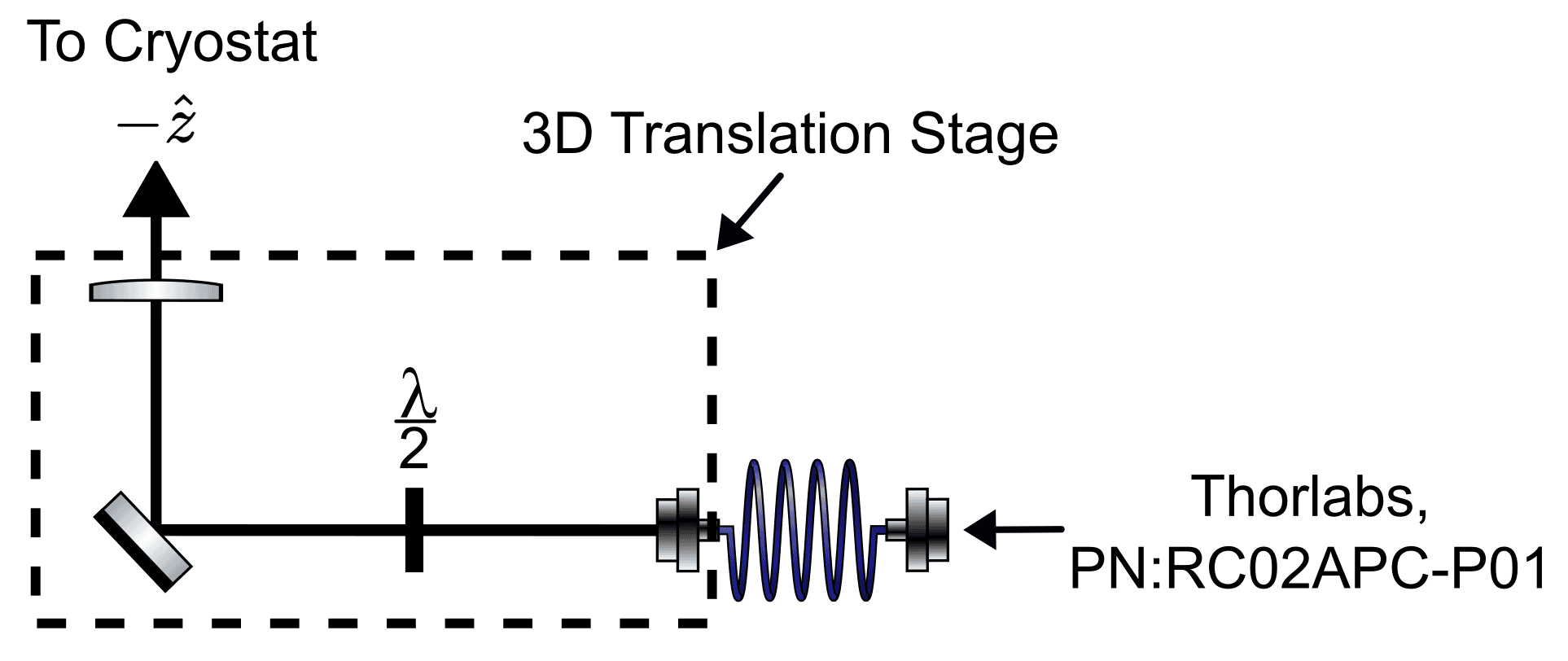}
\caption[Repumper Optomechanics]{Optomechanics to send repumper beams to ions along the $z$ axis. }
\label{fig:repumper-optomechanics-z}
\end{figure}

An optomechanics system was built to send additional repumpers to the ions along the $z$ axis and is shown in Figure \ref{fig:repumper-optomechanics-z}. A broadband silver reflective collimator (Thorlabs, PN: RC02APC-P01) is used to couple 935 nm, 976 nm, 1345 nm, or 1650 nm light to a single fiber. This fiber is routed to a 3D translation stage with a wave plate, mirror, and focusing lens to deliver the light to the ions. The collimator, mirror, wave plate, and focusing lens at the 3D translation stage are chosen so that there is a compromise for all wavelengths to sufficiently repump the ion.

\subsection{Imaging and Detection Breadboard}
\label{subsec:imaging-and-detection-breadboard}

\begin{figure}[!!h]
\centering
\includegraphics[width=0.75\textwidth]{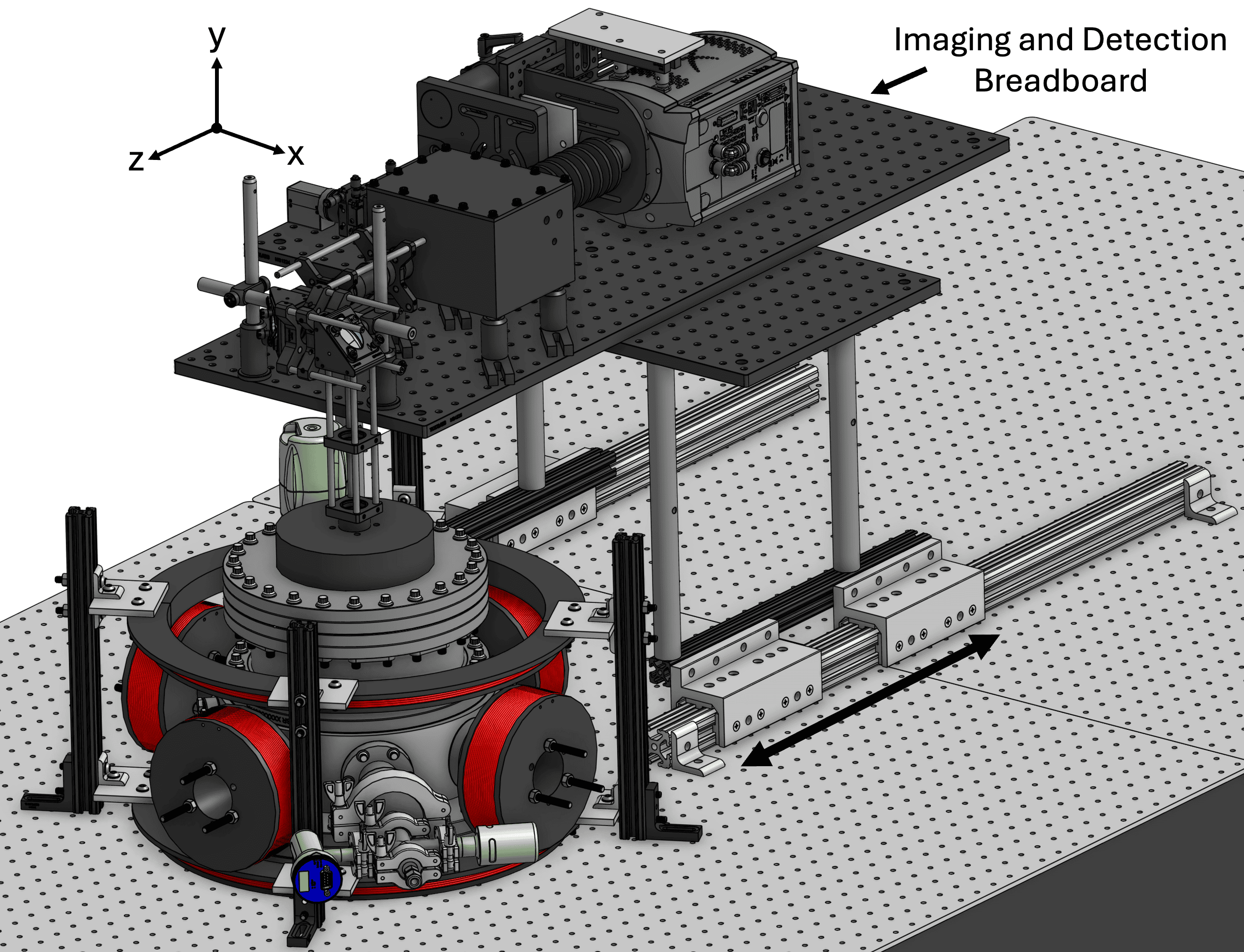}
\caption[Imaging and Detection Breadboard]{CAD drawing of optomechanical system on sliding breadboard to collect photons from the ions for imaging and detection. The black double arrow depicts how the breadboard can slide relative to the cryostat.}
\label{fig:imaging-and-detection-breadboard}
\end{figure}

An optomechanical system was built to collect photons from the ion for imaging and detection. First, an in-vacuum objective images the ions and then this image is relayed and magnified to the PMT and camera using ambient optics that are mounted on a sliding breadboard as shown in Figure \ref{fig:imaging-and-detection-breadboard}. The details about the design and construction of the imaging and detection subsystem are covered in Chapter \ref{chap:imaging-and-detection}. The breadboard can be slid away from the cryostat when in-vacuum maintenance needs to be performed as discussed in Section \ref{sec:in-vacuum-maintenance-system}.

\subsection{Vibration Probe, Repumper, and Laser Ablation Breadboard}
\label{subsec:vibration-probe-repumper-and-laser-ablation-breadboard}

\begin{figure}[!!h]
\centering
\includegraphics[width=0.49\textwidth]{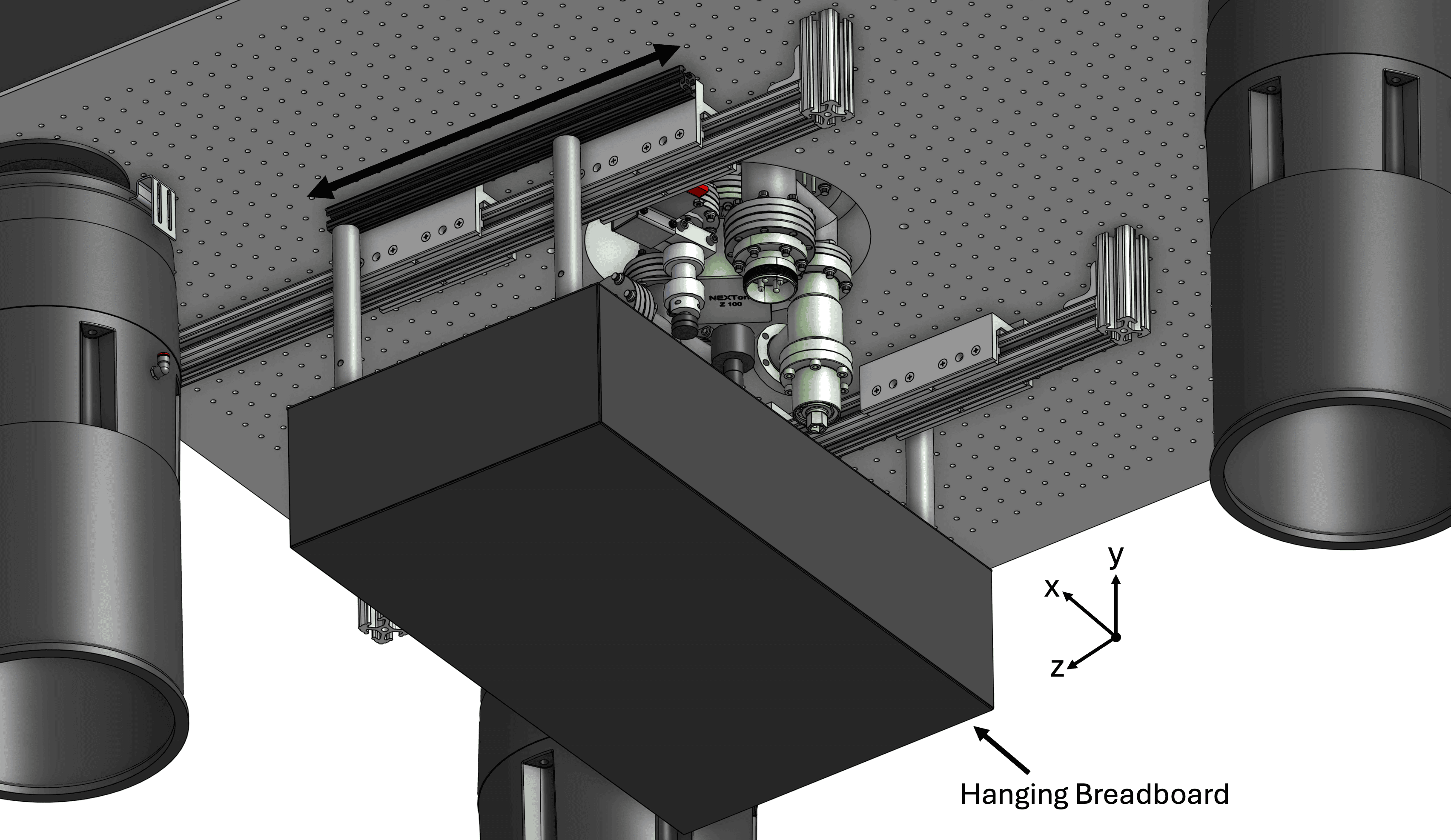}
\includegraphics[width=0.49\textwidth]{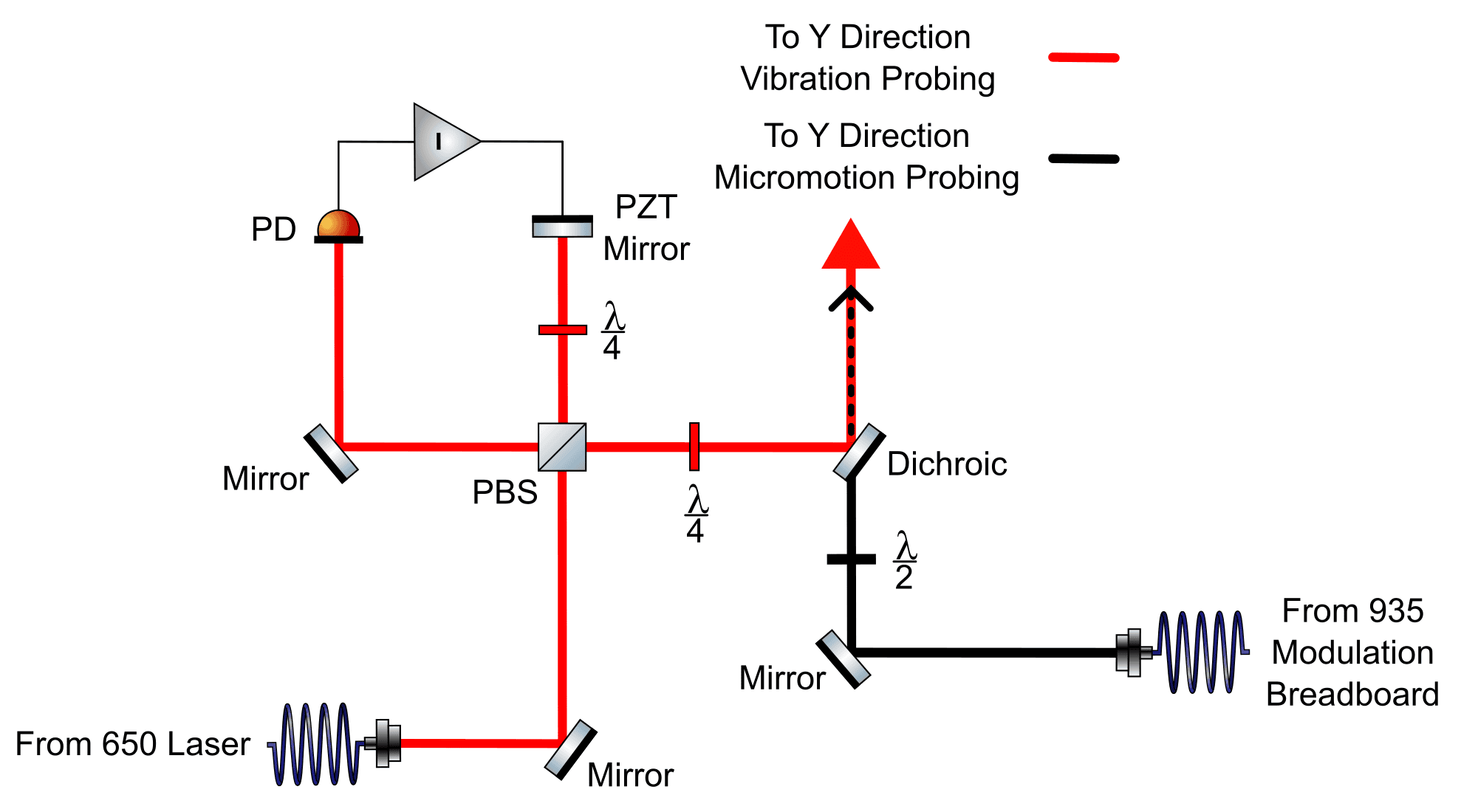}
\caption[Vibration Probe, Repumper, and Laser Ablation Breadboard]{(Left) CAD drawing of optomechanical system on hanging breadboard. The black double arrow depicts how the breadboard can slide relative to the cryostat. (Right) Schematic of optical setup on hanging breadboard to probe vibrations and micromotion along the $y$ axis and send the future ablation loading beam to the 4 K stage.}
\label{fig:vibration-probe-repumper-and-laser-ablation-breadboard}
\end{figure}

An optomechanical system was built to probe the vibrations of the 4 K stage along the $y$ axis, probe micromotion along the $y$ axis, and send a high intensity laser beam to the 4 K stage for future ablation loading. I ended up using an oven to load ions in this work (see Chapter \ref{chap:atomic-source}) instead of ablation loading. This optomechanical system consists of a sliding breadboard that hangs beneath the optics table as shown in Figure \ref{fig:vibration-probe-repumper-and-laser-ablation-breadboard}. This breadboard houses a interferometer for probing vibrations along the $y$ axis (see Subsection \ref{subsec:experimental-setup-vibration-measurement}) and optics for routing the 935 nm repumper beam to the ions for probing micromotion along the $y$ axis. The vibration probe beam and 935 nm beam are nearly overlapped using a dichroic. The breadboard can be slid away from the cryostat when in-vacuum maintenance needs to be performed as discussed in Section \ref{sec:in-vacuum-maintenance-system}.

\subsection{Ion-Cavity Optomechanics}
\label{subsec:ion-cavity-optomechanics}

All the cavity optomechanics were designed and built by Wance Wang and are documented in his thesis. In short, the necessary optics are mounted on a breadboard so that necessary beams can be aligned along the free space side of the cavity axis ($x$-axis).

\section{In-Vacuum Maintenance System}
\label{sec:in-vacuum-maintenance-system}

\begin{figure}[!!h]
\centering
\includegraphics[width=\textwidth]{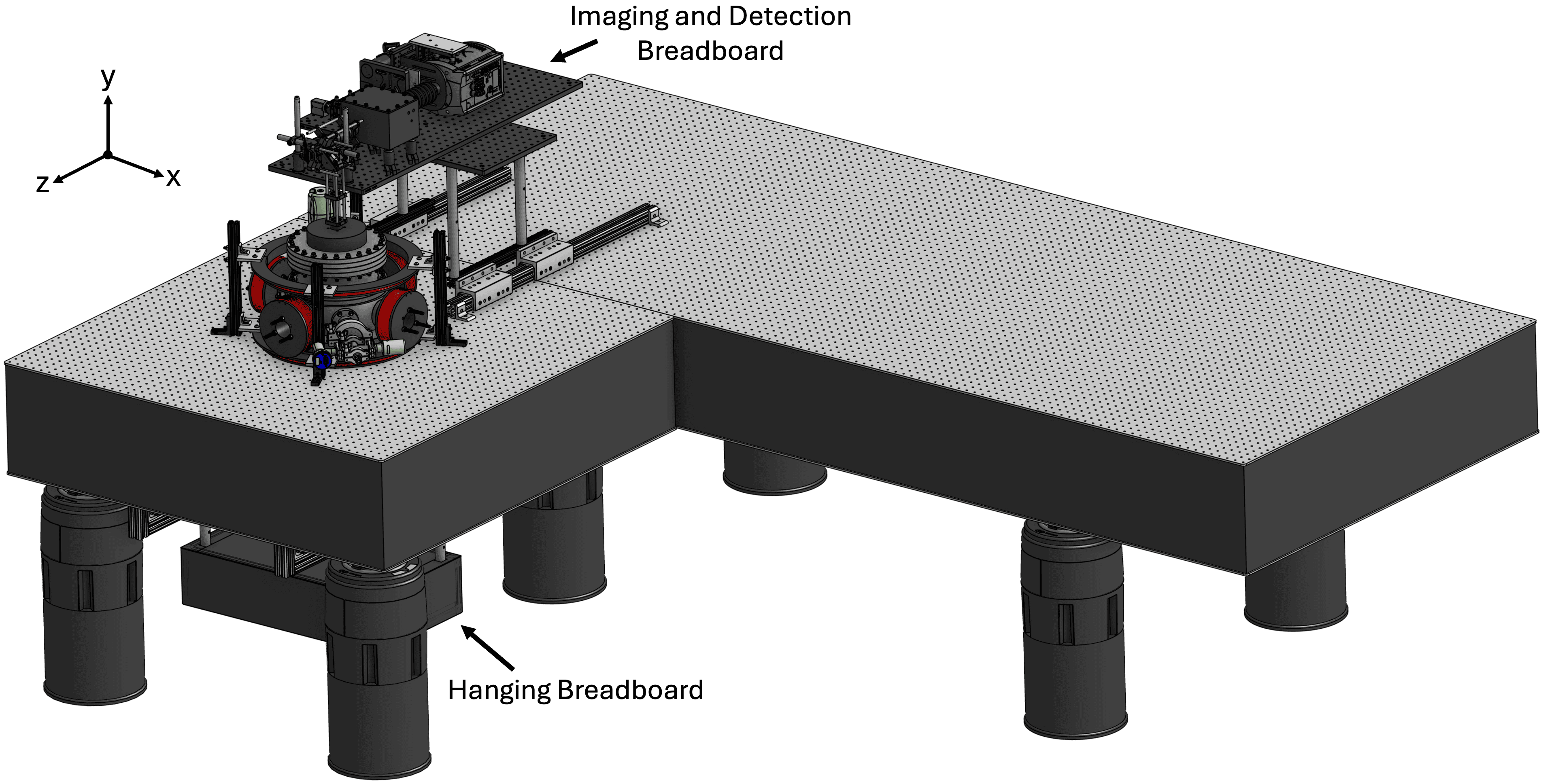}
\newline
\includegraphics[width=\textwidth]{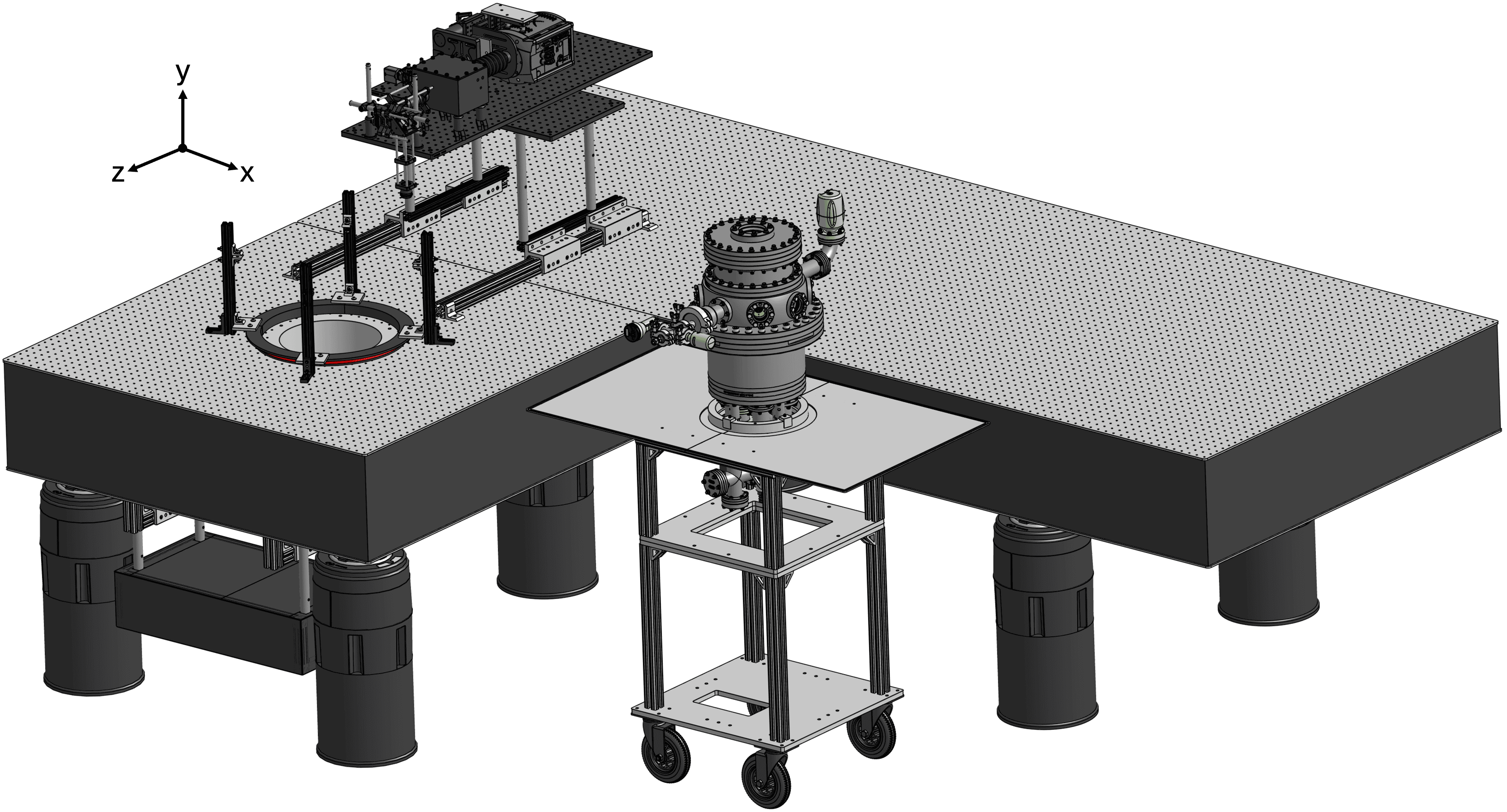}
\caption[In-Vacuum Maintenance System]{(Top) CAD drawing of cryo0 positioned in the optics table ready for ion trapping. (Bottom) CAD drawing of cryo0 lifted out of the optics table and positioned on the cryo cart for in-vacuum maintenance. The imaging and detection and hanging breadboards are shown to be slid away from the bore in the optics table enabling the cryostat to be lifted out of the table. The bottom magnetic field coil is kept in place to guide the cryostat back into position on the table after in-vacuum maintenance.}
\label{fig:in-vacuum-maintenance-system}
\end{figure}

The cryostat was designed to enable easy access to the 4 K stage to conduct in-vacuum maintenance on the ion trap and optical cavity. The are two modes of in-vacuum maintenance for this cryostat: 1.) simple adjustments to parts at the 4 K stage and 2.) more complicated/invasive adjustments to parts at the 4 K stage or the entire vacuum space. In mode 1, the imaging and detection breadboard needs to be removed so that the CF-10 to 4.5 zero-length reducer flange or CF-10 straight nipple can be removed. In-vacuum maintenance can then be performed while the cryostat remains on the optics table. In mode 2, both the imaging and detection breadboard and the vibration probe, repumper, and laser ablation breadboard need to be removed so that the cryostat can be lifted out of the optics table using the hoist, see Subsection \ref{subsec:hoist}. In-vacuum maintenance can then be performed while the cryostat is mounted on the cryo cart (see Subsection \ref{subsec:cryo-cart}) as shown in Figure \ref{fig:in-vacuum-maintenance-system}. Realignment of these breadboards relative to the cryostat after either mode of in-vacuum maintenance is achieved by simply sliding them back into position, preserving the local alignment of the optics on each breadboard. 

\subsection{Hoist}
\label{subsec:hoist}

\begin{figure}[!!h]
\centering
\includegraphics[width=\textwidth]{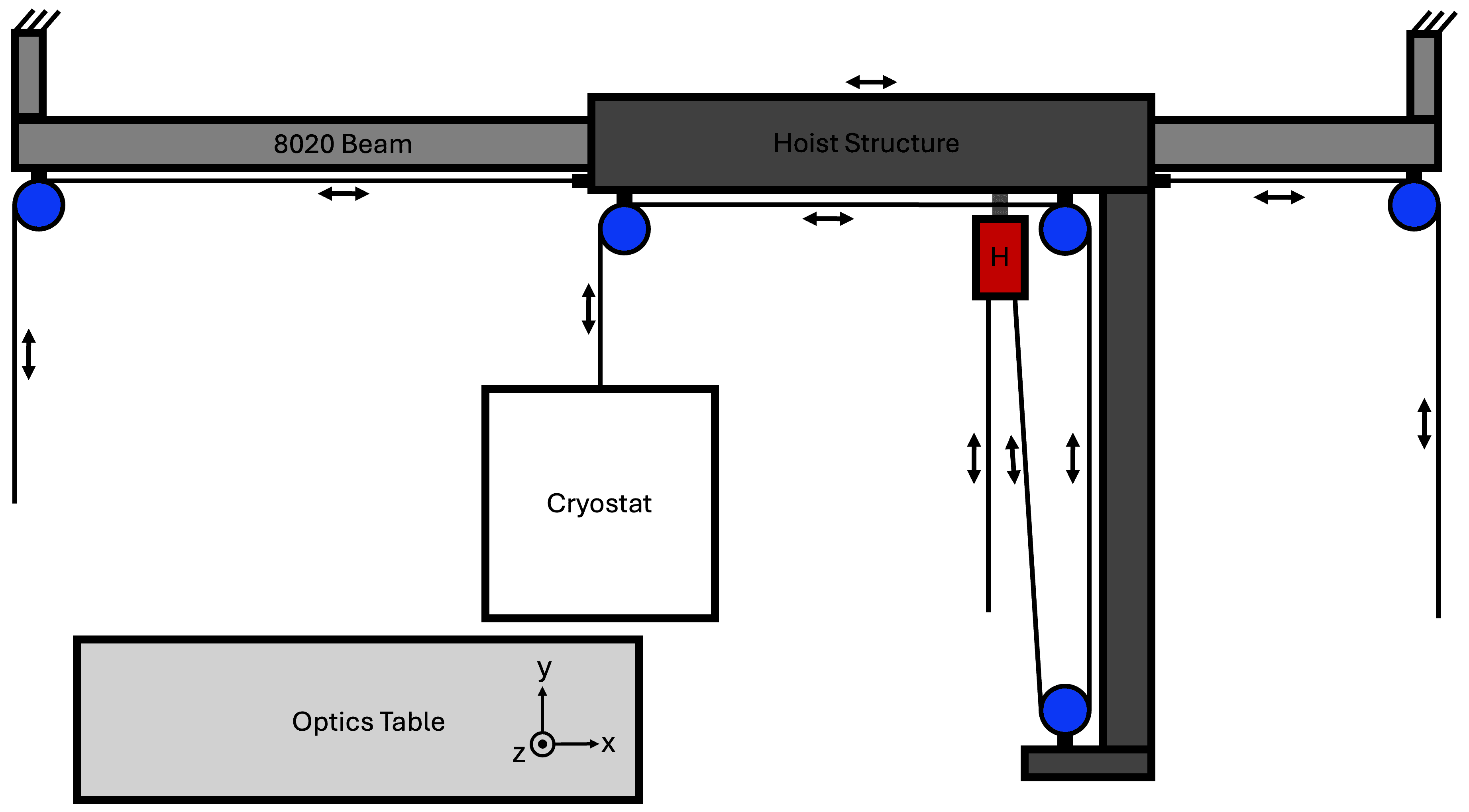}
\caption[Cryostat Hoist]{Schematic of hoist. The ropes and pulleys are depicted with black lines and blue circles respectively. A hand chain hoist, depicted in red, is used to lift the cryostat while the hoist structure, depicted in dark gray, can be slid along the $x$ direction by pulling ropes attached to each end of the structure.}
\label{fig:cryosat-hoist}
\end{figure}

\begin{figure}[!!h]
\centering
\includegraphics[width=0.5\textwidth]{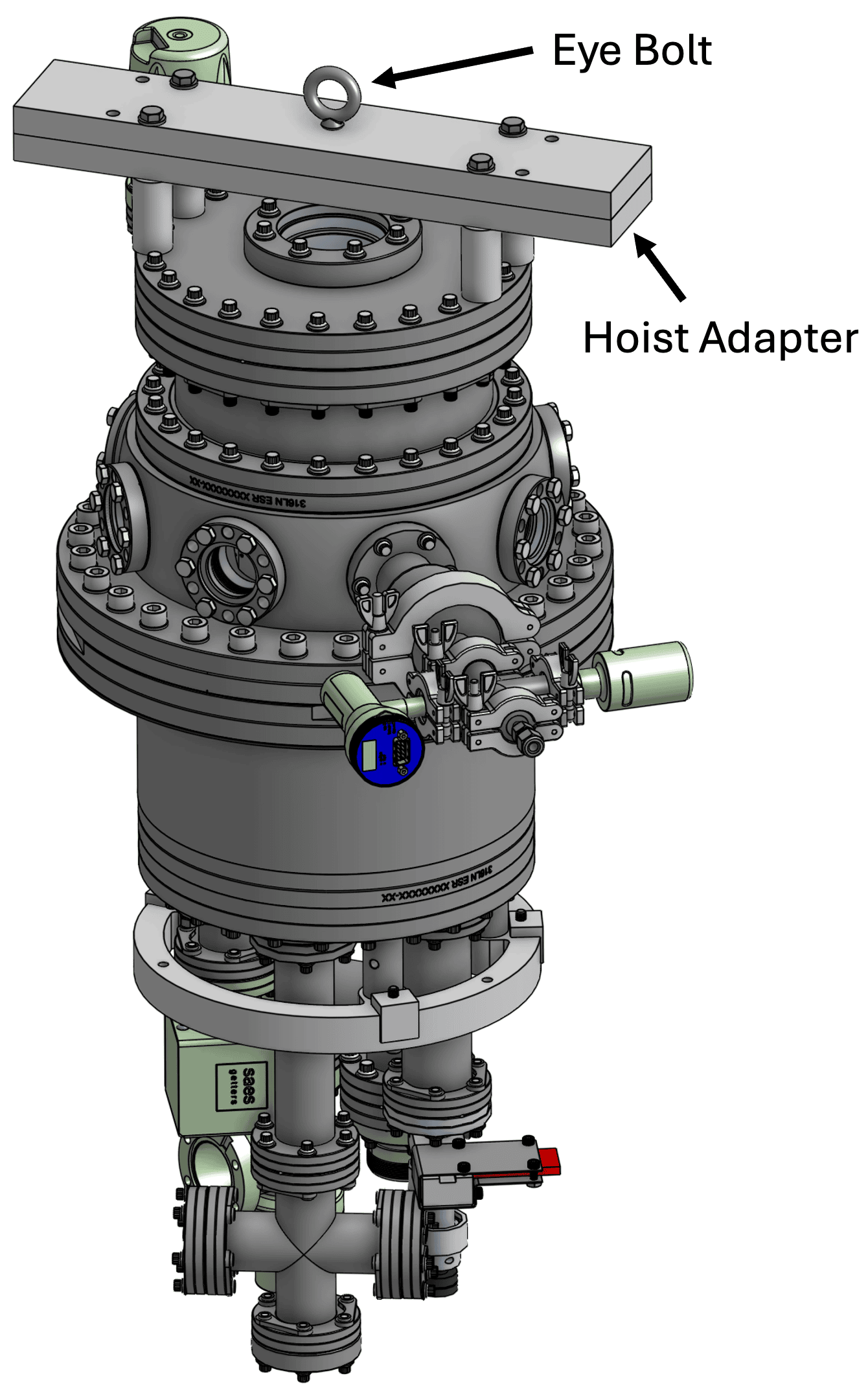}
\caption[Cryostat Hoist Adapter]{CAD drawing of cryostat with hoist adapter.}
\label{fig:cryosat-hoist-adapter}
\end{figure}

A hoist was built to lift the cryostat in and out of the optics table as shown in Figure \ref{fig:cryosat-hoist}. The design consists of a hand chain hoist (McMaster Carr, PN: 3094T14), rope (McMaster Carr, PN: 2222T45), and pulleys (McMaster Carr, PN: 3099T38) that are rigidly mounted to a monolithic structure. The rope is attached to the hand chain hoist and an eye bolt of an adapter that snares the CF-10 straight nipple of the cryostat as shown in Figure \ref{fig:cryosat-hoist-adapter}. The hand chain hoist provides a 133:1 mechanical advantage while the rope redirects the weight of the cryostat to the hand chain hoist via three pulleys. i.e., I only have to pull 1.5 lbs to lift the 206 lbs cryostat. The hand chain hoist has an auto locking mechanism and can operate in both lifting and lowering directions. The entire hoist structure is attached to a 8020 beam positioned above the optics table via a sleeve bearing (McMaster Carr, PN: 47065T956). Thus, the hoist structure that lifts the cryostat can be slid along the $x$ direction by pulling ropes attached to each end of the structure. Here, the mechanical advantage is 1:1, but the friction between the Teflon of the sleeve bearing and the 8020 beam is small. All the parts for the hoist structure and adapter were designed and machined by myself at UMD. A duplicate hoist structure was built for the second cryostat (cryo1).

\subsection{Cryo Cart}
\label{subsec:cryo-cart}

\begin{figure}[!!h]
\centering
\includegraphics[width=0.3\textwidth]{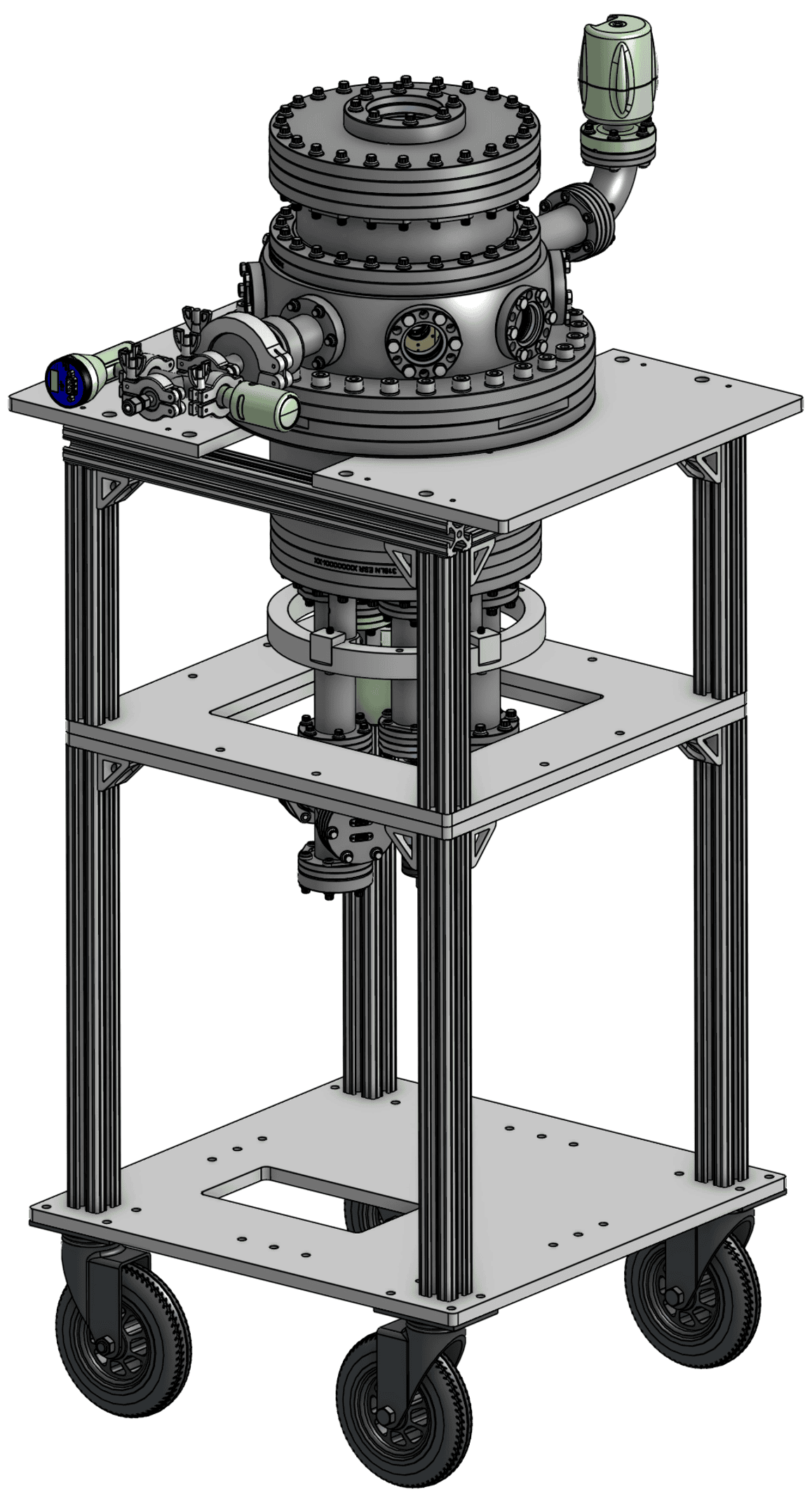}
\includegraphics[width=0.3\textwidth]{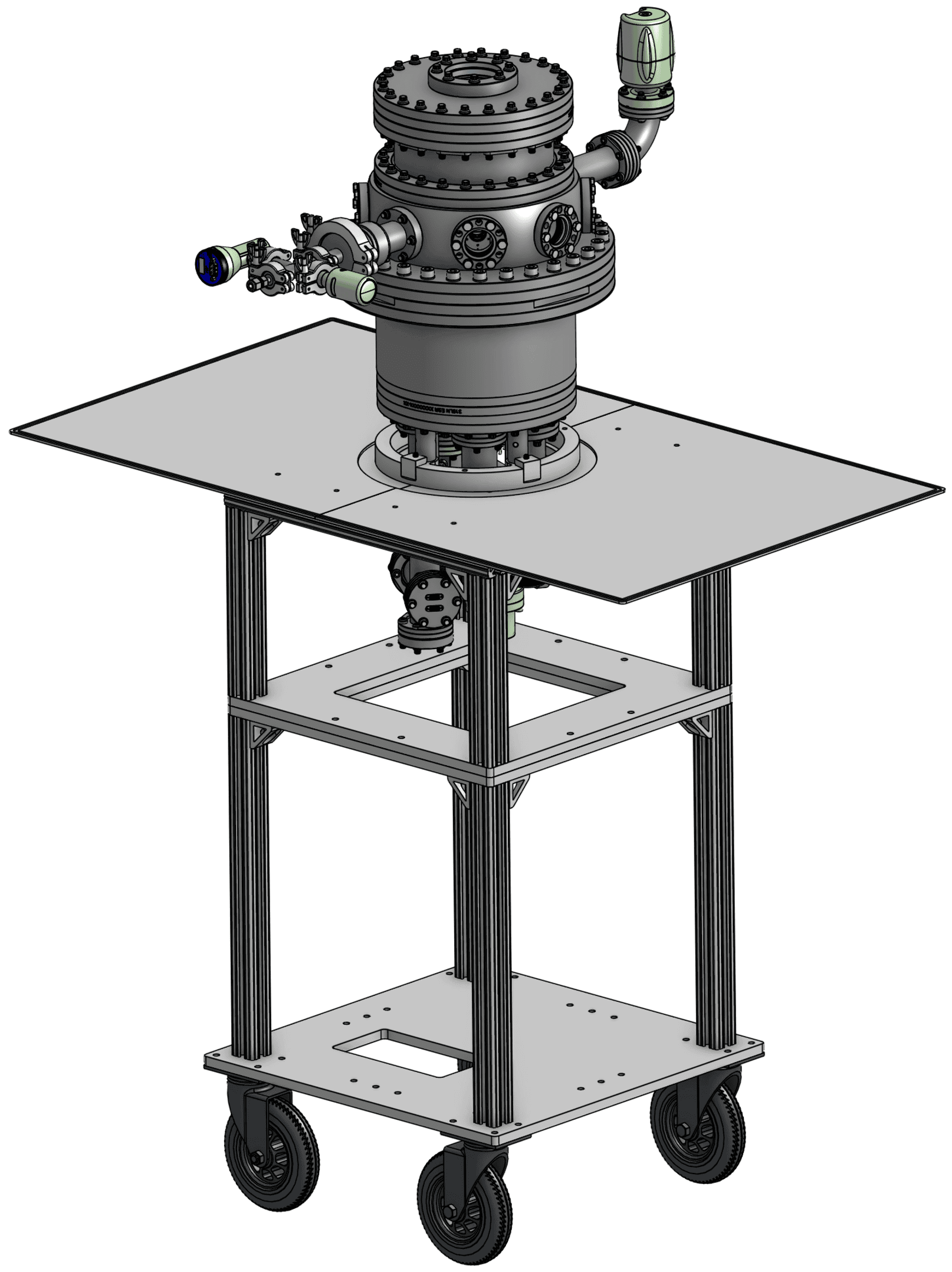}
\includegraphics[width=0.3\textwidth]{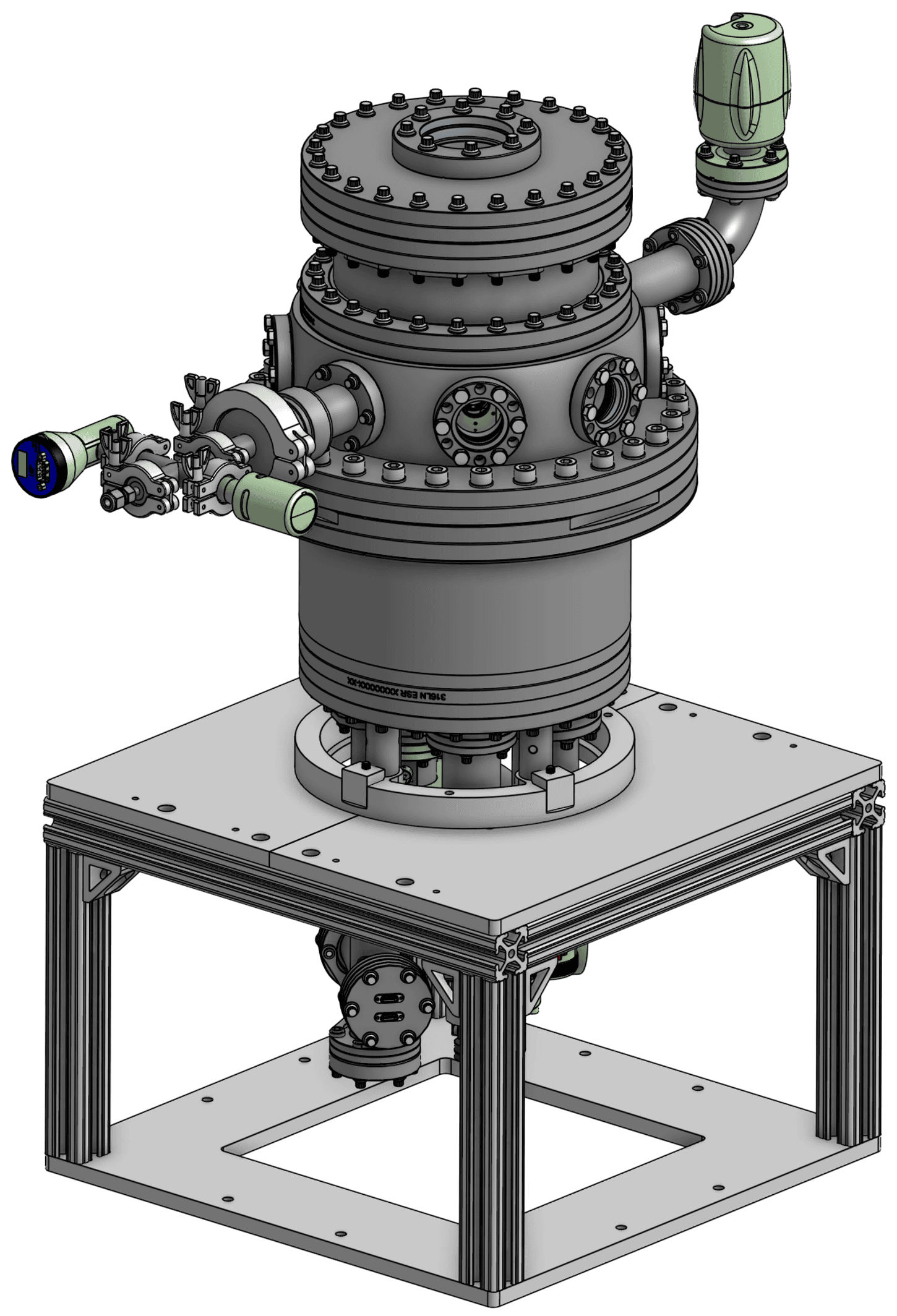}
\caption[Cryo Cart]{CAD drawings of cryo cart modes. The left drawing depicts modes 1 and 2 which provide enough clearance to install the Stinger bayonet to perform a cooldown and to conduct in-vacuum maintenance. The center drawing depicts mode 2 for conducting in-vacuum maintenance. The right drawing depicts mode 3 for performing a bakeout of the cryostat.}
\label{fig:cryo-cart}
\end{figure}

A cart (the cryo cart) was built to hold the cryostat while it's not positioned on the optics table and is shown in Figure \ref{fig:cryo-cart}. There are three modes of operation for the cryo cart: 1.) Cooling down the cryostat to perform diagnostic testing, 2.) perform in-vacuum maintenance and 3.) bakeout the cryostat. For mode 1, the cryostat can be cooled down on the cart as there is sufficient clearance to install the Stinger bayonet. This was done, e.g., after installing the cavity for the first time and checking for cavity misalignment after a cooldown before positioning the cryostat into the optics table. For mode 2, the cryostat can be wheeled to a clean environment where the flanges can be removed to perform in-vacuum maintenance. For mode 3, the cryostat can be wheeled over the bakeout station (see Appendix \ref{sec:bakeout-station}) where the top half of the cart holding the cryostat can be positioned inside the kiln. The carts were designed and built by myself at UMD and are made out of aluminum 8020 and Al6061-T6 plates and solid rubber wheels (McMaster Carr, PN: 4941T21).

\ifSubfilesClassLoaded{%
  \bibliographystyle{unsrt}
  \bibliography{Dissertation}
}{}
\end{document}

%% file: Chapter5.tex
\chapter{Ion Trap}
\label{chap:ion-trap}

\begin{figure}[!!h]
\centering
\includegraphics[width=0.6\textwidth]{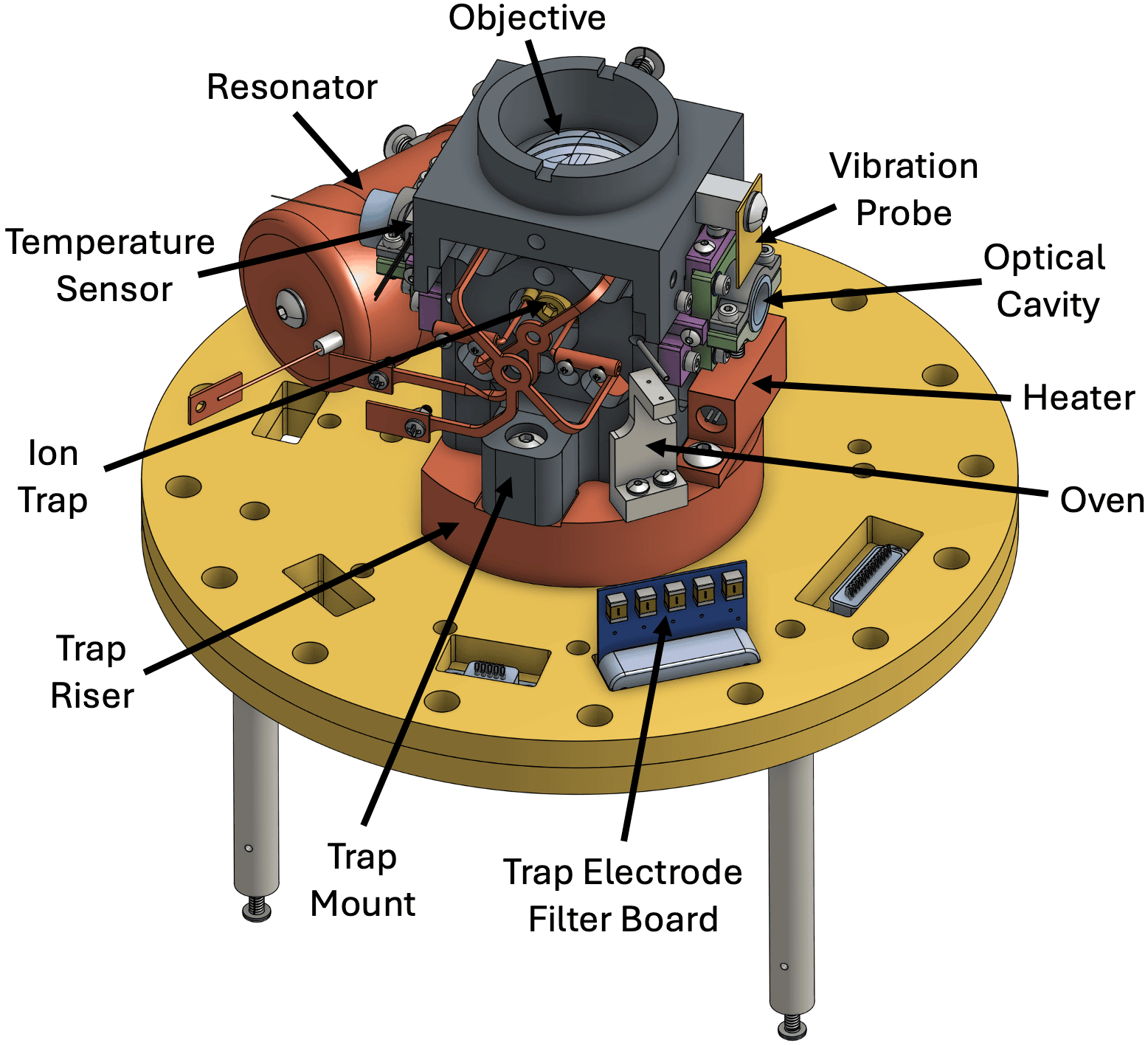}
\caption[Ion Trap Mounted at the 4 K Stage]{CAD drawing of ion trap mounted at the 4 K stage.}
\label{fig:ion-trap-mounted-at-4K-stage}
\end{figure}

In this Chapter, I'll discuss the details of ion trapping in this cryostat. The ion trap is mounted at the 4 K stage experiment plate as shown in Figure \ref{fig:ion-trap-mounted-at-4K-stage}.

\section{Linear RF Paul Trap Basics}
\label{sec:linear-rf-paul-trap-basics}

In order to trap an ion, an electric potential is generated at the trap center by applying DC and RF voltages to the trap electrodes. The electric potential of an ion trap can be decomposed into a time-dependent and time-independent potential as shown below
\begin{equation}
    \Phi(x,y,z,t)=\frac{U}{2}(\alpha x^{2}+\beta y^{2}+\gamma z^{2})+\frac{\tilde{U}\cos(\omega_{\text{RF}}t)+U_{r}}{2}(\alpha'x^{2}+\beta'y^{2}+\gamma'z^{2})
    \label{eq:ion-trap-potential-without-constraints}
\end{equation}
This potential has to satisfy the Laplace equation, $\nabla^{2}\Phi=0$ that imposes the following constraints on the coefficients $\alpha+\beta+\gamma=0$ and $\alpha'+\beta'+\gamma'=0$ \cite{leibfriedQuantumDynamicsSingle2003}. By choosing $-(\alpha+\beta)=\gamma>0$ and $\alpha'=-\beta'$ with $\gamma'=0$, the total potential takes the form
\begin{equation}
    \Phi(x,y,z,t)=\frac{U}{2}(\gamma z^{2}-\alpha x^{2}-\beta y^{2})+\frac{\tilde{U}\cos(\omega_{\text{RF}}t)+U_{r}}{2}(\alpha'x^{2}-\beta'y^{2})
    \label{eq:ion-trap-potential-with-constraints}
\end{equation}
which creates confinement along the $z$ axis (the axial direction) and anti-confinement along the $x,y$ axes (the radial directions) due to static potentials (DC confinement). Confinement along the $x,y$ axes is created by a dynamic quadrupole potential (RF confinement). To understand how the dynamic field generates confinement, we will consider just the dynamic potential along the $x$ direction in Equation \eqref{eq:ion-trap-potential-with-constraints}. The electric field along the $x$ direction is given by
\begin{equation}
    E_{x}(t)=-\frac{\partial\Phi(t)}{\partial x}=-\tilde{U}\alpha'x(t)\cos(\omega_{\text{RF}}t)
    \label{eq:inhomogeneous-E-field}
\end{equation}
where the radial DC bias potential $U_{r}=0$ and the equation of motion for an ion of mass $m$ and charge $e$\footnote{In general the charge of a particle is $Z|e|$where $|e|=1.6\times10^{-19}\text{ C}$ and $Z=+1$ is the charge number for a singly ionized atom. For notation simplicity, $e\equiv+1|e|$ for this work.} experiencing a force $F_{x}(t)=eE_{x}(t)$ in this field is 
\begin{equation}
    \ddot{x}(t)=-\frac{e\tilde{U}\alpha'}{m}x(t)\cos(\omega_{\text{RF}}t)
    \label{eq:EOM-of-ion-in-dynamic-quadrupole}
\end{equation}
When the RF drive frequency is sufficiently high that the ion's displacement during a single RF cycle is small, the ion's motion can be expressed as the sum of a slowly varying equilibrium position, $x_0$, and a harmonic oscillation about that position with amplitude $A=\frac{e\tilde{U}\alpha'x_{0}}{m\omega_{\text{RF}}^{2}}$,
\begin{equation}
    x(t)=x_{0}+A\cos(\omega_{\text{RF}}t)=x_{0}+\frac{e\tilde{U}\alpha'x_{0}}{m\omega_{\text{RF}}^{2}}\cos(\omega_{\text{RF}}t)
    \label{eq:position-of-ion-from-EOM}
\end{equation}
The electric force felt by the ion is,
\begin{equation}
\begin{aligned}
    F_{x}(t)&=eE_{x}(t)\\
    &=-e\tilde{U}\alpha'x(t)\cos(\omega_{\text{RF}}t) \\
    &=-e\tilde{U}\alpha'\cos(\omega_{\text{RF}}t)\left(x_{0}+\frac{e\tilde{U}\alpha'x_{0}} {m\omega_{\text{RF}}^{2}}\cos(\omega_{\text{RF}}t)\right) \\
    &=-e\tilde{U}\alpha'x_{0}\cos(\omega_{\text{RF}}t)-\frac{e^{2}\tilde{U}^{2}\alpha'^{2}x_{0}}{m\omega_{\text{RF}}^{2}}\cos^{2}(\omega_{\text{RF}}t)
\end{aligned}
\label{eq:electric-force-on-ion}
\end{equation}
And averaging over one RF period,
\begin{equation}
    \bar{F}_{x}=-\frac{e^{2}\tilde{U}^{2}\alpha'^{2}}{2m\omega_{\text{RF}}^{2}}x_{0}
    \label{eq:average-force-on-ion-over-RF-period}
\end{equation}
Thus, averaging over the fast motion (micromotion) of the ion reveals an effective restoring force acting on its slow motion, $x_{0}$. This restoring force arises from the ponderomotive effect, which causes charged particles to migrate toward regions of lower electric-field amplitude in an inhomogeneous oscillating field. The resulting slow oscillation is known as the secular motion and is well approximated as harmonic motion with secular frequency
\begin{equation}
    \omega_{\text{sec}}=\frac{e\tilde{U}\alpha'}{\sqrt{2}m\omega_{\text{RF}}}
    \label{eq:general-secular-frequency}
\end{equation}
with $\bar{F}_{x}=-\omega_{\text{sec}}^{2}mx_{0}$. The motion due to the dynamic quadrupole along the $y$ axis can be treated in a similar manner. A more careful analysis to obtain the equation of motion of the ion is presented below including both the DC and RF contributions of the confinement. 

The equation of motion of a particle with mass $m$ and charge $e$ subject to the potential given by Equation \eqref{eq:ion-trap-potential-without-constraints} along the $x$ axis is
\begin{equation}
    \ddot{x}=-\frac{e}{m}\frac{\partial\Phi}{\partial x}=-\frac{e}{m}\left(U\alpha+U_{r}\alpha'+\tilde{U}\alpha'\cos(\omega_{\text{RF}}t)\right)x
    \label{eq:EOM-FULL}
\end{equation}
By making a change of variables, $\xi=\frac{\omega_{\text{RF}}t}{2}$, we get that
\begin{equation}
    \frac{d^{2}x}{d\xi^{2}}=-\frac{4e}{m\omega_{\text{RF}}^{2}}\left(U\alpha+U_{r}\alpha'+\tilde{U}\alpha'\cos(2\xi)\right)x
    \label{eq:EOM-FULL-Change-of-variables}
\end{equation}
We then define the following parameters $a_{x}$ and $q_{x}$ as follows
\begin{equation}
    a_{x}=\frac{4e}{m\omega_{\text{RF}}^{2}}(U\alpha+U_{r}\alpha'), q_{x}=\frac{2e\tilde{U}\alpha'}{m\omega_{\text{RF}}^{2}}
    \label{eq:define-ax-qx}
\end{equation}
Here, $a_{x}$ and $q_{x}$ represent unitless parameters that determine the stability of the confinement. Again, we impose the constraints that $-(\alpha+\beta)=\gamma>0$ and $\alpha'=-\beta'$ with $\gamma'=0$. Since $\alpha'=-\beta'$ and $\gamma'=0$, it must be true that $q_{x}=-q_{y}$ and $q_{z}=0$. This differential equation is in the form of the Mathieu equation as shown below
\begin{equation}
    \frac{d^{2}x}{d\xi^{2}}+\left(a_{x}-2q_{x}\cos(2\xi)\right)x=0
    \label{eq:mathieu-equation}
\end{equation}
The solution to the Mathieu equation is,
\begin{equation}
    x(t)\approx\left[x_{0}+C_{x}\cos\left(\beta_{x}\frac{\omega_{\text{RF}}}{2}t\right)\right]\left[1-\frac{q_{x}}{2}\cos(\omega_{\text{RF}}t)\right]
    \label{eq:solution-to-mathieu-equation}
\end{equation}
where $\beta_{x}\approx\sqrt{a_{x}+\frac{q_{x}^{2}}{2}}$, $|a_{x}|$, $q_{x}^{2}\ll1$, the secular frequency is $\omega_{x}=\beta_{x}\frac{\omega_{\text{RF}}}{2}$ with amplitude $C_{x}$, and $x_{0}$ indicates the equilibrium offset position from the RF null \cite{leibfriedQuantumDynamicsSingle2003}. Unlike the simplified solution given by Equation \eqref{eq:position-of-ion-from-EOM}, the Mathieu solution contains a term that describes motion at the RF frequency applied to the trap electrodes, known as the micromotion of the ion. When the ion is at the RF null, $x_{0}=0$, and $q_{x}\ll1$, the motion of the ion is well approximated by harmonic motion. However, if there are stray electric fields that displace the ion from the RF null, $x_{0}\neq0$, then micromotion becomes magnified by the displacement and can quickly dominate the secular motion. Projection of this motion along the direction of a probe laser beam causes phase modulation of the laser spectrum, complicating ion-laser interactions. Therefore, it is imperative to minimize excess micromotion by applying compensation fields with additional electrodes until $x_{0}\approx0$ - and similarly along the $y,z$ axes. A more detailed summary of the secular frequencies along the axial and radial directions is presented in the following two sections.

\subsection{Axial Secular Frequency}
\label{subsec:axial-secular-frequency}
 The harmonic motion along the $z$ axis of the trap (axial direction) is summarized here. Since $\gamma'=0$, then $q_{z}=0$, and so $\beta_{z}\approx\sqrt{a_{z}}$. The axial secular frequency is,
 \begin{equation}
     \omega_{z}=\beta_{z}\frac{\omega_{\text{RF}}}{2}=\sqrt{a_{z}}\frac{\omega_{\text{RF}}}{2}=\sqrt{\frac{4eU\gamma}{m\omega_{\text{RF}}^{2}}}\frac{\omega_{\text{RF}}}{2}=\sqrt{\frac{eU\gamma}{m}}
     \label{eq:axial-secular-frequency}
 \end{equation}
and the equation of motion is 
\begin{equation}
    z(t)\approx C_{z}\cos(\omega_{z}t)
    \label{eq:axial-secular-frequency-EOM}
\end{equation}
where the equilibrium offset position is taken to be at the RF null, $z_{0}=0$.

\subsection{Radial Secular Frequencies}
\label{subsec:radial-secular-frequencies}

The harmonic motion along the $x,y$ axes of the trap (radial directions) is modeled here. Since $\alpha'=-\beta'$ and $\gamma'=0$, then $q_{z}=0$, and $\beta_{x}\approx\sqrt{a_{x}+\frac{q_{x}^{2}}{2}}$ and $\beta_{y}\approx\sqrt{a_{y}+\frac{q_{y}^{2}}{2}}$. The radial secular frequencies are,
\begin{equation}
\begin{aligned}
    \omega_{x}&=\beta_{x}\frac{\omega_{\text{RF}}}{2}=\sqrt{a_{x}+\frac{q_{x}^{2}}{2}}\frac{\omega_{\text{RF}}}{2}=\sqrt{\frac{4e}{m\omega_{\text{RF}}^{2}}(U\alpha+U_{r}\alpha')+2\left(\frac{e\tilde{U}\alpha'}{m\omega_{\text{RF}}^{2}}\right)^{2}}\frac{\omega_{\text{RF}}}{2} \\
    \omega_{y}&=\beta_{y}\frac{\omega_{\text{RF}}}{2}=\sqrt{a_{y}+\frac{q_{y}^{2}}{2}}\frac{\omega_{\text{RF}}}{2}=\sqrt{\frac{4e}{m\omega_{\text{RF}}^{2}}(U\beta+U_{r}\beta')+2\left(\frac{e\tilde{U}\beta'}{m\omega_{\text{RF}}^{2}}\right)^{2}}\frac{\omega_{\text{RF}}}{2}
    \label{eq:radial-secular-frequencies-x,y}
\end{aligned}
\end{equation}
and the equations of motion are
\begin{equation}
\begin{aligned}
    x(t)&\approx C_{x}\cos(\omega_{x}t)\left(1-\frac{q_{x}}{2}\cos(\omega_{\text{RF}}t)\right) \\
    y(t)&\approx C_{y}\cos(\omega_{y}t)\left(1-\frac{q_{y}}{2}\cos(\omega_{\text{RF}}t)\right)
    \label{eq:radial-secular-frequency-EOM}
\end{aligned}
\end{equation}
where the equilibrium offset position is taken to be at the RF null, $x_{0}$, $y_{0}=0$. Note that if $\alpha=\beta$, $\alpha'=-\beta'$, and $U_{r}=0\text{ V}$, then the radial secular frequencies are equal $\omega_{x}=\omega_{y}$. 

\section{Alpine Quantum Technologies (AQT) Pine Trap}
\label{sec:apline-quantum-technologies-pine-trap}

\begin{figure}[!!h]
\centering
\includegraphics[width=0.49\textwidth]{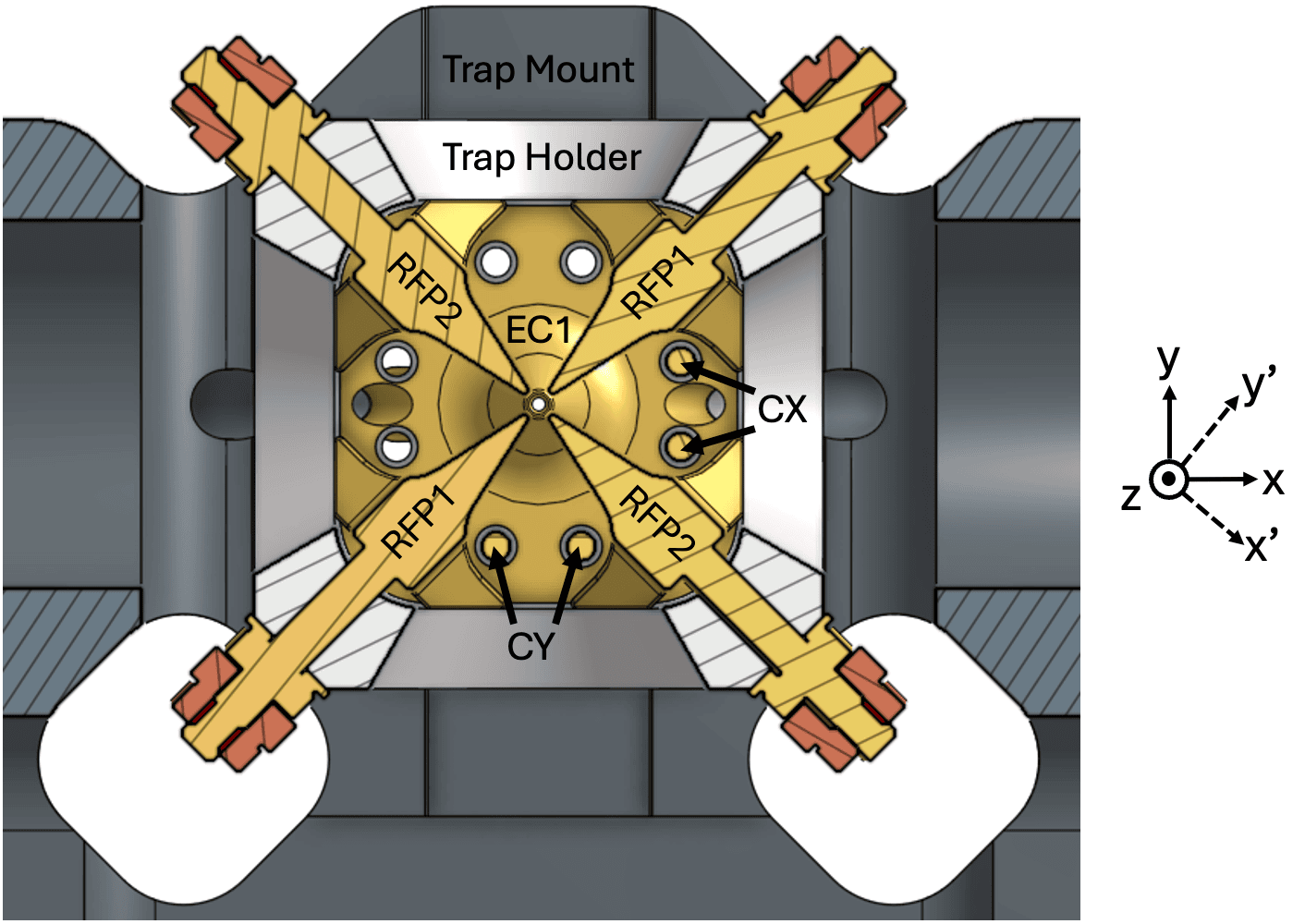}
\includegraphics[width=0.49\textwidth]{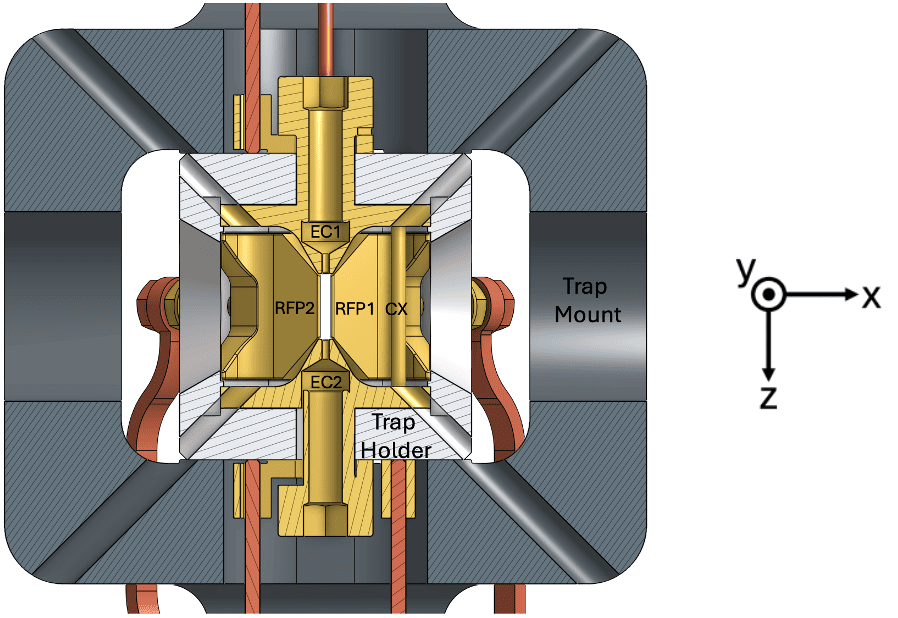}
\caption[AQT Trap]{AQT trap cross sections. For notational simplicity in this thesis, I'll refer to the radial secular frequencies as $\omega_x,\omega_y$.}
\label{fig:aqt-trap}
\end{figure}

We used a linear RF Paul trap in this work designed and built by Alpine Quantum Technologies (AQT) - the Pine Trap. This trap is colloquially called a blade trap as it consists of 4 RF electrodes (blades) for radial confinement, 2 DC electrodes (endcaps) for axial confinement, and an additional 2 DC electrodes (compensation) to null excess micromotion, see Figure \ref{fig:aqt-trap}. The blades positioned opposite to each relative to the trap center are electrically shorted together and are denoted RF pair 1 (RFP1) and RF pair 2 (RFP2). In this work, RF voltage is applied to RFP1 via a helical resonator, a DC bias voltage is applied to RFP2, and the endcaps (EC1, EC2) and compensation electrodes (CX, CY) are all independently biased with DC voltages. Capacitors on a printed circuit board (PCB) near the trap are wired in parallel to all the DC electrodes to reduce RF pickup from RFP1, see Section \ref{sec:trap-electrode-filtering-circuits}. The ion-electrode distance is $577\text{ \ensuremath{\mu}m}$ and the distance between the two endcaps is $4.5\text{ mm}$. All of the electrodes are made of titanium (grade 5) with a $1.75\text{ \ensuremath{\mu}m}$ gold plating. The electrodes are held and insulated from the titanium (grade 5) trap mount using an alumina (AL998) holder. The trap mount is bolted to the 4 K stage of the cryostat. 

\subsection{Simulation and Measurement of Secular Frequencies}
\label{subsec:simulation-and-measurement-of-secular-frequencies}

\begin{figure}[!!h]
\centering
\includegraphics[width=0.49\textwidth]{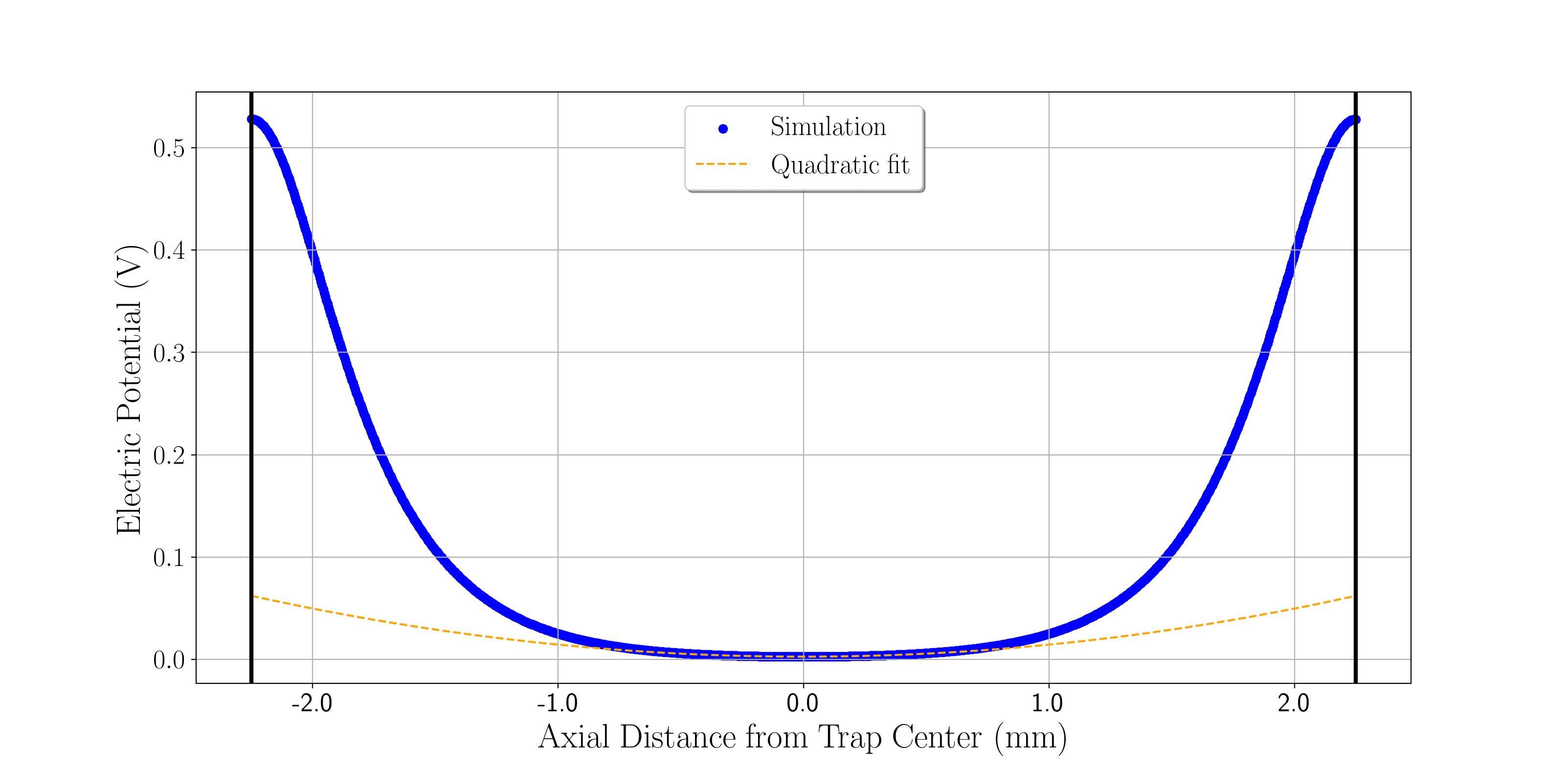}
\includegraphics[width=0.49\textwidth]{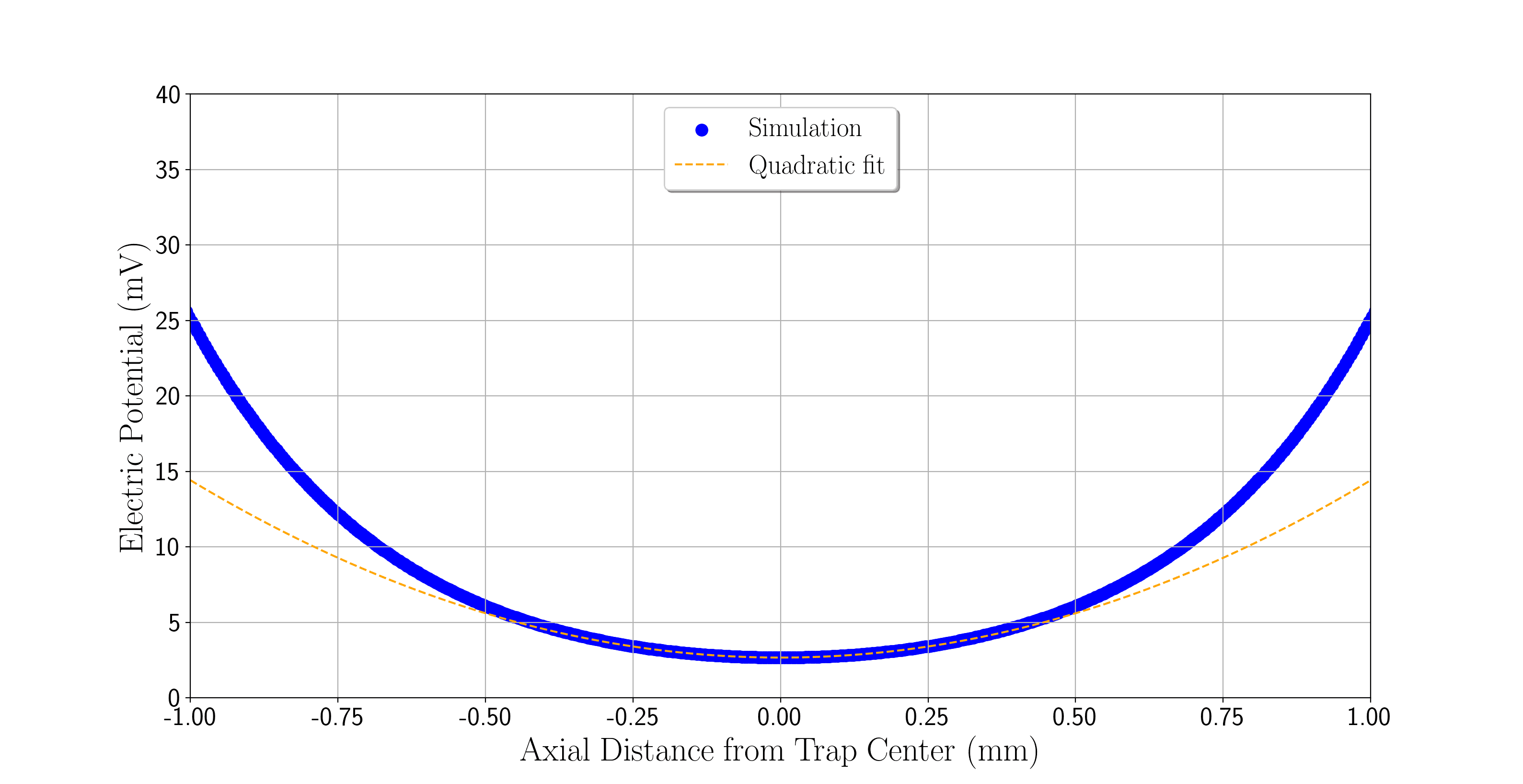}
\caption[Axial FEM Simulation of Trap Potential]{(Left) FEM simulation data of the axial electrode potential, with $5\text{ \ensuremath{\mu}m}$ resolution, as a function of the distance from the trap center with 1 V applied to each endcap electrode with all other electrodes grounded. The vertical black lines indicate the position of the endcaps. A quadratic fit is made for the region $\pm200\text{ \ensuremath{\mu}m}$ about the trap center to extract the geometric prefactor $\gamma=2.35\times10^{4}\text{ }\frac{1}{\text{m}^{2}}$. (Right) The same FEM simulation and fit plotted in the region $\pm1\text{ mm}$ about the trap center.}
\label{fig:axial-FEM-simulation-potential}
\end{figure}

\begin{figure}[!!h]
\centering
\includegraphics[width=\textwidth]{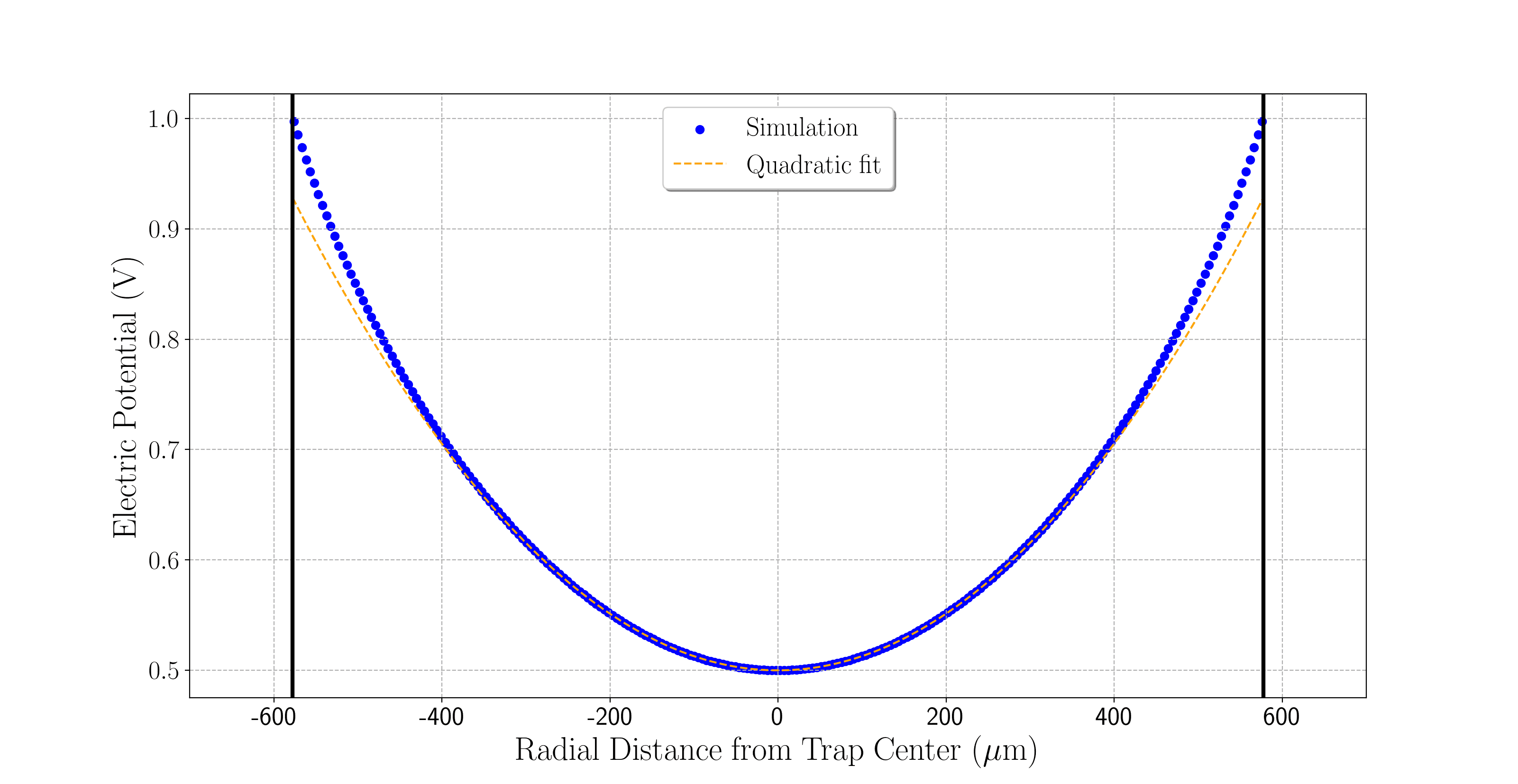}
\caption[Radial FEM Simulation of Trap Potential]{FEM simulation data of the radial electric potential, with $5\text{ \ensuremath{\mu}m}$ resolution, as a function of the distance from the trap center with 1 V applied to one pair of RF blades, RFP1. The vertical black lines indicate the position the RF electrodes. A quadratic fit is made for the region $\pm200\text{ \ensuremath{\mu}m}$ about the trap center to extract the geometric prefactor $\alpha'=2.57\times10^{6}\text{ }\frac{1}{\text{m}^{2}}$.}
\label{fig:radial-FEM-simulation-potential}
\end{figure}

In the limit of small excursions around the trap center, the motion of the ion is approximately harmonic. To predict the secular frequencies, a static Finite Element Analysis (FEM) simulation was performed using COMSOL, with $5\text{ \ensuremath{\mu}m}$ resolution, to extract the potential curvatures in each spatial direction via quadratic fits. For the axial direction, 1 V is applied to both endcaps while all other electrodes are grounded. For the radial direction, 1 V is applied to one the pairs of RF blades, $\text{RFP1}$, while all other electrodes are grounded. For each direction, quadratic fits were made to a range of $\pm200\text{ \ensuremath{\mu}m}$ around the trap center to extract the geometric prefactors in Equation \eqref{eq:ion-trap-potential-with-constraints}. The FEM simulation data and quadratic fits for the axial and radial potentials are shown in Figures \ref{fig:axial-FEM-simulation-potential} and \ref{fig:radial-FEM-simulation-potential} respectively. 

\begin{figure}[!!h]
\centering
\includegraphics[width=\textwidth]{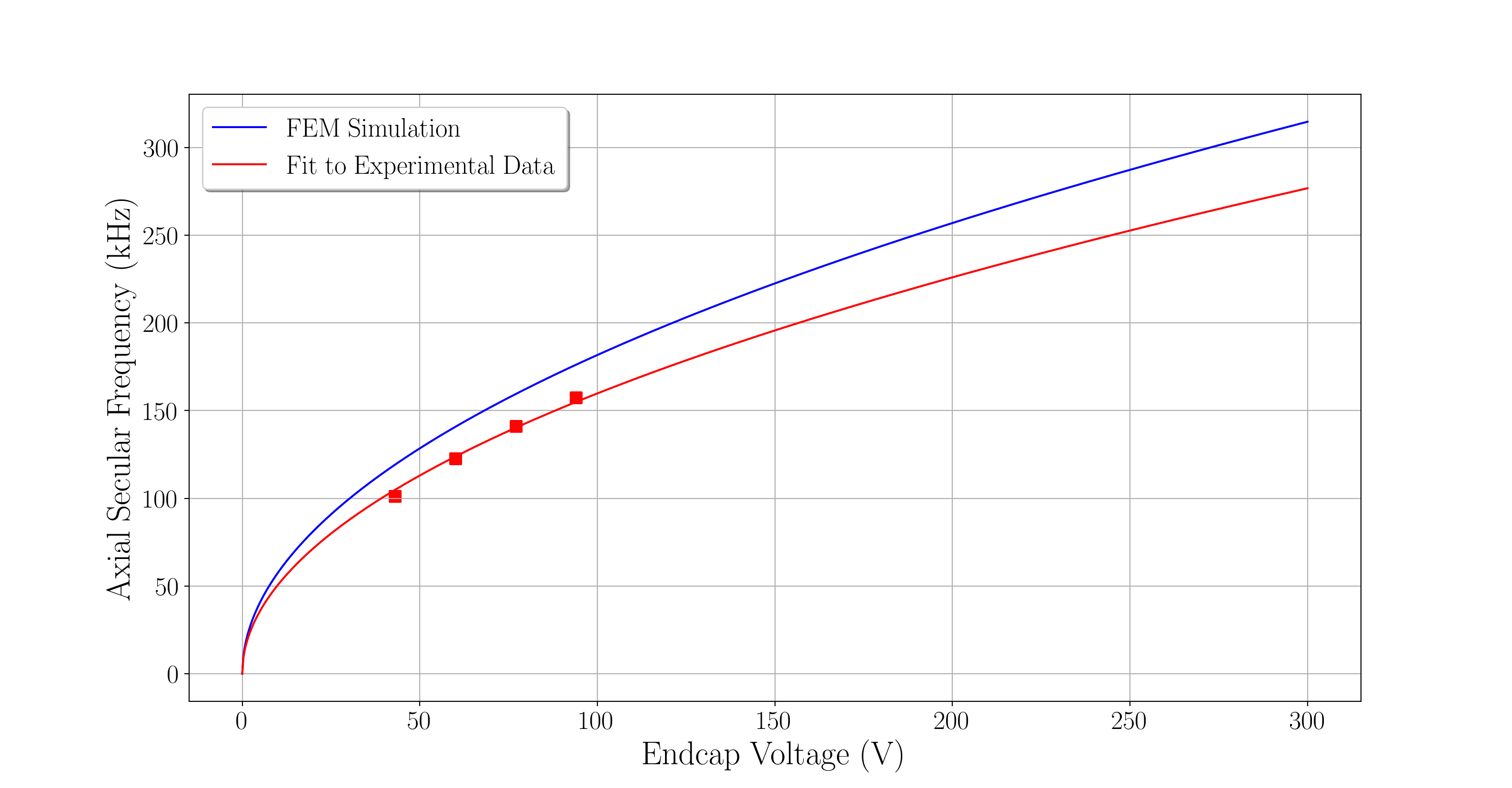}
\caption[Axial Secular Frequency vs. Endcap Voltage]{Axial secular frequency as a function of the endcap voltage in from the FEM simulation and experiment. The geometric prefactor extracted from the experimental data is $\gamma=1.82\times10^{4}\text{ }\frac{1}{\text{m}^{2}}$ and thus $\omega_{z}=2\pi\times16.0\text{ }\frac{\text{kHz}}{\sqrt{\text{V}}}\sqrt{U_{EC}}$. The maximum axial secular frequency given by the experiment data is $\omega_{z}=2\pi\times277\text{ kHz}$ constrained by the 300 V voltage rating of the DC feedthroughs. The maximum axial secular frequency for $^{174}\text{Yb}^+$ in this trap is $\omega_{z}=2\pi\times506.0\text{ kHz}$ given the 1 kV voltage rating as per the AQT manual.}
\label{fig:axial-secular-frequency-vs-voltage}
\end{figure}

The geometric prefactor $\gamma=2.35\times10^{4}\text{ }\frac{1}{\text{m}^{2}}$ is extracted from the quadratic fit shown in Figure \ref{fig:axial-FEM-simulation-potential}. The axial secular frequency can now be predicted using Equation \eqref{eq:axial-secular-frequency} and is plotted together with the measured secular frequencies as a function of the endcap voltage in Figure \ref{fig:axial-secular-frequency-vs-voltage}. A nonlinear least squares fit is made to the experimental data to extract an experimental geometric prefactor $\gamma=1.82\times10^{4}\text{ }\frac{1}{\text{m}^{2}}$ where we can write the axial secular frequency using the following equation.
\begin{equation}
    \omega_{z}=2\pi\times16.0\text{ }\frac{\text{kHz}}{\sqrt{\text{V}}}\sqrt{U_{EC}}
    \label{experimental-axial-secular-frequency}
\end{equation}
The difference in measured versus simulated secular frequencies could be due to errors in the simulation but also due to the fact the simulation is performed at 300 K. Titanium contracts 3 times more than alumina from 300 K to 4 K, and since the electrodes are held in place by the alumina trap holder, the endcap distance could increase by $\sim30\text{ \ensuremath{\mu}m}$ at 4 K. This was not simulated in this work. 

\begin{figure}[!!h]
\centering
\includegraphics[width=\textwidth]{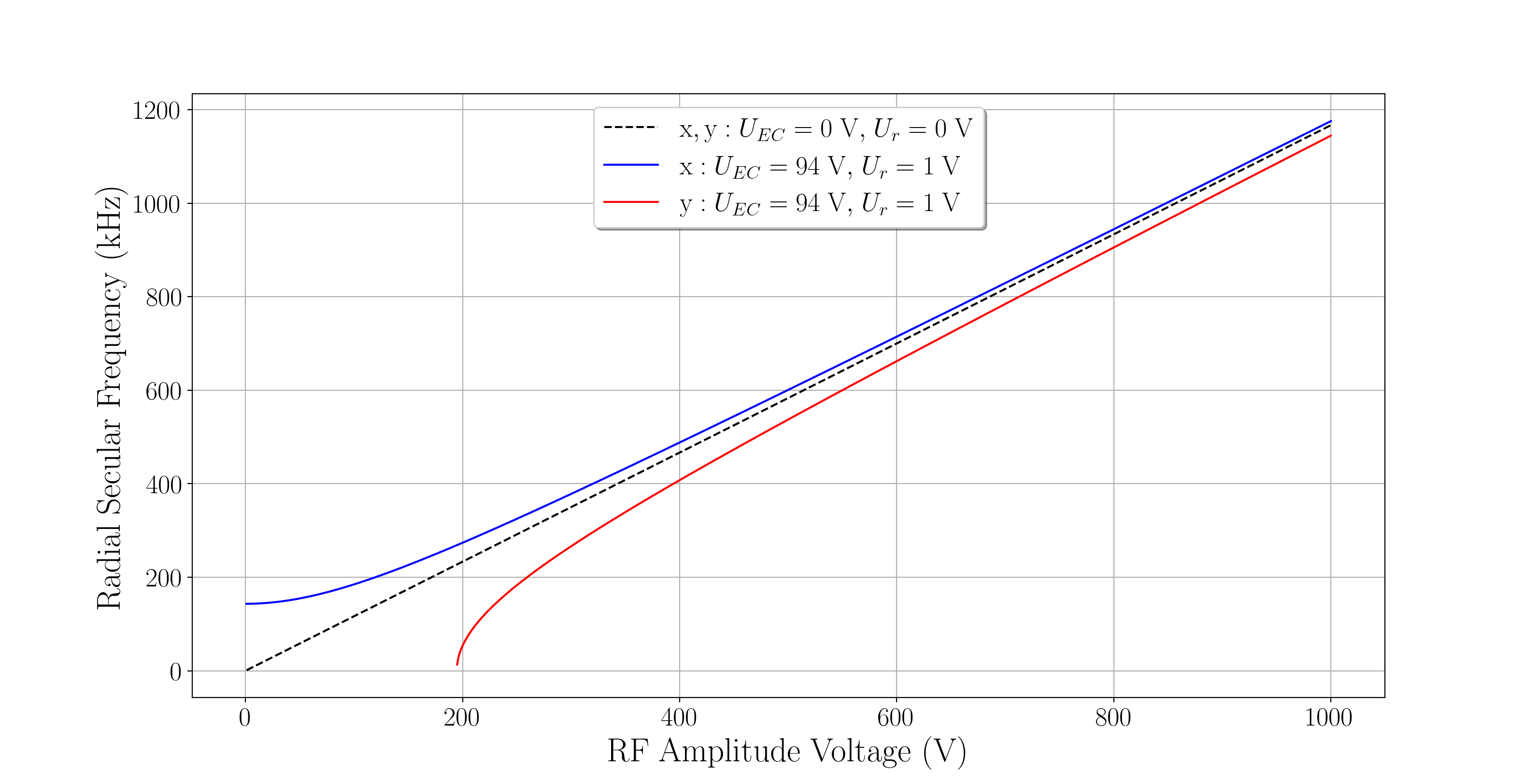}
\caption[Radial Secular Frequency vs. Blade Voltage]{Radial secular frequency as a function of the blade voltage from the FEM simulation. The geometric prefactor extracted from the FEM simulation is $\alpha^{\prime}=-\beta^{\prime}=1.82\times10^{4}\text{ }\frac{1}{\text{m}^{2}}$.}
\label{fig:radial-secular-frequency-vs-voltage}
\end{figure}

\begin{figure}[!!h]
\centering
\includegraphics[width=\textwidth]{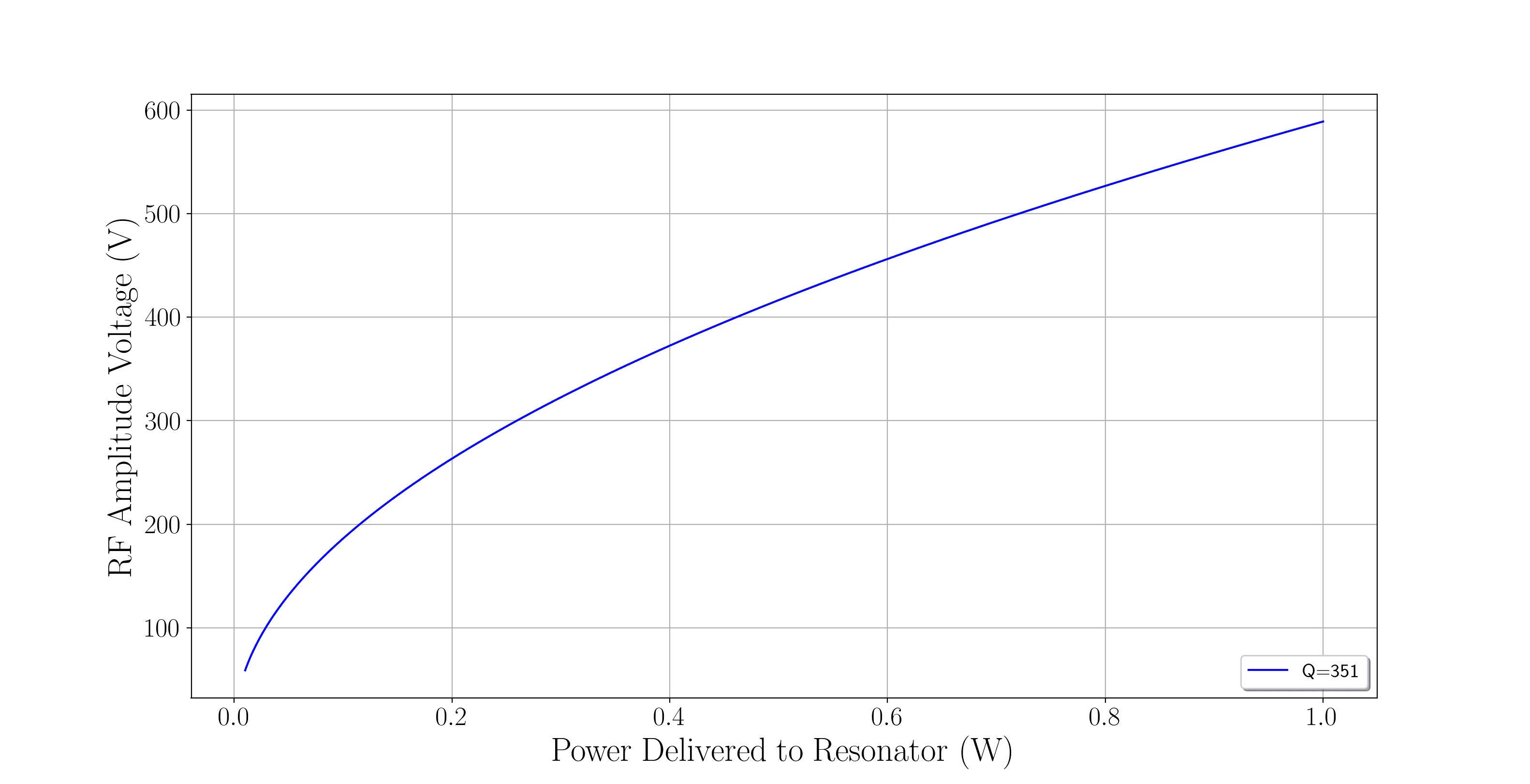}
\caption[Resonator RF Amplitude Voltage vs. Power]{Resonator RF amplitude voltage versus input power simulation.}
\label{fig:resonator-RF-amplitude-voltage-vs-power}
\end{figure}

For the radials, the geometric prefactor extracted from the FEM simulation is $\alpha^{\prime}=-\beta^{\prime}=1.82\times10^{4}\text{ }\frac{1}{\text{m}^{2}}$. Thus, the radial secular frequencies can be predicted using Equation \eqref{eq:radial-secular-frequencies-x,y}. In this work, the voltage was not directly measured at the blades as discussed in Section \ref{sec:helical-resonator}. Instead the voltage is predicted as in \cite{sivernsApplicationRadioFrequency2012} shown in Figure \ref{fig:resonator-RF-amplitude-voltage-vs-power}. In this work, the operating input power to the resonator is $448\text{ mW}$ with endcap and blade DC offset voltages of 94 V and 1 V respectively creates measured radial secular frequencies of $\omega_{x}=2\pi\times500\text{ kHz}$ and $\omega_{y}=2\pi\times570\text{ kHz}$. From the simulation in Figure \ref{fig:radial-secular-frequency-vs-voltage}, this corresponds to an RF amplitude voltage of 470 V.

\section{Helical Resonator}
\label{sec:helical-resonator}

\begin{figure}[!!h]
\centering
\includegraphics[width=0.6\textwidth]{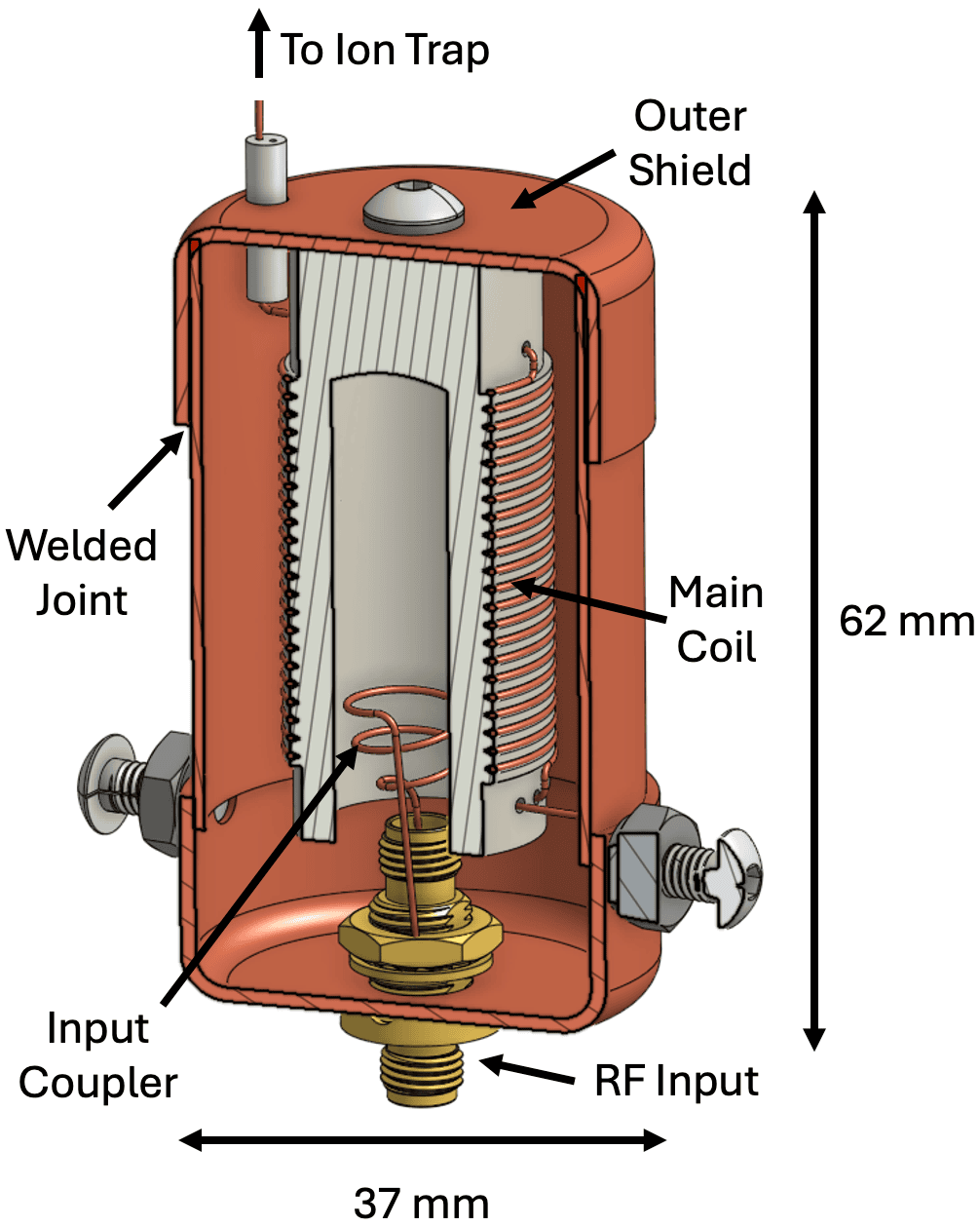}
\caption[Helical Resonator Assembly]{CAD drawing of the helical resonator.}
\label{fig:helical-resonator-assembly}
\end{figure}

We generate the radial confinement potential of the ion trap using an in-vacuum helical resonator mounted on the 4 K stage. The helical resonator consists of an adjustable input coupler, a main helical coil, and an outer shield as shown in Figure \ref{fig:helical-resonator-assembly}. The input coupler consists of an SMA bulkhead connector (Pasternack, PN: PE9184NH) soldered to an antenna fabricated from 24 AWG C10100 wire. The antenna is mounted on a holder that clamps to the outer shield, allowing its position relative to the main coil to be adjusted by sliding the holder along the shield before securing it with \#8-32 SS316 bolts and nuts. To ensure that the main coil remains mechanically rigid, a helical groove with a pitch of 17.5 threads per inch was machined into a PEEKGF30 cylinder, around which 24 AWG C10100 wire was wound. The completed coil is then connected directly to the ion trap RFP1 electrode.

As reported in \cite{paganoCryogenicTrappedionSystem2018a}, cooling the resonator to cryogenic temperatures substantially increases its unloaded quality factor, $Q_{0}$, due to the reduced resistivity of copper. Since the input antenna geometry determines the external quality factor, $Q_{\text{e}}$, and cooling increases the unloaded quality factor, $Q_{0}$, the impedance matching between the input coupler and the resonator changes for the same antenna position. Consequently, the antenna must be retracted slightly from the main coil to restore critical coupling at cryogenic temperatures. At room temperature, the optimum antenna position produces a resonance frequency of 21.624689 MHz with a loaded quality factor of $Q_\text{L}=117$ and a coupling of $S11(\omega_{\text{res}})=-3.8\text{ dB}$, corresponding to  $\sim$59\% of the incident RF power being delivered to the resonator. After cooling to 5.4 K, the optimum coupling of $S11(\omega_{\text{res}})=-38.8\text{ dB}$ occurs at a resonance frequency of 21.854224 MHz with a loaded quality factor of $Q_\text{L}=351$, corresponding to  $\sim$99.9\% power transfer, see Figure \ref{fig:helical-resonator-S11-300K-and-cryo}. The increase in resonance frequency is attributed to the slight reduction in the resonator inductance and capacitance resulting from the thermal contraction of the copper conductors and the PEEKGF30 support. 

\begin{figure}[!!h]
\centering
\includegraphics[width=0.49\textwidth]{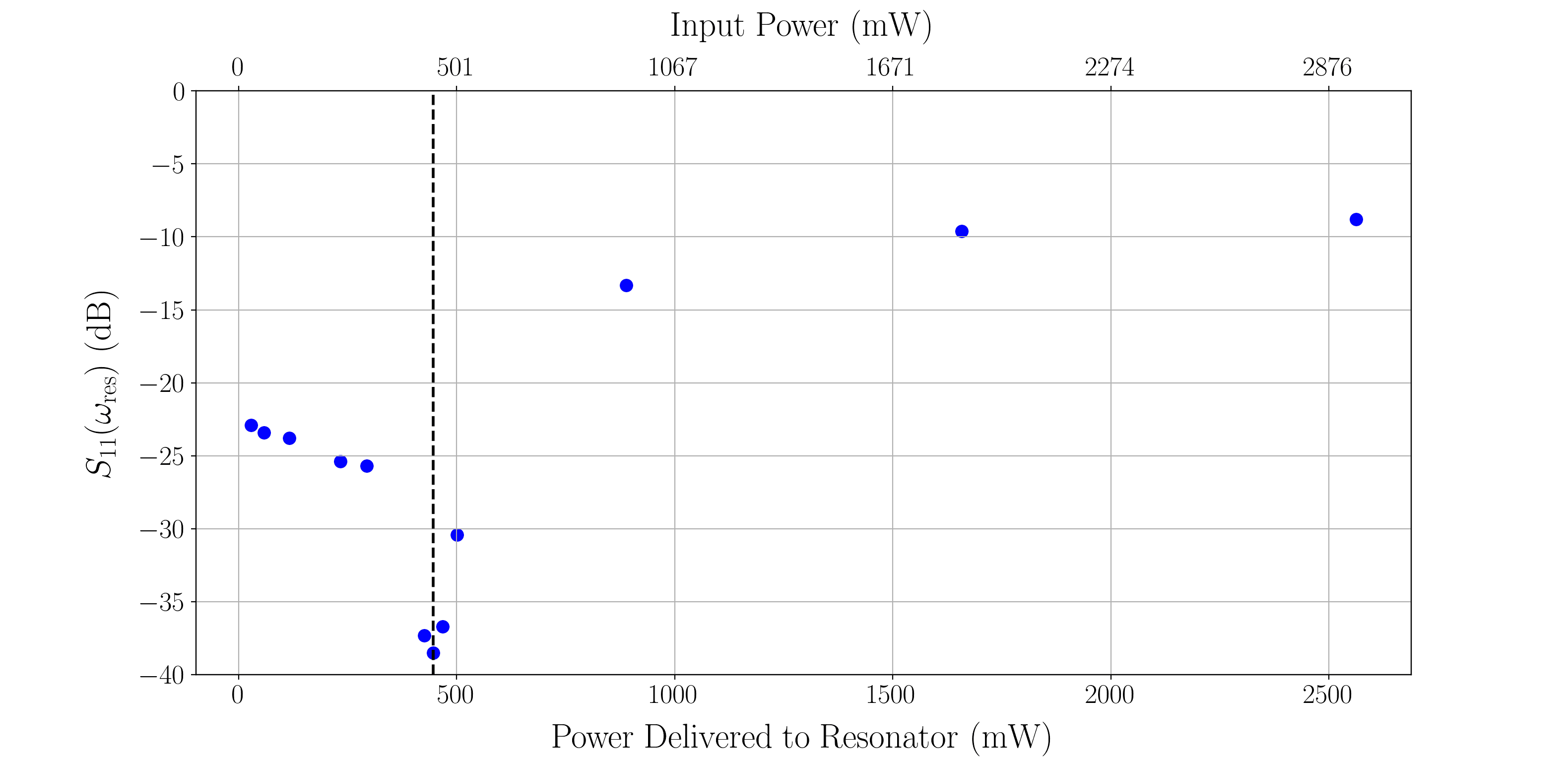}
\includegraphics[width=0.49\textwidth]{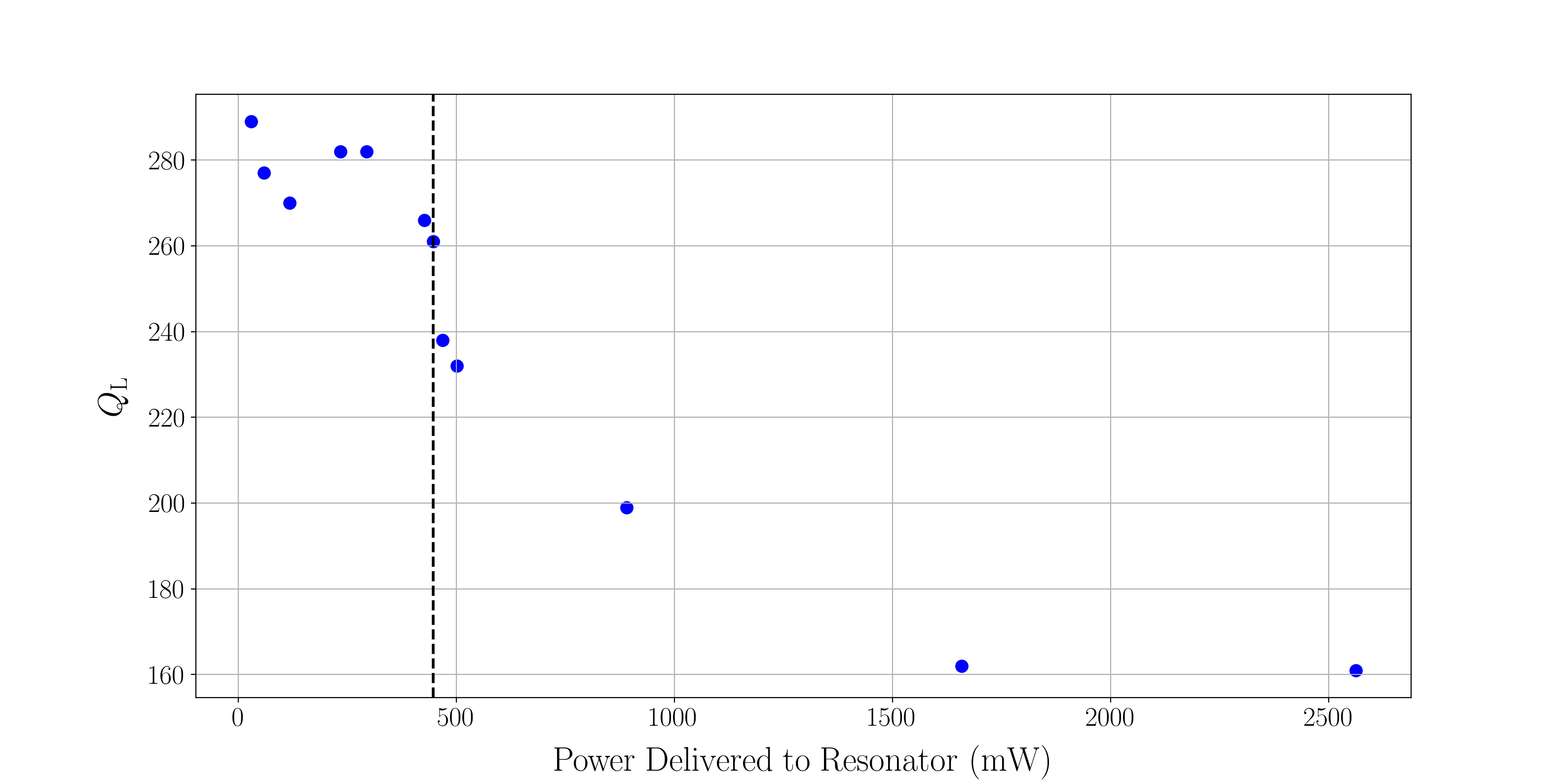}
\newline
\includegraphics[width=0.49\textwidth]{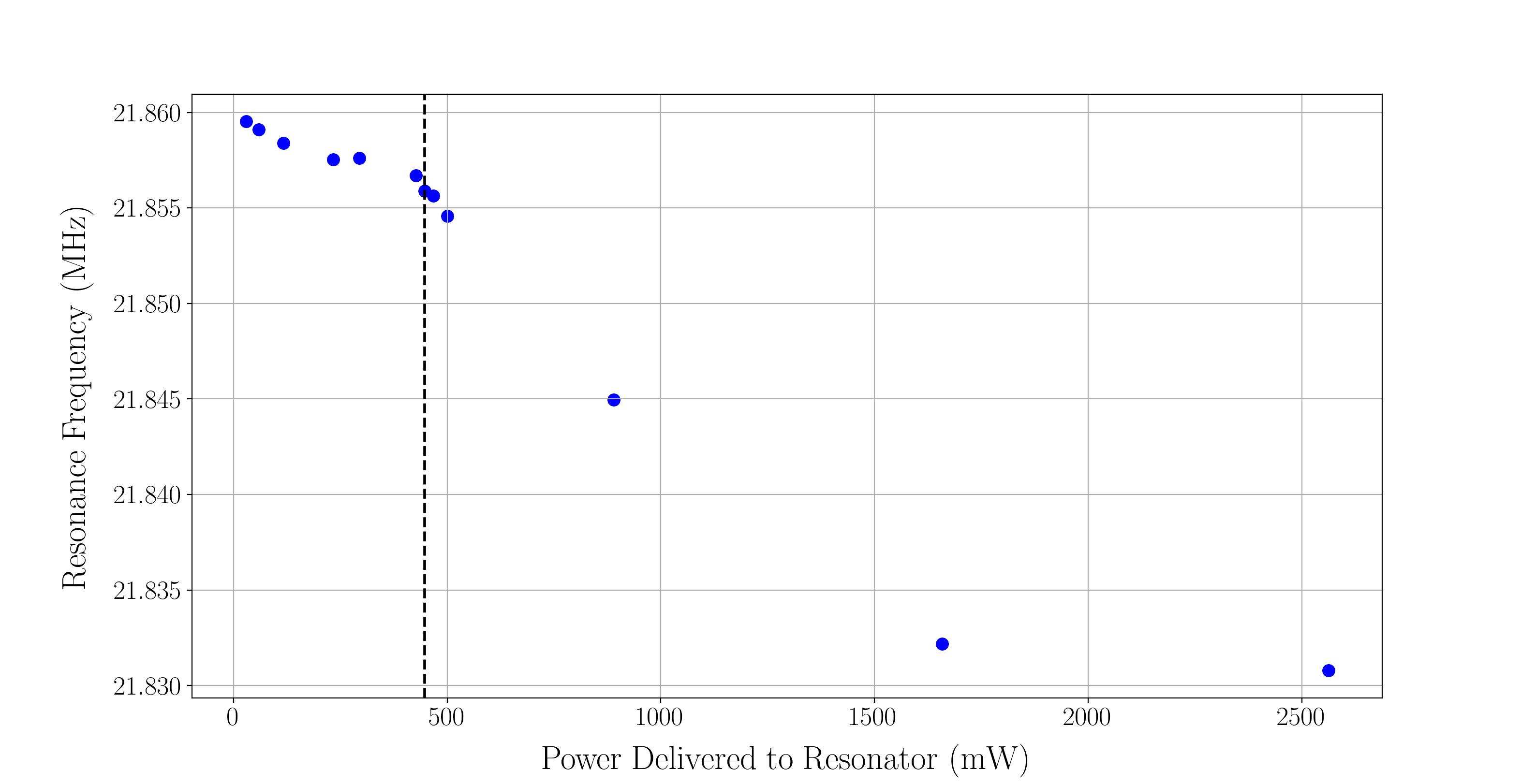}
\includegraphics[width=0.49\textwidth]{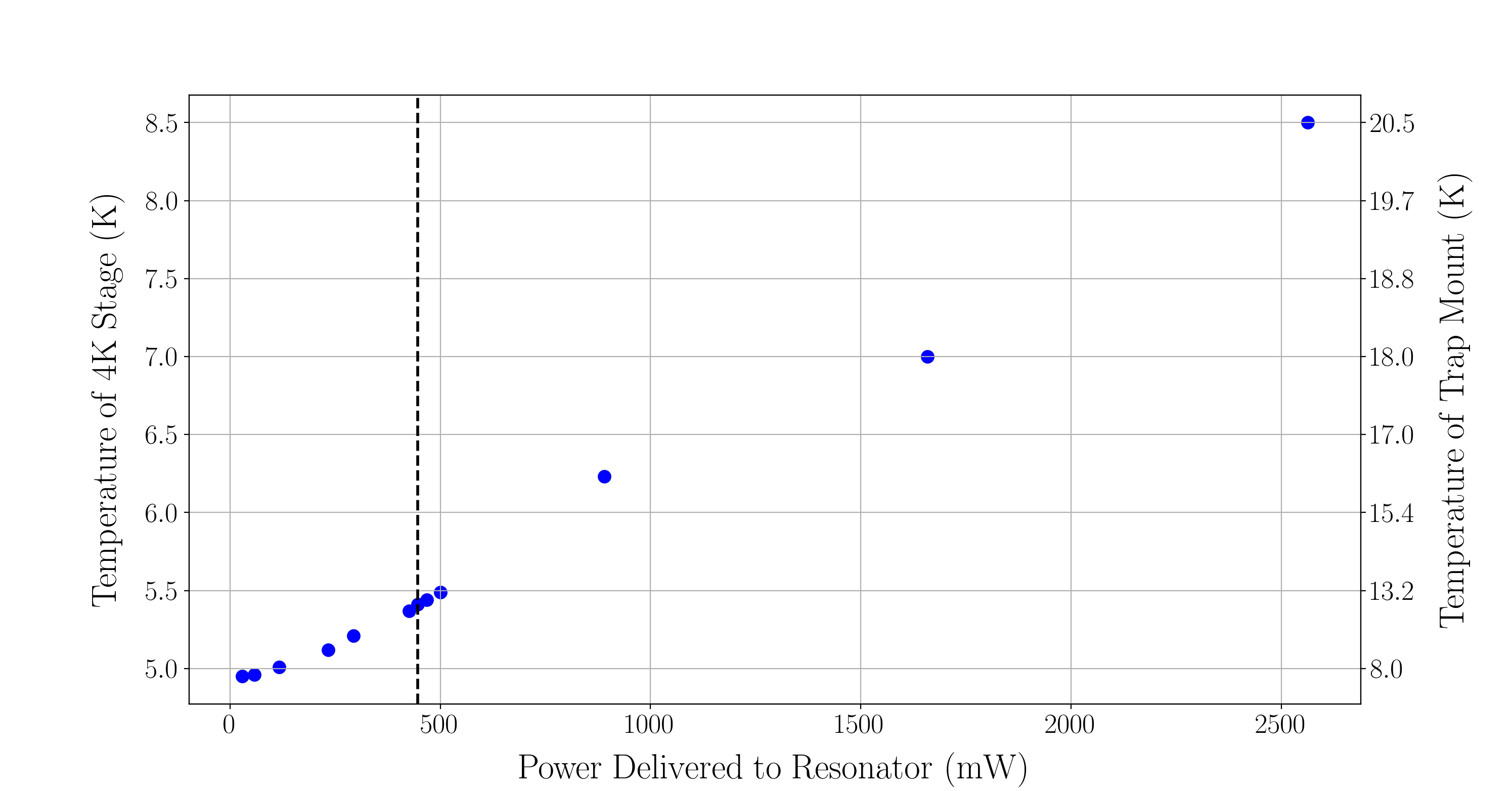}
\caption[Helical Resonator Response at Cryogenic Temperatures]{Resonator response at cryogenic temperatures. All data points are taken at optimal coupling once the system has equilibrated after $\sim1\text{ hr}$. (Top Left) $S11(\omega_{\text{res}})$ vs. power, (Top Right) $Q_{\text{L}}$ vs. power, (Bottom Left) $F_{\text{res}}$ vs. power, and (Bottom Right) Temperature of the 4 K stage and Trap Mount vs. power where the conductance between the trap electrodes and 4 K stage is extracted to be $\sim 0.06\,\frac{\text{W}}{\text{K}}$, which is consistent with the AQT manual.}
\label{fig:helical-resonator-response-at-cryogenic-temperatures}
\end{figure}

The resonator response at cryogenic temperatures is shown in Figure \ref{fig:helical-resonator-response-at-cryogenic-temperatures} including $S11(\omega_{\text{res}}), Q_{\text{L}}$, $F_{\text{res}}$, and 4 K stage and Trap Mount temperatures versus power delivered to the resonator. The coupling of a one-port helical resonator is characterized by the coupling coefficient $\beta=\frac{Q_{0}}{Q_{\text{e}}}$ where $ Q_{0}$ is the unloaded quality factor and $Q_{\text{e}}$ is the external quality factor determined by the input coupler. The loaded quality factor is related to the unloaded quality factor by $Q_{\text{L}}=\frac{Q_{0}}{1+\beta}$ and the reflection coefficient measured at resonance is $|S11(\omega_{\text{res}})|=\frac{|1-\beta|}{|1+\beta|}$ \cite{pozar2012microwave}. During these measurements, the antenna position was fixed so that $Q_{\text{e}}$ remained approximately constant. Consequently, changes in the unloaded quality factor, $Q_{0}$, alter the effective coupling coefficient, $\beta$, and therefore the reflection coefficient, $S11(\omega_{\text{res}})$. As shown in Figure \ref{fig:helical-resonator-response-at-cryogenic-temperatures}, increasing the applied RF power increases the thermal load on the 4 K stage, causing a modest increase in resonator temperature. Although the corresponding shift in resonance frequency is small, indicating only a weak temperature dependence of the resonator inductance and capacitance, the resonator losses changes with temperature, modifying the unloaded quality factor, $Q_{0}$. As a result, the coupling coefficient changes with RF power, causing the resonator to move toward or away from critical coupling. In the present system, the reflected power reaches a minimum near an input power of 460 mW, indicating that the resonator is closest to critical coupling. At critical coupling, $\beta=1$ so that $Q_{\text{L}}=\frac{Q_{0}}{2}$ and $S11(\omega_{\text{res}})\rightarrow-\infty$.

\subsection{Experimental Setup}
\label{subsec:experimental-setup-helical-resonator}

\begin{figure}[!!h]
\centering
\includegraphics[width=\textwidth]{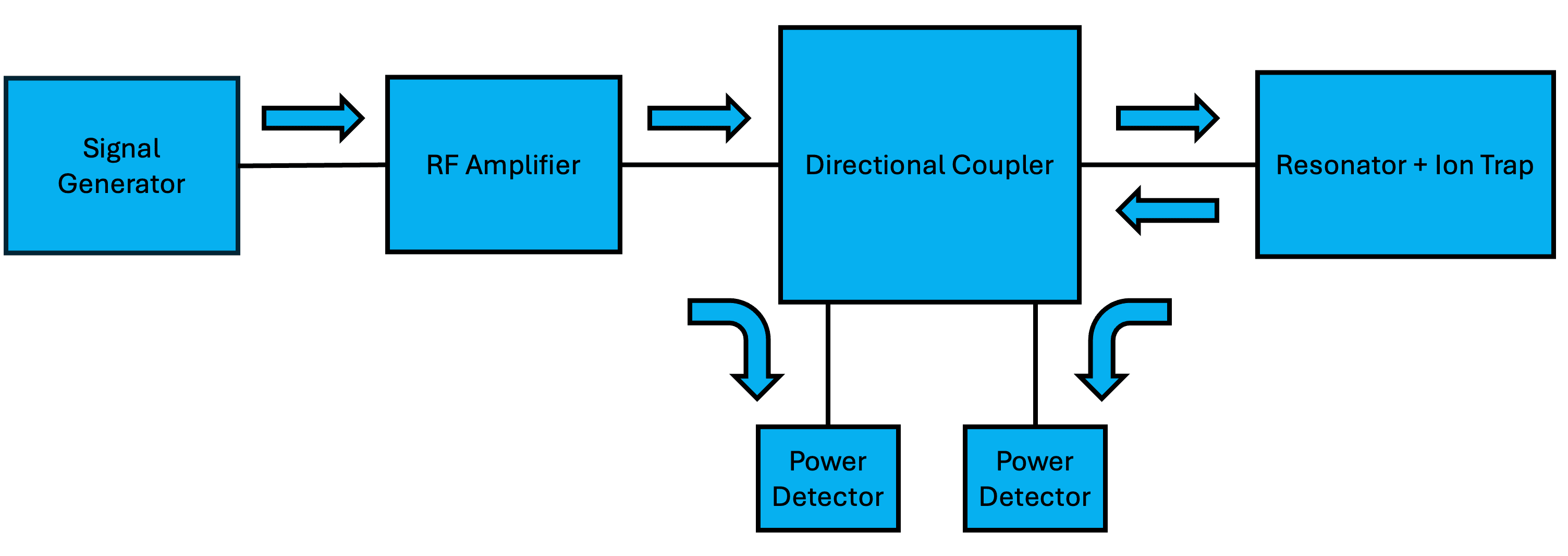}
\caption[Helical Resonator Circuit Diagram]{Resonator RF circuit diagram.}
\label{fig:helical-resonator-RF-circuit-diagram}
\end{figure}

The RF drive is delivered by a signal generator (Rigol, PN: DG4162) and amplified using an amplifier (Mini-circuits, PN: ZHL-1-2W-S+). To monitor S11 of the resonator a directional coupler (Mini-circuits, PN: ZFBDC20-62HP-S+) is placed in between the amplifier and resonator. The RF power measured at the coupled ports of the directional coupler is monitored using power detectors (Mini-circuits, PN: ZX47-60LN-S+) as shown in Figure \ref{fig:helical-resonator-RF-circuit-diagram}. 

\begin{figure}[!!h]
\centering
\includegraphics[width=\textwidth]{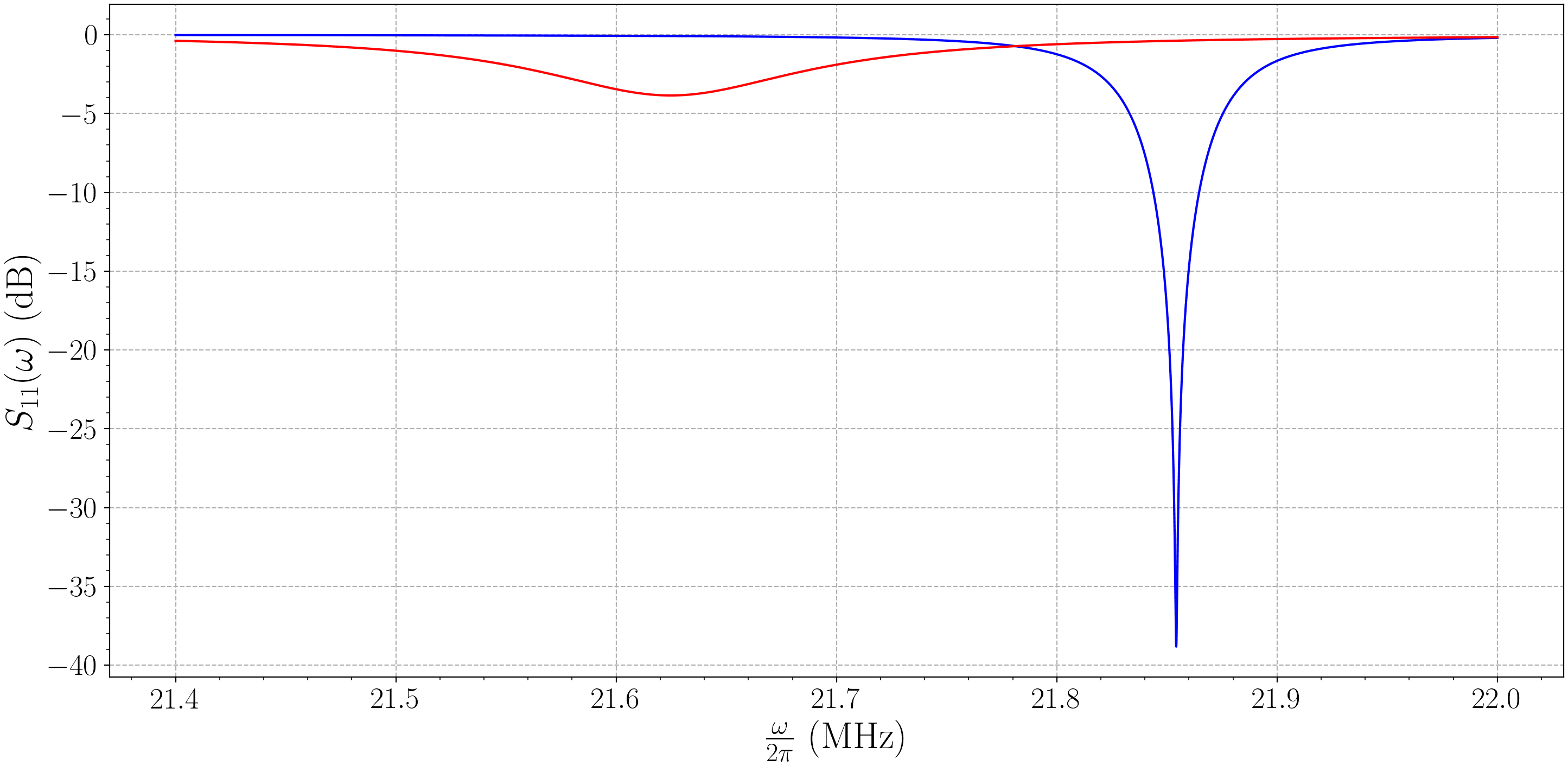}
\caption[Helical Resonator S11 at 300 K and at 5.4 K]{A plot of of the resonator's $S_{11}(\omega)$ measured at 300 K (red) and 5.4 K (blue). At 300K, the resonance frequency is 21.624689 MHz, the loaded quality factor is $Q_\text{L}=117$, and the coupling at resonance is $S11(\omega_{\text{res}})=-3.8\text{ dB}$ corresponding to $\sim$59\% of the power being delivered to the resonator. At 5.4 K, the resonance frequency is 21.854224 MHz, the loaded quality factor is $Q_\text{L}=351$, and the coupling at resonance is $S11(\omega_{\text{res}})=-38.8\text{ dB}$ corresponding to $\sim$99.9\% of the power is delivered to the resonator.}
\label{fig:helical-resonator-S11-300K-and-cryo}
\end{figure}

To extract the quality factor, $S11(\omega)$, and resonance frequency of the resonator, we model the resonator as a series RLC circuit with an input impedance, 
\begin{equation}
    Z_{\text{in}}=R+i\omega L+\frac{1}{i\omega C}
    \label{eq:input-impedance-RLC-circuit}
\end{equation}
where $R$ is the total resistive losses in the resonator and ion trap in $\text{\ensuremath{\Omega}}$, $L$ is the inductance of the resonator and ion trap in $\text{H}$, and $C$ is the capacitance of the resonator and ion trap in $\text{F}$. The reflection coefficient at the input to the RLC circuit is 
\begin{equation}
    \Gamma=\frac{Z_{\text{in}}-Z_{0}}{Z_{\text{in}}+Z_{0}}
    \label{eq:reflection-coefficient-RLC}
\end{equation}
where $Z_{\text{in}}$ is the input impedance given by Equation \eqref{eq:input-impedance-RLC-circuit} and $Z_{0}=50\text{ \ensuremath{\Omega}}$ is the RF source impedance. The $S_{11}$ and return loss (RL) referenced to power is given by,
\begin{equation}
    S_{11}=-RL=10\log(|\Gamma|^{2})=10\log\left(\frac{P_{\text{ref}}}{P_{\text{in}}}\right)=P_{\text{ref}}[\text{dBm}]-P_{\text{in}}[\text{dBm}]
    \label{eq:general-definition-S11}
\end{equation}
where $P_{\text{in}}$ and $P_{\text{ref}}$ are the input and reflected powers of the RLC circuit respectively. By combining Equations \eqref{eq:reflection-coefficient-RLC} and \eqref{eq:general-definition-S11} we can write,
\begin{equation}
    S_{11}(\omega)=10\log\left(\frac{(\gamma-\gamma_{\text{s}})^{2}\omega^{2}+(\omega^{2}-\omega_{0}^{2})^{2}}{(\gamma+\gamma_{\text{s}})^{2}\omega^{2}+(\omega^{2}-\omega_{0}^{2})^{2}}\right)
    \label{eq:S11-vs-omega}
\end{equation}
where $\gamma=\frac{R}{L}$ is the frequency width of the unloaded resonator\footnote{The unloaded Q of a series RLC circuit is $Q_{0}=\frac{\omega_{0}L}{R}$.}, $\gamma_{\text{s}}=\frac{Z_{0}}{L}$ relates the source impedance and inductance of circuit, $\omega_{0}=\frac{1}{\sqrt{LC}}$ is the resonance frequency, and $Q_{\text{L}}=\frac{\omega_{0}}{\gamma+\gamma_{\text{s}}}$ is the loaded Q-factor of the RLC circuit.\footnote{The external coupling due to the $Z_{0}=50\text{ Ohm}$ source is $Q_{\text{e}}=\frac{\omega_{0}L}{Z_{0}}$ and so the loaded Q is $\frac{1}{Q_{\text{L}}}=\frac{1}{Q_{\text{e}}}+\frac{1}{Q_{0}}$.}\cite{dubielzigUltralowvibrationClosedcycleCryogenic2021}. Thus, we can infer the coupling of the RF source to the resonator by measuring $P_{\text{in}}$ and $P_{\text{ref}}$ while sweeping the RF drive frequency to get $S_{11}(\omega)$, and then fit Equation \eqref{eq:S11-vs-omega} to the data to extract $\gamma$, $\gamma_{\text{s}}$, $\omega_{0}$, and $Q$. Knowing the coupling, $S_{11}(\omega)$, we can calculate the percent of the input power that is delivered to the resonator and ion trap using $\frac{P_{\text{load}}}{P_{\text{in}}}=\left(1-10^{S_{11}(\omega)/10}\right)$. A plot of of the resonator quality factor, $S_{11}(\omega)$, and resonance frequency measured at 300 K and 5.4 K is shown in Figure \ref{fig:helical-resonator-S11-300K-and-cryo}. 

\section{Trap Electrode Filtering Circuits}
\label{sec:trap-electrode-filtering-circuits}

\subsection{Circuit Analysis}
\label{subsec:trap-electrode-filtering-circuit-analysis}

\begin{figure}[!!h]
\centering
\includegraphics[width=\textwidth]{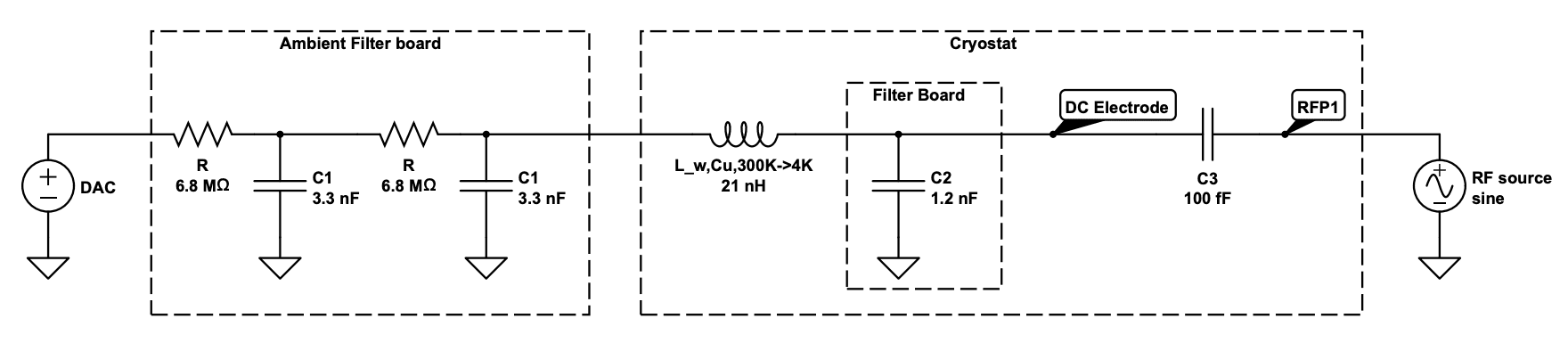}
\caption[DC Electrode Filtering Circuits with In-Vacuum Capacitor]{DC electrode circuit diagram. The 2-pole RC filter network is positioned in the ambient and connected in series to the in-vacuum copper wire, modeled as $L_{\text{w}}$. The in-vacuum filter board is positioned at the 4 K stage near the DC electrodes. RF coupling between RFP1 and the DC electrodes is modeled as $C_{3}$.}
\label{fig:DC-electrode-circuit-with-RF-coupling-in-vacuum-capacitor}
\end{figure}

In this section, we discuss the engineering solution to reduce electronic noise and RF pickup on the DC electrodes. We use a digital-to-analog converter (DAC) in series with an amplifier to deliver voltage to the DC electrodes. However, these DACs have electronic noise in the frequency range of 0-10 MHz. If this noise is at the ion's secular frequency, it will coherently drive the ion which causes motional heating and decoherence. To suppress the noise from the DAC and amplifier at the DC electrodes, we built a 2-pole RC low-pass filter network that is connected in series with the DC electrodes as shown in Figure \ref{fig:DC-electrode-circuit-with-RF-coupling-in-vacuum-capacitor}, where $R=6.8\text{ M\ensuremath{\Omega}}$, $C_{1}=3.3\text{ nF}$, and the cutoff frequency is $f_{\text{cutoff}}=\frac{1}{2\pi RC_{1}}=7\text{ Hz}$. Additionally, the RF voltage presented at RFP1 via an RF source and helical resonator can couple to the DC electrodes where this coupling is modeled as a capacitor, $C_{3}=0.1\text{ pF}$. To suppress this coupling, capacitors, $C_{2}=1.2\text{ nF}$, soldered on an in-vacuum PCB near the trap are wired in parallel to all the DC electrodes. The details of the in-vacuum filter board are discussed in Subsection \ref{subsec:filter-board-manufacturing}. The in-vacuum 34 AWG copper wire (Accu-Glass, PN:110824) used to deliver the voltage from the 300 K stage to the 4 K stage is modeled as an inductor, $L_{\text{w}}$, with self-inductance given by,
\begin{equation}
    L_{\text{w}}=\frac{\mu_{0}L}{2\pi}\left[\ln\left(\frac{2L}{r}\right)-1\right]
    \label{eq:self-inductance-wire}
\end{equation}
where $\mu_{0}=4\pi\times10^{-7}\text{ }\frac{\text{H}}{\text{m}}$ is the permeability of free space, $L$ is the length of the wire in $\text{m}$, and $r$ is the radius of the wire in m \cite{rosaSelfMutualinductancesLinear1908}. The total impedance of the copper wire is the sum of the resistance of the wire, $R_{\text{w}}=\bar{\rho}\frac{L}{A}$, and the impedance of the wire inductance, $Z_{L\text{w}}=i\omega L_{\text{w}}$, given as $Z_{\text{w}}=R_{\text{w}}+i\omega L_{\text{w}}$. The impedances of the capacitors are given as $Z_{C}=\frac{1}{i\omega C}$. 

\begin{figure}[!!h]
\centering
\includegraphics[width=\textwidth]{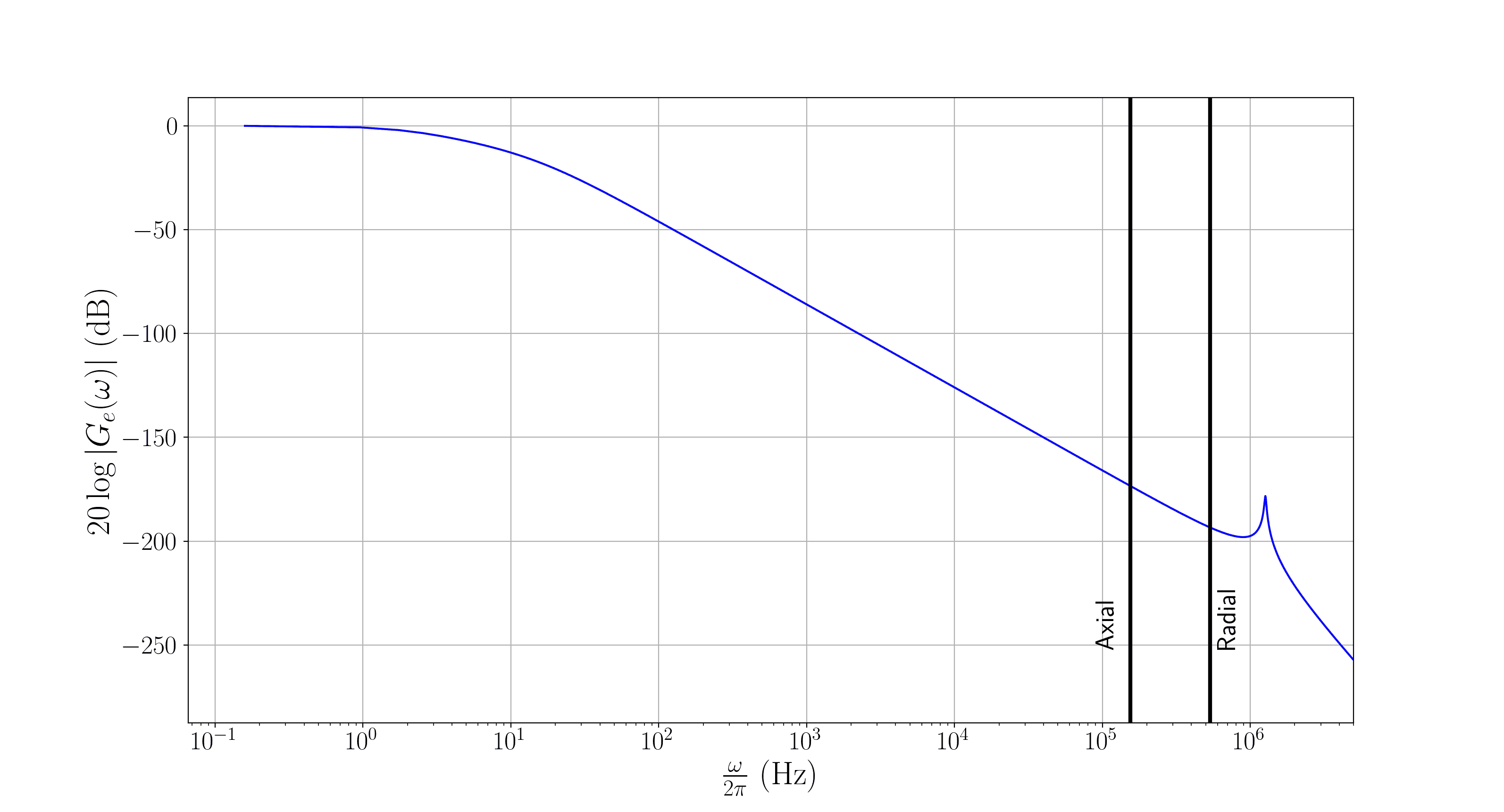}
\caption[Frequency Response of Gain at DC Electrode]{Frequency response of the gain $|G_{\text{e}}(\omega)|$ at the DC electrode relative to the DAC. The cutoff frequency of the 2-pole RC network is $f_{\text{cutoff}}=\frac{1}{2\pi RC_{1}}=7\text{ Hz}$ and the attenuation decreases with a slope of $-40\text{ }\frac{\text{dB}}{\text{decade}}$. There is a resonant frequency between $L_{\text{w}}$ and $C_{2}\parallel C_{3}$ at $f_{\text{res}}=\frac{1}{2\pi\sqrt{L_{\text{w}}C_{2}\parallel C_{3}}}=1.3\text{ MHz}$ indicated by the pole in the frequency response. The secular frequencies are marked with black lines. The suppression at the axial and radial secular frequencies is $20\log|G_{\text{e}}(2\pi\times155\text{ kHz})|=-173\text{ dB}$ and $20\log|G_{\text{e}}(2\pi\times535\text{ kHz})|=-194\text{ dB}$ respectively.}
\label{fig:frequency-response-of-gain-at-DC-electrode-with-capacitor}
\end{figure}

Analyzing the circuit from left-to-right, the frequency response of the gain at the DC electrode relative to the DAC is,
\begin{equation}
    G_{\text{e}}(\omega)=\frac{U_{\text{e}}(\omega)}{U_{\text{DAC}}(\omega)}=\frac{Z_{C2}}{(Z_{\text{w}}+Z_{C2})}\frac{Z_{C1}}{(R+Z_{C1})}\frac{Z_{\text{F}}}{Z_{\text{tot}}}
    \label{eq:gain-at-trap-elecrode-with-capacitor}
\end{equation}
where $Z_{\text{tot}}=R+Z_{\text{F}}$ and
\begin{equation}
    Z_{\text{F}}=\left(\frac{1}{Z_{C1}}+\frac{1}{R+\left(\frac{1}{Z_{C1}}+\frac{1}{Z_{\text{w}}+\left(\frac{1}{Z_{C2}}+\frac{1}{Z_{C3}}\right)^{-1}}\right)^{-1}}\right)^{-1}
\end{equation}
which is complex. The magnitude of the gain, $|G_{\text{e}}(\omega)|$, together with the location of the secular frequencies is plotted in Figure \ref{fig:frequency-response-of-gain-at-DC-electrode-with-capacitor}. Thus, the RC filtering circuit provides suppression at the axial and radial secular frequencies of $20\log|G_{\text{e}}(2\pi\times155\text{ kHz})|=-173\text{ dB}$ and $20\log|G_{\text{e}}(2\pi\times1\text{ MHz})|=-197\text{ dB}$ respectively. It is also important to note that we do not need to change the voltages to the DC electrodes at rates $>$ 7 Hz for the current experiment.

Analyzing the circuit in Figure \ref{fig:DC-electrode-circuit-with-RF-coupling-in-vacuum-capacitor} from right-to-left, the RF pickup gain at the DC electrode can be estimated by ignoring the RC filter network and wire inductance and analyzing just the capacitive divider between $C_{2}$ and $C_{3}$,
\begin{equation}
    G_{\text{e}}\approx\frac{C_{3}}{C_{2}+C_{3}}
    \label{eq:gain-at-DC-electrode-capacitive-divider}
\end{equation}
The suppression is $20\log G_{\text{e}}=-82\text{ dB}$. This is a good approximation as it differs from the frequency dependent gain considering the full circuit by only 0.6\%. This is due to the fact that at the drive frequency of $2\pi\times22\text{ MHz}$, the impedance of the RC network and wire inductance looking from the right is high relative to the parallel impedance of capacitors $C_{2}$ and $C_{3}$. 

\begin{figure}[!!h]
\centering
\includegraphics[width=\textwidth]{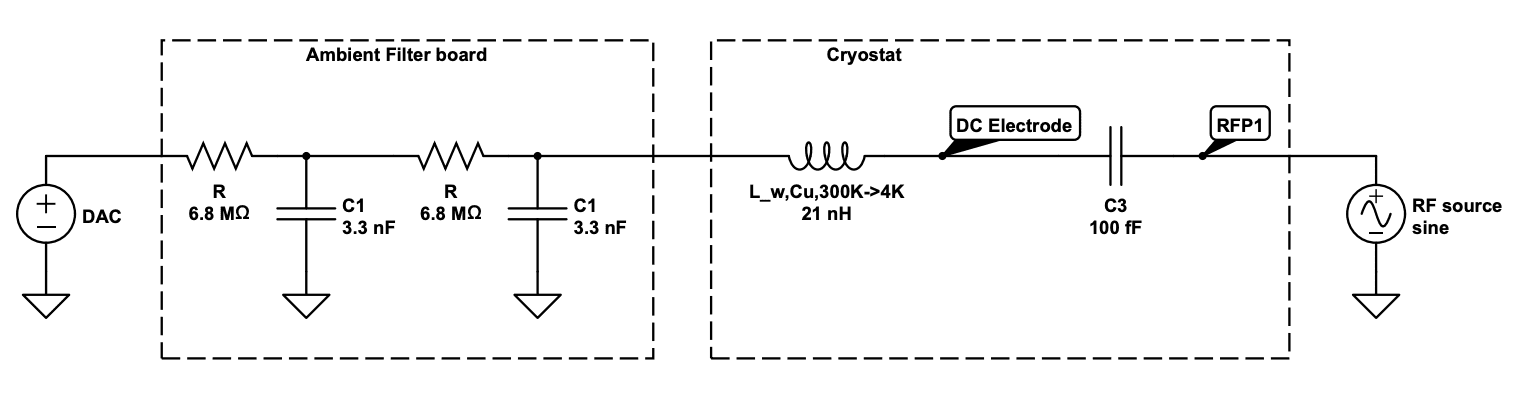}
\caption[DC Electrode Filtering Circuits without In-Vacuum Capacitor]{DC electrode circuit diagram without the in-vacuum filter board. The 2-pole RC filter network is positioned in the ambient and connected in series to the in-vacuum copper wire, modeled as $L_{\text{w}}$. RF coupling between RFP1 and the DC electrodes is modeled as $C_{3}$.}
\label{fig:DC-electrode-circuit-without-RF-coupling-in-vacuum-capacitor}
\end{figure}

To quantify the effect of including the capacitor $C_{2}$ on the gain $G_{\text{e}}$, we consider the DC electrode circuit without the in-vacuum filter board shown in Figure \ref{fig:DC-electrode-circuit-without-RF-coupling-in-vacuum-capacitor}. Analyzing this circuit from right-to-left, the RF pickup gain at the DC electrode can be estimated by ignoring the RC filter network and analyzing just the voltage divider between $L_{\text{w}}$ and $C_{3}$,
\begin{equation}
    |G_{\text{e}}(\omega)|\approx\left|\frac{Z_{\text{w}}}{Z_{C3}+Z_{\text{w}}}\right|
    \label{eq:gain-at-DC-electrode-voltage-divider-between-C3-and-Lw}
\end{equation}
The suppression is $20\log|G_{\text{e}}(2\pi\times22\text{ MHz})|=-32\text{ dB}$. This is a good approximation as it differs from the gain considering the full circuit by only 0.08\%. This is due to the fact that at the drive frequency of $2\pi\times22\text{ MHz}$, the impedance of the RC network looking from the right is high relative to the series impedance of $L_{\text{w}}$ and $C_{3}$. Thus, by having the capacitor, $C_{2}$, the suppression is 50 dB higher - a linear factor of 316. 

To achieve a suppression of $20\log|G_{\text{e}}(2\pi\times22\text{ MHz})|=-82\text{ dB}$ using Equation \eqref{eq:gain-at-DC-electrode-voltage-divider-between-C3-and-Lw} for the circuit in Figure \ref{fig:DC-electrode-circuit-without-RF-coupling-in-vacuum-capacitor} assuming $C_{3}$ does not change, the length of the copper wire would need to be 30 mm which is impractical even for room temperatures systems. This is especially impractical for cryogenic systems, as the leads to the 4 K stage relative to the 300 K stage need to be long to reduce thermal loading. To increase the suppression even more using Equation \eqref{eq:gain-at-DC-electrode-capacitive-divider} for the circuit in Figure \ref{fig:DC-electrode-circuit-with-RF-coupling-in-vacuum-capacitor}, the capacitance of $C_{2}$ must increase. This is challenging as there are few readily large capacitance capacitors $>$ 1.2 nF commercially available that are UHV compatible, bakeable to 500 K, cryogenically compatible, high voltage rated (2 kV), and are low profile.

\subsection{Filter Board Manufacturing}
\label{subsec:filter-board-manufacturing}

\begin{figure}[!!h]
\centering
\includegraphics[width=0.49\textwidth]{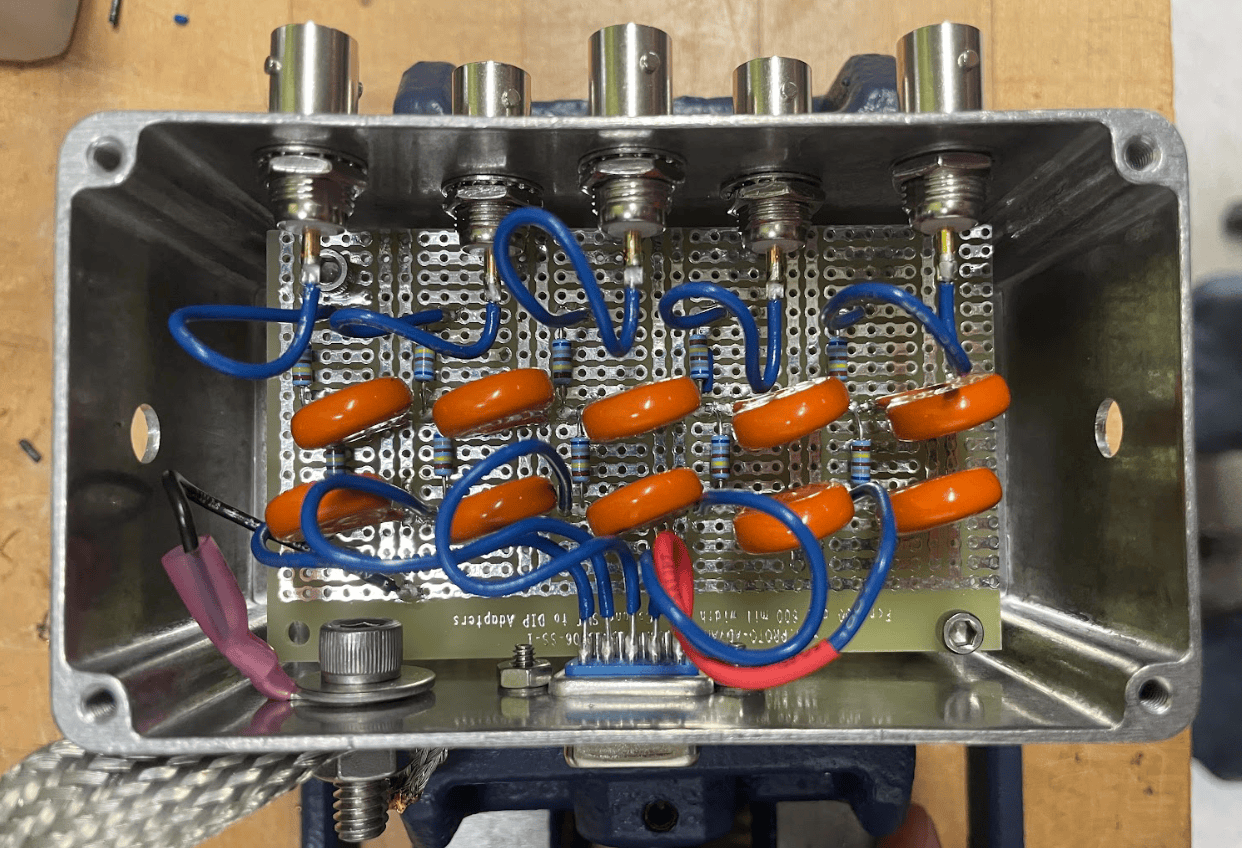}
\includegraphics[width=0.49\textwidth]{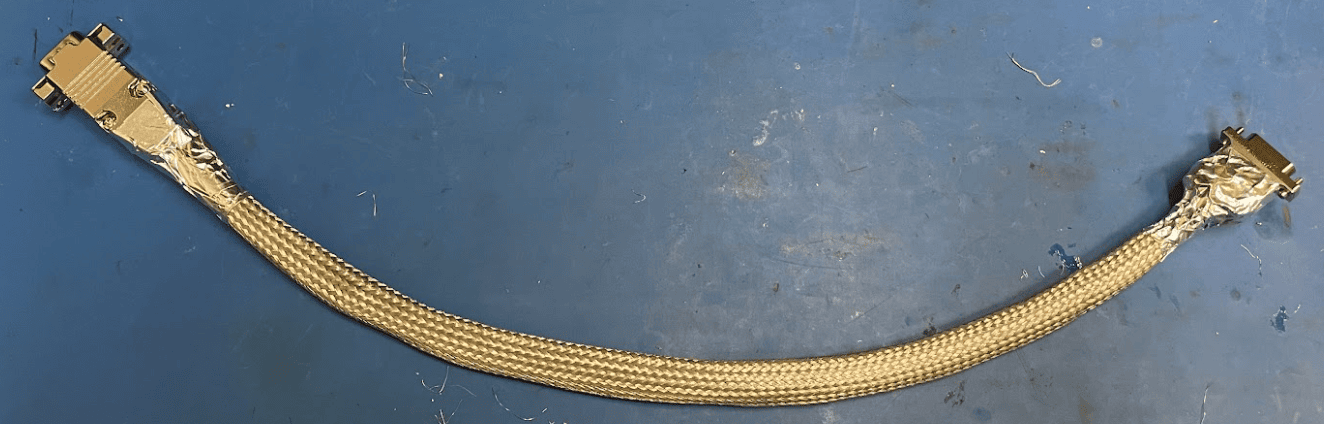}
\caption[Ambient DC Electrode Filter Board]{(Top) Ambient filter board housed in a closed steel box. Input to the board is via BNC connecters. Output of the board is via a DB9 connector. The ground of the circuit is tied to the box and cryostat chamber using a copper braid. (Bottom) Shielded cable assembly connected the output of the filter board to the cryostat chamber feedthrough.}
\label{fig:ambient-DC-electrode-filter-board}
\end{figure}

The ambient filter board was made in-house using standard resistors, ceramic capacitors (Vishay, PN: 564R60GAD33), and an FR4 circuit board with through-hole pads. The circuit board is mounted inside a closed steel box to reduce electromagnetic interference (EMI) where the input connections are BNCs and the output connection is a DB9 connector, see Figure \ref{fig:ambient-DC-electrode-filter-board}. A shielded cable assembly was made to connect the DB9 connector output to to a Micro-D feedthrough at the cryostat chamber. The ground of the circuit is tied to the box and cryostat chamber using a copper braid. 

\begin{figure}[!!h]
\centering
\includegraphics[width=0.49\textwidth]{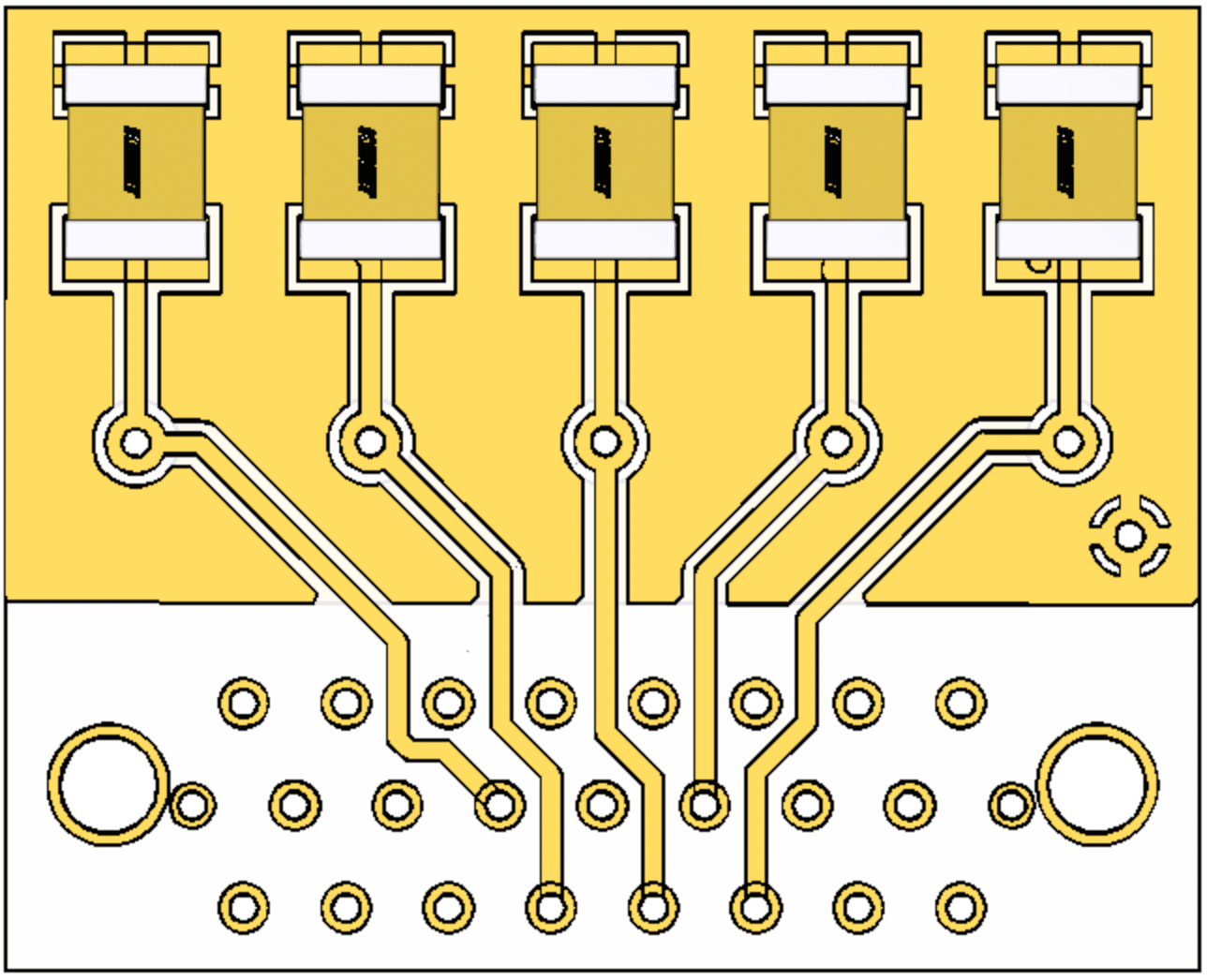}
\includegraphics[width=0.49\textwidth]{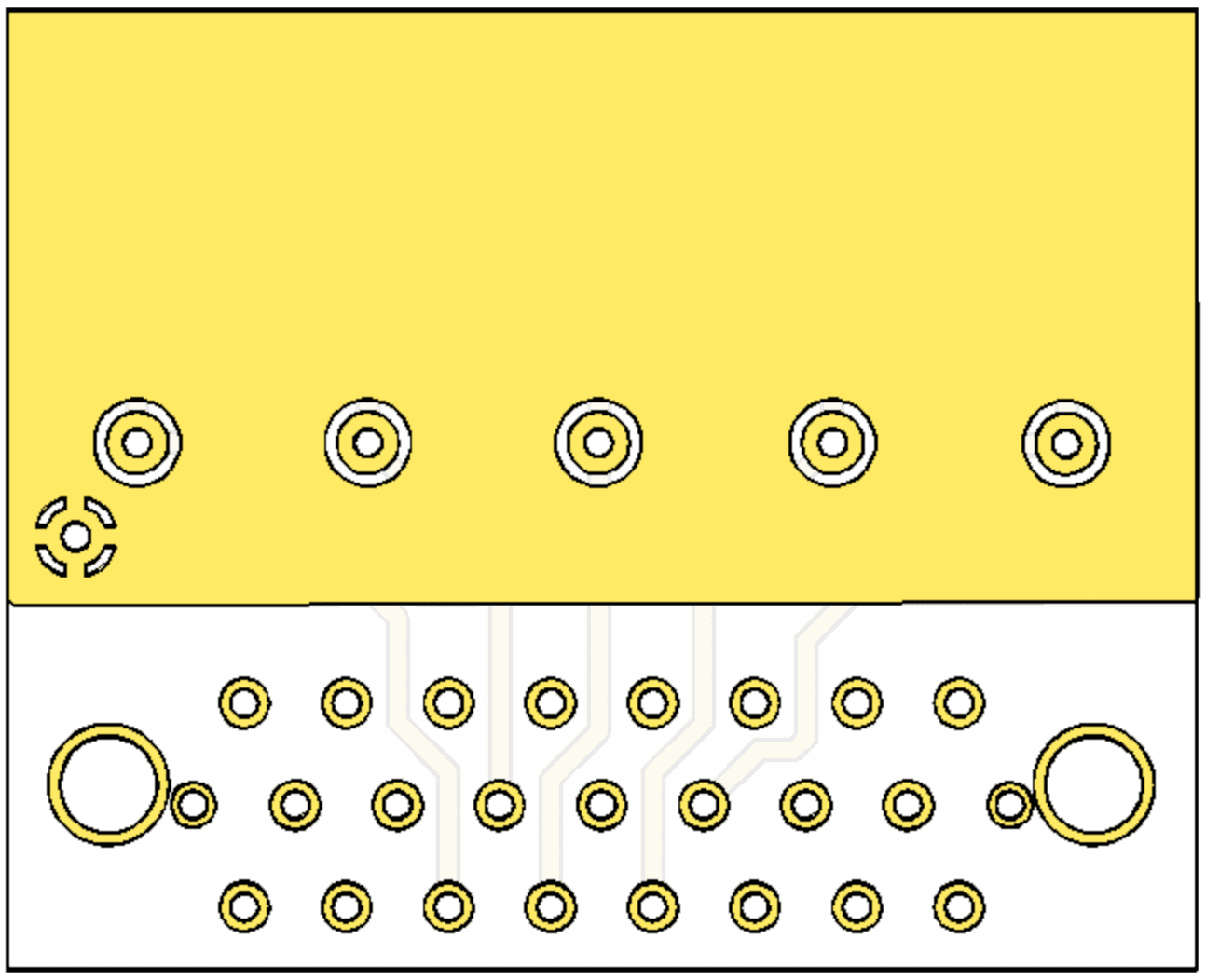}
\newline
\includegraphics[width=0.49\textwidth]{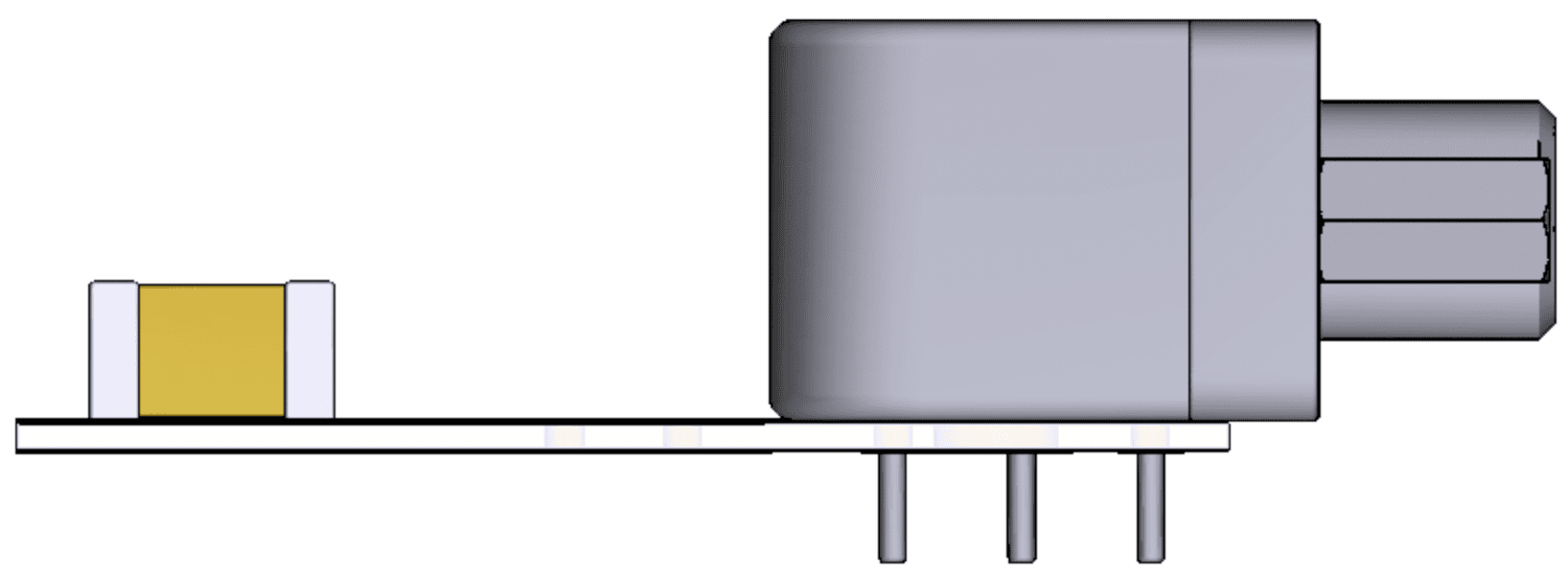}
\includegraphics[width=0.49\textwidth]{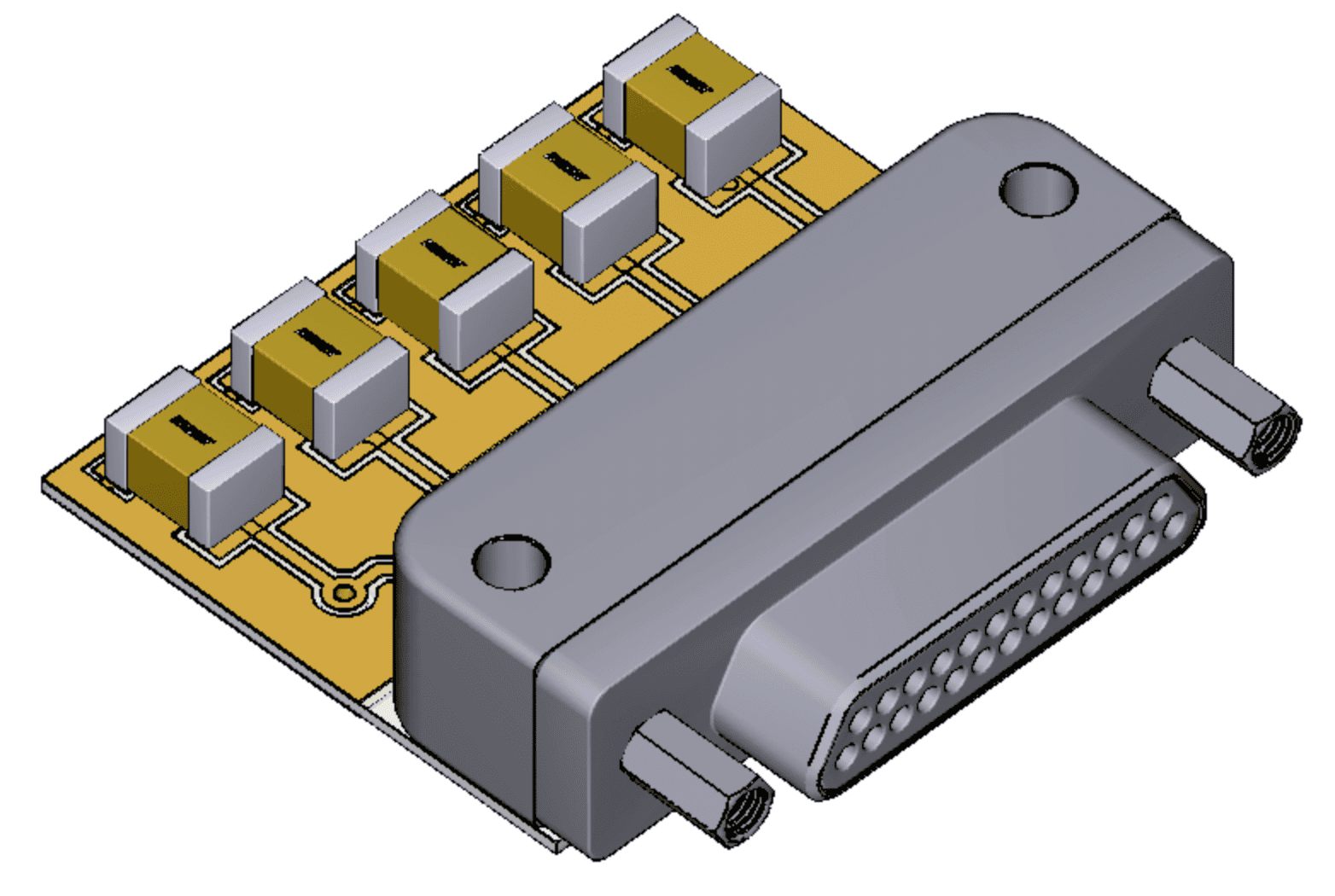}
\caption[In-Vacuum DC Electrode Filter Board]{CAD drawing of in-vacuum filter board with capacitors and Micro-D header. Gold surfaces are the electroless nickel immersion hold (ENIG) finish over the copper clad and the white surfaces are the dielectric. Wires soldered to the through-hole pads are not shown. (Top Left) top layer without header. (Top Right) Bottom layer without header. (Bottom Left) Side view of assembly showing pins of Micro-D header. (Bottom Right) Orthographic view of assembly.}
\label{fig:in-vacuum-DC-electrode-filter-board}
\end{figure}

The in-vacuum filter board consists of a two-layer PCB with milled-out traces and plated through holes, 5 capacitors, and a header that mates with the DC electrode Micro-D connector at the 4 K stage, see Figure \ref{fig:in-vacuum-DC-electrode-filter-board}. The pins of the header are soldered to the PCB at through-hole pads. The capacitors are soldered onto pads on the top layer of the PCB - one pad is connected to the trace and the other pad is connected to the ground plane on the top layer. The trap electrodes are connected to the capacitors and traces on the PCB with wires that are soldered at through-holes pads. The cryostat ground (copper of 4 K stage) is connected to the ground planes of the top and bottom layers of the PCB with a wire that is soldered at another through-hole pad. 

The two-layer board used here is Rogers, PN: R04350B with a thickness of 0.020 in and copper cladding of 0.5 oz. While a single layer board could be used here as the circuit is simple, having a 2-layer board is better as it prevents the PCB from bending due to the mismatch of the copper and dielectric CTEs as it goes through temperature cycling from 300 K to 4 K and 500 K. The Rogers R04350B series dielectric has a low dissipation factor of $3.7\times10^{-3}$ @ 10 GHz, is bakeable to 500 K, and has a low outgassing rate as it has been used in other room-temperature ion trap systems \cite{guiseInvacuumActiveElectronics2014,tanakaDesignCharacterizationPlanar2009a,shappertSpatiallyUniformSinglequbit2013}. It also can go through similar fabrication processes as FR4, as compared to Rogers RT/Duroid 6002 series, which makes it cheaper to make. An electroless nickel immersion gold (ENIG) finish was applied over the copper clad as this enables easier soldering. All the through holes are plated which allows more solder to be applied at each through hole pad joint. The PCB was designed using Altium and manufactured by Cirexx, International. The capacitors, wires, and header were soldered by myself at UMD. 

The solder used was Accu-Glass UHV solder, PN:110796 which contains 85\% gold and 15\% proprietary alloy. The capacitors used here were KEMET, PN: C1812H122JGGAFT050 which are COG (NPO) ceramic capacitors. COG ceramic capacitors are imperative for this work as they experience a negligible change in capacitance as a function of temperature as compared to X7R and X5R ceramic capacitors. The Micro-D header used was Milnec, PN: MDP4-N25X3-H which mates with the Omnetics Micro-D feedthroughs used for the cryostat wiring.

\section{Magnetic Field Coils}
\label{sec:magnetic-field-coils}

The ion resides in a magnetic field of 5 G along the $y$-axis of the cryostat for definition of a quantization axis and to avoid coherent population trapping \cite{berkelandDestabilizationDarkStates2002}. We built three pairs of a field coils along the $x,y,z\text{-axes}$ of the cryostat that can easily generate 5 G in an any direction in three space. The coils consist of 12 AWG insulated copper wire (MWS Wire Industries, PN: 40236) wound in nylon mounts that were 3-D printed at UMD. All the materials and fasteners used to secure the coils to the cryostat and optics table are non-magnetic. The magnetic field generated by each coil was measured using a Teslameter (Lake Shore Cryotronics, PN: F41) in a space absent of magnetic materials. The magnitude of the magnetic field generated along the axis of two coils separated by a distance s is given as follows,
\begin{equation}
    B(z)=\frac{\mu_{0}IR^{2}N}{2}\left(\frac{1}{((z-\frac{s}{2})^{2}+R^{2})^{3/2}}+\frac{1}{((z+\frac{s}{2})^{2}+R^{2})^{3/2}}\right)
    \label{eq:magnitude-of-B-field-due-to-two-coils}
\end{equation}
where $\mu_{0}=4\pi\times10^{7}\text{ }\frac{\text{T-m}}{\text{A}}$ is the permeability of free space, $I$ is the current through the coils in $\text{A}$, $R$ is the radius of each coil in $\text{m}$ , $N$ is the number of turns of the copper wire wound in the coil mount, and the coils are aligned along the $z\text{-axis}$ and centered about $z=0$. The magnitude of the magnetic field generated by two coils aligned along the $x$ or $y$ axes is computed by substituting $x$ or $y$ for $z$ and adjusting $R$ and $s$ accordingly in Equation \eqref{eq:magnitude-of-B-field-due-to-two-coils}. 

\begin{figure}[!!h]
\centering
\includegraphics[width=\textwidth]{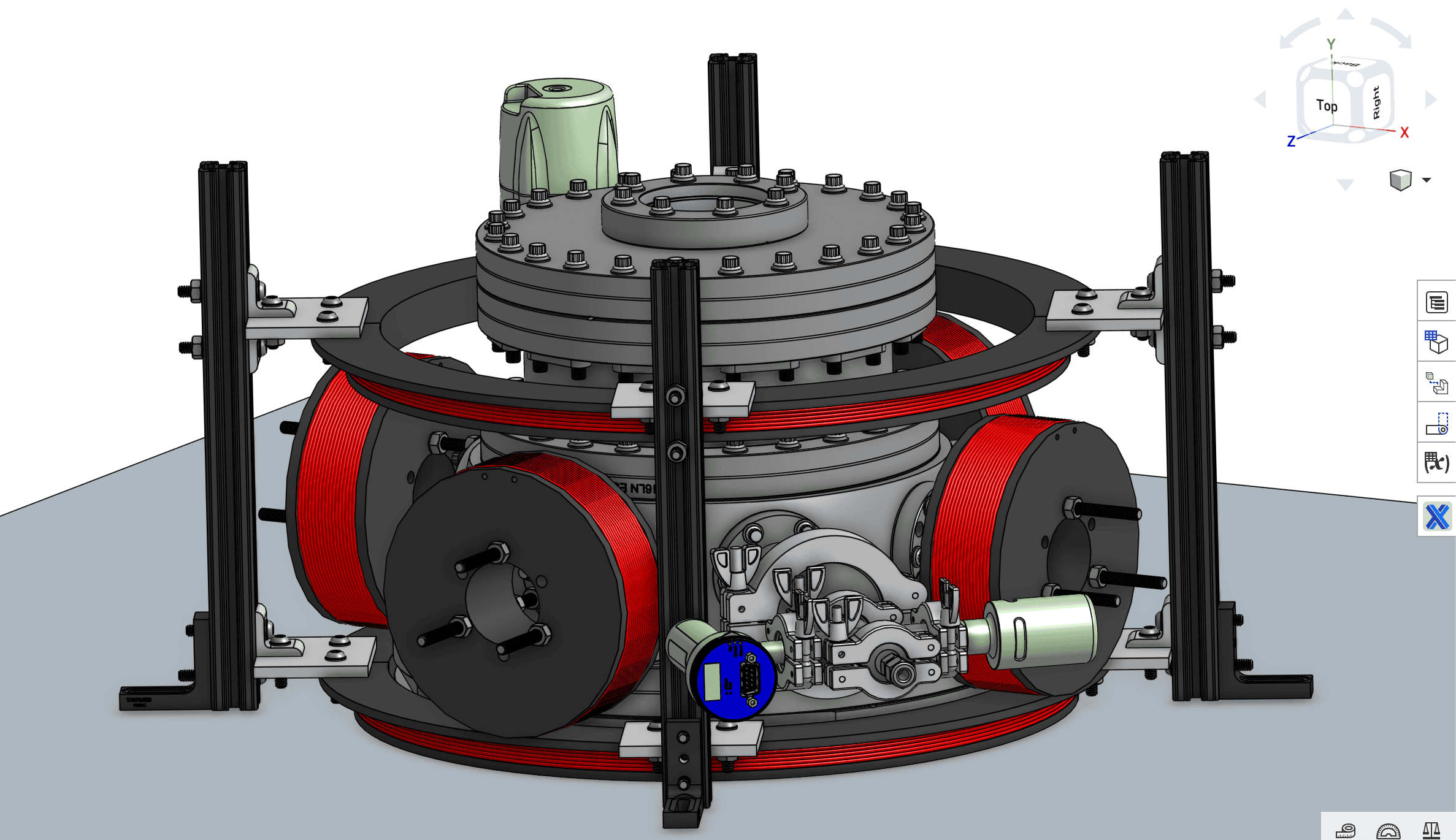}
\caption[B-Field Coils of the Cryostat]{CAD drawing of the cryostat showing the three pairs of magnetic field coils along the $x,y,z\text{-axes}$.}
\label{fig:cryostat-with-B-field-coils}
\end{figure}

\begin{figure}[!!h]
\centering
\includegraphics[width=\textwidth]{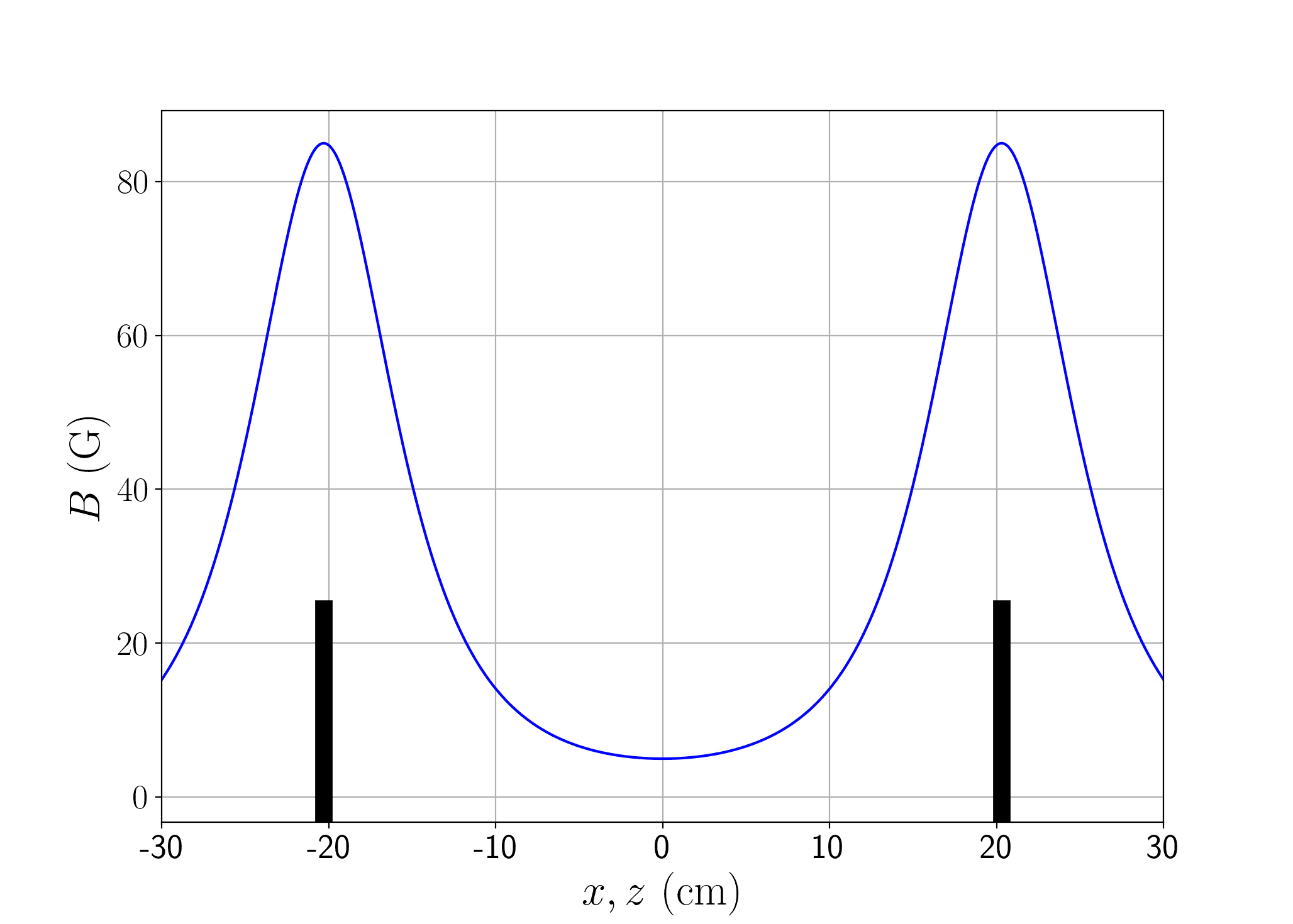}
\caption[B-Field Magnitude Along the $x,z$-axes]{Magnitude of the magnetic field along the $x,z\text{-axes}$ due to two coils each positioned about the trap center at $x,z=0$. The positions of the coils are marked by the solid black lines. Running 8.1 A through each coil with 110 turns of copper wire creates a magnetic field strength of 5 G at the trap center.}
\label{fig:B-field-magnitude-along-xz}
\end{figure}

\begin{figure}[!!h]
\centering
\includegraphics[width=\textwidth]{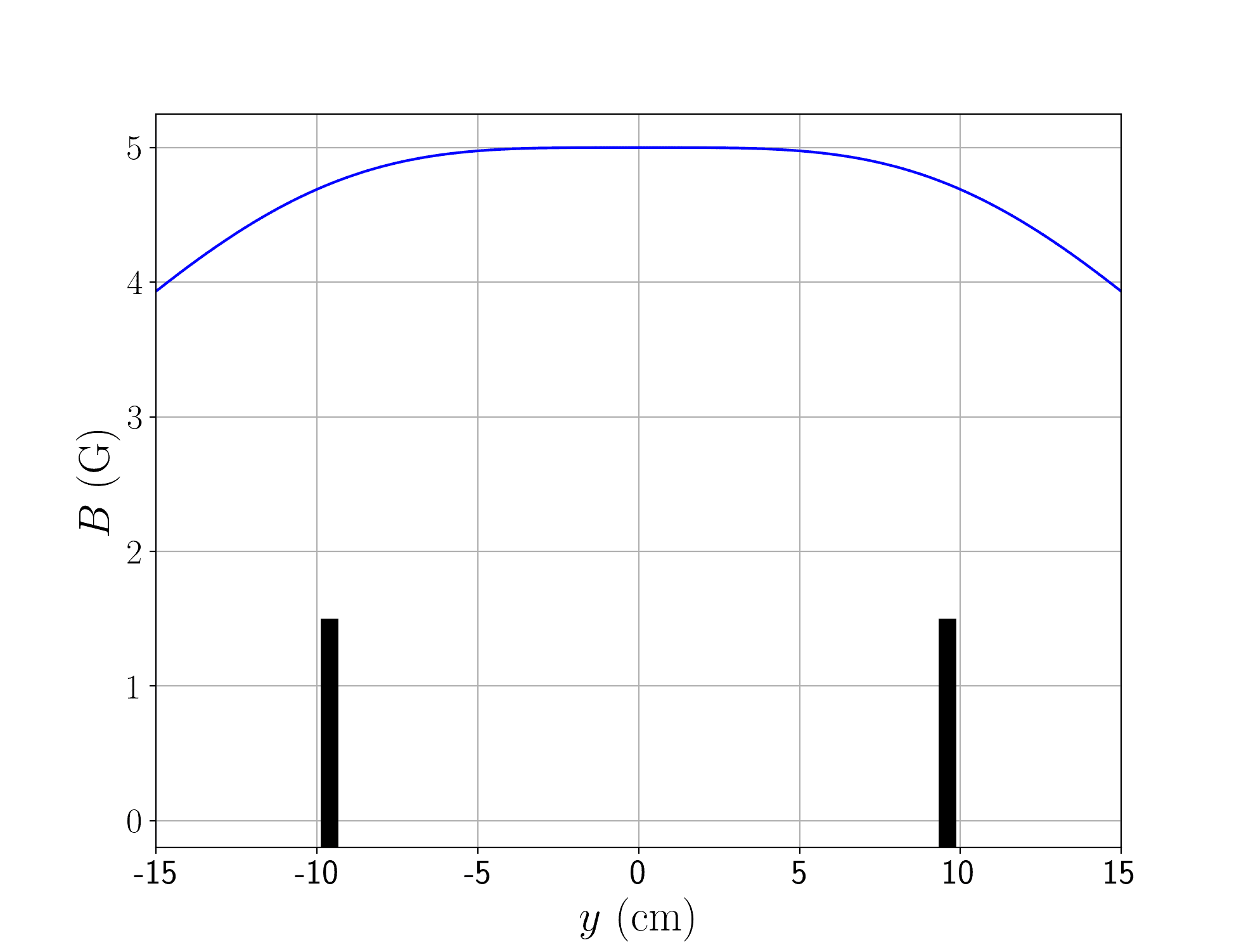}
\caption[B-Field Magnitude Along the $y$-axis]{Magnitude of the magnetic field along the $y\text{-axis}$ due to two coils positioned in Helmholtz configuration about the trap center at $y=0$. The positions of the coils are marked by the solid black lines. Running 3.6 A through each coil with 30 turns of copper wire creates a magnetic field strength of 5 G at the trap center.}
\label{fig:B-field-magnitude-along-y}
\end{figure}

The field coils aligned along the $x$ and $z$ axes are directly mounted to the cryostat chamber using threaded rods shown in Figure \ref{fig:cryostat-with-B-field-coils}. The threaded rods are screwed into taps in the chamber that are accessible because the window flanges are machined with double density through holes. To generate 5\text{ G} at the trap center, 8.1\text{ A} is run through each pair of coils along the $x$ and $z$ axes as shown in Figure \ref{fig:B-field-magnitude-along-xz}. The field coils aligned along the $y$ axis are mounted to the optical table. To ensure the coils are collinear, a mount made of 8020 and aluminum angle brackets was built to hold the coils in place. To generate 5 G at the trap center, 3.6 A is run through the pair of coils along the $y$\text{-axis} as shown in Figure \ref{fig:B-field-magnitude-along-y}.

\section{Ion Trapping Basics}
\label{sec:ion-trapping-basics}

\subsection{Doppler Cooling}
\label{subsec:doppler-cooling}

Unlike neutral atoms in free space, a trapped ion requires only a single Doppler cooling beam because the trapping forces provide confinement along all three spatial directions, while the radiation pressure exerted by the laser is comparatively weak. The cooling process is ultimately limited by heating due to photon recoil. To cool all three motional degrees of freedom, the laser wave vector must have a nonzero projection onto each of the trap's principal axes. This requirement also implies that the three secular frequencies cannot be degenerate. If two or more frequencies are degenerate, the corresponding principal axes are not uniquely defined and may rotate such that one axis becomes perpendicular to the laser wave vector, preventing Doppler cooling along that direction.

\begin{figure}[!!h]
\centering
\includegraphics[width=0.75\textwidth]{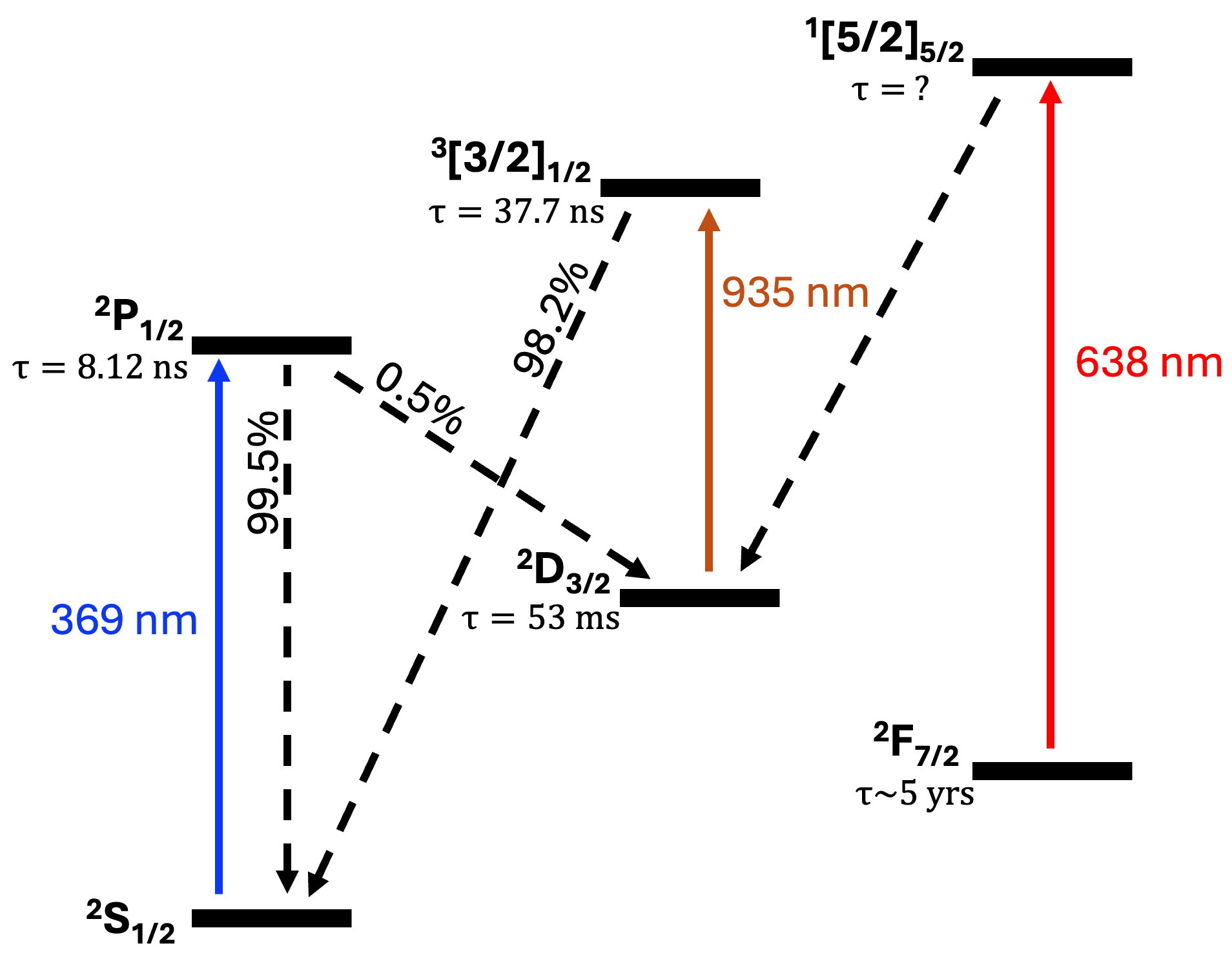}
\caption[$^{174}\text{Yb}^{+}$ Doppler Cooling Level Diagram 1]{$^{174}\text{Yb}^{+}$ level diagram relevant for laser cooling. The cooling beams 369 nm and 935 nm are only required unless collision-induced shelving into the $^{2}\text{F}_{7/2}$ occurs at which point the 638 nm repumps the ion back into the cooling cycle. The branching ratio from $^{1}[5/2]_{5/2}$ to $^{2}D_{3/2}$ is unknown.}
\label{fig:Doppler-cooling-174Yb-level-diagram-ion-trap}
\end{figure}

Doppler cooling of $^{174}\text{Yb}^{+}$ ions is on the $^{2}\text{S}_{1/2}-^{2}\text{P}_{1/2}$ transition at 369 nm, but the optically excited $^{171}\text{Yb}^{+*}$ ions in the $^{2}\text{P}_{1/2}$ state leak into the $^{2}\text{D}_{3/2}$ state and are repumped back to $^{2}\text{S}_{1/2}$ via the $^{2}\text{D}_{3/2}-^{3}[3/2]_{1/2}$ transition at 935 nm, as shown in Figure \ref{fig:Doppler-cooling-174Yb-level-diagram-ion-trap}. Optically excited $^{171}\text{Yb}^{+*}$ ions in the $^{2}\text{P}_{1/2}$ and $^{2}\text{D}_{3/2}$ states can undergo hydrogen collision-induced shelving to the $^{2}\text{F}_{7/2}$ state or simply decay to the the $^{2}\text{F}_{7/2}$ state from the $^{2}\text{D}_{5/2}$. These ions can be repumped back to the $^{2}\text{S}_{1/2}$ state via the $^{2}\text{F}_{3/2}-^{1}[5/2]_{5/2}$ transition at 638 nm. 

In the low-intensity limit of a two-level atom interacting with a red-detuned traveling-wave laser field, Doppler cooling reaches a minimum temperature, known as the Doppler limit, $T_{\text{D}}=\frac{\hbar\Gamma}{2k_{\text{B}}}$, which occurs for a laser detuning of $\Delta=-\Gamma/2$. For $^{174}\mathrm{Yb}^{+}$, the Doppler cooling transition is the $(^{2}\text{S}_{1/2}-{}^{2}\text{P}_{1/2})$ transition at $369\text{ nm}$, which has a natural linewidth of $\Gamma/2\pi=19.6,\mathrm{MHz}$. This corresponds to a Doppler temperature of $T_{\text{D}}=\frac{\hbar\Gamma}{2k_{\text{B}}}\approx470~\mu\mathrm{K}$.

The average motional occupation of a harmonic trap at the Doppler limit is approximately $\bar{n}\approx\frac{k_{B}T_{D}}{\hbar\omega}=\frac{\Gamma}{2\omega}$, where $\omega$  is the trap secular frequency. For typical axial secular frequencies in this work of $\omega_{z}/2\pi=155~\mathrm{kHz}$, $\omega_{r1}/2\pi=500~\mathrm{kHz}$, and $\omega_{r2}/2\pi=530~\mathrm{kHz}$, this yields $\bar{n}_{z}\approx63.2$, $\bar{n}_{r1}\approx19.6$, and $\bar{n}_{r2}\approx18.5$. The coupling between the ion's motion and the cooling laser is characterized by the Lamb-Dicke parameter, $\eta_{i}=k_{i}z_{0,i}=\left(\mathbf{k}\cdot\hat{\mathbf{e}}_{i}\right)\sqrt{\frac{\hbar}{2m\omega_{i}}}$, where $m$ is the mass of $^{174}\mathrm{Yb}^{+}$, $\hat{\mathbf{e}}_{i}$ is the unit vector along the $i$-th principal axis of the trap, and $z_{0,i}=\sqrt{\frac{\hbar}{2m\omega_{i}}}$ is the ground-state spatial extent of the harmonic oscillator along that axis. In this trap, we have a single cooling beam $k=\frac{\hat{x}-\hat{z}}{\sqrt{2}}$ and principal axes $\hat{e}_{z}=\hat{z}$, $\hat{e}_{r1}=\frac{\hat{x}+\hat{y}}{\sqrt{2}}$, and $\hat{e}_{r2}=\frac{\hat{x}-\hat{y}}{\sqrt{2}}$ with projections $|\hat{k}\cdot\hat{e}_{z}|=\frac{1}{\sqrt{2}}$ and $|\hat{k}\cdot\hat{e}_{r1,2}|=\frac{1}{2}$. For a trap frequencies of $\omega_{z}/2\pi=155~\mathrm{kHz}$, $\omega_{r1}/2\pi=500~\mathrm{kHz}$, and $\omega_{r2}/2\pi=530~\mathrm{kHz}$, the ground-state spatial extent is approximately $z_{0,z}\approx14~\mathrm{nm}$, $z_{0,r1}\approx8~\mathrm{nm}$, and $z_{0,r2}\approx7~\mathrm{nm}$ respectively corresponding to Lamb-Dicke parameters of $\eta_{z}=0.165$, $\eta_{r1}=0.065$, and $\eta_{r2}=0.064$ for the $369~\mathrm{nm}$ transition. The thermal wavepacket extent of the ion is given as $z_{\text{RMS}}=z_{0}\sqrt{2\bar{n}+1}$, and thus for Doppler cooled ions we get $z_{\text{RMS,z}}=154\text{ nm}$, $z_{\text{RMS,r1}}=48\text{ nm}$, and $z_{\text{RMS,r2}}=45\text{ nm}$.

The Lamb-Dicke regime is reached when $\eta_{i}^{2}\left(2\bar{n}+1\right)\ll1$, such that the ion's spatial wavepacket is much smaller than the optical wavelength and photon recoil only weakly couples the internal and motional degrees of freedom. For these secular frequencies, $\eta_{i}^{2}\left(2\bar{n}+1\right)$ is 3.5, 0.17, and 0.15 for $\omega_{z}/2\pi=155~\mathrm{kHz}$, $\omega_{r1}/2\pi=500~\mathrm{kHz}$, and $\omega_{r2}/2\pi=530~\mathrm{kHz}$ respectively. Thus, Doppler cooling alone of the axial mode does not satisfy the Lamb-Dicke criterion, motivating the use of higher trap frequencies or resolved-sideband cooling for experiments requiring preparation near the motional ground state.

\subsection{Micromotion Compensation}
\label{subsec:micromotion-compensation}

If the RF pseudopotential is nonzero at an ion's equilibrium position, the ion undergoes a driven oscillation at the trap drive frequency, $\Omega_{\mathrm{RF}}$, known as micromotion. Micromotion arises through two distinct mechanisms: intrinsic micromotion and excess micromotion.

Intrinsic micromotion is a consequence of the trap geometry. In an ideal quadrupole Paul trap, the pseudopotential minimum coincides with the RF electric-field null. However, practical ion traps often incorporate more complex electrode geometries that produce a finite RF field at the pseudopotential minimum, particularly along the trap axis. As a result, the ion experiences micromotion even when located at its equilibrium position. Because this effect is determined by the electrode geometry, it cannot be eliminated by applying compensation voltages and is referred to as intrinsic micromotion \cite{berkelandMinimizationIonMicromotion1998}.

Excess micromotion, in contrast, arises from stray static electric fields that displace the ion away from the RF field null. The displaced ion experiences the oscillating RF field and is therefore driven at the trap frequency. Unlike intrinsic micromotion, excess micromotion can be minimized by applying appropriate compensation voltages to the DC control electrodes \cite{berkelandMinimizationIonMicromotion1998}.

Both intrinsic and excess micromotion modify the interaction between the ion and incident laser fields, affecting Doppler cooling, fluorescence detection, Raman transitions, and other coherent operations. As the ion oscillates, it experiences a time-dependent Doppler shift, causing the laser field in the ion's rest frame to become frequency modulated. This modulation produces sidebands at integer multiples of the trap drive frequency, $\Omega_{\mathrm{RF}}$, while reducing the optical power contained in the carrier. The strength of the modulation is characterized by the modulation index,

\begin{equation}
    \beta=\frac{2\pi x_{\mu\mathrm{m}}}{\lambda}\cos\theta,
    \label{eq:micromotion-modulation-index}
\end{equation}
where $x_{\mu\mathrm{m}}$ is the micromotion amplitude, $\lambda$  is the laser wavelength, and $\theta$  is the angle between the laser wave vector and the direction of micromotion. As the modulation index increases, less optical power remains in the carrier frequency, causing the resonant fluorescence to decrease and, in the limit of large micromotion, potentially disappear.

\begin{figure}[!!h]
\centering
\includegraphics[width=\textwidth]{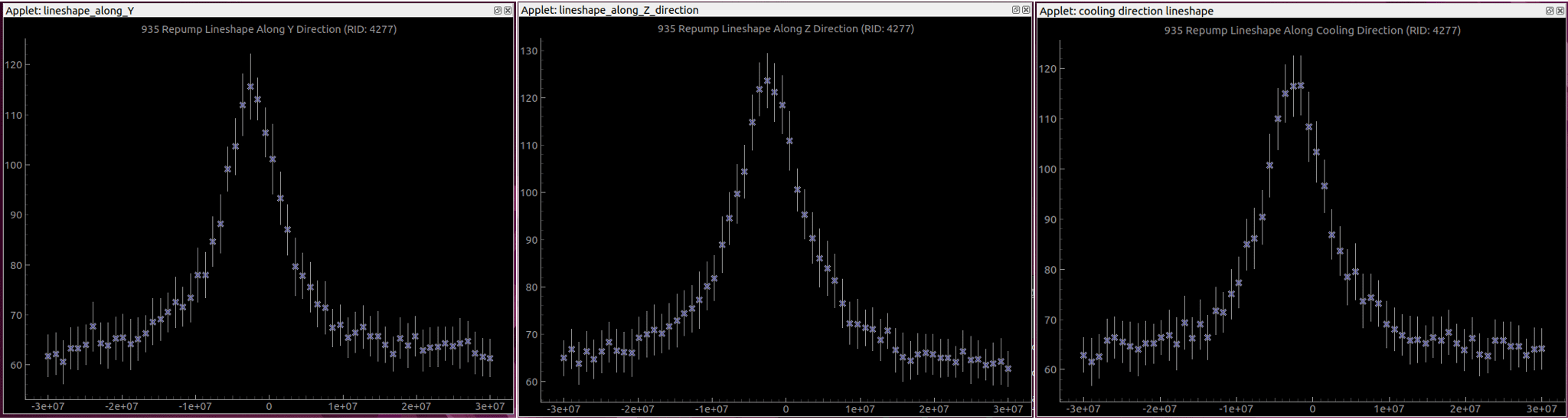}
\caption[Micromotion Compensation]{Coarse micromotion nulling using the 935 nm repumper. DC electrode voltages are as such: EC1, EC2=94 V, CX=9 V, CY=55 V, RFP2=1 V. }
\label{fig:micromotion-compensation}
\end{figure}

To probe the micromotion in the AQT trap, we perform a spectroscopy trace on the $^{2}D_{3/2}-^{3}[3/2]_{2}$ 935 nm transition while varying the voltages to the compensation electrodes as in \cite{gohamResolvedsidebandMicromotionSensing2022}. As the natural linewidth of the $^{2}D_{3/2}-^{3}[3/2]_{2}$ is $\Gamma=2\pi\times4.2\text{ MHz}$ and the trap drive is 22 MHz, we can easily resolve the micromotion sidebands. Excess micromotion was to nulled to $\beta=0.4$ along the $\frac{\hat{x}-\hat{z}}{\sqrt{2}}$ (cooling direction), $z$, and $y$ axes of the trap as shown in Figure \ref{fig:micromotion-compensation}. 

\subsection{Measuring Secular Frequencies}
\label{subsec:measureing-secular-frequencies}

\begin{figure}[!!h]
\centering
\includegraphics[width=0.3\textwidth]{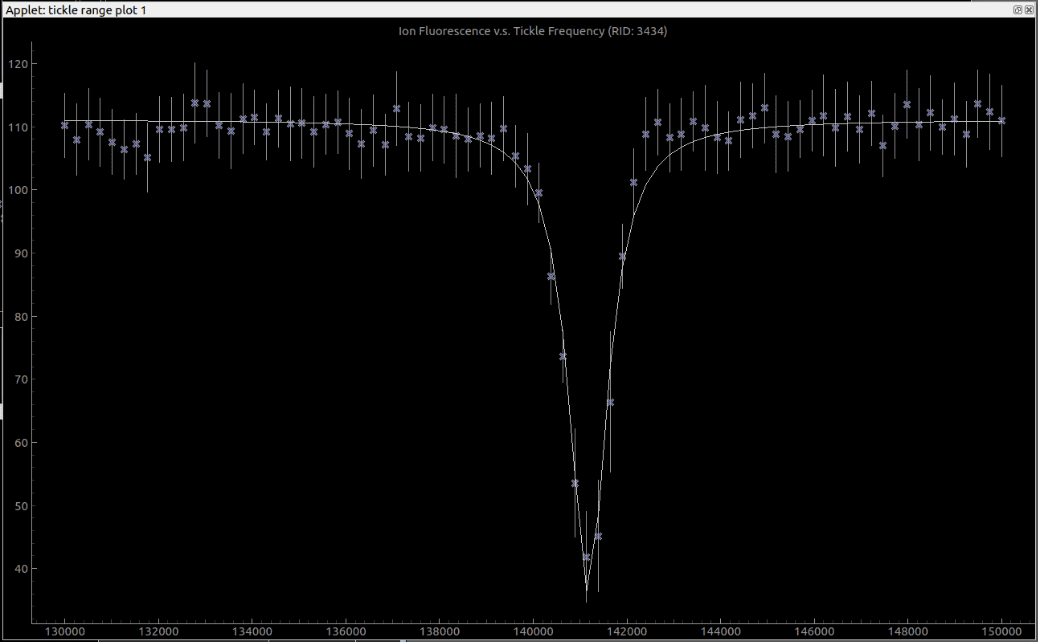}
\includegraphics[width=0.3\textwidth]{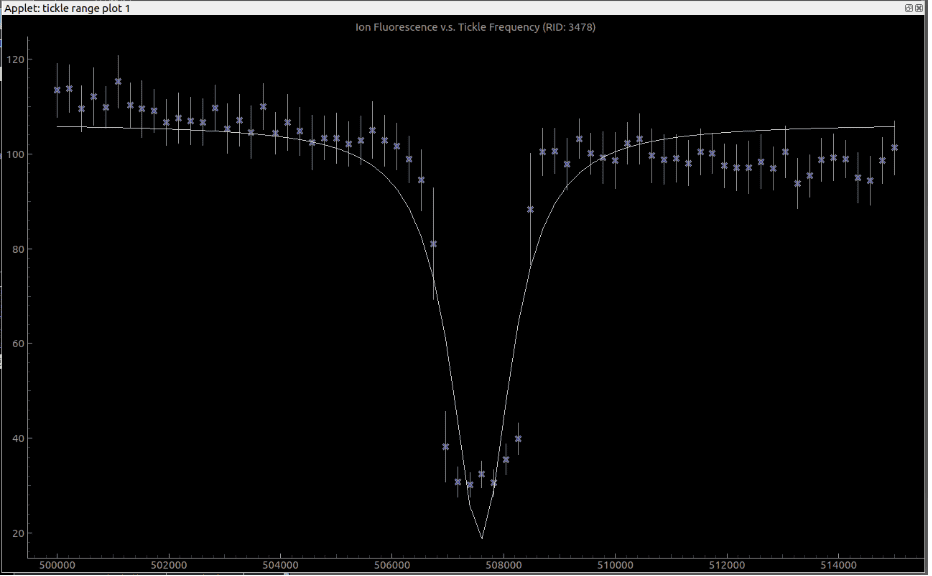}
\includegraphics[width=0.3\textwidth]{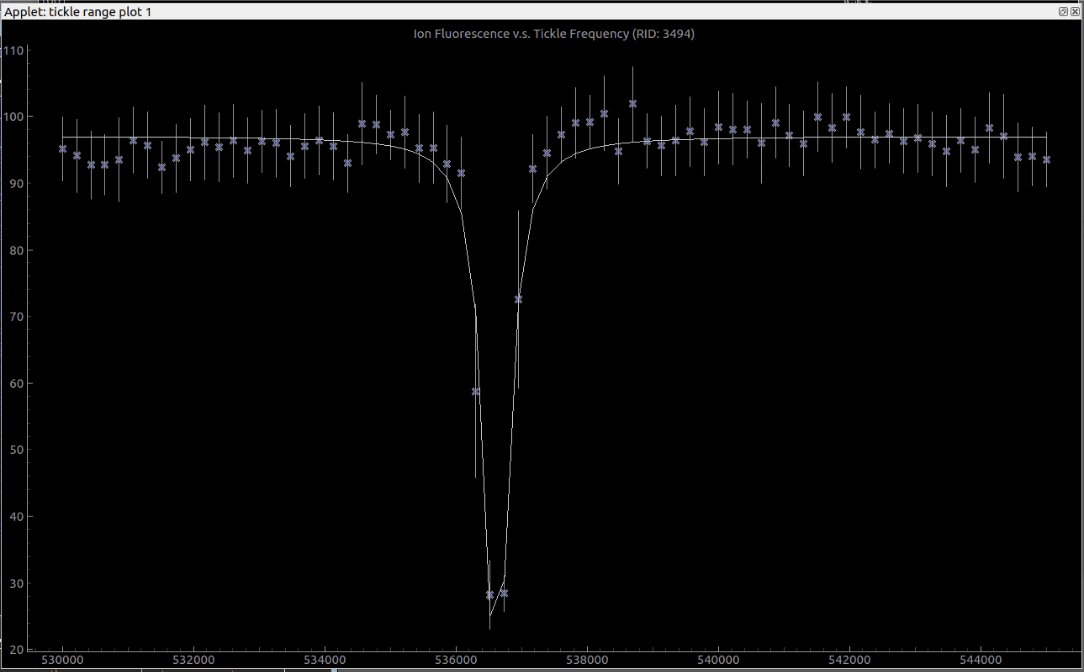}
\caption[Measuring Secular Frequencies]{RF tickle measurement of secular frequencies. Axial secular frequency is 141 kHz (Left), radial 1 is 508 kHz (Center), and radial 2 is 537 kHz (Right). Electrode configuration is as such: EC1, EC2=77 V, CX=9 V, CY=55,RFP2=0.151 V.}
\label{fig:measuring-secular-frequencies}
\end{figure}

\begin{figure}[!!h]
\centering
\includegraphics[width=0.49\textwidth]{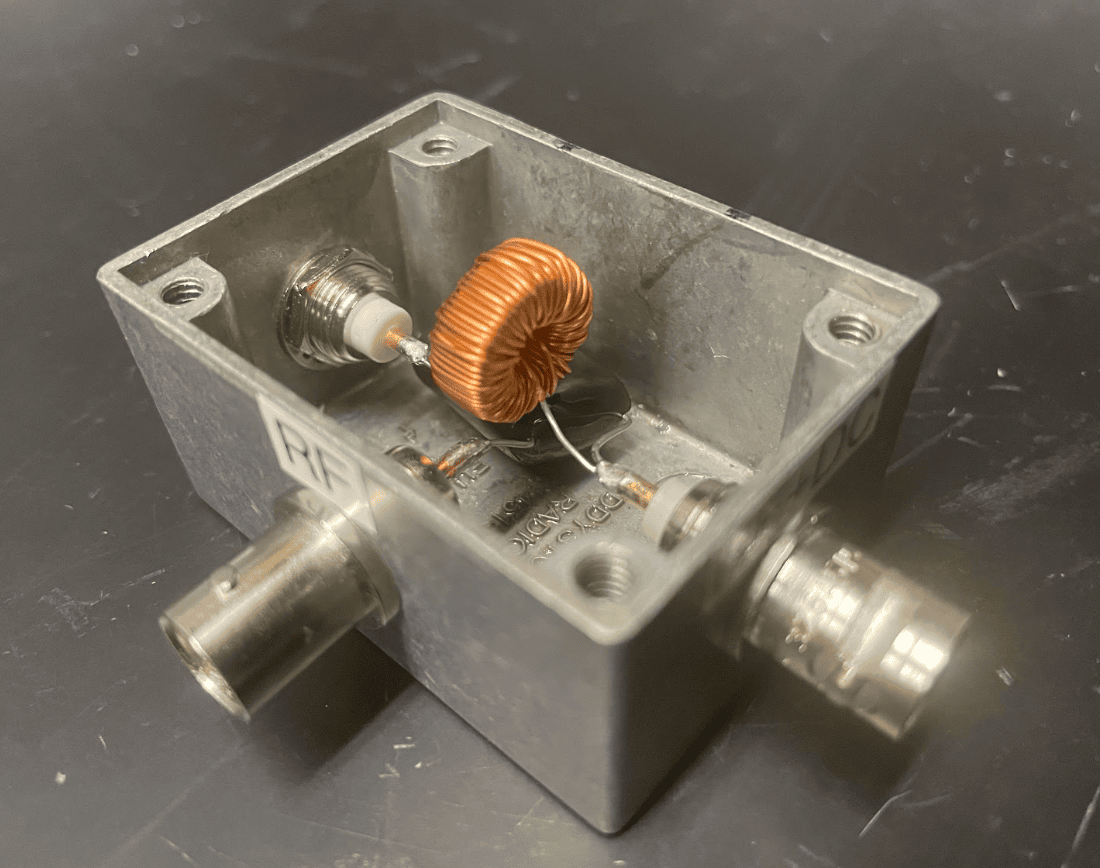}
\includegraphics[width=0.49\textwidth]{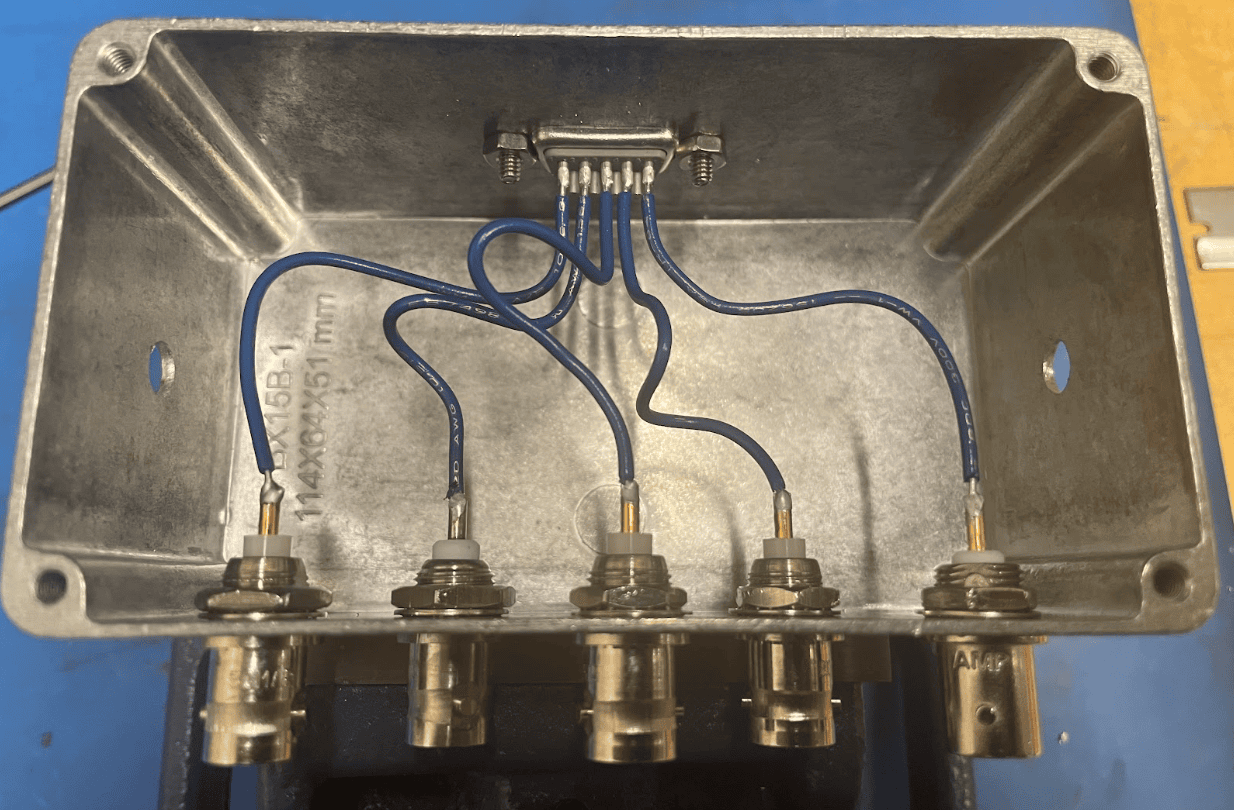}
\caption[Bias Tee]{(Left) RF bias tee. (Right) RF tickle box bypassing the ambient filter board.}
\label{fig:bias-tee}
\end{figure}

Measuring secular frequencies involves applying a weak AC drive to the trap electrodes and sweeping its frequency until it matches the ion's natural motional frequency. This resonant condition coherently drives the ion, dramatically increasing its oscillation amplitude and velocity. The resulting large velocity cycles cause the laser light to become heavily Doppler-shifted out of atomic resonance, creating a sharp, measurable drop in the observed fluorescence as shown in Figure \ref{fig:measuring-secular-frequencies}. To conduct this measurement, we apply AC voltage to the endcaps or compensation electrodes using a bias tee, bypassing the ambient filter board as shown in Figure \ref{fig:bias-tee}. 

\begin{figure}[!!h]
\centering
\includegraphics[width=\textwidth]{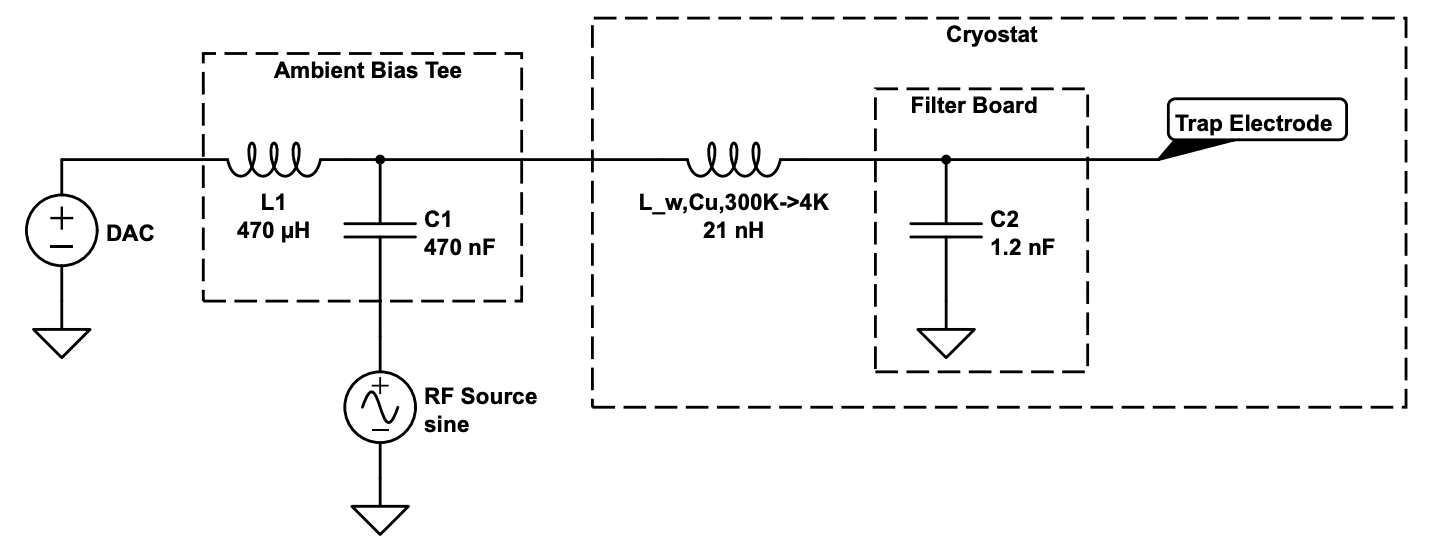}
\caption[Bias Tee Circuit Connected to DC Electrode Circuit Diagram]{Ambient bias tee circuit connected to DC electrode circuit diagram.}
\label{fig:bias-tee-circuit}
\end{figure}

\begin{figure}[!!h]
\centering
\includegraphics[width=\textwidth]{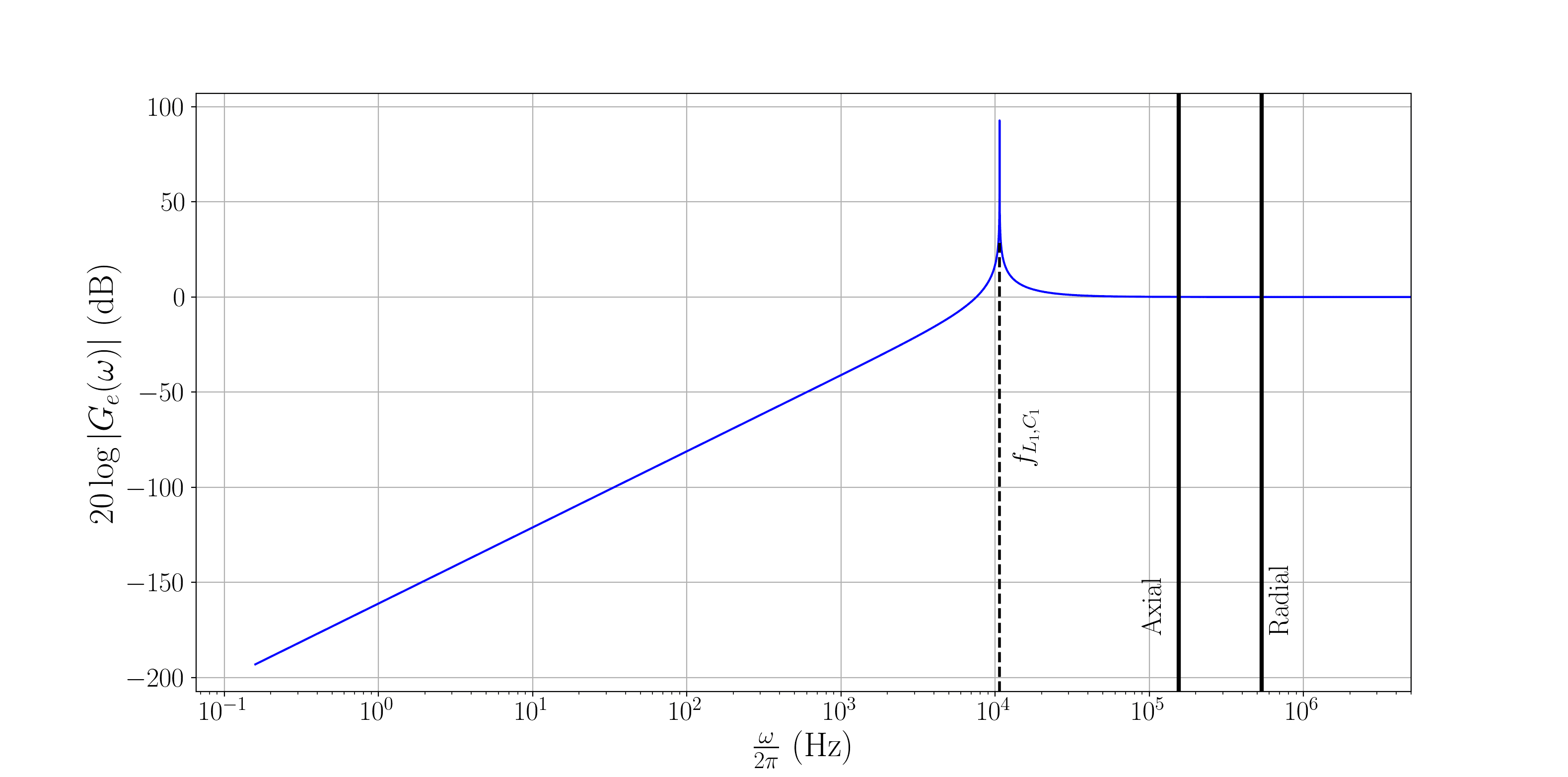}
\caption[Frequency Response of Gain at DC Electrode Relative to Input Bias Tee]{Frequency response of the gain at the DC electrode relative to the input bias tee.}
\label{fig:frequency-response-of-gain-at-DC-bias-tee}
\end{figure}

A circuit diagram of the bias tee connected to the trap electrodes is shown in Figure \ref{fig:bias-tee-circuit}. Analyzing the circuit from left-to-right, the frequency response of the gain at the DC electrode relative to the RF source is,
\begin{equation}
    G_{\text{e}}(\omega)=\frac{U_{\text{e}}(\omega)}{U_{\text{RF}}(\omega)}=\frac{Z_{C2}}{Z_{L\text{w}}+Z_{C2}}\frac{Z_{\text{F}}}{Z_{\text{tot}}}
    \label{eq:gain-at-DC-electrode-due-to-RF-source-bias-tee}
\end{equation}
where $Z_{\text{tot}}=Z_{C1}+Z_{\text{F}}$ and
\begin{equation}
    Z_{\text{F}}=\left(\frac{1}{Z_{L1}}+\frac{1}{Z_{L\text{w}}+Z_{C2}}\right)^{-1}
\end{equation}
which is complex. The magnitude of the gain, $|G_{e}(\omega)|$, together with the location of the secular frequencies and $L$, $C$ resonances of the circuit is plotted in Figure \ref{fig:frequency-response-of-gain-at-DC-bias-tee}. Additionally, the frequency response is plotted in the same figure assuming a negligible wire inductance $L_{\text{w}}$.

\subsection{Ion Lifetime}
\label{subsec:ion-lifetime}

We can trap an ion without laser cooling for 5 days. To conduct this measurement, I loaded an ion and then turned off the cooling beams. I then came back 5 days later, turned the cooling beams on as well as the 638 nm repump, and observed that the ion was still trapped.

\section{Optical Cavity}
\label{sec:optical-cavity}

In this work, I installed the optical cavity at the 4 K stage together with Wance Wang. In this section I'll present the experiment we conducted together to map the ion position with the cavity mode at target transitions. 

\subsection{Ion-Cavity Mode Overlap}
\label{subsec:ion-cavity-mode-overlap}

\begin{figure}[!!h]
\centering
\includegraphics[width=0.75\textwidth]{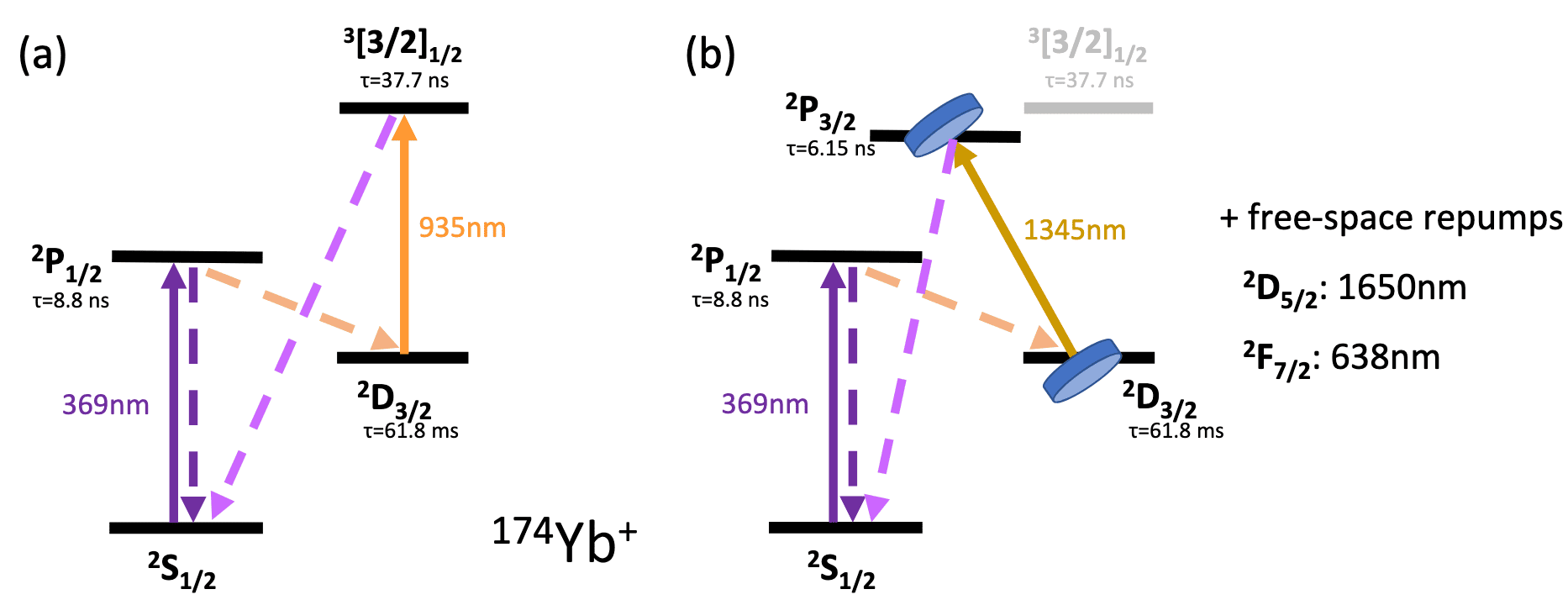}
\newline
\includegraphics[width=0.75\textwidth]{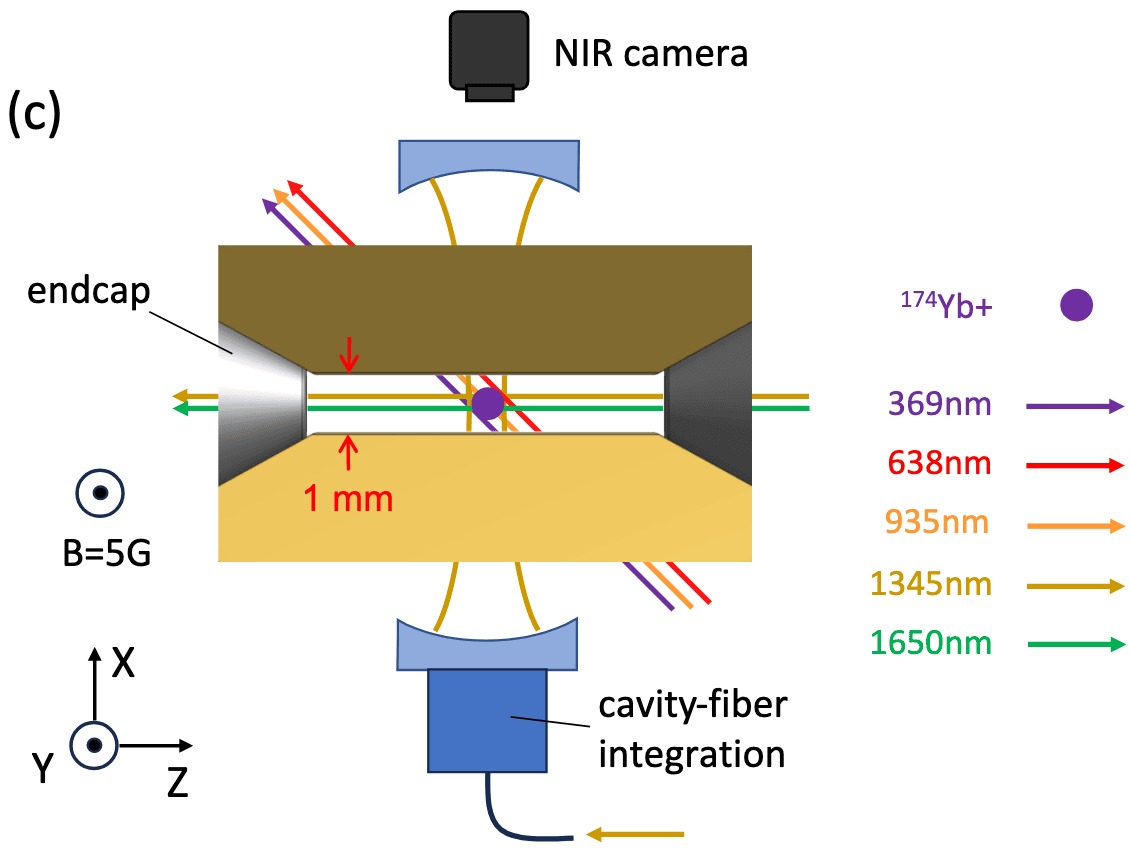}
\caption[Ion-Cavity Mode Overlap Experimental Setup]{(a) Generation of 369 nm fluorescence by repumping the $^{2}\text{D}_{3/2}$ state with 935 nm light. The use of $^{174}\text{Yb}^{+}$ eliminates the need address hyperfine structure. (b) Experimental scheme for verifying spatial overlap between the trapped ion and the 1345 nm cavity mode. Additional free-space repump lasers at 1650 nm and 638 nm return population from the $^{2}\text{D}_{5/2}$ and $^{2}\text{F}_{7/2}$ states, respectively. (c) Laser beam orientations relative to the ion trap.}
\label{fig:ion-cavity-overlap-experimental-setup}
\end{figure}

The cavity mirrors are coated for both 1345 nm and 1650 nm, so spatially overlapping the ion with a cavity mode at one wavelength also overlaps it with the corresponding mode at the other wavelength. To demonstrate ion-cavity overlap, we replaced the conventional 935 nm repump laser with a cavity mode resonant with the $^{2}\text{D}{3/2}-^{2}[P]{3/2}$ transition at 1345 nm, as shown in Figure \ref{fig:ion-cavity-overlap-experimental-setup}. When the ion overlaps the cavity mode, population is repumped from the $^{2}\text{D}_{3/2}$ state and the ion fluoresces; otherwise, the ion remains dark. The 369 nm, 935 nm, and 638 nm beams propagate along the cooling direction, $\frac{\hat{x}-\hat{z}}{\sqrt{2}}$, while the 1650 nm beam propagates along the cavity axis through apertures in the endcap electrodes. The 1345 nm light is coupled into the cavity through the integrated single-mode fiber and 369 nm fluorescence is collected along the $y$ direction using the in-vacuum objective, see Chapter \ref{chap:imaging-and-detection}. We perform this experiment using $^{174}\text{Yb}^{+}$ to avoid the additional complexity associated with hyperfine structure.

\begin{figure}[!!h]
\centering
\includegraphics[width=\textwidth]{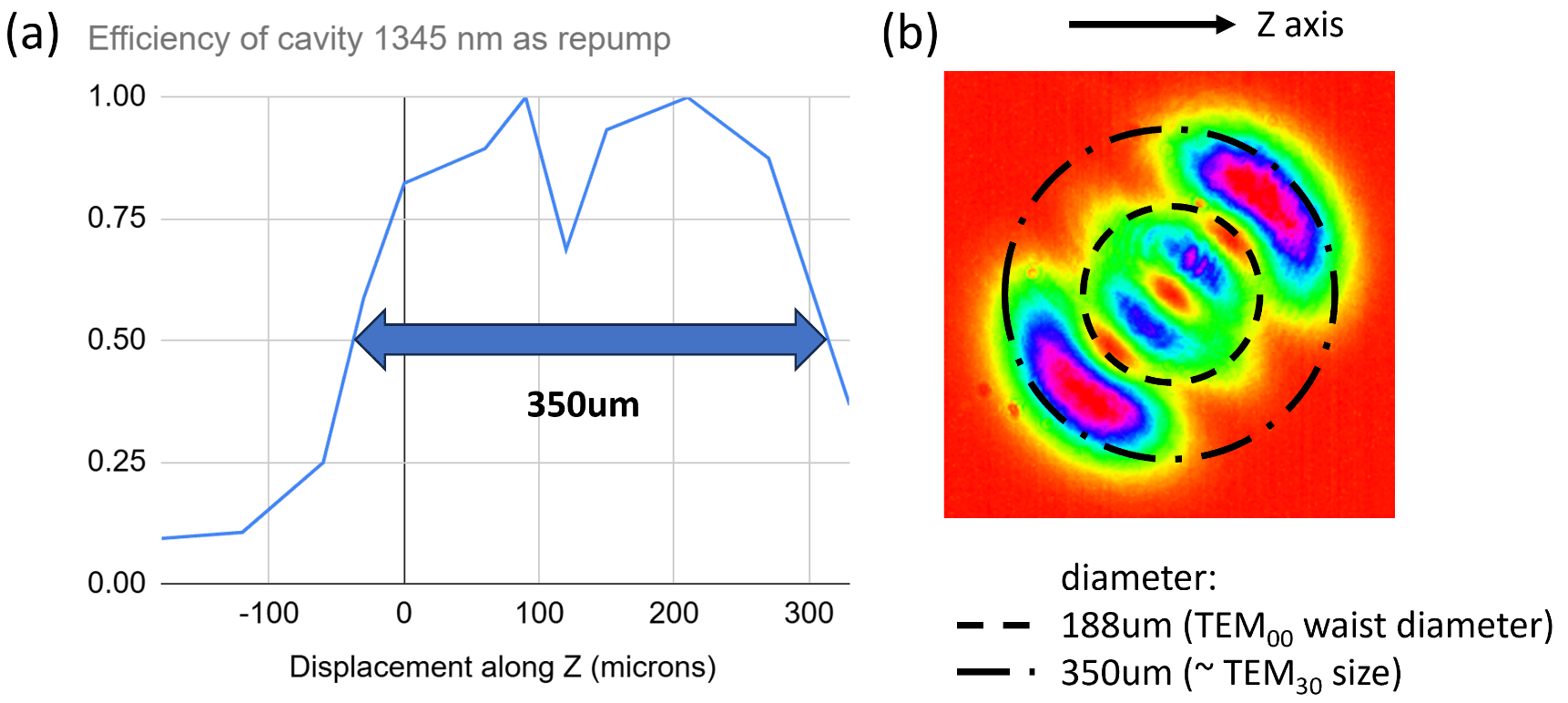}
\caption[Ion-Cavity Mode Overlap Experimental Result]{(a) Normalized fluorescence as a function of ion displacement. The fluorescence obtained using the 1345 nm cavity-mode repump is normalized to that obtained using the conventional 935 nm free-space repump. (b) NIR camera image of the 1345 nm cavity $\text{TEM}_{30}$ mode. The labeled circle has a diameter of $350\text{ \ensuremath{\mu}m}$. For comparison, the cavity $\text{TEM}_{00}$ mode has a diameter of $188\text{ \ensuremath{\mu}m}$.}
\label{fig:ion-cavity-mode-overlap-result}
\end{figure}

For the overlap measurement, the 1345 nm laser frequency was stabilized using the wavemeter while the cavity resonance was scanned with the piezoelectric actuators rather than actively locking the cavity to the laser frequency. During this experiment, the strongest accessible cavity mode was the $\mathrm{TEM}{30}$ mode shown in Figure \ref{fig:ion-cavity-mode-overlap-result}. The fundamental $\mathrm{TEM}{00}$ mode could not be reached within the available PZT tuning range because the cavity fiber was not yet optimally aligned (see Wance Wang's thesis).

To map the cavity mode, we translated a single $^{174}\text{Yb}^{+}$ ion along the cavity axis by varying the endcap voltage. At each ion position, the fluorescence obtained using conventional free-space repumping (935 nm) was used to normalize the fluorescence measured when repumping was provided by the 1345 nm cavity mode together with the free-space 1650 nm laser. The resulting normalized fluorescence profile is shown in Figure \ref{fig:ion-cavity-mode-overlap-result}. The measured profile has a width of approximately $350~\mu\mathrm{m}$, consistent with the spatial extent of the $\mathrm{TEM}{30}$ cavity mode. The reduced visibility of the intensity minima is attributed primarily to the cavity mode waist of approximately $25~\mu\mathrm{m}$ rather than the ion's thermal motion, since the Doppler-cooled axial wavepacket has an rms extent of only $z_{\mathrm{RMS},z}=154~\mathrm{nm}$. This measurement demonstrates successful spatial overlap between a trapped ion and a cavity mode, providing an important milestone toward achieving optimal coupling to the fundamental $\mathrm{TEM}_{00}$ mode.

\ifSubfilesClassLoaded{%
  \bibliographystyle{unsrt}
  \bibliography{Dissertation}
}{}
\end{document}

%% file: Chapter6.tex
\chapter{Atomic Source}
\label{chap:atomic-source}

Loading an ion trap requires ionizing neutral atoms within the trapping region, typically using a two-stage photoionization process. Several methods are used to produce sufficient atomic density in the trapping region, including collimated thermal atomic beams, laser ablation \cite{hendricksAllopticalIonloadingTechnique2007a}, and cold beams from magneto-optical traps (MOTs) \cite{sageLoadingSurfaceelectrodeIon2012}. Collimated thermal atomic beams are generated by vaporizing a sample of the desired species held in a small tube using a resistive heating element, colloquially named an oven. This is a popular approach due to its design simplicity, but requires high-current feedthroughs and in-vacuum wires to deliver the necessary power to the oven. There is also a warm up time for the oven which limits the overall loading rate and the hot oven evolves gas that degrades vacuum for a brief period. Laser ablation is advantageous due to its near-instantaneous loading rate and lack of high-current feedthroughs and in-vacuum wires, but requires a high-intensity pulsed laser source and careful optomechanical design to direct the light to a sample of the desired species near the trap \cite{leibrandtLaserAblationLoading2007,olmschenkLaserAblationProduction2017}. Additionally, the ablation plume typically emits a combination of neutral atoms and directly ionized atoms which complicates isotope-selective loading to the trap. MOT-based loading provides a well controlled source of atoms to the trap but requires considerable experimental overhead making it infrequently used in ion trap systems.

\begin{figure}[!!h]
\centering
\includegraphics[width=\textwidth]{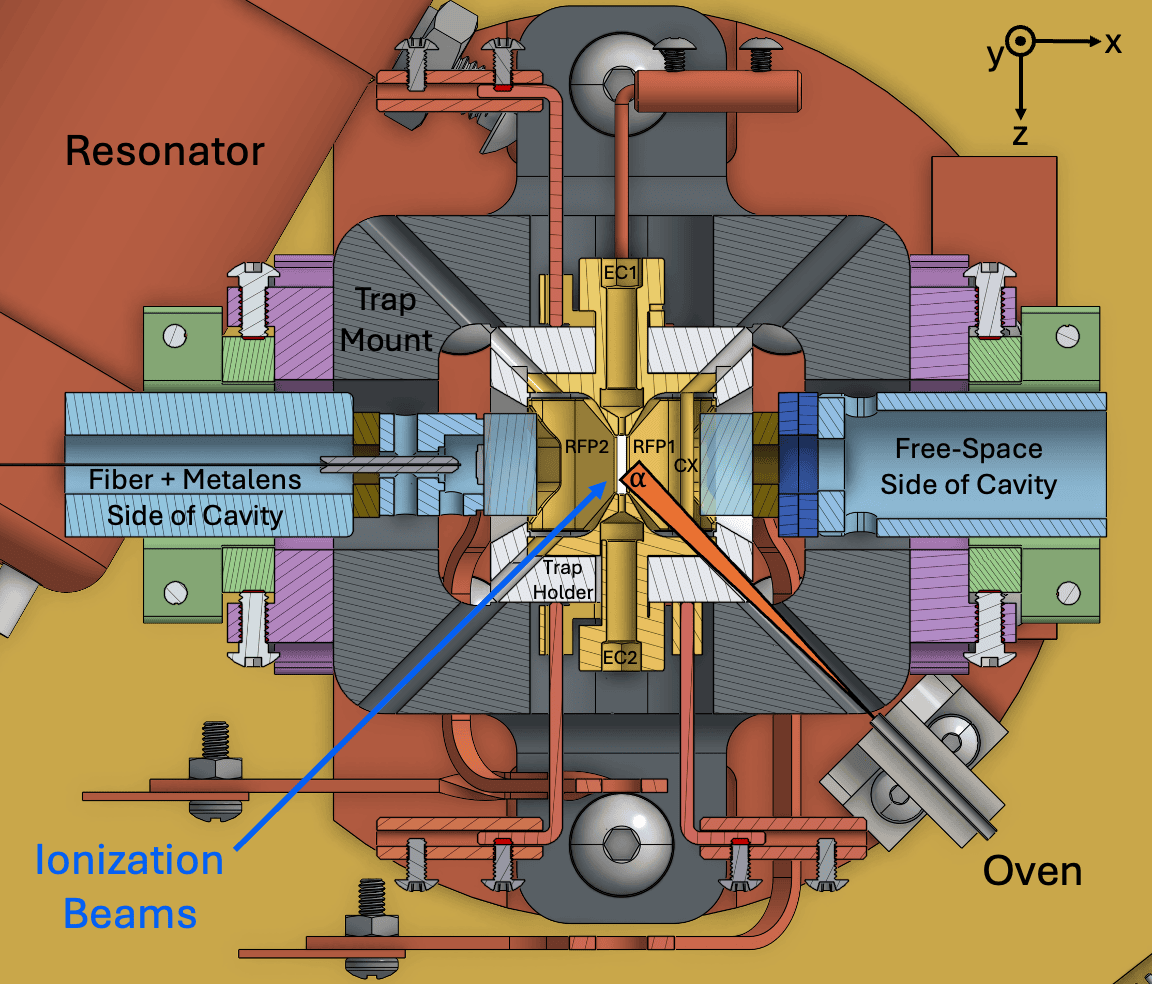}
\caption[Ion Loading Schematic]{Schematic of the ion trap at the 4 K stage showing that the atomic beam is oriented at $\sim90^{\circ}$ relative to the ionization beams and $\sim45^{\circ}$ relative to the trap axis. The trap electrodes are depicted in gold: EC1, EC2, RFP1, RFP2, and CX (CY is not in view). The atomic beam is collimated by the aspect ratio of the aperture in the trap mount and trap holder generating an atomic beam with an angular spread of $\alpha=0.08\text{ radians}$ at the trap center.}
\label{fig:Ion-loading-Schematic}
\end{figure}

In this work, we generate neutral $\text{Yb}$ atoms by vaporizing a sample of natural abundance $\text{Yb}$ using an oven and form a collimated thermal atomic beam that is aligned to the center of the trap. The $\text{Yb}$ metal is held in a stainless steel tube of length $12.7\text{ mm}$, $1.1\text{ mm}$ outer diameter, and $0.86\text{ mm}$ inner diameter. One end of the tube is closed while the other end is kept open and is pointed towards the trap. Once the oven tube is heated, $\text{Yb}$ atoms inside the oven tube have random thermal motion, but only those moving towards the open end of the tube escape. The inner diameter of the tube, $0.86\text{ mm}$, is much smaller than the mean free path of the hot $\text{Yb}$ atoms inside the oven tube, $\bar{\lambda}=\frac{k_{\text{B}}T}{\sqrt{2}\pi d_{0}^{2}P}\approx3.3\text{ km}$, so we can assume the atoms do not collide with each other as they escape - this is called an effusive atomic beam.\footnote{The mean free path, $\bar{\lambda}=\frac{k_{\text{B}}T}{\sqrt{2}\pi d_{0}^{2}P}$, was computed using a $\text{Yb}$ molecular diameter $d_{0}=4.5\times10^{-10}\text{ m}$, the simulated temperature of the oven tube $T\approx481\text{ K}$, and the vapor pressure of $\text{Yb}$ at that temperature $P_{\text{vap}}(481\text{\,K)}\approx2.2\times10^{-6}\text{ Pa}$.} The beam is then collimated by the aspect ratio of an aperture machined into the trap structure creating a small angular spread of $\alpha=0.08\text{ radians}$ as shown in Figure \ref{fig:Ion-loading-Schematic}. We then ionize $^{174}\text{Yb}$ atoms at the trap center using two lasers. The details about the design and the physics of the oven heating are discussed in Sections \ref{sec:oven-engineering} and \ref{sec:thermal-model-oven}. 

\section{Photoionization of Yb}
\label{sec:photoionization-of-Yb}

Yb is a lanthanide, a solid at room temperature with a vapor pressure of $5.7\times10^{-17}\text{ Pa}$ and a melting point of $1097\text{ K}$ \cite{alcockVapourPressureEquations1984}. It does not oxidize in air which simplifies the manufacturing and operation of an oven. The oven holds a sample of natural abundance $\text{Yb}$ so the atomic beam consists of its naturally occurring isotopes. The relative abundances of $\text{Yb}$ isotopes are listed in Table \ref{tab:natural-abundance-of-Yb-isotopes-and-shifts}. We use $^{174}\text{Yb}$ as it is the most abundant isotope and has no complicated hyperfine structure. 

Photoionization is the process of removing a single electron from an atom using a laser with a photon of wavelength, $\lambda$, and energy, $\frac{hc}{\lambda}$, that provides sufficient energy to ionize the atom. The ionization energy of $\text{Yb}$ is $7.31\text{ eV}$ $(170\text{ nm})$. There are Free-electron lasers (FELs) and Excimer lasers that can produce this wavelength, but they are expensive, have optics limitations, and are typically operated in vacuum because photons in this deep ultra-violet regime are strongly absorbed by air. A more convenient photoionization approach is implemented in this work using two photons from External Cavity Diode Lasers (ECDLs). First, the electron in the neutral atom is resonantly excited from $^{1}\text{S}_{0}$ to $^{1}\text{P}_{1}$ by a $399\text{ nm}$ laser. The natural linewidth of the $^{1}\text{S}_{0}-^{1}\text{P}_{1}$ transition is $\Gamma\approx2\pi\times28\text{ MHz}$ \cite{kleinertMeasurementYb02016}. From this excited state, the continuum limit for ionizing the electron is $\approx394\text{ nm}$. We use the Doppler cooling laser at $369\text{ nm}$ for this ionization process. To load a particular isotope, we need to detune the $399\text{ nm}$ 369 nm laser accordingly given the isotope shift of this transition listed in Table \ref{tab:natural-abundance-of-Yb-isotopes-and-shifts}.

\begin{table}[!!h]
\centering
\tablefont
\caption[Natural Abundance of Yb Isotopes and Isotope Shifts]{Natural abundance of $\text{Yb}$ isotopes and isotope shifts, $\Delta f$, of $^{1}\text{S}_{0}-^{1}\text{P}_{1}$ relative to $^{174}\text{Yb}$. The natural linewidth of the $^{1}\text{S}_{0}-^{1}\text{P}_{1}$ transition is $\Gamma\approx2\pi\times28\text{ MHz}$ \cite{kleinertMeasurementYb02016}.}
\label{tab:natural-abundance-of-Yb-isotopes-and-shifts}
\begin{tabularx}{\textwidth}{>{\centering\arraybackslash}p{0.3\textwidth}>{\centering\arraybackslash}p{0.3\textwidth}c}
\toprule
\textbf{Isotope} & \textbf{Abundance (\%)} & \textbf{Isotope Shift,$\Delta f$ (MHz)} \\
\midrule
168 & 0.1 & 1889 \\
\midrule
170 & 3.1 & 1190 \\
\midrule
171 & 14.3 & 835 \\
\midrule
172 & 21.9 & 531 \\
\midrule
173 & 16.1 & -250 \\
\midrule
174 & 31.8 & 0 \\
\midrule
176 & 12.7 & -509 \\
\bottomrule
\end{tabularx}
\end{table}

For a dilute atomic beam there are three key types of broadening of an atomic transition: its natural linewidth, power broadening, and Doppler broadening.\footnote{For the $^{1}\text{S}_{0}-^{1}\text{P}_{1}$ transition in $\text{Yb}$, collisional, transit-time, and second-order Doppler broadening are insignificant relative to the natural linewidth.} To obtain isotope-selective photoionization we require these shifts to be small relative to the isotope shift - for ionizing $^{174}\text{Yb}$, the nearest isotope is $^{173}\text{Yb}$ with $\Delta f=250\text{ MHz}$. The neutral excitation and ionizing beams are overlapped with the center of the trap and are oriented perpendicular to the $\text{Yb}$ atomic beam as shown in Figure \ref{fig:Ion-loading-Schematic}. This is important because the neutral atoms are thermally distributed with a most probable speed, $v_{\text{mp}}=\sqrt{\frac{2k_{\text{B}}T}{m_{\text{Yb}}}}=216\text{ \ensuremath{\frac{\text{m}}{\text{s}}}}$, derived from the Maxwell-Boltzmann distribution where $k_{\text{B}}=1.38\times10^{-23}\text{ }\frac{\text{m}^{2}\text{-kg}}{\text{s}^{2}\text{-K}}$ is the Boltzmann constant, $m_{\text{Yb}}=2.84\times10^{-25}\text{ kg}$ is the mass of a $\text{Yb}$ atom, and $T=481\text{ K}$ is the temperature of the atoms. We will define the component of the atomic velocity of the atoms along the axis of the laser to be $v_{\text{mp,laser}}=v_{\text{mp}}\sin\frac{\alpha}{2}$ where $\alpha$  is the angular spread of the atomic beam as shown in Figure \ref{fig:Ion-loading-Schematic}. If the neutral excitation laser beam was oriented parallel to the atomic beam, $\alpha=\pi\text{ radians}$, the Doppler-broadened linewidth of the $^{1}\text{S}_{0}-^{1}\text{P}_{1}$ transition becomes $\Delta\omega_{\text{D}}=2\pi\times\frac{2\sqrt{\ln2}}{\lambda_{0}}v_{\text{mp,laser}}=2\pi\times\frac{2\sqrt{\ln2}}{\lambda_{0}}\sqrt{\frac{2k_{\text{B}}T}{m}}\sin(\frac{\pi}{2})\approx2\pi\times900\text{ MHz}$ where $\lambda_{0}=399\text{ nm}$ is the wavelength of the transition, making isotope selective loading impossible.\footnote{The Maxwell-Boltzmann distribution can be rewritten in terms of frequency considering the Doppler effect on the atomic beam-laser interaction. This distribution is also Gaussian where the full width at half maximum (FWHM) is given as $\text{FWHM}=2\sqrt{2\ln2}\sigma$  with standard deviation $\sigma=\frac{2\pi}{\lambda_{0}}\sqrt{\frac{k_{\text{B}}T}{m}}$.} By orienting the neutral excitation laser beam perpendicular to the atomic beam with only a small angular spread, $\alpha=0.08\text{ radians}$ as shown in Figure \ref{fig:Ion-loading-Schematic}, the Doppler-broadened linewidth of the $^{1}\text{S}_{0}-^{1}\text{P}_{1}$ transition becomes $\Delta\omega_{\text{D}}=2\pi\times\frac{2\sqrt{\ln2}}{\lambda_{0}}v_{\text{mp,laser}}=2\pi\times\frac{2\sqrt{\ln2}}{\lambda_{0}}\sqrt{\frac{2k_{\text{B}}T}{m}}\sin(\frac{0.08}{2})\approx2\pi\times70\text{ MHz}$ which is small relative to the nearest isotope shift $\Delta f=250\text{ MHz}$ \cite{footAtomicPhysics2005,gaoOpticallyheatedAtomicSource2020,lucasIsotopeselectivePhotoionizationCalcium2004}. 

So given the current oven design, we are limited by Doppler broadening as the natural linewidth of the $^{1}\text{S}_{0}-^{1}\text{P}_{1}$ transition is $\Gamma\approx2\pi\times28\text{ MHz}$. Consequently, we tune the intensity of excitation laser such that the power-broadened linewidth of the $^{1}\text{S}_{0}-^{1}\text{P}_{1}$ transition is commensurate with the Doppler-broadened linewidth $\Delta\omega_{\text{D}}\approx2\pi\times70\text{ MHz}$ - increasing the excitation rate. The intensity of the excitation laser is tuned to $3056\text{ }\frac{\text{W}}{\text{m}^{2}}$ which corresponds to a laser power of $3\text{ \ensuremath{\mu}W}$ with a beam waist of $25\text{ \ensuremath{\mu}m}$.\footnote{The intensity of a Gaussian beam is $I=\frac{2P}{\pi w^{2}}$ where $P$ is the power of the laser beam in $\text{W}$ and $w$ is the waist of the beam in $\text{m}$.} With this intensity, the power-broadened linewidth of the $^{1}\text{S}_{0}-^{1}\text{P}_{1}$ transition becomes $\Delta\omega_{\text{P}}=\Gamma\sqrt{1+\frac{I}{I_{\text{sat}}}}\approx2\pi\times70\text{ MHz}$ where $\Gamma\approx2\pi\times28\text{ MHz}$ is the natural linewidth of the transition, $I$ is the intensity of the laser in $\text{\ensuremath{\frac{\text{W}}{\text{m}^{2}}}}$, and $I_{\text{sat}}=\frac{\pi hc\Gamma}{3\lambda^{3}}=575\text{ }\frac{\text{W}}{\text{m}^{2}}$ is the saturation intensity of the transition \cite{footAtomicPhysics2005}. The $369\text{ nm}$ ionizing beam is limited to $\sim60\text{ \ensuremath{\mu}W}$ due to the available laser power - this is an intensity of $61115\text{ }\frac{\text{W}}{\text{m}^{2}}$ at the ions assuming a $25\text{ \ensuremath{\mu}m}$ beam waist. As there is no power broadening of this second absorption to the continuum, a faster ionization rate could be achieved by using the Raman $355\text{ nm}$ laser or another laser closer to $394\text{ nm}$ and/or with more power. 

\section{Oven Engineering}
\label{sec:oven-engineering}

The first design goal is to determine the minimum oven temperature required to create sufficient atom density at the center of the trap. In principle, if the density of $\text{Yb }$ atoms at the center of the trap to ionize a single atom is known, the required vaporization rate can be calculated - placing a bound on the temperature of the natural abundance $\text{Yb}$ oven as discussed in Section \ref{sec:number-density-at-trap-center}. In this work, the target temperature of the $\text{Yb}$ oven was estimated by relating it to the known target temperature of a $\text{Ca}$ oven from Roland Hablutzel's Master's Thesis \cite{hablutzelNovelAtomSources2012}. He observed plating of $\text{Ca}$ on a glass slide - an opaque spot - in a vacuum chamber when the temperature of the oven, measured using a thermocouple, was $\approx706\text{ K}$. We can then estimate the temperature of $\text{Yb}$ that yields an equivalent vaporization rate as that observed for $\text{Ca}$. The theoretical principles of vaporization are briefly explained below. 

\begin{figure}[!!h]
\centering
\includegraphics[width=\textwidth]{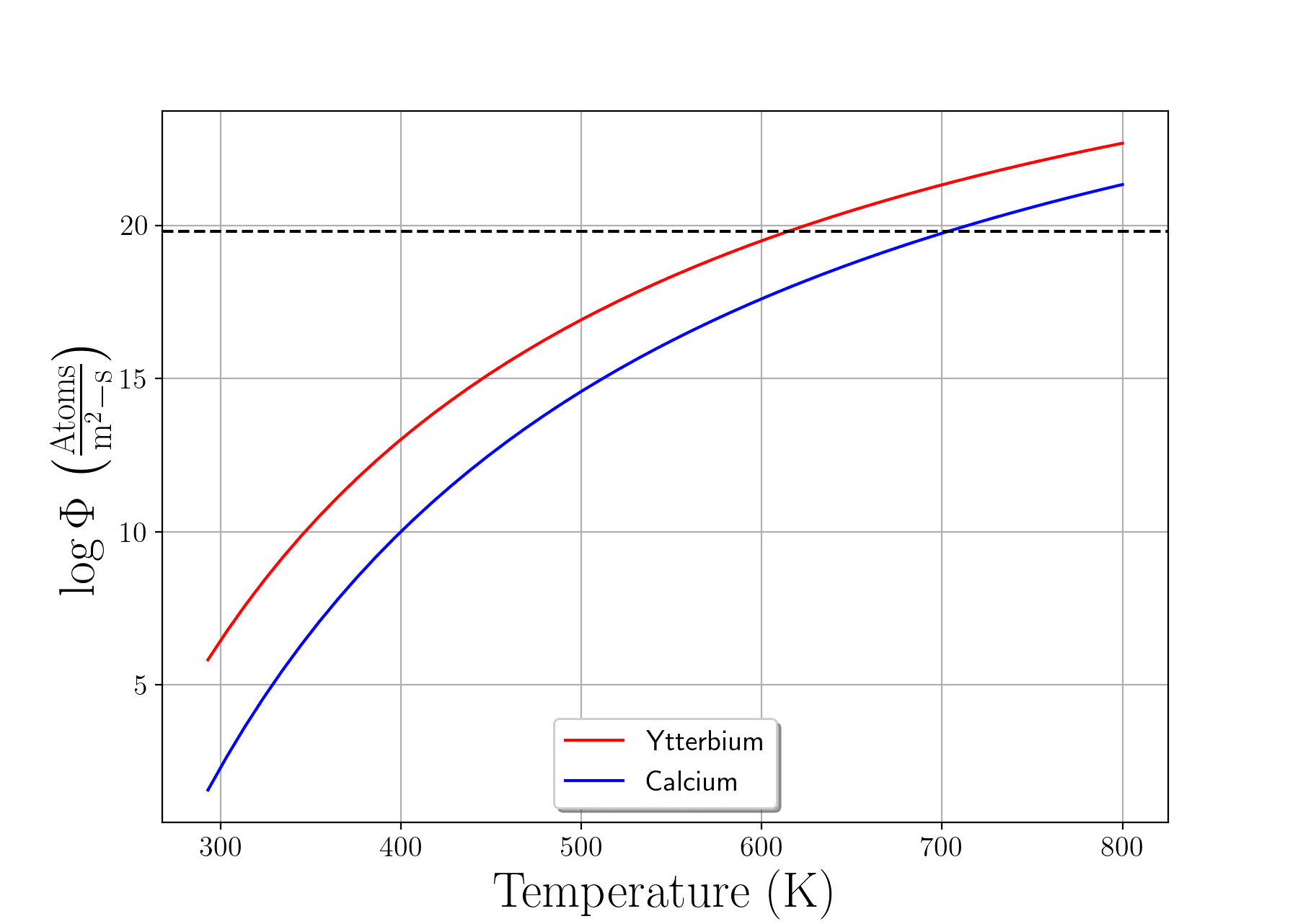}
\caption[Vaporization Rates of Yb and Ca as a Function of Temperature]{Vaporization rates of Yb and Ca are plotted as a function of temperature. The dashed line marks the vaporization rate of $\Phi=10^{19.8}\frac{\text{Atoms}}{\text{m}^{2}-\text{s}}$ that corresponds to the onset of $\text{Ca}$ plating on a glass slide \cite{hablutzelNovelAtomSources2012}.}
\label{fig:vaporization-rate-of-Yb-and-Ca}
\end{figure}

Vaporization is the thermally driven entry of molecules into the vapor phase. The system is in dynamic equilibrium when the rate at which molecules leave the surface of a solid equals the rate at which they arrive at the surface. The pressure of the vapor over the surface in dynamic equilibrium is the vapor pressure of the solid, provided the solid and the vapor are at the same temperature.\footnote{Vapor pressure can also be defined for a liquid but is not relevant in this context.} To generate atomic flux, we intentionally heat the $\text{Yb}$ sample so that it is not in equilibrium and thus vaporization dominates. The maximum rate of vaporization of a solid in $\frac{\text{atoms}}{\text{m}^{2}\text{-s}}$ is given as follows,
\begin{equation}
    \Phi(T)=2.63\times10^{24}\frac{P_{\text{vap}}(T)}{(MT)^{1/2}}
    \label{eq:vaporization-rate}
\end{equation}
where $P_{\text{vap}}(T)$ is the vapor pressure in $\text{Pa}$, $M$ is the molar mass in $\frac{\text{kg}}{\text{kmol}}$, and $T$ is the temperature in $\text{K}$. The vapor pressure of most metallic elements is well approximated by the formula $\log(P_{\text{vap}})=5.006+A+BT^{-1}+C\log(T)+DT^{-3}$ where $T$ is the temperature and $A,\text{ }B,\text{ }C,\text{ and }D$ are empirical coefficients \cite{alcockVapourPressureEquations1984}. Using 706 K as the $\text{Ca}$ target temperature, an equivalent vaporization rate of $\text{Yb}$ is achieved at 615 K as shown in Figure \ref{fig:vaporization-rate-of-Yb-and-Ca}. In general, experiments using atomic species with low vapor pressures will need to operate their ovens at higher temperatures.

The second design goal is to determine an oven geometry that can heat up to 615 K in a reasonable amount of time, $\approx1\text{ min}$, with a current $<3\text{ A}$ (set by the feedthroughs). Our approach uses a 316 stainless steel tube that holds the Yb metal. One end of the tube is flattened and then sealed by welding the end shut. The other end is kept open and is pointed towards our trap. Two thin 316 stainless steel wires are welded to each end of the tube. Two Kapton insulated OFHC copper wires (Accu-Glass, PN: 112615) are welded to each stainless steel wire and are routed to the feedthrough in the vacuum system. All of the welds were made using a Pulse Arc and Resistance welder that is discussed in Appendix \ref{chap:pulse-arc-and-resistance-welder}. Two copper wires are then used on the ambient side and connected to a power supply. As current passes through the circuit, the resistive heating occurs at the stainless steel wires because it is more resistive than the tube and the copper wire. The heat generated in the stainless steel wire is transferred to the oven tube via conduction through the weld - this heats up the oven tube. To determine the current required to heat up the oven tube to $615\text{ K}$, three geometries of the stainless steel parts of this design were tested in a test setup and are summarized in the following Subsections, \ref{subsec:Designs-1-and-2} and \ref{subsec:design-3}.

\subsection{Designs 1 and 2}
\label{subsec:Designs-1-and-2}

The components of Designs 1 and 2 are summarized below in Table \ref{tab:components-of-oven-designs-1-and-2}.

\begin{table}[!!h]
\centering
\tablefont
\caption[Components of Oven Designs 1 and 2.]{Components of oven Designs 1 and 2.}
\label{tab:components-of-oven-designs-1-and-2}
\begin{tabularx}{\textwidth}{>{\centering\arraybackslash}p{0.3\textwidth}>{\centering\arraybackslash}p{0.3\textwidth}c}
\toprule
\textbf{Part} & \textbf{Design 1} & \textbf{Design 2} \\
\midrule
Oven Tube & \makecell{SS 316 Tube\\(McMaster, PN: 1800T511)\\OD: 3.18 mm\\ID: 2.92 mm\\L: 25.4 mm} & \makecell{SS 316 Tube\\(McMaster, PN: 1800T511)\\OD: 3.18 mm\\ID: 2.92 mm\\L: 25.4 mm} \\
\midrule
Wire & \makecell{SS 316 Wire\\(McMaster, PN: 92705K11)\\OD: 0.38 mm\\L: 12.7 mm} & \makecell{SS 316 Wire\\(Accu-Glass wire, PN: 112746)\\OD: 0.51 mm\\L: 12.7 mm} \\
\bottomrule
\end{tabularx}
\end{table}
To quickly validate this approach, Designs 1 and 2 were built and heated up in air. The test setup consisted of a power supply connected to the test ovens with a thermocouple attached to the oven tube. The time for the oven tube to reach 500 K was measured for each using two different currents and is summarized in Table \ref{tab:time-for-oven-tube-designs-12-to-reach-500K}. This is an upper bound on the required current as there will not be air mediated conduction and convention in a vacuum environment.

\begin{table}[!!h]
\centering
\tablefont
\caption[Time for the Oven Tube, Designs 1 and 2, to Reach 500\text{ K}.]{Time for the oven tube, Designs 1 and 2, to reach 500\text{ K}.}
\label{tab:time-for-oven-tube-designs-12-to-reach-500K}
\begin{tabularx}{\textwidth}{>{\centering\arraybackslash}p{0.3\textwidth}>{\centering\arraybackslash}p{0.3\textwidth}c}
\toprule
\textbf{Design} & \textbf{Time to reach 500 K with 6 A (s)} & \textbf{Time to reach 500 K with 8 A (s)} \\
\midrule
Design 1 & 20 & Melted Wire\\
\midrule
Design 2 & 40 & 20 \\
\bottomrule
\end{tabularx}
\end{table}

The results of this experiment clearly show that the oven tube will heat up and that the design goal was satisfied. Design 1 was used for the next step in testing as it heated up faster. This is due to the fact that outer diameter of the stainless steel wire in Design 1 is smaller than that in Design 2 by a factor of 1.3 which is a factor of 1.7 larger in power dissipated and a factor of 1.7 lower in thermal conductance. The results also show that there is a maximum current at which the Design 1 fails due to the stainless steel melting, which occurs at $\approx1600\text{ K}$. 

\begin{figure}[!!h]
\centering
\includegraphics[width=0.3\textwidth]{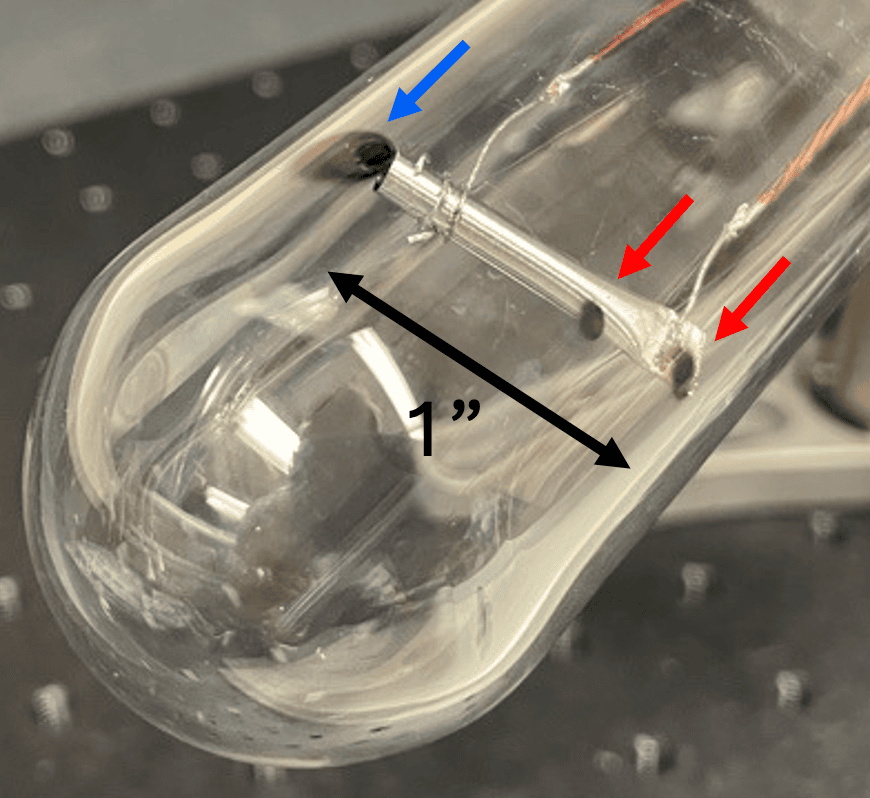}
\includegraphics[width=0.3\textwidth]{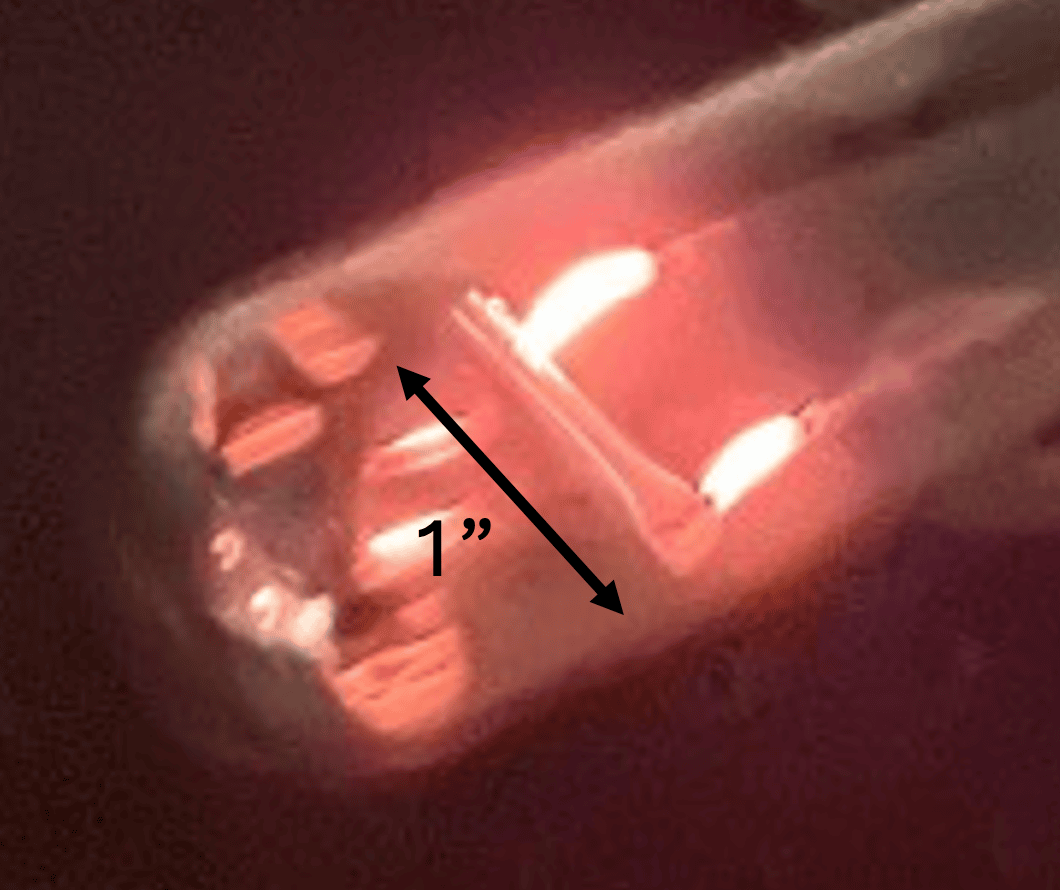}
\includegraphics[width=0.3\textwidth]{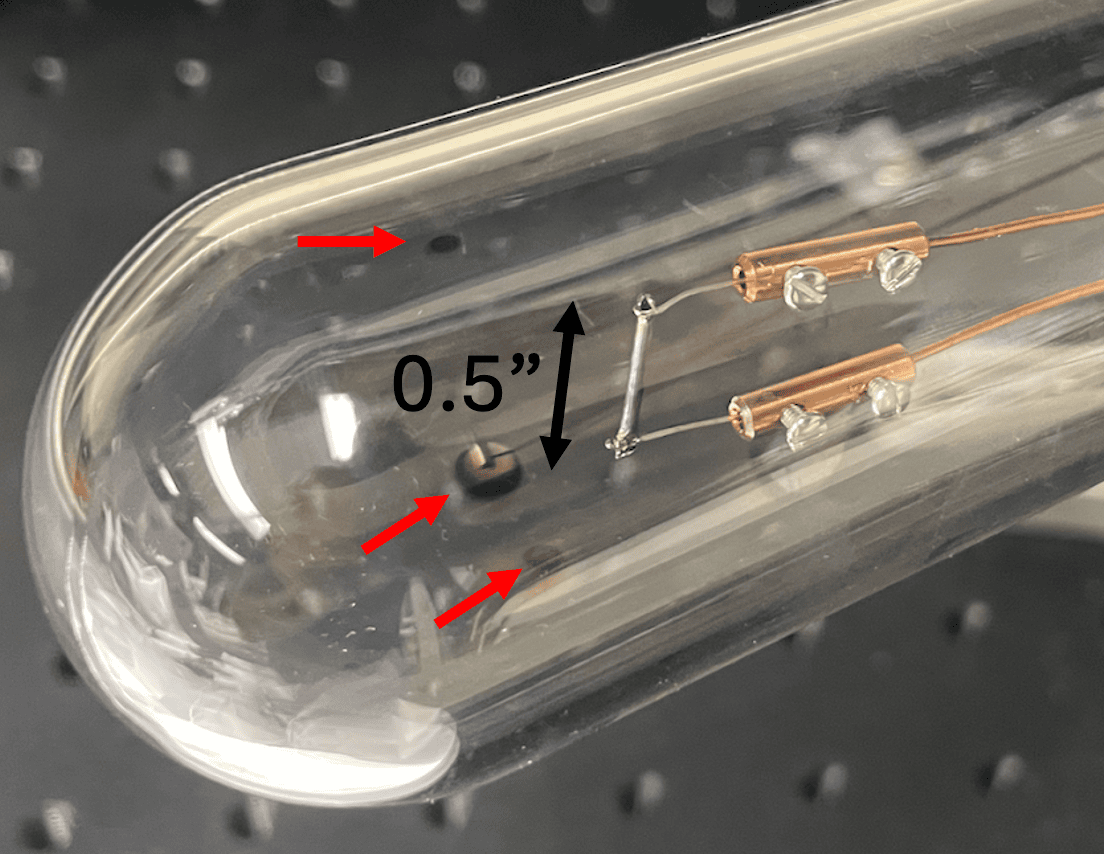}
\caption[Oven Designs 1 and 3 Plating Test Results]{(Left) Oven test setup for Design 1. The black spots on the bell jar is confirmation of Yb plating. The majority of the plating is observed at the open end of the oven tube indicated by the blue arrow. There is also plating at the opposite side of the tube, indicated by the red arrows, that is supposed to be welded closed. This shows that there were small holes at this end of the tube created during the welding process. (Center) Oven test setup for Design 1. This is picture of the oven running with $4\text{ A}$ simply to show the resistive heating of the stainless steel wires and that the stainless steel wire is at a higher temperature than the oven tube inferred from visible radiation. (Right) Oven test setup for Design 3. A picture showing the successful plating of Yb for this design was not taken. This picture is used to show how much smaller the oven tube is relative to Design 1. The plating spots, indicated by the red arrows, are left over from prior testing of Design 1 (not cleaned off).}
\label{fig:oven-design-1-3-plating-test-results}
\end{figure}

Design 1 was then assembled in a vacuum chamber consisting of a feedthrough and a bell jar connected to a turbo molecular pump as shown in Figure \ref{fig:oven-design-1-3-plating-test-results}. While pumping on the chamber, the current to the oven was increased until plating on the bell jar was observed. The voltage, current, and confirmation of plating is summarized in Table \ref{tab:oven-design-1-plating-test-results}. 

\begin{table}[!!h]
\centering
\tablefont
\caption[Oven Design 1 Plating Test Results.]{Voltage, current, and confirmation of plating for Design 1 in test setup. Plating was observed using a current of $3.1\text{ A}$ after $2\text{ min}$.}
\label{tab:oven-design-1-plating-test-results}
\begin{tabularx}{\textwidth}{>{\centering\arraybackslash}p{0.3\textwidth}>{\centering\arraybackslash}p{0.3\textwidth}c}
\toprule
\textbf{Voltage (V)} & \textbf{Current (A)} & \textbf{Plating (yes/no)} \\
\midrule
1.08 & 2.50 & no\\
\midrule
1.24 & 2.80 & no \\
\midrule
1.46 & 3.00 & no \\
\midrule
1.49 & 3.10 & yes (after 2 min) \\
\bottomrule
\end{tabularx}
\end{table}

The results of this experiment show that oven tube can be heated up enough to observe plating of Yb on the bell jar after $2\text{ min}$. As a sanity check, the time it takes for an opaque metal film, $\approx10\text{ nm}$, to form on the bell jar was estimated.\footnote{Using the Beer-Lambert's Law, $>99$\% attenuation is reasonable for a $10\text{ nm}$ film \cite{axelevitchInvestigationOpticalTransmission2012}.}. We assume a constant Yb vaporization rate of $\Phi(615\text{ K})=10^{19.8}\text{ }\frac{\text{Atoms}}{\text{m}^{2}-\text{s}}$ and that 100\% of the atoms are adsorbed on the bell jar. The growth rate of the film is then $R_{\text{film}}=\Phi V_{Yb}$ where $V_{\text{Yb}}=\frac{M_{\text{Yb}}}{\rho_{\text{Yb}}N_{\text{A}}}$ is the atomic volume of Yb where $M_{\text{Yb}}=173\text{ }\frac{\text{kg}}{\text{kmol}}$ is the molar mass, $\rho_{\text{Yb}}=6900\text{ \ensuremath{\frac{\text{kg}}{\text{m}^{3}}}}$ is the bulk density, and $N_{\text{A}}=6.02\times10^{26}\text{ }\frac{\text{Atoms}}{\text{kmol}}$ is Avogadro's Number. The time it takes to form an opaque metal film is $t_{\text{film}}=\frac{10\text{ nm}}{R_{\text{film}}}=4\text{ s}$. To make sense of this result, we will consider two edge cases: 1.) the oven tube takes $\approx2\text{ min}$ to heat up to $615\text{ K}$ and then plating occurs after $4\text{ s}$ and 2.) the oven tube instantly heats up to $550\text{ K}$ generating a constant $\text{Yb}$ vaporization rate of $\Phi(550\text{ K})=10^{18.3}\frac{\text{Atoms}}{\text{m}^{2}-\text{s}}$ that takes $\approx2\text{ min}$ to form an opaque film. This puts the final temperature of the oven tube where Yb plating is observed somewhere in the temperature range of 550 K-615 K. This is a rough estimate of the temperature range because the onset of plating was observed by eye. A better test would have been to monitor the temperature of the oven tube using an in-vacuum thermocouple. 

The experiment also showed that the manufacturing of this oven was poor because holes were created in the tube during the welding process as indicated by plating observed at the closed end of the oven tube shown in Figure \ref{fig:oven-design-1-3-plating-test-results}. Subsequent ovens built were examined under a microscope to visually inspect the tube for holes. Finally, the results showed that $\approx3.1\text{ A}$ is required to observe plating. However, the feedthroughs used in the cryostat have a current limit of $3\text{ A}$, so Design 1 was abandoned. This motivated a final Design 3 summarized in Subsection \ref{subsec:design-3}. 

\subsection{Design 3}
\label{subsec:design-3}

The components of Design 3 are summarized below in Table \ref{tab:components-of-oven-design-3}.

\begin{table}[!!h]
\centering
\tablefont
\caption[Components of Oven Design 3.]{Components of oven Design 3.}
\label{tab:components-of-oven-design-3}
\begin{tabularx}{\textwidth}{>{\centering\arraybackslash}p{0.5\textwidth}c}
\toprule
\textbf{Part} & \textbf{Design 3} \\
\midrule
Oven Tube & \makecell{SS 316 Tube\\(McMaster, PN: 89935K69)\\OD: 1.1 mm\\ID: 0.86 mm\\L: 12.7 mm} \\
\midrule
Wire & \makecell{SS 316 Wire\\(McMaster, PN: 92705K11)\\OD: 0.38 mm\\L: 12.7 mm} \\
\bottomrule
\end{tabularx}
\end{table}

Design 3 consists of a smaller oven tube that is a factor of 10 lighter in mass - decreasing its heat capacity by a factor of 10 - so that a smaller current is required to observed plating. This design was built and tested using the same chamber as shown in Figure \ref{fig:oven-design-1-3-plating-test-results}. The voltage, current, and confirmation of plating is summarized in Table \ref{tab:oven-design-3-plating-test-results}.

\begin{table}[!!h]
\centering
\tablefont
\caption[Oven Design 3 Plating Test Results.]{Voltage, current, and confirmation of plating for Design 3 in test setup. Plating was observed after $2\text{ min}$ using a current of $1.9\text{ A}$.}
\label{tab:oven-design-3-plating-test-results}
\begin{tabularx}{\textwidth}{>{\centering\arraybackslash}p{0.3\textwidth}>{\centering\arraybackslash}p{0.3\textwidth}c}
\toprule
\textbf{Voltage (V)} & \textbf{Current (A)} & \textbf{Plating (yes/no)} \\
\midrule
0.58 & 1.5 & no \\
\midrule
0.65 & 1.7 & no \\
0.74 & 1.9 & yes (after 2 min) \\
\bottomrule
\end{tabularx}
\end{table}

The results of this experiment show that oven tube can be heated up enough to observe plating of Yb on the bell jar after $2\text{ min}$. The current required to observe plating in Design 3 is $\approx1.9\text{ A}$ which is well below the $3\text{ A}$ current limit of the feedthroughs. Thus, Design 3 was implemented into the cryostat and is discussed in the following sections.

\subsection{40 K Stage Oven}
\label{subsec:40K-stage-oven}

It is advantageous to position the oven near the trap because it simplifies the alignment of the atomic flux to the center of trap. This design goal is more challenging to implement in a cryostat than in a room temperature system for two reasons. First, high-current feedthroughs and in-vacuum wires are required to deliver the current to the oven which creates a design compromise between having wires with low thermal conductance between each cold stage and wires that have an overall low resistance. Additionally, $\approx1.4\text{ W}$ is dissipated in the oven of this size which is on the order of the cooling power for most 4 K cryostats. Taking this into consideration, we chose to house the oven at the 40 K stage that has a larger cooling power.

\begin{figure}[!!h]
\centering
\includegraphics[width=0.49\textwidth]{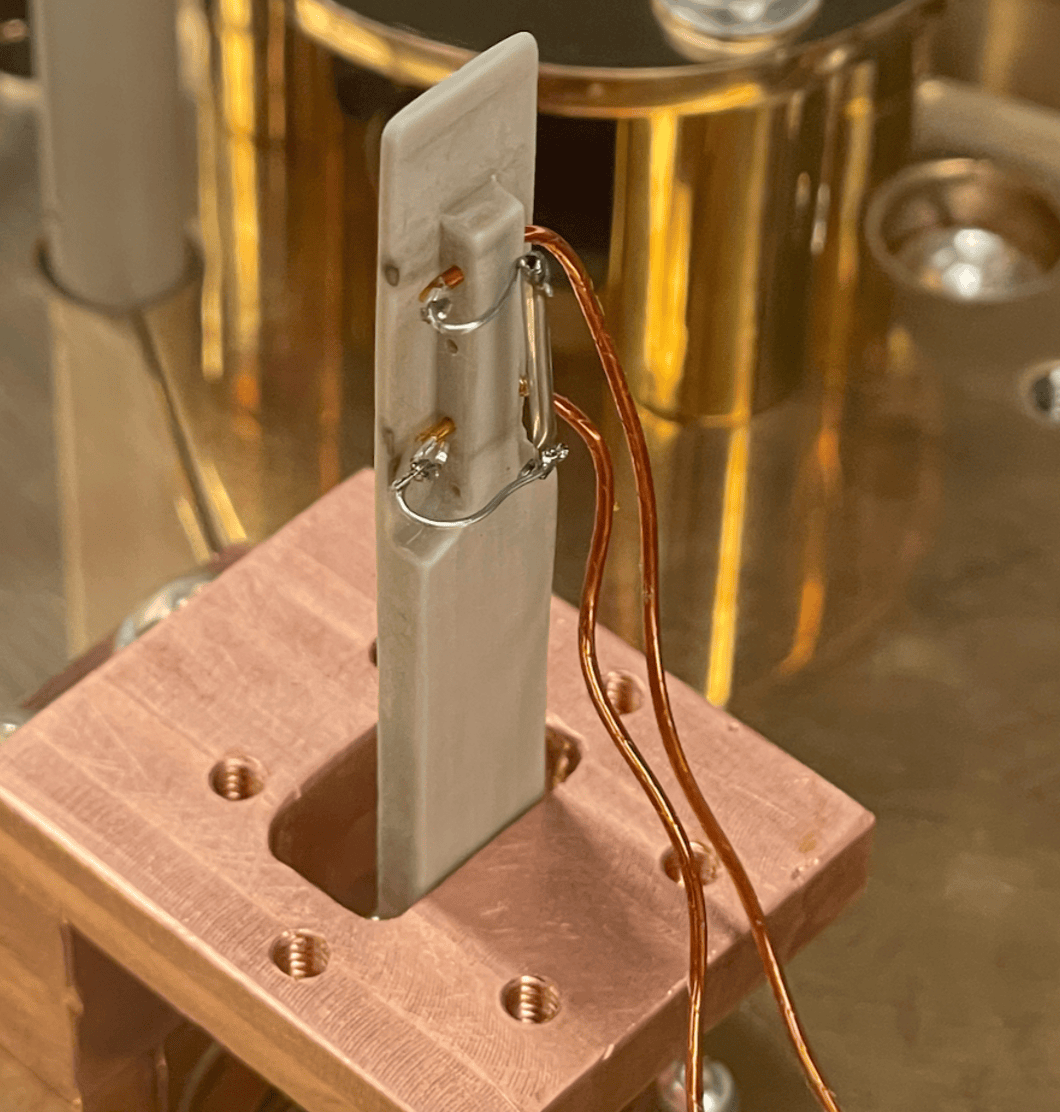}
\includegraphics[width=0.49\textwidth]{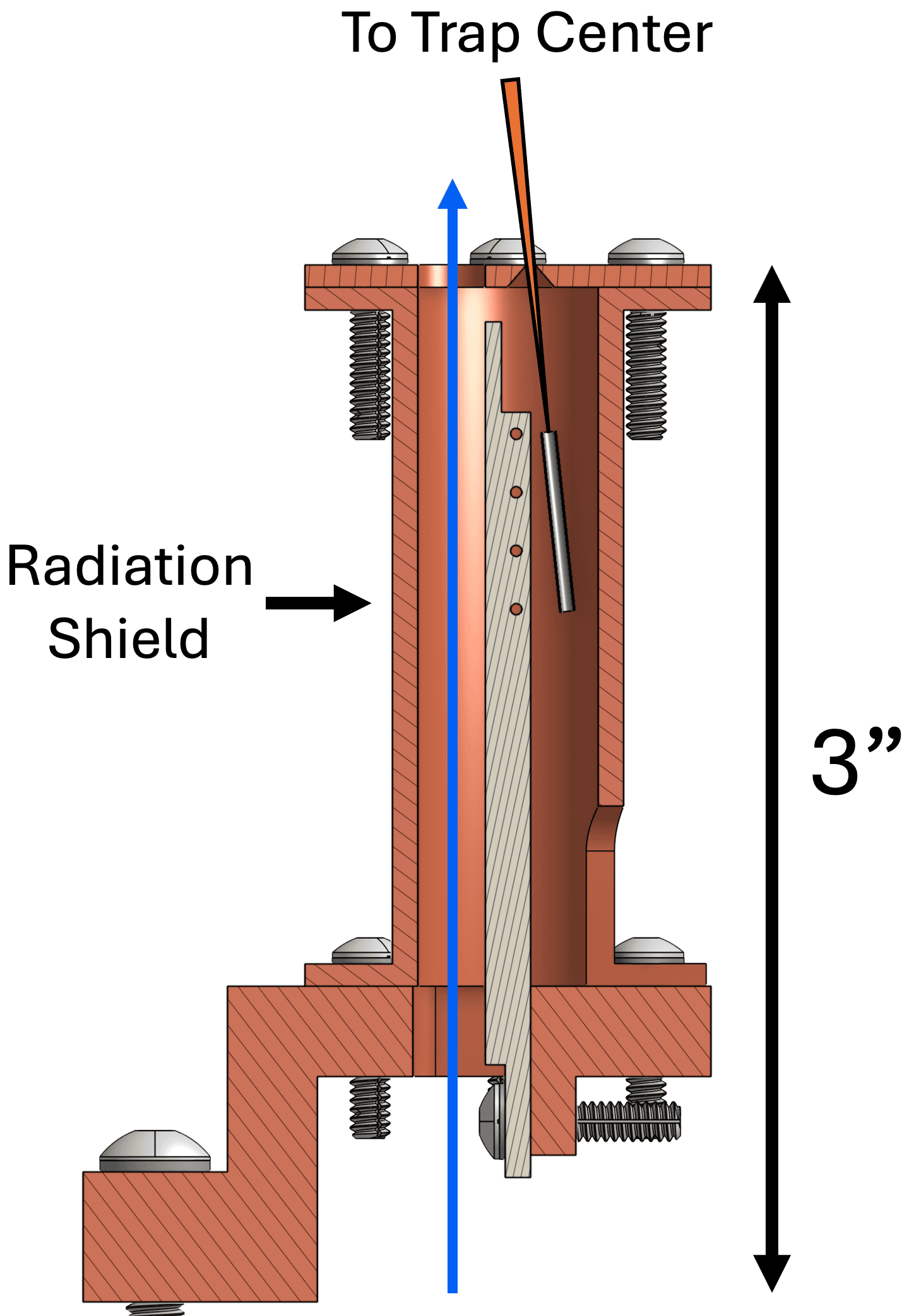}
\caption[40 K Stage Oven Assembly]{(Left) Picture of the 40 K oven assembly showing how the copper wires are press fit into the PEEK holder. (Right) CAD drawing of 40 K oven assembly showing how the atomic beam is collimated by the aperture in the radiation shield lid. The blue arrow depicts optical access to the ions from the bottom of the cryostat.}
\label{fig:40K-stage-oven-assembly}
\end{figure}

The 40 K stage oven assembly consists of Design 3 of the oven held by press-fitting the copper wire leads to through holes in a PEEK mount - machined by myself at UMD - that insulates the hot oven from the 40 K stage as shown in Figure \ref{fig:40K-stage-oven-assembly}. PEEK is used because it has a low thermal conductivity and can be easily machined. A radiation shield encapsulates the oven and PEEK holder to conduct the radiative heat lost from the oven to the 40 K stage. The assembly sits below the ion trap at the 4 K stage and the atomic beam is collimated by an aperture in the radiation shield lid. The size of the aperture was chosen such that the atomic beam will only intersect the center of the trap and not get deposited on the electrodes or alumina trap holder. 

Loading was unsuccessful with this oven as the atomic beam was misaligned to the trap. However, the oven was successfully manufactured as plating was only found on the bottom of the 4 K stage indicating no leakage metal out of unwanted places in the oven tube. Additionally, the temperatures of the 4 K and 40 K stages increased by 100 mK and 3 K respectively which is commensurate with the measured cooling capacity indicating good thermal isolation. In retrospect, the alignment should have been carried out by aligning a thin metal rod between the oven and the trap center using a mock trap mount, as the distance between trap electrodes is $1.5\text{ mm}$. This oven was abandoned and we decided to mount a new oven at the 4 K stage as discussed in the next section. This was a good compromise to get started as we wanted to simply load an ion as soon as possible.

\subsection{4 K Stage Oven}
\label{subsec:4K-stage-oven}

\begin{figure}[!!h]
\centering
\includegraphics[width=0.49\textwidth]{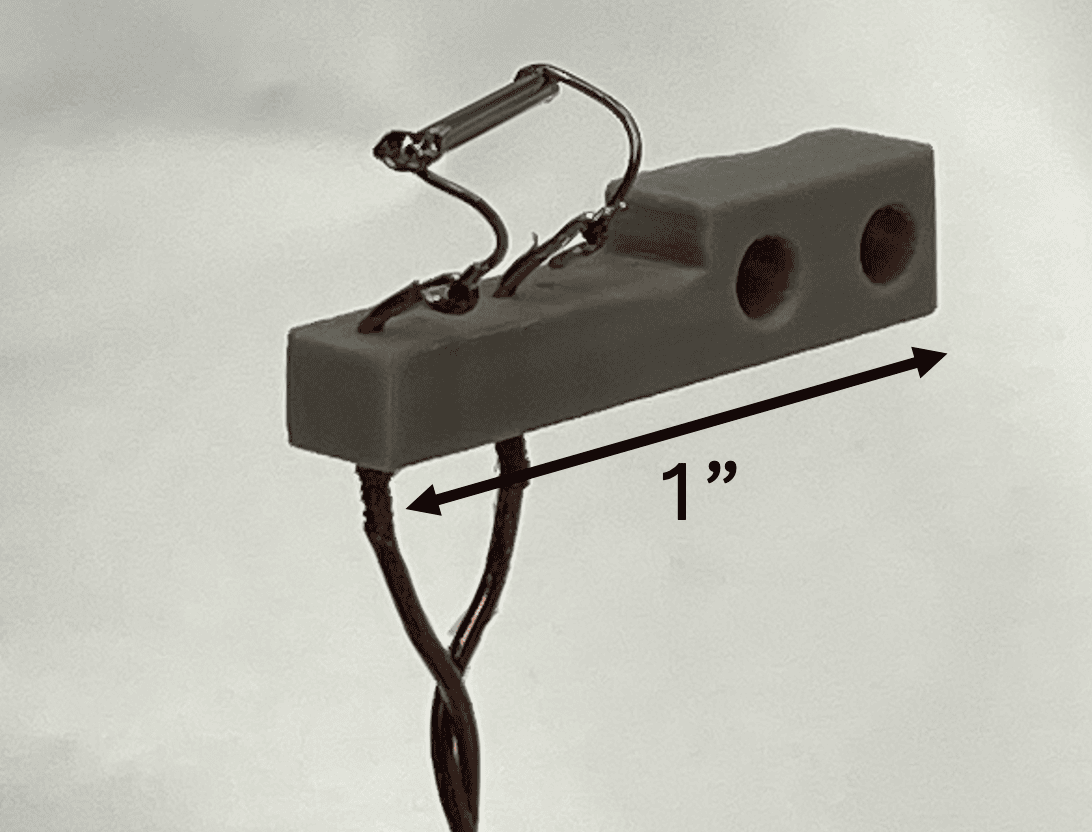}
\includegraphics[width=0.49\textwidth]{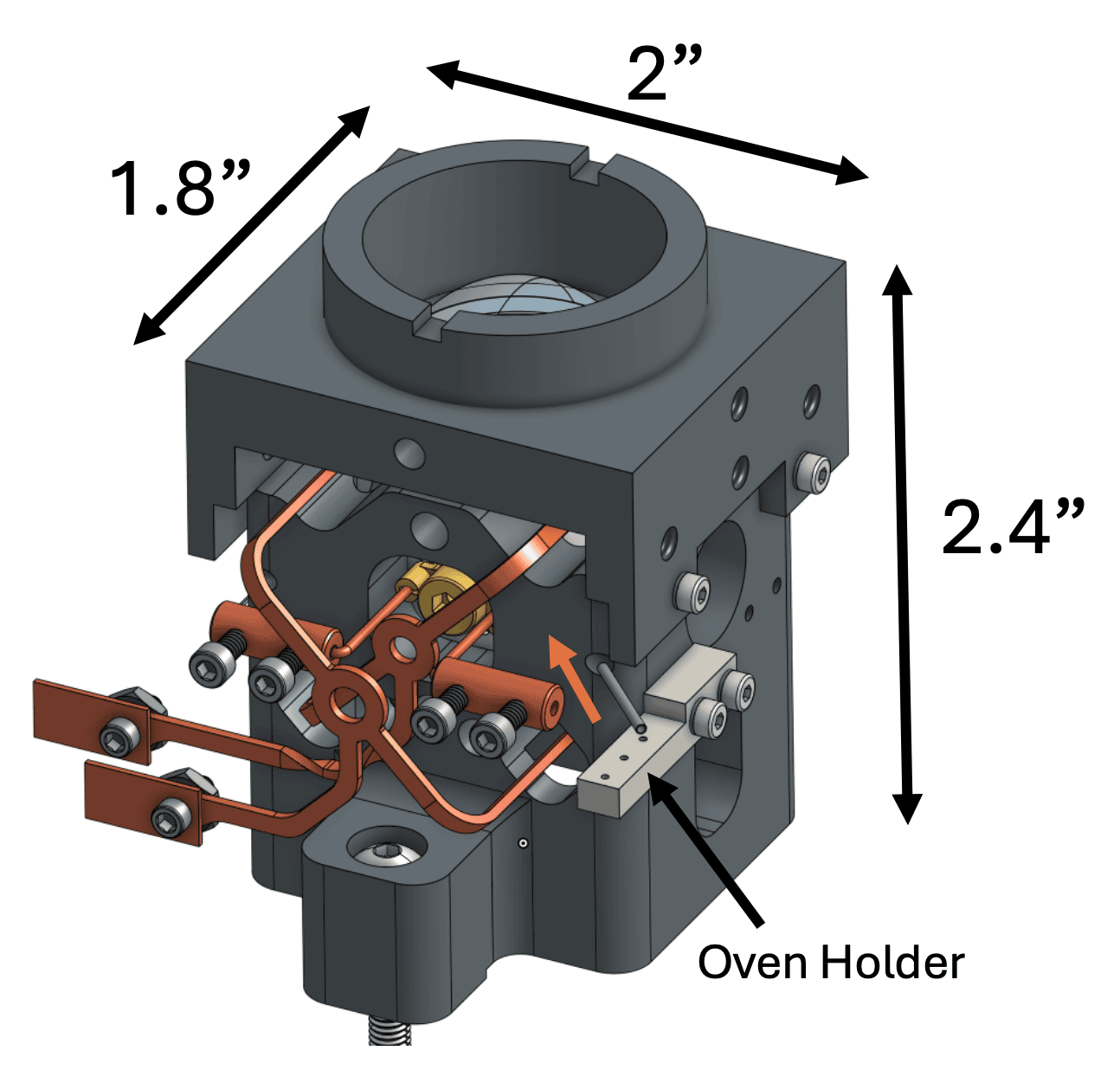}
\caption[4 K Stage Oven Version 1 Assembly]{(Left) Picture of version 1 of the 4 K stage oven assembly. (Right) CAD drawing of version 1 of the 4 K stage oven assembly mounted to the trap mount. The orange arrow indicates the direction of the atomic beam towards the trap center through the aperture in the trap mount.}
\label{fig:4K-stage-oven-version-1-assembly}
\end{figure}

\begin{figure}[!!h]
\centering
\includegraphics[width=0.49\textwidth]{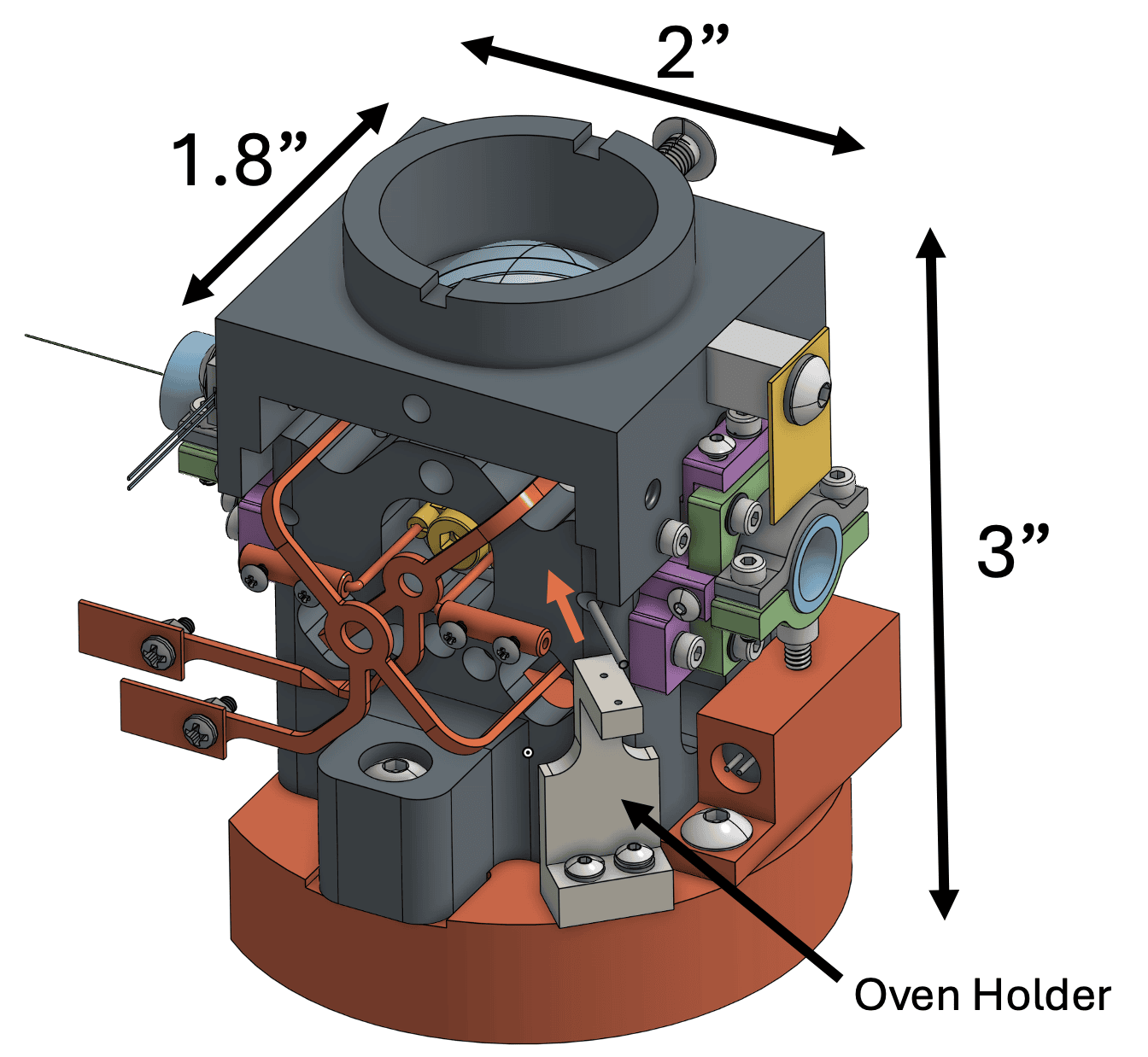}
\caption[4 K Stage Oven Version 2 Assembly]{CAD drawing of version 2 of the 4 K stage oven assembly mounted to the 4 K stage to accommodate the optical cavity. The orange arrow indicates the direction of the atomic flux towards the trap center through the aperture in the trap mount. See Figure \ref{fig:Ion-loading-Schematic} for a more detailed picture of the atomic beam intersecting the trap center as this is the final assembly.}
\label{fig:4K-stage-oven-version-2-assembly}
\end{figure}

The 4 K stage oven assembly consists of a Design 3 oven held by press-fitting the copper wire leads to through holes in a PEEK mount that insulates the hot oven from the 4 K stage. The oven is positioned near the trap mount so that the atomic beam is collimated by the aspect ratio of the trap aperture creating a small angular spread of $\alpha=0.08\text{ radians}$ at the trap center as shown in Figure \ref{fig:Ion-loading-Schematic}. Two versions of this assembly were implemented in the cryostat. In version 1, the PEEK holder was bolted to the trap mount as shown in Figure \ref{fig:4K-stage-oven-version-1-assembly}. Ions were successfully loaded for the first time using version 1. To accommodate the optical cavity, version 2 was implemented where the PEEK holder is bolted directly to the 4 K stage as shown in Figure \ref{fig:4K-stage-oven-version-2-assembly}. 

\section{Thermal Model}
\label{sec:thermal-model-oven}

\begin{figure}[!!h]
\centering
\includegraphics[width=\textwidth]{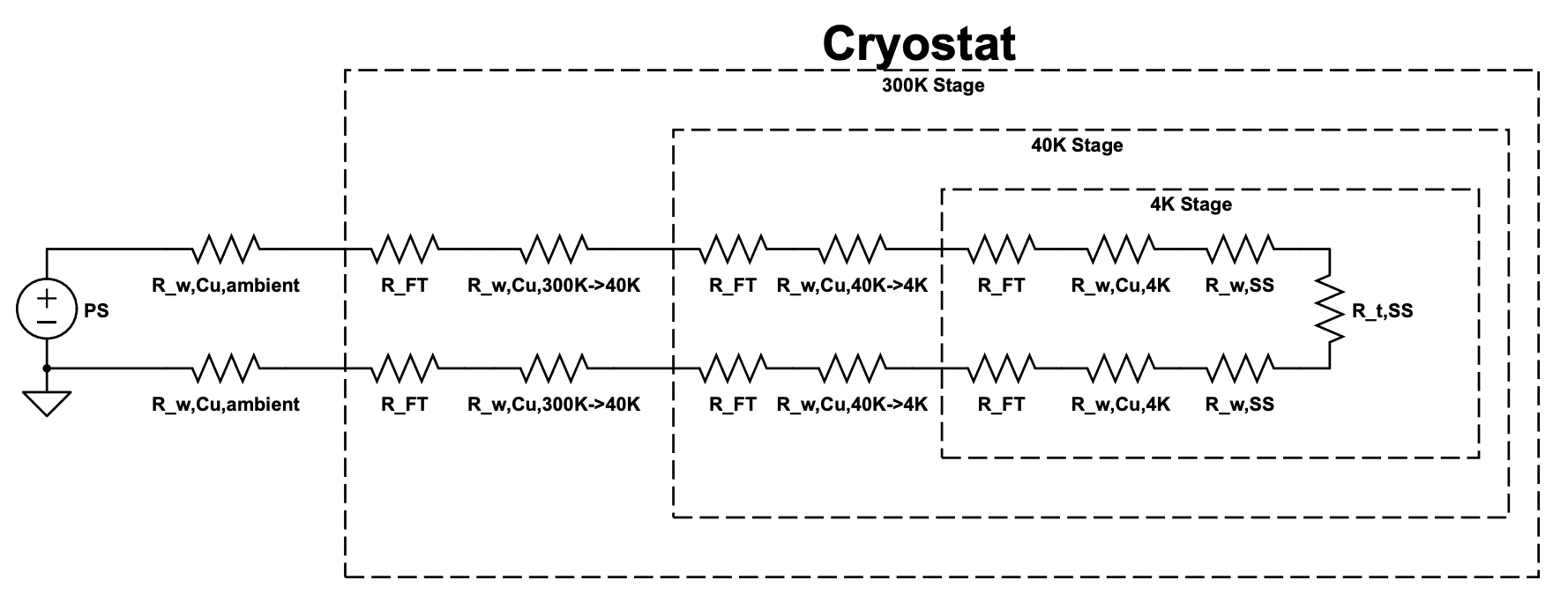}
\caption[Oven Electrical Circuit Diagram]{Oven electrical circuit. The relevant circuit elements are the resistances of the copper wires, $R_{\text{w}_{\text{Cu}}}$, at the different temperature stages in the cryostat, the resistance of the stainless steel wires, $R_{\text{w}_{\text{SS}}}$, the resistance of the oven tube, $R_{\text{t}_{\text{SS}}}$, and the resistance of the feedthroughs, $R_{\text{FT}}$, heat sunk at each temperature stage. The power dissipation of each circuit element is sumamrized in Table \ref{tab:power-dissipation-in-oven-circuit}}
\label{fig:oven-electrical-circuit-diagram}
\end{figure}

A simple model is presented here to understand how the oven tube heats up - the assembly modeled here is version 2 of the 4 K stage oven assembly (oven Design 3) displayed in Figure \ref{fig:4K-stage-oven-version-2-assembly}. First, the oven circuit diagram is studied to understand how the power is dissipated, see Figure \ref{fig:oven-electrical-circuit-diagram}. Current is supplied to the oven using a power supply with a $5\text{ m}$ long 12 AWG copper wire in series with $0.46\text{ m}$ long 26 AWG copper wire that is connected to a feedthrough at the vacuum chamber at the 300 K stage. Inside the cryostat, the current is first routed to a feedthrough at the 40 K stage using $2\text{ m}$ 30 AWG copper wire (Accu-Glass, PN:100670) and then to another feedthrough at the 4 K stage using another $2\text{ m}$ of the same wire. At the 4 K stage, the current is routed from the feedthrough to the $0.0127\text{ m}$ stainless steel wire (McMaster, PN: 92705K11) and oven tube (McMaster, PN: 1800T511) using a $0.15\text{ m}$ 30 AWG copper wire (Accu-Glass, PN:100670). The current is then returned to the power supply via an identical wire path. 

The electrical resistance of the wires is calculated using $R=\bar{\rho}\frac{L}{A}$ where $\bar{\rho}$ is the mean resistivity of the material in $\text{\ensuremath{\Omega}-\text{m}}$, $L$ is the length of the wire in $\text{m}$, and $A$ is the cross-sectional area of the wire in $\text{m}^{2}$. The mean resistivity is used here because the resistivity, $\rho(T)$, is temperature dependent and is given as $\bar{\rho}=\frac{1}{\Delta T}\int_{T_{1}}^{T_{2}}\rho(T)dT$ where $\Delta T=T_{2}-T_{1}$. The resistance of the oven tube is estimated using $R_{\text{t}}=\frac{\bar{\rho}}{2\pi L}\ln\frac{b}{a}$ where $\bar{\rho}$ is the mean resistivity of the material in $\text{\ensuremath{\Omega}-\text{m}}$, $L$ is the length of the wire in $\text{m}$, and $a$ and $b$ are the inner and outer radii of the tube respectively in m. The contact resistance of the Micro-D feedthroughs, $R_{\text{FT}}$, is specced to be $26\times10^{-3}\text{ \ensuremath{\Omega}}$ from the manufacturer, Omnetics, and is assumed to be temperature independent. The resistances of the circuit elements when the oven has reached the desired temperature for routinely loading ions are listed in Table \ref{tab:power-dissipation-in-oven-circuit}. The resistance of the ambient copper wire, $R_{\text{w}_{\text{Cu}}\text{,ambient}}$, is estimated using a resistivity of $\rho(300\text{ K})=1.7\times10^{-8}\text{ \ensuremath{\Omega}-m}$. The resistance of the in-vacuum copper leads routed from the 300 K stage to the 40 K stage, $R_{\text{w}_{\text{Cu}}\text{,300\,K-40\,K}}$, is estimated using a mean resistivity of $\bar{\rho}=8.78\times10^{-9}\text{ \ensuremath{\Omega}-m}$. The resistance of the in-vacuum copper leads routed from the 40 K stage to the 4 K stage, $R_{\text{w}_{\text{Cu}}\text{,40\,K-4\,K}}$, is estimated using a mean resistivity of $\bar{\rho}=4.08\times10^{-10}\text{ \ensuremath{\Omega}-m}$. The resistance of the copper leads routed from the feedthroughs at the 4 K stage to the stainless steel wire, $R_{\text{w}_{\text{Cu}}\text{,4\,K}}$, is estimated using a mean resistivity of $\bar{\rho}=1.57\times10^{-8}\text{ \ensuremath{\Omega}-m}$ assuming a final stainless steel wire temperature of $537\text{ K}$ as discussed in the next section. The resistances of the stainless steel wire and oven tube, $R_{\text{w}_{\text{SS}}}$ and $R_{\text{t}_{\text{SS}}}$, are estimated using mean resistivity of $\bar{\rho}=9.39\times10^{-7}\text{ \ensuremath{\Omega}-m}$ assuming a final temperature of $537\text{ K}$. Finally, the resistive heating for each element is estimated using $\dot{Q}_{\text{res}}=I^{2}R$ where $\dot{Q}_{\text{res}}$ is the power dissipated in $\text{W}$, $I$ is the current in $\text{A}$, and $R$ is the resistance in $\text{\ensuremath{\Omega}}$. 

A current of $1.8\text{ A}$ is passed through the circuit and the total power dissipation is $4.42\text{ W}$, see Table \ref{tab:power-dissipation-in-oven-circuit}. In summary, only 15\% of the power in the circuit is dissipated in the stainless steel wires while practically no power is dissipated in the oven tube. Thus, the resistive heating of the stainless steel wire heats the oven tube via conduction as explained using a thermal model in the next section. 51\% of the power is dissipated in the 30 AWG wires between the 300 K and 40 K stages. This is a design compromise of limiting the thermal conductance of the wires between the cold stages while also keeping the total resistance of the wires low. It is important to keep the resistances of the wires low so that you can deliver the desired voltage to a load without voltage limiting the power supply of the circuit and to mitigate resistive heating in the wires that could be comparable to the cooling power of the cryostat cold stages. It is an okay compromise here because this power is dissipated at the 40 K stage which has plenty of cooling power. An alternate design would be to replace the 30 AWG wires with 24 AWG wires so that the resistive heating is decreased by a factor of 4, but this comes at the cost of organizing bulkier and more springy wire in-vacuum which can be challenging. 

\begin{table}[!!h]
\centering
\tablefont
\caption[Power Dissipation in Oven Circuit]{The resistance for each circuit element in Figure \ref{fig:oven-electrical-circuit-diagram} the oven has reached its target temperature is listed along with its total power dissipation, $\dot{Q}_{\text{res}}$, in the circuit. The current through the circuit is $1.8\text{ A}$ and the net power dissipated in the circuit is $4.42\text{ W}$. $680\text{ mW}$ is dissipated in the stainless steel wires at the 4 K stage which is 15\% of the total power dissipated in the circuit.}
\label{tab:power-dissipation-in-oven-circuit}
\begin{tabularx}{\textwidth}{>{\centering\arraybackslash}p{0.24\textwidth}>{\centering\arraybackslash}p{0.24\textwidth}cc}
\toprule
\textbf{Circuit Element} & \textbf{Mean Resistivity ($\text{\ensuremath{\mathbf{\Omega}}-m}$)} & \textbf{Resistance ($\mathbf{\Omega}$)} & \textbf{Total Power Dissipation (W)}\\
\midrule
$R_{\text{w}_{\text{Cu}}\text{,ambient}}$ & $1.7\times10^{-8}$ & 0.09 & $2\times0.29=0.58$ \\
\midrule
$R_{\text{w}_{\text{Cu}}\text{,300\,K-40\,K}}$ & $8.78\times10^{-9}$ & 0.35 & $2\times1.13=2.25$ \\
\midrule
$R_{\text{w}_{\text{Cu}}\text{,40\,K-4\,K}}$ & $4.08\times10^{-10}$ & 0.02 & $2\times0.05=0.10$ \\
\midrule
$R_{\text{w}_{\text{Cu}}\text{,4\,K}}$ & $1.57\times10^{-8}$ & 0.05 & $2\times0.15=0.30$ \\
\midrule
$R_{\text{FT}}$ & - & 0.03 & $6\times0.085=0.51$ \\
\midrule
$R_{\text{w}_{\text{SS}}}$ & $9.39\times10^{-7}$ & 0.12 & $2\times0.34=0.68$ \\
\midrule
$R_{\text{t}_{\text{SS}}}$ & $9.39\times10^{-7}$ & $2.4\times10^{-6}$ & $1\times7.9\times10^{-6}=7.9\times10^{-6}$ \\
\bottomrule
\end{tabularx}
\end{table}

A simple thermal model is presented here that simulates how the temperatures of the stainless steel wire, oven tube, trap electrodes, trap mount, and 4 K stage change with time due to the resistive heating of the stainless wire. Each of these elements exchange heat with each other via conduction and radiation 
\begin{equation}
    \dot{Q}=\dot{Q}_{\text{cond}}+\dot{Q}_{\text{rad}}
    \label{eq:cryogenic-heat-transfer-oven}
\end{equation}
and we assume that the heat transferred via conduction and convection through the vacuum is negligible and is not included in the model. The heat transferred via conduction through a solid material is given as 
\begin{equation}
    \dot{Q}_{\text{cond}}=\bar{k}\frac{A}{L}\Delta T
    \label{eq:Fouriers-law-oven}
\end{equation}
where $\bar{k}$ is the mean thermal conductivity of the material in $\frac{\text{W}}{\text{m-K}}$, $A$ is the cross-sectional area of the material in $\text{m}^{2}$, $L$ is the length of the material in $\text{m}$, and $\Delta T$ is the temperature gradient through the material in $\text{K}$. The mean thermal conductivity is used here because the thermal conductivity, $k(T)$, is temperature dependent and is given as $\bar{k}=\frac{1}{\Delta T}\int_{T_{1}}^{T_{2}}k(T)dT$ where $\Delta T=T_{2}-T_{1}$. It is convenient to define the conductance $G=\bar{k}\frac{A}{L}$ in $\frac{\text{W}}{\text{K}}$ so now we can write $\dot{Q}_{\text{cond}}=G\Delta T$. The radiative heat transfer between two materials 1 and 2 is given as 
\begin{equation}
    \dot{Q}_{\text{rad,1-2}}=\frac{A_{1}\sigma(T_{1}^{4}-T_{2}^{4})}{\frac{1-\varepsilon_{1}}{\varepsilon_{1}}+\frac{1}{F_{12}}+\frac{1-\varepsilon_{2}}{\varepsilon_{2}}\frac{A_{1}}{A_{2}}}
    \label{eq:radiative-heat-transfer-oven}
\end{equation}
 where $\sigma=5.67\times10^{-8}\frac{\text{W}}{\text{m}^{2}\text{-K}^{4}}$ is the Stefan-Bolztmann constant, $T_{1}$ and $T_{2}$ are the temperatures of materials 1 and 2 respectively in K, $A_{1}$ and $A_{2}$ are the surface areas of materials 1 and 2 respectively in $\text{m}^{2}$, $\epsilon_{1}$ and $\epsilon_{2}$ are the emissivities of materials 1 and 2, and $F_{12}$ is the fraction of radiative heat due to material 1 that is communicated to material 2, denoted the view factor which is dependent on solid angle. It is useful to define the radiative heat transfer coefficient $h_{\text{1-2}}=\frac{\sigma(T_{1}^{2}+T_{2}^{2})(T_{1}+T_{2})}{\frac{1-\varepsilon_{1}}{\varepsilon_{1}}+\frac{1}{F_{12}}+\frac{1-\varepsilon_{2}}{\varepsilon_{2}}\frac{A_{1}}{A_{2}}}$ in $\frac{\text{W}}{\text{m}^{2}\text{-K}}$ so that we can write $\dot{Q}_{\text{1-2}}=h_{\text{1-2}}A_{1}(T_{1}-T_{2})$. 

A lumped system analysis is used in this thermal model which assumes that temperature gradient within each lump - the stainless steel wire, oven tube, trap electrodes, trap mount, and 4 K stage - is negligible. The validity of this lumped system analysis is characterized by the value of the dimensionless quantity called the Biot number $\text{Bi}=\frac{h_{\text{1-2}}}{k}L_{\text{c}}$ where $h_{\text{1-2}}$ is the radiative heat transfer coefficient between materials 1 and 2 in $\frac{\text{W}}{\text{m}^{2}\text{-K}}$, $k$ is the thermal conductivity of the material in $\frac{\text{W}}{\text{m-K}}$, and $L_{c}=\frac{V}{A}$ is the characteristic length in m where $V$ is the volume of the material in $\text{m}^{3}$ and $A$ is the surface area of the material in $\text{m}^{2}$. A Biot number that is $\ll1$ indicates that the temperature inside a body does not vary significantly in space when the body is heated or cooled by a heat flux at its surface. We only consider radiative heat transfer at the surface since convection is negligible in vacuum. The Biot numbers of all the lumps are $\ll1$ so we can assume that temperature gradients within each lump are small. The problem is then reduced to energy balances for each lump that are summarized below. 

The energy balance for one of the stainless steel wires is given below in Equation \eqref{eq:stainless-steel-wire-energy-balance}. The heat delivered to the stainless steel wire is due to the resistive heating of the wire itself, $\dot{Q}_{\text{res,w}}$. The heat delivered to the stainless steel wire due to the resistive heating in the copper leads in negligible. The wire loses heat via conduction and radiation. The conductive losses are to the oven tube via a contact conductance of the weld, $\dot{Q}_{\text{cond,w-t}}$ and to the 4 K stage via the copper wire and PEEK oven holder, $\dot{Q}_{\text{cond,w-4\,K}}$. The radiative losses are to the 4 K stage, $\dot{Q}_{\text{rad,w-4\,K}}$, and to the trap mount, $\dot{Q}_{\text{rad,w-m}}$. We assume that there is no radiation exchange between the stainless steel wire and the oven tube. We also assume there is no radiation exchange between the stainless steel wire and the trap electrodes because there is no direct line of sight. The heat capacity of the stainless steel wire is calculated using the mass of the wire, $m_{\text{w}}$, and the specific heat of stainless steel, $C_{\text{SS}}(T_{\text{w}})$. Since the specific heat of 316 stainless steel changes from $2\text{ }\frac{\text{\text{J}}}{\text{kg-K}}$ at 4 K to $525\text{ }\frac{\text{J}}{\text{kg-K}}$ at 500 K - a factor of 263 - we use the temperature dependent specific heat equation extracted from experimental data \cite{kimThermophysicalPropertiesStainless1975}.

\begin{equation}
\begin{aligned}
    \dot{Q}_{\text{res,w}}-\dot{Q}_{\text{cond,w-t}}-\dot{Q}_{\text{cond,w-4\,K}}-\dot{Q}_{\text{rad,w-4\,K}}-\dot{Q}_{\text{rad,w-m}}&=m_{\text{w}}C_{\text{SS}}(T_{\text{w}})\frac{\text{d}T_{\text{w}}}{\text{d}t} \\
    I^{2}R_{\text{w}}(T_{\text{w}})-G_{\text{w-t}}(T_{\text{w}}-T_{\text{t}})-G_{\text{w-4\,K}}(T_{\text{w}}-T_{\text{4\,K}})- \\
    h_{\text{w-4\,K}}A_{\text{w}}(T_{\text{w}}-T_{\text{4\,K}})-h_{\text{w-m}}A_{\text{w}}(T_{\text{w}}-T_{\text{m}})&=m_{\text{w}}C_{\text{SS}}(T_{\text{w}})\frac{\text{d}T_{\text{w}}}{\text{d}t}
    \label{eq:stainless-steel-wire-energy-balance}
\end{aligned}
\end{equation}

The resistance of the stainless steel wire, $R_{\text{w}}(T_{\text{w}})=\rho_{\text{SS}}(T)\frac{A_{\text{w}_{\text{SS}}}}{L_{\text{w}_{\text{SS}}}}$, using the temperature dependent resistivity of stainless steel extracted from experimental data \cite{hoElectricalResistivityThermal}. The conductance of the weld between the stainless steel wire and oven tube, $G_{\text{w-t}}=\bar{k}_{\text{SS}}\frac{A_{\text{w}_{\text{SS}}}}{L_{\text{w}_{\text{SS}}}}$, is estimated by assuming a $1\text{ mm}$ length of stainless steel that has the same cross-sectional area of the wire and mean thermal conductivity $\bar{k}_{SS}=13\text{ }\frac{\text{W}}{\text{m-K}}$ \cite{kimThermophysicalPropertiesStainless1975}. The total conductance between the stainless steel wire and 4 K stage is the sum of the conductance of the weld between the stainless steel and copper wire and the conductances of the copper wire and the PEEK oven holder in series $G_{\text{w-4K}}=(\frac{1}{G_{\text{w}_{\text{SS}}\text{-w}_{\text{Cu}}}}+\frac{1}{G_{\text{Cu}}+G_{\text{P}}})^{-1}$. The conductance of the copper wire, $G_{\text{Cu}}=\bar{k}_{\text{Cu}}\frac{A_{\text{Cu}}}{L_{\text{Cu}}}$, is estimated using a length of $0.15\text{ m}$, a cross-sectional area of $2\times10^{-7}\text{ m}^{2}$, and a mean thermal conductivity of $\bar{k}_{\text{Cu}}=481\text{ }\frac{\text{W}}{\text{m-K}}$ \cite{simonPropertiesCopperCopper1992}. The conductance of the PEEK holder, $G_{\text{P}}=\bar{k}_{\text{P}}\frac{A_{\text{P}}}{L_{\text{P}}}$, is estimated using a length of $0.29\text{ m}$, a cross-sectional area of $3.3\times10^{-5}\text{ m}^{2}$, and a mean thermal conductivity $\bar{k}_{\text{P}}=0.24\text{ }\frac{\text{W}}{\text{m-K}}$ \cite{ruleLowtemperatureThermalConductivity1989}. The conductance of the weld between the stainless steel wire and copper wire, $G_{\text{w}_{\text{SS}}\text{-w}_{\text{Cu}}}=(\frac{L_{\text{w}_{\text{SS}}}}{\bar{k}_{\text{SS}}A_{\text{w}_{\text{SS}}}}+\frac{L_{\text{w}_{\text{Cu}}}}{\bar{k}_{\text{Cu}}A_{\text{w}_{\text{Cu}}}})^{-1}$, is estimated by assuming a $1\text{ mm}$ length of stainless steel and copper wire and using the same cross-sectional areas and mean thermal conductivities as stated above. The heat lost via radiation to the 4 K stage, $h_{\text{w-4\,K}}A_{\text{w}}(T_{\text{w}}-T_{\text{4\,K}})$, is estimated by assuming the emissivities of the stainless steel wire and 4 K stage to be 1 and 0.02 respectively. An emissivity of 1 for the stainless steel wire is assumed because the surface treatment is unknown. The emissivity of 0.02 is assumed because the 4 K stage mainly consists of gold-plated copper \cite{ekinExperimentalTechniquesLowtemperature2006a}. The heat lost via radiation to the trap mount, $h_{\text{w-m}}A_{\text{w}}(T_{\text{w}}-T_{\text{m}})$, is estimated by assuming the emissivities of both the stainless steel wire and trap mount to be 1 because the surface treatment of these materials is unknown. We assume the view factor between the stainless steel wire and 4 K stage to be 0.75 and the view factor between the stainless steel wire and trap mount to be 0.25 - i.e., 25\% of the heat radiated from the stainless steel wire is communicated to the trap mount and the remaining 75\% is communicated to the 4 K stage. 

The energy balance of the oven tube is given below in Equation \eqref{eq:stainless-steel-tube-energy-balance}. Heat is delivered to the oven tube via conduction through the contact between the stainless steel wire and oven tube, $\dot{Q}_{\text{cond,w-t}}$. The factor of 2 in front of $\dot{Q}_{\text{cond,w-t}}$ is needed to account for the second stainless steel wire that is also resistively heating with a temperature dependence given by Equation \eqref{eq:stainless-steel-wire-energy-balance}. The tube loses heat via radiation to the 4 K stage, $\dot{Q}_{\text{rad,t-4\,K}}$, to the trap mount, $\dot{Q}_{\text{rad,t-m}}$, and to the trap electrodes, $\dot{Q}_{\text{rad,t-e}}$. The heat capacity of the stainless steel tube is calculated using the mass of the tube, $m_{\text{t}}$, and the specific heat of stainless steel, $C_{\text{SS}}(T_{\text{t}})$. 

\begin{equation}
\begin{aligned}
    2\dot{Q}_{\text{cond,w-t}}-\dot{Q}_{\text{rad,t-4\,K}}-\dot{Q}_{\text{rad,t-m}}-\dot{Q}_{\text{rad,t-e}}&=m_{\text{t}}C_{\text{SS}}(T_{\text{t}})\frac{\text{d}T_{\text{t}}}{\text{d}t} \\
    2G_{\text{w-t}}(T_{\text{w}}-T_{\text{t}})-h_{\text{t-4\,K}}A_{\text{t}}(T_{\text{t}}-T_{\text{4\,K}})- \\
    h_{\text{rad,t-m}}A_{\text{t}}(T_{\text{t}}-T_{\text{m}})-h_{\text{rad,t-e}}A_{\text{t}}(T_{\text{t}}-T_{\text{e}})=&m_{\text{t}}C_{\text{SS}}(T_{\text{t}})\frac{\text{d}T_{\text{t}}}{\text{d}t}
    \label{eq:stainless-steel-tube-energy-balance}
\end{aligned}
\end{equation}

The heat lost via radiation to the 4 K stage, $h_{\text{t-4\,K}}A_{\text{t}}(T_{\text{t}}-T_{\text{4\,K}})$, is estimated by assuming the emissivities of the stainless steel tube and 4 K stage to be 1 and 0.02 respectively. An emissivity of 1 is used for the tube because the surface treatment is unknown. The heat lost via radiation to the trap mount, $h_{\text{rad,t-m}}A_{\text{t}}(T_{\text{t}}-T_{\text{m}})$, is estimated by assuming the emissivities of both the stainless steel tube and trap mount to be 1. The heat lost via radiation to the trap electrodes, $h_{\text{rad,t-e}}A_{\text{t}}(T_{\text{t}}-T_{\text{e}})$, is estimated by assuming the emissivities of the stainless steel tube and trap electrodes to be 1 and 0.02 respectively. The emissivity of 0.02 is assumed because the trap electrodes are made of gold-plated titanium. The view factor from the tube to the trap electrodes is estimated by assuming only the radiation that is emitting from one end of the tube of area, $A_{\text{end}}$, hits the electrodes, i.e., $\frac{A_{\text{end}}}{A_{\text{total}}}=0.02$, where $A_{\text{total}}$ is the total surface area of the tube. As with the stainless steel wire, we assume 75\% of the radiation is emitted to the 4 K stage and so the view factor between the oven tube and trap mount is constrained to be $1-0.75-0.02=0.23$.

The energy balance of the trap electrodes is given below in Equation \eqref{eq:trap-electrode-energy-balance}. Heat is delivered to the trap electrodes due to the radiative heat load from the oven tube, $\dot{Q}_{\text{rad,t-e}}$. Heat is lost via conduction through the trap holder made of alumina to the trap mount, $\dot{Q}_{\text{cond,h}}$. The radiative heat lost to the 4 K stage, trap holder, and trap mount is negligible because the electrodes do not heat up significantly. The heat capacity of the titanium electrodes is calculated using the mass of the electrodes, $m_{\text{e}}$, and the specific heat of titanium, $C_{\text{Ti}}$. Since the electrodes do not heat up significantly, we can assume a constant specific heat $C_{\text{Ti}}(4\text{\,K})=0.064\text{ }\frac{\text{J}}{\text{kg-K}}$ \cite{zieglerSpecificHeatThermal1963}. 

\begin{equation}
\begin{aligned}
    \dot{Q}_{\text{rad,t-e}}-\dot{Q}_{\text{cond,h}}&=m_{\text{e}}C_{\text{Ti}}\frac{\text{d}T_{\text{e}}}{\text{d}t} \\
    h_{\text{t-e}}A_{\text{t}}(T_{\text{t}}-T_{\text{e}})-G_{\text{h}}(T_{\text{e}}-T_{\text{m}})&=m_{\text{e}}C_{\text{Ti}}\frac{\text{d}T_{\text{e}}}{\text{d}t}
    \label{eq:trap-electrode-energy-balance}
\end{aligned}
\end{equation}
The conductance of the trap holder is taken to be $G_\text{h}=0.045\,\frac{\text{W}}{\text{K}}$ as per the AQT trap manual.

The energy balance of the trap mount is given below in Equation \eqref{eq:trap-mount-energy-balance}. Heat is delivered to the trap mount due to the radiative heat load from the oven tube, $\dot{Q}_{\text{rad,t-m}}$, and the stainless steel wires, $\dot{Q}_{\text{rad,w-m}}$. Heat is lost via conduction through the trap holder, $\dot{Q}_{\text{cond,h}}$, to the trap electrodes and through the contact conductance between the trap mount and the 4 K stage, $\dot{Q}_{\text{cond,m-4\,K}}$. 

\begin{equation}
\begin{aligned}
    \dot{Q}_{\text{rad,t-m}}+\dot{Q}_{\text{rad,w-m}}-\dot{Q}_{\text{cond,h}}-\dot{Q}_{\text{cond,m-4\,K}}&=m_{\text{m}}C_{\text{Ti}}\frac{\text{d}T_{\text{m}}}{\text{d}t} \\
    h_{\text{t-m}}A_{\text{t}}(T_{\text{t}}-T_{\text{m}})+h_{\text{w-m}}A_{\text{w}}(T_{\text{w}}-T_{\text{m}})- \\
    G_{\text{h}}(T_{\text{m}}-T_{\text{e}})-G_{\text{m-4\,K}}(T_{\text{m}}-T_{\text{4\,K}})&=m_{\text{m}}C_{\text{Ti}}\frac{\text{d}T_{\text{m}}}{\text{d}t}
    \label{eq:trap-mount-energy-balance}
\end{aligned}
\end{equation}

The contact conductance between the trap mount and the 4 K stage, $G_{\text{m-4\,K}}=0.31\text{ \ensuremath{\frac{\text{W}}{\text{K}}}}$, was determined empirically by applying a known heat load to the 4 K stage and measuring the equilibration temperature, see Subsection\ref{subsec:current-system}.

Lastly, the energy balance of the 4 K stage is given below in Equation \eqref{eq:4K-stage-energy-balance}. Heat is delivered to the 4 K stage via conduction through the copper wire and the PEEK holder, $\dot{Q}_{\text{cond,w-4\,K}}$, and through the contact conductance between the trap mount and the 4 K stage, $\dot{Q}_{\text{\text{cond,m-4\,K}}}$. Heat is also delivered due to the radiative heat load from the stainless steel wires, $\dot{Q}_{\text{rad,w-4\,K}}$, and the oven tube, $\dot{Q}_{\text{rad,t-4\,K}}$. Heat is taken away from the 4 K stage via conduction to the cold helium flowing through the stage, $\dot{Q}_{\text{cond,cool}}$. The heat capacity of the 4 K stage is estimated using the mass of the stage, $m_{\text{4\,K}}$, and the specific heat of copper, $C_{\text{Cu}}$. Since the 4 K stage does not heat up significantly, we can assume a constant specific heat $C_{\text{Cu}}(4\text{\,K})=0.099\text{ }\frac{\text{J}}{\text{kg-K}}$ \cite{zieglerSpecificHeatThermal1963}. 

\begin{equation}
\begin{aligned}
    \dot{Q}_{\text{cond,w-4\,K}}+\dot{Q}_{\text{cond,m-4\,K}}+\dot{Q}_{\text{rad,w-4\,K}}+\dot{Q}_{\text{rad,t-4\,K}}-\dot{Q}_{\text{cond,cool}}&=m_{\text{4\,K}}C_{\text{Cu}}\frac{\text{d}T_{\text{4\,K}}}{\text{d}t} \\
    G_{\text{w-4\,K}}(T_{\text{w}}-T_{\text{4\,K}})+G_{\text{m-4\,K}}(T_{\text{m}}-T_{\text{4\,K}})+h_{\text{w-4\,K}}A_{\text{w}}(T_{\text{w}}-T_{\text{4\,K}})+ \\
    h_{\text{t-4\,K}}A_{\text{t}}(T_{\text{t}}-T_{\text{4\,K}})-G_{\text{cool}}(T_{\text{4\,K}}-4\text{\,K})&=m_{\text{4\,K}}C_{\text{Cu}}\frac{\text{d}T_{\text{4\,K}}}{\text{d}t}
    \label{eq:4K-stage-energy-balance}
\end{aligned}
\end{equation}
The cooing capacity of the 4 K stage characterized by the conductance, $G_{\text{cool}}=0.39\text{ }\frac{\text{W}}{\text{K}}$, was determined empirically by applying a known heat load to the 4 K stage and measuring the equilibration temperature, see Subsection \ref{subsec:current-system}. 

\begin{figure}[!!h]
\centering
\includegraphics[width=0.49\textwidth]{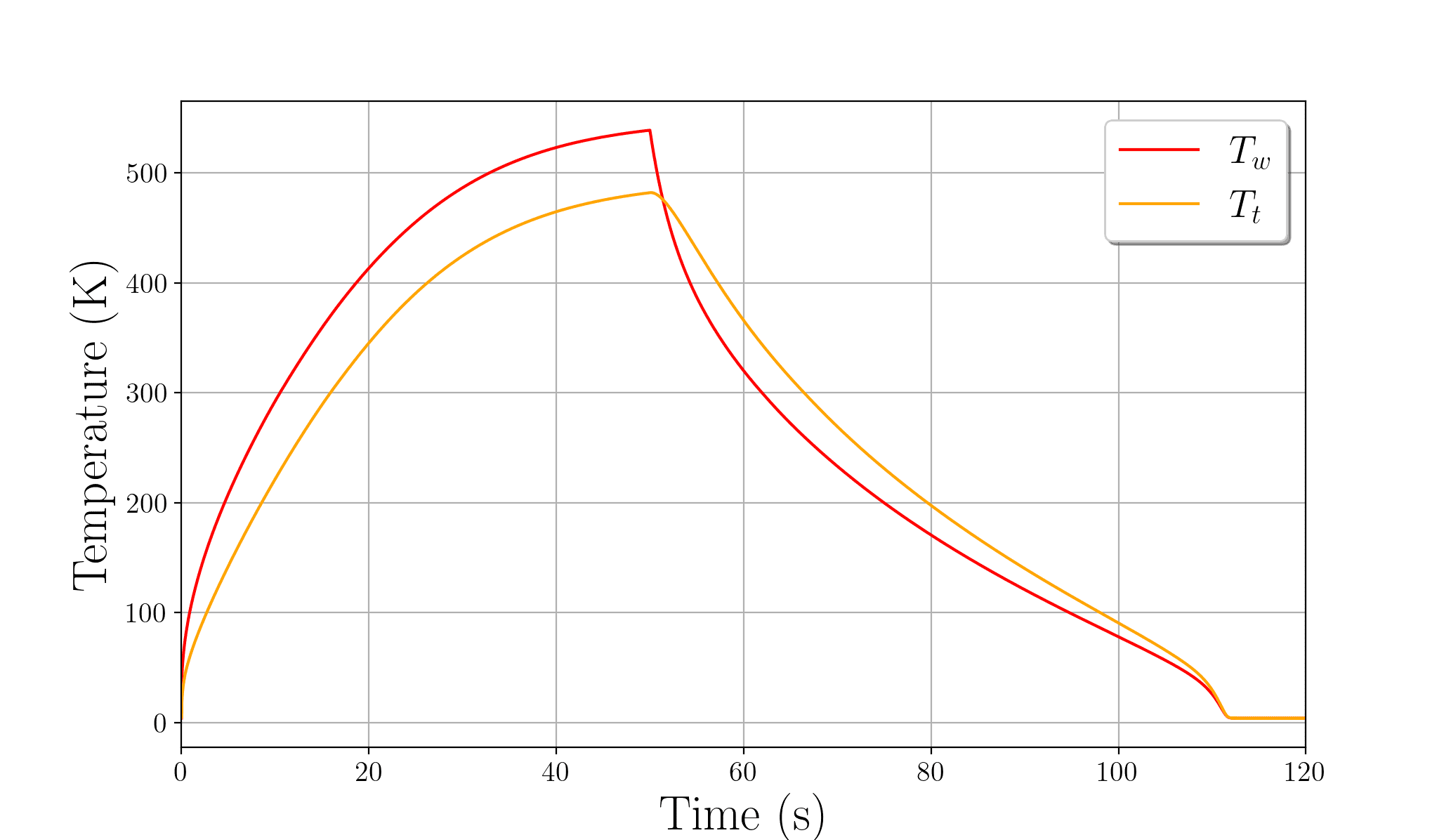}
\includegraphics[width=0.49\textwidth]{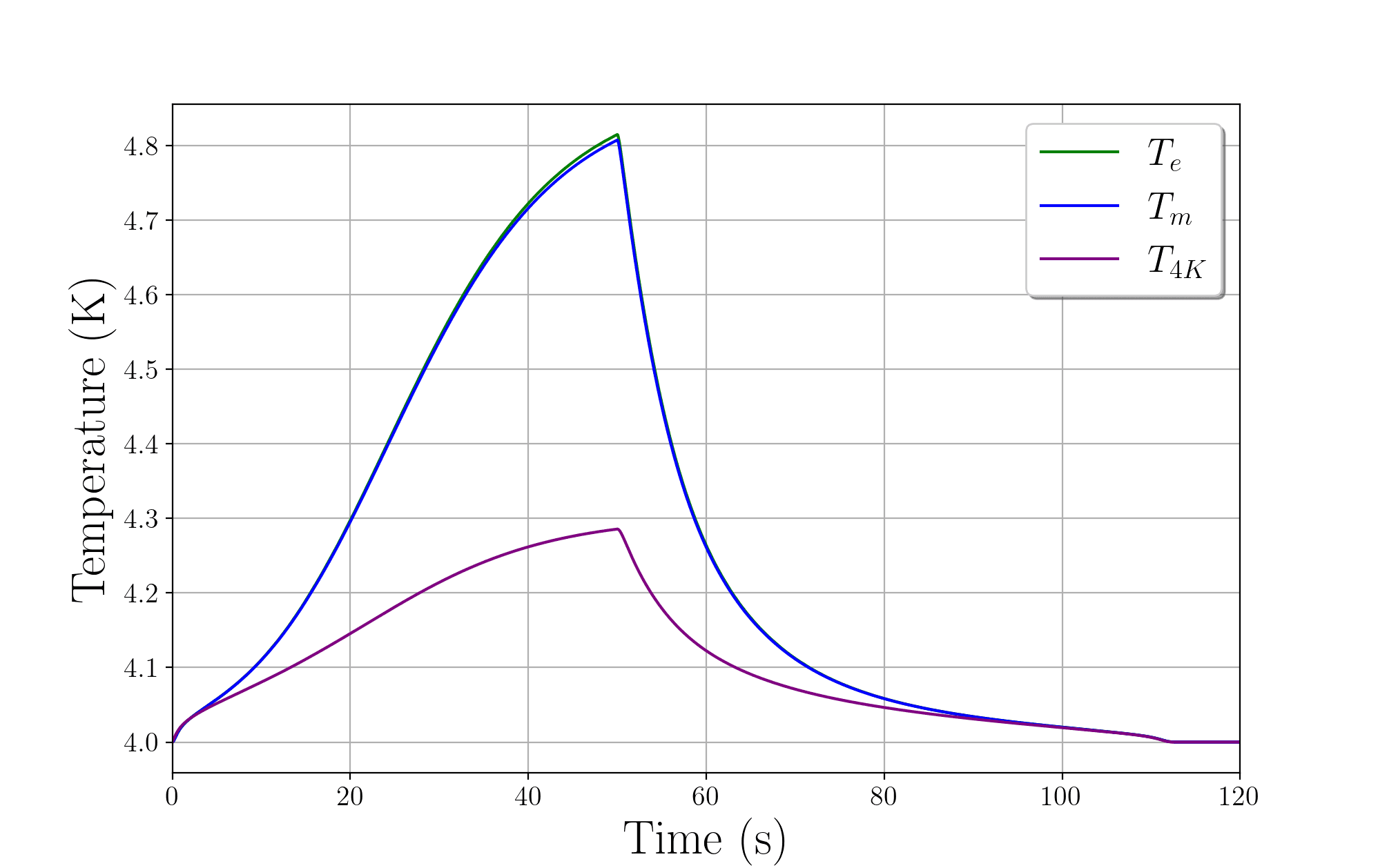}
\caption[Simulation of Oven Design 3]{(Left) The simulated temperature of the stainless steel wire, $T_{\text{w}}(t)$, and oven tube, $T_{\text{t}}(t)$, as a function of time for oven Design 3. (Right) The temperature of the trap electrodes, $T_{\text{e}}(t)$, trap mount, $T_{\text{m}}(t)$, and 4 K stage, $T_{\text{4\,K}}(t)$, as a function of time for oven Design 3. A current of $1.8\text{ A}$ is supplied to the circuit for 50 s and then turned off, simulating the time it takes to routinely load an ion. The final temperatures for the stainless wire, oven tube, trap electrodes, and trap mount are 537 K, 481 K, 4.81 K, 4.80 K, and 4.29 K respectively. The temperatures of each lump take $55\text{ s}$ to re-equilibrate to 4 K.}
\label{fig:simulation-of-oven-design-3}
\end{figure}

This system of coupled ordinary differential equations - \eqref{eq:stainless-steel-wire-energy-balance}, \eqref{eq:stainless-steel-tube-energy-balance}, \eqref{eq:trap-electrode-energy-balance}, \eqref{eq:trap-mount-energy-balance}, and \eqref{eq:4K-stage-energy-balance}- can be solved numerically using the $\text{RKF}45$ method to obtain $T_{\text{w}}(t)$, $T_{\text{t}}(t)$, $T_{\text{e}}(t)$, $T_{\text{m}}(t)$, and $T_{\text{4\,K}}(t)$. We assume the initial temperatures of the all the lumps in the model to be at 4 K. To routinely load an ion, current is supplied to the oven circuit for $\approx50\text{ s}$ so a piecewise function for this current is used in the numerical method, 
\begin{equation}
    I(t)=\begin{cases}
1.8\text{ A} & 0\text{ s}<t<50\text{ s}\\
0 & \text{other times}
\label{eq:piecwise-function-for-numerical-method-oven}
\end{cases}
\end{equation}
The plots of $T_{\text{w}}(t)$, $T_{\text{t}}(t)$, $T_{\text{e}}(t)$, $T_{\text{m}}(t)$, and $T_{\text{4\,K}}(t)$ for oven Design 3 are shown in Figure \ref{fig:simulation-of-oven-design-3}. 

The results of the simulation show that the oven tube heats up to 481 K in 50 s using $1.8\text{ A}$ of current. An ion is routinely loaded within a coupled of seconds after the current to the oven is turned off. We will assume the temperature of the oven tube and thus the Yb atomic beam to load an ion is $\approx481\text{ K}$. The temperature gradient between the stainless steel wire and the oven tube is due to the assumed contact conductance of the weld in the thermal model. This is consistent with the observation of hotter stainless steel wires in the oven test setup in Figure \ref{fig:oven-design-1-3-plating-test-results}. The simulated temperature increase of the trap mount during loading is $800\text{ mK}$ while the measured temperature increase is $1.7\text{ K}$ confirmed by a temperature sensor heat sunk to the trap mount. The temperature of increase of the trap electrodes is estimated to be $800\text{ mK}$ according to the simulation. The simulated temperature increase of the 4 K stage is $300\text{ mK}$ which is more consistent with the measured temperature increase $250\text{ mK}$ confirmed by a temperature sensor heat sunk to the 4 K stage. Additionally, I measured an equilibration time of 30 minutes, not $\sim1 \text{ min}$ as in the simulation. 

There are limitations in this model that make it hard to predict the final temperatures of each lump. The thermal contact conductances for the welds were estimated to allow for temperature gradients within the oven circuit, but the exact conductance is unknown. The copper wire, PEEK holder, and trap mount are all assumed to be thermally continuous to the 4 K stage but there really is an unknown contact conductance. Additionally, the emissivities for the trap mount, stainless steel wire, and oven tube were assumed to be 1 but are unknown. Lastly, while the lumped-element approximation is appropriate for the oven and stainless-steel heater wire, it breaks down for the much larger 4 K stage. The stage exhibits a thermal equilibration time of approximately 30 minutes, indicating that heat diffusion within the structure is comparable to the transient heating timescale. As a result, the temperature cannot be assumed to be spatially uniform, and the transient response must instead be modeled by solving the heat equation. The purpose of this thermal model and simulation was to understand how the oven tube heats up and how it increases the temperatures of the 4 K stage and trap mount as observed empirically. 

\section{Number Density at Trap Center}
\label{sec:number-density-at-trap-center}

Using the temperature of the Yb atomic beam to be $481\text{ K}$, we can estimate the number density, $n(T)$ in $\text{m}^{3}$, at the center of the trap that is present when loading. The flux of Yb atoms out of the oven tube is estimated to be $\Phi(481\text{ K})=10^{16.3}\frac{\text{Atoms}}{\text{m}^{2}\text{-s}}$ using Equation \eqref{eq:vaporization-rate}. The flux on the sphere $r$ away from the oven tube is $\phi(r,\theta)=\phi_{0}(r)\cos\theta$  where $\theta$  is the angle to the atomic beam axis\footnote{The flux for an effusive source varies as $\cos\theta$.} and $\phi_{0}(r)$ is the flux along the atomic beam axis. We can then relate $\Phi(T)$ to $\phi_{0}(r)$ by applying conservation of flux out of the oven tube onto the half sphere $\Phi(T)A_{\text{oven}}=\phi_{0}(r)2\pi r^{2}\int_{0}^{\frac{\pi}{2}}\cos\theta\sin\theta d\theta=\phi_{0}(r)\pi r^{2}$ where $A_{\text{oven}}$ is the inner diameter of the oven tube. We can now calculate the number density of the atoms at the center of the trap as follows,
\begin{equation}
    n(T)=\frac{\phi_{0}(r_{\text{t}})}{\bar{v}(T)}=\frac{\Phi(T)A_{\text{oven}}}{\bar{v}(T)\pi r_{\text{t}}^{2}}
    \label{eq:number-density-vs-atomic-flux}
\end{equation}
where $r_{\text{t}}=27\text{ mm}$ is the distance from the oven tube to the center of the trap, $\phi_{0}(r_{\text{t}})$ is the flux at the center of the trap, and $\bar{v}(T)=\sqrt{\frac{8k_{\text{B}}T}{\pi m}}$ is the mean speed of the Yb atoms in the atomic beam. The number density is estimated to be $n(481\text{\,K})=2.0\times10^{4}\text{ }\frac{\text{Atoms}}{\text{cm}^{3}}$. This is comparable to predicted number densities sufficient to achieve ion loading within a few seconds \cite{gaoOpticallyheatedAtomicSource2020,lucasIsotopeselectivePhotoionizationCalcium2004}. 

If the oven were mounted at the 40 K stage, the distance from the oven tube to the trap center would be $r_{\text{t}}=60\text{ mm}$ - a factor of 2.2 lower than that at the 4 K stage. Consequently, the oven tube and atomic beam would need to be at $502\text{ K}$ to generate the same number density at the trap center. If the oven were mounted at the 300 K stage, the distance from the oven tube to the trap center would be $r_{\text{t}}=115\text{ mm}$ - a factor of 4.3 than that at the 4 K stage. Consequently, the oven tube and atomic beam would need to be at $522\text{ K}$ to generate the same number density at the trap center. In conclusion, Design 3 of the oven could be used with currents in the range of 1.8-3 A at each stage of the cryostat because the target temperatures are only within 40 K. This is due to the fact that the vaporization rate, $\Phi(T)$, is a strong exponential function of temperature even though the number density scales as $\frac{1}{r_{\text{t}}^{2}}$.

The number density can also be estimated using Equation \eqref{eq:number-density-vs-photon-count-rate} by measuring the photon count rate due to scattering of the $^{1}\text{S}_{0}-^{1}\text{P}_{1}$ transition of an ensemble of $^{174}\text{Yb}$ atoms 
\begin{equation}
    n(T)=\frac{C}{R_{\text{scat}}\eta}
    \label{eq:number-density-vs-photon-count-rate}
\end{equation}
where $C$ is the photon count rate, $R_{\text{scat}}\sim\frac{\Gamma}{2}\frac{s}{1+s+(\frac{2\Delta}{\Gamma})^{2}}$ is the scattering rate for a single atom, and $\eta$  is the total detection efficiency.\footnote{Here the total detection efficiency includes the efficiency of the PMT, the losses in the optics, and the collection efficiency of the in-vacuum objective.} The temperature of the atomic beam could then be estimated by using Equation \eqref{eq:number-density-vs-atomic-flux}. This was not measured but would be the easiest way to confirm the temperature without installing a thermocouple on the oven tube.

\section{Oven Lifetime}
\label{sec:oven-lifetime}

Every time the oven is run, mass of Yb in the oven tube is lost. The initial mass of the Yb inside the oven tube is $\approx50\text{ mg}$ which is $2\times10^{20}\text{ Atoms}$. The rate of Yb atoms leaving the oven at $481\text{ K}$ is $\Phi(481\text{\,K})A_{\text{oven}}=1.2\times10^{11}\text{ }\frac{\text{Atoms}}{\text{s}}$. Thus, the lifetime of the oven running continuously at 481 K is $\approx500\text{ years}$.

\ifSubfilesClassLoaded{%
  \bibliographystyle{unsrt}
  \bibliography{Dissertation}
}{}
\end{document}

%% file: Chapter7.tex
\chapter{Imaging and Detection}
\label{chap:imaging-and-detection}

A combined imaging and detection system consisting of an Electron-Multiplying Charge-Coupled Device (EMCCD), colloquially known as a camera, and a photomultiplier tube (PMT) is essential in trapped ion experiments as the two detectors serve fundamentally different roles. While the camera provides spatially resolved fluorescence imaging that is critical for initial alignment, diagnostic testing, debugging, trap loading, and monitoring single ions or ion chains, the PMT provides high-speed photon counting with extremely low dark noise enabling high fidelity qubit state detection. Consequently, both cameras and PMTs are incorporated in most if not all trapped ion experiments.

Designing an imaging and detection system for trapped ion fluorescence at cryogenic temperatures places competing demands on numerical aperture (NA), working distance, and mechanical integration. A high NA objective is essential for maximizing photon collection efficiency and achieving sub-micron spatial resolution, yet cryogenic architectures such as radiation shields impose strict geometric constraints that force large working distances $>$ 35 mm for objectives placed outside the vacuum chamber \cite{paganoCryogenicTrappedionSystem2018a,todaroScalableStateDetection2020,mickeClosedcycleLowvibration42019}. Even if a NA of 0.5 is attainable at these working distances, objectives with outer diameters of $>$ 40 mm will be required and sending ion fluorescence through radiation shield and chamber windows introduces significant spherical aberration that can be difficult to correct without multi-element optical stacks which can be bulky and expensive. 

These constraints motivate the use of in-vacuum single-element objectives with working distances in the range of 5-20 mm in an infinite-conjugate configuration. At these working distances, single-element high NA objectives with outer diameters $<$ 40 mm can be positioned inside the radiation shields so that light exits the cryostat already collimated and largely free of window-induced aberrations \cite{dubielzigUltralowvibrationClosedcycleCryogenic2021,brandlCryogenicSetupTrapped2016a,spiveyHighStabilityCryogenicSystem2022}. However, integrating brittle optical substrates (e.g., fused silica, sapphire, N-BK7) into a cryogenic environment is nontrivial, as differential thermal contraction between materials can induce mechanical stress, leading to cracking or strain-induced birefringence that introduces polarization-dependent aberrations. Additionally, high NA single-element objectives suffer from a small diffraction-limited field of view and depth of focus demanding tight axial positioning tolerances of $<1\text{ \ensuremath{\mu}m}$ to minimize geometric and defocus aberration. As a result, in-vacuum piezo actuators (Smaract, JPE, Attocube) that can provide step sizes of $\sim1\text{ nm}$ are required, but are limited to $\approx400\text{ K}$ bakeout temperatures and necessitate additional wiring to the 4 K stage. It is also important to note that if a high NA $>$ 0.5 and a large diffraction-limited field of view of $500\text{ \ensuremath{\mu}m}$ is desired for imaging and detection of hundreds of ions in a cryogenic Paul trap \cite{paganoCryogenicTrappedionSystem2018a} or Penning trap \cite{ballSiteresolvedImagingBeryllium2019}, a multi-element in-vacuum objective would be required to cancel the aberrations across the field of view. Lastly, decreasing the working distance too far comes at the cost of potentially increasing dielectric-induced heating - the ion motional heating due to proximity to a dielectric \cite{tellerHeatingTrappedIon2021}. Together, these factors define the engineering motivation and constraints behind the development of a robust imaging and detection system.

\section{Design of Optical System}
\label{sec:design-of-optical-system}

\begin{figure}[!!h]
\centering
\includegraphics[width=\textwidth]{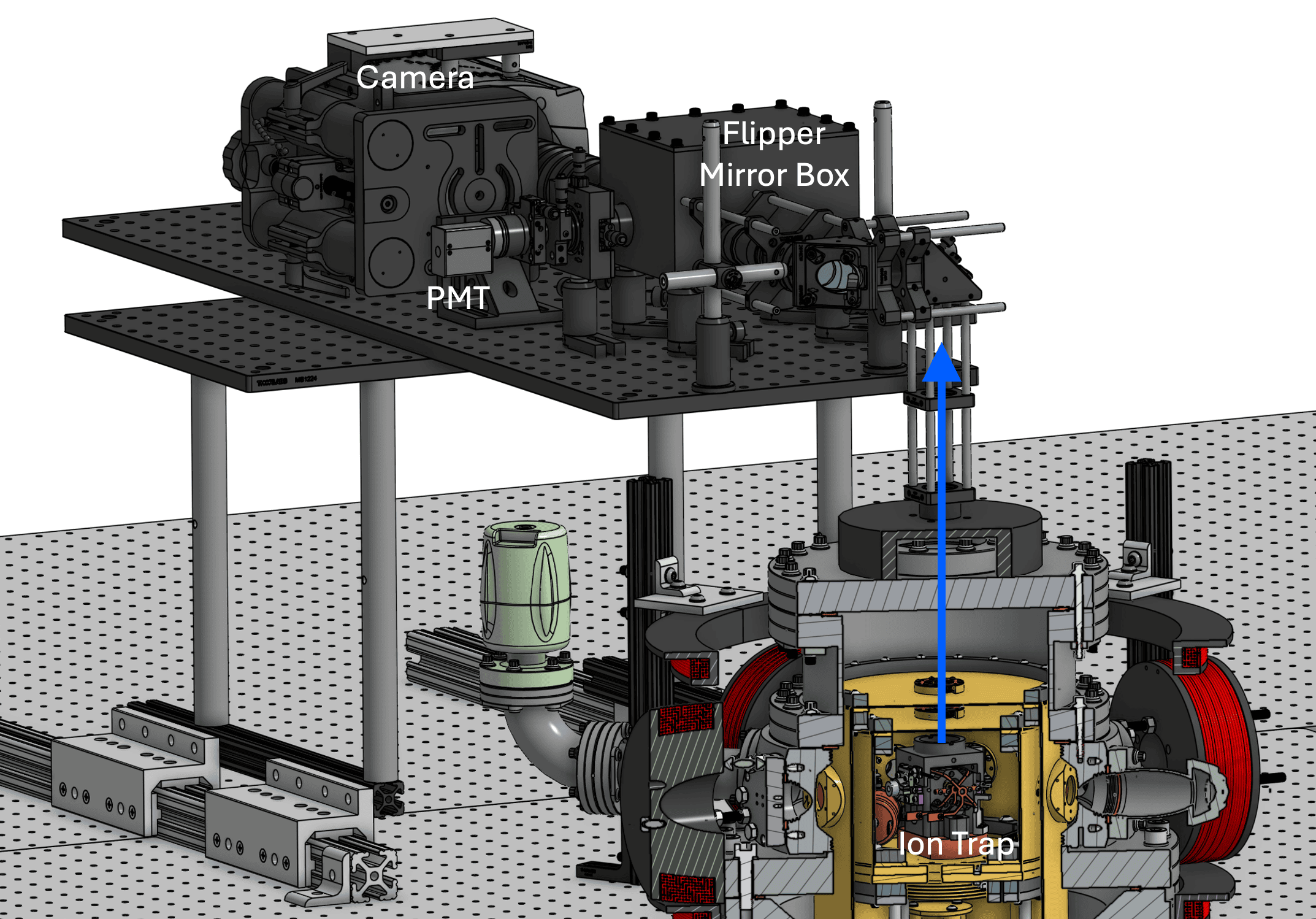}
\caption[Imaging and Detection System Assembly]{CAD drawing showing how the imaging and detection optical system interfaces with the cryostat. The blue arrow depicts the ion fluorescence collected by the in-vacuum objective.}
\label{fig:imaging-and-detection-system-assmebly}
\end{figure}

In this work, an in-vacuum aspheric lens (Edmund Optics, PN: 84-338) was positioned vertically above the center of the trap in an adjustable lens tube, described in in Subsection \ref{subsec:in-vacuum-aspheric-lens-optomechanics}, with a working distance of 19.052 mm. The lens is a finite corrected objective where the ion is positioned 1.09 mm away from the focus of the lens forming a real image 559 mm away from the lens outside the cryostat on an adjustable optics breadboard. At this intermediate image, an iris is inserted to prevent stray light and scattered light off of the trap electrodes to propagate through the rest of the optical system - this is known as a Lyot stop. At this point, the ion image can be either relayed to the PMT for detection or the camera for imaging. For the PMT, the light at the intermediate image is collimated with a plano-convex spherical lens (Thorlabs, PN: LA4148-A-ML) with a focal length of 50 mm, then directed to another plano-convex spherical lens (Thorlabs, PN: LA4380-A) with a focal length of 100 mm via a flipper mirror where it is imaged at the PMT with a magnification of $\approx46$. There is also a 369 nm bandpass filter (Semrock, PN: FF01-370/6-25) just before the PMT to block stray light from other lasers sent to the ion and ambient light from the lab. For the camera, the light at the intermediate image is collimated with a plano-convex spherical lens (Thorlabs, PN: LA4148-A-ML) with a focal length of 50 mm, then directed to another plano-convex spherical lens (Thorlabs, PN: LA4158-A) with a focal length of 250 mm where the light is imaged at the camera with a magnification of $\approx100$. Another 369 nm bandpass filter is placed just before the camera. CAD drawings of the imaging and detection system, in-vacuum objective optomechanics, and flipper mirror box are displayed in Figures \ref{fig:imaging-and-detection-system-assmebly}, \ref{fig:in-vacuum-aspheric-lens-assembly}, and \ref{fig:flipper-mirror-box-assembly} respectively. A schematic showing the ray tracing from the ions to the camera is displayed in Figure \ref{fig:finite-conjugate-imaging-ray-tracing}.

\subsection{In-vacuum Aspheric Lens Optomechanics}
\label{subsec:in-vacuum-aspheric-lens-optomechanics}

\begin{figure}[!!h]
\centering
\includegraphics[width=0.59\textwidth]{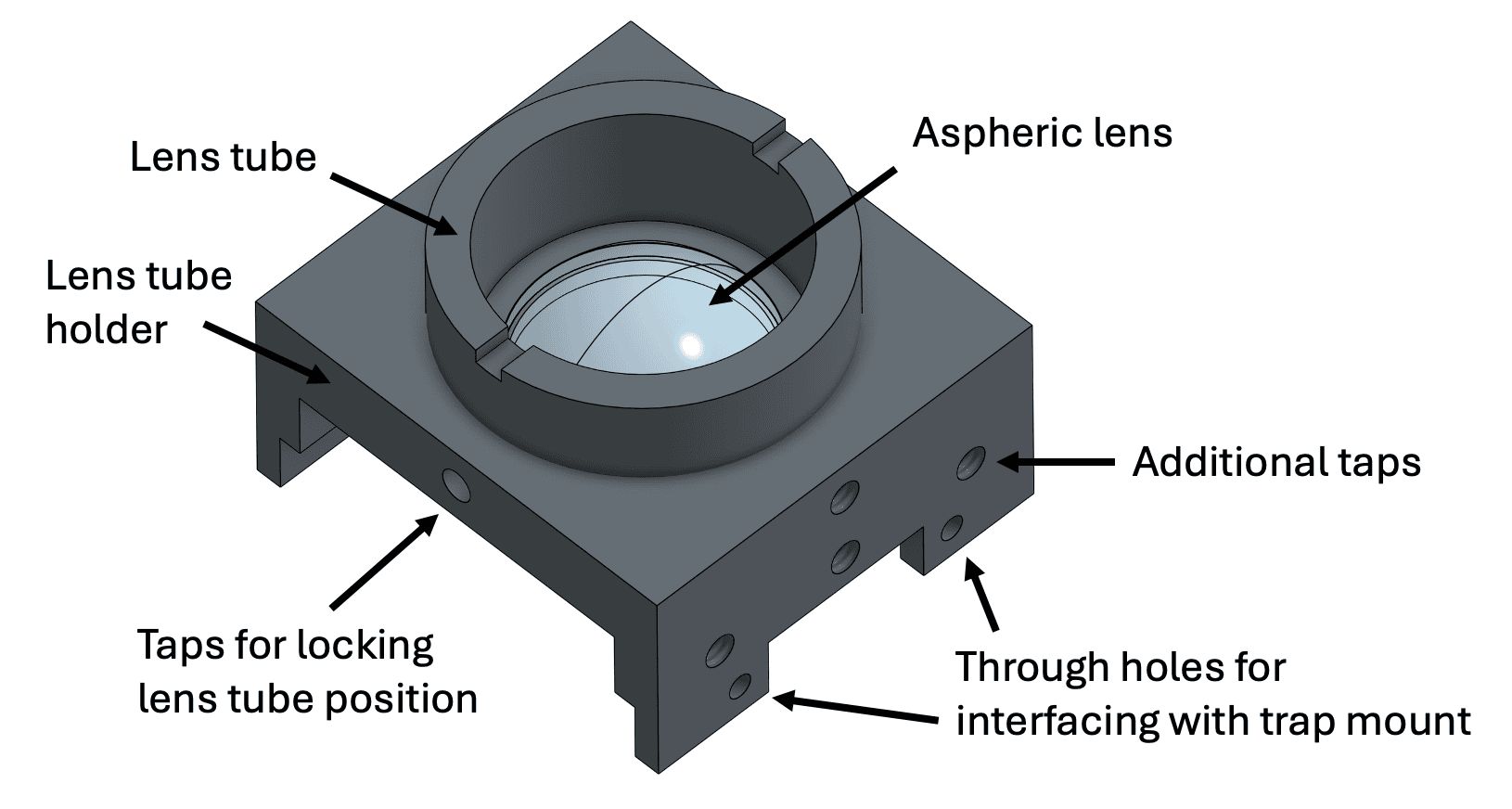}
\includegraphics[width=0.39\textwidth]{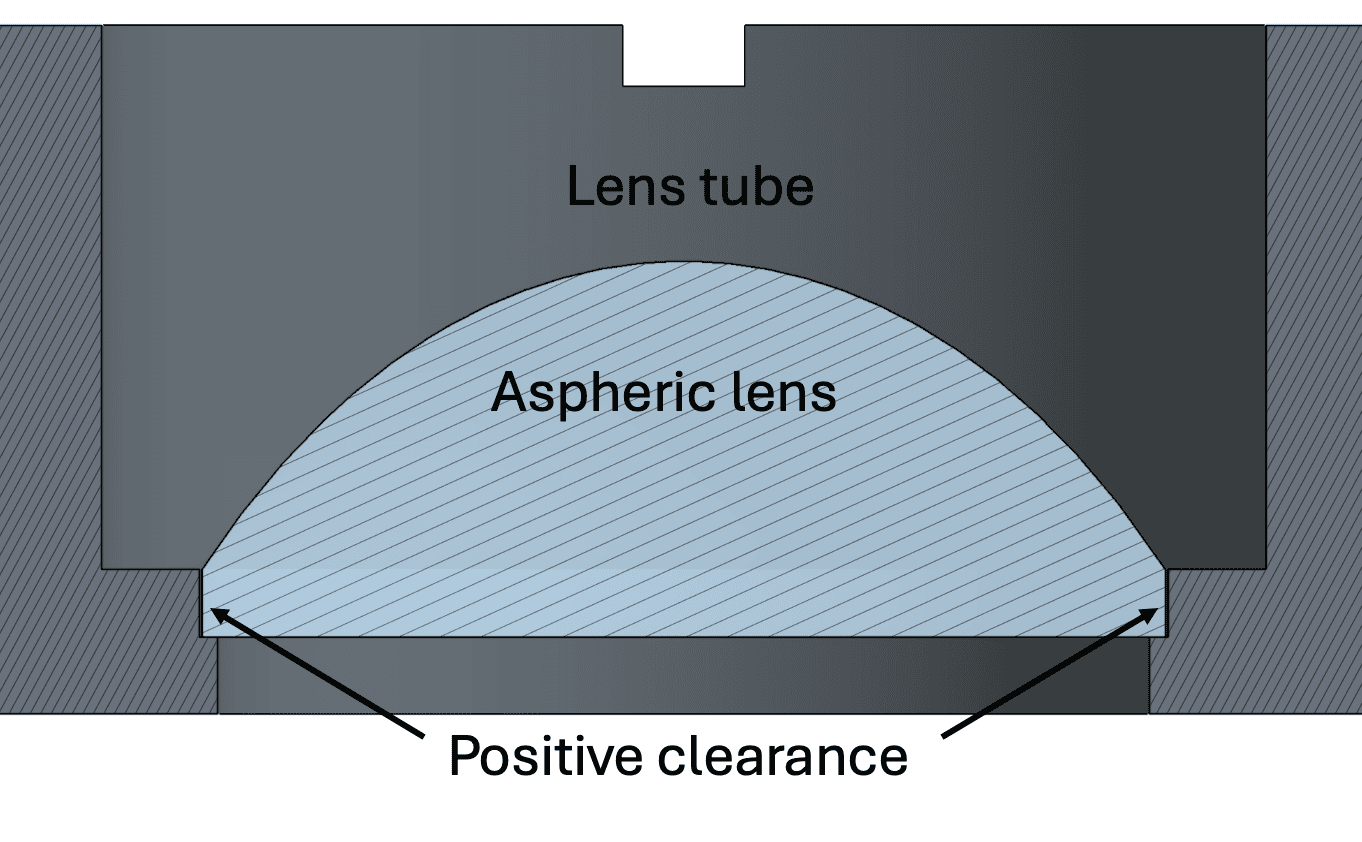}
\caption[In-Vacuum Aspheric Lens Assembly]{Aspheric lens tube and holder made out of grade 2 titanium. The external threads on the lens tube are 1.5-12 UNF. (Left) CAD drawing of asphere held in lens tube with holder. (Right) Asphere mounted in lens tube showing the clearance between the inner sidewalls of lens tube and asphere.}
\label{fig:in-vacuum-aspheric-lens-assembly}
\end{figure}

The aspheric lens is mounted in a lens tube machined with external threads that mates with a holder attached to the trap mount such that the position of the lens tube relative to the ions can be adjusted as shown in Figure \ref{fig:in-vacuum-aspheric-lens-assembly}. The lens tube thread pitch is 1.5-12 UNF resulting in a 2.12 mm change in working distance per revolution of the lens tube - or $130\text{ \ensuremath{\mu}m}$ per 1/16 revolution. The lens tube and holder were machined out of titanium (grade 2) because it is a non-magnetic material\footnote{The relative permeability of grade 2 titanium is $\mu_{\text{r}}=\frac{\mu}{\mu_{0}}(1+\chi)=1.00005$.} and so that there is minimal coefficient of thermal expansion (CTE) mismatch between the holder and trap mount (machined out of titanium grade 5). The asphere is held in place on the lens tube by gravity instead of clamping it since there is a large CTE mismatch between titanium and fused silica (the substrate of the asphere). To account for this CTE mismatch the inner diameter of the lens tube was machined such that there is a 0.001 in positive clearance between the lens tube inner wall and asphere at cryogenic temperatures. The thermal contractions of the titanium inner diameter and the asphere outer diameter are computed using the following equation, $D_{\text{5\,K}}=D_{\text{300\,K}}(1+\epsilon)$, where $D_{\text{300\,K}}$ and $D_{\text{5\,K}}$ are the diameters of the material at 300 K and 5 K respectively and $\epsilon=\frac{\Delta L}{L}=\int_{\text{5\,K}}^{\text{300\,K}}\alpha(T)dT$ is the strain of the material with thermal expansion coefficient $\alpha(T)$ from 300 K to 5 K. The strains of titanium and fused silica from 300 K to 5 K are $\epsilon_{\text{Ti}}=-1.51\times10^{-3}$ and $\epsilon_{\text{FS}}=-5.14\times10^{-5}$ respectively \cite{ekinExperimentalTechniquesLowtemperature2006a,okajiLaserInterferometricDilatometer1995}. Lastly, the position of the lens tube relative to the ions is locked with \#8-32 stainless steel bolts as shown in Figure \ref{fig:in-vacuum-aspheric-lens-assembly}. The lens tube and holder were designed at UMD and machined by ZeroHour Parts.

\subsection{Numerical Aperture and Depth of Focus}
\label{subsec:numerical-aperture-and-depth-of-focus}

\begin{figure}[!!h]
\centering
\includegraphics[width=0.75\textwidth]{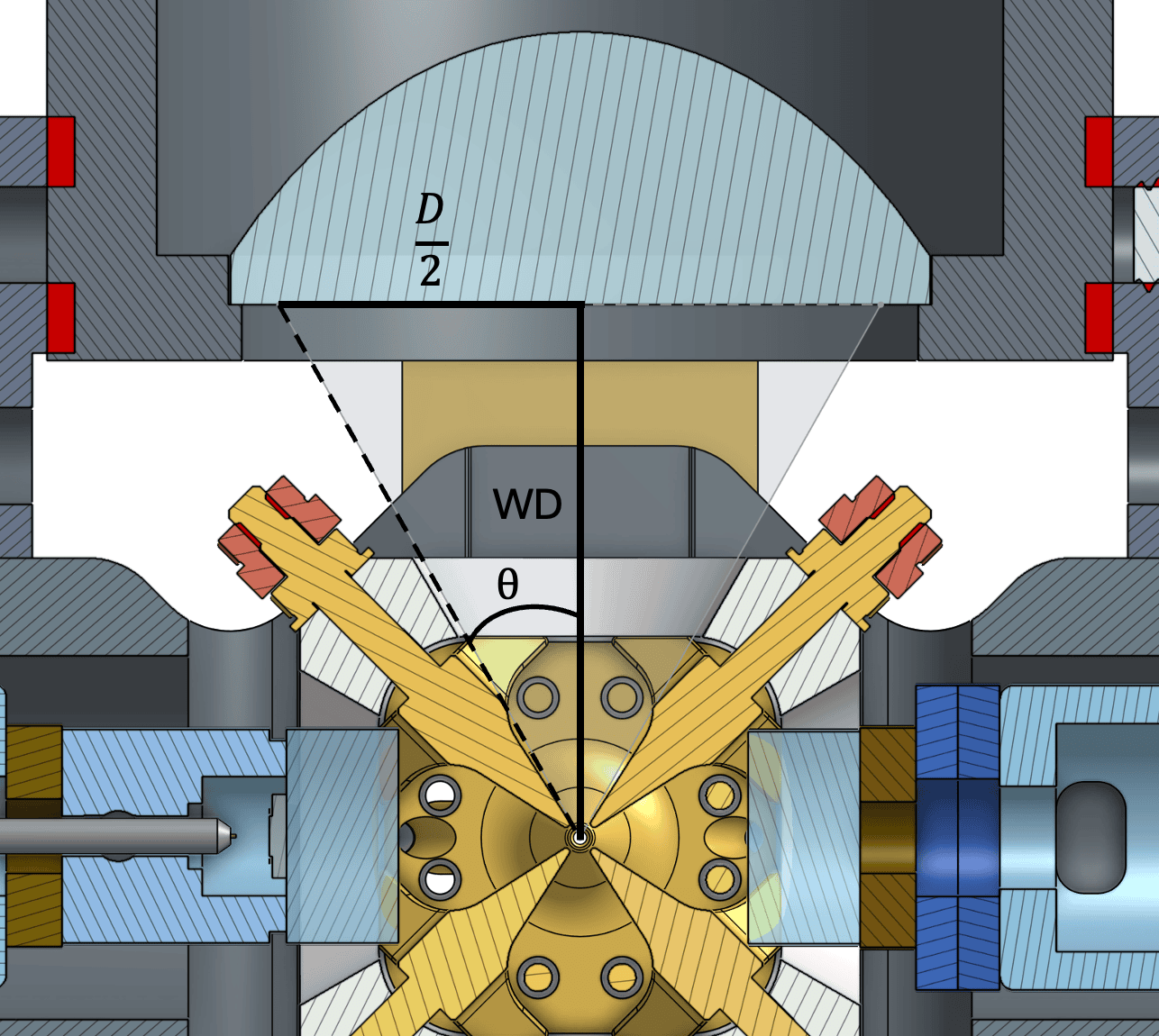}
\caption[Numerical Aperture of In-Vacuum Aspheric Lens]{Cross sectional view of the ion trap showing the NA of the in-vacuum objective. The NA is constrained by the effective focal length of the lens and the aspect ratio of the ion trap. The working distance is $\mathit{WD}=19.052\text{ mm}$ and the stop diameter is $D=21.480\text{ mm}$ resulting with a numerical aperture of $\mathit{NA}=0.49.$}
\label{fig:numerical-aperture-of-in-vacuum-aspheric-lens}
\end{figure}

\begin{figure}[!!h]
\centering
\includegraphics[width=0.5\textwidth]{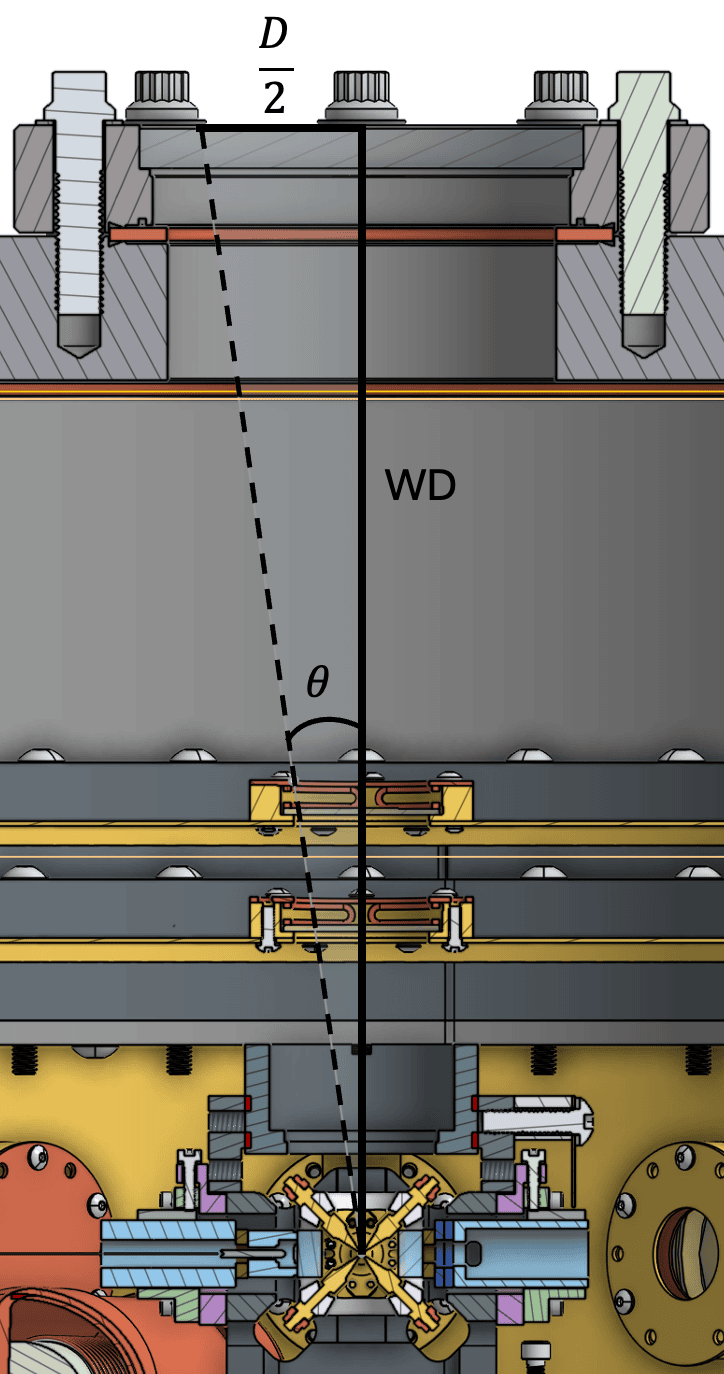}
\caption[Numerical Aperture to Ions for Imaging and Detection Without In-Vacuum Optics]{Cross sectional view of the ion trap showing the NA of the objective if it was placed just outside the cryostat window. Here the NA is constrained by the sizes of the radiation shield and chamber windows. The working distance is $\mathit{WD}=184\text{ mm}$ and the stop diameter is $D=53\text{ mm}$ resulting in an $\mathit{NA}=0.21$.}
\label{fig:numerical-to-ions-without-in-vacuum-optics}
\end{figure}

The numerical aperture (NA) for the in-vacuum objective is calculated using the working distance and the stop diameter. The working distance $(\mathit{WD})$ is defined as the distance from the object (the ion at the center of the trap) to the first physical surface of the lens (the flat, plano surface). The stop diameter is defined as the aperture diameter of the lens that limits the marginal rays as seen from the ion. The marginal ray is the ray that forms the maximum possible ray angle with respect to the optical axis that gets collected by the lens. The aperture diameter in this case is limited by the aspect ratio of the trap blades as shown in Figure \ref{fig:numerical-aperture-of-in-vacuum-aspheric-lens}. The numerical aperture is estimated to be $\mathit{NA}=0.49$ using the following equation 
\begin{equation}
    \mathit{NA}=n_{\text{vac}}\sin\theta=n_{\text{vac}}\frac{D/2}{\sqrt{\mathit{WD}^{2}+(D/2)^{2}}}
    \label{eq:numerical-aperture}
\end{equation}
where the index of refraction of vacuum is $n_{\text{vac}}=1$, the stop diameter is $D=21.480\text{ mm}$, and the working distance is $\mathit{WD}=19.052\text{ mm}$. The stop diameter here is less than the clear aperture of 22.5 mm specced by the manufacturer so we can guarantee that light from the ion gathered by the aspheric lens given an $\mathit{NA}=0.49$ will form an image. The diffraction-limited NA for this geometry was not calculated for this work, but can be calculated using non-paraxial ray tracing analysis tools like Zeemax. However, assuming that the NA for this geometry is diffraction-limited we can estimate the diffraction-limited ion spot size by calculating the Airy disk radius - a fundamental wave-optics limit independent of imperfections in the lens and aberrations. When light with wavelength $\lambda$  goes through a lens with a circular aperture with some NA, the intensity distribution at the focal plane (the Point Spread Function (PSF)) produced by the circular aperture is known as the Airy pattern. The PSF has a central peak with the intensity dropping to zero at a distance given by 
\begin{equation}
    r_{\text{airy}}=0.61\frac{\lambda}{\mathit{NA}}
    \label{eq:airy-disc-radius}
\end{equation}
where the factor of 0.61 parametrizes the distance between the central intensity peak and the first zero. The Airy disk radius for a $\text{Yb}^+$ ion scattering light at wavelength $\lambda=369\text{ nm}$ that is imaged using a circular lens with an $\mathit{NA}=0.49$ is $r_{\text{airy}}=459\text{ nm}$ - i.e., the diffraction limited ion spot size. This Airy disk radius is also the minimum distance two $\text{Yb}^+$ ions can be separated for each ion to be `just resolved' in the image plane. However, if the objective were placed just outside the cryostat as shown in Figure \ref{fig:numerical-to-ions-without-in-vacuum-optics}, the working distance and stop diameter would increase to $\mathit{WD}=184\text{ mm}$ and $D=53\text{ mm}$ respectively resulting in a large outer diameter objective $>$ 53 mm, a maximum NA of 0.21, and a diffraction-limited spot size of $1\text{ \ensuremath{\mu}m}$.

Another fundamental wave-optics limit to consider here is the diffraction-limited depth of focus (DOF). If the ion is displaced along the optical axis away from the focal plane of the aspheric lens, there will be wavefront curvature error along the ion plane known as defocus aberration. The Rayleigh $\lambda/4$ criterion states that the image of the ion will remain diffraction limited as long as the defocus causes no more than $\lambda/4$ wavefront error across the ion plane - i.e., the maximum optical path length difference of the wavefront across the ion plane is $\leq\lambda/4$.\footnote{There is zero optical path length difference of the wavefront across the ion plane when there is zero defocus and the PSF reduces to the Airy pattern.} The DOF can be parameterized in terms of NA using the following equation \cite{salehFundamentalsPhotonics}
\begin{equation}
    \mathit{DOF}=\pm\frac{\lambda}{\mathit{NA}^{2}}
    \label{eq:depth-of-focus}
\end{equation}
For this objective with $\mathit{NA}=0.49$ collecting light from a $\text{Yb}^+$ ion scattering light at wavelength $\lambda=369\text{ nm}$, the DOF is $\pm1.5\text{ \ensuremath{\mu}m}$. The DOF establishes the necessary tolerance for axial positioning of the lens tube, ensuring that defocus aberration remains within acceptable limits. This objective was positioned in-vacuum to increase the NA and is only used for imaging and state detection while the 1650 nm photons are collected from the cavity into a single-mode (SM) optical fiber. However, for diffraction limiting imaging and/or for free-space collection of 1650 nm photons with this lens (without additional ambient correcting optics) to then be coupled into SM fibers for quantum networking, axial positioning less than the $\mathit{DOF}(1650\text{\,nm})=\pm6.87\text{ \ensuremath{\mu}m}$ would be required as defocus aberration can greatly effect coupling into SM fibers \cite{carterIonTrapInvacuum2024}. This resolution would require thread pitches of $\approx500\text{ \ensuremath{\frac{\text{threads}}{\text{in}}}}$ which will be expensive to machine and potentially challenging to integrate into a vacuum and cryogenic environment. There are vendors that can machine 3/16 in outer diameter screws with these fine pitches\footnote{kozakmicro.com.}, but this trajectory was not explored in this work. Of course, translational stages that have $1\text{ \ensuremath{\mu}m}$ resolution are routinely made by optics vendors such as Thorlabs and can be used for the ambient optics, but these are not vacuum compatible. Alternatively, vacuum and cryogenic compatible piezo stages (Smaract, JPE, Attocube) could be used to provide the fine axial positioning with step sizes of $\approx1\text{ nm}$ for these high NA lenses. 

\subsection{Flipper Mirror Box}
\label{subsec:flipper-mirror-box}

\begin{figure}[!!h]
\centering
\includegraphics[width=0.49\textwidth]{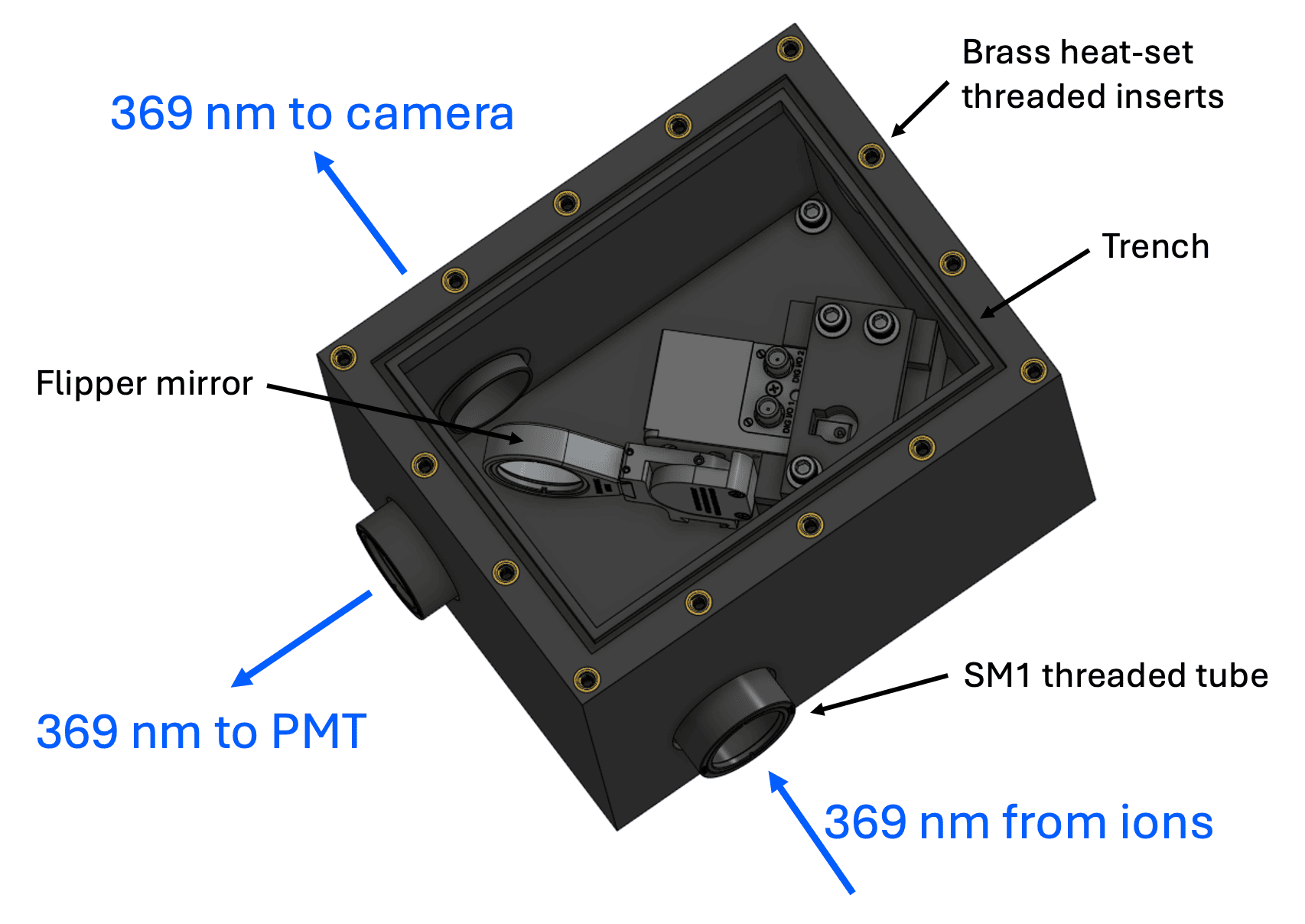}
\includegraphics[width=0.49\textwidth]{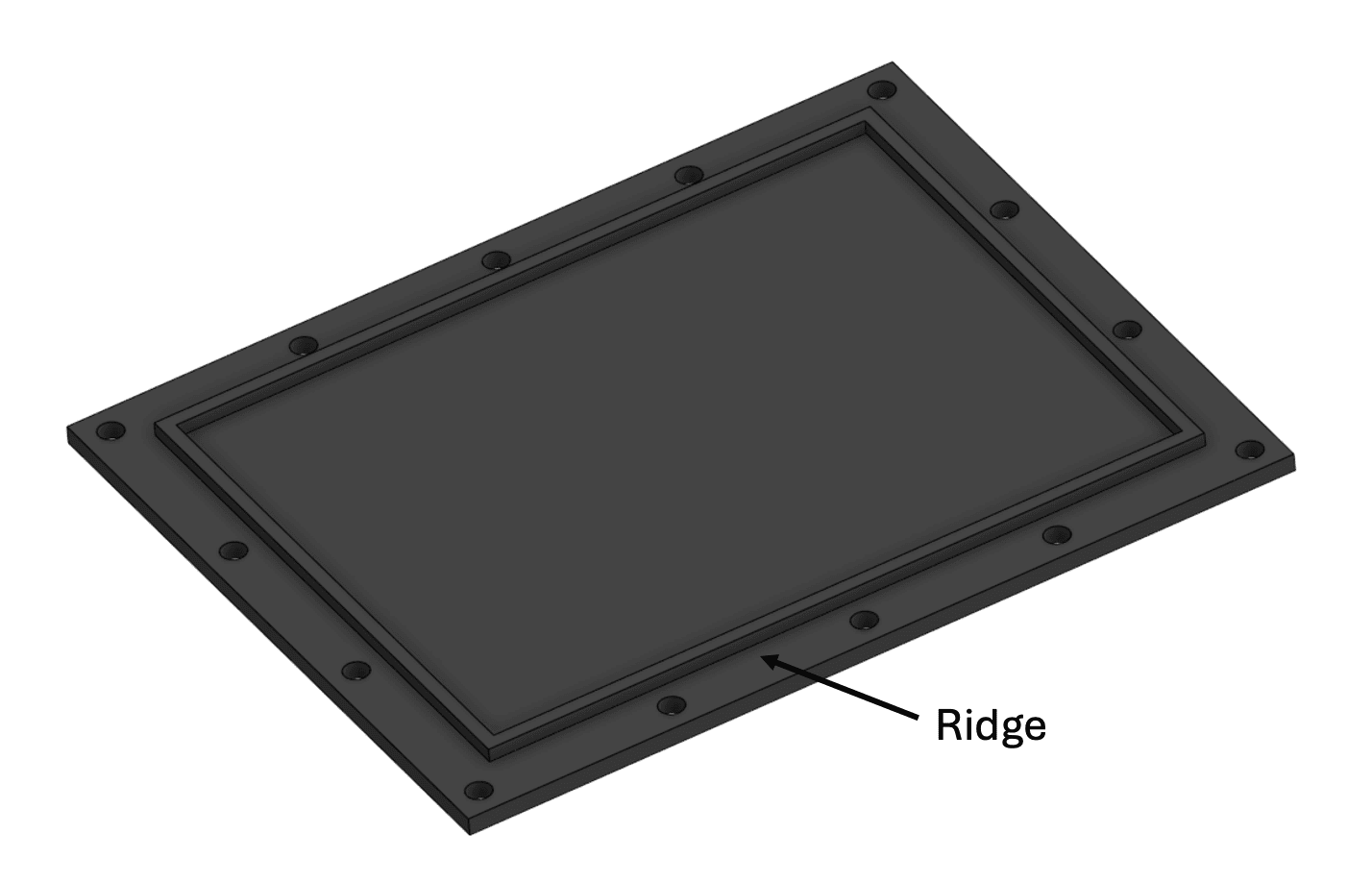}
\caption[Flipper Mirror Box Assembly]{(Left) Light-tight flipper mirror box 3D printed out of nylon. (Right) Lid of box also 3D printed out of nylon. Thorlabs SM1 threaded tubes are mounted to the box to interface with the imaging and detection optics. A light-tight seal is made between the box and lid that consists of a mating trench and ridge. The lid is bolted to the box with \#8-32 taps that mate with heat-set threaded inserts. The threaded inserts are positioned with high density around the trench to maintain the light-tight seal.}
\label{fig:flipper-mirror-box-assembly}
\end{figure}

To seamlessly switch between imaging and state detection, a flipper mirror (Thorlabs, PN: MFF101) is placed in the optical setup inside a light-tight box to preserve signal to noise. The flipper mirror box was 3D printed out of nylon at UMD as shown in Figure \ref{fig:flipper-mirror-box-assembly} and is integrated into the image and detection optical system as shown in Figures \ref{fig:finite-conjugate-imaging-ray-tracing} and \ref{fig:infinite-conjugate-imaging-ray-tracing}.

\section{Modeling of Optical System}
\label{sec:modeling-of-optical-system}

\subsection{Estimate focal length of aspheric lens}
\label{subsec:estimate-focal-length-of-aspheric-lens}

The effective focal length for this lens (Edmund Optics, PN: 84-338) is 25 mm at 587.6 nm, but is $<$ 25 mm at 369 nm due to chromatic aberration - $n\propto\frac{1}{\lambda}$. The index of refraction for Corning 7980 fused silica at 293 K - the substrate used for this lens - is $n(369\text{\,nm})=1.47397$. Using the convex surface radius for the lens of 11.462 mm we can estimate the focal length at 369 nm using the thick lens equation 
\begin{equation}
    f(\lambda)=\frac{1}{(n(\lambda)-1)}\left[\frac{1}{R_{1}}-\frac{1}{R_{2}}+\frac{(n(\lambda)-1)d}{n(\lambda)R_{1}R_{2}}\right]^{-1}
    \label{eq:thick-lens-equeation}
\end{equation}
where $R_{1}=11.462\text{\text{ mm}}$ is the convex surface radius of the lens, $R_{2}=\infty$ is the flat surface of the lens, and $d=9.75\text{ mm}$ is the center thickness \cite{hechtOptics2017}. Since $R_{2}=\infty$, Equation \eqref{eq:thick-lens-equeation} reduces to
\begin{equation}
    f(\lambda)=\frac{R_{1}}{n(\lambda)-1}
    \label{eq:simplified-thin-lens-equation}
\end{equation}
and the effective focal length is $f(369\text{\,nm})=24.183\text{ mm}$. A separate effect occurs which is that the index of refraction isotropically decreases as a function of temperature of the fused silica. The change in index of refraction due to a temperature change of $\Delta T=293\text{ K}-30\text{ K}$ for Corning 7980 fused silica at 400 nm is $\Delta n=-1.6\times10^{-3}$ \cite{levitonTemperaturedependentAbsoluteRefractive2006}. Using the convex surface radius $R_{1}=11.462\text{\text{ mm}}$ and the index of refraction at 293 K $n(369\text{\,nm})=1.47397$, the change in focal length is $\Delta f=\frac{-R_{1}}{(n(\lambda)-1)^{2}}\Delta n\approx80\text{ \ensuremath{\mu}m}$. So the net effect is that the effective focal length for 369 nm at 5 K is 24.264 mm.

\subsection{ABCD Matrices}
\label{subseq:ABCD-matrices}

To calculate the location and magnification of the image of the ions we use the ABCD matrix formalism for geometric optics. Here a light ray is described as a column vector, called the ray vector, that contains the height, $h$, and slope, $\theta$, relative to the optical axis. Optical elements in the optical system that effect the light ray are described by $\text{2\ensuremath{\times}2}$ matrices, called ray matrices. The optical element's effect on the light ray is found by multiplying by its ray vector as shown below in Equation \eqref{eq:general-abcd-ray-matrix}. This linear mapping is only possible assuming the slope, $\theta$, relative to the optical axis is small, known as the paraxial approximation. 

\begin{equation}
\begin{aligned}
    \left(\begin{array}{c}
h_{\text{f}}\\
\theta_{\text{f}}
\end{array}\right)&=\left(\begin{array}{cc}
A & B\\
C & D
\end{array}\right)\left(\begin{array}{c}
h_{\text{i}}\\
\theta_{\text{i}}
\end{array}\right) \\
&=\left(\begin{array}{cc}
\frac{\partial h_{\text{f}}}{\partial h_{\text{i}}} & \frac{\partial h_{\text{f}}}{\partial\theta_{\text{i}}}\\
\frac{\partial\theta_{\text{f}}}{\partial h_{\text{i}}} & \frac{\partial\theta_{\text{f}}}{\partial\theta_{\text{i}}}
\end{array}\right)\left(\begin{array}{c}
h_{\text{i}}\\
\theta_{\text{i}}
\end{array}\right)
\label{eq:general-abcd-ray-matrix}
\end{aligned}
\end{equation}
    
The optical elements considered for this system are listed below in Table \ref{tab:ray-matrices}. Modeling the aspheric lens using the thins ray matrix is sufficient here as we are only using the ABCD matrix formalism to calculate the image/object distances and spatial magnifications. To study the geometric (spherical, coma, astigmatic) aberrations as well as numerical aperture (NA) diffraction limit analysis we would need to use non-paraxial ray tracing analysis tools like Zeemax. Lastly, the important criterion we will assert here is that for finite object distances, it must be true that the image height does not depend on the ray input angle, i.e., $B=\frac{\partial h_{\text{f}}}{\partial\theta_{\text{i}}}=0$ and that the spatial magnification is $A=\frac{\partial h_{\text{f}}}{\partial h_{\text{i}}}$ \cite{salehFundamentalsPhotonics}.

\begin{table}[!!h]
\centering
\tablefont
\caption[Ray Matrices]{Table of relevant optical elements and their corresponding ray matrices \cite{salehFundamentalsPhotonics}.}
\label{tab:ray-matrices}
\begin{tabularx}{\textwidth}{>{\centering\arraybackslash}p{0.4\textwidth}cc}
\toprule
\textbf{Optical Element} & \textbf{Ray Matrix} & \textbf{Comments} \\
\midrule
Propagation, $M_{l}$ & $\left(\begin{array}{cc}
1 & l\\
0 & 1
\end{array}\right)$ & \makecell{Propagation in free pace\\or in a medium of constant\\index of refraction} \\
\midrule
Refraction at a Flat Interface, $M_{\text{int}}$ & $\left(\begin{array}{cc}
1 & 0\\
0 & \frac{n_{1}}{n_{2}}
\end{array}\right)$ & \makecell{$n_{1}=\text{initial refractive index}$\\$n_{2}=\text{final refractive index}$} \\
\midrule
Reflection from a flat Mirror, $M_{\text{m}}$ & $\left(\begin{array}{cc}
1 & 0\\
0 & 1
\end{array}\right)$ & \makecell{Valid for flat mirrors\\oriented at any angle to the\\incoming beam. Both the ray\\and the optical axis are reflected equally,\\so there is no change in height or slope} \\
\midrule
Thin Lens, $M_{f}$ & $\left(\begin{array}{cc}
1 & 0\\
-\frac{1}{f} & 1
\end{array}\right)$ & \makecell{$f=\text{focal length of lens}$\\$f>0\text{ for convex/positive converging lens}$} \\
\bottomrule
\end{tabularx}
\end{table}

\subsection{Finite Conjugate Imaging of Ions}
\label{subsec:finite-conjugate-imaging-of-ions}

The focus of the in-vacuum aspheric lens was intentionally positioned $\approx1\text{ mm}$ away from the trap center so that we could image the ions and determine the object distance at 369 nm. Once the object distance is determined, the position of the lens can be adjusted accordingly when the cryostat is at 300 K such that the focus of the lens is aligned to the trap center at $5\text{ K}$ within $\pm130\text{ \ensuremath{\mu}m}$ - infinite conjugate configuration. The real image location was measured to be $559\text{ mm}$. Using the ABCD matrix formalism, we estimated that the object distance is $25.371\text{ mm}$ and the magnification is $-22$. The object distance is greater than the focal length which is consistent with the fact that a real image of the ions is formed and the minus sign in the magnification indicates that the image is inverted. 

\begin{figure}[!!h]
\centering
\includegraphics[width=\textwidth]{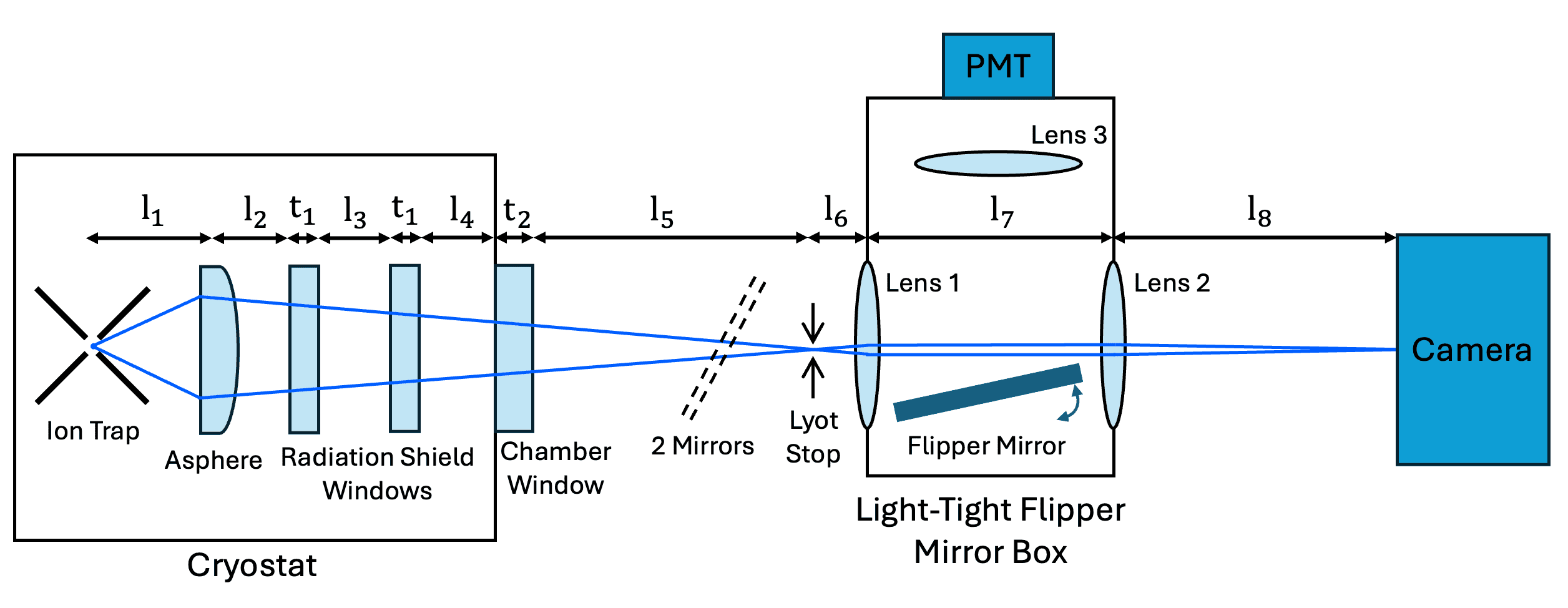}
\caption[Finite Conjugate Imaging Ray Tracing Diagram]{Finite conjugate imaging and detection configuration showing marginal ray tracing from the ion to the camera - schematic is not to scale. The two mirrors positioned between the cryostat and the Lyot stop are depicted with dashed lines for convenience.}
\label{fig:finite-conjugate-imaging-ray-tracing}
\end{figure}

The location of this image was used as the position of the Lyot stop in the imaging and detection system. Here the image is either relayed to the PMT for detection or to the camera for imaging as shown in Figure \ref{fig:finite-conjugate-imaging-ray-tracing}. We can compute the effect on a ray vector from the ion, $\left(\begin{array}{c}
h_{\text{i}}\\
\theta_{\text{i}}
\end{array}\right)$, through the optical system to calculate the location and magnification of the image at the PMT or camera. The ray propagation is shown below for imaging at the camera where the ray matrices for the mirrors are not included as they have no effect on the calculation.

\begin{equation}
\begin{aligned}
    \left(\begin{array}{c}
h_{\text{f}}\\
\theta_{\text{f}}
\end{array}\right)&=M_{l_{8}}M_{f_{2}}M_{l_{7}}M_{f_{1}}M_{l_{6}}M_{l_{5}}M_{\text{int}_{\text{FS-air}}}M_{t_{2}}M_{\text{int}_{\text{vac-FS}}}M_{l_{4}}M_{\text{int}_{\text{FS-vac}}} \\
&M_{t_{1}}M_{\text{int}_{\text{vac-FS}}}M_{l_{3}}M_{\text{int}_{\text{FS-vac}}}M_{t_{1}}M_{\text{int}_{\text{vac-FS}}}M_{l_{2}}M_{f_\text{asphere}}M_{l_{1}}\left(\begin{array}{c}
h_{\text{i}}\\
\theta_{\text{i}}
\end{array}\right) \\
&=M_{\text{tot}}\left(\begin{array}{c}
h_{\text{i}}\\
\theta_{\text{i}}
\end{array}\right)
\label{eq:abcd-ray-matrix-finite-conjugate}
\end{aligned}
\end{equation}
Here, the focal lengths of the asphere, lens 1, and lens 2 are $f_{\text{asphere}}=24.264\text{ mm}$, $f_{1}=50\text{ mm}$, and $f_{2}=250\text{ mm}$ respectively. The thicknesses $t_{1}$ and $t_{2}$ represent the propagation distances through radiation shield window and chamber window respectively. The ray matrix representing light going through an interface from medium 1 to medium 2 is given as $M_{\text{int}_{\text{1-2}}}$. Once the total ray matrix $M_{\text{tot}}$ for the optical system is computed, we can impose the constraint that $B=\frac{\partial h_{\text{f}}}{\partial\theta_{\text{i}}}=0$ and solve for the unknown $l_{8}$ and compute the spatial magnification $A=\frac{\partial h_{\text{f}}}{\partial h_{\text{i}}}$. For this optical system, the distance between the flipper mirror box and camera is $l_{8}=200.467\text{ mm}$, the image location relative to the in-vacuum aspheric lens is 1011.510 mm, and the total magnification at the camera is 101. The same analysis can be done for the ray propagation from the ions to the PMT.

\subsection{Infinite Conjugate Imaging of Ions}
\label{subsec:infinite-conjugate-imaging-of-ions}

\begin{figure}[!!h]
\centering
\includegraphics[width=\textwidth]{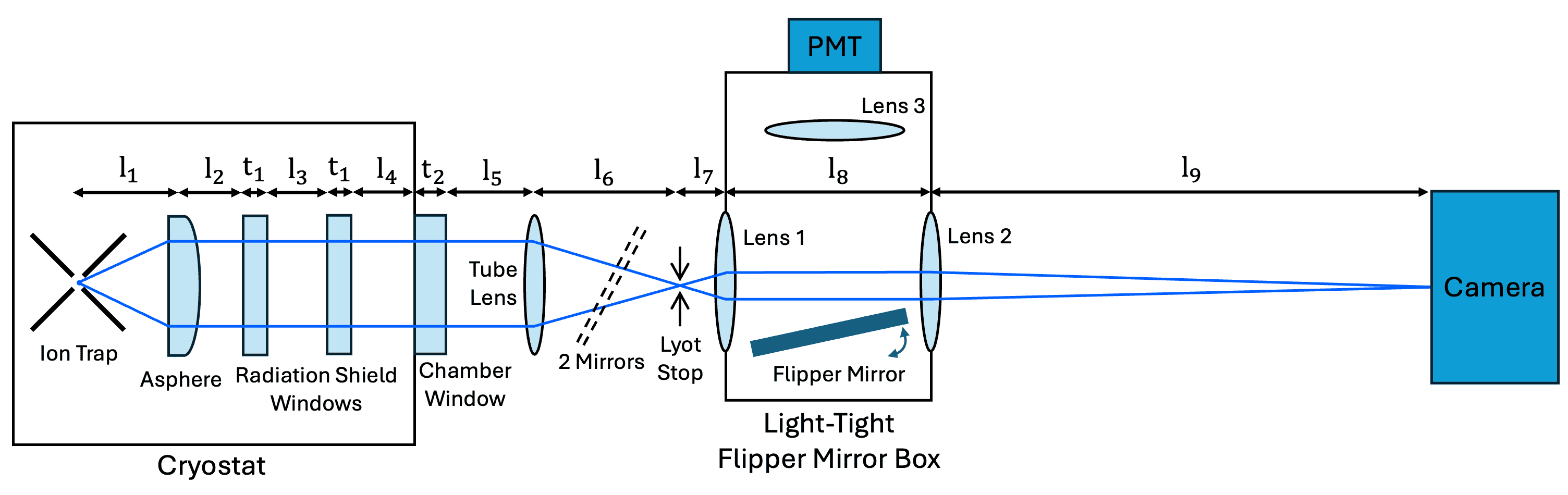}
\caption[Infinite Conjugate Imaging Ray Tracing Diagram]{Infinite conjugate imaging and detection configuration showing marginal ray tracing from the ion to the camera - schematic is not to scale. The two mirrors positioned between the cryostat and the Lyot stop are depicted with dashed lines for convenience.}
\label{fig:infinite-conjugate-imaging-ray-tracing}
\end{figure}

The in-vacuum objective position was never moved during this work, but the imaging and detection system was designed to accommodate infinite conjugate imaging. However, if the aspheric lens position is adjusted such that the focus of the lens is aligned to the trap center at 5 K within $130\text{ \ensuremath{\mu}m}$, an additional lens, called a tube lens, can be placed outside of the cryostat at a distance from the position of the Lyot stop equal to the focal length of the lens. Thus the position of the flipper mirror box doesn't need to change as shown in Figure \ref{fig:infinite-conjugate-imaging-ray-tracing}. The ray propagation is shown below for imaging at the camera where the ray matrices for the mirrors are not included as they have no effect on the calculation.

\begin{equation}
\begin{aligned}
    \left(\begin{array}{c}
h_{\text{f}}\\
\theta_{\text{f}}
\end{array}\right)&=M_{l_{9}}M_{f_{2}}M_{l_{8}}M_{f_{1}}M_{l_{7}}M_{l_{6}}M_{f_{\text{tube}}}M_{l_{5}}M_{\text{int}_{\text{FS-air}}}M_{t_{2}}M_{\text{int}_{\text{vac-FS}}}M_{l_{4}}M_{\text{int}_{\text{FS-vac}}} \\ 
&M_{t_{1}}M_{\text{int}_{\text{vac-FS}}}M_{l_{3}}M_{\text{int}_{\text{FS-vac}}}M_{t_{1}}M_{\text{int}_{\text{vac-FS}}}M_{l_{2}}M_{f_{\text{asphere}}}M_{l_{1}}\left(\begin{array}{c}
h_{\text{i}}\\
\theta_{\text{i}}
\end{array}\right) \\
&=M_{\text{tot}}\left(\begin{array}{c}
h_{\text{i}}\\
\theta_{\text{i}}
\end{array}\right)
\label{eq:abcd-ray-matrix-infinite-conjugate}
\end{aligned}
\end{equation}
Here the focal lengths of the asphere, tube lens, lens 1, and lens 2 are $f_{\text{asphere}}=24.264\text{ mm}$, $f_{\text{tube}}=300\text{ mm}$, $f_{1}=50\text{ mm}$, and $f_{2}=500\text{ mm}$ respectively. The thicknesses $t_{1}$ and $t_{2}$ represent the propagation distances through radiation shield window and chamber window respectively. The ray matrix representing light going through an interface from medium 1 to medium 2 is given as $M_{\text{int}_{\text{1-2}}}$. Once the total ray matrix $M_{\text{tot}}$ for the optical system is computed, we can impose the constraint that $B=\frac{\partial h_{\text{f}}}{\partial\theta_{\text{i}}}=0$ and solve for the unknown $l_{9}$ and compute the spatial magnification $A=\frac{\partial h_{\text{f}}}{\partial h_{\text{i}}}$. For this configuration, the distance between the flipper mirror box and camera is $l_{9}=500\text{ mm}$, the image location relative to the aspheric lens 1311.040 mm and the total magnification at the camera is 124. There is enough room on the imaging breadboard for the camera to be translated another 300 mm away from the flipper mirror box. The same analysis can be done for the ray propagation from the ions to the PMT. 

\subsection{Spatial Resolution of Imaging System}
\label{subsec:spatial-resolution-of-imagin-system}

\begin{figure}[!!h]
\centering
\includegraphics[width=0.75\textwidth]{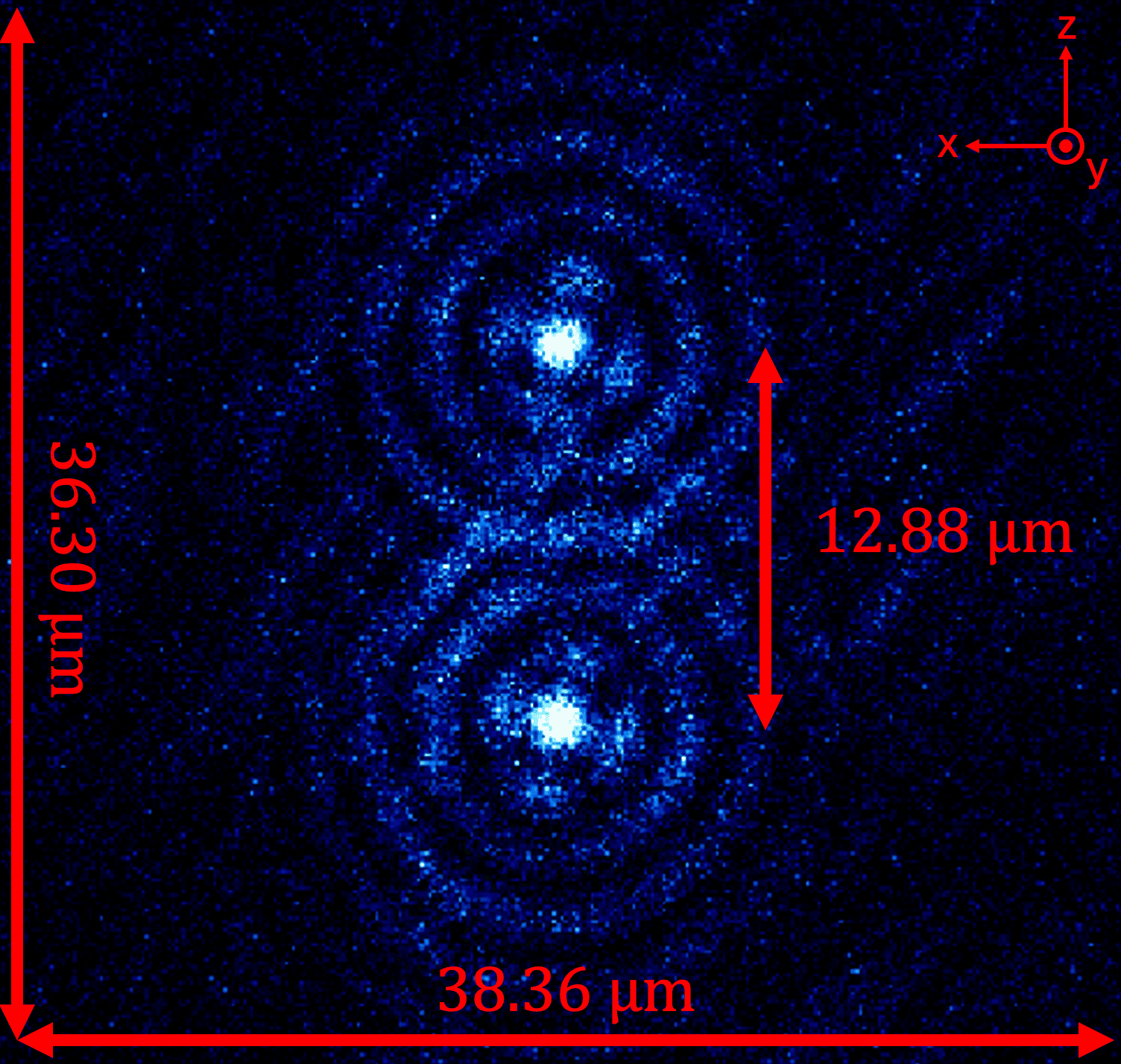}
\caption[Spatial Resolution of Imaging System]{Camera image of a two ion chain with COM mode $\omega_{z}=2\pi\times155.9\text{ kHz}$ giving a separation of $z_{\text{min}}(2\text{\,ions})=12.88\text{ \ensuremath{\mu}m}$. The field of view in this image is $298\text{ px}\times282\text{ px} (38.36\text{ \ensuremath{\mu}m}\times36.30\text{ \ensuremath{\mu}m})$. The maximum field of view not shown here is $1024\text{ px}\times1024\text{ px} (131.80\text{ \ensuremath{\mu}m}\times131.80\text{ \ensuremath{\mu}m})$.}
\label{fig:camera-image-of-two-ions-endcap-95p21-V}
\end{figure}

We calculated the spatial resolution of the imaging system by relating the real space distance between two trapped $^{174}\text{Yb}^+$ ions to the distance in pixels of the camera image. Numerical calculations of the minimum inter-ion spacing for different ion chain sizes is given by the following equation
\begin{equation}
    z_{\text{min}}(N)=\left(\frac{e^{2}}{4\pi\epsilon_{0}m\omega^{2}}\right)^{1/3}\frac{2.018}{N^{0.559}}
    \label{eq:minimum-inter-ion-spacing}
\end{equation}
where $N$ is the number of ions in the chain, $\omega$  is the axial center of mass (COM) mode frequency in $\text{s}^{-1}$, $m$ is the mass of an ion in kg, $e=1.6\times10^{-19}\text{ C}$, and $\epsilon_{0}=8.85\times10^{-12}\text{ }\frac{\text{C}^{2}-\text{s}^{2}}{\text{kg}-\text{m}^{3}}$ is the permittivity of free space \cite{jamesQuantumDynamicsCold1998}. The spatial resolution and magnification are calculated in Table \ref{tab:spatial-resolution-of-imgaging-system} using two $^{174}\text{Yb}^+$ ions loaded in the trap with an endcap voltage of 95.21 V. Camera images of the two ions are presented in Figure \ref{fig:camera-image-of-two-ions-endcap-95p21-V}. The magnification of 101 computed here is consistent with the magnification calculated using the ABCD matrices in Section \ref{subsec:finite-conjugate-imaging-of-ions}. 

\begin{table}[!!h]
\centering
\tablefont
\caption[Spatial Resolution of Imaging System]{Parameters used to calculate the spatial resolution and magnification of the imaging system. Two $^{174}\text{Yb}^+$ ions are trapped with an endcap voltage of $95.21\text{ V}$ which corresponds to a COM mode frequency of $\omega_{z}=2\pi\times155.9\text{ kHz}$ using Equation \eqref{eq:axial-secular-frequency}. The minimum real space inter-ion distance of $12.88\text{ \ensuremath{\mu}m}$ is calculated using Equation \eqref{eq:minimum-inter-ion-spacing} and compared to the inter-ion distance of 100 px at the image plane of the camera yields a spatial resolution of $129\text{ }\frac{\text{nm}}{\text{px}}$. The magnification of 101 is calculated by comparing the spatial resolution to the $13\text{ \ensuremath{\mu}m}$ pixel size of Andor iXon Ultra camera.}
\label{tab:spatial-resolution-of-imgaging-system}
\begin{tabularx}{\textwidth}{>{\centering\arraybackslash}p{0.49\textwidth}>{\centering\arraybackslash}p{0.49\textwidth}}
\toprule
\textbf{Parameter} & \textbf{Value} \\
\midrule
Endcap Voltage (V) & 95.21 \\
\midrule
Axial COM Mode $\frac{\omega}{2\pi}$ (kHz) & 155.9 \\
\midrule
$z_{\text{min}}(2\text{ ions}) (\text{\ensuremath{\mu}m})$ & 12.88 \\
\midrule
$z_{\text{min}}(2\text{ ions}) (\text{px})$ & 100 \\
\midrule
Spatial Resolution $(\frac{\text{nm}}{\text{px}})$ & 129 \\
\midrule
Pixel Size $(\text{\ensuremath{\mu}m})$ & 13 \\
\midrule
Magnification & 101 \\
\bottomrule
\end{tabularx}
\end{table}

\subsection{Ion Image Analysis}
\label{subsec:ion-image-analysis}

\begin{figure}[!!h]
\centering
\includegraphics[width=0.75\textwidth]{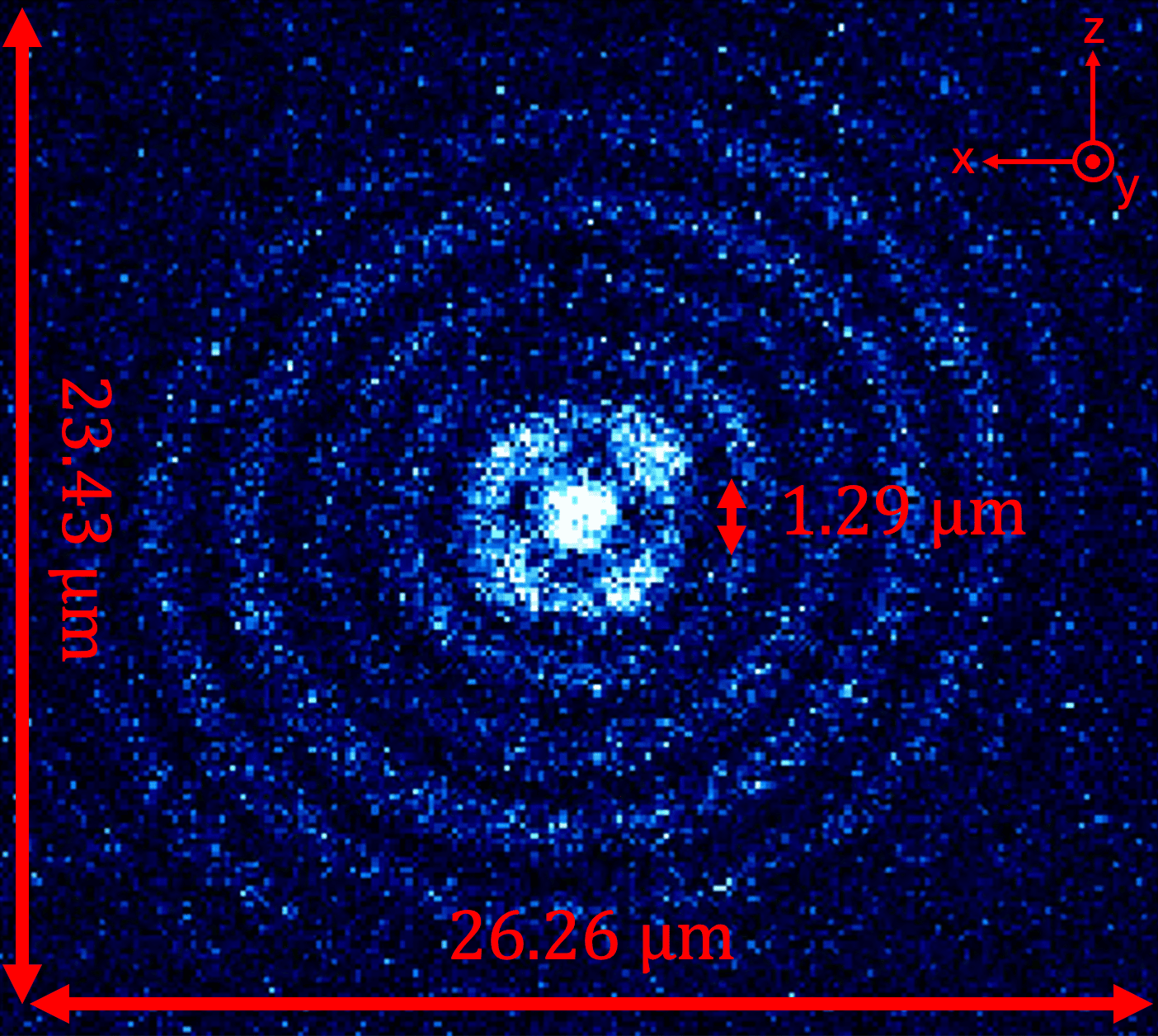}
\caption[Camera Image Analysis]{Camera image of a single ion with a field of view of $204\text{ px}\times182\text{ px}$ and a spatial resolution of $129\text{ }\frac{\text{nm}}{\text{px}}$. The ion is imaged 1.109 mm away from best focus, far outside the $\pm1.5\text{ \ensuremath{\mu}m}$ diffraction-limited depth of field. As a result, the PSF exhibits a pronounced ring surrounding the central intensity peak, most likely characteristic of strong defocus with no definitive signatures of coma, astigmatism, or spherical aberration. The measured distance between the central maximum and the first intensity minimum is approximately $1.29\text{ \ensuremath{\mu}m}$, which sets the effective minimum resolvable separation for this finite-conjugate imaging configuration.}
\label{fig:camera-image-analysis}
\end{figure}

The quality of the ion image is evaluated in this section by examining the point-spread function (PSF) measured with the camera shown in Figure \ref{fig:camera-image-analysis}. We know that the ion is displaced from best focus by 1.109 mm, which is far larger than the diffraction-limited depth of field of $\pm1.5\text{ \ensuremath{\mu}m}$ so we expect that the PSF should not resemble an Airy pattern. This is confirmed by the measured distance between the central peak and the first intensity minimum, which is $1.29\text{ \ensuremath{\mu}m}$ - approximately three times larger than the ideal diffraction-limited Airy radius. In addition, the presence of a bright symmetric donut surrounding the core and several subsequent rings of comparable intensity most likely indicates a PSF dominated by strong defocus with no definitive signatures of coma, astigmatism, or spherical aberration. It is likely that these aberrations are small compared to the imposed defocus and cannot be quantitatively assessed from this measurement alone.

\subsection{Ion Position vs. Time}
\label{subsec:ion-position-vs-time}

The position of a single $^{174}\text{Yb}^+$ ion, the temperature of the cold stages, and the temperature of the imaging and detection breadboard were monitored over 5 hours. The 369 nm, 935 nm, and 638 nm lasers were kept on for the duration of the experiment to prevent the ion from going dark. Exposures were taken with the camera every 2 s monitoring the ion position vs. time along the $x,z$ axes as shown in Figure \ref{fig:camera-image-analysis}. The temperatures of the cold stages were sampled every 15 ms while the temperature of the imaging and detection breadboard was sampled every 10 s (the lowest sampling rate for the Elitech sensor, PN: RC0-5+). The drift along the $y$-axis manifests as a change in the PSF (measured by the camera) due to defocus aberration. A more detailed analysis of the PSF versus time would need to be done to infer the drift along the $y$-axis and was not explored in this work. However, the ion position in the $x,z$-plane was measured to vary $<400\text{ nm}$ (about 3 px) over 5 hours. The results of this experiment are summarized in Table \ref{tab:ion-position-vs-time}. 

\begin{table}[!!h]
\centering
\tablefont
\caption[Ion Position vs. Time]{Results of ion position vs. time experiment.}
\label{tab:ion-position-vs-time}
\begin{tabularx}{\textwidth}{>{\centering\arraybackslash}p{0.3\textwidth}cccccc}
\toprule
\textbf{Average Temperature of Trap Mount (K)} & \makecell{\textbf{Average Temperature of}\\ \textbf{Imaging and Detection Breadboard (K)}} & \makecell{\textbf{Ion Position Drift}\\\textbf{Along $\mathbf{x,z}$ Directions (nm)}}\\
\midrule
$5.87\pm0.01$ & $297.34\pm0.03$ & $<400$ \\
\bottomrule
\end{tabularx}
\end{table}

\subsection{End-to-End Detection Efficiency}
\label{subsec:end-to-end-detection-efficiency}

\begin{figure}[!!h]
\centering
\includegraphics[width=0.75\textwidth]{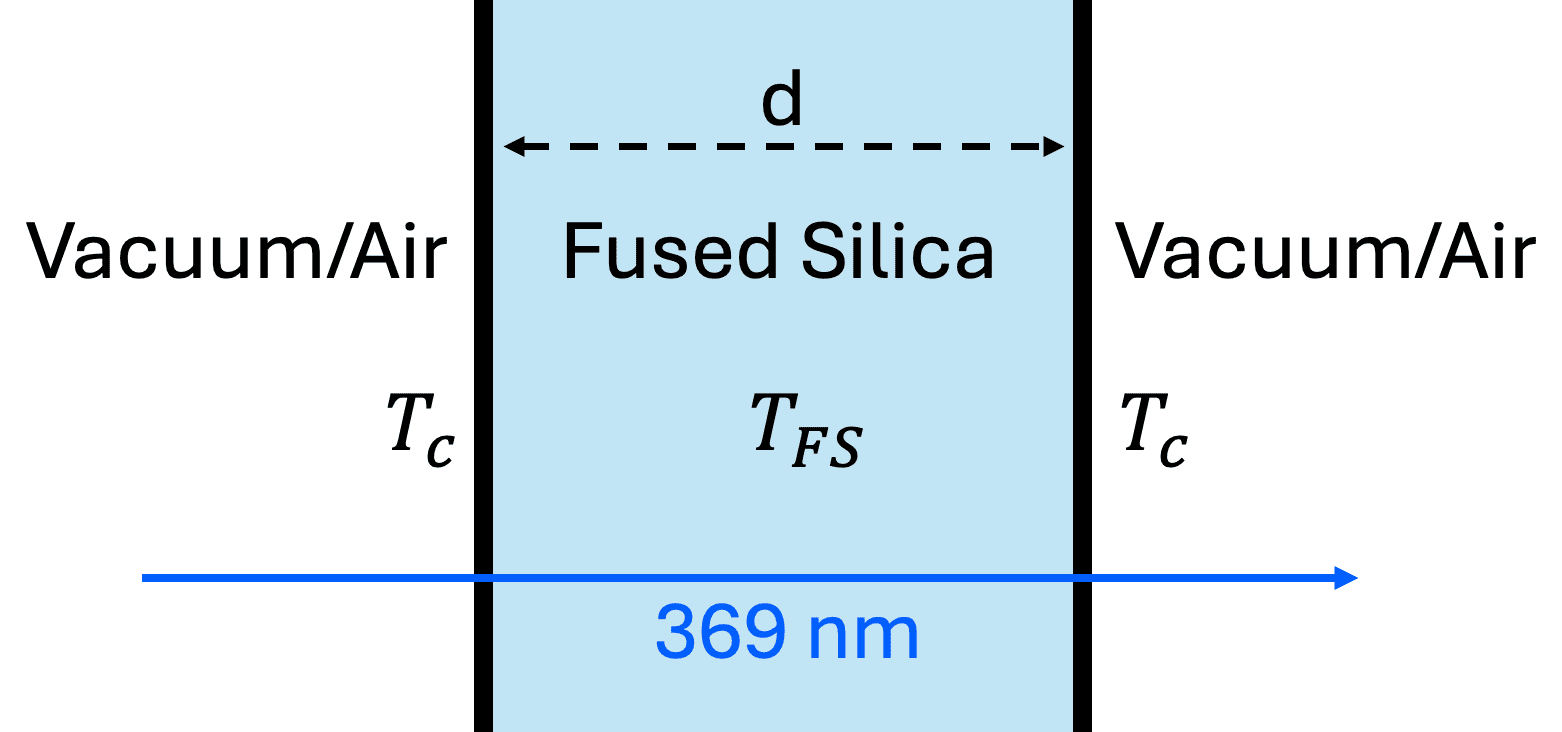}
\caption[Transmission Through Optically Coated Substrate]{Schematic of 369 nm light propagating through an optically coated fused silica substrate in vacuum or air where the coatings are marked with vertical solid lines. The total transmission through the optic is $T_{\text{tot}}=T_{\text{c}}^{2}T_{\text{FS}}$ where $T_{\text{c}}$ is the transmission through the optical coating and $T_{\text{FS}}=e^{-\mu d}$ is the transmissions through a Corning 7980 fused silica substrate with optical density $\mu(369\text{\,nm})=7.334\text{ m}^{-1}$ and thickness $d$.}
\label{fig:transmission-through-optically-coated-substrate}
\end{figure}

The end-to-end detection efficiency depends on the collection efficiency of the objective, the transmission through the optical system, and the quantum efficiency of the photomultiplier (PMT). The collection efficiency is estimated as follows,
\begin{equation}
\begin{aligned}
    \eta_{\text{c}}&=\frac{\Omega}{4\pi} \\
    &=\frac{1}{4\pi}\left[\int_{0}^{2\pi}\int_{0}^{\theta}\sin\theta d\theta d\phi\right] \\
    &=\frac{1}{2}\left[1-\cos(\arcsin(\text{NA}))\right]
    \label{eq:objective-collection-efficiency}
\end{aligned}
\end{equation}
where $\Omega$  is the solid angle subtended by a cone of half angle $\theta$  (the angle between the cone's axis and its side) and the numerical aperture is $\mathit{NA}=0.49$. Thus, the collection efficiency of the objective is $\eta_{\text{c}}=6.41\text{\%}$. The transmission through the optical system, $\eta_{\text{o}}$, is the net transmission through each optical element taking into account the coatings and optical substrates and is summarized in Table \ref{tab:total-transmission-through-optical-system}. The total transmission through the optically coated fused silica substrates in this optical system (windows and lenses) can be calculated using the following equation, $T_{\text{tot}}=T_{\text{c}}^{2}T_{\text{FS}}$, where $T_{\text{c}}$ is the transmission through the coating and $T_{\text{FS}}=e^{-\mu d}$ is the transmission through a fused silica substrate with absorption coefficient $\mu=\frac{4\pi k}{\lambda}$ and thickness $d$ using the Beer-Lambert law \cite{hechtOptics2017}, see Figure \ref{fig:transmission-through-optically-coated-substrate}. The absorption coefficient for Corning 7980 fused silica at 369 nm is $\mu(369\text{\,nm})=7.334\text{ m}^{-1}$.

\begin{table}[!!h]
\centering
\tablefont
\caption[Total Transmission Through Optical System]{Total transmission through the optical system, $\eta_{o}$.}
\label{tab:total-transmission-through-optical-system}
\begin{tabularx}{\textwidth}{>{\centering\arraybackslash}p{0.49\textwidth}>{\centering\arraybackslash}p{0.49\textwidth}}
\toprule
\textbf{Optical Element} & \textbf{Transmission (\%)} \\
\midrule
In-Vacuum Objective & 92.2 \\
\midrule
Radiation Shield Windows & 89.2 \\
\midrule
300 K Window & 94.4 \\
\midrule
3 Ambient Mirrors & 98.6\footnote{This is reflectance.} \\
\midrule
2 Ambient Lenses & 92.1 \\
\midrule
369 nm Bandpass Filter & 91 \\
\midrule
Total, $\eta_{o}$ & 64 \\
\bottomrule
\end{tabularx}
\end{table}

The quantum efficiency of the PMT is estimated using Equation \eqref{eq:quantum-efficiency-of-PMT} to be $\eta_{\text{QE}}=23\text{\%}$,
\begin{equation}
    \eta_{\text{QE}}=\frac{S_{369}}{\Phi_{\text{ph}}}
    \label{eq:quantum-efficiency-of-PMT}
\end{equation}
where the $S_{\text{369}}=4.3\times10^{17}\text{ }\frac{\text{counts}}{\text{s-W}}$ is the count sensitivity specified for the PMT (Hamamatsu, PN: H10682-210) at 369 nm and $\Phi_{\text{ph}}=\frac{1\text{W}}{hc/\lambda}=1.86\times10^{18}\text{ \ensuremath{\frac{\text{photons}}{\text{s}}}}$ is the photon flux per Watt where $h=6.626\times10^{-34}\text{ J-s}$ is Planck's constant, $c=3\times10^{8}\text{ }\frac{\text{m}}{\text{s}}$ is the speed of light and $\lambda=369\text{ nm}$ \cite{hamamatsuPhotomultiplierTubesBasics2006,foxQuantumOpticsIntroduction2006}. The end-to-end detection efficiency is estimated using Equation \eqref{eq:end-to-end-detection-efficiency} to be $\eta=0.8\text{\%}$.
\begin{equation}
    \eta=\eta_{\text{c}}\eta_{\text{o}}\eta_{\text{QE}}
    \label{eq:end-to-end-detection-efficiency}
\end{equation}

\subsection{Photon Count Rate and Signal to Noise}
\label{subsec:photon-count-rate-and-signal-to-noise}

The photon count rate at the PMT while scattering photons off the $^{2}\text{S}_{1/2}-^{2}\text{P}_{1/2}$ transition of a single $^{174}\text{Yb}^+$ ion is estimated to be $352\times10^{3}\text{ }\frac{\text{counts}}{\text{s}}$ using Equation \eqref{eq:photon-count-rate-at-PMT}, 
\begin{equation}
    C=\eta R_{\text{scat}}
    \label{eq:photon-count-rate-at-PMT}
\end{equation}
where $R_{\text{scat}}\sim\frac{\Gamma}{2}\frac{s}{1+s+(\frac{2\delta}{\Gamma})^{2}}=44\times10^{6}\text{ }\frac{\text{photons}}{\text{s}}$ is the scattering rate and $\eta=0.8\text{\%}$ is the detection efficiency from Equation \eqref{eq:end-to-end-detection-efficiency}. The natural linewidth of the $^{2}\text{S}_{1/2}-^{2}\text{P}_{1/2}$ transition is $\Gamma=2\pi\times19.6\text{ MHz}$, the saturation intensity is $I_{\text{sat}}=\frac{\pi hc\Gamma}{3\lambda^{3}}=510\text{ }\frac{\text{W}}{\text{m}^{2}}$, the detuning is $\delta=-\frac{\Gamma}{2}$, and the intensity of the 369 nm beam is $2546\text{ }\frac{\text{W}}{\text{m}^{2}}$ giving a saturation parameter $s=\frac{I}{I_{\text{sat}}}=5$. The photon count rate of the PMT was first measured using a frequency counter (BK precision, PN: 1823 A) to be $30\times10^{3}\text{ \ensuremath{\frac{\text{counts}}{\text{s}}}}$, a factor of $\approx12$ lower than expected. I'm not entirely sure why this discrepancy is so large. There could be bad alignment between the center of the trap and radiation shield lids creating an unwanted stop and effectively reducing the achievable NA or higher-than-specced absorption of optics. The signal-to-noise (SNR) was measured by comparing the counts measured at the PMT when the resonant 369 nm beam was on and off. The SNR was measured to be 75 and the data is displayed in Table \ref{tab:signal-to-noise-of-PMT}.

\begin{table}[!!h]
\centering
\tablefont
\caption[Signal-to-Noise of PMT]{Signal-to-noise (SNR) of PMT.}
\label{tab:signal-to-noise-of-PMT}
\begin{tabularx}{\textwidth}{>{\centering\arraybackslash}p{0.3\textwidth}>{\centering\arraybackslash}p{0.3\textwidth}c}
\toprule
\textbf{Measurement Configuration} & \textbf{Counts ($\mathbf{s^{-1}}$)} & \textbf{Comments} \\
\midrule
Signal & $30\times10^{3}$ & \makecell{Resonant 369 On\\Repumpers On} \\
\midrule
Noise & 400 & \makecell{Resonant 369 Blocked\\Repumpers On} \\
\midrule
Dark Counts & 2 & Light Directed to Camera \\
\bottomrule
\end{tabularx}
\end{table}

\ifSubfilesClassLoaded{%
  \bibliographystyle{unsrt}
  \bibliography{Dissertation}
}{}
\end{document}

%% file: Chapter8.tex
\chapter{Cryostat Characterization}
\label{chap:cryostat-characterization}

\section{Cooldown and Warmup}
\label{sec:cooldown-and-warmup}

\subsection{Cooldown}
\label{subsec:cooldown}

\begin{figure}[!!h]
\centering
\includegraphics[width=\textwidth]{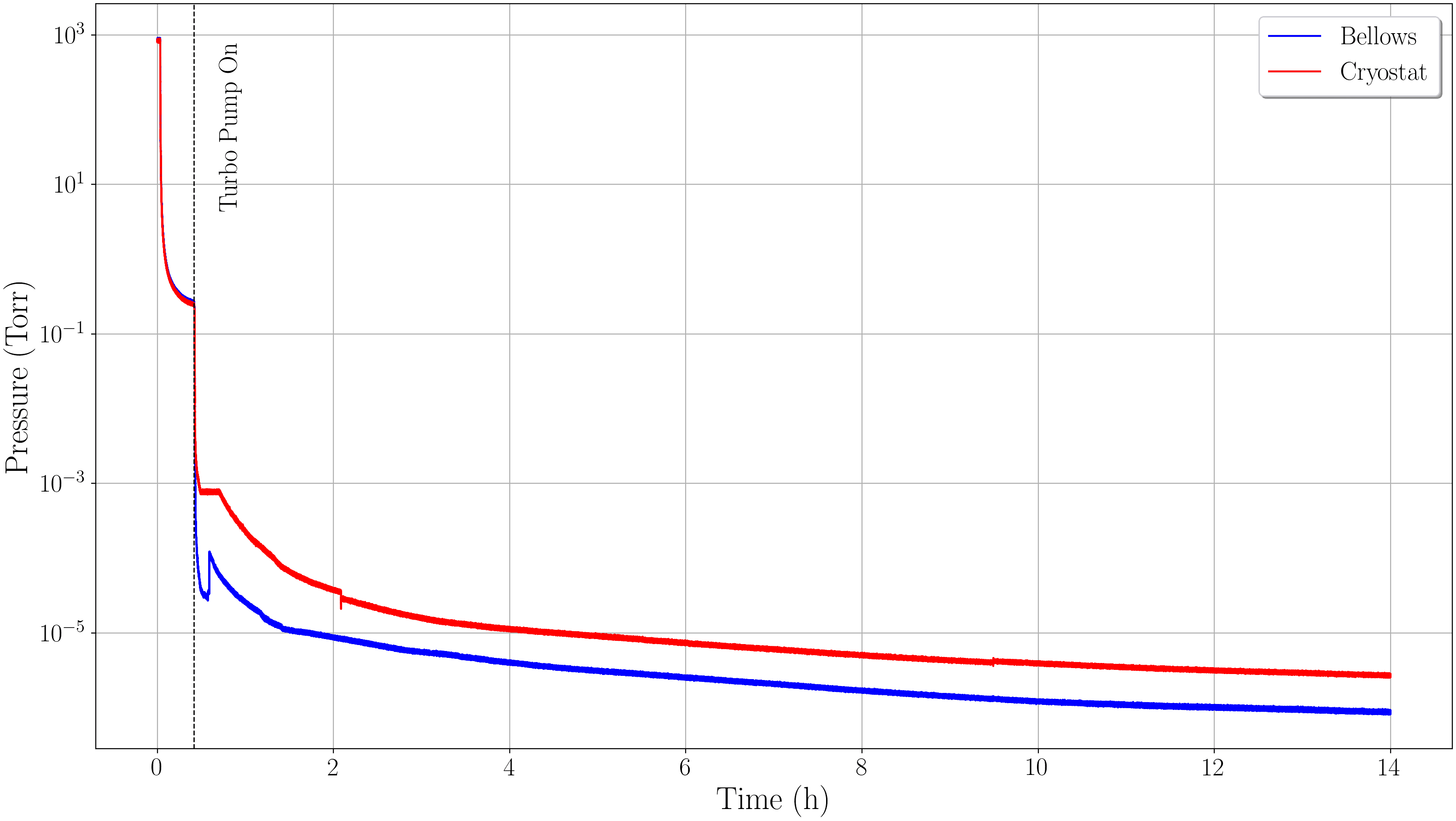}
\caption[Cryostat Vacuum Pump Down]{Plot of the pressure of the cryostat and bellows vs. time prior to cooldown. The base pressure of the cryostat and bellows after pumping with just the scroll pump for 25 minutes is $4.2\times10^{-1}\text{ Torr}$. The base pressure of the cryostat and bellows after pumping with the turbo pump for 14 hours is $2.6\times10^{-6}\text{ Torr}$ and $8.3\times10^{-7}\text{ Torr}$ respectively.}
\label{fig:cryostat-vacuum-pump-down}
\end{figure}

Prior to initiating a cooldown, the cryostat volume is evacuated using a turbo pump (Agilent, TPS-Compact, PN: X3580A) equipped with a backing scroll pump (Agilent, PN: IDP-3) as shown in Figure \ref{fig:cryostat-vacuum-pump-down}. First, just the scroll pump is turned on to evacuate the initial large gas load for about 25 minutes achieving a base pressure of $4.2\times10^{-1}\text{ Torr}$. The turbo pump is then turned on where the cryostat pressure reaches $2.6\times10^{-6}\text{ Torr}$ in 14 hours. The pressure at the bellows is also monitored and reaches $8.3\times10^{-7}\text{ Torr}$ in 14 hours. The vacuum conductance, $C$, between the turbo pump and the cryostat can be calculated using the following equation

\begin{equation}
    C=\frac{Q}{\Delta P}=\frac{P_{\text{bellows}}S_{\text{pump}}}{P_{\text{cryostat}}-P_{\text{bellows}}}
    \label{eq:vacuum-conductance-of-cryostat}
\end{equation}
where Q is the gas throughput in $\frac{\text{Torr-L}}{\text{s}}$, $S_{\text{pump}}=48\text{ \ensuremath{\frac{\text{L}}{\text{s}}}}$ is the pumping speed of the turbo pump, and $P_{\text{cryostat}}$ and $P_{\text{bellows}}$ are the vacuum pressure of the cryostat and bellows respectively in Torr. Since the throughput is constant at every position along the conductance path in a steady state, we can simply compute it using the pressure measured at the bellows which is near the turbo pump. The vacuum conductance using the base pressures of the cryostat and bellows after 14 hours is $23\,\frac{\text{L}}{\text{s}}$. This initial evacuation is imperative for two reasons: 1.) it reduces the thermal conduction through the residual gas and limits the accumulation of adsorbed gas layers on internal surfaces (increasing the surface's emissivity), both of which would otherwise increase the thermal load on the cold stages through conductive and radiative heat transfer, and 2.) it reduces the hydrogen gas load eventually seen at the 4 K stage, the least efficiently cryopumped gas\footnote{Helium is the least efficiently pumped gas, but is assumed to be at a negligible partial pressure in this system.}  as discussed in more detail in Appendix \ref{sec:vacuum-fundamentals}.

The conduction through residual gas in the cryostat is presented here for different vacuum regimes parametrized using the Knudsen number $\text{Kn}=\frac{\text{\ensuremath{\bar{\lambda}}}}{L}$ where $\bar{\lambda}=\frac{k_{\text{B}}T}{\sqrt{2}\pi d_{0}^{2}P}$ is the mean-free-path of a gas molecule and $L$ is the relevant distance scale for problem: the distance between the inner surface of the 300 K stage and the 40 K radiation shield side walls, $L=16\text{ mm}$. To analyze the conduction at different vacuum regimes, we will assume that the total pressure of the cryostat is approximately equal to the partial pressure of nitrogen. This is a good approximation at ambient pressure of $760\text{ Torr}$ and the scroll pump pressure of $10^{-1}\text{ Torr}$ as the vacuum composition is still air-like. In this regime $\text{Kn}<1$ meaning gas-gas collisions dominate over gas-surface collisions which is known as the Hydrodynamic vacuum regime and gas conduction follows Fourier's law
\begin{equation}
    \dot{Q}_{\text{cond,hydro}}=k_{\text{gas}}\frac{A}{L}\Delta T
    \label{eq:gas-conduction-hydrodynamic-regime}
\end{equation}
and is independent of pressure where $k_{\text{gas}}=0.025\,\frac{\text{W}}{\text{m-K}}$ is the thermal conductivity of nitrogen at 300 K, $A$ is the area of the surfaces in $\text{m}^{2}$, and $\Delta T$ is the temperature difference between the surfaces in K \cite{ekinExperimentalTechniquesLowtemperature2006a}. The conductance between the 300 K stage and the 40 K stage for this regime is summarized in Table \ref{tab:thermal-conductance-through-N2-gas}. 

At the base pressure of $2.6\times10^{-6}\text{ Torr}$ the vacuum composition is more likely to be dominated by water, carbon dioxide, and hydrogen but we will assume it be nitrogen to simplify the analysis. In this regime, $\text{Kn}>1$ meaning gas-surface collisions dominate over gas-gas collisions which is known as the Free-Molecular vacuum regime and gas conduction becomes linear with pressure and independent of stage separation as follows,
\begin{equation}
    \dot{Q}_{\text{cond,mol}}=\kappa a_{0}AP\Delta T
    \label{eq:gas-conduction-free-molecular-regime}
\end{equation}
where $\kappa=1.2$ is a constant for nitrogen that is dependent on its adiabatic index $\gamma=\frac{C_{\text{p}}}{C_{\text{v}}}$ and its mass, $a_{0}=0.33$ is the accommodation coefficient which characterizes how effectively gas molecules exchange energy with a surface\footnote{If $a_{0}=1$, the molecule leaves the collision in thermal equilibrium with the surface. If $a_{0}=0$, the molecule collides elastically, keeping its original energy.}, $A$ is the area of the surfaces in $\text{m}^{2}$, $P$ is the pressure of nitrogen in Pa, and $\Delta T$ is the temperature difference between the surfaces in K \cite{ekinExperimentalTechniquesLowtemperature2006a}. The conductance between the 300 K stage and the 40 K stage for this regime is summarized in Table \ref{tab:thermal-conductance-through-N2-gas}.
 
\begin{table}[!!h]
\centering
\tablefont
\caption[Thermal Conductance Through Nitrogen Gas in the Cryostat]{Thermal conductance through nitrogen gas at different vacuum regimes in the cryostat.}
\label{tab:thermal-conductance-through-N2-gas}
\begin{tabularx}{\textwidth}{>{\centering\arraybackslash}p{0.24\textwidth}>{\centering\arraybackslash}p{0.2\textwidth}>{\centering\arraybackslash}p{0.2\textwidth}c}
\toprule
\textbf{$\mathbf{N_2}$ Pressure (Torr)} & \textbf{$\mathbf{\bar{\lambda}}$ (m)} & \textbf{Kn} & \textbf{Conductance $\mathbf{\frac{\dot{Q}_{\text{cond}}}{\Delta T}\,\left(\frac{\text{W}}{\text{K}}\right)}$}\\
\midrule
760 & $7.1\times10^{-8}$ & $4.4\times10^{-6}$ & 1.0 \\
\midrule
$1\times10^{-1}$ & $5.4\times10^{-4}$ & $3.4\times10^{-2}$ & 1.0 \\
\midrule
$1\times10^{-4}$ & 0.55 & 35 & $3\times10^{-3}$ \\
\midrule
$2.6\times10^{-6}$ & 21 & 1296 & $8.7\times10^{-5}$ \\
\bottomrule
\end{tabularx}
\end{table}

In summary, residual gas conduction can impose a problematic thermal load if a cooldown is initiated at insufficiently low pressure. Using nitrogen as the representative gas and the stage separation, $L=16\text{ mm}$, the effective gas conduction thermal load between the 300 K stage and 40 K stage at $1\times10^{-1}\text{ Torr}$ is $\dot{Q}_{\text{gas}}\approx1\,\frac{\text{W}}{\text{K}}\times(300\,\text{K}-40\,\text{K})\approx260\text{ W}$ which far exceeds the available cooling power of the 40 K stage of 4 W at 40 K discussed in subsection \ref{subsec:current-system}. Consequently, the 40 K stage would stall at an elevated temperature and the system would be unable to cool into the cryogenic regime. By contrast, at $2.6\times10^{-6}\text{ Torr}$, the corresponding thermal load on the 40 K stage would be $\dot{Q}_{\text{gas}}\approx8.7\times10^{-5}\,\frac{\text{W}}{\text{K}}\times(300\,\text{K}-40\,\text{K})\approx20\text{ mW}$ making gas conduction negligible compared to other thermal loads. This is also an overestimate as the nitrogen partial pressure is decreasing over time as the system cools down, but the effect is made clear.

\begin{figure}[!!h]
\centering
\includegraphics[width=\textwidth]{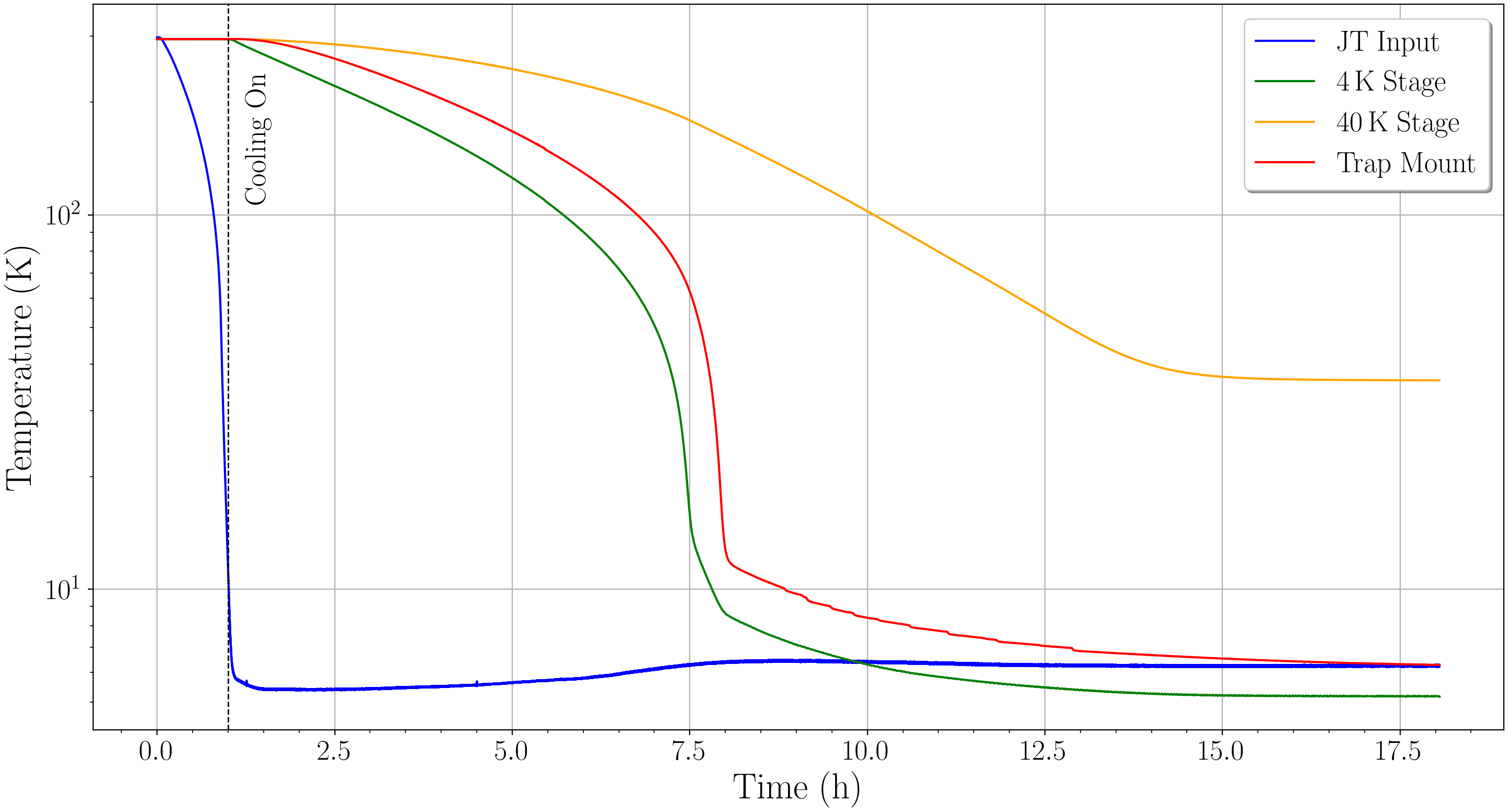}
\caption[Cryostat Cold Stages Temperature vs. Time During Cooldown]{Cooldown time of cryostat stages. Cooling at the cryostat begins after about 1 hour once the GM cryocooler cools the recirculating helium to $\sim42\text{ K}$, the maximum JT inversion temperature of helium at 110 psig \cite{sciverHeliumCryogenics2012,bellPurePseudopureFluid2014CoolProp,lemmon2010nistREFPROP}. The final temperatures of the JT input, 4 K stage, 40 K stage, and Trap Mount in this plot are 6.25 K, 5.18 K. 36.16 K, and 6.29 K respectively.}
\label{fig:cryostat-temperature-vs-time-during-cooldown}
\end{figure}

\begin{figure}[!!h]
\centering
\includegraphics[width=\textwidth]{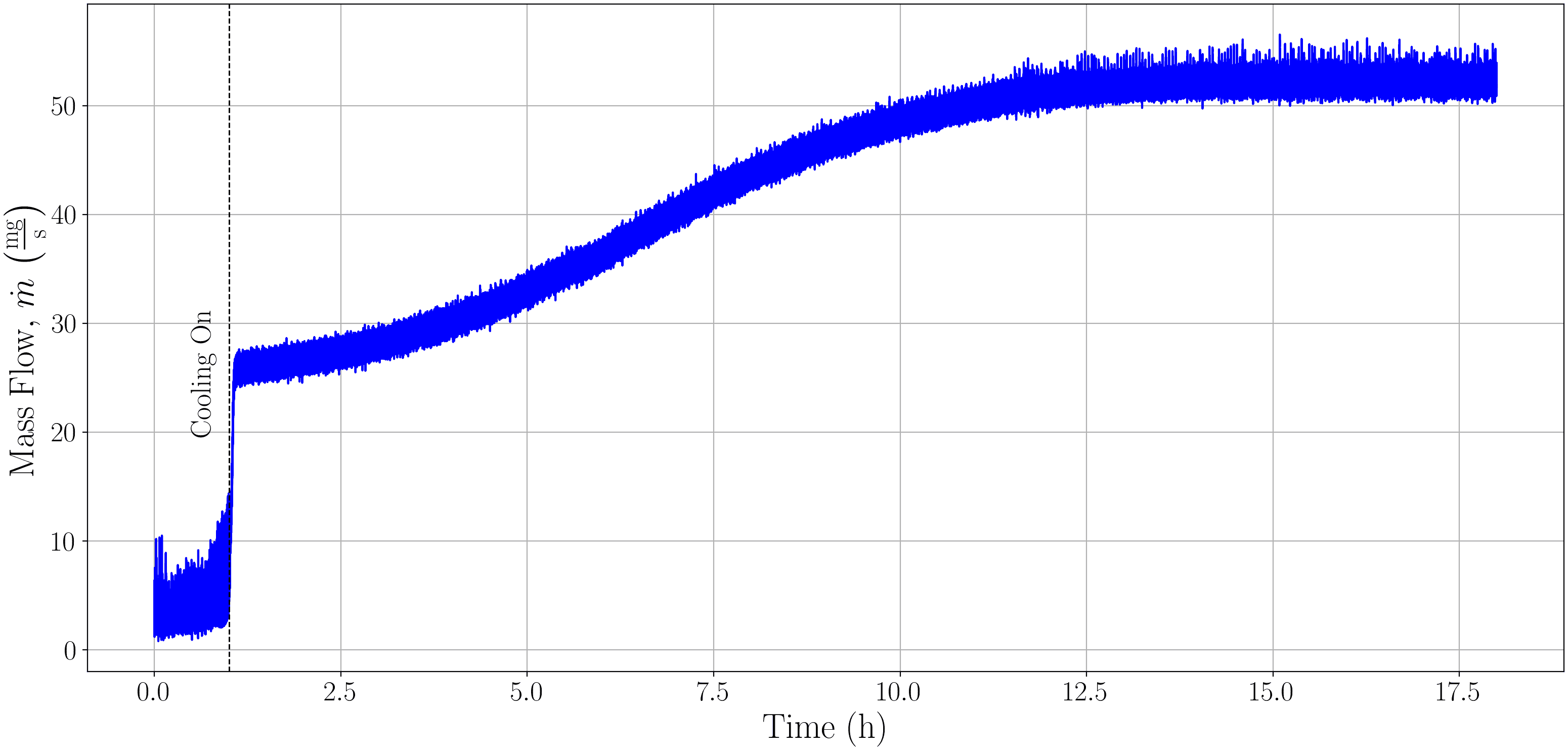}
\caption[Cryostat Helium Mass Flow vs. Time During Cooldown]{Mass flow of helium during cooldown. Cooling at the cryostat begins after about 1 hour once the GM cryocooler cools the recirculating helium to $\sim42\text{ K}$, the maximum JT inversion temperature of helium at 110 psig \cite{sciverHeliumCryogenics2012,bellPurePseudopureFluid2014CoolProp,lemmon2010nistREFPROP}. At this point the flow increases to $25\,\frac{\text{mg}}{\text{s}}$ and continues to increase to $52\,\frac{\text{mg}}{\text{s}}$ when the cryostat cold stages reach their base temperatures.}
\label{fig:cryostat-helium-mass-flow-vs-time-during-cooldown}
\end{figure}

Following evacuation, the GM cryocooler and helium recirculating loop are engaged to begin the cooldown. Cooling at the cryostat begins after about 1 hour once the GM cryocooler cools the recirculating helium to $\sim42\text{ K}$, the maximum JT inversion temperature of helium at 110 psig \cite{sciverHeliumCryogenics2012,bellPurePseudopureFluid2014CoolProp,lemmon2010nistREFPROP}. A temperature sensor located near HX2 in Figure \ref{fig:Schematic-of-CET-Stinger-Cyrogenic-system} monitors the JT input temperature. The complete cooldown requires $\approx17\text{ hours}$, and the temperatures of the JT input, 4 K stage, 40 K stage, and Trap Mount throughout the cooldown are shown in Figure \ref{fig:cryostat-temperature-vs-time-during-cooldown}. As shown in Figure \ref{fig:cryostat-helium-mass-flow-vs-time-during-cooldown}, the helium mass flow increases throughout the cooldown. This occurs because the viscosity of helium decreases as its temperature falls, reducing the hydraulic resistance of the JT capillary and associated cold-side plumbing \cite{sciverHeliumCryogenics2012}. Consequently, for a fixed supply pressure, the mass flow rate increases as the cryostat approaches its base operating temperature. The flow is monitored using a mass flow meter discussed in Subsection \ref{subsec:Gas-Manifold}.

\begin{figure}[!!h]
\centering
\includegraphics[width=\textwidth]{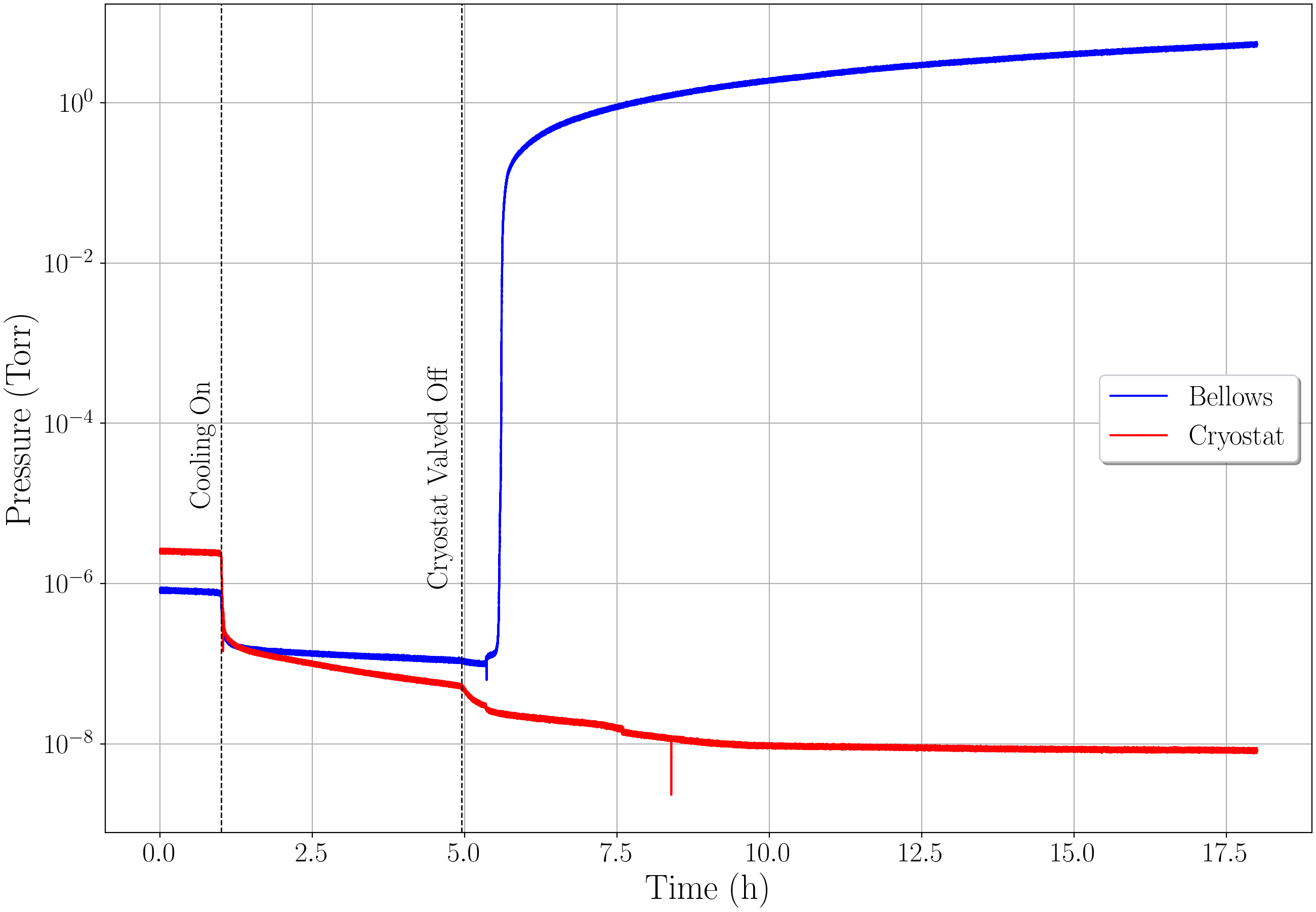}
\caption[Cryostat Vacuum Pressure vs. Time During Cooldown]{Vacuum pressure of the cryostat during cooldown. Cooling at the cryostat begins after about 1 hour once the GM cryocooler cools the recirculating helium to $\sim42\text{ K}$, the maximum JT inversion temperature of helium at 110 psig \cite{sciverHeliumCryogenics2012,bellPurePseudopureFluid2014CoolProp,lemmon2010nistREFPROP}. The cryostat is valved off after $\approx5\text{ hours}$ and the turbo pump is turned off shortly after that. Base pressure of cryostat is $8.1\times10^{-9}\text{ Torr}$.}
\label{fig:cryostat-vacuum-pressure-vs-time-during-cooldown}
\end{figure}

A residual gas analyzer (RGA) was not installed on the cryostat so the exact partial pressures of all the gases are not known, but some high-level statements about the vacuum are presented here. As soon as the helium flow increases and the JT temperature is 6 K after about 1 hour during the cooldown, the pressure suddenly deceases by an order of magnitude in about 3 minutes as shown in Figure \ref{fig:cryostat-vacuum-pressure-vs-time-during-cooldown}. This occurs because as the cold helium enters the cryostat it creates a 6 K cold finger at the in-vacuum helium lines. This 6 K surface acts as a powerful cryopump, rapidly condensing water, carbon dioxide, nitrogen, oxygen, and organics. However, the turbo pump is kept on as it is not known how effectively these gases are cryopumped as the effective cryopumping speed of these gases depends on the vacuum conductance to the 6 K cold finger, $C$, and the cryopumping speed of the cold finger, $S$, $S_{\text{eff}}=\frac{CS}{C+S}$, which is discussed more in Section \ref{sec:vacuum-characterization}. After this point, the pressure continues to decrease gradually as the residual gas pressure drops linearly with temperature, the surface outgassing rates fall with temperature, the cold surfaces increasingly act as cryopumps, and the turbo pump remains connected which continues to pump the residual non-condensed gases. The cryostat is valved off when the 4 K stage temperature reaches $\approx120\text{ K}$ because at this point the cryopumping surpasses the turbo pumping. At this point, water, carbon dioxide, and most organics are able to be cryopumped onto the 4 K stage. After this, the chamber pressure continues to decrease as cold stages provide stronger cryopumping of nitrogen and oxygen. At the base temperature, the residual gas at the 4 K stage is assumed to be completely hydrogen. If there is a small leak in the in-vacuum recirculating loop there could be trace amounts of helium at the 4 K stage, but we cannot know without an RGA. 

During a cooldown, the clamping mechanism at the JT interface loosens and slightly limits the cooling power at the cryostat. This has occurred ever since the system was delivered to UMD. This is most likely due to the the CTE mismatch between the Teflon seal and the metals of the bayonet and Stinger interface. My standard operating procedure is as follows: after a successful cooldown determined by measuring similar cold stage temperatures, helium flow, and vacuum pressure outlined in Figures \ref{fig:cryostat-temperature-vs-time-during-cooldown}, \ref{fig:cryostat-helium-mass-flow-vs-time-during-cooldown}, and \ref{fig:cryostat-vacuum-pressure-vs-time-during-cooldown} the Stinger interface clamp is tightened a bit further. After tightening the clamp and waiting for the cryostat to equilibrate, the final cold stage temperatures, helium flow, and vacuum pressure were measured and are summarized in Table \ref{tab:final-temp-pressure-helium-mass-flow-of-cryostat}. The larger standard deviation for the Stinger temperature is due to due fact that it oscillates at $\approx200\text{ mHz}$ which is most likely correlated with the GM compression-expansion cycle frequency of $\approx1\text{ Hz}$. The temperatures of the other stages do not oscillate an any particular frequency.

\begin{table}[!!h]
\centering
\tablefont
\caption[Final Temperatures, Pressures, and Helium Mass Flow of the Cryostat]{Final temperatures, pressures, and flow of the cryostat. The average values with standard deviations are recorded here over 12 min. }
\label{tab:final-temp-pressure-helium-mass-flow-of-cryostat}
\begin{tabularx}{\textwidth}{>{\centering\arraybackslash}p{0.49\textwidth}>{\centering\arraybackslash}p{0.49\textwidth}}
\toprule
\textbf{Parameter} & \textbf{Value} \\
\midrule
JT Input (K) & $6.185\pm0.022$ \\
\midrule
4 K Stage (K) & $4.936\pm0.006$ \\
\midrule
Trap Mount (K) & $5.976\pm0.001$ \\
\midrule
40 K Stage (K) & $27.229\pm0.003$ \\
\midrule
Helium Flow $\left(\frac{\text{mg}}{\text{s}}\right)$ & $51.3\pm0.4$ \\
\midrule
Helium Pressure at Gas Manifold (psig) & $105\pm2$ \\
\midrule
Vacuum Pressure (Torr) & $6.3\pm0.1\times10^{-9}$ \\
\bottomrule
\end{tabularx}
\end{table}

\subsection{Warmup}
\label{subsec:warmup}

To warm the system, the GM cryocooler and compressors are shut down and the helium remaining in the recirculating loop between the gas manifold and cryostat is evacuated. The helium contained between the gas manifold and compressors is isolated and retained within the recirculating loop. Without active heating, the cryostat warms passively to room temperature in approximately 72 hours because of its large thermal mass and excellent thermal isolation from the 300 K vacuum chamber. This warmup time was reduced to $\approx5\text{ hours}$ by applying power via heaters mounted at the 4 K and 40 K stages. We routinely apply 20 W to the 4 K stage using one heater and 40 W to the 40 K stage using two heaters (20 W each). The second heater on the 4 K stage is unusable due to a lead breaking while performing in-vacuum work on the cryostat. Currently, the circuit is limited to 20 W per heater by the power supplies, but more power could be applied in the future.

\begin{figure}[!!h]
\centering
\includegraphics[width=0.75\textwidth]{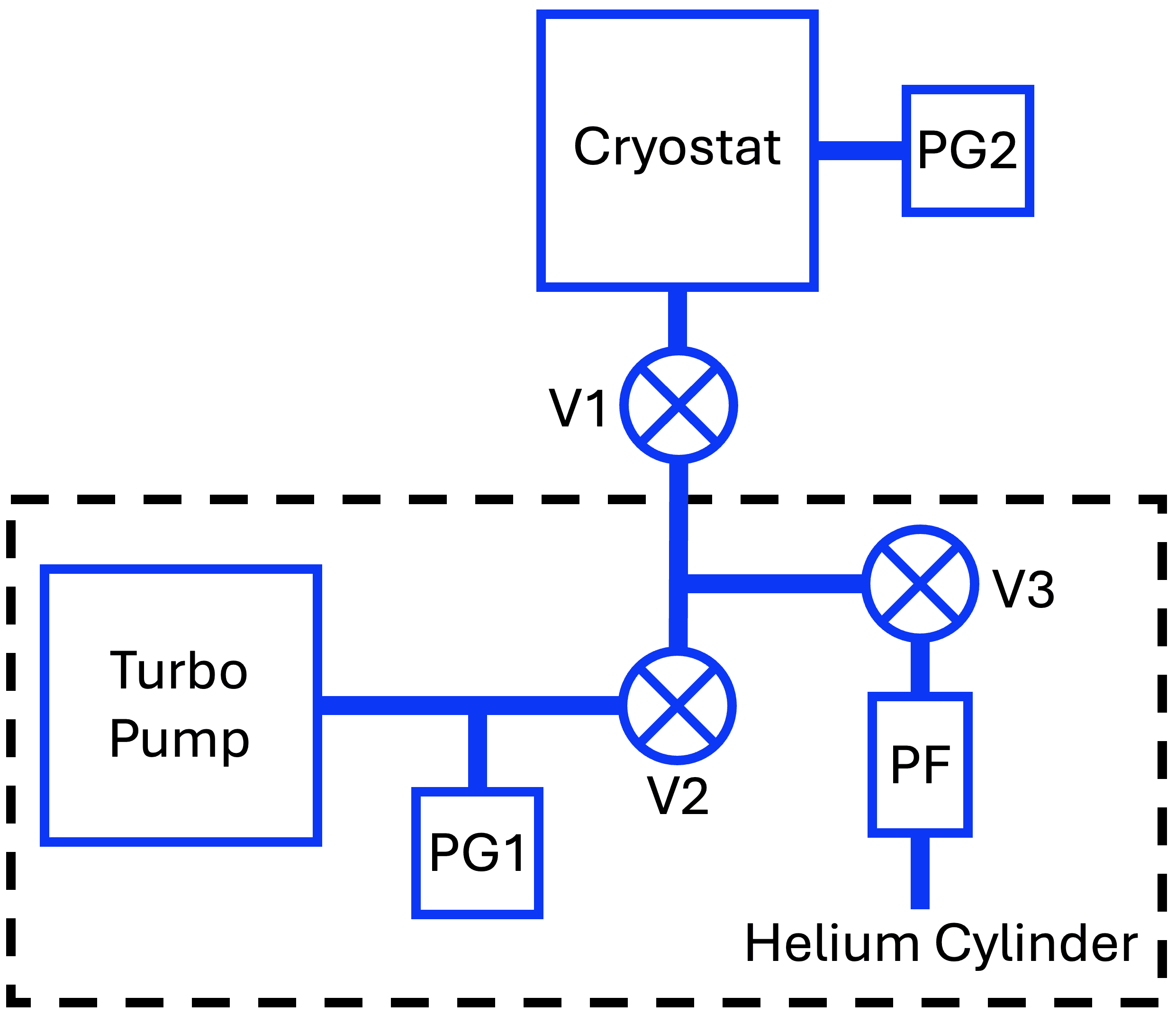}
\caption[Gas Injection Vacuum Space]{Gas injection vacuum space for bleeding gas into the cryostat vacuum space depicted inside the dashed black box. The particulate filter (PF) and needle valve (V3) were connected with Swagelok-1/4 fittings, while the rest of the parts were connected with KF-50 and CF-2.75 components. Parts list is given in Table \ref{tab:parts-list-for-gas-injection-vacuum-space}.}
\label{fig:gas-injection-vacuum-space}
\end{figure}

Another way to reduce the passive warmup time is to bleed gas into the cryostat which creates a thermal short between the cold stages and the 300 K stages via gas conduction. This was performed once using helium when there were no usable heaters available at the 4 K stage. A gas injection vacuum space was built to bleed helium gas into the cryostat and is shown in Figure \ref{fig:gas-injection-vacuum-space}. While keeping valve 1 closed, the gas injection vacuum space was evacuated with a turbo pump to a base pressure of $\approx10^{-6}\text{ Torr}$, measured using gauge 1. Valve 1 was then opened so that the gas injection vacuum space is continuous with the cryostat vacuum space. This was done when the cryostat cold stages were at $\approx100\text{ K}$. Valve 2 was then closed, isolating the turbo pump, and helium was carefully bled into the cryostat vacuum space by slowing opening valve 3 until the pressure reached $\approx7.5\times10^{-3}\text{ Torr}$, measured using gauge 2. Taking the partial pressure of helium to be $\approx7.5\times10^{-3}\text{ Torr}$ at 100 K, the Knudsen number is $<1$ (Hydrodynamic regime) so the gas conductance between the cold stages and 300 K stage can be using Equation \eqref{eq:gas-conduction-hydrodynamic-regime}. Taking the mean thermal conductivity for helium to $0.11\,\frac{\text{W}}{\text{m-K}}$, this yields a conductance of $\frac{\dot{Q}_{\text{cond,hydro}}}{\Delta T}\approx4.4\,\frac{\text{W}}{\text{K}}$ which reduced the warmup time to $\approx10\text{ hours}$. 

Helium is a good gas to use because it has one of the highest thermal conductivities, will not condense at any point during the warmup, and is safe to use in a lab environment. A list of thermal conductivities for various alternate gases are presented in Table \ref{tab:list-of-mean-thermal-conductivities-of-various-gases}. Hydrogen could be used as it has a slightly higher thermal conductivity than helium, but would immediately become a fire hazard if there are leaks in the gas injection vacuum space to atmosphere. Nitrogen, argon, or neon could alternatively be used which have reasonable thermal conductivities and do not pose any detrimental lab hazards. The cheapest solution would be to use nitrogen gas which would provide a conductance of $G_{\text{cond,hydro}}\approx0.7\,\frac{\text{W}}{\text{K}}$ using the same operating conditions which is worse than helium, but still much better than the estimated conductance of vacuum $\frac{\dot{Q}_{\text{cond,mol}}}{\Delta T}\approx8.7\times10^{-5}\:\frac{\text{W}}{\text{K}}$ calculated in Table \ref{tab:thermal-conductance-through-N2-gas}. 

\begin{table}[!!h]
\centering
\tablefont
\caption[Parts List for Gas Injection Vacuum Space]{Parts list for gas injection vacuum space in Figure \ref{fig:gas-injection-vacuum-space}.}
\label{tab:parts-list-for-gas-injection-vacuum-space}
\begin{tabularx}{\textwidth}{>{\centering\arraybackslash}p{0.49\textwidth}c}
\toprule
\textbf{Schematic Label} & \textbf{Part} \\
\midrule
V1, V2 & \makecell{UHV Valve\\VAT, PN: 54132-GE02} \\
\midrule
V3 & \makecell{Needle Valve\\Swagelok, PN: SS-4MG} \\
\midrule
PF & \makecell{Particulate Filter\\Swagelok, PN: SS-4F-05} \\
\midrule
PG1, PG2 & \makecell{Pressure Gauge\\Edwards Vacuum, PN: D3G0021100} \\
\bottomrule
\end{tabularx}
\end{table}

\begin{table}[!!h]
\centering
\tablefont
\caption[List of Mean Thermal Conductivities of Various Gases]{List of mean thermal conductivities of various gases in the temperature range of 100 K-300 K \cite{flynnCryogenicEngineering2005}.}
\label{tab:list-of-mean-thermal-conductivities-of-various-gases}
\begin{tabularx}{\textwidth}{>{\centering\arraybackslash}p{0.5\textwidth}>{\centering\arraybackslash}p{0.4\textwidth}}
\toprule
\textbf{Gas} & $\mathbf{\bar{k}}\text{ } \mathbf{\left(\frac{\text{W}}{\text{m-K}}\right)}$ \\
\midrule
He & 0.11 \\
\midrule
$\text{N}_2$ & 0.02 \\
\midrule
Ne & 0.04 \\
\midrule
$\text{H}_2$ & 0.13 \\
\midrule
Ar & 0.012 \\
\bottomrule
\end{tabularx}
\end{table}

\section{Cooling Performance}
\label{sec:cooling-performance}

The cooling performance of the cold stages is measured by applying a heat load to each stage using resistive heaters and measuring the equilibration temperatures of the stages. This is a useful metric as it gives the possible operating temperatures for the system given a heat load at each stage. This measurement was first conducted by CET before delivering the cryostat to UMD and is discussed in Subsection \ref{subsec:system-delivered-by-CET}. The measurement was conducted again at UMD after making many changes to the cryostat and is discussed in Subsection \ref{subsec:current-system}.

\subsection{System Delivered by CET}
\label{subsec:system-delivered-by-CET}

\begin{figure}[!!h]
\centering
\includegraphics[width=\textwidth]{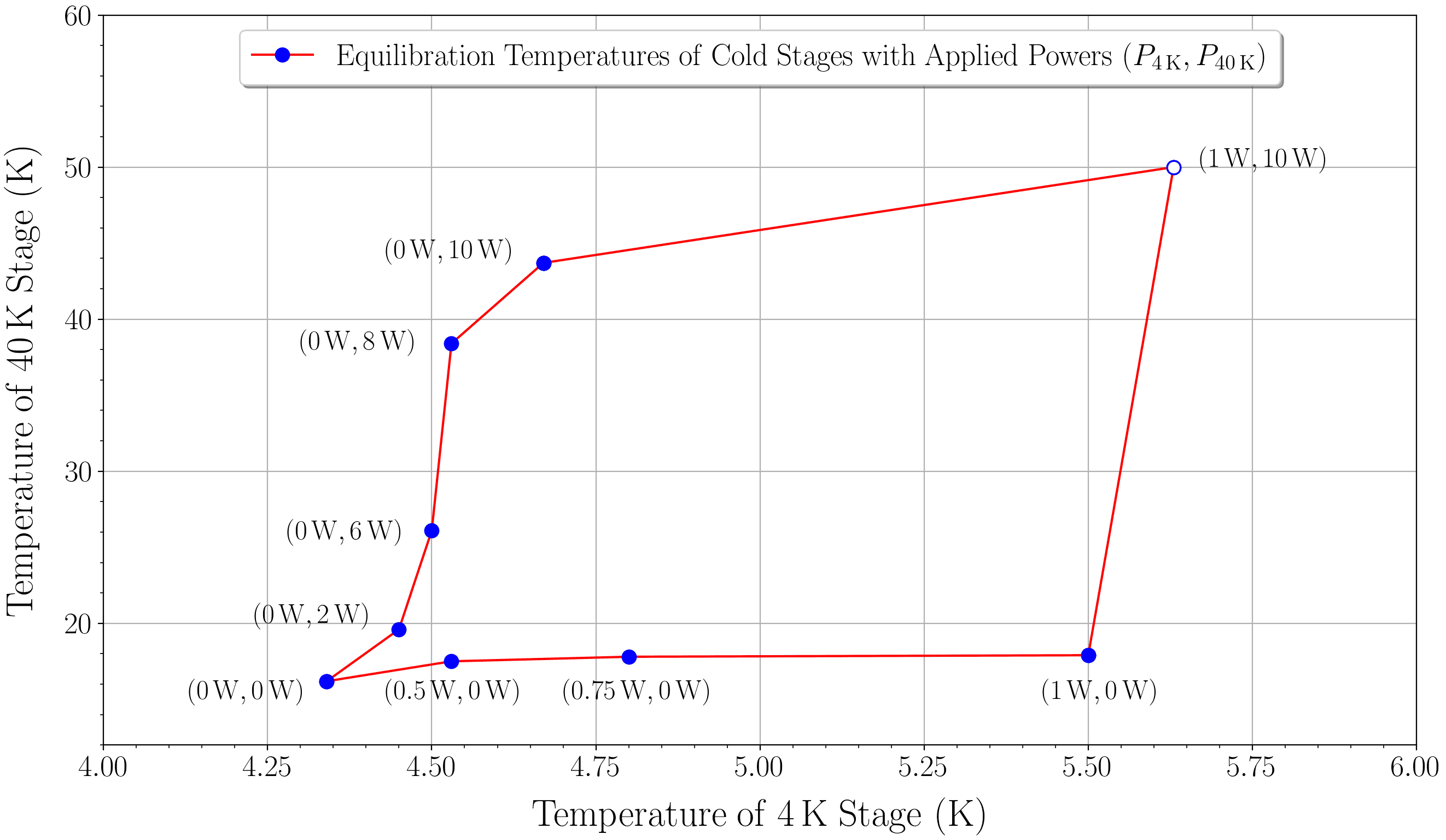}
\caption[Cooling Map of Cryostat Measured by CET]{Cooling map of cryostat measured by CET. The 4 K and 40 K stage equilibration temperatures are plotted, indicated by the blue points, when heat loads $(P_{\text{4\,K}},P_{\text{40\,K}})$ are applied to each stage respectively. Equilibration times are unknown as this was measured by CET. The unfilled blue point at (1 W, 10 W) was not measured but is extrapolated from the data. The red lines are used to outline the possible operating temperatures of the cryostat.}
\label{fig:cooling-map-of-cryostat-CET}
\end{figure}

CET characterized the overall cooling performance of the cryostat by generating a heat map displayed in Figure \ref{fig:cooling-map-of-cryostat-CET}. In this configuration, there were only 4 leads to each cold stage: 2 for temperature sensors and 2 for heaters. The radiation shield windows had copper blanks, there were no 300 K windows, the PEEK support rods were thinner, there were no PEEK tabs for vibration damping, and there was no additional load at the cold stages. The cooldown time of the cryostat in this configuration was $\approx12\text{ hours}$.

\subsection{Current System}
\label{subsec:current-system}

Many changes were made to the system once it arrived at UMD, including completely re-engineering the wiring, installing the ion trap, resonator, oven, optical cavity, and radiation shield windows, and also installing PEEK tabs on the radiation shields to provide addition passive vibration isolation from the GM cryocooler. 

\begin{figure}[!!h]
\centering
\includegraphics[width=0.49\textwidth]{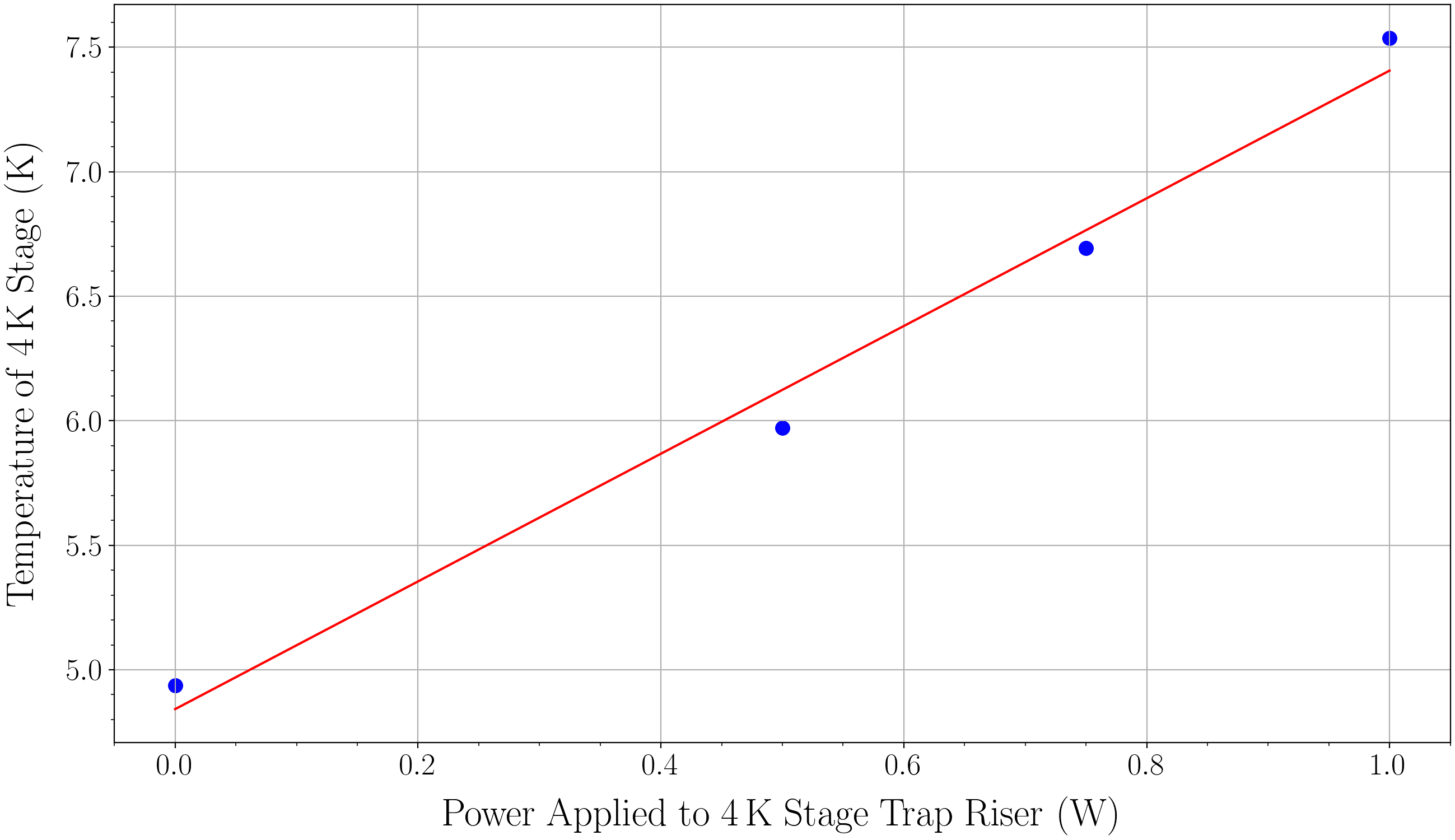}
\includegraphics[width=0.49\textwidth]{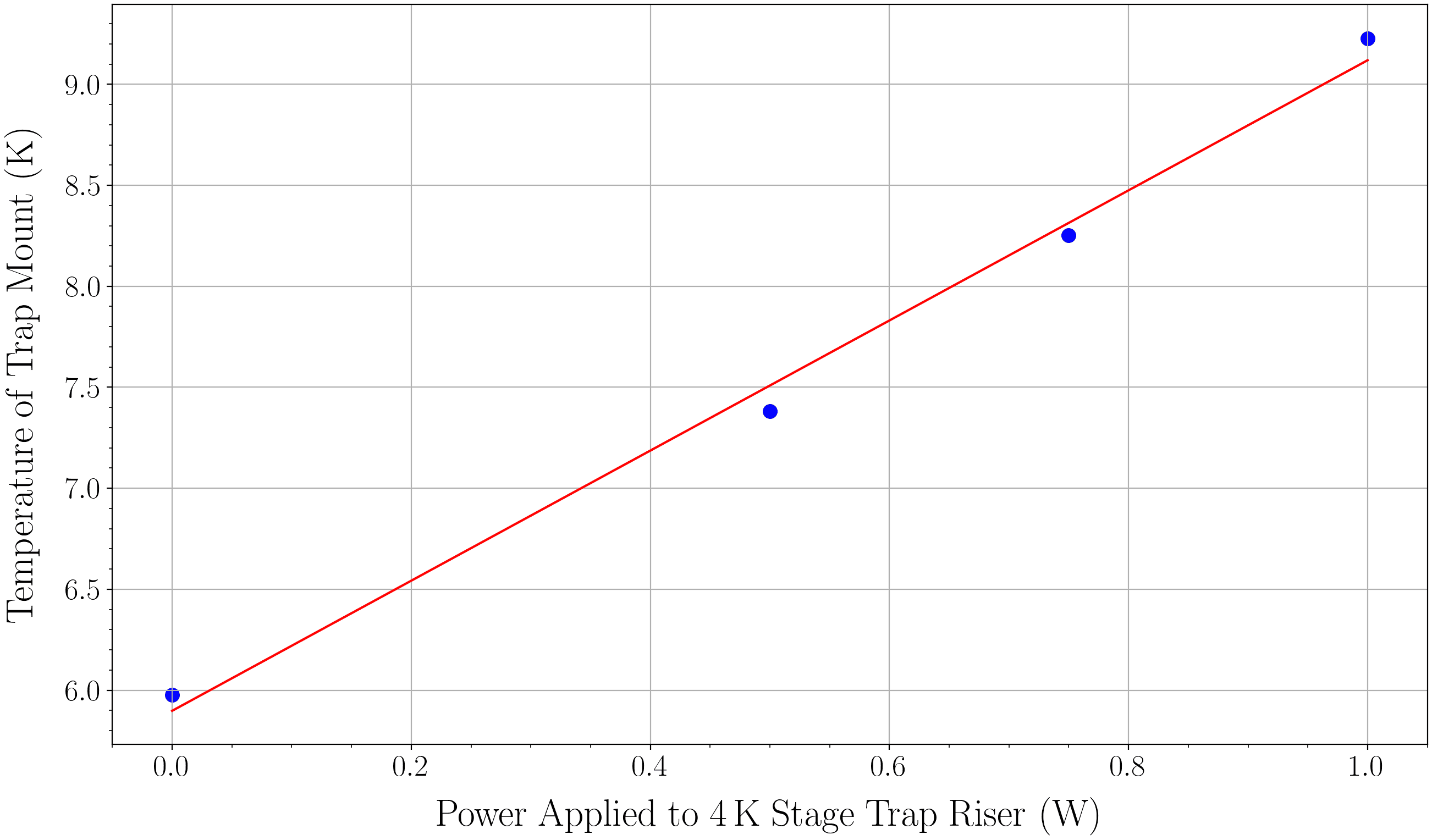}
\caption[Cooling Power of 4 K Stage and Trap Mount]{(Left) Equilibration temperatures of 4 K stage as a function of power applied to the 4 K stage trap riser. A linear fit to the data yields an apparent thermal conductance $G_{\text{app,4\,K}}=0.390\pm0.034\text{ \ensuremath{\frac{\text{W}}{\text{K}}}}$. (Right) Equilibration temperatures of trap mount as a function of power applied to the 4 K stage trap riser. A linear fit to the data yields an apparent thermal conductance $G_{\text{app,m}}=0.311\pm0.018\text{ \ensuremath{\frac{\text{W}}{\text{K}}}}$. Equilibration times are $\approx2$ hours.}
\label{fig:cooling-capacity-4K-Stage-Trap-mount}
\end{figure}

\begin{figure}[!!h]
\centering
\includegraphics[width=0.6\textwidth]{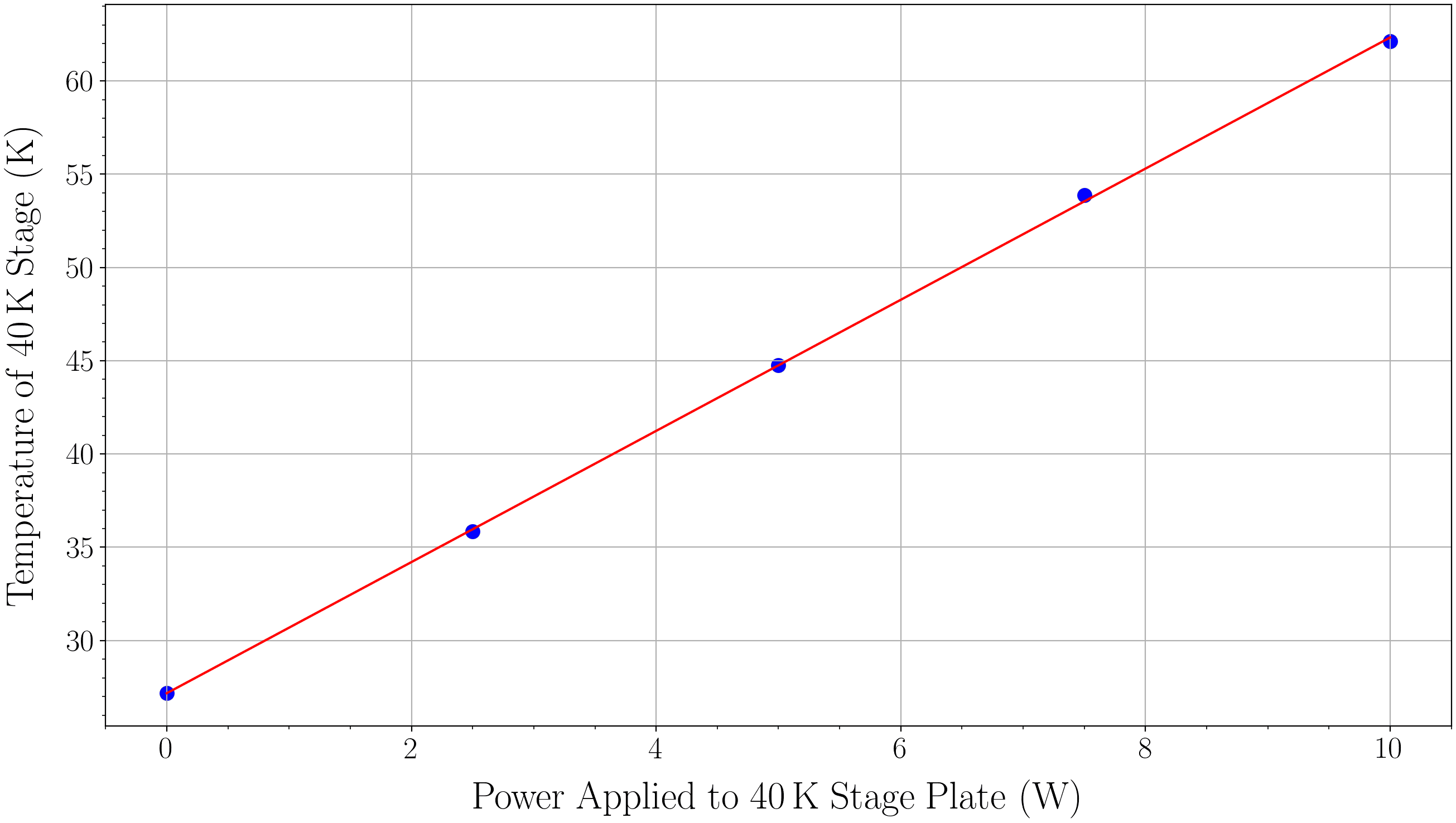}
\caption[Cooling Capacity of 40 K Stage]{Equilibration temperatures of 40 K stage as a function of power applied to the 40 K stage plate A linear fit to the data yields an apparent thermal conductance $G_{\text{app,40\,K}}=0.283\pm0.002\text{ \ensuremath{\frac{\text{W}}{\text{K}}}}$. Equilibration times are $\approx2$ hours. }
\label{fig:cooling-capacity-40K-Stage}
\end{figure}

Power is incrementally applied to the 4 K stage trap riser while no power is applied to the 40 K stage to determine the equilibration temperatures of the 4 K stage and the trap mount as shown in Figure \ref{fig:cooling-capacity-4K-Stage-Trap-mount}. Similarly, power is incrementally applied to the 40 K stage plate while no power is applied to 4 K stage to determine the equilibration temperature of the 40 K stage as shown in Figure \ref{fig:cooling-capacity-40K-Stage}. A linear least squares fit to the data in in Figures \ref{fig:cooling-capacity-4K-Stage-Trap-mount} and \ref{fig:cooling-capacity-40K-Stage} is performed to extract the apparent thermal conductances, $G_{\text{app}}=\left(\frac{\text{d}T}{\text{d}P}\right)^{-1}$, which represents the incremental heat removal capability of the cryostat as seen at 4 K stage, trap mount and 40 K stage. This reflects the aggregate thermal response of the full system rather than individual interface/material conductances and are summarized in Table \ref{tab:apparent-thermal-conductances-4K-40K-trap-mount}. 

\begin{table}[!!h]
\centering
\tablefont
\caption[Apparent Thermal Conductances of 4 K Stage, 40 K Stage, and Trap Mount]{The incremental heat removal capability of the cryostat as seen at the 4 K stage, trap mount, and 40 K stages.}
\label{tab:apparent-thermal-conductances-4K-40K-trap-mount}
\begin{tabularx}{\textwidth}{>{\centering\arraybackslash}p{0.25\textwidth}>{\centering\arraybackslash}p{0.2\textwidth}cc}
\toprule
\textbf{Element} & \makecell{\textbf{Power Applied to}\\\textbf{4 K Trap Riser}} & \makecell{\textbf{Power Applied to}\\\textbf{40 K Plate}} & \makecell{\textbf{Apparent Thermal}\\\textbf{Conductance, }$\mathbf{G_{\text{app}}\left(\frac{\text{W}}{\text{K}}\right)}$} \\
\midrule
4 K Stage & Yes & No & $0.390\pm0.034$ \\
\midrule
Trap Mount & Yes & No & $0.311\pm0.018$ \\
\midrule
40 K Stage & No & Yes & $0.283\pm0.002$ \\
\bottomrule
\end{tabularx}
\end{table}

\begin{figure}[!!h]
\centering
\includegraphics[width=0.6\textwidth]{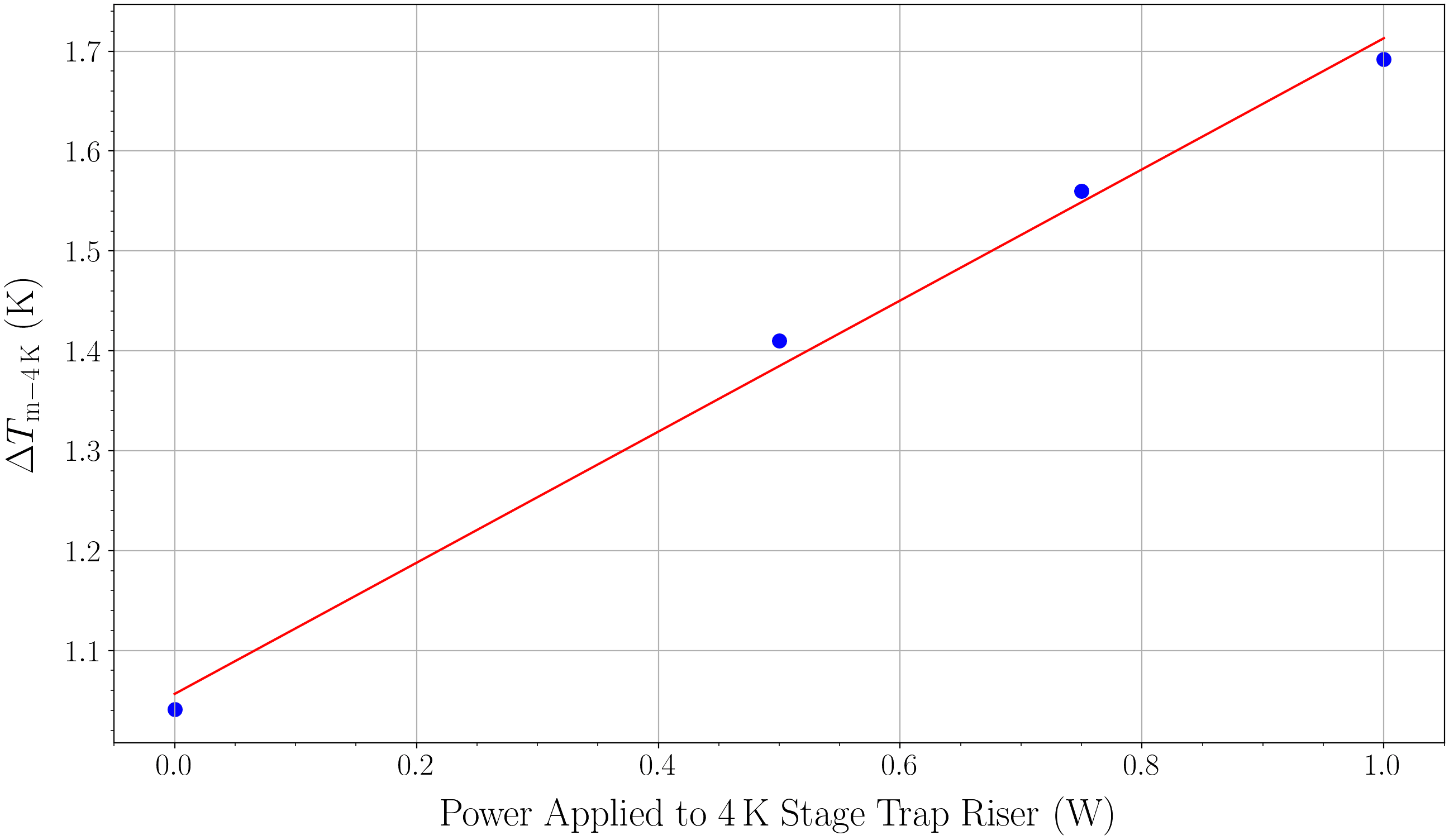}
\caption[Steady State Temperature Gradient Between Trap Mount and 4 K Stage]{Steady state temperature gradient between trap mount and 4 K stage as a function of power applied to the 4 K trap riser. A linear fit to the data yields a thermal conductance $G_{\text{m-4\,K}}=1.523\pm0.084\text{ \ensuremath{\frac{\text{W}}{\text{K}}}}$. Equilibration times are $\approx2$ hours.}
\label{fig:steady-state-temperature-gradient-Trap-mount-4K-stage}
\end{figure}

The thermal conductance between the trap mount and the 4 K stage is also calculated by studying the steady state temperature gradient between the trap mount and 4 K stage as a function of applied power at the trap riser as shown in Figure \ref{fig:steady-state-temperature-gradient-Trap-mount-4K-stage}. This data is generated using the equilibration temperatures of the trap and 4 K stage and applied powers in Figure \ref{fig:cooling-capacity-4K-Stage-Trap-mount}. A linear least squares fit to the data in in Figure \ref{fig:steady-state-temperature-gradient-Trap-mount-4K-stage} is performed to extract the thermal conductance, $G_{\text{m-4\,K}}=\left(\frac{\text{d}T}{\text{d}P}\right)^{-1}=1.523\pm0.084\,\frac{\text{W}}{\text{K}}$, which represents the aggregate thermal transport capability of the material stack and interfaces between the trap mount and 4 K stage under the measured operating conditions. It includes contributions from bulk material thermal conductivities, geometric cross-sections, and contact conductances between parts in the stack. The non-zero temperature gradient observed at $P_{\text{4\,K}}=0\text{ W}$ indicates the presence of a baseline parasitic heat load into the trap mount most likely due to conduction through the resonator coaxial cable from the 300 K stage and direct line-of-sight radiation from the 300 K stage to the trap mount. 

\begin{figure}[!!h]
\centering
\includegraphics[width=\textwidth]{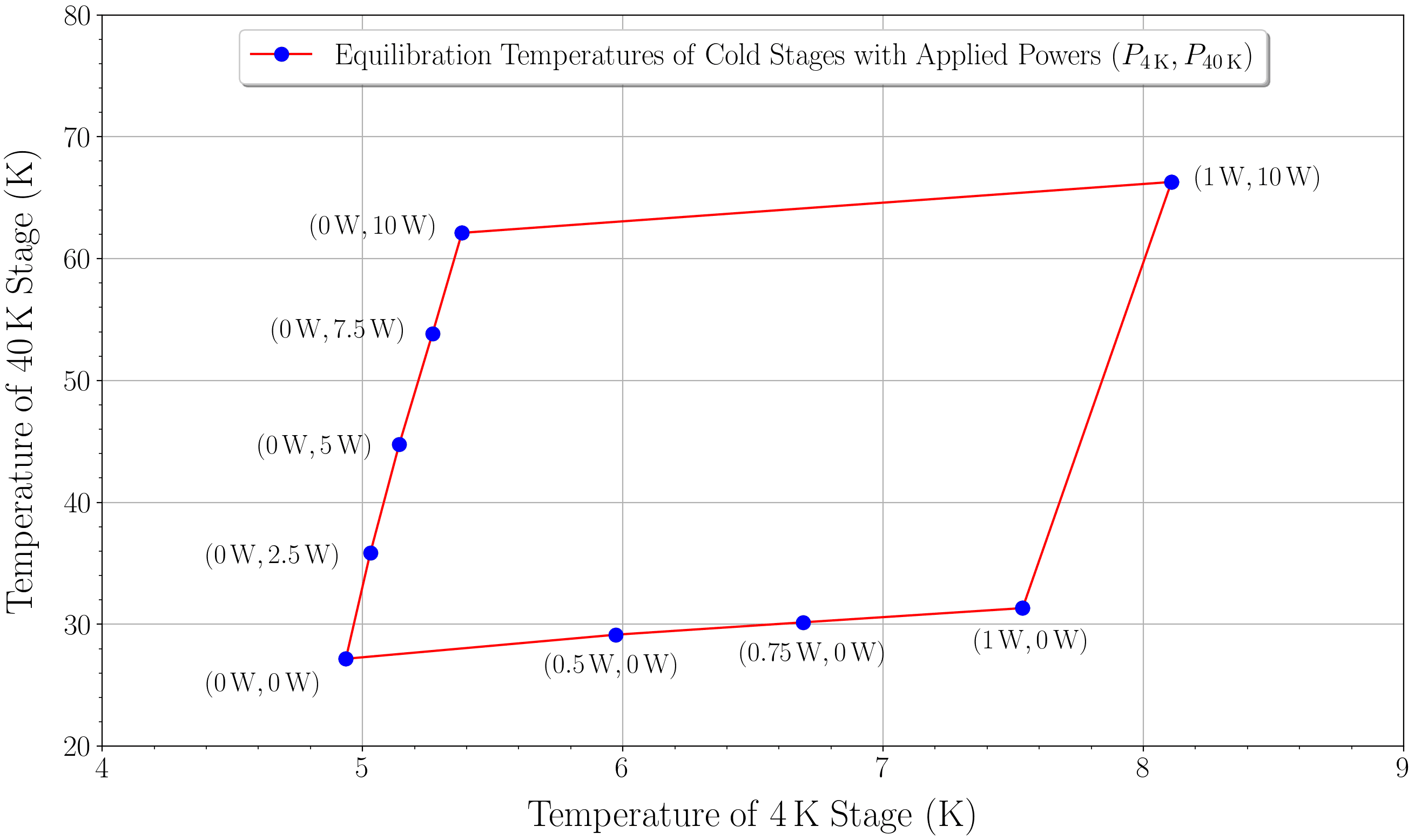}
\caption[Cooling Map of Cryostat Measured at UMD]{Cooling map of cryostat. The 4 K and 40 K stage equilibration temperatures are plotted, indicated by the blue points, when heat loads $(P_{\text{4\,K}},P_{\text{40\,K}})$ are applied to each stage respectively. Equilibration times are $\approx2$ hours. The red lines are used to outline the possible operating temperatures of the cryostat. }
\label{fig:cooling-map-UMD}
\end{figure}

We can completely characterize the overall cooling performance of the cryostat by generating a heat map displayed in Figure \ref{fig:cooling-map-UMD}. This data is generated using the equilibration temperatures of the 4 stage, trap mount, and 40 K stage and applied powers in Figures \ref{fig:cooling-capacity-4K-Stage-Trap-mount} and \ref{fig:cooling-capacity-40K-Stage}. Comparing the base temperatures in this cooling map to the baseline cooling map measured by CET in Figure \ref{fig:cooling-map-of-cryostat-CET}, we can see that there is approximately 6 W of additional power added to the 40 K stage and 0.75 W of additional power added to the 4 K stage. In addition, the cooldown time increased by $\approx5\text{ hours}$. This is most likely due to the following: PEEK tabs creating thermal link between each stage, DC wiring that are touching the cold stages in unwanted locations, thicker PEEK support rods, windows at all stages, direct line of sight thermal radiation through cavity feedthroughs in the cold stages, and unanticipated higher thermal loading from RF wiring. 

\section{Vacuum Characterization}
\label{sec:vacuum-characterization}

The vacuum quality of a trapped ion system directly influences the ion lifetime and operational reliability. Collisions between trapped ions and residual background gas molecules can transfer sufficient energy to heat, reorder, or eject ions from the trap, interrupting quantum operations. Consequently, trapped-ion quantum computers typically require ultra-high vacuum (UHV) conditions on the order of $10^{-11}$ Torr to maintain stable ion chains over experimentally relevant timescales. For well baked room temperature systems and cryogenic systems operating at $<20\text{ K}$, the residual background gas is nominally hydrogen $(\text{H}_{2})$. 

Extensive work has gone into estimating the residual hydrogen pressure near trapped ions by studying the rate of hydrogen-ion collisions, modeled using classical Langevin theory \cite{paganoCryogenicTrappedionSystem2018a,wangProbingExtremehighVacuum2025,mickeClosedcycleLowvibration42019,spiveyHighStabilityCryogenicSystem2022,aikyoVacuumCharacterizationCompact2020,hankinSystematicUncertaintyDue2019,hahnRoomTemperatureExtremeHigh2025,dubielzigUltralowvibrationClosedcycleCryogenic2021}. In this work, I will first present a hydrogen pressure estimate at the 4 K stage of the cryostat using cryopumping speed and cryosorption capacity and is covered in Subsection \ref{subsec:pressure-estimate-using-cryopuming-speed-and-cryosorption-capacity}. Next, I will present the ion-neutral molecule interaction and show how a hydrogen-ion collision can be modeled using classical Langevin theory and is covered in Subsection \ref{subsec:pressure-estimate-using-langevin-collisions}. Finally, I will present the experimental setup and measurements performed to estimate the hydrogen pressure near the ions which is covered in Subsections \ref{subsec:experimental-setup-pressure-measurement}, \ref{subsec:pressure-inference-dark-ion-rate}, and \ref{subsec:pressure-inference-from-reordering-rate}.

\subsection{Pressure Estimate Using Cryopumping Speed and Cryosorption Capacity}
\label{subsec:pressure-estimate-using-cryopuming-speed-and-cryosorption-capacity}

An estimate of the pressure at the 4 K stage for this cryostat is made by measuring the pressure at the 300 K stage and calculating the cryopumping speed at the 4 K stage and the vacuum conductance between the 300 K pressure gauge and 4 K stage. As the base temperatures of the 4 K stage and ion trap while trapping an ion is less than 15 K, we can assume the residual background gas is hydrogen $(\text{H}_{2})$. We define the hydrogen gas flow (throughput) from the 300 K stage to the 4 K stage as $Q=C(P_{\text{300\,K}}-P_{\text{4\,K}})$ where $C$ is the vacuum conductance between the stages and the gas cryopumped onto the 4 K stage to be $Q=SP_{\text{4\,K}}$ where $S$ is the cryopumping speed. Setting them equal and rearranging we get,
\begin{equation}
    P_{\text{4\,K}}=\frac{C}{S+C}P_{\text{300\,K}}
    \label{eq:pressure-4K-vs-C-S-300K}
\end{equation}
Plugging Equation \eqref{eq:pressure-4K-vs-C-S-300K} back into $Q=SP_{\text{4\,K}}$, we get that $Q=S_{\text{eff}}P_{\text{300\,K}}$ where $S_{\text{eff}}$ is the effective pumping speed seen at the 300 K stage,
\begin{equation}
    S_{\text{eff}}=\frac{CS}{C+S}
    \label{eq:effective-pumping-speed-at-300K}
\end{equation}
As the the cryopumping speed is almost always greater than the vacuum conductance, $S\gg C$, the effective speed seen at the 300 K stage will be $S_{\text{eff}}\approx C$. 

In this cryostat, the conductance path to the 4 K stage consists of the nipple between the 300 K pressure gauge and chamber followed by multiple apertures in the 40 K and 4 K stage annular plates and window holders, as discussed in Subsection \ref{subsec:300K-chamber} and Section \ref{sec:4K-and-40K-stages}. The conductance was estimated using the molecular flow conductances for a long tube and aperture taking the gas flow species to be 300 K hydrogen. The conductance through a long tube $(d\ll l)$ is given as $C_{\text{tube}}=\frac{\pi}{12}\bar{v}\frac{d^{3}}{l}$ where $\bar{v}$ is the mean speed of 300 K hydrogen in $\frac{\text{m}}{\text{s}}$, $d$ is the diameter of the tube in m, and $l$ is the length of the tube in m. The conductance through an orifice is given as $C=\frac{1}{4}\bar{v}A_{\text{a}}$ where $\bar{v}$ is the mean speed of 300 K hydrogen in $\frac{\text{m}}{\text{s}}$ and $A_{\text{a}}$ is the area of the aperture in $\text{m}^{2}$. Summing the conductances of the nipple and apertures appropriately yields a total conductance of $C=0.06\text{ \ensuremath{\frac{\text{m}^{3}}{\text{s}}}} \left(60\text{ }\frac{\text{L}}{\text{s}}\right)$. The cryopumping speed of hydrogen at the 4 K stage can be estimated using the following equation,
\begin{equation}
    S=\frac{1}{4}A_{\text{4\,K}}\bar{v}\left(1-\frac{P_{\text{vap}}}{P}\right)
    \label{eq:cryopumping-speed}
\end{equation}
where $\bar{v}$ is the mean speed of 4 K hydrogen, $A_{\text{4\,K}}$ is the surface area of the 4 K stage, $P_{\text{vap}}$ is the vapor pressure of hydrogen at 4 K, and $P$ is the initial pressure of hydrogen in the chamber before cooldown \cite{baglinCryopumpingVacuumSystems2020}.\footnote{A capture probability of 1 is assumed here for simplicity.} Taking $A_{\text{4\,K}}=0.7\text{ m}^{2}$, $\bar{v}=205\text{ \ensuremath{\frac{\text{m}}{\text{s}}}}$, $P_{\text{vap}}=10^{-7}\text{ Torr}$, and $P=10^{-6}\text{ Torr}$\footnote{Hydrogen outgassing from the in-vacuum metal surfaces in not included here, so this is a lower bound of the hydrogen pressure.} we get a cryopumping speed of $S=32\text{ }\frac{\text{m}^{3}}{\text{s}} \left(32000\text{ }\frac{\text{L}}{\text{s}}\right)$. Thus, the hydrogen pressure at the 4 K stage using Equation \eqref{eq:pressure-4K-vs-C-S-300K} with $C=0.06\text{ \ensuremath{\frac{\text{m}^{3}}{\text{s}}}}$, $S=32\text{ }\frac{\text{m}^{3}}{\text{s}}$, and $P_{\text{300\,K}}=6.3\times10^{-9}\text{ Torr}$ (base pressure when the system is cold) is $10^{-11}\text{ Torr}$. However, this this pressure estimate is limited by the errors in the conductances and capture probability. In particular, the conductance between the 300 K chamber and cold stages as well the conductance between each cold stage is not well modeled using the simple equations introduced here. Consequently, the actual conductance is most likely lower than $C=0.06\text{ \ensuremath{\frac{\text{m}^{3}}{\text{s}}}}$. A more rigorous simulation using the Monte Carlo numerical method could be used to estimate the total conductance and capture probability, but was not done in this work.\footnote{MolFlow simulation software: molflow.docs.cern.ch.}

Another way of estimating the hydrogen pressure is by calculating the cryosorption capacity of hydrogen on the 4 K stage. A simple calculation is presented here and more information about adsorption surface physics is covered in Appendix \ref{sec:vacuum-fundamentals}. The hydrogen monolayer capacity on a 4 K stainless steel surface has been reported to be $3\times10^{19}\,\frac{\text{H}_{2}}{\text{m}^{2}}$ with an equilibrium pressure of $4\times10^{-11}\text{ Torr}$. Once the surface coverage increases to $10^{20}\,\frac{\text{H}_{2}}{\text{m}^{2}}$, the equilibration pressure saturates at the vapor pressure of $10^{-7}\text{ Torr}$ \cite{baglinCryopumpingVacuumSystems2020,benvenutiINFLUENCETHERMALRADIATION}. However, a sub-monolayer surface coverage of $10^{19}\text{ }\frac{\text{H}_{2}}{\text{m}^{2}}$ yielded an equilibrium pressure of $3\times10^{-12}\text{ Torr}$. So the ultimate pressure is a function of the surface coverage which can be minimized by either having a small gas load or generating a large surface area. Thus, by knowing the initial hydrogen load in the cryostat before cooldown, $P_{\text{H}_{2}}V_{\text{cryostat}}$, we can calculate the required surface area of the 4 K stage to cryosorb hydrogen to an equilibrium pressure of $3\times10^{-12}\text{ Torr}$ as follows,
\begin{equation}
    A_{\text{4\,K}}=\frac{P_{\text{H}_{2}}V_{\text{cryostat}}}{k_{\text{B}}T\times10^{19}\,\frac{\text{H}_{2}}{\text{m}^{2}}}
    \label{eq:4K-SA-to-cryosorb-H2}
\end{equation}
Taking $P_{\text{H}_{2}}=1.3\times10^{-4}\text{ Pa} (10^{-6}\text{ Torr})$, $V_{\text{cryostat}}=0.018\text{ m}^{3} (18 \text{ L})$, and $T=300\text{ K}$, the required surface area of the 4 K stage is $A_{4\,\text{K}}=6\times10^{-5}\text{ m}^{2}$. The 4 K stage surface area in this cryostat is $\approx0.7\text{ m}^{2}$ so there is sufficient surface area to enable cryosorption at 4 K using this model. As a corollary to this, the sticking coefficient of physisorbed hydrogen is most likely less than 1 meaning that it might take a couple of bounces off the 4 K stage before it sticks. Thus, it is desirable to have a 4 K enclosure such that there is more cold surface area and so the hydrogen can bounce multiple times and thus increase the probability of sticking. However, this estimate is also limited by the error in the sorption capacity as this was reported for a stainless steel surface and the majority of the surface area of the 4 K stage for this cryostat is gold-plated copper. To my knowledge, the adsorption isotherm of hydrogen on a gold surface at 4 K has not been measured. 

\subsection{Pressure Estimate Using Langevin Collisions}
\label{subsec:pressure-estimate-using-langevin-collisions}

The most direct way of inferring the pressure near the ions at the 4 K stage is to measure the collision rate between a background gas molecule and a trapped ytterbium ion. As the base temperatures of the 4 K stage while trapping an ion is less than 15 K, we can assume the residual background gas is hydrogen $(\text{H}_{2})$. In general, the ion-neutral molecule interaction is between the electric potential of the induced dipole moment of the neutral molecule by the electric field of the ion and the charge of the ion. In this case, the electric field of a $\text{Yb}^+$ ion, $E_{\text{ion}}=\frac{1}{4\pi\epsilon_{0}}\frac{e}{r^{2}}$, polarizes a neutral hydrogen molecule creating an induced dipole moment, $p=\alpha_{\text{H}_{2}}E_{\text{ion}}$, where $\alpha_{\text{H}_{2}}=\left(4\pi\epsilon_{0}\right)\alpha_{\text{V,H}_{2}}=\left(4\pi\epsilon_{0}\right)\times8.06\times10^{-31}\text{ m}^{3}$ and $\alpha_{\text{V,H}_{2}}$ is the polarizability volume \cite{akindinovaMolecularPolarizabilityQuantum2009}. This induced dipole has an electrical potential given by $V_{\text{dipole}}=-\frac{\alpha e}{2(4\pi\epsilon_{0})^{2}}\frac{1}{r^{4}}$ and the interaction between this dipole and the charge of the $\text{Yb}^+$ ion is
\begin{equation}
    U(r)=-\frac{C_{4}}{r^{4}}
    \label{eq:ion-neutral-molecule-interaction}
\end{equation}
where $C_{4}=\frac{\alpha_{\text{H}_{2}}e^{2}}{2(4\pi\epsilon_{0})^{2}}$ \cite{griffithsIntroductionElectrodynamics2013}. 

\begin{figure}[!!h]
\centering
\includegraphics[width=\textwidth]{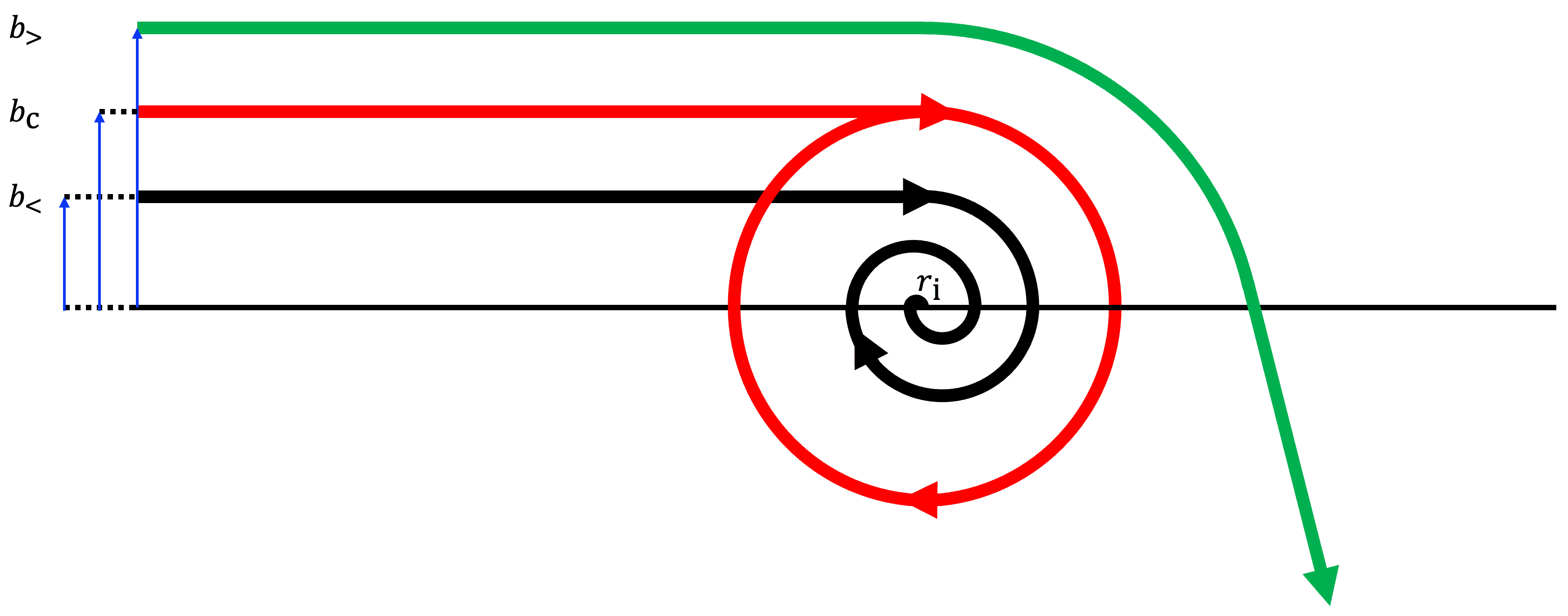}
\caption[Classical Langevin Collision Schematic]{Schematic of the collision between hydrogen and a $\text{Yb}^+$ ion, $\mathbf{r}_{\text{COM}}\approx\mathbf{r}_{\text{i}}$, showing the classical hydrogen trajectories for three impact parameters. For small values of impact parameter $(b_{<})$, the hydrogen follows the trajectory indicated by the black curve spiraling into the ion - these are Langevin collisions. For large values of impact parameter $(b_{>})$, the hydrogen follows the trajectory indicated by the green curve and is deflected away from the ion. At the critical impact parameter $(b_{\text{c}})$, the hydrogen follows the trajectory indicated by the red curve where it orbits the ion. Schematic was inspired from \cite{simbotinQuantumCorrectionLangevin2025}.}
\label{fig:classical-langevin-collision-schematic}
\end{figure}

We will now derive the classical Langevin cross section and describe the nature of the collisions that can occur. It is convenient to treat this collision in the center of mass frame where the location of the center of mass is $\mathbf{r}_{\text{COM}}=\frac{m_{\text{m}}\mathbf{r}_{\text{m}}+m_{\text{i}}\mathbf{r}_{\text{i}}}{m_{\text{m}}+m_{\text{i}}}$ and the reduced mass is $\mu=\frac{m_{\text{m}}m_{\text{i}}}{m_{\text{m}}+m_{\text{i}}}$ where $\mathbf{r}_{\text{m}}, m_{\text{m}}$ and $\mathbf{r}_{\text{i}}$, $m_{\text{i}}$ are the position and mass of the hydrogen molecule and ion respectively. The position of the hydrogen relative to the ion is $\mathbf{r}=\mathbf{r}_{\text{m}}-\mathbf{r}_{\text{i}}$. In this case with $m_{\text{m}}=m_{\text{H}_{2}}$ and $m_{\text{i}}=m_{^{174}\text{Yb}^{+}}$, $m_{\text{m}}\ll m_{\text{i}}$ and so $\mu\approx m_{\text{m}}$, $\mathbf{r}_{\text{COM}}\approx\mathbf{r}_{\text{i}}$, and $\mathbf{r}\approx\mathbf{r}_{\text{m}}$. The effective potential energy of this collision in the center of mass frame is written as follows, 
\begin{equation}
    U_{\text{eff}}(|\mathbf{r}|)=\frac{L^{2}}{2\mu|\mathbf{r}|^{2}}-\frac{C_{4}}{|\mathbf{r}|^{4}}
    \label{eq:effective-potential-energy-ion-neutral}
\end{equation}
where the angular momentum is $L=\mu|\dot{r}|b$ and $b$ is the impact parameter \cite{taylor2005classical}. To find the centrifugal barrier height, we impose $\frac{\text{d}U_{\text{eff}}}{\text{d}r}=0$, solve for $r$, and find that maximum of the effective potential is $U_{\text{eff,max}}=\frac{L^{4}}{16C_{4}\mu^{2}}$. By equating the initial energy, $E_{\text{m}}=\frac{1}{2}\mu|\dot{\mathbf{r}}|^{2}$, to the maximum effective potential energy we can solve for the critical impact parameter $b_{\text{c}}=\left(\frac{8C_{4}}{\mu|\mathbf{\dot{r}}|^{2}}\right)^{1/4}$. Taking $|\mathbf{\dot{r}}|=\bar{v}=205\text{ }\frac{\text{m}}{\text{s}}$ where $\bar{v}$ is the mean speed of 4 K hydrogen, following a Boltzmann distribution, the critical impact parameter is $b_{\text{c}}=1.5\text{ nm}$. Thus, there are long range (glancing) collisions that never cross the centrifugal barrier, $b>b_{\text{c}}$, and short range Langevin collisions, $b<b_{\text{c}}$, where the ion and hydrogen spiral inward towards each other as depicted in Figure \ref{fig:classical-langevin-collision-schematic}. We can write the classical Langevin cross section as follows, 
\begin{equation}
    \sigma_{\text{L}}=\pi b_{\text{c}}^{2}=\pi\sqrt{\frac{8C_{4}}{\mu|\mathbf{\dot{r}}|^{2}}}
    \label{eq:langevin-cross-section}
\end{equation}
and the scattering rate of these collisions assuming a hydrogen density $n=\frac{P_{\text{H}_{2}}}{k_{\text{B}}T}$ in $\frac{\text{H}_{2}}{\text{m}^{3}}$ as
\begin{equation}
    \gamma=n|\mathbf{\dot{r}}|\sigma_{\text{L}}=\frac{P_{\text{H}_{2}}e}{k_{\text{B}}T2\epsilon_{0}}\sqrt{\frac{\alpha_{\text{H}_{2}}}{\mu}}
    \label{eq:langevin-collision-rate}
\end{equation}
Thus, we expect the hydrogen density, or pressure, in the vicinity of the ion to be related to the Langevin collision rate. 

The validity of assuming a classical Langevin cross section is determined by comparing the initial energy of the hydrogen molecule to the centrifugal barrier corresponding to the onset of the quantum threshold regime where only a few partial waves contribute to the scattering. In general, this energy is commonly derived for collision processes by setting the quantum mechanical kinetic energy scale $\frac{\hbar^{2}}{2\mu|\mathbf{r}|^{2}}$ equal to the interaction potential which in this case is $\frac{C_{4}}{|\mathbf{r}|^{4}}$ from Equation \eqref{eq:ion-neutral-molecule-interaction} to get a characteristic length given by, 
\begin{equation}
    R_{4}=\sqrt{\frac{2\mu C_{4}}{\hbar^{2}}}=7.4\text{ nm}
    \label{eq:ion-neutral-molecule-interaction-characteristic-length}
\end{equation}
Thus, the corresponding characteristic energy is given by,
\begin{equation}
    E_{4}=\frac{\hbar^{2}}{2\mu R_{4}^{2}}=190\text{ neV}
    \label{eq:ion-neutral-molecule-interaction-characteristic-energy}
\end{equation}
In general, the impact parameter $b$ can be related to the quantum mechanical angular momentum $l$ by $kb=\sqrt{l(l+1)}$, where $k(E)=\sqrt{\frac{2\mu E}{\hbar^{2}}}$. Taking the impact parameter to be the characteristic length $b=R_{4}$ and evaluating the wavenumber at the characteristic energy $k(E_{4})$ we get that,
\begin{equation}
    k(E_{4})R_{4}=1\approx l_{\text{max}}
    \label{eq:characteristic-angular-momentum}
\end{equation}
Thus, the characteristic length and energy are defined such that the maximum angular momentum is of the order 1 by convention \cite{jachymskiQuantumTheoryReactive2013a}. This is implies $s$-wave $(l=0)$ and $p$-wave $(l=1)$ partial wave contributions and is the onset of the quantum threshold regime. 

The initial energy of a hydrogen molecule at 4 K in this collision can be taken to be its average energy following a Boltzmann distribution $E_{\text{m}}=\frac{3}{2}k_{\text{B}}\times4\,\text{K}=0.5\text{ meV}$. Thus, the system lies far above the quantum threshold regime as $k(E_{\text{m}})R_{4}=\sqrt{\frac{E_{\text{m}}}{E_{4}}}\approx52$ implying that many partial waves contribute in general. As adjacent partial waves differ by $\Delta l=1$, the corresponding change in impact parameter is $\Delta b=\frac{1}{k}$, so the number of impact parameter bins in this case is $\approx52$ and is on the onset of the continuous classical continuum of trajectories and not the few partial wave quantum threshold regime. Therefore, a classical Langevin description of the short range collision rate is a reasonable approximation. However, if information about the exact angular distribution of the scattered hydrogen is desired, a full quantum mechanical partial wave treatment is required. 

These Langevin collisions are either elastic or inelastic. For elastic collisions, $\Delta\mathit{KE}_{\text{tot}}=0$, kinetic energy and momentum are redistributed between the ion and hydrogen. When a background gas molecule collides elastically with one ion, the ion receives a sudden momentum kick that displaces it in the trap potential. This energy initially appears as increased secular motion of that ion but rapidly redistributes throughout the chain via the Coulomb coupling, heating the collective motional modes of the crystal. If the ions are being cooled, the excess energy is gradually removed and the crystal returns to equilibrium. If collisions occur infrequently, the crystal briefly heats and then re-cools; however, at higher collision rates the repeated energy kicks can disrupt the ordered structure of the chain, leading to increased motional heating or even melting of the Coulomb crystal. The benefit of being cryogenic is that the 4 K hydrogen collisions are factor of 75 lower than the 300 K collisions due to the temperature dependence of the average energy, following a Boltzmann distribution, making crystal melting less likely. 

\begin{figure}[!!h]
\centering
\includegraphics[width=0.75\textwidth]{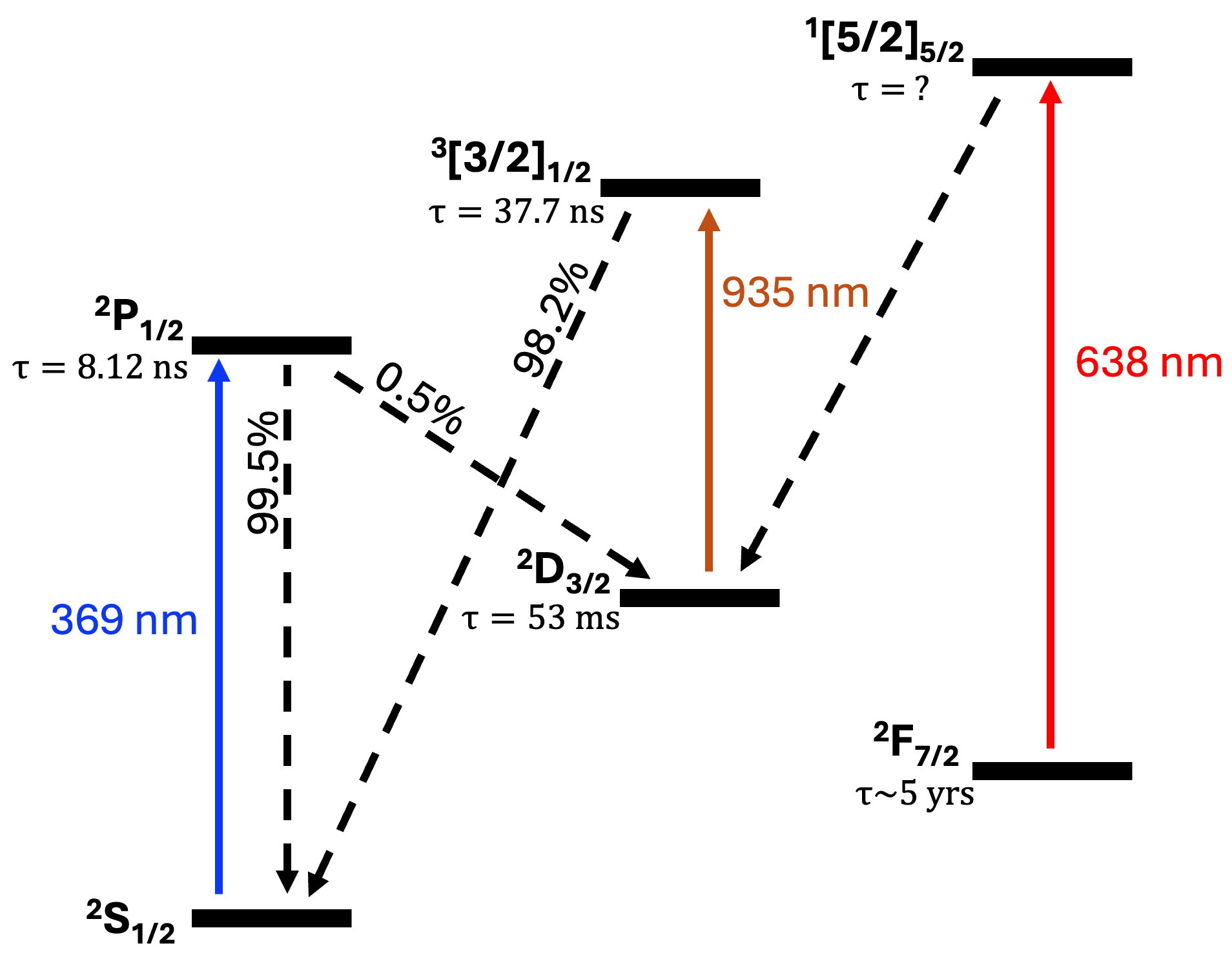}
\caption[$^{174}\text{Yb}^{+}$ Doppler Cooling Level Diagram 2]{$^{174}\text{Yb}^+$ ion level diagram relevant for laser cooling. The cooling beams 369 nm and 935 nm are only required unless collision-induced shelving into the $^{2}\text{F}_{7/2}$ occurs at which point the 638 nm repumps the ion back into the cooling cycle. The branching ratio from $^{1}[5/2]_{5/2}$ to $^{2}D_{3/2}$ is unknown. }
\label{fig:174Yb-ion-doppler-cooling-level-diagram-2}
\end{figure}

Alternatively, for inelastic collisions, $\Delta\mathit{KE}_{\text{tot}}\neq0$, the internal state of the ion and/or hydrogen changes or a chemical reaction occurs changing the identity of the ion and molecule. In particular while laser cooling $^{174}\text{Yb}^+$ ions, the optically excited $^{174}\text{Yb}^{+*}$ ions spend time in the $^{2}\text{P}_{1/2}$ and $^{2}\text{D}_{3/2}$ states as shown in Figure \ref{fig:174Yb-ion-doppler-cooling-level-diagram-2}. This is relevant, as it has been reported that shelving into the $^{2}\text{F}_{7/2}$ state and molecule association of $\text{YbH}^{+}$ occur due collisions between optically excited $^{174}\text{Yb}^{+*}$ ions in the $^{2}\text{P}_{1/2}$ and $^{2}\text{D}_{3/2}$ states and hydrogen \cite{lehmitzPopulationTrappingExcited1989,sugiyamaProductionYbHChemical1997}. For the former, the collision with hydrogen perturbs the ion's electronic structure making states that are normally weakly coupled or forbidden by selection rules partially mixed. This collisional mixing effectively relaxes the selection rules allowing population in the excited $^{2}\text{P}_{1/2}$ and $^{2}\text{D}_{3/2}$ manifolds to be transferred into higher-lying states that subsequently decay into the $^{2}\text{F}_{7/2}$ state. As the $^{2}\text{F}_{7/2}$ state is extremely long-lived, $\sim5$ years, the ions remain trapped there until they are repumped using the 638 nm laser. For the latter, the collision with hydrogen has enough energy to induce a chemical reaction forming $\text{YbH}^{+}$. In both cases, these dark ions remain trapped and can remain cold via sympathetic cooling between other bright (laser-cooled) ions.

\subsection{Experimental Setup}
\label{subsec:experimental-setup-pressure-measurement}

\begin{figure}[!!h]
\centering
\includegraphics[width=\textwidth]{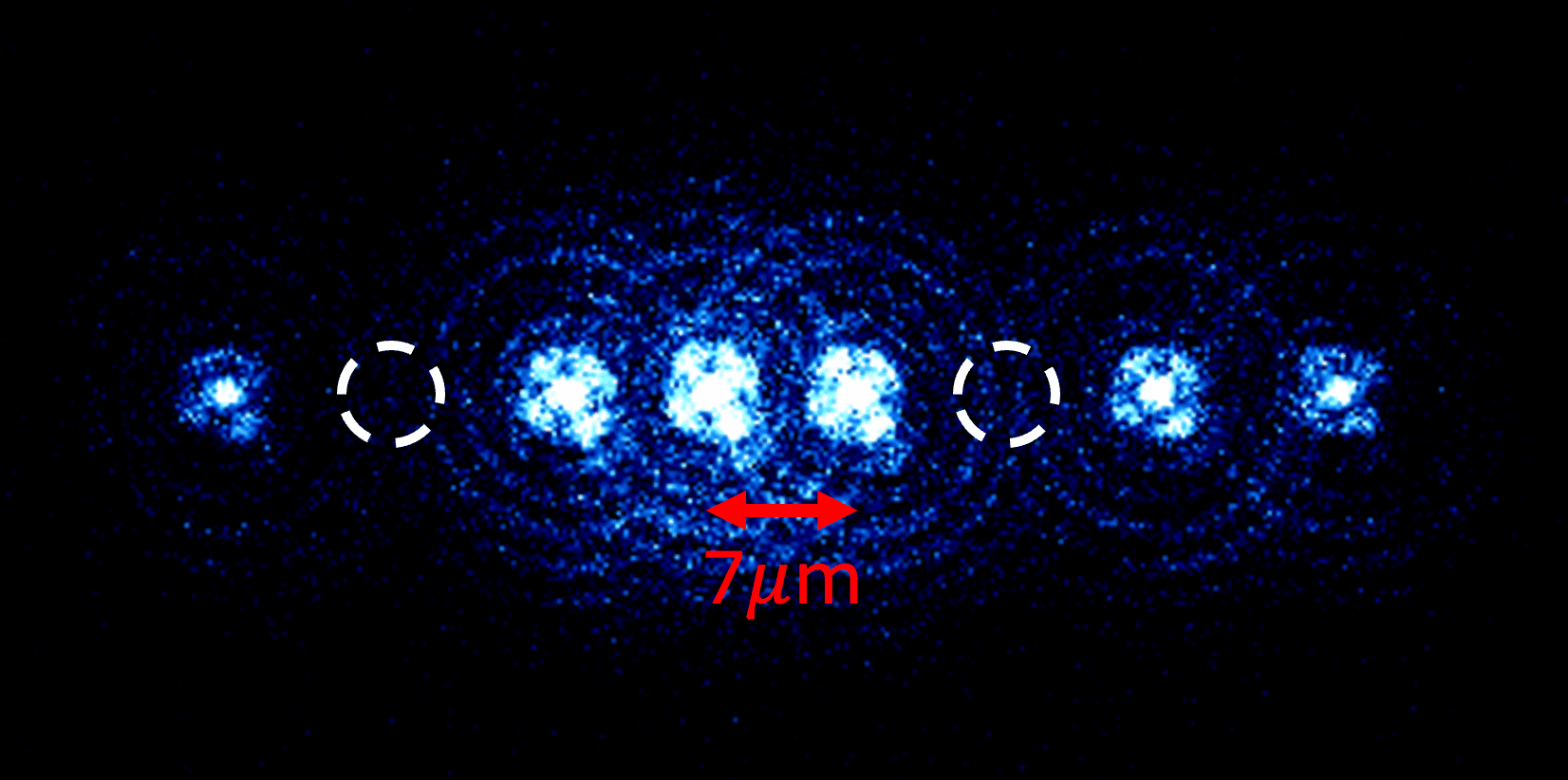}
\caption[Camera Image of 8 Trapped $\text{Yb}^+$ Ions]{Camera image of 8 trapped $\text{Yb}^{+}$ ions: $6\times^{174}\text{Yb}^+$ bright ions, $1\times^{174}\text{Yb}^+$ dark ion, and another unknown isotope. The positions of the dark ions are marked with white circles. The trap secular frequencies are $\left(\frac{\omega_{x}}{2\pi},\frac{\omega_{y}}{2\pi},\frac{\omega_{z}}{2\pi}\right)=\left(570,500,120\right)\text{ kHz}$.}
\label{fig:camera-image-of-ions-pressure-measurement}
\end{figure}

In this experiment, we measured the rate at which trapped $^{174}\text{Yb}^+$ ions go dark and the rate of reordering events in the ion chain, where we assume these events are contributed to collisions between the residual hydrogen and the trapped $^{174}\text{Yb}^+$ ions at the 4 K stage. We assume that the dark ion rate is related to the inelastic collision rate, $\gamma_{\text{dark}}=p_{\text{in}}\gamma$, while the reordering rate is related to the elastic collision rate, $\gamma_{\text{reorder}}=p_{\text{el}}\gamma$, where $p_{\text{in}}$ and $p_{\text{el}}$ are the probabilities that an inelastic and elastic collision occur respectively and $\gamma$ is the Langevin collision rate given by Equation \eqref{eq:langevin-collision-rate}. To conduct this experiment, we first loaded 8 $\text{Yb}^+$ ions into the trap: $7\times ^{174}\text{Yb}^+$ ions and 1 unknown isotope. This was done by keeping the 399 nm laser near-resonant on the $^{174}\text{Yb}$ $^1\text{S}_0-^1\text{P}_1$ transition and increasing the optical power to increase the broadening, which is covered in Section \ref{sec:photoionization-of-Yb}. The Doppler cooling beams 369 nm and 935 nm were applied and the fluorescence captured on the camera provided information about the locations of the ions, while the locations of the dark ions can be inferred from the spacing between the bright ions as shown in Figure \ref{fig:camera-image-of-ions-pressure-measurement}. We kept the Doppler cooling beams 369 nm and 935 nm on with a saturation parameter $s=9$ and the $^{2}\text{F}_{7/2}$ state repump beam 638 nm off during the duration of the measurement while monitoring the dark ion rate and reordering rate. We measured these rates with trap secular frequencies $\omega_{x}=2\pi\times570\text{ kHz}$, $\omega_{y}=2\pi\times500\text{ kHz}$, and $\omega_{z}=2\pi\times120\text{ kHz}$ in two temperature configurations: 4 K stage = 5.7 K, Trap Mount = 13.1 K and 4 K stage = 10.0 K, Trap Mount = 16.2 K where 1.6 W was applied to the Trap Riser using a heater. 

The images of the ion chain were taken every 10 s with a 5 s exposure over a total time $T_{\text{total}}$. Assuming that these events are equally likely at any time and uncorrelated, we can write the dark ion rate, reordering rate, and uncertainty as,
\begin{equation}
    \gamma_{\text{dark,reorder}}=\frac{N_{\text{obs}}}{T}\pm\frac{\sqrt{N_{\text{obs}}}}{T_{\text{total}}}
    \label{eq:scattering-rate-analysis-uncertainty}
\end{equation}
where $N_{\text{obs}}$ is the number of dark ion or reordering events observed during the observation window $T_{\text{tot}}$. 

\subsection{Pressure Inference from Dark Ion Rate}
\label{subsec:pressure-inference-dark-ion-rate}

To relate the measured dark ion rate to the hydrogen background pressure, we need to use a relative measurement as the probability of an inelastic collision, $p_{\text{in}}$, is unknown. In \cite{paganoCryogenicTrappedionSystem2018a} our colleague measured the dark ion rate in $^{171}\text{Yb}^{+}$ in an ion trap at 300 K where the ion gauge pressure was $P_{\text{H}_{2}}^{\text{300\,K}}=10^{-11}\text{ Torr}$ and the dark ion rate was $\gamma_{\text{dark}}^{300\,\text{K}}=2\times10^{-4}\text{ s}^{-1}$ with a saturation parameter for the Doppler cooling laser of $\approx1$. While we don't know the probability of an inelastic scattering event, $P_{\text{in}}$, we assume it is temperature independent which allows us to eliminate $p_{\text{in}}$ in Equation \eqref{eq:langevin-collision-rate} and compute the hydrogen pressure $P_{\text{H}_{2}}^{T}$ as follows,
\begin{equation}
    P_{\text{H}_{2}}^{T}=\frac{P_{\text{H}_{2}}^{\text{300\,K}}\gamma_{\text{dark}}^{T}T}{\gamma_{\text{dark}}^{300\,\text{K}}\times300\,\text{K}}C_{s}
    \label{eq:H2-pressure-inelastic}
\end{equation}
where $T$ is the temperature of the Trap Mount in K while measuring the dark ion rate and $C_{s}=\frac{\rho_{^{2}\text{P}_{1/2}}^{\text{300\,K}}(1)}{\rho_{^{2}\text{P}_{1/2}}^{T}(9)}=0.4$ is an additional correction factor to account for the different saturation parameters, as the excited state population is a function of the saturation parameter, $\rho_{^{2}\text{P}_{1/2}}(s)=\frac{s}{2(2+s)}$ with $\Delta=-\frac{\Gamma}{2}$ \cite{footAtomicPhysics2005}. 

For the temperature configuration with the 4 K stage = 5.7 K, Trap Mount = 13.1 K, the number of dark ion events observed was $N_{\text{obs}}=5$ over a time duration of 10.7 hours $T_{\text{tot}}=38660\text{ s}$ yielding an average dark ion rate of $\gamma_{\text{dark}}^{\text{13.1\,K}}=2.6\times10^{-4}\pm1.2\times10^{-4}\text{ s}^{-1}$. Using $\gamma_{\text{dark}}^{\text{13.1\,K}}$ and $T=13.1\text{ K}$ in Equation \eqref{eq:H2-pressure-inelastic}, we infer a hydrogen pressure of $P_{\text{H}_{2}}^{\text{13.1\,K}}\sim2\times10^{-13}\text{ Torr}$. For the temperature configuration with the 4 K stage = 10.0 K, Trap Mount = 16.2 K, the number of dark ion events observed was $N_{\text{obs}}=5$ over a time duration of 4.2 hours $T_{\text{tot}}=15200\text{ s}$ yielding an average dark ion rate of $\gamma_{\text{dark}}^{\text{16.2\,K}}=6.6\times10^{-4}\pm2.9\times10^{-4}\text{ s}^{-1}$. Using $\gamma_{\text{dark}}^{\text{16.2\,K}}$ and $T=16.2\text{ K}$ in Equation \eqref{eq:H2-pressure-inelastic}, we infer a hydrogen pressure of $P_{\text{H}_{2}}^{\text{16.2\,K}}\sim7\times10^{-13}\text{ Torr}$. However, the saturation parameter and dark ion rate reported at 300 K in \cite{paganoCryogenicTrappedionSystem2018a} have an unknown uncertainty and thus we take this pressure inference using Equation \eqref{eq:H2-pressure-inelastic} as just an estimate. The dark ion rates and corresponding pressures for each configuration are summarized in Table \ref{tab:dark-ion-rate-pressure-inference}. 

\begin{table}[!!h]
\centering
\tablefont
\caption[Pressure Inference from Dark Ion Rate]{Reported dark ion rates and inferred pressures computed using Equation \eqref{eq:H2-pressure-inelastic} for each temperature configuration.}
\label{tab:dark-ion-rate-pressure-inference}
\begin{tabularx}{\textwidth}{>{\centering\arraybackslash}p{0.2\textwidth}>{\centering\arraybackslash}p{0.2\textwidth}cc}
\toprule
\makecell{\textbf{Temperature of}\\\textbf{4 K Stage}} & \makecell{\textbf{Temperature of}\\\textbf{Trap Mount}} & \textbf{Dark Ion Rate, }$\mathbf{\gamma_{\textbf{dark}}\text{ }(\textbf{s}^{-1})}$ & \textbf{Pressure, }$\mathbf{P_{\textbf{H}_2}^T}$\textbf{ (Torr)}\\
\midrule
5.7 & 13.1 & $(2.6\pm1.2)\times10^{-4}$ & $2\times10^{-13}$ est \\
\midrule
10.0 & 16.2 & $(6.6\pm2.9)\times10^{-4}$ & $7\times10^{-13}$ est \\
\bottomrule
\end{tabularx}
\end{table}

\subsection{Pressure Inference from Reordering Rate}
\label{subsec:pressure-inference-from-reordering-rate}

\begin{figure}[!!h]
\centering
\includegraphics[width=\textwidth]{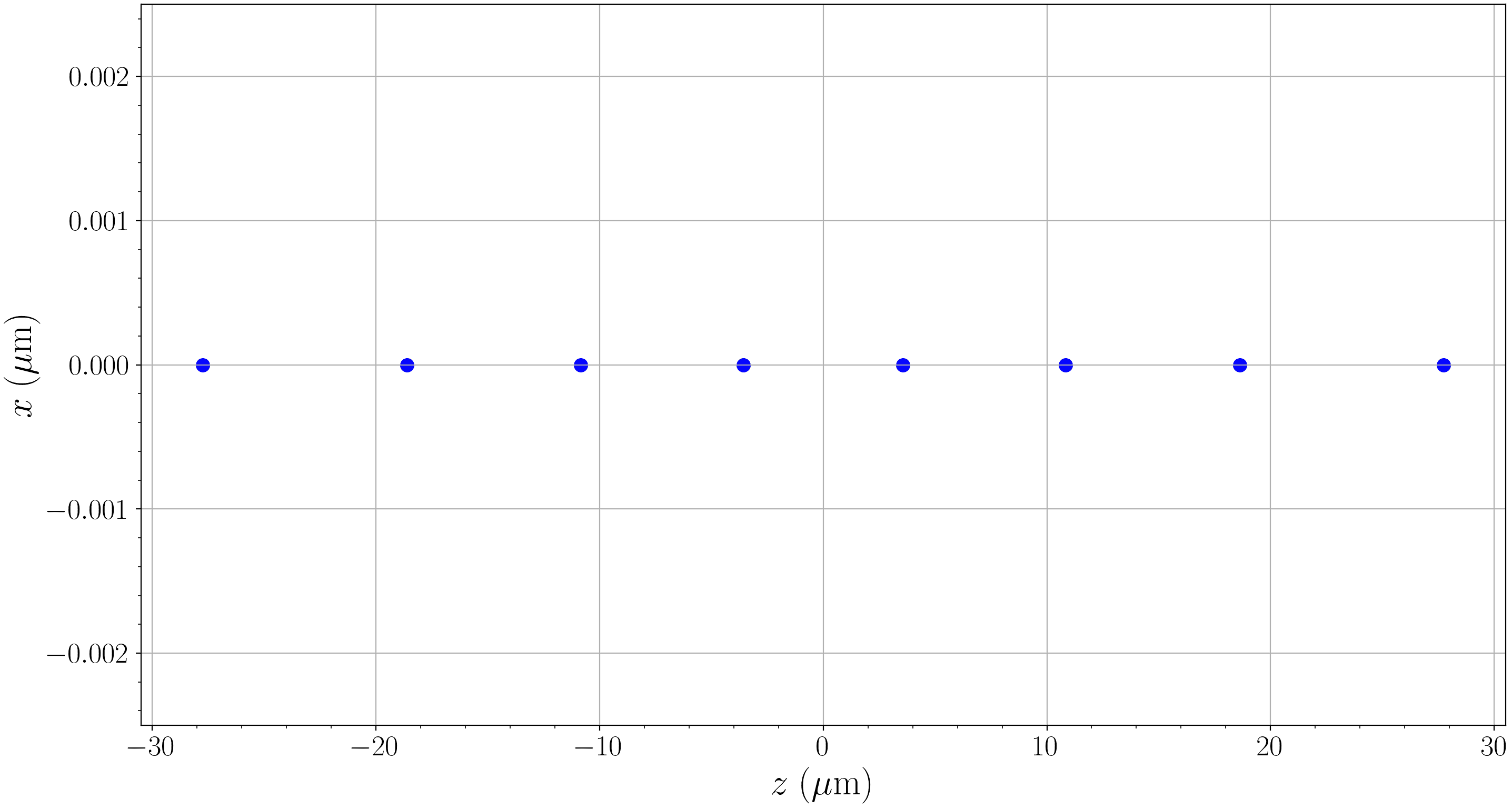}
\caption[Ion Chain Equilibrium Positions]{Plot of ion chain equilibrium positions in the $x,z$ plane that minimize the total energy given by Equation \eqref{eq:ion-harmonic-trap-plus-Coulombg-potential-energy}. The minimized total energy is $U=3.8\text{ meV}$ and the equilibrium positions are in agreement with the positions on the camera image in Figure \ref{fig:camera-image-of-ions-pressure-measurement} given a spatial resolution of $129\text{ }\frac{\text{nm}}{\text{px}}$ computed in Subsection \ref{subsec:spatial-resolution-of-imagin-system}.}
\label{fig:Ion-chain-equilibrium-positions}
\end{figure}

To relate the measured reordering rate to the hydrogen background pressure, we first need to calculate the probability that an elastic collision occurs, i.e., that a reordering event occurs, $p_{\text{el}}=p_{\text{reorder}}$. This was done by solving for the total energy of the ion chain during an ion position swapping event following the work done in \cite{aikyoVacuumCharacterizationCompact2020,hahnRoomTemperatureExtremeHigh2025}. The total energy of an ion crystal is the sum of the potential due to the trap and the potential due to the Coulomb interaction between each ion in the chain,
\begin{equation}
    U=\frac{1}{2}m\sum_{i=1}^{N}\left(\omega_{x}^{2}x_{i}^{2}+\omega_{y}^{2}y_{i}^{2}+\omega_{z}^{2}z_{i}^{2}\right)+\frac{e^{2}}{4\pi\epsilon_{0}}\sum_{i<j}\frac{1}{\left|\mathbf{r}_{i}-\mathbf{r}_{j}\right|}
    \label{eq:ion-harmonic-trap-plus-Coulombg-potential-energy}
\end{equation}
where $\omega_{x}=2\pi\times570\text{ kHz}$, $\omega_{y}=2\pi\times500\text{ kHz}$ (radial), and $\omega_{z}=2\pi\times120\text{ kHz}$ (axial) are the trap secular frequencies, the position of each ion in the chain is $\mathbf{r}_{i}=(x_{i},y_{i},z_{i})$, and the relative position between each ion is $\left|\mathbf{r}_{i}-\mathbf{r}_{j}\right|$. We first compute the equilibrium positions that minimize the total energy of the chain to $U=3.8\text{ meV}$ as shown in Figure \ref{fig:Ion-chain-equilibrium-positions}. We then constrain the neighboring ions such that they have the same axial positions, $z_{i}=z_{i+1}$, and then freely vary the $3N-2$ coordinates to compute the equilibrium positions that minimize the total energy, $U(z_{i}=z_{i+1})$. Thus, we can assert that the energy difference $E_{\text{barrier,}i}=U(z_{i}=z_{i+1})-U$ is the minimum energy required to cause a reordering event at the $i^{\text{th}}$ ion provided it happens slow relative to a secular period - adiabatic limit. 

\begin{figure}[!!h]
\centering
\includegraphics[width=\textwidth]{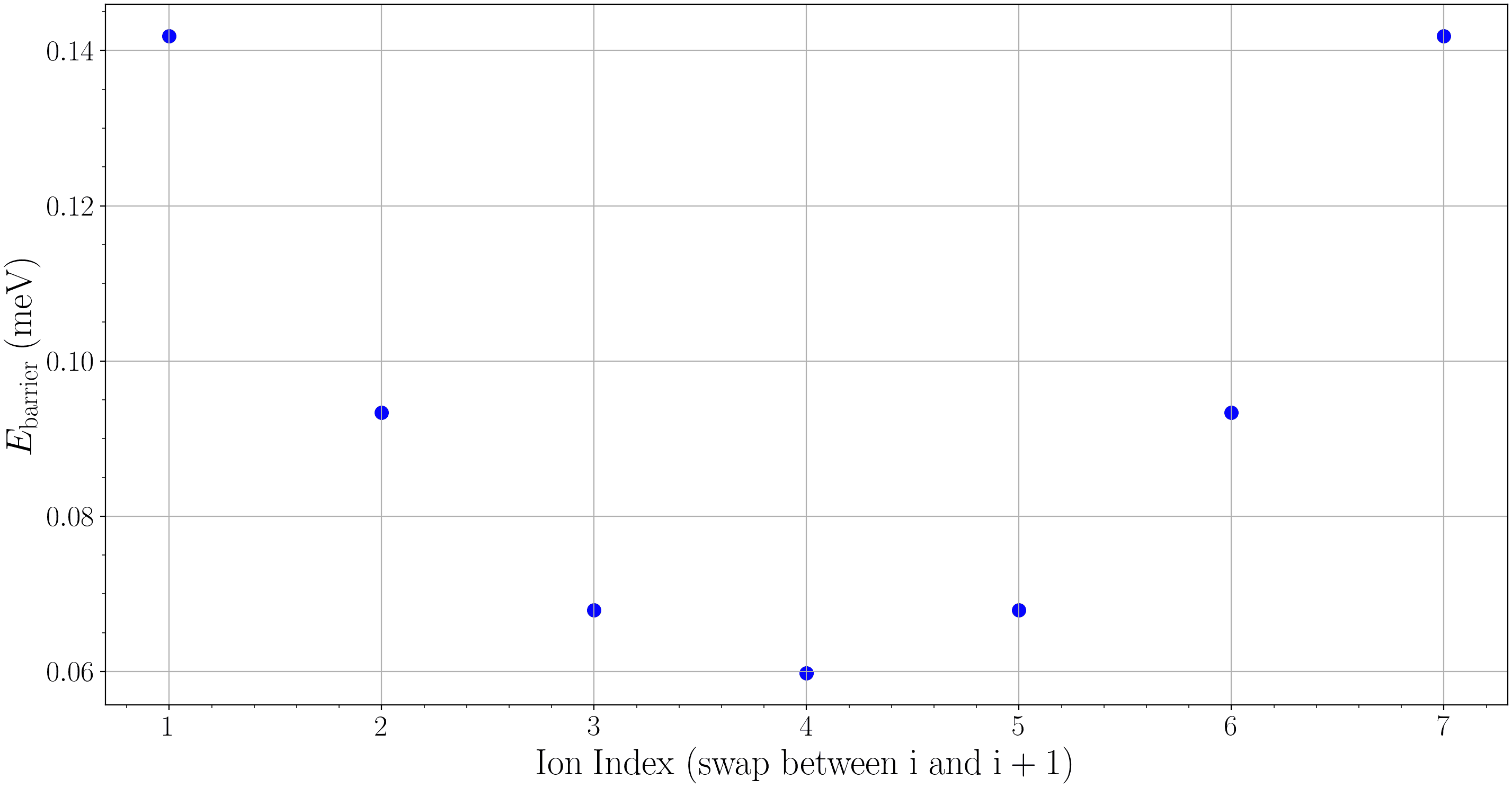}
\caption[Reordering Event Energy Barrier vs. Ion Position Index in Chain]{Plot of energy barrier versus ion position index in chain. For each point, we compute the equilibrium positions that minimize the total energy $U(z_{i}=z_{i+1})$ by constraining neighboring ions such they have the same axial position.}
\label{fig:reordering-event-barrier-energy-vs-ion-index}
\end{figure}

To solve this problem, we first define a characteristic length scale following the work in \cite{jamesQuantumDynamicsCold1998} to be,
\begin{equation}
    l_{0}=\left(\frac{e^{2}}{4\pi\epsilon_{0}m\omega_{z}^{2}}\right)^{1/3}
    \label{eq:characterisitc-length-ion-sim}
\end{equation}
and rescale the coordinates as $\mathbf{r}_{i}=l_{0}\mathbf{\mathbf{\rho}}_{i}$ such that the total energy in Equation \eqref{eq:ion-harmonic-trap-plus-Coulombg-potential-energy} can be written in dimensionless form as,
\begin{equation}
    U(\mathbf{\mathbf{\rho}})=\frac{1}{2}\sum_{i=1}^{N}\left(\alpha_{x}\rho_{i,x}^{2}+\alpha_{y}\rho_{i,y}^{2}+\rho_{i,z}^{2}\right)+\sum_{i<j}\frac{1}{|\mathbf{\rho}_{i}-\mathbf{\rho}_{j}|}
    \label{eq:dimensionless-ion-sim-energy}
\end{equation}
where $\alpha_{x}=\left(\frac{\omega_{x}}{\omega_{z}}\right)^{2}$, $\alpha_{y}=\left(\frac{\omega_{y}}{\omega_{z}}\right)^{2}$, and the corresponding physical energy is related to the dimensionless energy by,
\begin{equation}
    U_{\text{physical}}=\frac{e^{2}}{4\pi\epsilon_{0}l_{0}}U(\mathbf{\rho})
    \label{eq:physical-energy-ion-sim}
\end{equation}
To estimate the ion reordering barrier for a selected neighboring pair $(i,j)$, we compute a constrained equilibrium configuration in which the axial separation is fixed to $\rho_{j,z}-\rho_{i,z}=\Delta z_{\text{target}}$. In the barrier calculation, this is evaluated with $\Delta z_{\text{target}}=0$ forcing the two ions into the same axial plane. We then introduce a quadratic penalty term and minimize the modified objective, 
\begin{equation}
    U_{\text{pen}}(\mathbf{\rho})=U(\mathbf{\rho})+\frac{\lambda}{2}(\rho_{j,z}-\rho_{i,z}-\Delta z_{\text{target}})^{2}
    \label{eq:penalty-objective-ion-sim}
\end{equation}
where $\lambda=10^{6}$ is a large positive penalty parameter. In the limit as $\lambda\rightarrow\infty$, the solution approaches the exact constrained minimum. Thus, we are letting the ion chain adiabatically relax to its equilibrium while holding two ions together along the axial direction with a stiff spring. To improve numerical stability, the axial coordinates of the selected pair are adjusted so that the initial guess satisfies the target separation: $\rho_{\text{mid}}=\frac{\rho_{i,z}+\rho_{j,z}}{2}$, $\rho_{i,z}\leftarrow\rho_{\text{mid}}-\frac{\Delta z_{\text{target}}}{2}$, and $\rho_{j,z}\leftarrow\rho_{\text{mid}}+\frac{\Delta z_{\text{target}}}{2}$. The penalized potential $U_{\text{pen}}(\mathbf{\rho})$ is then minimized and the resulting coordinates correspond to a configuration that minimizes the original potential while approximately satisfying the axial constraint. The reordering barrier for each pair $(i,j)$ is defined as $E_{\text{barrier,}i}=U(z_{i}=z_{i+1})-U$ and is plotted in Figure \ref{fig:reordering-event-barrier-energy-vs-ion-index}. We find that the maximum of all reordering energies is $E_{\text{barrier,max}}=\text{max\{}E_{\text{barrier,}i}\}=0.142\text{ meV}$ and use this as the threshold energy required to cause a reordering event. 

The average energy gained by the ion during an elastic collision with hydrogen is calculated using basic kinematics as done in \cite{aikyoVacuumCharacterizationCompact2020,hahnRoomTemperatureExtremeHigh2025,spiveyHighStabilityCryogenicSystem2022}. I will analyze this collision by taking the center of mass to be at rest, then compute the final velocity of the ion in the lab frame, and finally compute the average energy gained by the ion during the collision as a function of the initial energy of the hydrogen. We can write the position of the hydrogen and ion in the lab frame as $\mathbf{r}_{\text{m}}=\mathbf{r}_{\text{COM}}+\frac{m_{\text{i}}}{M}\mathbf{r}$ and $\mathbf{r}_{\text{i}}=\mathbf{r}_{\text{COM}}-\frac{m_{\text{m}}}{M}\mathbf{r}$ where $M=m_{\text{i}}+m_{\text{m}}$ and $\mathbf{r}=\mathbf{r}_{\text{m}}-\mathbf{r}_{\text{i}}$ is the position of the hydrogen relative to the ion. Similarly, the velocities of the hydrogen and the ion in lab frame are $\mathbf{\dot{r}}_{\text{m}}=\mathbf{\dot{r}}_{\text{COM}}+\frac{m_{\text{i}}}{M}\mathbf{\dot{r}}$ and $\mathbf{\dot{r}}_{\text{i}}=\mathbf{\dot{r}}_{\text{COM}}-\frac{m_{\text{m}}}{M}\mathbf{\dot{r}}$. Before the collision, the velocity of the ion in the lab frame is $\mathbf{\dot{r}}_{\text{i}}=0$ and thus the velocity of the center of mass in the lab frame is $\mathbf{\dot{r}}_{\text{COM}}=\frac{m_{\text{m}}}{M}\mathbf{\dot{r}}$. Additionally, the momenta of each particle before the collision in the center of mass frame are equal but in opposite directions with magnitude $\mu\mathbf{|\dot{r}}|$ where $\mu=\frac{m_{\text{m}}m_{\text{i}}}{m_{\text{m}}+m_{\text{i}}}$ is the reduced mass. After the collision the magnitude of the momentum stays the same but the direction rotates by the scattering angle $\theta$, $\mu|\mathbf{\dot{r}}|\left(\cos\theta\hat{x}-\sin\theta\hat{y}\right)$ where the sign in the $y$ component is simply dependent on the choice of coordinate system \cite{taylor2005classical}. Thus, we can now write the final velocity of the ion in the lab frame to be,
\begin{equation}
    \mathbf{\dot{r}}_{\text{i}}^{\prime}=\frac{m_{\text{m}}|\mathbf{\dot{r}}|}{m_{\text{m}}+m_{\text{i}}}\left(1-\cos\theta\right)\hat{x}+\frac{m_{\text{m}}|\mathbf{\dot{r}}|}{m_{\text{m}}+m_{\text{i}}}\sin\theta\hat{y}
    \label{eq:final-velocity-of-ion-H2-collision}
\end{equation}
and the energy gained by the ion is,
\begin{equation}
    E_{\text{i}}^{\prime}(\theta)=\frac{1}{2}m_{\text{i}}|\mathbf{\dot{r}}_{\text{i}}^{\prime}|^{2}=\frac{m_{\text{i}}m_{\text{m}}^{2}|\mathbf{\dot{r}}|^{2}}{\left(m_{\text{m}}+m_{\text{i}}\right)^{2}}\left(1-\cos\theta\right)
    \label{eq:ion-energy-increase-H2_collision}
\end{equation}
Assuming isotropic scattering over the entire sphere, we get that the probability density for a solid angle $\Omega$ is $P(\Omega)=\frac{1}{4\pi}$ so that $\int_{\Omega}P(\Omega)\text{d}\Omega =1$. To get the probability density in $\theta$  alone, we integrate over $\phi$  to such that $\int_{\Omega}P(\Omega)\text{d}\Omega=\int_{0}^{2\pi}\int_{0}^{\pi}P(\Omega)\sin\theta\text{d}\theta\text{d}\phi=\frac{1}{4\pi}\int_{0}^{2\pi}\int_{0}^{\pi}\sin\theta\text{d}\theta\text{d}\phi=\int_{0}^{\pi}\frac{\sin\theta}{2}\text{d}\theta=1$, and so the probability distribution in $\theta$  is $P(\theta)=\frac{\sin\theta}{2}$. We can now compute the average energy gained by the ion averaging over all angles assuming isotropic scattering to be,
\begin{equation}
    \left\langle E_{\text{i}}^{\prime}\right\rangle _{\theta}=\int_{0}^{\pi}E_{\text{i}}^{\prime}(\theta)P(\theta)\text{d}\theta=\frac{m_{\text{i}}m_{\text{m}}^{2}|\mathbf{\dot{r}}|^{2}}{\left(m_{\text{m}}+m_{\text{i}}\right)^{2}}\approx\frac{2m_{\text{i}}m_{\text{m}}}{\left(m_{\text{m}}+m_{\text{i}}\right)^{2}}E_{\text{m}}
    \label{eq:average-ion-energy-increase-H2_collision}
\end{equation}
where we we take the initial energy of hydrogen to be $E_{m}\approx\frac{1}{2}m_{\text{m}}|\mathbf{\dot{r}_{\text{m}}}|^{2}$ since $\mu\approx m_{\text{m}}$, $\mathbf{r}_{\text{COM}}\approx\mathbf{r}_{\text{i}}$, and $\mathbf{r}\approx\mathbf{r}_{\text{m}}$. Modeling the energy of the residual hydrogen as a Boltzmann distribution, as done in \cite{aikyoVacuumCharacterizationCompact2020,hahnRoomTemperatureExtremeHigh2025,spiveyHighStabilityCryogenicSystem2022}, we can calculate the probability that a hydrogen-ion collision causes a reordering event $p_{\text{reorder}}$. This is equivalent to calculating fraction of hydrogen molecules at temperature $T$ that have an energy of at least $E_{\text{m}}=\frac{E_{\text{barrier,max}}}{\frac{2m_{\text{i}}m_{\text{m}}}{\left(m_{\text{m}}+m_{\text{i}}\right)^{2}}}=6.3\text{ meV}$ using Equation \eqref{eq:average-ion-energy-increase-H2_collision} and is given as follows, 
\begin{equation}
    p_{\text{reorder}}(E_{\text{m}},T)=\frac{N([E_{\text{m}},\infty],T)}{N_{\text{total}}}=\int_{E_{\text{m}}}^{\infty}\frac{2}{\sqrt{\pi}(k_{\text{B}}T)^{3/2}}\sqrt{E}e^{-\frac{E}{k_{\text{B}}T}}dE
    \label{eq:probability-of-reordering-event}
\end{equation}
To compute the integral in Equation \eqref{eq:probability-of-reordering-event}, we can define $x=\frac{E_{\text{m}}}{k_{\text{B}}T}$ and rewrite the integral as follows,
\begin{equation}
    \text{erfc}(\sqrt{x})+\frac{2}{\sqrt{\pi}}\sqrt{x}e^{-x}
    \label{eq:probability-of-reordering-event-integral-as-erfc}
\end{equation}
where $\text{erfc}(z)=\frac{2}{\sqrt{\pi}}\int_{z}^{\infty}e^{-t^{2}}dt$ is the complementary error function.

We also need to calculate the probability that a reordering event is distinguishable on the camera image as done in \cite{hahnRoomTemperatureExtremeHigh2025}. There are $\binom{N}{n}=\frac{N!}{n!(N-n)!}$ ways to place $n$ dark ions among $N$ positions in an $N$-ion chain. After a collision, all configurations are equally likely. As there are $\binom{N}{n}$ possible configurations and only 1 of them looks identical to the original configuration before the collision, the probability that the system looks unchanged is $1/\binom{N}{n}$. Thus, the probability that you see a reordering event is 
\begin{equation}
    p_{\text{obs}}=1-\frac{1}{\binom{N}{n}}=1-\frac{n!(N-n)!}{N!}
    \label{eq:probability-of-reordering-distinguishable}
\end{equation}
Finally, using $\gamma_{\text{reorder}}=p_{\text{reorder}}p_{\text{obs}}\gamma$  where $\gamma$ is the Langevin collision rate from Equation \eqref{eq:langevin-collision-rate}, we can compute the hydrogen pressure in the vicinity of the ions as follows,
\begin{equation}
    P_{\text{H}_{2}}=\frac{\gamma_{\text{reorder}}k_{\text{B}}T2\epsilon_{0}}{p_{\text{reorder}}p_{\text{obs}}e}\sqrt{\frac{\mu}{\alpha_{\text{H}_{2}}}}
    \label{eq:H2-pressure-elastic}
\end{equation}

For the temperature configuration with the 4 K stage = 5.7 K, Trap Mount = 13.1 K, there were no reordering events observed $N_{\text{obs}}=0$ over a time duration of 10.7 hours $T_{\text{tot}}=38660\text{ s}$ yielding an average reordering rate of $\gamma_{\text{reorder}}^{\text{13.1\,K}}=0\text{ s}^{-1}$ and so the hydrogen pressure cannot be estimated for this configuration using Equation \eqref{eq:H2-pressure-elastic}. However, the probability of observing no events $N_{\text{obs}}$ given the Poisson distribution $p(N_{\text{obs}})=\frac{\lambda^{N_{\text{obs}}}e^{-\lambda}}{N_{\text{obs}}!}$ is $p(N_{\text{obs}}=0)=e^{-\lambda}=e^{-\gamma T_{\text{total}}}$ where $\gamma$  is the true reordering rate. Thus, the true reordering rate for this configuration is most likely small relative to the observation time $T_{\text{total}}$. 

For the temperature configuration with the 4 K stage = 10.0 K, Trap Mount = 16.2 K, the number of dark ion events was reported for 5 different ion chain configurations as the dark ion rate was simultaneously being measured as shown in Table \ref{tab:reordering-rate-pressure-inference} - as the number of dark ions increases the probability of observing a reordering event $p_{\text{obs}}$ changes. Using $\gamma_{\text{reorder}}$ and $p_{\text{obs}}$ for each configuration together with $T=16.2\text{ K}$ and $p_{\text{reorder}}(6.3\text{ meV},T=16.2\text{ K})=0.03$, we inferred the hydrogen pressures $P_{\text{H}_{2}}$ for each ion chain configuration. $p_{\text{reorder}}$ is so small in this case because the average energy of hydrogen at $T=16.2\text{ K}$, following a Boltzmann distribution, is $E_{\text{m}}=\frac{1}{2}m_{\text{m}}|\mathbf{\dot{r}_{\text{m}}}|^{2}=\frac{3}{2}k_{\text{B}}\times16.2\,K=2\text{ meV}$ and so on average, the energy gained by the ion due to a hydrogen-ion collision is only $\left\langle E_{\text{i}}^{\prime}\right\rangle _{\theta}=47\text{ \ensuremath{\mu}eV}<E_{\text{barrier,max}}=0.142\text{ meV}$ using Equation \eqref{eq:average-ion-energy-increase-H2_collision}. For the ion chain configuration with 6 bright ions and 2 dark ions, there was no observed reordering rate and so it is assumed that the actual reordering rate for this configuration is small relative to the observation time $T_{\text{total}}$. By performing a weighted average over all pressures for each ion chain configuration we calculate an average pressure of $(2.2\pm0.2)\times10^{-10}\text{ Torr}$, inferred from ion reordering. 

\begin{table}[!!h]
\centering
\tablefont
\caption[Pressure Inference from Ion Reordering Rate]{Reported reordering rates and inferred pressures computed using Equation \eqref{eq:H2-pressure-elastic} for each ion chain configuration for the temperature configuration with the 4 K stage = 10.0 K, Trap Mount = 16.2 K.}
\label{tab:reordering-rate-pressure-inference}
\begin{tabularx}{\textwidth}{>{\centering\arraybackslash}p{0.14\textwidth}>{\centering\arraybackslash}p{0.12\textwidth}>{\centering\arraybackslash}p{0.12\textwidth}>{\centering\arraybackslash}p{0.12\textwidth}>{\centering\arraybackslash}p{0.15\textwidth}>{\centering\arraybackslash}p{0.2\textwidth}}
\toprule
\makecell{\textbf{Ion Chain}\\\textbf{Configuration}} & $p_\text{obs}$ & $T_\text{total}$ & $N_\text{obs}$ & $\mathbf{\gamma_{\text{reorder}}\text{ }\left(\text{s}^{-1}\right)}$ & $P_{\text{H}_2}$ \\
\midrule
\makecell{7 bright ions\\1 dark} & 0.88 & 2930 & 17 & $(5.8\pm1.4)\times10^{-3}$ & $(2.4\pm0.6)\times10^{-10}$ \\
\midrule
\makecell{6 bright ions\\2 dark} & 0.96 & 560 & 0 & 0 & - \\ 
\midrule
\makecell{5 bright ions\\3 dark} & 0.98 & 2980 & 12 & $(4.0\pm1.2)\times10^{-3}$ & $(1.5\pm0.4)\times10^{-10}$ \\
\midrule
\makecell{4 bright ions\\4 dark} & 0.99 & 2220 & 18 & $(8.1\pm1.9)\times10^{-3}$ & $(3.0\pm0.7)\times10^{-10}$ \\
\midrule
\makecell{3 bright ions\\5 dark} & 0.98 & 6510 & 39 & $(6.0\pm1.0)\times10^{-3}$ & $(2.2\pm0.4)\times10^{-10}$ \\
\bottomrule
\end{tabularx}
\end{table}

\subsection{Summary}
\label{subsec:summary-pressure-measurement}
The residual hydrogen pressure was inferred by measuring ion reordering and dark ion events. In both cases, the pressure inference assumes that the observed event rate is proportional to the Langevin collision rate, which accounts only for close-range collisions with impact parameters $b<b_{\text{c}}$. Furthermore, only a fraction of Langevin collisions are expected to produce a measurable observable, such as ion-chain reordering or collisional quenching into a dark state.

For the temperature configuration with the 4 K stage = 10.0 K, Trap Mount = 16.2 K, the reordering probability is $p_{\text{reorder}}=0.03$ and the pressure inferred from ion reordering is $(2.2\pm0.2)\times10^{-10}\text{ Torr}$, while the pressure inferred from the dark ion rate is $P_{\mathrm{\text{H}_{2}}}^{\text{dark}}\sim7\times10^{-13}\ \mathrm{Torr}$. The substantially lower pressure inferred from the dark-ion rate is due to the unknown uncertainties of the saturation parameter and dark ion rate reported at 300 K in \cite{paganoCryogenicTrappedionSystem2018a}, thus we take this pressure inference using Equation \eqref{eq:H2-pressure-inelastic} as just an estimate.

For the temperature configuration with the 4 K stage = 5.7 K, Trap Mount = 13.1 K, the reordering probability decreases to $p_{\text{reorder}}=0.01$ and no reordering events were observed, while the dark ion rate yielded an estimated pressure of $\sim2\times10^{-13}$ Torr. However, the reduction in the reordering rate probability is only a factor a 3, so the corresponding reordering rate should only decrease to $\sim2\times10^{-3}\text{ s}^{-1}$ which would still be experimentally observable, assuming the hydrogen density remains constant. Thus, the hydrogen density at this temperature configuration must be significantly lower. This is plausible because even though the 4 K stage and trap mount are not below 3 K where complete hydrogen cryocondensation occurs, the 4 K stage surface area is large enough relative to the hydrogen density enabling good cryosorption pumping as discussed earlier in Subsection \ref{subsec:pressure-estimate-using-cryopuming-speed-and-cryosorption-capacity}. This conclusion is further supported by the observation that ions remained trapped for 5 days without laser cooling at this temperature configuration. In summary, the reordering rate for this temperature configuration is most likely very small relative to the observation time. For example, if we assume that the probability of a reordering event occurring in the observation time $T_{\text{tot}}=38660\text{ s}$ is 50\%, this would correspond to a reordering rate of $\gamma_\text{reorder}\sim1\times10^{-5}\text{ s}^{-1}$ yielding a pressure of $1\times10^{-12}\text{ Torr}$ using Equation \eqref{eq:H2-pressure-elastic}.

A limitation of this ion chain reordering technique is that only a small fraction of Langevin capture collisions produce an observable event because sufficient momentum must be transferred to overcome the ion chain reordering barrier. A more sensitive collision-based pressure diagnostic could instead exploit the two degenerate zig-zag crystal configurations separated by an energy barrier demonstrated in \cite{paganoCryogenicTrappedionSystem2018a,spiveyHighStabilityCryogenicSystem2022}. In this geometry, elastic collisions induce transitions between the two zig-zag crystal configurations rather than requiring complete ion-chain reordering. Since this zig-zag transition barrier is lower than the linear ion chain reordering barrier, a significantly larger fraction of elastic collisions generate a measurable event, improving both the sensitivity and statistical precision of the pressure inference, but was not done in this work. 

\section{Vibration Characterization}
\label{sec:vibration-characterization}

Closed-cycle cryocoolers (typically Gifford-McMahon (GM), Pulse Tube, or Stirling) are mechanical refrigeration systems that achieve cryogenic temperatures without using liquid cryogens, utilizing closed-loop gas circulation, but produce significant vibrations on the order of $10\text{ \ensuremath{\mu}m}$ \cite{tomaruVibrationAnalysisCryocoolers2004}. Consequently, a critical design goal for interfacing atomic systems with these cryocoolers is to decouple their vibrations from the atomic systems. These vibrations effect laser-atom interactions by creating relative motion between the atoms and interrogating lasers producing intensity fluctuations, a Doppler shift of the laser frequency in the rest frame of the atom, and a time varying magnetic field at the atom if there is a field gradient. In particular for atom cavity interactions, these vibrations can perturb the cavity length which effectively broadens the cavity linewidth and degrades the ion-cavity cooperativity.

Extensive work has gone into building low vibration cryogenic ion trap and neutral atom atomic systems using closed-cycle cryocoolers. Many cryogenic ion trap systems have been built utilizing passive vibration isolation techniques where the thermal link between the ion trap and the cryocooler is a helium gas exchange region \cite{paganoCryogenicTrappedionSystem2018a,dubielzigUltralowvibrationClosedcycleCryogenic2021,moritzCryogenicPaulTrap2025,hartsellDesignCharacterizationCryogenic2025}, a flexible copper braid \cite{wangProbingExtremehighVacuum2025,niedermayrCryogenicSurfaceIon2014b}, or a long copper inertial pendulum \cite{qinQinEvaluationSuppressionVibrational2025,mickeClosedcycleLowvibration42019}. Other ion traps have been installed in cryostats built by Montana Instruments Corporation and Four 9 Design, but the exact cryocooler vibration decoupling technique is not publicly known \cite{spiveyHighStabilityCryogenicSystem2022}. For this work and the neutral atom systems in \cite{zhangHighOpticalAccess2025,jinExtendedRydbergLifetimes2026}, a ColdEdge Technologies (CET) Stinger system was used that consists of a closed-cycle recirculating helium loop that is cooled by a GM cryocooler. In this case, the thermal link between the atomic system and the cryocooler is the recirculating helium via a long flexible transfer line which provides passive vibration isolation. 

Separate engineering work has gone into passive and active vibration isolation of close-cycle GM, Pulse Tube, and Stirling cryocoolers and is summarized in \cite{caoVibrationControlMechanical2022}. Some of the passive isolation techniques consist of a well damped structure and absorber mass connecting the device under test to the cryocooler. Thus, vibrations of the device under test can be suppressed at the driving of the cryocooler by tuning the resonance frequency of the absorber mass to the driving frequency. The active isolation techniques consist of servoing the position of the device under test so as to cancel the vibrations at driving frequency of the cryocooler. In this work using the CET Stinger system, only passive vibration isolation techniques without a tuned absorber were implemented. 

I will first present some basic theory on passive vibration isolation which is covered in Subsection \ref{subsec:passive-vibration-isolation-theory}. I will then present the passive vibration isolation design goals for this cryostat and an eigenfrequency simulation of the cryostat cold stages which is covered in Subsections \ref{subsec:two-passive-vibration-isolation-problems} and \ref{subsec:eigenfrequency-analysis-of-cold-stages}. Finally, the experimental setup to measure the vibrations of the 4 K stage of this cryostat relative to the optics table, results, and discussion of how the vibration level effects laser-ion interactions and cavity length fluctuations is presented in Subsections \ref{subsec:experimental-setup-vibration-measurement}, \ref{subsec:results-vibration-measurement}, and \ref{subsec:vibration-impact-on-ion-laser-interactions-and-cavity-length-fluctuations}.

\subsection{Passive Vibration Isolation Theory}
\label{subsec:passive-vibration-isolation-theory}

\begin{figure}[!!h]
\centering
\includegraphics[width=0.3\textwidth]{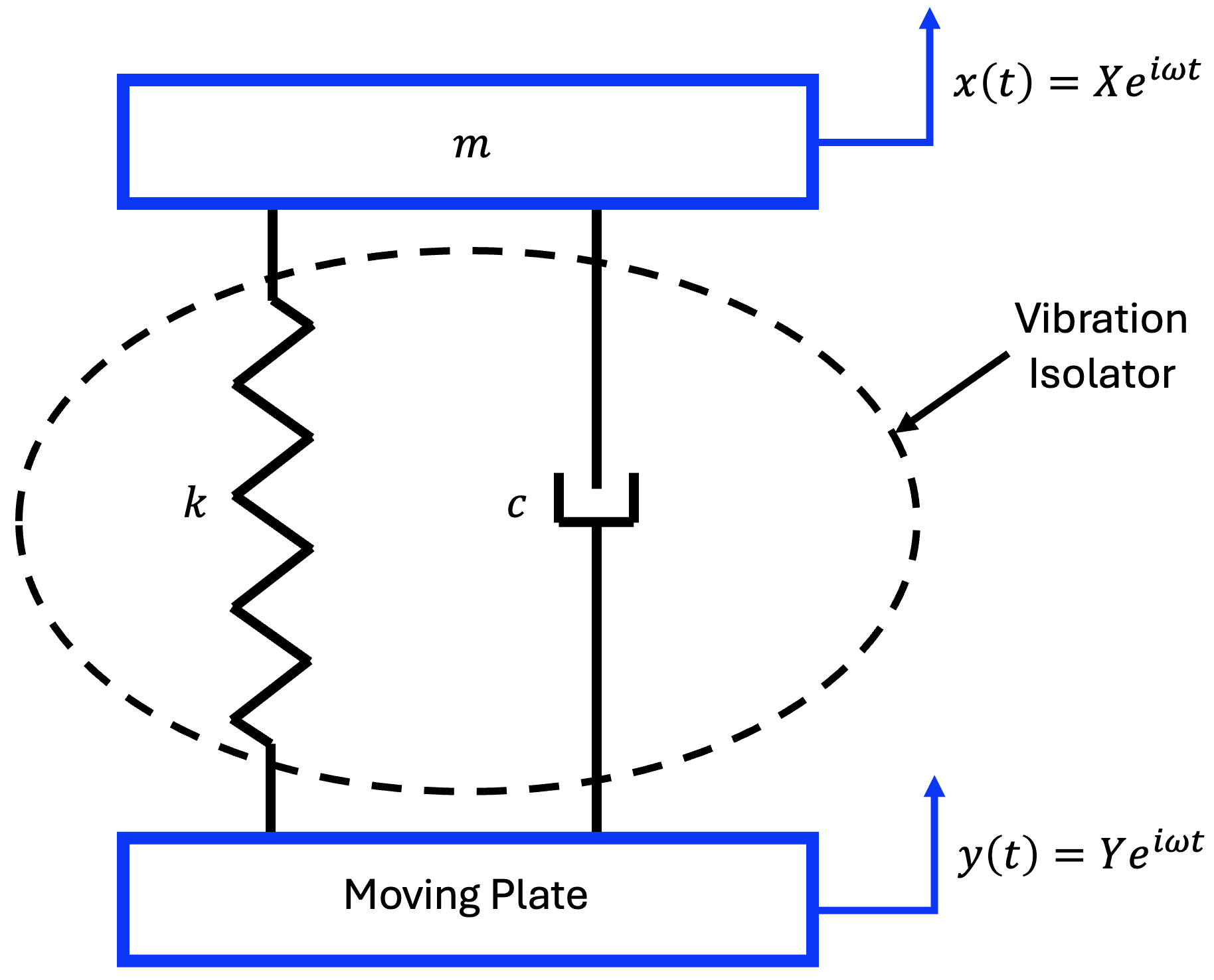}
\includegraphics[width=0.68\textwidth]{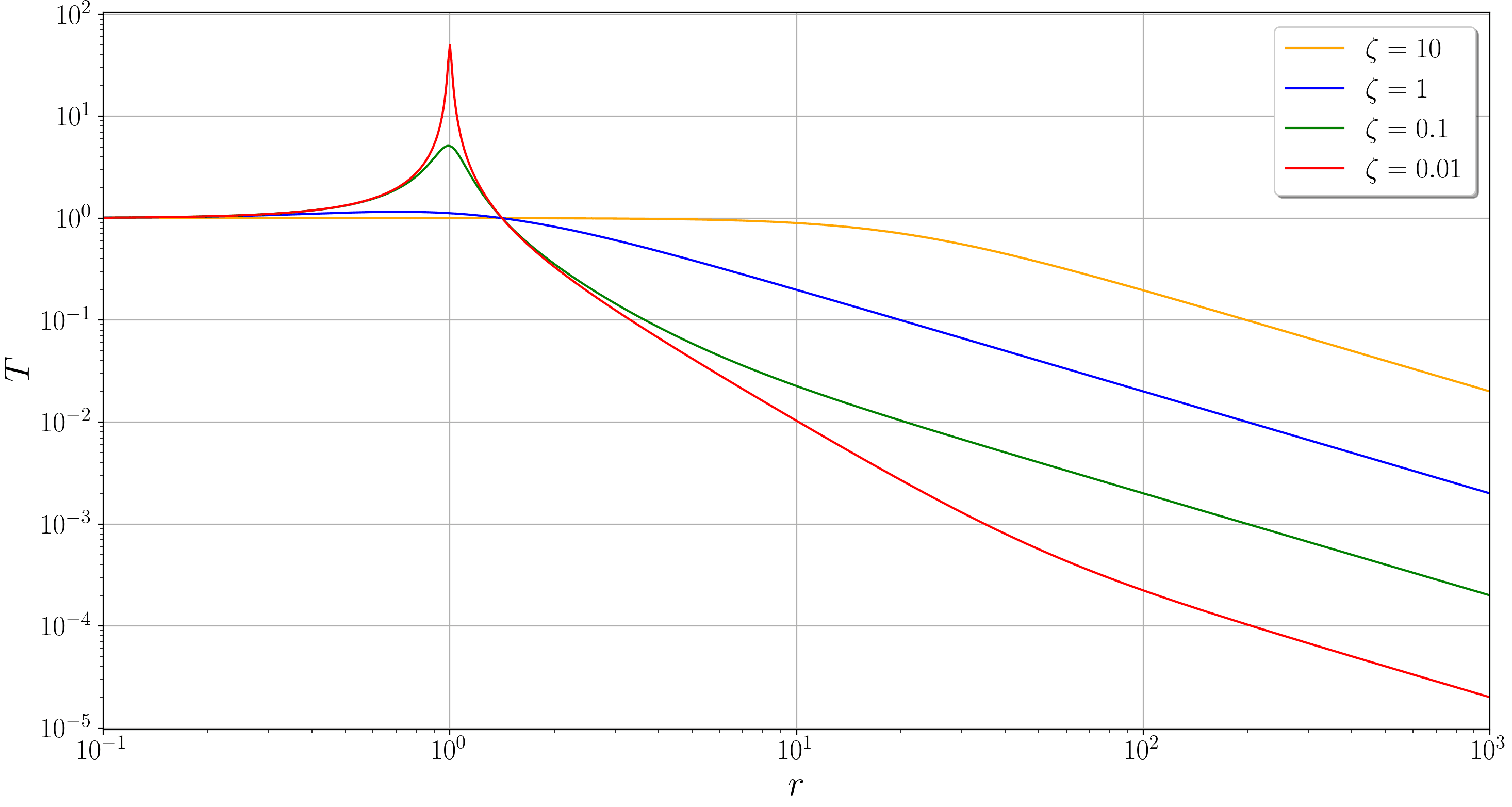}
\caption[Ion Chain Equilibrium Positions]{(Top) Schematic of stead-state motion of the mass $x(t)$ and plate $y(t)$ (the source of vibration). (Bottom) Displacement transmissibility as a function of the frequency ratio $r=\frac{\omega}{\omega_{\text{n}}}$ from Equation \eqref{eq:displacement-transmissibility} plotted with damping ratios $\zeta=[0.01,0.1,1,10]$.}
\label{fig:Displacement-transmissibility}
\end{figure}

I will first present how to model the problem of passive vibration isolation between coupled mechanical systems without an absorber. The equation of motion of a mass $m$ with displacement $x(t)$ connected to a moving plate $y(t)$ (the source of vibration) through a spring of stiffness $k$ and damping $c$ is 
\begin{equation}
    m\ddot{x}+c(\dot{x}-\dot{y})+k(x-y)=0
    \label{eq:displacement-transmissibility-EOM}
\end{equation}
Assuming steady-state harmonic motion of the plate at frequency $\omega$, the motion of the plate and mass are $y(t)=Ye^{i\omega t}$ and $x(t)=Xe^{i\omega t}$ and a schematic of this motion is shown in Figure \ref{fig:Displacement-transmissibility}. After substituting into Equation \eqref{eq:displacement-transmissibility-EOM} and rearranging we get that the displacement transmissibility is
\begin{equation}
    T=\left|\frac{X}{Y}\right|=\sqrt{\frac{1+(2\zeta r)^{2}}{(1-r^{2})^{2}+(2\zeta r)^{2}}}
    \label{eq:displacement-transmissibility}
\end{equation}
where $r=\frac{\omega}{\omega_{\text{n}}}$ is the frequency ratio, $\omega_{\text{n}}=\sqrt{\frac{k}{m}}$ is the natural frequency of the mass, and $\zeta=\frac{c}{2m\omega_{\text{n}}}$ is the damping ratio. The damping ratio can be parameterized in terms of the intrinsic damping known as the loss factor $\eta=2\zeta$  which represents the fraction of vibrational energy dissipated inside the material due to internal friction \cite{Inman_engineering_vibration}. A plot of the displacement transmissibility as a function of the frequency ratio for various damping ratios is shown in Figure \ref{fig:Displacement-transmissibility}. In general, for all frequency ratios the mass $X$ oscillates at the frequency $\omega$  but when $r\ll1$ the $X$ motion follows the $Y$ motion, when $r=1$ resonance occurs causing amplification of the $X$ motion, and when $r>\sqrt{2}$ isolation occurs causing suppression of the $X$ motion. Another effect is that higher damping is more effective on resonance and less effective for isolation. 

\subsection{Two Passive Vibration Isolation Problems}
\label{subsec:two-passive-vibration-isolation-problems}

\begin{figure}[!!h]
\centering
\includegraphics[width=0.49\textwidth]{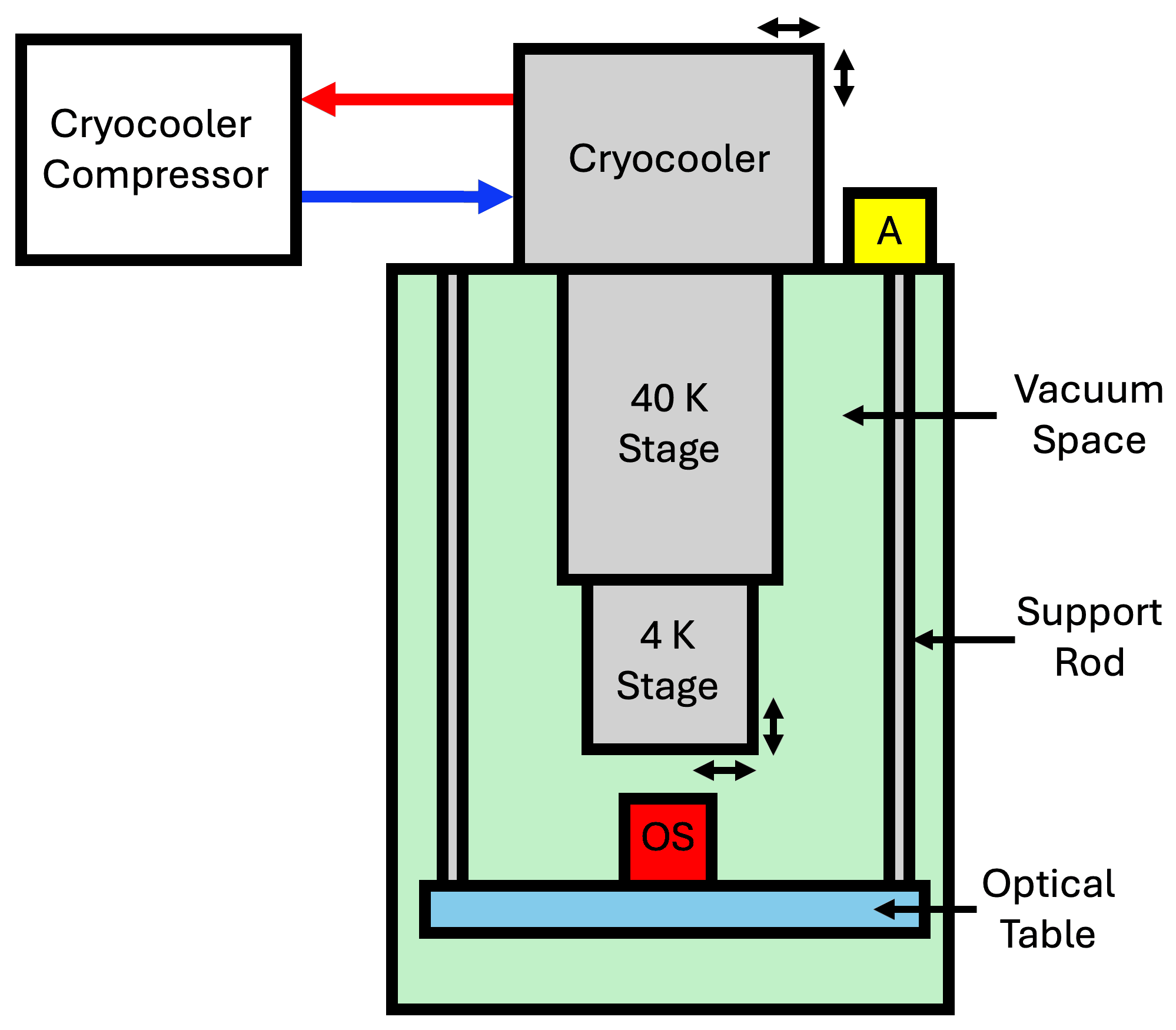}
\includegraphics[width=0.49\textwidth]{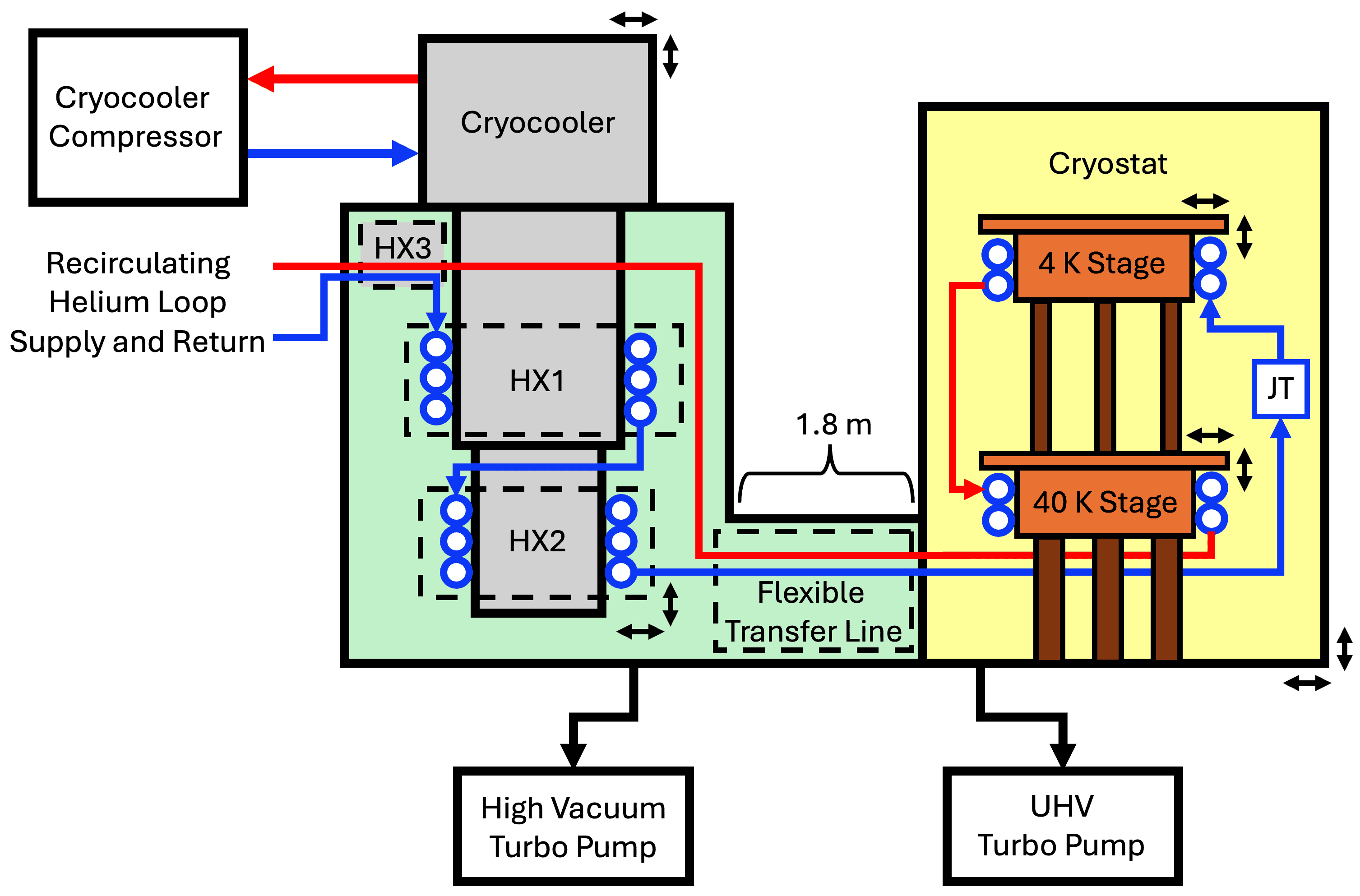}
\caption[GM Cryocooler Assembly Vibrations]{(Left) Schematic for probing the vibrations of the GM cryocooler assembly in \cite{tomaruVibrationAnalysisCryocoolers2004}. The vibrations of the cryocooler were measured using an accelerometer indicated by the yellow box. The vibrations of the cold stage relative to the cryocooler were measured using an optical sensor located at an in-vacuum optical table indicated in red. Axial and lateral vibrations were measured for both sensors depicted with the double arrows. (Right) Simplified schematic of CET Stinger system from Figure \ref{fig:Schematic-of-CET-Stinger-Cyrogenic-system} showing how vibrations from the GM cryocooler are communicated to the cryostat. The cryostat cold stage support rods are depicted with the brown rectangles and vibrations are depicted with the double arrows.}
\label{fig:GM-cryocooler-assmebly-vibrations}
\end{figure}

Here, I will present two vibration isolation problems for our cryostat, cooled using the Stinger CET system. GM cryocooler assemblies consist of the mechanical cryocooler itself, its cold stages, and a vacuum enclosure. There are vibrations of the cryocooler and vibrations due to the mechanical resonances of its cold stages. Additional vibrations due to the compressor get coupled in through the flexible helium lines. I will refer to all these vibrations as the cryocooler vibrations for simplicity. For the CET Stinger system in this work, a recirculating helium line is pre-cooled at the cold stages of a GM cryocooler and then transported to the cryostat via a transfer line where it undergoes Joule-Thomson (JT) expansion and cools the cryostat cold stages explained in more detail in Chapter \ref{chap:Cryocooler}. Thus, the vibrations due to the GM cryocooler need to be isolated from the cryostat. This boils down to two vibration isolation problems: 1.) vibrations from the GM cryocooler transmitting to the cryostat and 2.) vibrations of the cryostat transmitting to its cold stages, see Figure \ref{fig:GM-cryocooler-assmebly-vibrations}. 

The vibration amplitude spectral density of a GM cryocooler was measured by Sumitomo in the frequency range of 0.1-800 Hz \cite{tomaruVibrationAnalysisCryocoolers2004}. A schematic for the cryocooler showing how the vibrations were measured is shown in Figure \ref{fig:GM-cryocooler-assmebly-vibrations}. The motion of the cryocooler was measured using an accelerometer while the motion of the 4 K stage relative to the cryocooler was measured using an optical sensor attached to an in-vacuum optical table \cite{tomaruVibrationAnalysisCryocoolers2004}. The dominant motion of the cryocooler is due to the driving frequency of the GM expansion/compression cycle of $\approx1\text{ Hz}$ with an amplitude of $12\text{ \ensuremath{\mu}m} \left(80\text{ \ensuremath{\frac{\text{\ensuremath{\mu}m}_{\text{RMS}}}{\sqrt{\text{Hz}}}}}\right)$ and harmonics of this frequency up to 10 Hz. The resonant frequency of its 4 K cold stage was measured to be 115 Hz but two orders of magnitude lower than the driving frequency. Additional vibrations due to the cryocooler as well as vibrations coupled in via the flexible helium lines due to the compressor were measured between 10-800 Hz, but were 3 orders of magnitude lower. As the CET Stinger system used in this work has a GM cryocooler, we can use the data taken by Sumitomo as an estimate of the vibration spectral density coming from the GM cryocooler. We will now discuss the two vibration isolation problems in detail with reference to this estimated vibration spectrum to formulate design goals.

\noindent\textbf{Problem 1: Vibrations from the GM Cryocooler Assembly Transmitting to the Cryostat}

\begin{figure}[!!h]
\centering
\includegraphics[width=0.5\textwidth]{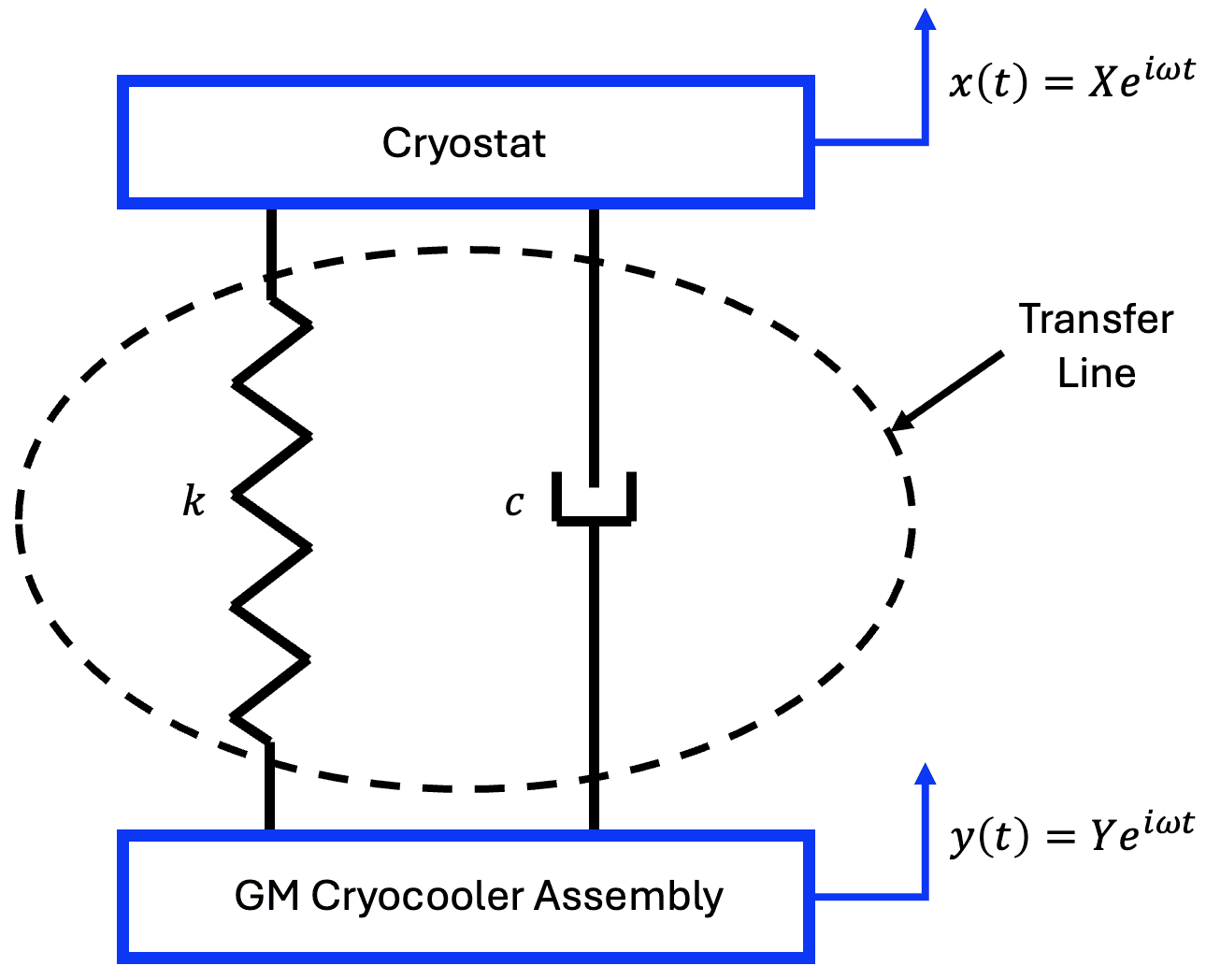}
\caption[Vibration Isolation Problem 1 Schematic]{Schematic of steady-state motion of the cryostat $x(t)$ and GM cryocooler assembly $y(t)$ coupled by the transfer line.}
\label{fig:vibration-isolation-problem-1-schematic}
\end{figure}

Casting this problem using Equation \eqref{eq:displacement-transmissibility-EOM}, $X$ is the cryostat motion, $Y$ is the GM cryocooler motion, and the stiffness $k$ and damping $c$ are that of the transfer line as shown in the schematic in Figure \ref{fig:vibration-isolation-problem-1-schematic}. So the design goal for isolating the GM cryocooler vibrations is to manufacture the transfer line such that its natural resonance frequency is below 1 Hz, $r>\sqrt{2}$. The CET Stinger transfer line is a flexible stainless steel edge welded bellows with a length of 2 m and radius of 10 mm that encloses the recirculating helium supply and return to the cryostat. We can estimate the natural frequency of this transfer line by modeling it as a fixed cantilever circular tube. The natural frequency of a fixed cantilever circular tube is $f_{\text{cant}}=\frac{1}{2\pi}\sqrt{\frac{EI}{\rho AL^{4}}}$ where $E$ is the Young's modulus, $I$ is the second moment of inertia of the circular tube wall $I=\pi r^{3}t$ where $t$ is the thickness of the tube wall, $\rho$  is the density, $A=2\pi rt$ is the cross-sectional area of the tube wall, and $L$ is the length of the tube \cite{Rao_Mechanical_vibrations}. Rearranging we get that the estimated natural frequency of the transfer line is,
\begin{equation}
    f_{\text{TL}}=\frac{1}{2\pi}\sqrt{\frac{E}{\rho}}\frac{r}{L^{2}}
    \label{eq:transfer-line-resonance-frequency}
\end{equation}
Taking $E=2\times10^{11}\text{ Pa}$, $\rho=8000\text{ \ensuremath{\frac{\text{kg}}{\text{m}^{3}}}}$, $r=10\text{ mm}$, and $L=2\text{ m}$, we get $f_{\text{TL}}=7\text{ Hz}$. Since this is near the dominant cryocooler vibrational frequencies [0-10 Hz], this implies $r\leq1$ so additional damping needs to be applied to reduce the transmissibility to the cryostat. However, the transfer line should provide isolation for the higher frequencies [10-800 Hz]. This is only a simple model and the damping $c$ is not known, but the motivation for engineering a long, thin flexible transfer line is made clear.

\noindent\textbf{Problem 2: Vibrations of the Cryostat Transmitting to its Cold Stages}

\begin{figure}[!!h]
\centering
\includegraphics[width=0.5\textwidth]{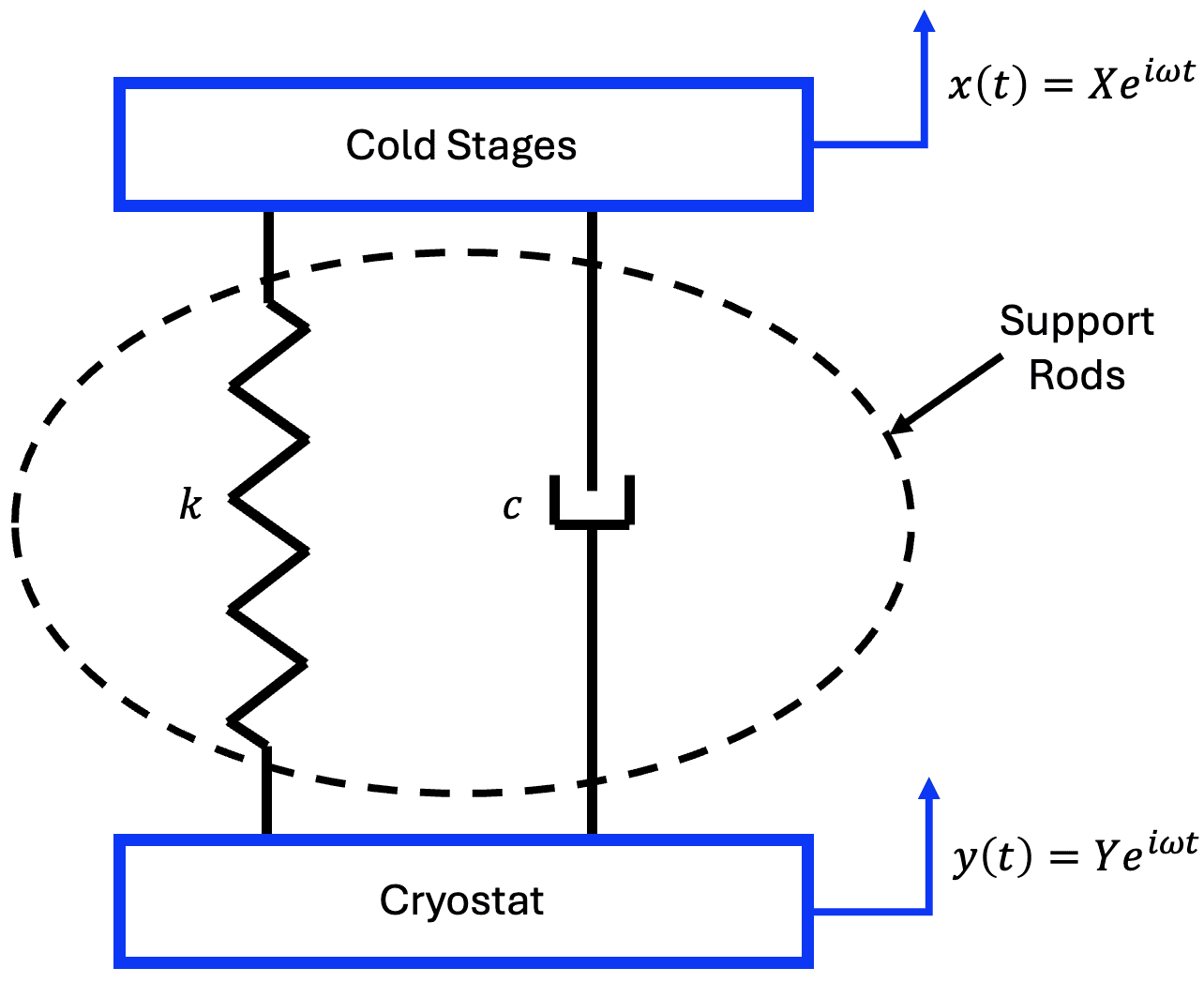}
\caption[Vibration Isolation Problem 2 Schematic]{Schematic of stead-state motion of the cryostat cold stages $x(t)$ and cryostat $y(t)$ coupled by the support rods.}
\label{fig:vibration-isolation-problem-2-schematic}
\end{figure}

Casting this problem using Equation \eqref{eq:displacement-transmissibility-EOM}, $X$ is the cryostat cold stages motion, $Y$ is the cryostat motion following the GM cryocooler motion, and the stiffness $k$ and damping $c$  are that of the cold stage support rods as shown in the schematic in Figure \ref{fig:vibration-isolation-problem-2-schematic}. Similar to problem 1, the design goal for isolating this vibration is to manufacture the cold stages such their natural resonance frequency is below 1 Hz, $r>\sqrt{2}$. However, this is challenging as the cryostat cold stages are typically small structures made of copper that are supported by insulating materials. Thus, the practical design goal is to make the support rods stiff (drive up resonance frequency) such that $r\ll1$ for all the vibrational frequencies transmitted to the cryostat and are made out of a material with a high loss factor $\eta$  so that $c$ is relatively large. 

In this cryostat, the cold stages are made of C10100 circular plates supported by PEEKGF30 support rods. PEEKGF30 is a good material for the support rods as it is stiff for a plastic $(E=18\times10^{9}\text{ Pa}$ at 20 K \cite{chuMechanicalThermalExpansion2010a}) and has a relatively high loss factor $\eta=0.03$ \cite{Inman_engineering_vibration,ekinExperimentalTechniquesLowtemperature2006a}. Also, PEEKGF30 has a Young's Modulus that is a factor of 2 higher than unfilled normal PEEK. The axial translation resonance frequency of the cold stages along the direction of the length of the support rods ($y$-axis) can be estimated as follows, 
\begin{equation}
    f_{\text{a}}=\frac{1}{2\pi}\sqrt{\frac{K_{\text{tot}}}{m}}=\frac{1}{2\pi}\sqrt{\frac{3EA}{Lm}}
    \label{eq:vertical-resonance-of-cold-stages}
\end{equation}
Here, $K_{\text{tot}}=3k_{\text{r,a}}$ is the total stiffness of three support rods each with stiffness $k_{\text{r,a}}=\frac{EA}{L}$, $E$ is the Young's modulus, $A$ is the cross sectional area of the rod, $L$ is the length of the rod, and $m$ is the mass of the cold stage \cite{Rao_Mechanical_vibrations,Inman_engineering_vibration}. Taking $A=5\times10^{-5}\text{ m}^{2}, L=76\text{ mm}$, $E=18\times10^{9}\text{ Pa}$, and $m=5.6\text{ kg}$, the axial translation resonance frequency of the 4 K stage is $f_{\text{a}}=401\text{ Hz}$. Taking $A=1.8\times10^{-4}\text{ m}^{2}$, $L=92\text{ mm}$, $E=18\times10^{9}\text{ Pa}$, and $m=8.4\text{ kg}$, the axial translation resonance frequency of the 40 K stage is $f_{\text{a}}=564\text{ Hz}$. 

The lateral translation resonance frequency of the cold stages transverse to the axial ($x,z$-axes) can be estimated as follows,
\begin{equation}
    f_{\text{l}}=\frac{1}{2\pi}\sqrt{\frac{K_{\text{tot}}}{m}}=\frac{1}{2\pi}\sqrt{\frac{36EI}{L^{3}m}}
    \label{eq:lateral-resonance-frequency-of-cold-stages}
\end{equation}
Here, $K_{\text{tot}}=3k_{\text{r,l}}$ is the total lateral stiffness of three support rods each with stiffness $k_{\text{r,l}}=\frac{12EI}{L^{3}}$ where $I=\frac{\pi r^{4}}{4}$ is the second moment of inertia of the rod, $E$ is the Young's modulus, $L$ is the length of the rod, and m is the mass of the cold stage \cite{Rao_Mechanical_vibrations,Inman_engineering_vibration}. Taking $r=4\text{ mm}$, $L=76\text{ mm}$, $E=18\times10^{9}\text{ Pa}$, and $m=5.6\text{ kg}$, the lateral translation resonance frequency of the 4 K stage is $f_{\text{l}}=37\text{ Hz}$. Taking $r=7.5\text{ mm}$, $L=92\text{ mm}$, $E=18\times10^{9}\text{ Pa}$, and $m=8.4\text{ kg}$, the lateral translation resonance frequency of the 40 K stage is $f_{\text{l}}=79\text{ Hz}$. 

In general, the axial and lateral translation resonance frequencies increase with less massive cold stages, shorter and thicker support rods, and a stiffer material choice for the support rods. However, shorter and thicker support rods creates a higher thermal conductance between the stages. This geometry is a reasonable compromise providing cold stage resonances greater than the dominant cryocooler vibrational frequencies at [0-10] Hz. The torsional and rocking resonance frequencies are more complicated to estimate with a simple analytical expression, but were simulated in the following section.

\subsection{Eigenfrequency Analysis of Cold Stages}
\label{subsec:eigenfrequency-analysis-of-cold-stages}

\begin{figure}[!!h]
\centering
\includegraphics[width=0.5\textwidth]{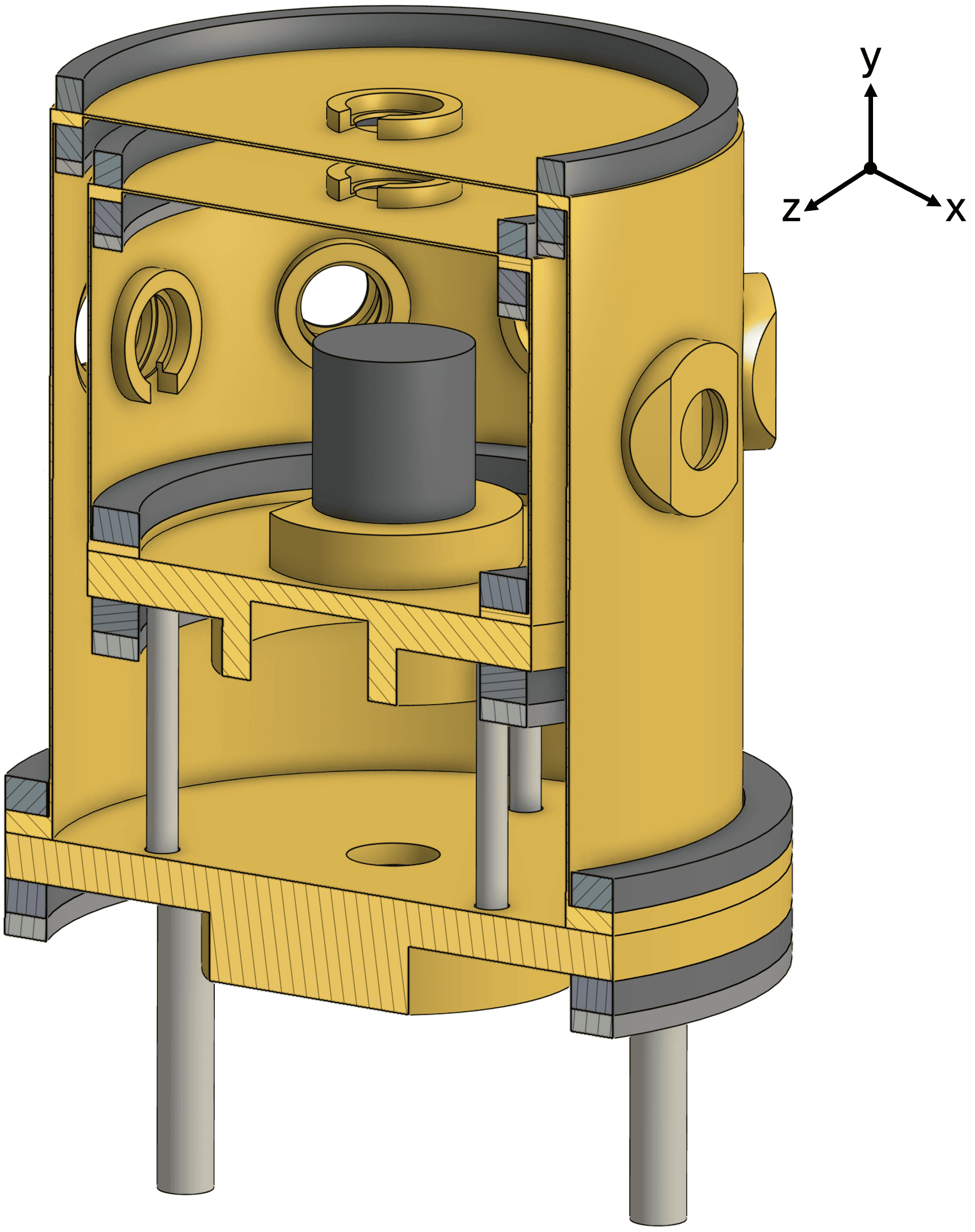}
\caption[CAD Drawing of Cold Stages for COMSOL Eigenfrequency Simulation]{Simplified CAD drawing of the cryostat cold stages used for the COMSOL eigenfrequency simulation.}
\label{fig:simplified-CAD-drawing-of-cold-stages-for-COMSOL-sim}
\end{figure}

An eigenfrequency analysis of the cryostat cold stages was simulated in COMSOL. To perform this simulation, the CAD drawing of the cold stages was simplified, removing intricate small features like through holes, taps, bolts, and small chamfers/fillets. The idea is to simplify the drawing to prevent unnecessarily fine meshing to run the simulation, but to ensure the total mass and distribution of the mass of the cold stages remains roughly the same. Thus, the ion trap was approximated by a titanium cylinder mounted to the trap riser such that it has the same mass as the real ion trap structure. The helical corrugated surface of the cold stage heat exchangers were approximated by copper C10100 cylinders and the resonator, wires, micro-D connectors, radiation shield windows and springs, and heaters were omitted from the drawing as they have negligible mass. The cold stages CAD drawing used for the COMSOL simulation is shown in Figure \ref{fig:simplified-CAD-drawing-of-cold-stages-for-COMSOL-sim} consisting of the C10100 cold stages, radiation shields, and window holders, TiG2 spacers and ion trap, SS316 plate nuts, and PEEKGF30 support rods.

The simulation was setup up by importing the CAD drawing as an assembly to create identity pairs (contacts between each parts) and assigning the material properties to each part. The densities of the materials were taken to be their 300 K values as the densities only change by $\sim1\%$ between 300 K and 4 K. The Young's Modulus of PEEKGF30 increases by a factor of 2 between 300 K and 20 K so the Young's modulus was taken to be its value at 20 K, $18\text{ GPa}$ \cite{chuMechanicalThermalExpansion2010a}. A fixed constraint was applied at the bases of the 40 K support rods and a continuity/bonded constraint was applied to the other identity pairs. In reality, there is some compliance at the support rod-cold stages joints so a continuity/bonded constraint (infinitely stiff) is an approximation. Consequently, this simulation will slightly overestimate the eigenfrequencies. Finer meshing, $\sim0.1\times D$, was applied at the support rods with diameter $D$ as these parts mainly govern the axial, lateral, torsional, and rocking eigenfrequencies of the cold stages. The simulation was solved for eigenfrequencies, their mode shapes, and participation factors in the frequency range of 0-1000 Hz and the results are summarized in Table \ref{tab:results-of-cold-stages-COMSOL-eigenfrequency-sim}. 

In summary, 28 eigenfrequencies were obtained from the simulation together with their mode shapes and participation factors. In short, the mode shape gives information about what part of the cold stages is moving and the participation factor represents how easily that mode can be excited. So the modes that will be the most sensitive to measuring vibrations of the 4 K stage relative to the optics table are the ones that involve movement of the 4 K stage with a relatively high participation factor. It is also important to note that lateral, torsional, and rocking modes can manifest as vibrations in $x,y,z$ and thus the eigenfrequencies of interest are [36.0, 36.4, 38.3, 76.3, 85.3, 85.5, 219.0, 250.0, 340.8, 361.5, 383.0, 686.2] Hz. The simulated lateral translation mode frequencies of the 4 K and 40 K stages roughly agree with the ones analytically obtained using Equation \eqref{eq:lateral-resonance-frequency-of-cold-stages}. However, the simulated axial translation mode frequencies of the 4 K and 40 K stages are very different from the analytically obtained ones using Equation \eqref{eq:vertical-resonance-of-cold-stages}. This discrepancy in the analytical frequencies is probably due to assuming the total mass distribution of the stages is located right at the support rods and not modeling the stages as connected to each other.

\begin{table}[!!h]
\centering
\tablefont
\caption[Results of Cold Stages COMSOL Eigenfrequency Simulation]{Summary of 28 cryostat cold stages eigenfrequencies, their mode shapes, and participation factors in the frequency range of 0-1000 Hz simulated in COMSOL.}
\label{tab:results-of-cold-stages-COMSOL-eigenfrequency-sim}
\begin{tabularx}{\textwidth}{>{\centering\arraybackslash}p{0.19\textwidth}>{\centering\arraybackslash}p{0.26\textwidth}cc}
\toprule
\textbf{Frequency (Hz)} & \textbf{Mode Type} & \textbf{Comments} & \makecell{\textbf{Participation}\\\textbf{Factor}}\\
\midrule
36.0, 36.4 & Lateral Translation in $z,x$ & 4 K stage & 2.9, 2.9 \\
\midrule
38.3 & Torsional About $y$ & 4 K stage & 0.3 \\
\midrule
76.3 & Torsional About $y$ & Both stages out of phase & 0.6 \\
\midrule
85.3, 85.4 & Lateral Translation in $x,z$ & 40 K stage & 2.2, 2.1 \\
\midrule
193.0 & Zero-Order Plate Mode in $y$ & 40 K stage shield lid & 0.4 \\
\midrule
219.0 & Rocking About $x$ & Both stages in phase & 0.3 \\
\midrule
247.0 & Zero-Order Plate Mode in $y$ & 4 K stage shield lid & 0.8 \\
\midrule
250.6 & \makecell{Rocking About $z$\\Zero-Order Plate Mode in $y$} & \makecell{Both stages in phase\\Both stages shield lids in phase} & 0.2 \\
\midrule
340.8 & \makecell{Axial Translation in $y$\\Zero-Order Plate Mode in $y$} & \makecell{4 K stage\\Both stages shield lids in phase} & 3.1 \\
\midrule
361.5 & \makecell{Rocking About $x$\\Higher-Order Plate Mode in $y$} & \makecell{Both stages out of phase\\Both stages shield lids in phase} & 0.1 \\
\midrule
383.0 & \makecell{Rocking About $z$\\Higher-Order Plate Mode in $y$} & \makecell{Both stages out of phase\\Both stages shield lids in phase} & 0.1 \\
\midrule
417.2, 423.2 & Higher-Order Plate Mode in $y$ & 40 K stage shield lid & 0.1, 0.01 \\
\midrule
526.9, 540.1 & Higher-Order Plate Mode in $y$ & 4 K stage shield lid & 0.003, 0.01 \\
\midrule
686.2 & \makecell{Axial Translation in $y$\\Higher-Order Plate Mode in $y$} & \makecell{40 K stage\\Both stages shield lids out of phase} & 1.6 \\
\midrule
739.5, 749.9, 750.0 & Higher-Order Plate Mode in $x,z$ & 40 K stage shield side wall & 0.02, 0.06, 0.03 \\
\midrule
848.0, 848.3 & Higher-Order Plate Mode in $x,y,z$ & 40 K stage shield lid and side wall & 0.01, 0.01 \\
\midrule
854.9, 858.5, 881.1 & Higher-Order Plate Mode in $y$ & 40 K stage shield lid & 0.0003, 0.003, 0.2 \\
\midrule
922.4, 1012.3 & Higher-Order Plate Mode in $x,z$ & 40 K stage shield side wall & 0.003, 0.001 \\
\bottomrule
\end{tabularx}
\end{table}

\subsection{Experimental Setup}
\label{subsec:experimental-setup-vibration-measurement}

\begin{figure}[!!h]
\centering
\includegraphics[width=0.39\textwidth]{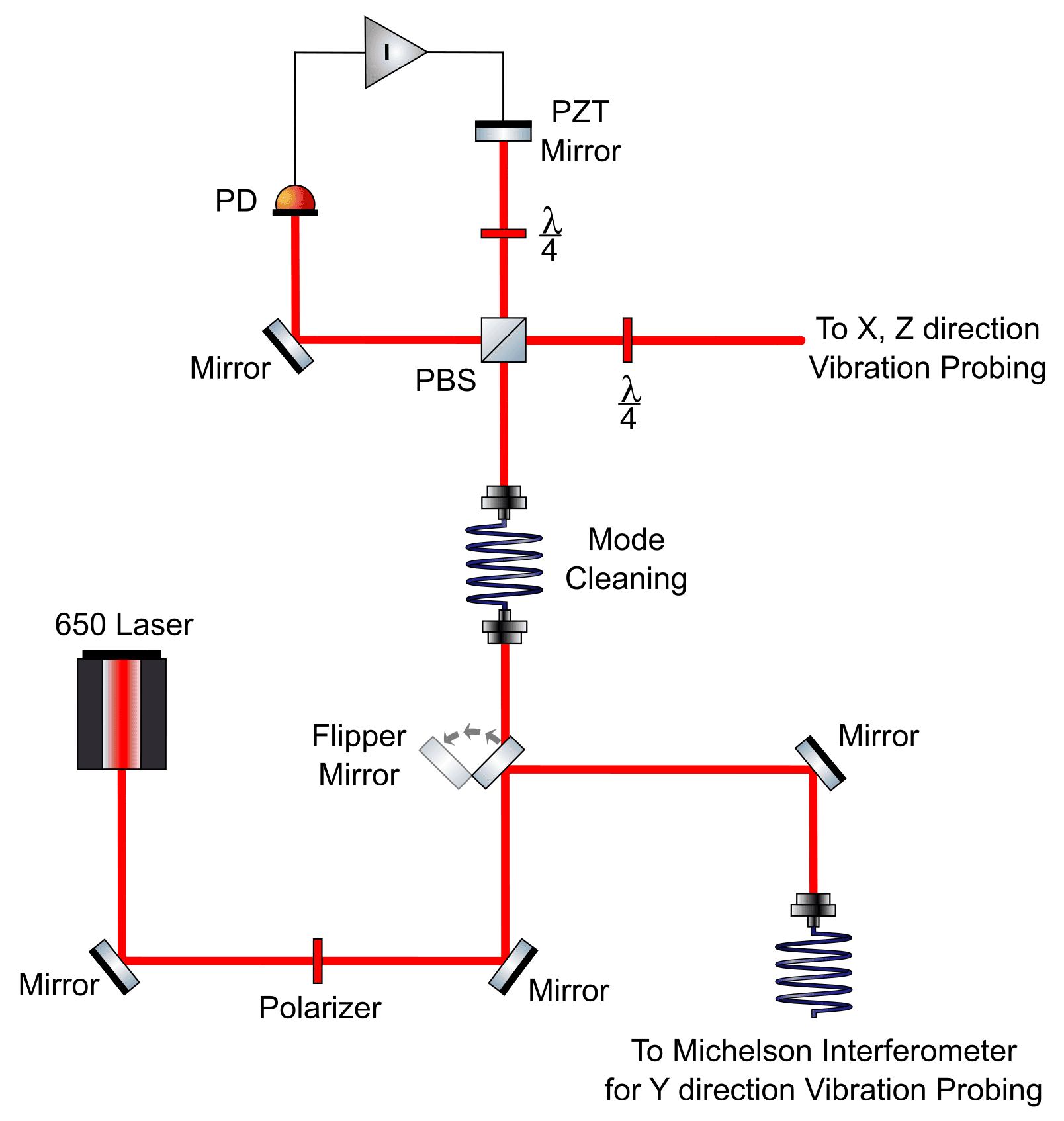}
\includegraphics[width=0.59\textwidth]{Figures/Cryostat_Characterization/Michelson_Interferometer_and_Micromotion_Y_Direction_optics_breadboard.png}
\caption[Michelson Interferometer Optical Setup]{Michelson interferometers. (Left) Michelson interferometer mounted on top of the optics table used to probe vibrations along the $x,z$-axes of the cryostat. (Right) Michelson interferometer mounted on a hanging breadboard below the optics table used to probe vibrations along the $y$-axis of the cryostat. Light is directed to this interferometer by a flipper mirror and optical fiber shown in the left schematic. The beam is then overlapped onto the 935 nm repump beam via a dichroic mirror.}
\label{fig:Michelson-interferometer-optical-setup}
\end{figure}

We measured the motion of the 4 K stage relative to the optics table in 3D using a Michelson interferometer. For each spatial direction, the Michelson interferometer consists of a 650 nm single-mode external diode laser (ECDL), a non-polarizing beamsplitter, a mirror mounted to a piezoelectric actuator (PZT mirror), a probe mirror, and a photodiode (PD). The probe mirror is mounted inside the cryostat at the 4 K stage along each spatial direction while the ECDL, beamsplitter, PZT mirror, and PD are mounted on the optics table as shown in Figure \ref{fig:Michelson-interferometer-optical-setup}. We used gold-plated C10100 tabs as the probe mirrors because they can be easily mounted at the 4 K stage and the surface finish of the plating and reflectance of 650 nm light off gold are sufficient to form a Michelson fringe. The reflectivity of gold is $\beta=0.94$ using Equation \eqref{eq:reflectivity-Fresnel-Drude-normal-indcidence} taking the optical constants to be $n(653\,\text{nm})=0.166$ and $k(653\,\text{nm})=3.15$ \cite{brewsterThermalRadiativeTransfer1992}. The arms of the interferometer differ by roughly 1 m, but a fringe can still be formed as the coherence length of the ECDL is $>$ 300 m. We used the Michelson interference fringe detected by the PD as an error signal, with the set point at the steepest slope of the fringe, and servoed the PZT mirror using a slow integrator. The slow feedback voltage applied to the PZT tracks low-frequency drifts in the optical path length, while fluctuations in the PD signal about the locked operating point measure the high-frequency vibration-induced path-length excursions. I will now discuss the experimental setup in more detail.

\begin{figure}[!!h]
\centering
\includegraphics[width=\textwidth]{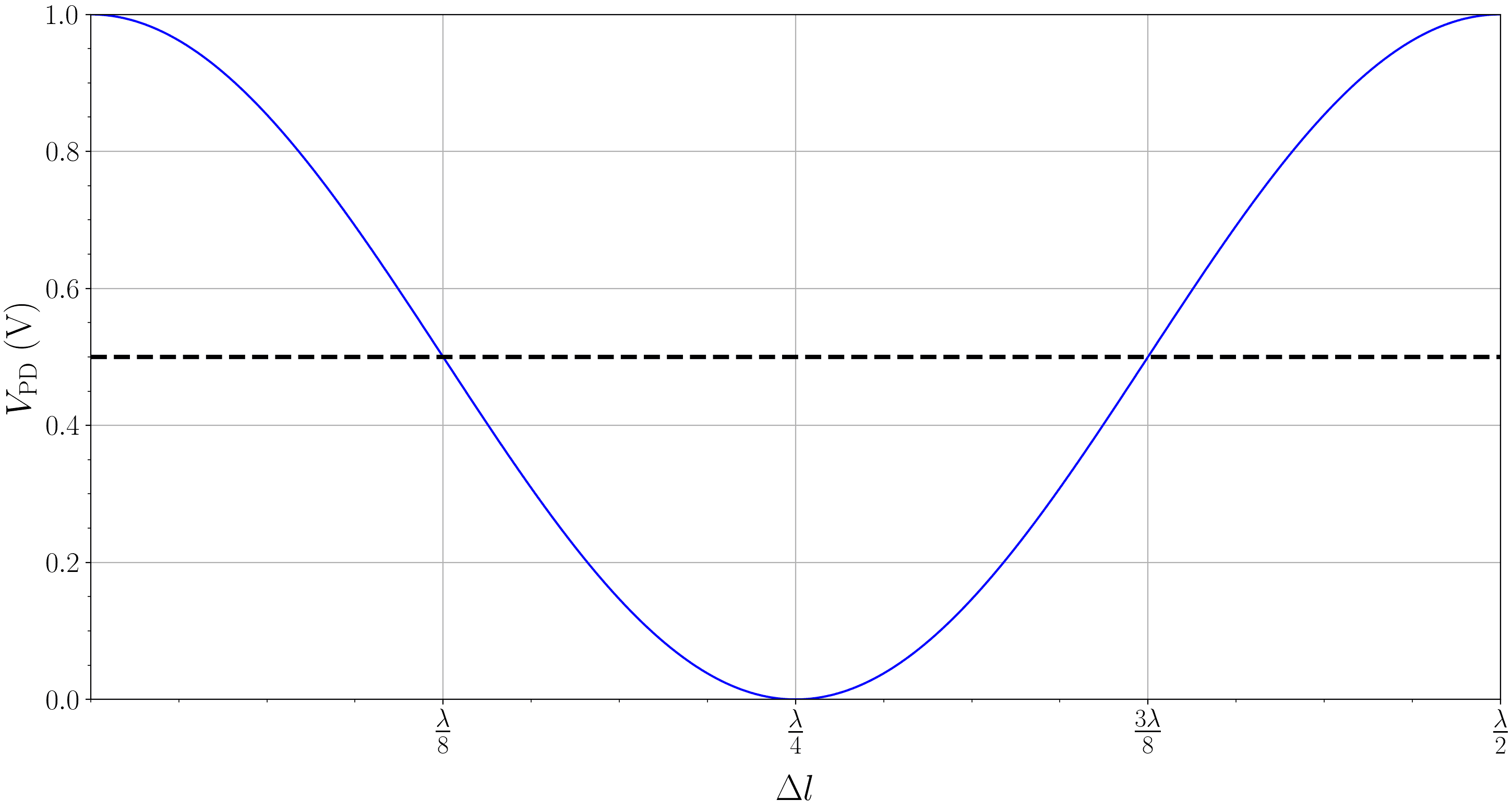}
\caption[Michelson Interferometer Fringe]{Michelson Fringe normalized to the PD voltage in Equation \eqref{eq:Michelson-fringe-equation} where $V_{\text{min}}=0$ and $V_{\text{max}}=1$ for clarity. The PZT is servoed at the steepest slope of the fringe indicted by the dashed horizontal line.}
\label{fig:Michelson-interferometer-fringe}
\end{figure}

The Michelson fringe normalized to the voltage measured at the PD (Thorlabs, PN:DET36A) can be written as,
\begin{equation}
    V_{\text{PD}}(\Delta l)=V_{\text{min}}+\frac{\left(V_{\text{max}}-V_{\text{min}}\right)}{2}\left(1+\cos2k\Delta l\right)
    \label{eq:Michelson-fringe-equation}
\end{equation}
where $V_{\text{min}}$ and $V_{\text{max}}$ are the minimum and maximum voltages of the fringe, $k=\frac{2\pi}{\lambda}$ is the wavenumber, and $\Delta l$ is the relative length change between the arms of the interferometer \cite{hechtOptics2017}. One fringe period corresponds to a relative length change of $\Delta l=\frac{\lambda}{2}$ as shown in Figure \ref{fig:Michelson-interferometer-fringe}. Thus, by scanning over multiple fringes the PZT (Thorlabs, PN: TA0505D024W) and interferometer sensitivities can be measured. The PZT sensitivities for the $x,z$ interferometer and the $y$ interferometer were measured to be $17\text{ }\frac{\mbox{nm}}{\mbox{V}}$ and $24\text{ }\frac{\mbox{nm}}{\mbox{V}}$ respectively, which only needs to be done once. On the other hand, the interferometer sensitivities were calibrated before each measurement as they depend on the optical power. At the steepest slope of the fringe in Equation \eqref{eq:Michelson-fringe-equation} where $\Delta l=(2n+1)\frac{\lambda}{8}$ with $n\in\mathbb{Z}$, the interferometer sensitivity is
\begin{equation}
    \left|\frac{\text{d}V_{\text{PD}}}{\text{d}\Delta l}\left((2n+1)\frac{\lambda}{8}\right)\right|=k\Delta V
    \label{eq:interferometer-sensitivity}
\end{equation}
and is reduced to simply calculating $\Delta V=V_{\text{max}}-V_{\text{min}}$ before each measurement. The interferometer sensitivity for these measurements using Equation \eqref{eq:interferometer-sensitivity} with 650 nm is $\left|\frac{\text{d}V_{\text{PD}}}{\text{d}\Delta l}\left((2n+1)\frac{\lambda}{8}\right)\right|^{-1}=\frac{650\,\text{nm}}{2\pi\Delta V}=\frac{103\,\text{nm}}{\Delta V}$.

\begin{figure}[!!h]
\centering
\includegraphics[width=0.4\textwidth]{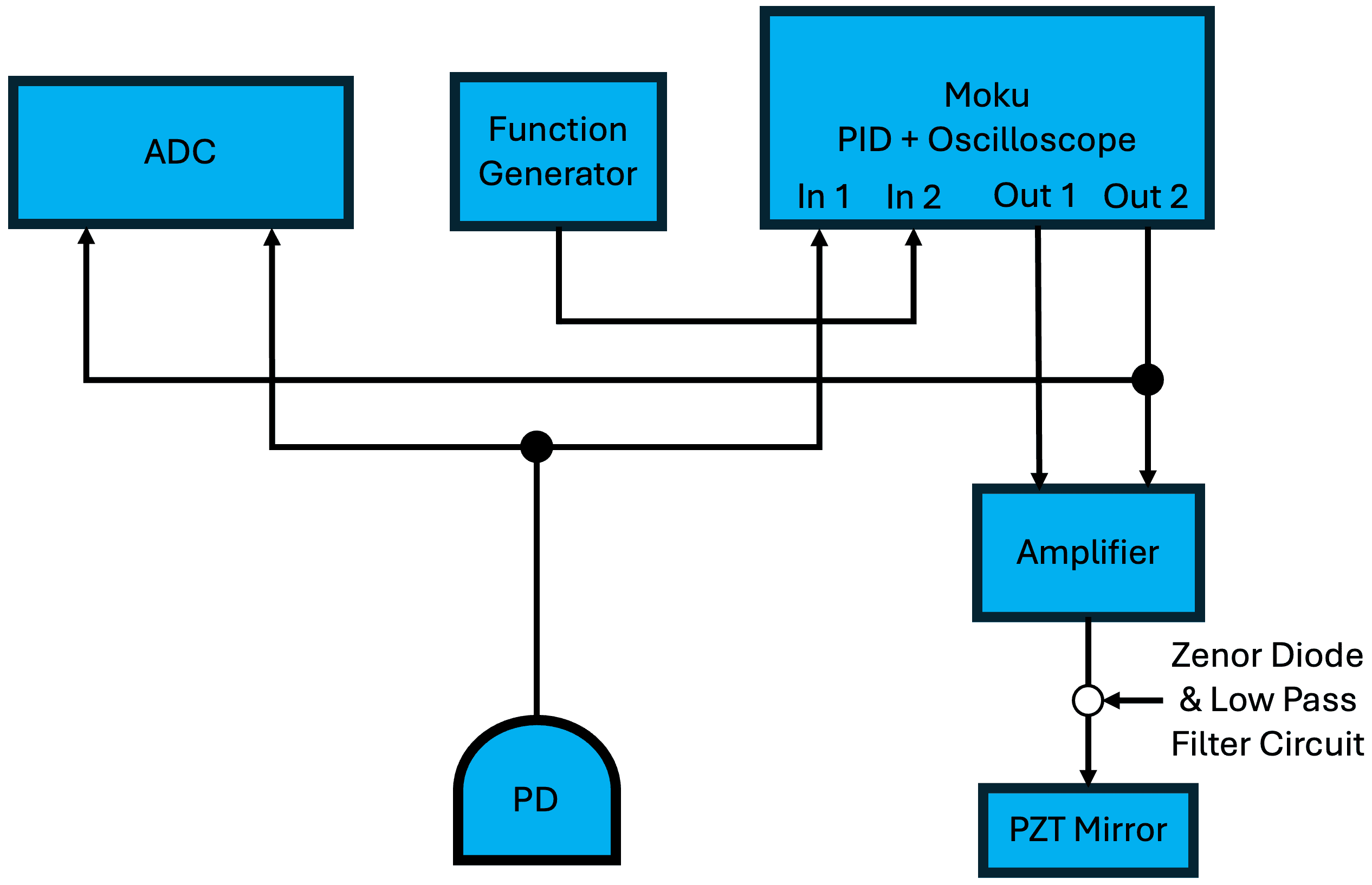}
\includegraphics[width=0.58\textwidth]{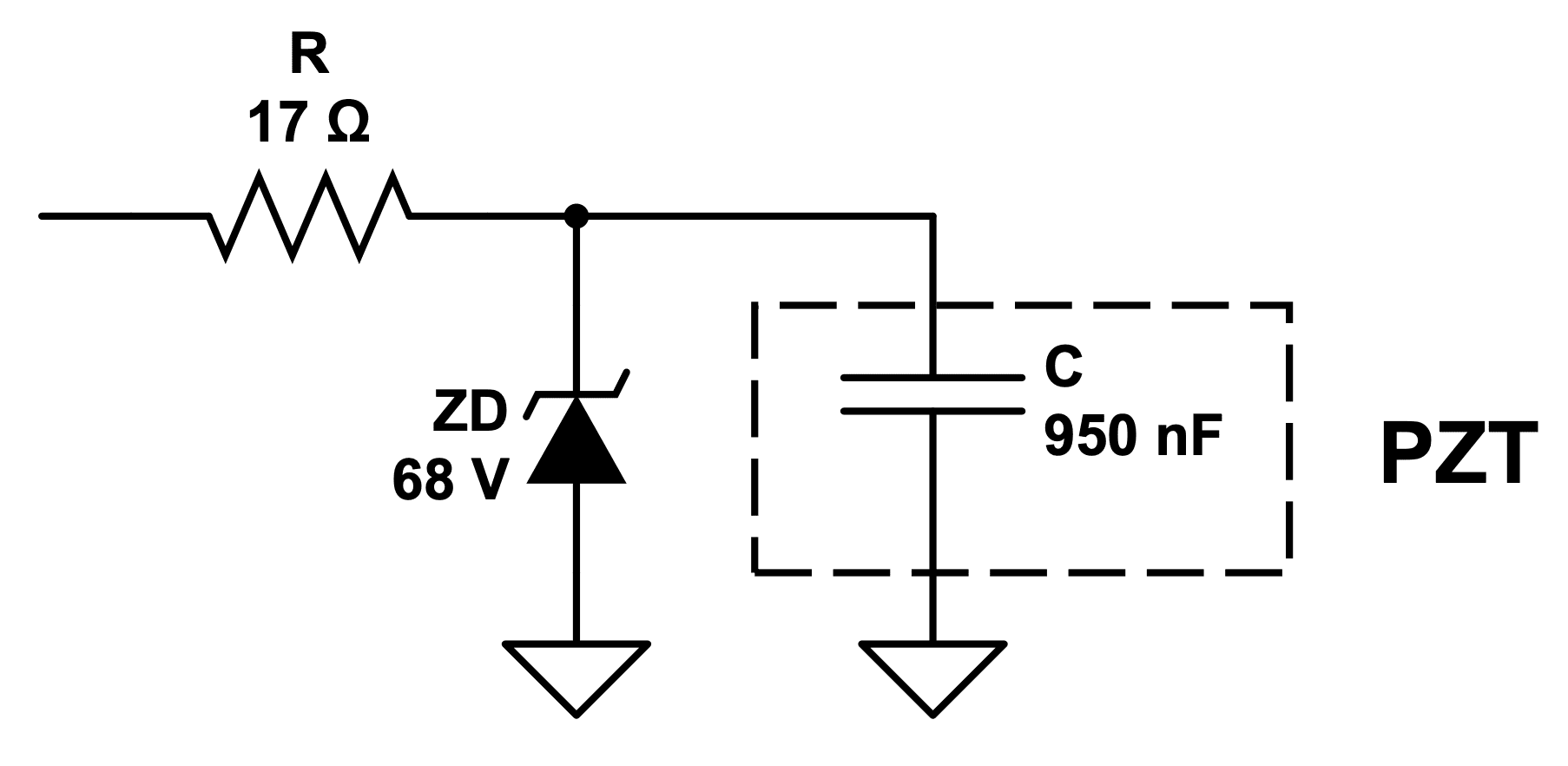}
\caption[Vibration Measurement Control Loop]{(Left) Schematic of control system where the arrows indicate electrical connections. The DC offset it set by sweeping over multiple fringes using the function generator (Rigol, PN:DG822), where the voltage is passed through the PID with unity gain to output 2. An amplifier (PiezoDrive, PN: PD32-V130,130) is used to amplify the voltage to the PZT with a gain of 9.5. A suitable error signal is found when the PZT voltage is roughly 34 V, provided by output 1, ensuring maximum dynamic range for the measurement. Once the lock is engaged the feedback voltage from output 2 is summed with output 1 to servo the PZT mirror. (Right) Equivalent circuit diagram of the low pass filter between $R$ and the capacitance of the PZT, $C$, where the voltage across the PZT is clamped to 0-68 V by the the Zenor diode. }
\label{fig:Vibration-measurement-control-loop}
\end{figure}

We used a hardware platform (Liquid Instruments, PN: Moku-Go-M2) as a PID and oscilloscope. The voltage from the PD is sent to input 1 where the DC offset is subtracted off creating an error signal, as shown in Figure \ref{fig:Vibration-measurement-control-loop}. This error signal is passed through an integrator in the Moku with a $500\text{ mHz}$ unity gain frequency and the output is feedback to the PZT mirror via output 2. After the amplifier, a 68 V Zenor diode is connected in parallel to the PZT mirror to clamp the voltage at 0-68 V, as the PZT has a operating voltage range of 0-70 V. A $17\text{ \ensuremath{\Omega}}$ resistor is placed before the Zenor diode creating a low pass filter as the PZT has an intrinsic capacitance of 950 nF. This low pass filter has a cutoff frequency of $f_{\text{3dB}}=\frac{1}{2\pi RC}=10\text{ kHz}$ which suppresses high frequency electronic noise from the PID. The dynamic range of the PZT feedback voltage is 0-68 V which corresponds to $1178\text{ \ensuremath{\mu}m}$ for the $x,z$ interferometer and $1632\text{ \ensuremath{\mu}m}$ for the $y$ interferometer. A suitable error signal is found when the PZT voltage is near 34 V, provided by output 1, ensuring maximum dynamic range for the measurement. Once the lock is engaged the feedback voltage from output 2 is summed with output 1 to servo the PZT mirror. 

We monitor the interferometer using an ADC independent of the Moku PID as shown in Figure \ref{fig:Vibration-measurement-control-loop}. The voltage measured at the PD and the feedback voltage to the PZT are digitized at a sampling frequency of $200\text{ \ensuremath{\frac{\text{kS}}{\text{s}}}}$ over a 0-10 V range using an analog-to-digital converter (ADC) (Measurement Computing Corporation, PN: USB-1808X). Thus, the feedback voltage to the PZT mirror measures the drift of the fringe at frequencies $<500\text{ mHz}$ while the PD voltage measures the fast excursions (vibrations) about the steepest slope on the fringe up to the Nyquist frequency $100\text{ kHz}$. This control system is convenient as the PD has a high bandwidth of $\approx20\text{ MHz}$ so it can easily detect these excursions. In addition, servoing the PZTs with a slow integrator ensures that the intrinsic resonances of the PZT as well as the PZT mirror assembly, in the range of 100 Hz-1 kHz, are not excited.

\subsection{Results}
\label{subsec:results-vibration-measurement}

\begin{figure}[!!h]
\centering
\includegraphics[width=\textwidth]{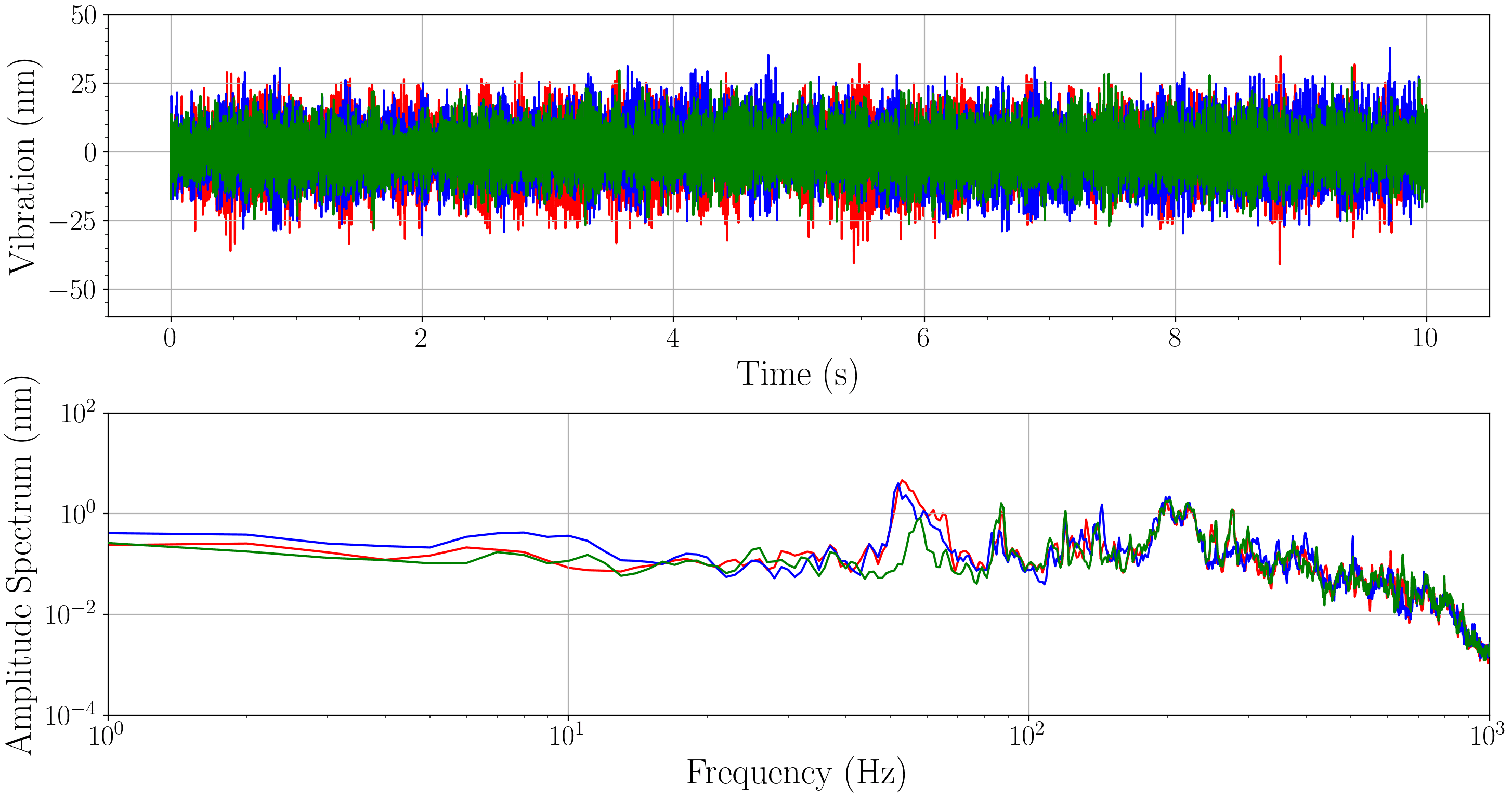}
\caption[$X$-axis Vibration Spectrum]{$X$-axis vibration analysis of the baseline (green), with lead shot (blue), and without lead shot (red). Vibration time series inferred from the PD over 10 s sampled at $f_{\text{s}}=200\text{ kHz}$ and corresponding amplitude spectrum. The frequency resolution for the measurement is $f_{\text{res}}=1\text{ Hz}$ which enables averaging over 20 segments, yielding an equivalent noise bandwidth (ENBW) of $1.5\text{ Hz}$. The amplitude spectrum is $<10^{-2}\text{ nm}_{\text{RMS}}$ for frequencies above 1 kHz and follows the noise floor - baseline configuration.}
\label{fig:x-axis-vibration-spectrum}
\end{figure}

\begin{figure}[!!h]
\centering
\includegraphics[width=\textwidth]{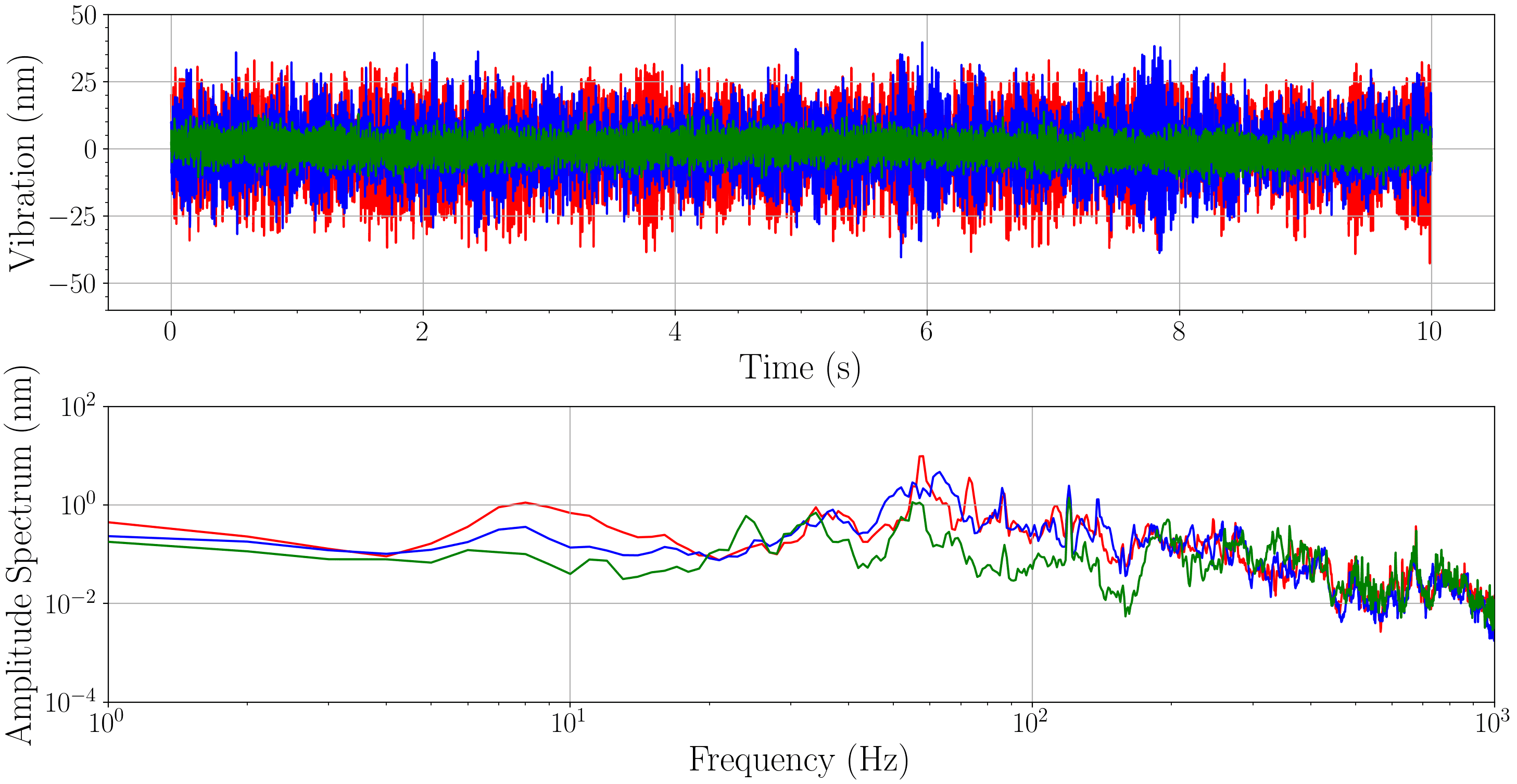}
\caption[$Y$-axis Vibration Spectrum]{$Y$-axis vibration analysis of the baseline (green), with lead shot (blue), and without lead shot (red). Vibration time series inferred from the PD over 10 s sampled at $f_{\text{s}}=200\text{ kHz}$ and corresponding amplitude spectrum. The frequency resolution for the measurement is $f_{\text{res}}=1\text{ Hz}$ which enables averaging over 20 segments, yielding an equivalent noise bandwidth (ENBW) of $1.5\text{ Hz}$. The amplitude spectrum is $<10^{-2}\text{ nm}_{\text{RMS}}$ for frequencies above 1 kHz and follows the noise floor - baseline configuration.}
\label{fig:y-axis-vibration-spectrum}
\end{figure}

\begin{figure}[!!h]
\centering
\includegraphics[width=\textwidth]{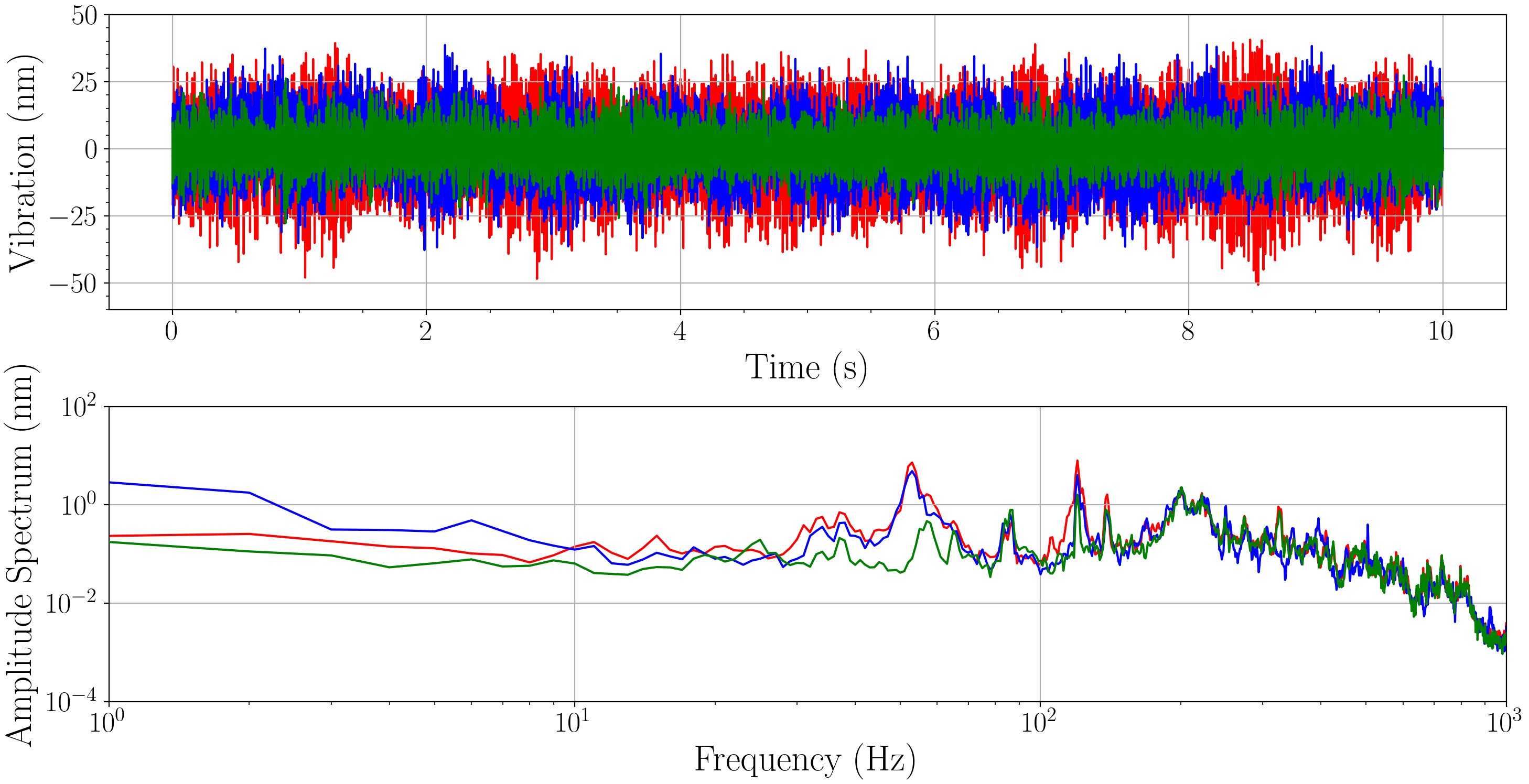}
\caption[$Z$-axis Vibration Spectrum]{$Z$-axis vibration analysis of the baseline (green), with lead shot (blue), and without lead shot (red). Vibration time series inferred from the PD over 10 s sampled at $f_{\text{s}}=200\text{ kHz}$ and corresponding amplitude spectrum. The frequency resolution for the measurement is $f_{\text{res}}=1\text{ Hz}$ which enables averaging over 20 segments, yielding an equivalent noise bandwidth (ENBW) of $1.5\text{ Hz}$. The amplitude spectrum is $<10^{-2}\text{ nm}_{\text{RMS}}$ for frequencies above 1 kHz and follows the noise floor - baseline configuration.}
\label{fig:z-axis-vibration-spectrum}
\end{figure}

The measured vibrations and corresponding amplitude spectra of the cryostat relative to the optics table in $x,y,z$ are presented Figures \ref{fig:x-axis-vibration-spectrum}, \ref{fig:y-axis-vibration-spectrum}, and \ref{fig:z-axis-vibration-spectrum} respectively. The data was measured in three configurations: baseline, lead shot, and no lead shot. In the baseline configuration, a probe mirror is placed just before the cryostat window along each direction and thus measures the vibrations of the optical components in the interferometer and is taken to be the noise floor of the measurement. In the no lead shot configuration, the flexible transfer line is resting on the floor with no bags of lead shot and the probe mirror is a gold-plated piece of copper at the 4 K stage. In the lead shot configuration, 4 bags of lead shot are placed over the flexible transfer line, clamping it to the floor and the probe mirror is a gold-plated piece of copper to the 4 K stage.\footnote{Our colleagues \cite{zhangHighOpticalAccess2025} observed that the vibrations of their cryostat were further reduced by placing bags of lead shot over the transfer line.} In addition, I installed PEEK spokes on the the radiation shields to increase the stiffness and damping of the cold stages. Thin PEEK tabs, 0.3 and 0.5 mm thick, were cut into rectangles and then folded back onto themselves to form a spring element. The ends of each tab were then bolted to the radiation shields, while the folded section acted as a spring that presses against either the chamber wall or the 40 K stage. Thus, the efficacy of the transfer line vibration isolation and increased stiffness and damping due to the PEEK spokes was measured. 

For each measurement, the time duration was 10 s and the frequency resolution was chosen to be $f_{\text{res}}=1\text{ Hz}$ which enables averaging over 19 segments, yielding an equivalent noise bandwidth (ENBW) of $1.5\text{ Hz}$. The final resolution was decided by comparing the RMS amplitude over various frequency bands where there appeared to be resonances using Equation \eqref{eq:RMS-motion-in-frequency-band-from-AS} while tuning $f_{\text{res}}$, and thus the averaging. I calculated that the RMS amplitude fluctuated within 10\% between $f_{\text{res}}=0.1\text{ Hz}$ (no averaging) and $f_{\text{res}}=1\text{ Hz}$ (20 segments). Thus, $f_{\text{res}}=1\text{ Hz}$ was used to reduce the noise variance. In hindsight, I should have increased the time duration to get lower frequency resolution with suitable averaging, but this was not done in this work. In addition, the amplitude spectra for each measurements was $<10^{-2}\text{ nm}_{\text{RMS}}$ for frequencies above 1 kHz and follows the noise floor (baseline).

In the $x$-direction, the vibrations in the frequency bands [1-20 Hz] and [130-150 Hz] were worse with the lead shot while the vibrations in the frequency band [43-70 Hz] were suppressed and is summarized in Table \ref{tab:x-axis-band-RMS}. 

\begin{table}[!!h]
\centering
\tablefont
\caption[$X$-axis RMS amplitude]{$X$-axis RMS amplitude in various frequency bands computed from the amplitude spectrum in Figure \ref{fig:x-axis-vibration-spectrum} using Equation \eqref{eq:RMS-motion-in-frequency-band-from-AS}.}
\label{tab:x-axis-band-RMS}
\begin{tabularx}{\textwidth}{>{\centering\arraybackslash}p{0.2\textwidth}>{\centering\arraybackslash}p{0.28\textwidth}cc}
\toprule
\textbf{Frequency Band (Hz)} & \textbf{Lead Shot, RMS (nm)} & \textbf{No Lead Shot, RMS (nm)} & \makecell{\textbf{Baseline, RMS (nm)}}\\
\midrule
1-4 & 0.7 & 0.4 & 0.4 \\
\midrule
5-16 & 1.0 & 0.4 & 0.4 \\
\midrule
28-35 & 0.3 & 0.4 & 0.3 \\
\midrule
43-70 & 6.6 & 9.0 & 1.4 \\
\midrule
130-150 & 2.5 & 1.6 & 1.3 \\
\bottomrule
\end{tabularx}
\end{table}

In the $y$-direction, the vibrations in the frequency bands [1-20 Hz], [28-80 Hz], and [100-130 Hz] were suppressed with the lead shot while the vibrations in the frequency bands [80-100 Hz] and [130-180 Hz] were worse and is summarized in Table \ref{tab:y-axis-band-RMS}. 

\begin{table}[!!h]
\centering
\tablefont
\caption[$Y$-axis RMS amplitude]{$Y$-axis RMS amplitude in various frequency bands computed from the amplitude spectrum in Figure \ref{fig:y-axis-vibration-spectrum} using Equation \eqref{eq:RMS-motion-in-frequency-band-from-AS}.}
\label{tab:y-axis-band-RMS}
\begin{tabularx}{\textwidth}{>{\centering\arraybackslash}p{0.2\textwidth}>{\centering\arraybackslash}p{0.28\textwidth}cc}
\toprule
\textbf{Frequency Band (Hz)} & \textbf{Lead Shot, RMS (nm)} & \textbf{No Lead Shot, RMS (nm)} & \makecell{\textbf{Baseline, RMS (nm)}}\\
\midrule
1-4 & 0.3 & 0.5 & 0.2 \\
\midrule
4-20 & 0.7 & 2.1 & 0.3 \\
\midrule
28-43 & 1.8 & 2.0 & 1.3 \\
\midrule
45-70 & 11.8 & 15.1 & 2.5 \\
\midrule
70-80 & 1.6 & 5.2 & 0.3 \\
\midrule
80-100 & 3.3 & 2.8 & 0.3 \\
\midrule
100-130 & 3.5 & 3.7 & 1.8 \\
\midrule
130-180 & 2.5 & 1.8 & 0.5 \\
\bottomrule
\end{tabularx}
\end{table}

In the $z$-direction, the vibrations in the frequency bands [1-10 Hz] were worse with the lead shot while the vibrations in the frequency bands [10-20 Hz] and [28-150 Hz] were suppressed and is summarized in Table \ref{tab:z-axis-band-RMS}.

\begin{table}[!!h]
\centering
\tablefont
\caption[$Z$-axis RMS amplitude]{$Z$-axis RMS amplitude in various frequency bands computed from the amplitude spectrum in Figure \ref{fig:z-axis-vibration-spectrum} using Equation \eqref{eq:RMS-motion-in-frequency-band-from-AS}.}
\label{tab:z-axis-band-RMS}
\begin{tabularx}{\textwidth}{>{\centering\arraybackslash}p{0.2\textwidth}>{\centering\arraybackslash}p{0.28\textwidth}cc}
\toprule
\textbf{Frequency Band (Hz)} & \textbf{Lead Shot, RMS (nm)} & \textbf{No Lead Shot, RMS (nm)} & \makecell{\textbf{Baseline, RMS (nm)}}\\
\midrule
1-4 & 3.4 & 0.4 & 0.2 \\
\midrule
4-10 & 0.8 & 0.3 & 0.2 \\
\midrule
10-20 & 0.3 & 0.5 & 0.2 \\
\midrule
28-45 & 1.0 & 1.6 & 0.4 \\
\midrule
45-70 & 8.1 & 11.4 & 1.0 \\
\midrule
100-130 & 5.2 & 10.5 & 2.1 \\
\midrule
130-150 & 1.3 & 2.5 & 1.5 \\
\bottomrule
\end{tabularx}
\end{table}

In summary, the vibrations in the frequency band of 1-10 Hz were measured at the cryostat. We attribute this to the dominant motion of the cryocooler is due to the driving frequency of the GM refrigeration cycle of $\approx1\text{ Hz}$ with an amplitude of $12\text{ \ensuremath{\mu}m} \left(80\text{ \ensuremath{\frac{\text{\ensuremath{\mu}m}_{\text{RMS}}}{\sqrt{\text{Hz}}}}}\right)$ and harmonics of this frequency up to 10 Hz \cite{tomaruVibrationAnalysisCryocoolers2004}. These vibrations are suppressed in the $y$ direction by clamping the transfer line to the floor with lead shot and are worse in the $x,z$ directions, but the amplitude is low in all directions at $\sim1\text{ nm}_{\text{RMS}}$. The lead shot suppressed vibrations particularly in the frequency band of 45-70 Hz for all three directions and 100-130 Hz in the $y$ direction yielding amplitudes $<12\text{ nm}_{\text{RMS}}$. The lowest COMSOL simulated resonances at 36.0 Hz, 36.4 Hz, and 38.3 Hz correspond to motion of the 4 K stage: lateral translation in $z,x$ plus a torsional mode about $y$. These are the most likely candidates for the dominant measured resonances in the frequency band of 45-70 Hz, shifted upward from the simplified simulated values by the stiffness of the PEEK tabs. The next simulated resonances at 76.3 Hz, 85.3 Hz, and 85.5 Hz involve out of phase lateral and torsional motion of the cold stages and are plausible contributors to the measured resonances in the frequency band of 80-180 Hz. The resonance around 120 Hz could be excited from the GM cryocooler resonances at 115 Hz \cite{tomaruVibrationAnalysisCryocoolers2004} and/or acoustic resonances from the cryocooler and the lab fans providing a positive pressure at the optics tables \cite{spiveyHighStabilityCryogenicSystem2022}. The simulated resonances at 219.0 Hz and 250 Hz involve rocking modes of the cold stages about $x$ and $z$ which do not appear to be excited in the measured $x$ and $z$ directions but are plausible contributors to the measured $y$ vibrations in the frequency band of 200-300 Hz. The simulated resonances at 340.8 Hz and 686.2 Hz involve axial translation in $y$ of the cold stages which do not appear to be excited in the measured $y$ directions. Lastly, the simulated resonances at 361.5 Hz 383.0 Hz correspond to rocking modes of the cold stages and are the most likely contributors to the weakly excited measured $x,z$ resonances in the 300-500 Hz frequency band. 

In short, it is difficult to accurately map the simulated and measured resonances as the PEEK tabs increased the stiffness of the cold stages and the measured amplitude spectra could also include harmonics of resonances. Additionally, acoustic frequencies in the lab can couple into the cryostat and excite the cold stage resonances. Another unknown contributor to these measured vibrations could be due to helium liquefaction at the JT capillary. Additionally, the vibrations were not measured with the cryocooler off. This would have provided a better control to show the efficacy of the transfer line vibration isolation as the cold stages could be excited via acoustic frequencies in the lab. 

The baseline configuration in the $y$ direction is much lower in frequency band of 80-200 Hz and 800-1000 Hz due to the fact that the probe mirror was mounted to the hanging breadboard and thus does not measure the resonances of the hanging breadboard relative to the optics table. The eigenfrequencies of the hanging breadboard were simulated in COMSOL resulting in dominant modes at 94.5 Hz, 105 Hz, 151 Hz, and 166 Hz and 13 higher order modes in the frequency band of 300-1000 Hz. Thus, the measured 4 K stage resonances in the $y$ direction are partly contributed to the hanging breadboard resonances. In the $x,z$ baseline configurations, dominant resonances around 200 Hz are assumed to be contributed to structural resonances of the optical components in the interferometer. The $x,z$ interferometer setup consisted of a 1 meter beam path to the cryostat via 3 mirrors with lens tubes installed on the mirror mounts to reduce air currents interfering with the measurement. This 200 Hz resonance does not appear in the $y$ baseline configuration as the beam path to the cryostat was more direct - the lens tube was attached to the box enclosing the interferometer and not the mirrors.

Looking forward, it might be possible to further decrease the resonance frequency of the transfer line by increasing its length and decreasing its radius as it scales with $\frac{r}{L^{2}}$, assuming the simple model using Equation \eqref{eq:transfer-line-resonance-frequency} - this might enable better better isolation $r>\sqrt{2}$. However, decreasing the radius is difficult as the transfer line encloses the supply and return of the recirculating helium. Additionally, making the transfer line longer will limit the cooling power at the cryostat due to increased parasitic thermal loading of the helium. Another potential improvement to this system would be to clamp the flexible helium line between the GM cryocooler and the compressor with bags of lead shot as this might help suppress vibrations from the compressor. The cryocooler could also be better acoustically isolated by installing more acoustic damping foam around the box it resides in. For the cryostat cold stages, the axial frequencies are reasonably high relative to the dominant vibrational frequencies of the GM cryocooler, but the lateral frequencies could be higher. The mass of the cold stages in this design are constrained as we want to have a large experimental region at the cold stages that are made out of C10100. In hindsight, the PEEKGF30 supports could have been shorter potentially by a factor of 3, which would increase the lateral resonances of the cold stages to above 200 Hz. In addition, a stiffer material such as carbon fiber could be used which can have Young's Modulus that is 40 times greater than PEEKGF30 depending on the class which would also increase the lateral resonances and would be easy to implement.

\subsection{Vibration Impact on Ion-Laser Interactions and Cavity Length Fluctuations}
\label{subsec:vibration-impact-on-ion-laser-interactions-and-cavity-length-fluctuations}

I will now discuss the impact on ion-laser interactions and cavity length fluctuations using the measured vibration amplitude spectra. For the ion-laser interactions, the vibration induced sources of error considered here are laser intensity errors and laser frequency (detuning) errors. Here, I will discuss these errors using the peak amplitudes of the dominant vibrational resonances measured in Subsection \ref{subsec:results-vibration-measurement}: $8.1\text{ nm}_{\text{RMS}}\times\sqrt{2}=12\text{ nm}$ at $53\text{ Hz}$ in the $x,z$ directions and $11.8\text{ nm}_{\text{RMS}}\times\sqrt{2}=17\text{ nm}$ at $62\text{ Hz}$ in the $y$ direction.

Laser intensity errors can be estimated assuming the intensity at the ions is that of a Gaussian laser beam which is given as follows, 
\begin{equation}
    I(r)=I_{0}e^{-\frac{2r^{2}}{w^{2}}}
    \label{eq:intensity-of-gaussian-beam}
\end{equation}
Here, $I_{0}$ is the peak intensity, $w$ is the beam waist at the ions, and $r$ is the radial distance from the ion \cite{salehFundamentalsPhotonics}. Taking the beam waist to be $25\text{ \ensuremath{\mu}m}$ at the ion and using the peak vibration in the $y$ direction of $17\text{ nm}$ at $62\text{ Hz}$, the fractional change in intensity at the ion is $9\times10^{-7}$. Fractional changes in intensity are sensitive to vibration in all directions, but I'll use the $y$ direction as its vibration level is the largest. Since the Rabi frequency scales with intensity as $\Omega\propto\sqrt{I}$, the fractional change in the Rabi frequency is $\frac{\text{d}\Omega}{\Omega}=\frac{1}{2}\frac{\text{d}I}{I}=4.5\times10^{-7}$. A change in Rabi frequency will cause the geometric phases accumulated in a two-qubit gate to differ from their ideal values. This gate error for a small Rabi frequency errors is written as follows,
\begin{equation}
    \epsilon_{\Omega}=\frac{\pi^{2}}{4}\left(\frac{\text{d}\Omega}{\Omega}\right)^{2}
    \label{eq:two-qubit-gate-error-rabi-error}
\end{equation}
where $\left(\frac{\text{d}\Omega}{\Omega}\right)$ is the fractional change in the Rabi frequency \cite{ballanceHighFidelityQuantumLogic2017a}. Thus, the vibration induced Rabi frequency error of $\left(\frac{\text{d}\Omega}{\Omega}\right)=4.5\times10^{-7}$ yields a two-qubit gate error of $\epsilon_{\Omega}=5\times10^{-13}$, which is acceptable. In summary, the maximum tolerable vibration level for a beam waist of $25\text{ \ensuremath{\mu}m}$ is $3\text{ }\mu\text{m}_\text{RMS}$ as this yields a two-qubit gate error of $\epsilon_{\Omega}=2\times10^{-3}$.

Detuning errors can be estimated by calculating the Doppler shift of a Raman beam at the ion. I will use the vibration amplitude of $12\text{ nm}$ at $53\text{ Hz}$ in $x,z$ directions as this defines the plane that the Raman beams would be aligned to the ions at the trap in this cryostat - $y$ vibrations will not create a Doppler shift. As the vibrational period of $19\text{ ms}$ is much longer than typical gate times $t_{\text{g}}=50\text{ \ensuremath{\mu}s}$, the ion's velocity remains nearly constant over the course of a single gate operation and can be modeled as a systematic detuning. An ion oscillating at $53\text{ Hz}$ with a peak amplitude of $12\text{ nm}$ has a maximum speed of $v_{\text{max}}=A_{\text{pk}}2\pi F=12\text{ nm}\times2\pi\times53\text{ Hz}=3.8\times10^{-6}\text{ }\frac{\text{m}}{\text{s}}$. This vibrating ion experiences a Doppler shifted Raman frequency of $\kappa=\frac{2\pi}{\lambda_{\text{Raman}}}v_{\text{max}}=\frac{2\pi}{355\text{ nm}}3.8\times10^{-6}\text{ }\frac{\text{m}}{\text{s}}=68\text{ Hz}$. A detuning error will prevent the loops in phase space to completely close, causing spin and motion to remain entangled at the end of a gate operation. The gate error for small detuning errors is written as follows, 
\begin{equation}
    \epsilon_{\kappa}=\frac{1+2K(1+2\bar{n})}{16K^{2}}(\kappa t_{\text{g}})^{2}
    \label{eq:two-qubit-gate-error-detuning-error}
\end{equation}
where $K$ is the number of loops, $\bar{n}$ is the mean thermal occupation of the motional mode, $\kappa$  is the detuning error, and $t_{\text{g}}$ is the gate time \cite{ballanceHighFidelityQuantumLogic2017a}. Assuming $K=1$, ideally ground state cooled ions $\bar{n}=0$, a gate time of $50\text{ \ensuremath{\mu}s}$, and a vibration induced detuning error of $\kappa=68\text{ Hz}$, the two-qubit gate error is $\epsilon_{\kappa}=2\times10^{-6}$, which is acceptable. In summary, the maximum tolerable vibration level is constrained to $300\text{ nm}_\text{RMS}$ as this yields a two-qubit gate error of $\epsilon_{\kappa}=3\times10^{-3}$.

The vibration impact on cavity length fluctuations was not done in this work, but I will at least briefly discuss a way that it could be simulated. Fluctuations of the cavity length degrade the ion-cavity cooperativity. The cavity linewidth is $\Delta L=\lambda/(2\mathcal{F})$ and the cavity operated in this work has $\lambda=1650$ nm and $\mathcal{F}=5.6\times10^{4}$. For Gaussian-distributed cavity length excursions of
RMS amplitude $\sigma$, the effective cooperativity degrades from its resonant value $C^{0}$ as follows \cite{fontanaMechanicallyStableTunable2021a}
\begin{equation}
C(\sigma)=C^{0}\sqrt{\frac{\pi}{8}}\frac{\Delta L}{\sigma}
\exp\!\left(\frac{\Delta L^{2}}{8\sigma^{2}}\right)
\operatorname{erfc}\!\left(\frac{\Delta L}{2\sqrt{2}\sigma}\right)
\end{equation}
The degradation is steep: at $\sigma=\Delta L/2$ the cooperativity has already fallen to approximately 66\% of $C^{0}$, and at $\sigma=\Delta L$ to approximately 44\%. With the resonant cooperativity of the 13 mm cavity in this work estimated at $C^{0}\approx2.1$, the RMS cavity length excursions must be $\sigma\ll 15$ pm.
Using the vibration time series or amplitude spectra measured in 3D in Subsection \ref{subsec:results-vibration-measurement} as a base excitation of the ion trap and cavity structure in COMSOL, the response to the cavity length could be simulated. This could be useful for design goals such as material choice and geometry of future cavity structures.

\ifSubfilesClassLoaded{%
  \bibliographystyle{unsrt}
  \bibliography{Dissertation}
}{}
\end{document}

%% file: Chapter9.tex
\chapter{Outlook}
\label{chap:outlook}

The work presented in this thesis established two low-vibration cryostats interfaced with the CET Stinger cryogenic system. The primary focus was the development and commissioning of cryo0, which integrates an Alpine Quantum Technologies (AQT) RF Paul trap and a Fabry-Pérot (FP) optical cavity at the 4 K stage. The optical cavity, designed by Wance Wang and described in his thesis, was incorporated into a cryogenic platform that I designed and assembled for trapped-ion experiments.

To realize this system, I developed the major cryogenic subsystems required for operation, including an in-vacuum 0.5 NA imaging objective, an Yb atomic source, an in-vacuum helical resonator, the in-vacuum wiring, and an in-vacuum DC electrode filter board. I also designed three-axis magnetic field coils to provide an arbitrary quantization axis, relaxing the optical access constraints for driving quadrupole transitions. Outside the vacuum system, I developed the laser delivery and imaging optics, implemented the experimental control system using ARTIQ, and characterized the trapped-ion performance through measurements of the secular frequencies, excess micromotion, photon collection efficiency, and ion lifetime. In parallel, I characterized the cryostat performance by measuring its cooling capacity, vibration, and vacuum level, and developed a helium purification system that increased continuous operating time from approximately 30 days to 72 days. Finally, Wance Wang and I integrated the Fabry-Pérot cavity with the AQT Pine ion trap and demonstrated spatial overlap and interaction between a trapped $^{174}\mathrm{Yb}^{+}$ ion and the cavity mode on the 1345 nm transition, establishing preliminary experimental validation of the integrated cryogenic ion-cavity structure.

A second cryostat, cryo1, was also constructed and commissioned to demonstrate the reproducibility of the overall design, although it was not used for the trapped-ion experiments presented in this thesis. In addition, I developed a dedicated bakeout station for vacuum processing of the cryostats and other ultra-high-vacuum systems.

The completed cryo0 platform achieved the primary design objectives established at the outset of this work. The system provides the high optical access and large number of RF and DC electrical connections required to implement our quantum networking architecture while maintaining vibration levels below $12~\mathrm{nm_{RMS}}$ in three dimensions, enabling both future high-fidelity quantum gates and stable optical cavity locking. The cryostat was constructed primarily from non-magnetic materials, although characterization of the residual magnetic field and field gradients remains future work. Cryogenic trapping of $^{174}\mathrm{Yb}^{+}$ ions was successfully demonstrated, with ion storage times exceeding five days without laser cooling. The system was also designed to withstand bakeout temperatures up to 500 K, providing a future path toward studying the effect of clean cryogenic surfaces on anomalous ion heating. Finally, the vacuum-continuous cold-stage architecture provides rapid access to the 4 K stage, greatly simplifying future modifications and experimental development.

Looking forward, the immediate next step is to bake the cryostat and measure trapped-ion heating rates under cryogenic conditions. Additionally, optical access along the $\frac{-\hat{x}-\hat{z}}{\sqrt{2}}$ and $\frac{\hat{x}+\hat{z}}{\sqrt{2}}$ directions could be enabled by installing a CF-2.75 tee at the 300 K Chamber ports together with appropriate windows. However, for this configuration the oven would need to be installed below the trap. More broadly, the experimental platform developed in this work provides the foundation for implementing cavity-mediated quantum networking between remote trapped-ion processors. Future experiments will focus on generating single 1650 nm photons into the cavity-coupled optical fiber, characterizing the fidelity of ion-photon entanglement, and ultimately demonstrating remote ion-ion entanglement through Hong-Ou-Mandel interference. These advances will move the system toward the long-term goal of distributing entanglement over distances exceeding 10 km between compute-capable $^{171}\mathrm{Yb}^{+}$ quantum nodes.

\ifSubfilesClassLoaded{%
  \bibliographystyle{unsrt}
  \bibliography{Dissertation}
}{}
\end{document}

%% file: AppendixA.tex
\ifSubfilesClassLoaded{\appendix}{}

\newpage

\chapter{Vacuum Processing}
\label{chap:vacuum-processing}

\section{Vacuum Fundamentals}
\label{sec:vacuum-fundamentals}
Gas molecules are bound on the inner surfaces of the vacuum chamber via adsorption (physisorption or chemisorption). Physisorbed molecules are bonded to the surfaces via weak Van Der Waals forces of energies $<0.4\,\frac{\text{eV}}{\text{molecule}}$. Adsorption at greater energies is known as chemisorption which is due to chemical bonds formed between the gas molecules and the surface \cite{grinhamReviewOutgassingMethods2017,ohanlonUsersGuideVacuum1989}. Sub-monolayer surface coverage can be modeled using the Polanyi-Wigner equation where the desorption rate is proportional to the concentration of molecules on the surface, $C$,
\begin{equation}
    \frac{\text{d}C(t)}{\text{d}t}=\nu e^{-E_{\text{des}}/k_{\text{B}}T}C(t)
    \label{eq:firs-order-desorption}
\end{equation}
where $E_{\text{des}}$ is the desorption energy in $\frac{\text{eV}}{\text{molecule}}$, $k_{\text{B}}$ is the Boltzmann constant, $T$ is the temperature of the surface, and $\nu$ is the vibrational frequency of a molecule in an adsorption site (typically $10^{13}\text{ Hz}$) \cite{ohanlonUsersGuideVacuum1989}. The average time an adsorbed molecule spends on the surface, called the residence time, is defined as follows,
\begin{equation}
    \tau_{\text{r}}=\frac{1}{\nu}e^{E_{\text{des}}/k_{\text{B}}T}
    \label{eq:residence-time-first-order-desorption}
\end{equation}
Residence times at different temperatures for relevant gas molecules in a vacuum system are listed in Table \ref{tab:residence-time-for-various-gas-molecules}. 

After a couple monolayers of surface coverage, molecules will begin bind to each other (instead of the surface) via Van Der Waals forces - i.e., condense - and the pressure of the molecules above the surface saturates, known as the vapor pressure. At this point, the adsorbed molecules behave like a bulk condensed phase at a temperature equal to the surface temperature and the heat of vaporization equals the binding energy between the molecules. This binding energy (heat of vaporization) is typically 2-3 times lower than the physisorbed and chemisorbed desorption energies \cite{baglinCryopumpingVacuumSystems2020}. The vapor pressure of a molecule can be modeled using the Clausius-Clapeyron equation
\begin{equation}
    \ln P_{\text{vap}}(T)=A-\frac{\Delta H_{\text{vap}}}{RT}
    \label{eq:clausius-clapeyron-equation}
\end{equation}
where $\Delta H_{\text{vap}}$ is the heat of vaporization of the molecule and $A$ is a constant \cite{baglinCryopumpingVacuumSystems2020}. 

We will now define a metric for when a gas is effectively cryocondensed on a surface. A surface held at temperature, $T_{\text{surface}}$, will begin to cryocondense a gas when the vapor pressure of the bulk gas, $P_{\text{vap,gas}}(T_{\text{surface}})$, is less than the partial partial pressure of the gas, $P_{\text{gas}}$, 
\begin{equation}
    P_{\text{vap,gas}}(T_{\text{surface}})<P_{\text{gas}}
    \label{eq:cryocondensing-metric}
\end{equation}
The vapor pressures of relevant molecules in well prepared vacuum systems are listed in Table \ref{tab:vapor-pressures-for-various-gas-molecules}. In summary, surfaces below 20 K will provide excellent bulk condensation of all the gases except for hydrogen. To reiterate, if the hydrogen partial pressure in the chamber is $10^{-6}\text{ Torr}$, bulk condensation will occur on a 4 K surface due to the criterion in Equation \eqref{eq:cryocondensing-metric}, but the hydrogen partial pressure in the chamber will saturate at $10^{-7}\text{ Torr}$. To pump hydrogen below its saturated vapor pressure at 4 K, another pumping mechanism is used known as cryosorption. 

The process of cryosorption is based on the fact that physisorbed hydrogen has a larger binding energy than condensed hydrogen, $0.05\,\frac{\text{eV}}{\text{molecule}}>0.009\,\frac{\text{eV}}{\text{molecule}}$, which corresponds to a longer residence time \cite{baglinCryopumpingVacuumSystems2020}. Consequently, the equilibration pressure of sub-monolayer physisorbed hydrogen is lower than that of condensed hydrogen. As a corollary to that, the ultimate pressure is a function of the surface coverage which can be minimized by having a small gas load or generating a large surface area. Additionally, the hydrogen adsorption capacity on an electroplated copper surface has been measured to increase with the addition of carbon dioxide cryodeposit, as it makes the surface more porous increasing the effective surface area \cite{wallenAdsorptionIsothermsH21996}. Another material used for cryosorption of hydrogen is activate charcoal at 4 K as it can provide a high desorption energy of $0.2\text{ }\frac{\text{eV}}{\text{molecule}}$ relative to metal surfaces and has a remarkable surface area to volume ratio $\approx1000\text{ }\frac{\text{m}^{2}}{\text{g}}$ \cite{baglinCryopumpingVacuumSystems2020,day-cern-cryopumping}. 

Another way to pump hydrogen below its saturated vapor pressure at 4 K is to utilize cryotrapping. This process involves injecting a gas species into the chamber that is condensable at a much high temperature. A porous cryodeposit layer is formed by the condensable gas, providing ample surface area for the hydrogen to adsorb. The adsorbed hydrogen then gets `trapped' by additional layers of the condensable gas onto the cold surface and this process continues. Cryotrapping of hydrogen by carbon dioxide has be shown to reduce the saturated equilibrium pressure of hydrogen by 2 orders of magnitude than that of cryosorption alone \cite{wallenAdsorptionIsothermsH21996}. Another gas typically used for cryotrapping hydrogen is argon \cite{jadejaStudyHydrogenPumping2012}. This is an effective pumping technique, but requires a more complicated vacuum design for bleeding in the `trapping' gas and pumping out the `trapping' gas when the system is warmed up.

\begin{table}[!!h]
\centering
\tablefont
\caption[Residence Times of Various Gas Molecules]{Residence times of various gas species at different temperatures rounded to the nearest order of magnitude. Desorption energies are between the gas molecules and a stainless steel surface \cite{ohanlonUsersGuideVacuum1989,day-cern-cryopumping,baglinCryopumpingVacuumSystems2020}.}
\label{tab:residence-time-for-various-gas-molecules}
\begin{tabularx}{\textwidth}{>{\centering\arraybackslash}p{0.07\textwidth}>{\centering\arraybackslash}p{0.2\textwidth}>{\centering\arraybackslash}p{0.1\textwidth}ccccc}
\toprule
\makecell{\textbf{Gas}\\\textbf{Species}} & \makecell{\textbf{Desorption Energy}\\$\mathbf{\left(\frac{\textbf{eV}}{\textbf{molecule}}\right)}$} & $\mathbf{\tau_\textbf{r}^{\textbf{4 K}}}$ \textbf{ (s)} & $\mathbf{\tau_\textbf{r}^{\textbf{77 K}}}$ \textbf{ (s)} & $\mathbf{\tau_\textbf{r}^{\textbf{120 K}}}$ \textbf{ (s)} & $\mathbf{\tau_\textbf{r}^{\textbf{300 K}}}$ \textbf{ (s)} & $\mathbf{\tau_\textbf{r}^{\textbf{400 K}}}$ \textbf{ (s)} & $\mathbf{\tau_\textbf{r}^{\textbf{500 K}}}$ \textbf{ (s)} \\
\midrule
$\text{H}_{2}\text{O}$ & 1.0 & $\infty$ & $10^{52}$ & $10^{29}$ & $10^{4}$ & $10^{-1}$ & $10^{-3}$ \\
\midrule
$\text{CO}_{2}$ & 0.2 & $10^{239}$ & $1$ & $10^{-5}$ & $10^{-10}$ & $10^{-11}$ & $10^{-11}$ \\
\midrule
$\mathbf{\text{O}_{2}}$ & 0.1 & $10^{112}$ & $10^{-7}$ & $10^{-9}$ & $10^{-11}$ & $10^{-12}$ & $10^{-12}$ \\
\midrule
$\mathbf{\text{N}_{2}}$ & 0.08 & $10^{87}$ & $10^{-8}$ & $10^{-10}$ & $10^{-12}$ & $10^{-12}$ & $10^{-13}$ \\
\midrule
$\mathbf{\text{H}_{2}}$ & 0.05 & $10^{50}$ & $10^{-10}$ & $10^{-11}$ & $10^{-12}$ & $10^{-13}$ & $10^{-13}$ \\
\bottomrule
\end{tabularx}
\end{table}

\begin{table}[!!h]
\centering
\tablefont
\caption[Vapor Pressures of Various Gas Molecules]{The vapor pressures of various molecules relevant for cryogenics and vacuum at different temperatures rounded to the nearest order of magnitude \cite{honigFairyTaleVapor1957,flynnCryogenicEngineering2005}. The vapor pressure is the pressure of the molecules that exists above a bulk solid or liquid (of that molecule) in thermodynamic equilibrium. The $-$ cells in the table means the vapor pressure is $<10^{-11}\text{ Torr}$. The $+$ cells in the table means the vapor pressure is $>10^{3}\text{ Torr}$. The cells with NA indicates that the vapor pressure is not defined at that temperature as its above the critical temperature of the molecule. }
\label{tab:vapor-pressures-for-various-gas-molecules}
\begin{tabularx}{\textwidth}{>{\centering\arraybackslash}p{0.08\textwidth}>{\centering\arraybackslash}p{0.08\textwidth}>{\centering\arraybackslash}p{0.09\textwidth}>{\centering\arraybackslash}p{0.09\textwidth}>{\centering\arraybackslash}p{0.09\textwidth}>{\centering\arraybackslash}p{0.09\textwidth}>{\centering\arraybackslash}p{0.09\textwidth}>{\centering\arraybackslash}p{0.09\textwidth}>{\centering\arraybackslash}p{0.09\textwidth}}
\toprule
\makecell{\textbf{Gas}\\\textbf{Species}} & \makecell{$\mathbf{P_\textbf{vap}^{\textbf{1 K}}}$\\ \textbf{ (Torr)}} & \makecell{$\mathbf{P_\textbf{vap}^{\textbf{3 K}}}$\\ \textbf{ (Torr)}} & \makecell{$\mathbf{P_\textbf{vap}^{\textbf{4 K}}}$\\ \textbf{ (Torr)}} & \makecell{$\mathbf{P_\textbf{vap}^{\textbf{10 K}}}$\\ \textbf{ (Torr)}} & \makecell{$\mathbf{P_\textbf{vap}^{\textbf{20 K}}}$\\ \textbf{ (Torr)}} & \makecell{$\mathbf{P_\textbf{vap}^{\textbf{77 K}}}$\\ \textbf{ (Torr)}} & \makecell{$\mathbf{P_\textbf{vap}^{\textbf{120 K}}}$\\ \textbf{ (Torr)}} & \makecell{$\mathbf{P_\textbf{vap}^{\textbf{200 K}}}$\\ \textbf{ (Torr)}}\\
\midrule
$\text{H}_{2}\text{O}$ & - & - & - & - & - & - & $10^{-11}$ & $10^{-3}$ \\
$\text{CO}_{2}$ & - & - & - & - & - & $10^{-8}$ & $10^{-1}$ & $10^3$ \\
$\text{CH}_{4}$ & - & - & - & - & - & $10$ & + & NA \\
$\text{O}_{2}$ & - & - & - & - & - & $10^2$ & + & NA \\
$\text{CO}$ & - & - & - & - & - & $10^2$ & + & NA \\
$\text{N}_{2}$ & - & - & - & - & $10^{11}$ & $10^3$ & + & NA \\
$\text{H}_{2}$ & - & $10^{-11}$ & $10^{-7}$ & 1 & $10^3$ & NA & NA & NA \\
$^4\text{He}$ & $10^{-1}$ & $10^{2}$ & $10^{3}$ & NA & NA & NA & NA & NA \\
$^3\text{He}$ & $10$ & $10^{3}$ & NA & NA & NA & NA & NA & NA \\
\bottomrule
\end{tabularx}
\end{table}
 
\section{Bakeout Station}
\label{sec:bakeout-station}

I designed and built a bakeout station to perform a bakeout of the cryostat and other UHV chambers. This system consists of a pump station (see Subsection \ref{subsec:pump-station}) that is connected to the UHV chamber which sits inside a kiln (see Subsection \ref{subsec:kiln}) and is shown in Figure \ref{fig:bakeout-station-assembly}. In short, the idea is to ramp up the temperature of the kiln to reduce the residence times of the residual gas in the chamber (see Section \ref{sec:vacuum-fundamentals}) while evacuating the gas using the pump station until the total pressure in the chamber reaches UHV. At this point, a pump (ion pump, NEG, TSP) local to the chamber is engaged and subsequently the chamber is isolated from the pump station using a vacuum valve local to the chamber. There are two main benefits of using a kiln: 1.) it ensures temperature uniformity over the chamber during the bakeout and 2.) the bakeout can be performed away from the optics table. This bakeout station is also suitable for vacuum annealing of metals. The idea is to elevate the temperature of the metal under vacuum to accelerate the diffusion and outgassing of impurities. Michael Straus and Henry Luo from Norbert Linke's lab helped me with with the vacuum cleaning, assembly of the parts, initial bakeout tests of the pump station, and initial temperature ramping of the kiln. 

\begin{figure}[!!h]
\centering
\includegraphics[width=0.75\textwidth]{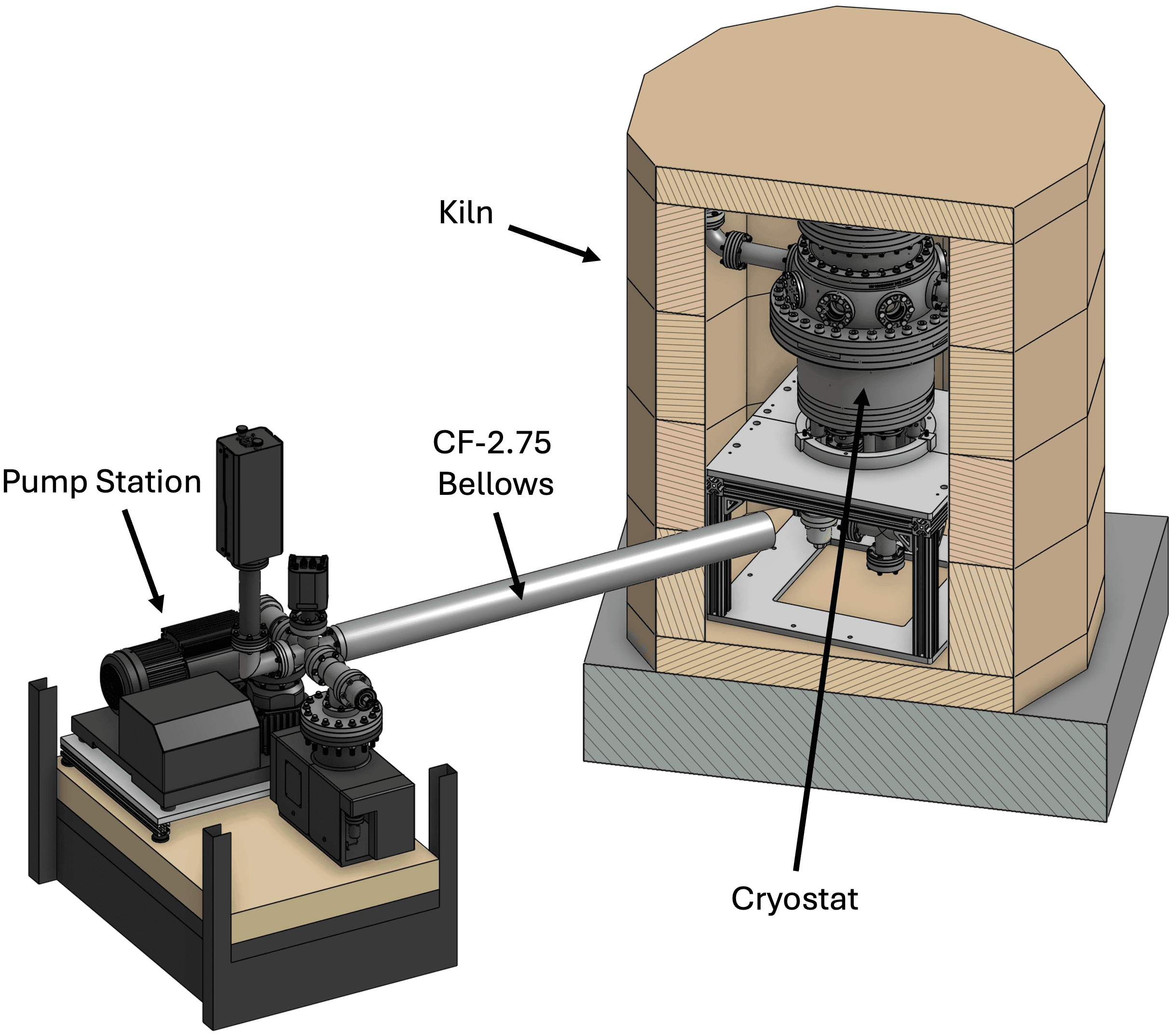}
\caption[Bakeout Station Assembly]{CAD drawing of bakeout station showing how the pump station interfaces with the UHV chamber (here, the cryostat) inside the kiln. The pump station is connected to the UHV chamber with a CF-2.75 bellows.}
\label{fig:bakeout-station-assembly}
\end{figure}

\subsection{Pump Station}
\label{subsec:pump-station}

The pump station consists of a turbo pump, ion pump, ion gauge, RGA, valves, and bellows as shown in Figure \ref{fig:pump-station-assembly}. A parts list of the pump station is summarize in Table \ref{tab:parts-list-for-pump-station}. To match the heights of the turbo pump and ion pump flanges, a riser was machined by Michael Straus and me at UMD. The pump station rests on top of a bed of fire bricks housed in a metal rack enabling a bakeout of the pump station as discussed in the next paragraph. Additional insulation between the pump station and fire bricks is provided by ceramic pedestals. The conflat parts were cleaned after which the pump station was assembled with a blank conflat flange installed on the end of the CF-2.75 bellows. The pump station was then evacuated using the turbo pump after which RGA traces were taken to determine its cleanliness. RGA traces for the unbaked pump station were recorded after 6 minutes and 12 hours and are shown in Figures \ref{fig:RGA-trace-of-pump-station-after-6-minutes} and \ref{fig:RGA-trace-of-pump-station-after-12-hours}. At this point, the pump station was helium-leak-checked using the RGA. 

A low temperature bakeout of the pump station using heater tape was performed as a final test of its cleanliness. During the bakeout, the ion pump was engaged and eventually the turbo pump was valved off. The turbo pump top flange could only be baked to 120 K due to the pump's temperature rating. The ion pump could only be baked to 120 C due to thermal coupling through the firebrick (Figure \ref{fig:pump-station-assembly}) and insufficient heater tape at the ion pump. The rest of the parts were baked to 500 K with the RGA and ion gauge electronics removed. In the future, better insulation and more heater tape can be installed to perform a higher temperature bakeout. The pressure inferred from the ion pump current during the bakeout plateaued to $7\times10^{-10}\text{ Torr}$ after two days. A final RGA trace was then taken after this bakeout and is shown in Figure \ref{fig:RGA-trace-of-pump-station-after-two-day-bakeout}.

\begin{figure}[!!h]
\centering
\includegraphics[width=0.75\textwidth]{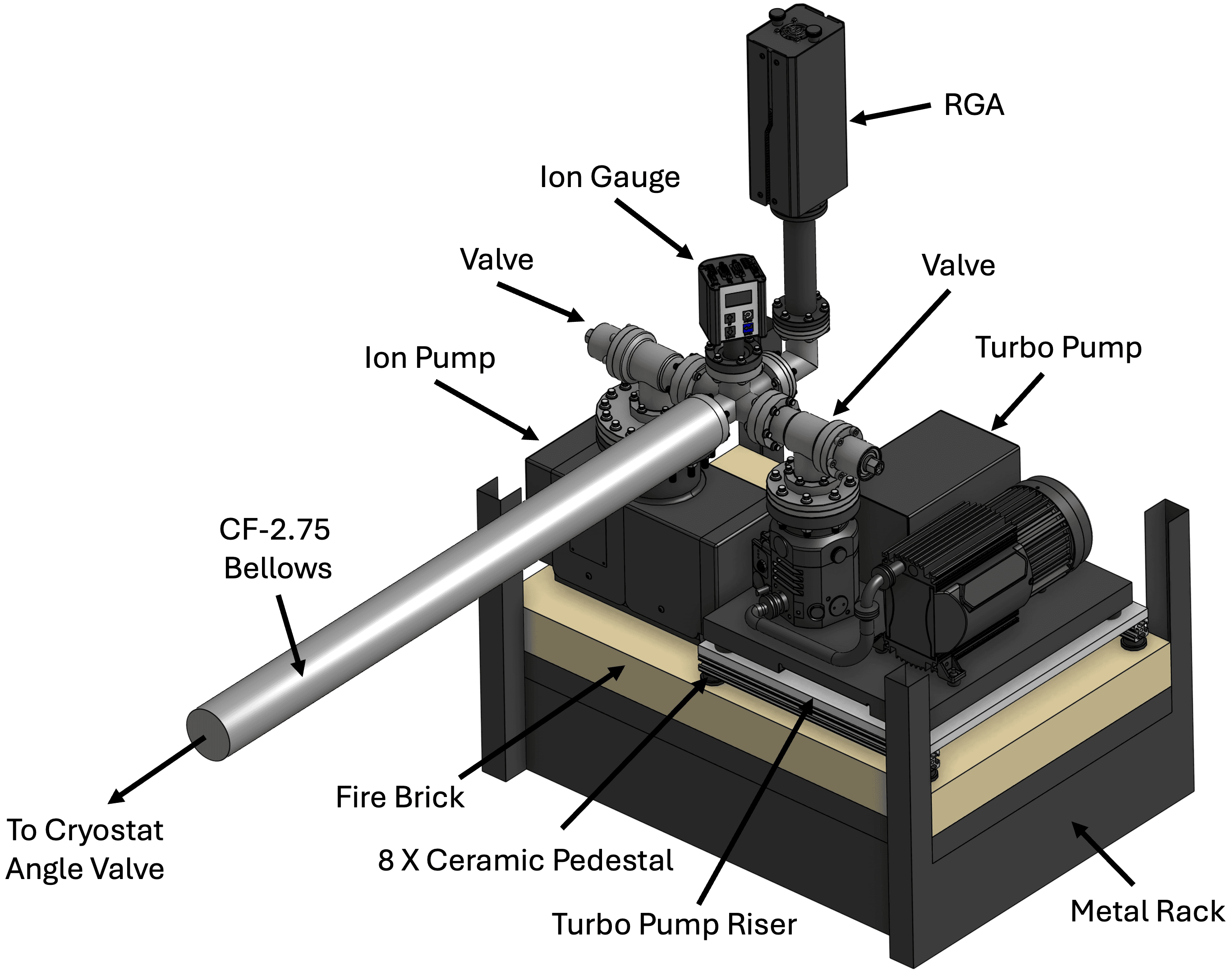}
\caption[Pump Station Assembly]{CAD drawing of pump station showing how the pump station is connected to the UHV chamber with a CF-2.75 bellows. The pump station rests on top of a bed of fire bricks housed in a metal rack. A part list is summarized in Table \ref{tab:parts-list-for-pump-station}.}
\label{fig:pump-station-assembly}
\end{figure}

\begin{table}[!!h]
\centering
\tablefont
\caption[Parts List for pump Station]{Parts list for pump station shown in Figure \ref{fig:pump-station-assembly}}
\label{tab:parts-list-for-pump-station}
\begin{tabularx}{\textwidth}{>{\centering\arraybackslash}p{0.5\textwidth}c}
\toprule
\textbf{Schematic Label} & \textbf{Part} \\
\midrule
Turbo Pump & Edwards Vacuum, PN: TS85D3002 \\
\midrule
Ion Pump & Agilent, PN: 9191410 \\
\midrule
Ion Gauge & Kurt J. Lesker, PN: KJLC392402YF \\
\midrule
RGA & Stanford Research Systems, PN: RGA200 \\
\midrule
Valve & VAT, PN: 54132-GE02 \\
\midrule
Fire Brick & McMast Carr, PN: 9355K2 \\
\midrule
Ceramic Pedestal & Thorlabs, PN: RS05PC \\
\bottomrule
\end{tabularx}
\end{table}

\begin{figure}[!!h]
\centering
\includegraphics[width=0.75\textwidth]{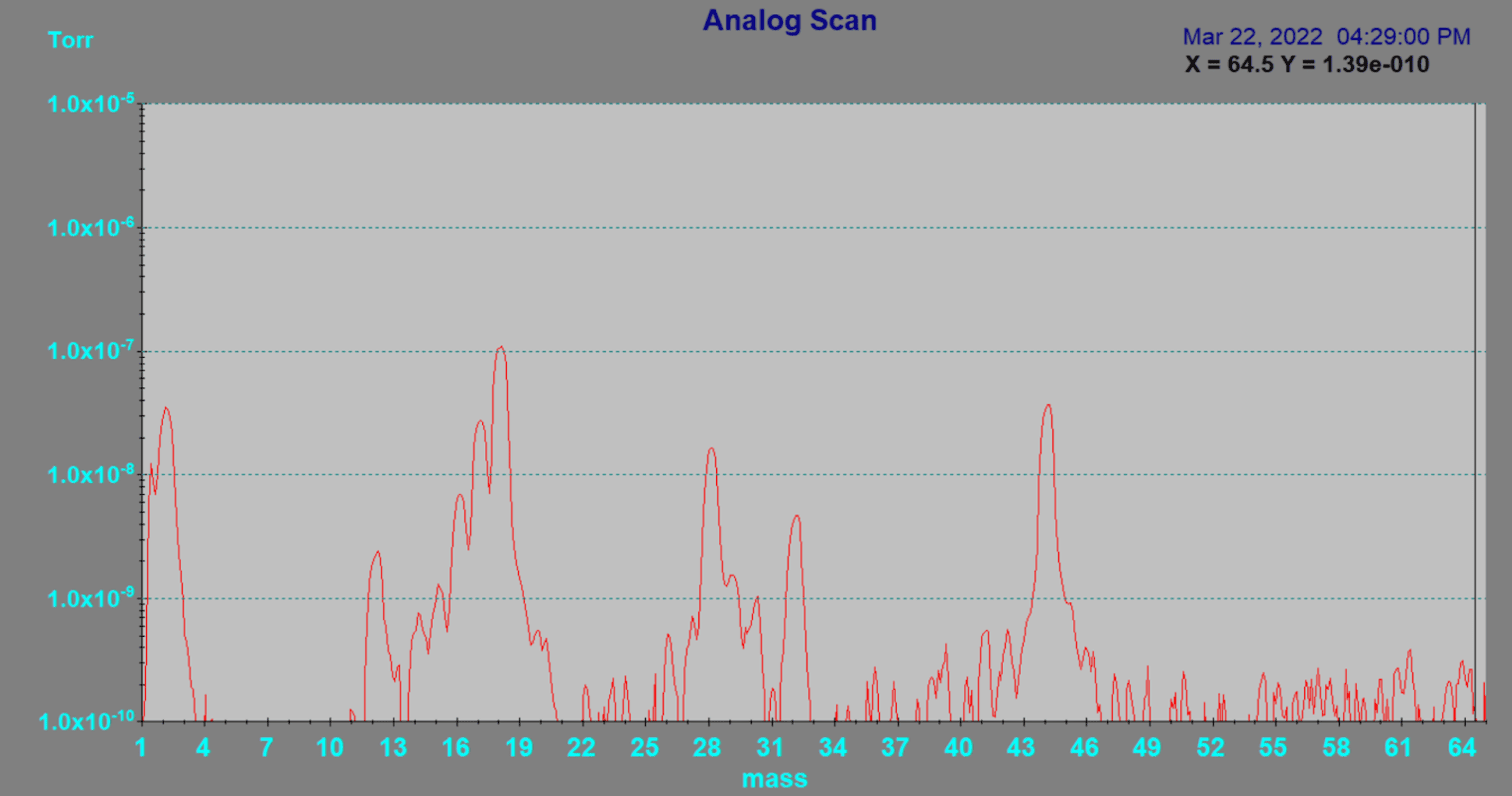}
\caption[RGA Trace of Pump Station After 6 Minutes of Turbo Pumping]{RGA trace of unbaked pump station after 6 minutes of turbo pumping. The partial pressures of the residual gases in this trace are summarized in Table \ref{tab:partial-pressures-of-pump-station-after-6-minutes}.}
\label{fig:RGA-trace-of-pump-station-after-6-minutes}
\end{figure}

\begin{table}[!!h]
\centering
\tablefont
\caption[Partial Pressures of Pump Station After 6 Minutes of Turbo Pumping]{Partial pressures of residual gases in unbaked pump station after 6 minutes of turbo pumping taken from the RGA trace in Figure \ref{fig:RGA-trace-of-pump-station-after-6-minutes}.}
\label{tab:partial-pressures-of-pump-station-after-6-minutes}
\begin{tabularx}{\textwidth}{>{\centering\arraybackslash}p{0.3\textwidth}>{\centering\arraybackslash}p{0.3\textwidth}>{\centering\arraybackslash}p{0.3\textwidth}}
\toprule
\textbf{Gas Species} & \textbf{Partial Pressure (Torr)} & \textbf{Percent of Total Pressure (\%)} \\
\midrule
$\text{H}_2\text{O}$ & $10^{-7}$ & 50 \\
\midrule
$\text{CO}_2$ & $4\times10^{-8}$ & 20 \\
\midrule
$\text{O}_2$ & $5\times10^{-9}$ & 3 \\
\midrule
$\text{N}_2$ & $2\times10^{-8}$ & 10 \\
\midrule
$\text{H}_2$ & $3.5\times10^{-8}$ & 18 \\
\bottomrule
\end{tabularx}
\end{table}

\begin{figure}[!!h]
\centering
\includegraphics[width=0.75\textwidth]{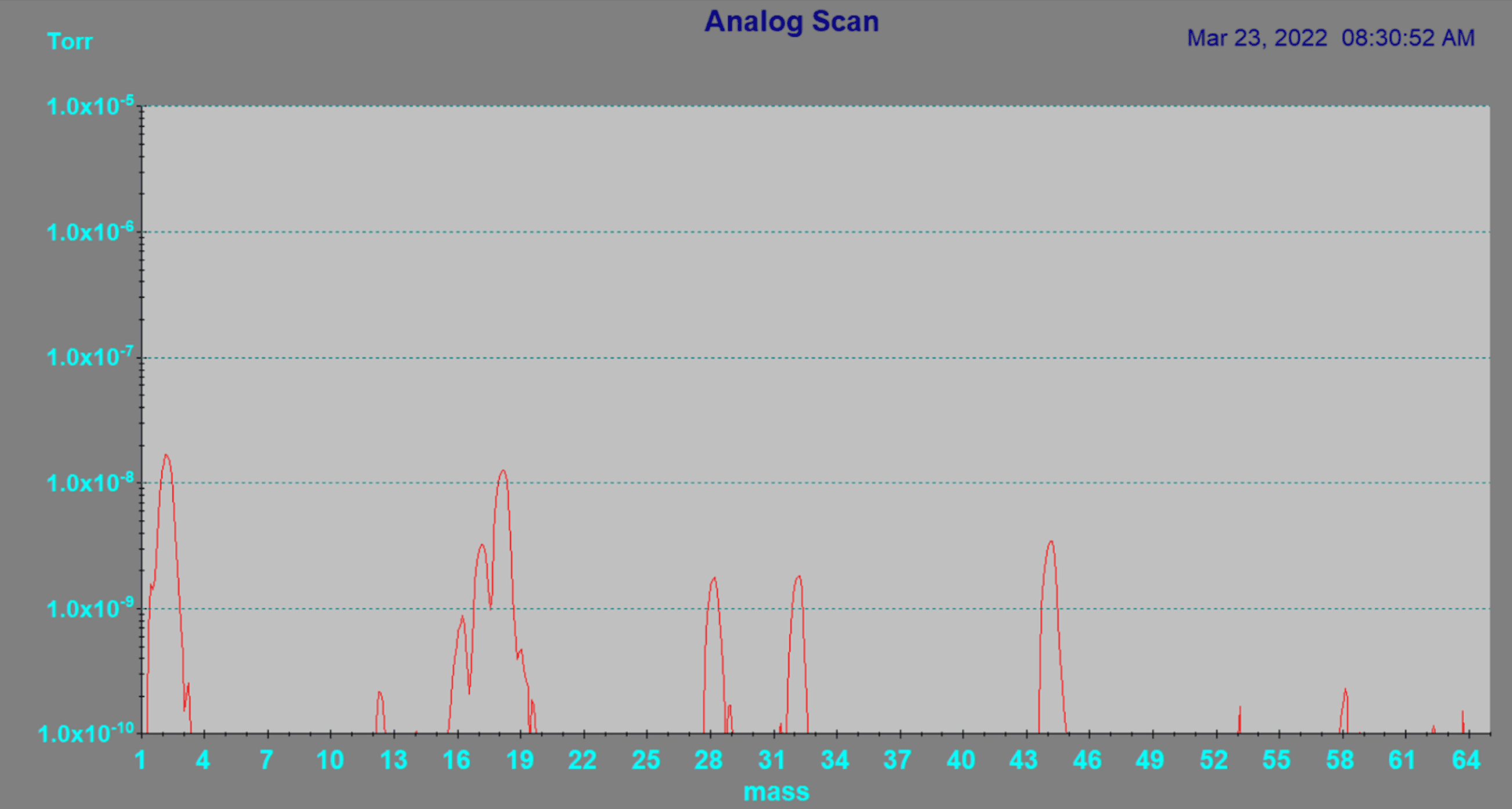}
\caption[RGA Trace of Pump Station After 12 Hours of Turbo Pumping]{RGA trace of unbaked pump station after 12 hours of turbo pumping. The partial pressures of the residual gases in this trace are summarized in Table \ref{tab:partial-pressures-of-pump-station-after-12-hours}.}
\label{fig:RGA-trace-of-pump-station-after-12-hours}
\end{figure}

\begin{table}[!!h]
\centering
\tablefont
\caption[Partial Pressures of Pump Station After 12 Hours of Turbo Pumping]{Partial pressures of residual gases in unbaked pump station after 12 hours of turbo pumping taken from the RGA trace in Figure \ref{fig:RGA-trace-of-pump-station-after-12-hours}.}
\label{tab:partial-pressures-of-pump-station-after-12-hours}
\begin{tabularx}{\textwidth}{>{\centering\arraybackslash}p{0.3\textwidth}>{\centering\arraybackslash}p{0.3\textwidth}>{\centering\arraybackslash}p{0.3\textwidth}}
\toprule
\textbf{Gas Species} & \textbf{Partial Pressure (Torr)} & \textbf{Percent of Total Pressure (\%)} \\
\midrule
$\text{H}_2\text{O}$ & $10^{-8}$ & 27 \\
\midrule
$\text{CO}_2$ & $3\times10^{-9}$ & 8 \\
\midrule
$\text{O}_2$ & $2\times10^{-9}$ & 5 \\
\midrule
$\text{N}_2$ & $2\times10^{-9}$ & 5 \\
\midrule
$\text{H}_2$ & $2\times10^{-8}$ & 54 \\
\bottomrule
\end{tabularx}
\end{table}

\begin{figure}[!!h]
\centering
\includegraphics[width=0.75\textwidth]{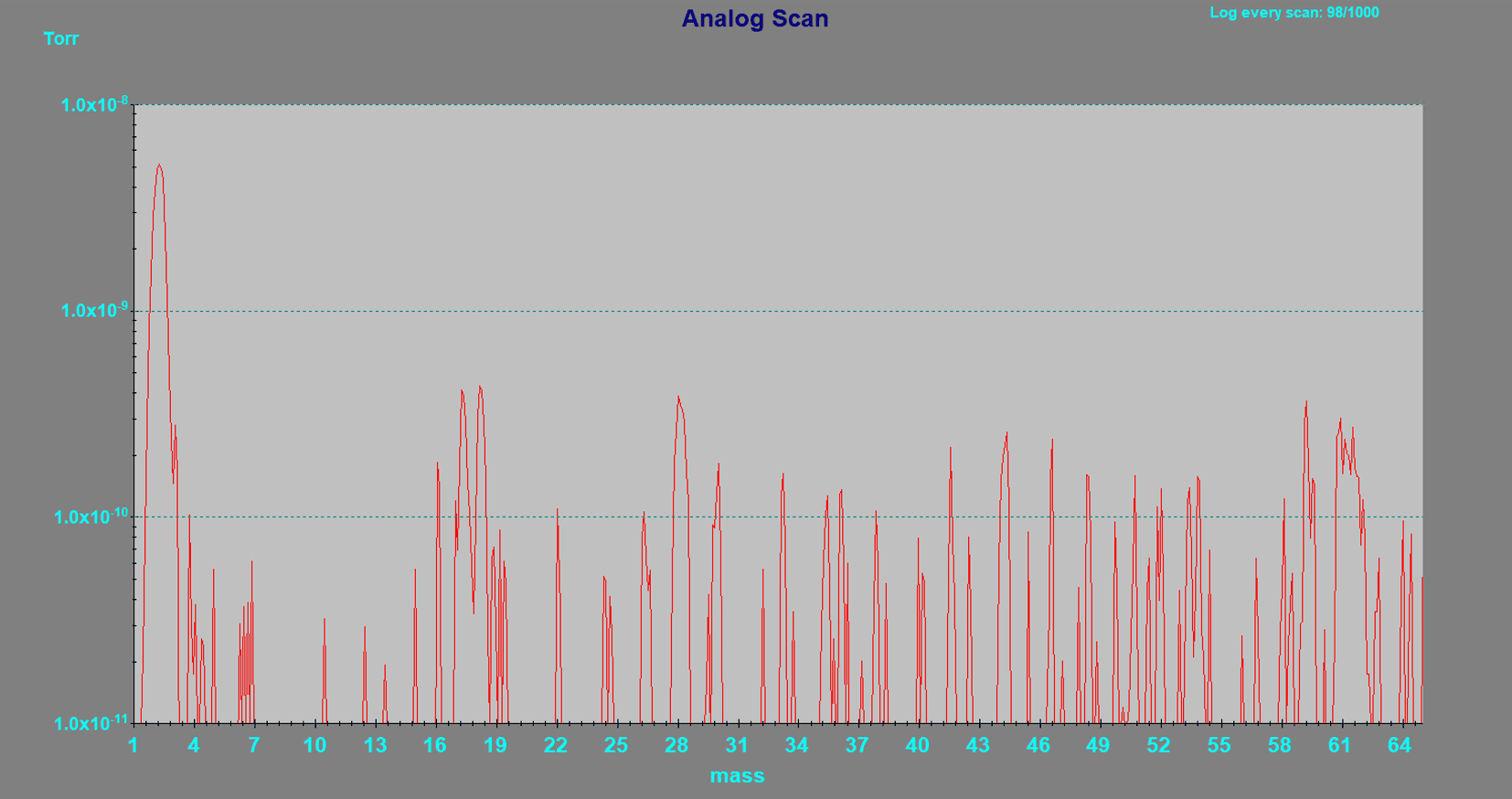}
\caption[RGA Trace of Pump Station After Two Day Bakeout]{RGA trace of pump station, measured at 300 K, after two day bakeout. The partial pressures of the residual gases in this trace are summarized in Table \ref{tab:partial-pressures-of-pump-station-after-two-day-bakeout}. Pumping is provided by the ion pump and the turbo pump is valved off, see Figure \ref{fig:pump-station-assembly}. The turbo pump top flange could only be baked to 120 K due to the pump's temperature rating. The ion pump could only be baked to 120 C due to thermal coupling through the firebrick (Figure \ref{fig:pump-station-assembly}) and insufficient heater tape at the ion pump. The rest of the parts were baked to 500 K with the RGA and ion gauge electronics removed.}
\label{fig:RGA-trace-of-pump-station-after-two-day-bakeout}
\end{figure}

\begin{table}[!!h]
\centering
\tablefont
\caption[Partial Pressures of Pump Station After Two Day Bakeout]{Partial pressures of residual gases in pump station, measured at 300 K, after two day bakeout taken from RGA trace in Figure \ref{fig:RGA-trace-of-pump-station-after-two-day-bakeout}.}
\label{tab:partial-pressures-of-pump-station-after-two-day-bakeout}
\begin{tabularx}{\textwidth}{>{\centering\arraybackslash}p{0.3\textwidth}>{\centering\arraybackslash}p{0.3\textwidth}>{\centering\arraybackslash}p{0.3\textwidth}}
\toprule
\textbf{Gas Species} & \textbf{Partial Pressure (Torr)} & \textbf{Percent of Total Pressure (\%)} \\
\midrule
$\text{H}_2\text{O}$ & $4\times10^-{10}$ & 7 \\
\midrule
$\text{CO}_2$ & $2.5\times10^{-10}$ & 4 \\
\midrule
$\text{O}_2$ & $5\times10^{-11}$ & 0.8 \\
\midrule
$\text{N}_2$ & $4\times10^{-10}$ & 7 \\
\midrule
$\text{H}_2$ & $5\times10^{-9}$ & 82 \\
\bottomrule
\end{tabularx}
\end{table}

\subsection{Kiln}
\label{subsec:kiln}

A kiln (L\&L Kilns, PN: J2927-3) was purchased to provide the oven environment for the bakeout of a UHV chamber and is shown in Figure \ref{fig:kiln-assembly}. The kiln is a fire brick enclosure that consists of three active layers containing a resistive heating element and a thermocouple used to servo the temperature inside the kiln. An additional passive layer without a heating element is used to provide a clearance hole for the CF-2.75 bellows - this hole was simply drilled at UMD using a hole saw. The kiln rests on top of a metal rack that has wheels making it possible to position the kiln relative to the pump station. 

\begin{figure}[!!h]
\centering
\includegraphics[width=0.75\textwidth]{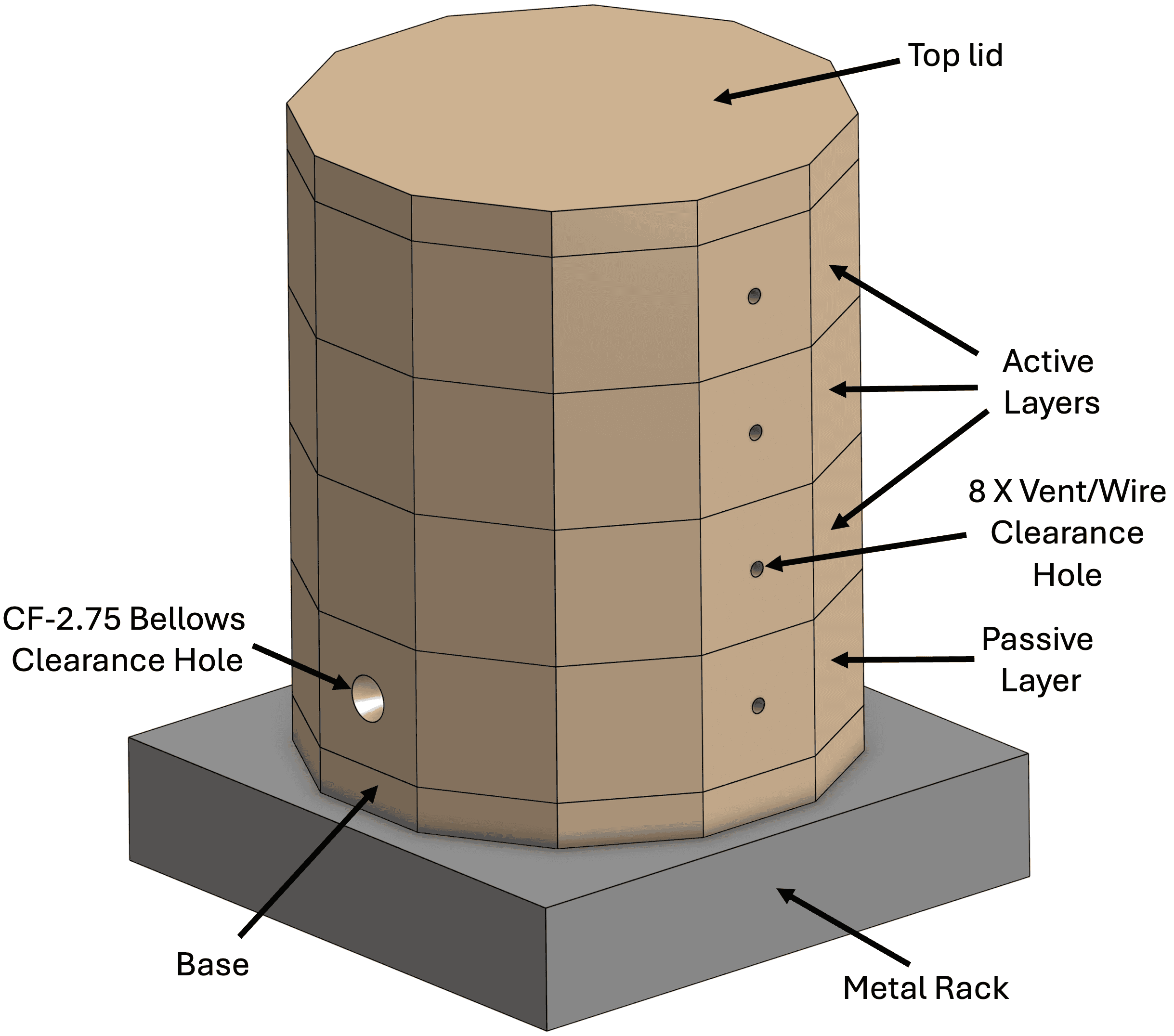}
\caption[Kiln Assembly]{CAD drawing of the kiln showing the different layers bakeout station and how the pump station interfaces with the UHV chamber inside the kiln. The pump station is connected to the UHV chamber with a CF-2.75 bellows.}
\label{fig:kiln-assembly}
\end{figure}

I'll now discuss the kiln control electronics setup by myself at UMD. Power is delivered to the heating elements of kiln using a three phase circuit. The three-phase power source for the kiln is a 208 Wye source from panel PP-2N-B0225-3 in ATL B0225 at UMD. The three active layers of the kiln each have three 31.2 $\Omega$  heating elements wired in parallel. Thus, the active layers are the load of the circuit and are wired in a delta configuration as shown in Figures \ref{fig:kiln-control-electronics-complete-circuit-diagram} and \ref{fig:kiln-control-electronics-effective-circuit-diagram}. Power to the kiln is controlled using a set of power relays and to servo the temperature inside the kiln, three PID controllers monitor the temperature of each layer and control the current to each layer using solid state relays. The relay circuit was provided by the L\&L Kilns while PID controllers and additional switches and fuses were purchases and integrated into the circuit by myself at UMD.

\begin{figure}[!!h]
\centering
\includegraphics[width=0.75\textwidth]{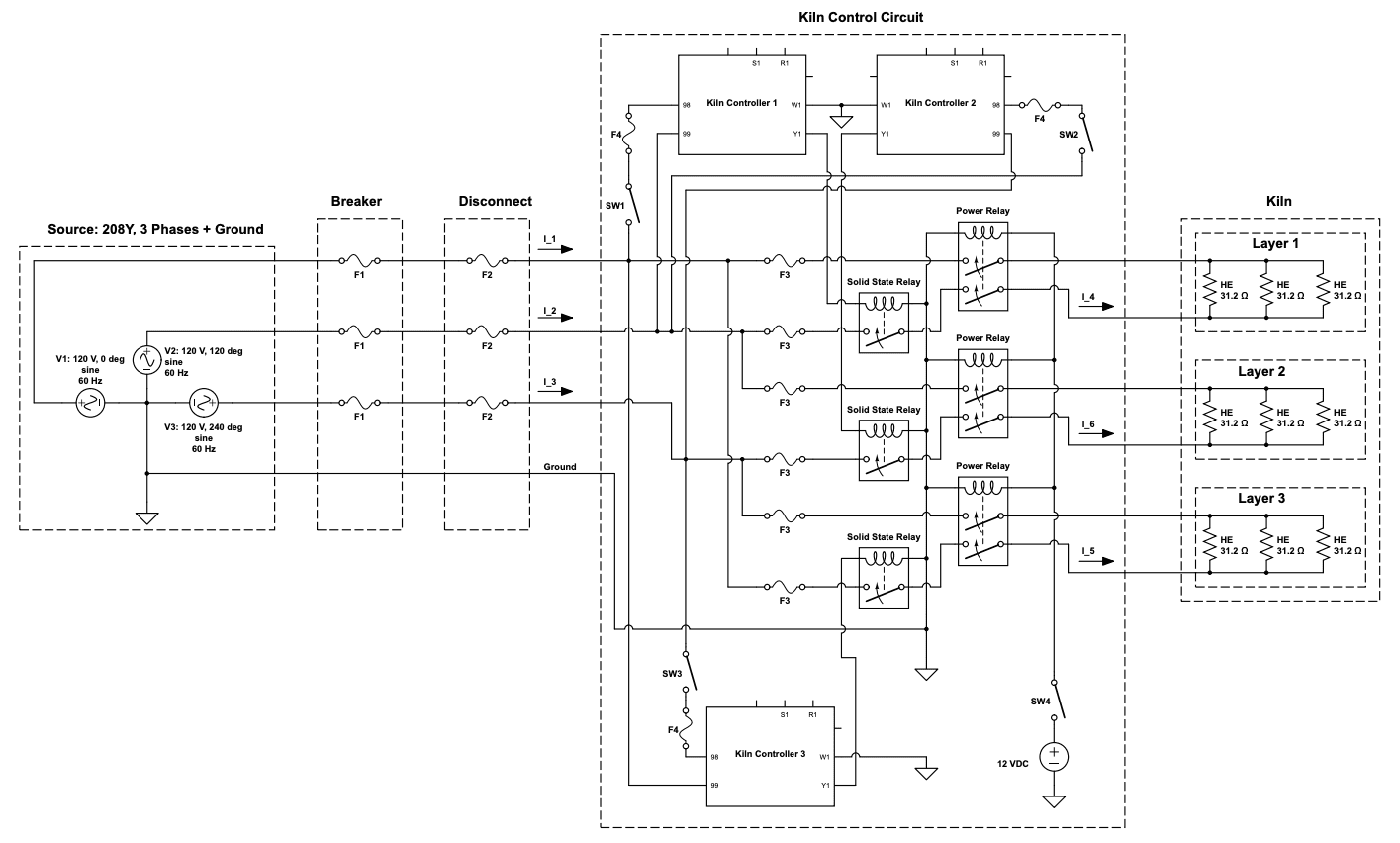}
\caption[Kiln Control Electronics Complete Circuit Diagram]{Kiln control electronics complete circuit diagram.}
\label{fig:kiln-control-electronics-complete-circuit-diagram}
\end{figure}

\begin{figure}[!!h]
\centering
\includegraphics[width=0.75\textwidth]{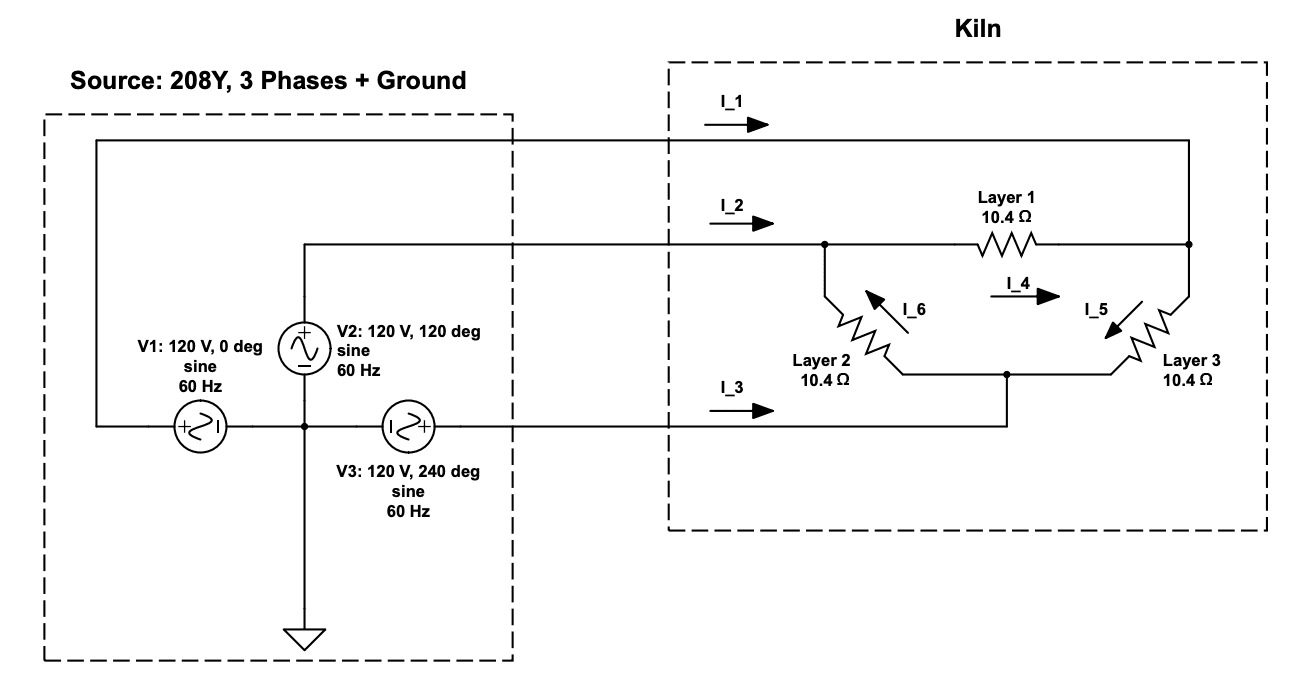}
\caption[Kiln Control Electronics Effective Circuit Diagram]{Effective kiln control circuit diagram. This is called a balanced Wye-Delta connection. The breaker is a Wye source and the kiln is a Delta load.}
\label{fig:kiln-control-electronics-effective-circuit-diagram}
\end{figure}

I'll now present an analysis of the kiln control circuit as it's not immediately obvious how to solve for the currents and voltages. All voltages and currents in this analysis are characterized by their frequency, root-mean-squared (RMS) amplitude, and phase. The conductors connected to the three points of a three-phase source or load are called lines. The three components comprising a three-phase source or load are called phases. The line voltage is defined as the voltage measured between any two lines in a three-phase circuit. The phase voltage is defined as the voltage measured across a single component in a three-phase source or load. The line current is the current through any one line between a three-phase source and load. The phase current is the current through any one component comprising a three-phase source or load \cite{FundamentalsofElectricCircuitsSadiku}.

\noindent\textbf{Phase Voltage of Source}

\noindent$V_{1}=120\text{V}\angle0^{\circ}, V_{2}=120\text{V}\angle120^{\circ} , V_{3}=120\text{V}\angle240^{\circ} $

\noindent\textbf{Phase Voltage of Load}

\noindent Layer 1

\noindent Apply Kirchhoff's voltage law: $V_{2}-\underbrace{(V_{2}-V_{1})}_{\text{Layer 1}}-V_{1}=0$
\begin{equation}
\begin{aligned}
    V_{2}-V_{1}&=120\text{V}\angle120^{\circ}-120\text{V}\angle0^{\circ} \\
    &=120\text{V}\cos(2\pi\times60\text{\,Hz\,}t+120^{\circ})-120\text{V}\cos(2\pi60\text{Hz}t) \\
    &=\text{Re}\left[120\text{V}e^{i2\pi60\text{Hz}t+120^{\circ}}-120\text{V}e^{i2\pi60\text{Hz}t}\right] \\
    &=\text{Re}\left[120\text{V}(e^{i120^{\circ}}-e^{i0^{\circ}})e^{i2\pi60\text{Hz}t}\right] \\
    &=\text{Re}\left[208\text{V}e^{i150^{\circ}}e^{i2\pi60\text{Hz}t}\right] \\
    &=208\text{V}\cos(2\pi60\text{Hz}t+150^{\circ})
\end{aligned}  
\end{equation}

\noindent Layer 2

\noindent Apply Kirchhoff's voltage law: $V_{3}-\underbrace{(V_{3}-V_{2})}_{\text{Layer 2}}-V_{2}=0$
\begin{equation}
\begin{aligned}
    V_{3}-V_{2}&=120\text{V}\angle240^{\circ}-120\text{V}\angle120^{\circ} \\
    &=208\text{V}\cos(2\pi60\text{Hz}t+270^{\circ})
\end{aligned}  
\end{equation}

\noindent Layer 3

\noindent Apply Kirchhoff's voltage law: $V_{1}-\underbrace{(V_{1}-V_{3})}_{\text{Layer 3}}-V_{3}=0$
\begin{equation}
\begin{aligned}
    V_{1}-V_{3}&=120\text{V}\angle0^{\circ}-120\text{V}\angle240^{\circ} \\
    &=208\text{V}\cos(2\pi60\text{Hz}t+30^{\circ})
\end{aligned}  
\end{equation}

\noindent\textbf{Phase Current of Load}

\noindent Layer 1

\noindent$I_{4}=\frac{V_{2}-V_{1}}{R_{\text{Layer 1}}}=\frac{208\text{V}\cos(2\pi60\text{Hz}t+150^{\circ})}{10.4\text{ \ensuremath{\Omega}}}=20A\cos(2\pi60\text{Hz}t+150^{\circ})$

\noindent Layer 2

\noindent$I_{6}=\frac{V_{3}-V_{2}}{R_{\text{Layer 2}}}=\frac{208\text{V}\cos(2\pi60\text{Hz}t+270^{\circ})}{10.4\text{ \ensuremath{\Omega}}}=20\text{A}\cos(2\pi60\text{Hz}t+270^{\circ})$

\noindent Layer 3

\noindent$I_{5}=\frac{V_{1}-V_{3}}{R_{\text{Layer 3}}}=\frac{208\text{V}\cos(2\pi60\text{Hz}t+30^{\circ})}{10.4\text{ \ensuremath{\Omega}}}=20\text{A}\cos(2\pi60\text{Hz}t+30^{\circ})$

\noindent\textbf{Line Current}

\noindent $I_{1}$

\begin{equation}
\begin{aligned}
    I_{1}&=I_{5}-I_{4} \\
    &=20\text{A}\cos(2\pi60\text{Hz}t+30^{\circ})-20\text{A}\cos(2\pi60\text{Hz}t+150^{\circ}) \\
    &=34.6\text{A}\cos(2\pi60\text{Hz}t)
\end{aligned}  
\end{equation}

\noindent $I_{2}$

\begin{equation}
\begin{aligned}
    I_{2}&=I_{4}-I_{6} \\
    &=20\text{A}\cos(2\pi60\text{Hz}t+150^{\circ})-20\text{A}\cos(2\pi60\text{Hz}t+270^{\circ}) \\
    &=34.6\text{A}\cos(2\pi60\text{Hz}t+120^{\circ})
\end{aligned}  
\end{equation}

\noindent $I_{3}$

\begin{equation}
\begin{aligned}
    I_{3}&=I_{6}-I_{5} \\
    &=20\text{A}\cos(2\pi60\text{Hz}t+270^{\circ})-20\text{A}\cos(2\pi60\text{Hz}t+30^{\circ}) \\
    &=34.6\text{A}\cos(2\pi60\text{Hz}t+240^{\circ})
\end{aligned}  
\end{equation}

\noindent\textbf{Sanity Check}

\noindent Apply Kirchhoff's current law at source. Show that $I_{1}+I_{2}+I_{3}=0$

\begin{equation}
\begin{aligned}
    I_{1}+I_{2}+I_{3}&=34.6\text{A}\cos(2\pi60\text{Hz}t)+34.6\text{A}\cos(2\pi60\text{Hz}t \\&+120^{\circ})+34.6\text{A}\cos(2\pi60\text{Hz}t+240^{\circ}) \\
    &=0
\end{aligned}  
\end{equation}

\noindent Apply Kirchhoff's voltage law in load loop. Show that $(V_{2}-V_{1})+(V_{1}-V_{3})+(V_{3}-V_{2})=0$

\begin{equation}
\begin{aligned}
    (V_{2}-V_{1})+(V_{1}-V_{3})+(V_{3}-V_{2})&=208\text{V}\cos(2\pi60\text{Hz}t+150^{\circ}) \\&+208\text{V}\cos(2\pi60\text{Hz}t+30^{\circ})+208\text{V}\cos(2\pi60\text{Hz}t+270^{\circ}) \\
    &=0
\end{aligned}  
\end{equation}

\noindent\textbf{Power Dissipation}

\noindent Layers 1, 2, 3

\noindent $P_{\text{Layer 1}}=I_{4}^{2}R_{\text{Layer 1}}=4.2\text{ kW}$ 

\noindent $P_{\text{Layer 2}}=I_{6}^{2}R_{\text{Layer 2}}=4.2\text{ kW}$ 

\noindent $P_{\text{Layer 3}}=I_{5}^{2}R_{\text{Layer 3}}=4.2\text{ kW}$ 

\noindent Total Power Dissipated in Kiln Circuit 

\begin{equation}
    P_{\text{total}}=P_{\text{Layer 1}}+P_{\text{Layer 2}}+P_{\text{Layer 3}}=12.6\text{kW}
\end{equation}

\ifSubfilesClassLoaded{%
  \bibliographystyle{unsrt}
  \bibliography{Dissertation}
}{}
\end{document}

%% file: AppendixB.tex
\ifSubfilesClassLoaded{\appendix}{}

\newpage

\chapter{Manufacturing}
\label{chap:manufacturing}

\section{Materials and Connectors}
\label{sec:materials-and-connectors}

\begin{table}[!!h]
\centering
\tablefont
\caption[List of Materials Used in the Cryostat.]{List of materials used in the Cryostat along the abbreviations used in this thesis.}
\label{tab:list-of-materials-used-in-the-cryostat}
\begin{tabularx}{\textwidth}{>{\centering\arraybackslash}p{0.49\textwidth}>{\centering\arraybackslash}p{0.49\textwidth}}
\toprule
\textbf{Material} & \textbf{Abbreviation} \\
\midrule
Stainless Steel 304 & SS304 \\
\midrule
Stainless Steel 316(L) & SS316(L) \\
\midrule
Oxygen-Free Electronic (OFE) Copper & C10100 \\
\midrule
Oxygen-Free (OF) Copper & C10200 \\
\midrule
Electrolytic Tough Pitch (ETP) Copper & C11000 \\
\midrule
Phosphorous Deoxidized (DHP) Copper & C12200 \\
\midrule
Titanium Grade 2 & TiG2 \\
\midrule
Titanium Grade 5 & TiG5 \\
\midrule
Aluminum 6061-T6 & Al6061T6 \\
\midrule
\makecell{Polyether Ether Ketone (PEEK)\\Glass-Filled 30\%} & PEEKGF30 \\
\bottomrule
\end{tabularx}
\end{table}

\begin{table}[!!h]
\centering
\tablefont
\caption[List of Connectors Used in the Cryostat]{List of connectors used in cryostat along the abbreviations used in this thesis.}
\label{tab:list-of-connectors-used-in-the-cryostat}
\begin{tabularx}{\textwidth}{>{\centering\arraybackslash}p{0.49\textwidth}>{\centering\arraybackslash}p{0.49\textwidth}}
\toprule
\textbf{Connector} & \textbf{Abbreviation} \\
\midrule
Conflat Flange & CF-[size] \\
\midrule
Klein Flange & KF-[size] \\
\midrule
Swagelok & Swagelok-[size] \\
\midrule
Vacuum Coupling Radiation & VCR-[size] \\
\midrule
Aeroquip & Aeroquip-[size] \\
\midrule
National Pipe Thread & NPT-[size] \\
\bottomrule
\end{tabularx}
\end{table}

\section{Machine Shops}
\label{se:machine-shops}

\begin{table}[!!h]
\centering
\tablefont
\caption[Machine Shops Used for Manufacturing the Cryostat]{Summary of machine shops used for manufacturing cryostat.}
\label{tab:machine-shops-used-for-manufacturing-the-cryostat}
\begin{tabularx}{\textwidth}{>{\centering\arraybackslash}p{0.49\textwidth}>{\centering\arraybackslash}p{0.49\textwidth}}
\toprule
\textbf{Machine Shop} & \textbf{Material} \\
\midrule
Zero Hour Parts & \makecell{Stainless Steel\\Titanium\\Aluminum} \\
\midrule
Amorphic Tech Ltd. & \makecell{Copper\\PEEK} \\
\midrule
Zimmerman Machine, Inc. & Copper \\
\midrule
Xometry & \makecell{Copper\\PEEK} \\
\midrule
UMD IREAP Shop & \makecell{PEEK\\Aluminum\\Titanium\\Stainless Steel\\Copper} \\
\bottomrule
\end{tabularx}
\end{table}

\ifSubfilesClassLoaded{%
  \bibliographystyle{unsrt}
  \bibliography{Dissertation}
}{}
\end{document}

%% file: AppendixC.tex
\ifSubfilesClassLoaded{\appendix}{}

\newpage

\chapter{Pulse Arc and Resistance Welder}
\label{chap:pulse-arc-and-resistance-welder}

An Orion S Series Pulse Arc and Resistance Welder (Sunstone, PN: Orion 150S) was used to weld small in-vacuum parts for the cryostat. A pulse arc welder is a specialized type of Tungsten Inert Gas (TIG) welder where a sharpened tungsten electrode is used in combination with electrical energy to create a high temperature plasma discharge - an arc. This high temperature plasma melts the metal in a small spot and can be used to melt two workpieces together. A capacitive discharge is used to create this pulse arc weld and argon is used as the shielding gas. The Orion can also be used as a resistance welder where a large electrical current is passed through two workpieces to be joined together. Resistive heating occurs at the contact between the two materials and with enough current the two materials can be melted together. 

\section{Welding Steps for Oven}
\label{sec:welding-steps-for-oven}

The steps used to weld oven Design 3 are discussed here

\begin{enumerate}
    \item Measure out 0.5” of SS316 tube and cut with steel rated wire cutter 
    \item Second item Flatten out 1/16” section from one end of tube with pliers 
    \item Third item Weld flat section closed. Weld Settings: Arc mode, Sharp tungsten electrode tip, Energy =2\text{ W-s}, Duration = 100\%, S+, argon pressure = 15 psi 
    \item Clamp SS316 wire to tube using tack mode welding pliers, leaving 1/4” extra to fold about the tube later. Tack weld wire to tube - this is a temporary weld to position the wire with the tube. Weld Settings: Tack mode, Energy =10\text{ W-s}, Pre-weld delay = medium
    \item Fold extra portion of wire onto other side of flattened tube using pliers
    \item Create permanent weld between wire and tube using the pulse arc mode. This nicely melts the wire onto the tube. Weld Settings: Arc mode, Sharp tungsten electrode tip, Energy =5\text{ W-s}, Duration = 100\%, S+, argon pressure = 15 psi
    \item Repeat steps 4 and 5 but on the others side of the tube
    \item Test welds in air by running 4 A through the oven 
\end{enumerate}

\ifSubfilesClassLoaded{%
  \bibliographystyle{unsrt}
  \bibliography{Dissertation}
}{}
\end{document}

%% file: AppendixD.tex
\ifSubfilesClassLoaded{\appendix}{}

\newpage

\chapter{Estimating Vibration Amplitude Spectra}
\label{chap:estimating-vibration-amplitude-spectra}

An amplitude spectrum is more useful than a simple Fourier transform when analyzing a vibration signal, as in \ref{subsec:results-vibration-measurement}, because it provides a physically interpretable measure of vibration amplitude at each frequency whose magnitudes do not depend on the record length (time duration of the measurement). We used Welch's method to estimate the amplitude spectra which averages multiple modified periodograms computed from overlapping, windowed segments of data which reduces noise variance. A Hahn window function was used which is optimized by having 50\% overlap between adjacent segments of the spectra \cite{heinzelSpectrumSpectralDensity}. The numerical method was implemented in Python using SciPy Welch. The amplitude spectrum, $\text{AS}[k]$ where $k$ is the frequency bin, is generated and can be related to the amplitude spectral density by $\text{ASD}[k]=\frac{\text{AS}[k]}{\sqrt{\text{ENBW}}}$. $\text{ENBW}=\frac{\sum w[j]}{\sum w[j]^{2}}$ where $w[j]$ with $j=0,\dots,N_{\text{nperseg}}$ is the window function and $N_{\text{nperseg }}$ is the number of bins in a segment. This is the equivalent noise bandwidth (ENBW) and is the width of a rectangular filter that passes the same noise power as the window function \cite{heinzelSpectrumSpectralDensity}. The RMS motion in a given frequency band, $[f_{1,}f_{2}]$, can be calculated from the amplitude spectrum as follows,
\begin{equation}
    d_{\text{RMS}}\left([f_{1,}f_{2}]\right)=\sqrt{\sum_{k\in[f_{1,}f_{2}]}\text{AS}[k]^{2}}
    \label{eq:RMS-motion-in-frequency-band-from-AS}
\end{equation}
To understand how to set up the measurement, I will discuss the constraints imposed on the time duration of the experiment. If the amplitude spectrum is calculated using a single periodogram (segment), then the frequency resolution is just $f_{\text{res}}=\frac{1}{T_{\text{run}}}=\frac{f_{\text{s}}}{N_{\text{total}}}$ where $T_{\text{run}}$ is the time duration of the measurement, $f_{\text{s}}$ is the sampling rate, and $N_{\text{total}}$ is the number of frequency bins. However, if the amplitude spectrum is computed by averaging over multiple segments, the frequency resolution scales in a different manner. Knowing the desired frequency resolution and number of segments to average, we can compute the required time duration of the experiment to be
\begin{equation}
    T_{\text{run}}(f_{\text{res}},\eta,N_{\text{seg}})=\frac{N_{\text{total}}}{f_{\text{s}}}=\frac{N_{\text{perseg}}+(N_{\text{seg}}-1)N_{\text{perseg}}(1-\eta)}{f_{\text{s}}}
    \label{eq:time-duration-welch}
\end{equation}
where $f_{\text{res}}$ is the desired frequency resolution, $\eta$  is the optimal overlap for a given window function, and $N_{\text{seg}}$ is the desired number of segments to average over. Here, the number of bins per segment is $N_{\text{perseg}}=\frac{f_{\text{s}}}{f_{\text{res}}}$ and the number of bins that are overlapped between segments is $N_{\text{overlap}}=N_{\text{perseg}}\text{\ensuremath{\eta}}$ where $\eta$  is the overlap percentage and so the number of new bins advanced between adjacent segments is $N_{\text{step}}=N_{\text{perseg}}(1-\eta)$. Thus the total number of bins required is $N_{\text{total}}=N_{\text{perseg}}+(N_{\text{seg}}-1)N_{\text{step}}=N_{\text{perseg}}+(N_{\text{seg}}-1)N_{\text{perseg}}(1-\eta)$. So if $100\text{ mHz}$ resolution is desired using a Hahn window, $\eta=0.5$, averaging over 20 segments, we get that the required time duration is 105 s, where as if only one segment is used with no averaging the time duration is 10 s.

\ifSubfilesClassLoaded{%
  \bibliographystyle{unsrt}
  \bibliography{Dissertation}
}{}
\end{document}